%% file: mainV1.tex
\documentclass[aps,onecolumn,prd,preprintnumbers,superscriptaddress,nofootinbib,bibnotes,floatfix,
longbibliography,
]{revtex4-2}

\pdfoutput=1
\usepackage{amsmath}
\usepackage{amsfonts}
\usepackage{amssymb}
\usepackage{mathrsfs}
\usepackage{graphicx}
\usepackage{color}
\usepackage{bbding}
\usepackage{tabularx}
\usepackage[usenames,dvipsnames,table]{xcolor}
\usepackage[normalem]{ulem}
\usepackage{makecell}
\usepackage{siunitx}
\usepackage{enumerate}
\usepackage{multirow}
\usepackage{makecell}
\usepackage{upgreek}
\usepackage{comment}
\usepackage{subfigure}
\usepackage{booktabs}
\usepackage{enumitem}

\usepackage{marvosym}
\definecolor{darkred}{rgb}{0.5,0,0}
\definecolor{darkblue}{rgb}{0,0,0.5}
\definecolor{firebrick}{rgb}{0.75,0.125,0.125}
\definecolor{darkgreen}{rgb}{0,0.5,0}
\usepackage[colorlinks=true,linkcolor=firebrick,citecolor=darkgreen,urlcolor=darkblue]{hyperref}
\setlistdepth{8}

\graphicspath{{figures/}}

\newcommand{\orcid}[1]{\href{https://orcid.org/#1}{\includegraphics[width=10pt]{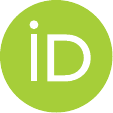}}}

\begin{document}

\title{Physics in Safety and Safety in Physics: Lessons from the LHC and its Underground Experiments}

\author{Alexander S.~Sakharov \orcid{0000-0001-6622-2923}}
\email{alexandre.sakharov@cern.ch}
\affiliation{Department of Mathematics and Physics, Manhattan University,\\
4513 Manhattan College Parkway, Riverdale, NY 10471, United States of America}
\affiliation{Experimental Physics Department, CERN, CH-1211 Gen\`eve 23, Switzerland}

\date{\today}

\begin{abstract}
Large underground particle-physics facilities are usually discussed by separating scientific performance from the systems that make their operation safe.  We develop a complementary view in which safety is treated as an intrinsic part of the physical and operational structure of an experiment.  Using the LHC, ATLAS, and CMS as principal case studies, we introduce the concepts of \emph{physics in safety} and \emph{safety in physics}: physical laws, measurements, and models determine hazards and protective barriers, while those barriers in turn define the machine, detector, access, and intervention states in which physics can be performed.  We formulate this relation through a \emph{safe operational phase space} and through admissible trajectories connecting different configurations, emphasizing that safety constrains not only states but also transitions between them.  The framework is applied to beam and magnet protection, cryogenic and oxygen-deficiency hazards, detector-specific risk configurations, personnel access, maintenance, and radiation protection, with particular attention to the different architectures of ATLAS and CMS.  We argue that validated protection does not merely restrict scientific activity but can enlarge the domain of sustainable operation.  This viewpoint provides a unified physical and operational framework for understanding safety at the LHC and for incorporating protection, maintainability, and recoverability into the design of future accelerator and underground facilities such as HL-LHC, FCC and muon collider.
\end{abstract}

\maketitle

\section{Introduction}
\label{sec:introduction}

Modern high-energy physics relies on experimental infrastructures whose
complexity extends far beyond the particle interactions that constitute
their primary scientific target.  The Large Hadron Collider (LHC) at CERN
is a particularly prominent example.  Its operation combines
ultrarelativistic particle beams, superconducting magnets, large stores of
electromagnetic and kinetic energy, cryogenic systems operating at
temperatures close to absolute zero, ultra-high vacuum, high-voltage
equipment, radiation fields, large underground structures, extensive
ventilation and electrical networks, and experimental detectors containing
many millions of individual components~\cite{Bruning:2004LHC,
Wenninger:2016MPO}.  The ATLAS~\cite{ATLAS:2008Detector} and CMS~\cite{CMS:2008Detector}
experiments add another level of
complexity through detector magnets, power-distribution systems, cooling
circuits, movable detector structures, gas systems, electronics,
radiation-sensitive components, and the simultaneous presence of large
numbers of personnel during installation, maintenance, and shutdown
activities.

Such facilities reveal a particularly close relationship between physics
and safety.  The same physical quantities that make frontier experiments
possible also determine their characteristic hazards.  Beam energy enables
high-energy particle collisions, but an uncontrolled beam loss can damage
accelerator components.  Large electrical currents generate the magnetic
fields required for beam steering and particle-momentum measurement, while
the corresponding stored magnetic energy must be safely controlled during
a quench or electrical failure.  Cryogenic helium permits the operation of
superconducting magnets, while an accidental release may generate
overpressure, very low temperatures, or oxygen deficiency hazard conditions in an
underground environment.  Ionizing radiation is an intrinsic consequence
of accelerator operation and particle interactions, while the same
interaction physics determines shielding requirements, activation,
dosimetry, and access conditions.

This relationship motivates the two complementary directions considered
in this work, which we refer to as \emph{physics in safety} and
\emph{safety in physics}.  By \emph{physics in safety}, we mean the use of
physical laws, quantitative models, instrumentation, and experimental
measurements to identify hazards, estimate their possible consequences,
and design systems capable of preventing or mitigating them.  By
\emph{safety in physics}, we mean the converse influence: safety
requirements define conditions under which accelerators and experiments
may be constructed, commissioned, accessed, maintained, and operated, and
therefore form part of the boundary conditions within which experimental
physics is performed.

The central premise of this paper can therefore be stated as follows:
\emph{at a large underground accelerator facility, safety is neither
external to physics nor merely administrative.  Physical laws determine
the hazards, measurements reveal the state of the system, protection
systems constrain its evolution, and operational experience feeds back
into improved physical and engineering models.  Conversely, these safety
constraints define the operational phase space within which experimental
physics can be performed.}

It is useful at the outset to distinguish several concepts that are
sometimes grouped together under the general term ``safety.''  Personnel
safety is concerned primarily with preventing unacceptable harm to people.
Machine protection is concerned with preventing or limiting damage to the
accelerator and its components.  Detector and equipment protection
similarly aims to prevent damage to experimental systems.  Radiation
protection introduces its own hierarchy of monitoring, zoning, access
control, shielding, and dose limitation.  Fire safety, electrical safety,
cryogenic safety, and occupational safety add further, partly overlapping
layers.  These domains strongly interact, but their objectives and
protection architectures need not be identical.  For example, the LHC
machine-protection system is designed to prevent damage from stored beam
and magnet energy, while the LHC Access Safety System provides collective
personnel protection by interlocking access conditions with the state of
the accelerator~\cite{Bordry:2001MPS,Wenninger:2016MPO,Ninin:2005LACS,Ladzinski:2009LHCAccess}.
Likewise, detector safety systems may automatically protect experimental
equipment without being identical to the systems responsible for personnel
protection~\cite{Luders:2003DSS,ATLAS:2008Detector,CMS:2008Detector}.

A useful schematic representation of the relation discussed in this paper
is
\begin{equation}
\text{physical state}
\;\longrightarrow\;
\text{hazard}
\;\longrightarrow\;
\text{risk}
\;\longrightarrow\;
\text{measurement}
\;\longrightarrow\;
\text{protection}
\;\longrightarrow\;
\text{operational constraint}.
\label{eq:physics_safety_chain}
\end{equation}
The sequence in Eq.~\eqref{eq:physics_safety_chain} should not be
interpreted as a strictly linear process.  In a real accelerator complex,
the different stages form a network of feedback loops.  Measurements
continuously update knowledge of the physical state; protection systems
may modify that state; operational restrictions change the accessible
machine configurations; and experience accumulated during commissioning,
operation, maintenance, anomalies, and incidents can lead to revisions of
both the hazard model and the associated protection strategy.

This motivates a second conceptual representation that will be
used throughout this work. Let the instantaneous physical and
operational state of an accelerator facility be represented
schematically by a multidimensional vector
\begin{equation}
{\boldsymbol X}(t)
=
\left\{
E_{\rm beam},
I_{\rm mag},
E_{\rm mag},
T,
P,
C_{O_2},
\dot{H}^{*},
V_{\rm HV},
\sigma_{\rm access},
\sigma_{\rm det},
\ldots
\right\},
\label{eq:facility_state_vector}
\end{equation}
where $E_{\rm beam}$ denotes stored beam energy,
$I_{\rm mag}$ and $E_{\rm mag}$ characterize magnet
currents and stored magnetic energy, $T$ and $P$
denote relevant temperatures and pressures,
$C_{O_2}$ is an oxygen concentration~\footnote{
Throughout this work, $C$ with an appropriate
species or constituent index generally denotes
a gas or airborne-constituent concentration,
including quantities relevant to smoke detection.
This notation is distinct from $\mathcal C$,
used for consequences, and from other explicitly
defined quantities such as diagnostic coverage.},
$\dot{H}^{*}$ represents an appropriate radiation
dose-rate quantity, and $V_{\rm HV}$ represents
high-voltage states. The variables
$\sigma_{\rm access}$ and $\sigma_{\rm det}$
denote discrete status variables describing,
respectively, personnel-access conditions and
the relevant detector operating configuration.
The precise choice of variables is system
dependent and may include many additional
quantities.

One may then introduce, at a conceptual level, a \emph{safe operational
phase space}
\begin{equation}
\Omega_{\rm safe}
\subset
\Omega_{\rm phys},
\label{eq:safe_phase_space}
\end{equation}
where $\Omega_{\rm phys}$ denotes the set of physically realizable machine
and experiment states, while $\Omega_{\rm safe}$ denotes those states,
transitions, and configurations for which the applicable protection
conditions are satisfied.  Safe operation may then be expressed
schematically as
\begin{equation}
{\boldsymbol X}(t)\in\Omega_{\rm safe}.
\label{eq:safe_condition}
\end{equation}
Approach to a relevant boundary,
\begin{equation}
{\boldsymbol X}(t)\rightarrow
\partial\Omega_{\rm safe},
\label{eq:safe_boundary}
\end{equation}
may trigger preventive or protective responses such as an interlock,
beam abort, power removal, magnet discharge, detector shutdown, alarm,
restriction of access, or evacuation.

Equations~\eqref{eq:facility_state_vector}--\eqref{eq:safe_boundary} are
not intended to replace the detailed hazard analyses, safety rules, or
engineering specifications applicable to a real installation.  In
particular, safety cannot in general be reduced to independent threshold
conditions on a small number of scalar variables.  Dependencies between
subsystems, transient conditions, common-cause failures, redundancy,
procedural states, human actions, and the history of a system can all be
important.  The construction nevertheless provides a useful physical
language for the present discussion: safety can be regarded as defining
an admissible subset of a much larger space of possible physical and
operational states.\footnote{The term ``safe'' is used here in the
engineering and operational sense of risk being controlled according to
the applicable protection requirements.  It does not imply the absolute
absence of hazard or the mathematical elimination of all possible adverse
events.}

This viewpoint is especially natural for the LHC because its operating
conditions are inseparable from machine protection.  A stored
ultrarelativistic proton beam may be described at the simplest level by
\begin{equation}
E_{\rm beam}
=
n_b N_b E_p,
\label{eq:lhc_beam_energy}
\end{equation}
where $n_b$ is the number of bunches, $N_b$ is the number of particles per
bunch, and $E_p$ is the energy per proton.  The scientific value of the
machine requires large values of these quantities, while the potential
consequence of an uncontrolled loss grows with the same stored energy.
Machine protection therefore does not merely protect a completed
accelerator from an external hazard; it enters directly into the
conditions under which beam intensity can be increased and machine
operation can proceed~\cite{Wenninger:2016MPO}.

A similar argument applies to superconducting systems.  To leading order,
the magnetic energy associated with an inductive circuit is
\begin{equation}
E_{\rm mag}
=
\frac{1}{2} L I^2,
\label{eq:inductive_energy}
\end{equation}
where $L$ is the relevant inductance and $I$ the current.  Large currents
and stored energies are indispensable for accelerator and detector
performance, but they simultaneously require quench detection, energy
extraction, electrical protection, cryogenic relief, and carefully defined
failure responses.  Thus, the scientific and safety aspects of the system
originate from the same underlying physical variables.

The LHC also provides an important historical illustration of why safety
models must evolve with empirical knowledge.  During powering tests on
19 September 2008, an electrical fault in sector 3--4 developed into an
arc, resulting in extensive mechanical and electrical damage and the
release of several tonnes of helium into the tunnel.  Subsequent analysis
showed that the helium flow associated with the event substantially
exceeded the maximum credible incident considered in an earlier cryogenic
risk analysis.  The experience led to a reassessment of the relevant
failure scenario, modifications to protection measures, and dedicated
studies of personnel safety in the LHC underground areas
\cite{Bajko:2009Incident,CryogenicRisk:2010,LHCSafetyTaskForce:2009,Wenninger:2016MPO}.
This episode is important for the present discussion not merely as an
accident case study, but as an example of a more general scientific
principle:
\begin{equation}
\text{model}
\rightarrow
\text{prediction}
\rightarrow
\text{observation}
\rightarrow
\text{model revision}
\rightarrow
\text{improved protection}.
\label{eq:safety_learning_cycle}
\end{equation}
In this sense, safety knowledge at the technological frontier contains an
empirical component.  The objective is, of course, not to learn primarily
through damaging events, but to extract information from commissioning
data, tests, simulations, operational anomalies, near misses, and
incidents and to incorporate that knowledge into progressively more robust
systems.

The underground environment adds another essential dimension.  Restricted
access and egress, the potential accumulation or displacement of gases,
long transport distances, radiation zoning, fire compartments, ventilation
requirements, and the presence of large experimental structures make the
state of the underground infrastructure itself part of the experimental
boundary conditions.  A cavern or accelerator tunnel is therefore not
simply a passive geometrical container in which physics is performed.  Its
ventilation, access, electrical, cryogenic, alarm, evacuation, and
radiation-protection systems participate directly in defining when and how
the accelerator or experiment may operate.

ATLAS and CMS provide complementary examples of this integration at the
detector scale.  Large experimental detectors must simultaneously achieve
high efficiency, stability, precision, and availability while remaining
protected against abnormal environmental and technical conditions.  The
distinction between detector control and detector protection is
particularly instructive.  A detector control system is primarily
concerned with bringing subsystems into the states required for operation
and physics data taking, monitoring their parameters, and coordinating
their configuration.  A detector safety or protection system, by contrast,
must be capable of overriding normal operation when predefined hazardous
conditions are detected.  In ATLAS, for example, the Detector Safety
System was designed as a dedicated and redundant protection layer capable
of initiating automatic actions to bring detector equipment into a safe
state independently of routine detector control
\cite{Luders:2003DSS,ATLAS:2008Detector}.  CMS employs its own hierarchy
of detector-control and subsystem-protection architectures adapted to its
detector design and operational requirements~\cite{CMS:2009DCS,DiCalafiori:2020CMS_ECAL}.
  The comparison is useful precisely because there is
no requirement for different experiments to implement protection in an
identical way; the relevant architecture follows from the physical system,
its failure modes, and the protection objectives.

The reverse influence---safety in physics---is equally important.
Accelerator and detector states cannot be chosen exclusively by asking
which configuration would maximize luminosity, efficiency, or data-taking
time.  Beam conditions, magnet currents, detector voltages, access modes,
maintenance activities, and commissioning sequences are constrained by
interlocks, permits, readiness conditions, radiation levels, equipment
states, and human-access requirements.  These constraints can affect
schedules, detector availability, commissioning strategies, and the rate
at which new operating regimes are approached.  It would therefore be
misleading to describe safety simply as a limitation imposed on scientific
activity.  In a highly energetic and complex installation, the ability to
establish and verify safe operating conditions is itself what makes
sustained high-performance operation possible.

This observation suggests a broader interpretation of protection systems.
Their purpose is not solely to react to failures.  They also define trusted
boundaries within which increasingly demanding scientific states can be
explored.  Redundancy, interlocks, monitoring, controlled access,
fail-safe design, testing, and operational procedures reduce the extent to
which safety depends on an unverified assumption that every component will
remain in its nominal state.  In this respect, a mature protection system
does not oppose scientific ambition; rather, it provides part of the
technical foundation on which such ambition can be exercised repeatedly
and reproducibly.

The aim of this paper is not to provide a comprehensive safety manual for
CERN, nor to reproduce the detailed regulatory framework applicable to
individual installations.  Instead, we use the LHC and the ATLAS and CMS
underground experiments as case studies for examining how physical
principles, quantitative hazard models, instrumentation, engineered
protection, operational rules, and human decision making become coupled in
a large scientific facility.  Particular attention is given to stored
beam and magnetic energy, machine protection, cryogenic hazards, radiation,
underground access, detector safety systems, and the feedback between
operational experience and subsequent protection strategies.

The novelty of the framework proposed here is not the introduction of
operational limits themselves, which are intrinsic to accelerator and
detector protection, but their formulation as a coupled
\emph{state-and-trajectory problem} for the scientific facility as a
whole.  Rather than treating machine protection, detector configuration,
cryogenics, infrastructure, personnel access, and intervention conditions
as independent safety domains, we regard them as components of a common
multidimensional operational state.
Safety therefore depends not only on whether the instantaneous values of
individual variables lie within acceptable limits, but also on whether
their combination represents an admissible global configuration and
whether the trajectory connecting two configurations remains admissible.
This unified phase-space description provides a common language for
machine protection, detector safety, cryogenic and environmental
conditions, access control, maintenance, commissioning, and operational
validation.

The discussion is organized as follows.  In
Sec.~\ref{sec:lhc_energetic_system}, we characterize the LHC as an
energetic underground physical system and identify the principal physical
quantities relevant to both performance and safety.
Section~\ref{sec:hazards_barriers} develops the transition from physical
variables to hazards, monitoring, protective barriers, and operational
constraints.  Stored beam and magnetic energy and their relation to
machine protection are discussed in Sec.~\ref{sec:machine_protection}.
Section~\ref{sec:cryogenic_safety} considers cryogenics, helium release,
oxygen-deficiency hazards, and underground personnel protection, including
lessons from the 2008 sector 3--4 incident.  In
Sec.~\ref{sec:atlas_cms_safety}, we examine ATLAS and CMS as complementary
examples of safety embedded within large detector systems.  The converse
problem---how safety requirements shape the actual practice of
experimental physics---is addressed in Sec.~\ref{sec:safety_in_physics}.
Section~\ref{sec:future_facilities} discusses implications for the
High-Luminosity LHC and future large accelerator and underground research
facilities.  Our conclusions are summarized in
Sec.~\ref{sec:conclusions}.

\section{The LHC as an energetic underground physical system}
\label{sec:lhc_energetic_system}

The LHC is naturally described as a particle accelerator,
but from the viewpoint developed in this work it is equally useful to regard
it as a spatially distributed system in which large amounts of energy are
stored, transported, transformed, and ultimately dissipated under highly
controlled conditions.  The relevant energy reservoirs are not limited to
the kinetic energy of the circulating beams.  They include the magnetic
energy stored in the superconducting accelerator and detector magnets,
electrical energy supplied to accelerator and experimental systems,
thermodynamic energy associated with the large cryogenic helium inventory,
mechanical and gravitational energy associated with massive detector
structures, and the energy subsequently deposited in matter through beam
losses and radiation fields.

The characteristic physical scales of these systems span many orders of
magnitude.  Proton beams circulate over a circumference of approximately
$26.7~{\rm km}$, while potentially damaging energy deposition may occur
within millimetre- or centimetre-scale regions.  Superconducting magnets
extend over kilometres of accelerator lattice while individual quenches
begin locally.  Liquid helium is distributed around the machine, whereas
an accidental release can influence the conditions within a comparatively
restricted tunnel volume.  The accelerator should therefore not be
characterized by a single ``stored energy,'' but by a hierarchy of energy
reservoirs, spatial distributions, characteristic release times, and
couplings between subsystems.

This distinction is central to the relation between physics and safety.
Large stored energy does not by itself define a hazard.  The consequences
of an abnormal event also depend on the rate at which the energy can be
released, the volume or mass into which it is transferred, and the
efficiency with which the energy reservoir couples to the affected system.
Schematically, a useful set of quantities is
\begin{equation}
\left\{
E_{\rm st},
\quad
\dot{E}_{\rm rel},
\quad
u_E,
\quad
\epsilon_{\rm dep},
\quad
\eta
\right\},
\label{eq:energetic_characteristics}
\end{equation}
where $E_{\rm st}$ denotes stored energy,
\begin{equation}
\dot{E}_{\rm rel}
=
\left|
\frac{dE_{\rm st}}{dt}
\right|
\label{eq:release_power}
\end{equation}
characterizes an energy-release rate,
\begin{equation}
u_E
=
\frac{E_{\rm st}}{V_{\rm eff}}
\label{eq:energy_density_generic}
\end{equation}
is an effective energy density, and
\begin{equation}
\epsilon_{\rm dep}
=
\frac{\eta E_{\rm st}}{m_{\rm eff}}
\label{eq:specific_deposition}
\end{equation}
represents a characteristic specific energy deposition into an affected
mass $m_{\rm eff}$.  The factor $\eta$ summarizes the fraction of the
available energy that can be transferred to the system under consideration.
Equations~\eqref{eq:energetic_characteristics}--\eqref{eq:specific_deposition}
are not intended as a substitute for detailed accident or protection
calculations.  They emphasize, however, why total energy alone is an
insufficient measure of the physical severity of a failure.

\subsection{Geometry and distributed character of the machine}
\label{subsec:lhc_geometry_energy}

The LHC is installed in the approximately $26.7~{\rm km}$ circumference
tunnel originally constructed for LEP and is divided into eight principal
sectors.  Its superconducting bending and focusing systems operate in
superfluid helium at approximately $1.9~{\rm K}$, while the principal
experimental interaction regions contain large underground caverns housing
ATLAS, CMS, ALICE, and LHCb~\cite{Bruning:2004LHC,Lebrun:1999Cryogenics}.
Two major experiments ATLAS and CMS are situated approximately $100~{\rm m}$ below ground and
constitute, in themselves, large technological infrastructures containing
superconducting magnets, detector services, electrical distribution,
cooling systems, gas installations, ventilation, access systems, and
radiation-protection infrastructure
\cite{ATLAS:2008Detector,CMS:2008Detector}.

The underground geometry has direct physical consequences.  Energy released
in a confined environment does not propagate or dissipate as it would in an
unbounded external space.  Gas expansion interacts with tunnel and cavern
geometry; ventilation determines transport and dilution; shielding modifies
radiation fields; access shafts and galleries determine evacuation paths;
and compartment boundaries can modify the propagation of smoke, pressure,
and heat.  The underground volume must therefore be considered part of the
physical system rather than merely a passive container surrounding the
accelerator.

At the same time, the longitudinal scale of the accelerator implies that
the relevant energies are highly distributed.  An order-$10~{\rm GJ}$
magnetic-energy inventory does not constitute a single localized
$10~{\rm GJ}$ source~\cite{Bruning:2004LHC,Wenninger:2016MPO}.  The accelerator is divided into electrical,
cryogenic, vacuum, and protection sectors, and the localization and
isolation of energy reservoirs are fundamental features of the machine
architecture.  Safety analysis consequently requires both a global view of
the facility and a local description of the physical subsystem in which an
abnormal transition occurs.

\subsection{Stored energy of the circulating beams}
\label{subsec:beam_energy_scale}

For a proton beam containing $n_b$ bunches, each with $N_b$ particles at
energy $E_p$, the total kinetic energy stored in the circulating beam is given by Eq.~\eqref{eq:lhc_beam_energy}.
At the nominal LHC design point,
\begin{equation}
n_b = 2808,
\qquad
N_b \simeq 1.15\times10^{11},
\qquad
E_p = 7~{\rm TeV},
\label{eq:lhc_nominal_beam_parameters}
\end{equation}
which gives
\begin{equation}
E_{\rm beam}^{\rm nom}
\simeq
3.62\times10^{8}~{\rm J}
=
362~{\rm MJ}
\label{eq:lhc_nominal_beam_energy}
\end{equation}
for each of the two beams
\cite{Bruning:2004LHC,Wenninger:2016MPO}.

Equation~\eqref{eq:lhc_nominal_beam_energy} illustrates an important
property of accelerator physics.  The energy of an individual
$7~{\rm TeV}$ proton is only
\begin{equation}
E_p
\simeq
1.12\times10^{-6}~{\rm J},
\label{eq:single_proton_joule}
\end{equation}
but the stored beam contains of order
$3\times10^{14}$ such particles.  A microscopic energy scale per particle
is therefore transformed, through particle number and beam coherence, into
a macroscopic energy reservoir.

The revolution period of the LHC is approximately
\begin{equation}
T_{\rm rev}
\simeq 89~\mu{\rm s}.
\label{eq:lhc_revolution_time}
\end{equation}
For purposes of illustrating the relevant timescale, the ratio
\begin{equation}
P_{\rm beam}^{(1\,{\rm turn})}
\equiv
\frac{E_{\rm beam}}{T_{\rm rev}}
\label{eq:beam_equivalent_power}
\end{equation}
is approximately
\begin{equation}
P_{\rm beam}^{(1\,{\rm turn})}
\simeq
4\times10^{12}~{\rm W}
\label{eq:beam_equivalent_power_number}
\end{equation}
at nominal beam energy.  This quantity should not be interpreted as a
continuous beam power: a stored circulating beam is not continuously
dissipating hundreds of megajoules.  Rather, it illustrates the enormous
instantaneous power scale that becomes relevant if a substantial fraction
of the beam is transferred to matter over a timescale comparable to one
revolution.

The physical consequences of beam loss are determined more directly by
energy density and energy deposition than by total beam energy alone.
High-energy protons penetrating matter initiate hadronic and
electromagnetic cascades, distributing energy over a volume determined by
the material, beam size, impact geometry, and the hydrodynamic response of
the target.  Detailed calculations have shown that the nominal LHC beam is
capable of causing extensive material damage if a sufficiently large
fraction is lost in an uncontrolled manner
\cite{Tahir:2005LHCBeam,Tahir:2009Graphite}.
The need to control where beam energy is allowed to be deposited is thus a
direct consequence of particle-interaction physics.

It is important for the present discussion that beam energy is also a
scientific resource.  Increasing particle energy, bunch population, or
number of bunches enhances the accessible collision rate or physics reach,
but simultaneously modifies the protection requirements.  Accelerator
performance and machine protection are therefore coupled already at the
level of Eq.~\eqref{eq:lhc_beam_energy}; they are not independent
considerations added at different stages of machine operation.

\subsection{Magnetic energy and the superconducting state}
\label{subsec:lhc_magnetic_energy}

The second major energy reservoir is contained in the superconducting
magnet system.  For an inductive circuit, the stored magnetic energy is given in Eq.~\eqref{eq:inductive_energy}.  The nominal field
of the LHC main dipoles is approximately $8.33~{\rm T}$, generated with a
current close to $12~{\rm kA}$.  A single main dipole stores approximately
$7~{\rm MJ}$ at nominal conditions, while the magnetic energy associated
with one LHC sector is of order
\begin{equation}
E_{\rm mag}^{\rm sector}
\simeq 1.1~{\rm GJ}.
\label{eq:lhc_sector_energy}
\end{equation}
The total magnetic-energy scale of the accelerator is consequently close
to $10~{\rm GJ}$~\cite{Wenninger:2016MPO}.

These values make superconductivity itself part of the safety problem.
During normal operation, the magnet windings have essentially vanishing
electrical resistance.  If a local region crosses the superconducting
boundary and becomes resistive, a quench develops.  The stored magnetic
energy must then be redistributed or extracted without producing
unacceptable local temperatures, voltages, mechanical stresses, or
cryogenic consequences.

The electromagnetic field also generates mechanical forces.  A useful
local scale is provided by the magnetic energy density
\begin{equation}
u_B
=
\frac{B^2}{2\mu_0},
\label{eq:magnetic_energy_density}
\end{equation}
which has the dimensions of pressure, where $\mu_0$ denotes the vacuum permeability.  One may therefore associate a
characteristic magnetic pressure
\begin{equation}
p_B
\sim
\frac{B^2}{2\mu_0}.
\label{eq:magnetic_pressure}
\end{equation}
For $B=8.33~{\rm T}$,
\begin{equation}
p_B
\simeq
2.8\times10^{7}~{\rm Pa}
\simeq
28~{\rm MPa}.
\label{eq:lhc_magnetic_pressure}
\end{equation}
The actual stresses in a magnet depend strongly on coil geometry,
mechanical supports, local field distribution, and structural constraints,
but Eq.~\eqref{eq:magnetic_pressure} demonstrates why high-field magnet
design is simultaneously an electromagnetic and mechanical problem.

This is another example of the two-way relation considered in this paper.
The large magnetic field required by accelerator physics generates the
forces and stored energies that protection systems must control, while the
mechanical, cryogenic, and electrical limits of the magnet system constrain
the admissible accelerator operating state.

\subsection{Cryogenics as a distributed thermodynamic system}
\label{subsec:lhc_cryo_energy}

The superconducting state of the main LHC magnets requires operation in
pressurized superfluid helium at approximately $1.9~{\rm K}$.  The LHC
cryogenic infrastructure is consequently a distributed thermodynamic
system extending around essentially the entire accelerator circumference
\cite{Lebrun:1999Cryogenics,Claudet:2010LHCcryo}.  The total helium
inventory associated with LHC operation is of order $130$ tonnes, although
this inventory is distributed among different parts and thermodynamic
states of the cryogenic system.

The safety significance of a cryogenic fluid cannot be characterized only
by its low temperature.  Helium undergoing depressurization and warming
can expand by a very large factor.  As a simple scale estimate, treating
helium as an ideal gas after warming to approximately room temperature
gives
\begin{equation}
V_{\rm He}
\simeq
\frac{m_{\rm He}}{M_{\rm He}}
\frac{RT}{P},
\label{eq:helium_expansion}
\end{equation}
where $M_{\rm He}$ and $R$ are the molar mass and gas constant of helium, respectively.  At
$T\simeq293~{\rm K}$ and $P\simeq1~{\rm bar}$,
\begin{equation}
V_{\rm He}
\simeq
6.1~{\rm m^3}
\left(
\frac{m_{\rm He}}{1~{\rm kg}}
\right)
\left(
\frac{T}{293~{\rm K}}
\right)
\left(
\frac{1~{\rm bar}}{P}
\right).
\label{eq:helium_expansion_numeric}
\end{equation}
This estimate is not an accident model: an actual cryogenic discharge
involves transient multiphase flow, heat transfer, pressure evolution,
ventilation, geometry, and non-equilibrium effects.  It nevertheless
illustrates why even a comparatively modest local helium inventory can
displace a substantial volume of air after warming and expansion.

For the LHC operating conditions, the relevant initial state is
superfluid helium at approximately $1.9~{\rm K}$, whose density is about
$\rho_{\rm He\,II}\simeq145~{\rm kg\,m^{-3}}$.  After vaporization and
warming to approximately room temperature at atmospheric pressure, one
kilogram of helium occupies a volume given by Eq.~\eqref{eq:helium_expansion_numeric}.
The corresponding volume-expansion factor is therefore
\begin{equation}
{\cal E}_{\rm He}
\equiv
\frac{V_{\rm He}^{\rm gas}}{V_{\rm He}^{\rm He\,II}}
=
\rho_{\rm He\,II}
\frac{V_{\rm He}^{\rm gas}}{m_{\rm He}}
\sim
10^{3}.
\label{eq:helium_expansion_factor}
\end{equation}
Thus, a comparatively small volume of superfluid helium can generate a
gas volume approximately three orders of magnitude larger after
vaporization and warming.  The precise expansion factor depends on the
initial thermodynamic state of the helium and on the final gas temperature
and pressure; the estimate in Eq.~\eqref{eq:helium_expansion_factor}
should therefore be understood as an order-of-magnitude value appropriate
for safety considerations.

The physical state of the cryogenic system is therefore naturally
described by several coupled quantities,
\begin{equation}
{\boldsymbol X}_{\rm cryo}
=
\left\{
T,
P,
m_{\rm He},
\dot{m}_{\rm He},
x_{\rm He},
\dot Q,
\ldots
\right\},
\label{eq:cryo_state_vector}
\end{equation}
where $\dot m_{\rm He}$ characterizes mass flow, $x_{\rm He}$ may denote a
local helium concentration, and $\dot Q$ represents relevant heat transfer.
A transition from a nominal cryogenic state to an abnormal one may then
couple magnet protection, pressure relief, ventilation, oxygen
concentration, and personnel access.  The 2008 LHC sector 3--4 incident,
discussed in detail in Sec.~\ref{sec:cryogenic_safety}, provides an
important example of precisely such cross-system coupling.

\subsection{Energetic systems of ATLAS and CMS}
\label{subsec:detector_energy_scales}

The large LHC experiments should themselves be regarded as energetic
physical systems rather than passive instruments placed around the
collision points.  In ATLAS and CMS, magnetic fields required for momentum
measurement are generated by superconducting magnet systems containing
gigajoule-scale stored energies.

ATLAS employs a central solenoid together with a barrel toroid and two
end-cap toroids.  At nominal currents, the corresponding stored magnetic
energies are approximately
\begin{equation}
E_{\rm sol}^{\rm ATLAS}
\simeq0.04~{\rm GJ},
\qquad
E_{\rm BT}^{\rm ATLAS}
\simeq1.08~{\rm GJ},
\qquad
E_{\rm ECT}^{\rm ATLAS}
\simeq2\times0.25~{\rm GJ},
\label{eq:atlas_magnet_energies}
\end{equation}
giving a total of approximately
\begin{equation}
E_{\rm mag}^{\rm ATLAS}
\simeq1.6~{\rm GJ}
\label{eq:atlas_total_magnet_energy}
\end{equation}
for the complete superconducting magnet system
\cite{ATLAS:2008Detector,tenKate:2005ATLASMagnets}.

CMS follows a very different detector and magnet philosophy.  Its large
superconducting solenoid was designed to produce a central field of
$4~{\rm T}$ and stores approximately
\begin{equation}
E_{\rm mag}^{\rm CMS}
\simeq2.6~{\rm GJ}
\label{eq:cms_magnet_energy}
\end{equation}
at the design field~\cite{CMS:2008Detector}.
Its normal operating field of approximately $3.8~{\rm T}$ is accompanied
by large electromagnetic forces transmitted to the coil and return-yoke
structures.

These detector-scale energy reservoirs are localized in experimental
caverns occupied by extensive detector services and, during appropriate
access states, by personnel.  Magnet operation must consequently be
integrated with cryogenic conditions, electrical states, detector
configuration, access, ventilation, and emergency systems.  The detector
is therefore simultaneously a measurement apparatus and a controlled
energetic installation.

Massive detector structures introduce another relevant energy scale.
For a component of mass $m$ undergoing a vertical displacement $h$, the
change in gravitational potential energy is
\begin{equation}
\Delta E_g
=
mgh.
\label{eq:gravitational_energy}
\end{equation}
During construction, installation, opening, closing, and maintenance,
detector components weighing tens to hundreds of tonnes---and in some
cases substantially more---must be moved under tightly controlled
conditions.  In this case the key physical quantities are not only total
potential or kinetic energy, but load distribution, centre of gravity,
structural deformation, acceleration, braking distance, support geometry,
and the integrity of lifting or transport systems.  Mechanical handling is
thus another example in which ordinary classical physics becomes directly
part of safety engineering.

\begin{table*}[t]
\caption{\label{tab:lhc_energy_scales}
Representative physical scales of the LHC and the ATLAS and CMS
experiments.  Values are given primarily to establish the hierarchy of
energies and operating conditions relevant to the discussion; actual
machine configurations vary with operating period and accelerator mode.
The beam values correspond to the nominal LHC design configuration.}
\begin{ruledtabular}
\begin{tabular}{lccc}
System
&
Physical quantity
&
Representative scale
&
Physical significance
\\
\hline
LHC proton beam
&
Energy per beam
&
$362~{\rm MJ}$
&
Potential localized energy deposition
\\
LHC proton beam
&
Proton energy
&
$6.8~{\rm TeV}$
&
Collision energy and shower development
\\
LHC main dipoles
&
Magnetic field
&
$8.33~{\rm T}$
&
Beam bending and electromagnetic stress
\\
LHC main dipoles
&
Operating temperature
&
$1.9~{\rm K}$
&
Superconducting state
\\
LHC dipole
&
Stored magnetic energy
&
$\sim7~{\rm MJ}$
&
Local quench/protection scale
\\
LHC sector
&
Stored magnetic energy
&
$\sim1.1~{\rm GJ}$
&
Circuit energy-extraction scale
\\
LHC magnet system
&
Stored magnetic energy
&
$\mathcal{O}(10~{\rm GJ})$
&
Distributed machine-scale energy reservoir
\\
LHC cryogenic system
&
Helium inventory
&
$\mathcal{O}(130~{\rm t})$
&
Cryogenic, pressure, and ODH considerations
\\
ATLAS magnet system
&
Stored magnetic energy
&
$\sim1.6~{\rm GJ}$
&
Detector magnet protection
\\
CMS solenoid
&
Stored magnetic energy
&
$\sim2.6~{\rm GJ}$ at $4~{\rm T}$
&
Detector magnet protection
\end{tabular}
\end{ruledtabular}
\end{table*}

\subsection{The underground environment as a dynamical subsystem}
\label{subsec:underground_environment_state}

The energetic description above becomes incomplete if the underground
environment is treated independently of the accelerator and detectors.
For personnel and infrastructure, quantities such as oxygen concentration,
air flow, temperature, pressure, radiation field, smoke concentration, and
access state form additional dynamical variables.  We may schematically
write
\begin{equation}
{\boldsymbol X}_{\rm env}
=
\left\{
C_{O_2},
C_{\rm He},
T_{\rm air},
P_{\rm air},
\dot V_{\rm vent},
\dot H^{*},
C_{\rm smoke},
{\sigma}_{\rm access},
\ldots
\right\},
\label{eq:environment_state_vector}
\end{equation}
where undefined earlier $\dot V_{\rm vent}$ denotes the effective
volumetric ventilation flow rate.

These environmental variables can now be incorporated into
the more general facility-state description of
Eq.~\eqref{eq:facility_state_vector}.  That equation represents the state
of the installation at the level of individual physical and operational
coordinates, such as beam energy, magnet current, temperature, pressure,
oxygen concentration, high voltage, and access state.  For the discussion
of coupled failure modes, however, it is useful to reorganize these
coordinates into groups associated with the principal physical subsystems.

Schematically, one may write
\begin{equation}
{\boldsymbol X}
=
\left(
{\boldsymbol X}_{\rm beam},
{\boldsymbol X}_{\rm mag},
{\boldsymbol X}_{\rm cryo},
{\boldsymbol X}_{\rm det},
{\boldsymbol X}_{\rm env},
{\boldsymbol X}_{\rm access}
\right),
\label{eq:lhc_coupled_state_vector}
\end{equation}
where the individual subsystem states may be represented as
\begin{align}
{\boldsymbol X}_{\rm beam}
&=
\left\{
E_{\rm beam},
{N}_{\rm beam},
{\boldsymbol x}_{\rm orbit},
\ldots
\right\},
\nonumber\\
{\boldsymbol X}_{\rm mag}
&=
\left\{
I_{\rm mag},
E_{\rm mag},
\sigma_{\rm quench},
\ldots
\right\},
\nonumber\\
{\boldsymbol X}_{\rm cryo}
&=
\left\{
T_{\rm cryo},
P_{\rm cryo},
m_{\rm He},
\dot m_{\rm He},
\ldots
\right\},
\nonumber\\
{\boldsymbol X}_{\rm det}
&=
\left\{
V_{\rm HV},
V_{\rm LV},
T_{\rm det},
\sigma_{\rm det},
\ldots
\right\},
\nonumber\\
{\boldsymbol X}_{\rm env}
&=
\left\{
C_{O_2},
C_{\rm He},
T_{\rm air},
P_{\rm air},
\dot V_{\rm vent},
\dot H^{*},
C_{\rm smoke},
\ldots
\right\},
\nonumber\\
{\boldsymbol X}_{\rm access}
&=
\left\{
\sigma_{\rm access},
\sigma_{\rm occupancy},
\ldots
\right\}.
\label{eq:lhc_subsystem_state_vectors}
\end{align}

Here, $E_{\rm beam}$ and $N_{\rm beam}$ denote the
stored beam energy and the total number of circulating particles
in the beam, respectively,
while ${\boldsymbol x}_{\rm orbit}$ represents the
beam-orbit coordinates. The magnet state is characterized
by the current $I_{\rm mag}$, stored magnetic energy
$E_{\rm mag}$, and discrete quench status
$\sigma_{\rm quench}$. The cryogenic quantities
$T_{\rm cryo}$ and $P_{\rm cryo}$ represent relevant
temperatures and pressures, whereas $m_{\rm He}$
and $\dot m_{\rm He}$ denote the helium mass
and mass-flow rate, respectively.
The detector state includes the high- and low-voltage
supply quantities $V_{\rm HV}$ and $V_{\rm LV}$,
the relevant detector temperatures $T_{\rm det}$,
and the discrete detector status $\sigma_{\rm det}$.
The environmental state is characterized by the
oxygen and helium concentrations $C_{O_2}$ and
$C_{\rm He}$, air temperature $T_{\rm air}$,
air pressure $P_{\rm air}$, effective volumetric
ventilation flow rate $\dot V_{\rm vent}$,
radiation dose-equivalent rate $\dot H^{*}$,
and smoke-related concentration $C_{\rm smoke}$.
Finally, $\sigma_{\rm access}$ and
$\sigma_{\rm occupancy}$ denote the discrete
personnel-access and occupancy statuses.

Thus, Eq.~\eqref{eq:lhc_coupled_state_vector}
should be understood as a block decomposition,
or coarse-grained reorganization, of the more
elementary state description introduced in
Eq.~\eqref{eq:facility_state_vector}, rather than
as a separate state definition. The grouping is
schematic and is not unique: some variables may
participate in more than one subsystem description,
and each block may include both continuous physical
quantities and discrete operational statuses.
For example, temperature and pressure may
characterize both cryogenic and environmental states,
while detector operation can depend simultaneously
on electrical, cooling, magnetic, and access conditions.

The usefulness of the block representation is that it makes the
inter-subsystem couplings explicit.  In general,
\begin{equation}
\frac{d{\boldsymbol X}_{i}}{dt}
=
{\boldsymbol F}_{i}
\left(
{\boldsymbol X}_{\rm beam},
{\boldsymbol X}_{\rm mag},
{\boldsymbol X}_{\rm cryo},
{\boldsymbol X}_{\rm det},
{\boldsymbol X}_{\rm env},
{\boldsymbol X}_{\rm access}
\right),
\label{eq:lhc_coupled_dynamics}
\end{equation}
so that the evolution of one subsystem need not depend only on its own
state.  The physical significance of this coupling becomes particularly
clear when an abnormal condition propagates from one subsystem to another.
A beam loss, for example, originates in beam dynamics but transfers energy
to accelerator components through particle--matter interactions.  The
resulting local energy deposition can heat material, generate secondary
radiation, and, when occurring in a superconducting magnet, induce a
quench.  A beam-dynamical perturbation can therefore evolve into a thermal,
electromagnetic, cryogenic, and radiation-protection problem.

Conversely, a magnet fault can propagate in the opposite direction.  A
quench begins as a transition of part of the conductor from the
superconducting to the resistive state, leading to redistribution of the
stored magnetic energy and heat deposition in the conductor, magnet
structure, and surrounding helium.  The resulting cryogenic response may
involve changes of temperature, pressure, helium mass flow, and, in a
sufficiently severe failure, loss of helium containment.  A cryogenic
release can then modify the underground environmental state by changing
the local pressure, temperature, gas composition, and oxygen
concentration.  These environmental changes may in turn alter the
permitted personnel-access state or activate alarm, ventilation, and
evacuation functions.  The complete sequence can therefore extend across
several subsystem boundaries,
\begin{equation}
{\boldsymbol X}_{\rm mag}
\rightarrow
{\boldsymbol X}_{\rm cryo}
\rightarrow
{\boldsymbol X}_{\rm env}
\rightarrow
{\boldsymbol X}_{\rm access}.
\label{eq:mag_cryo_env_access_chain}
\end{equation}

The coupling is not limited to failure sequences involving the accelerator
magnets.  Detector systems provide additional examples.  A loss of cooling
can cause the temperature of detector electronics or other sensitive
components to rise, requiring an automatic reduction or removal of
electrical power.  An electrical fault may similarly initiate a transition
of the detector from its nominal operating configuration to a protected
state even when the beam, magnets, and cryogenic systems of the accelerator
remain nominal.  Schematically, such a sequence may be represented as
\begin{equation}
{\boldsymbol X}_{\rm cooling}
\rightarrow
{\boldsymbol X}_{\rm det}
\rightarrow
{\boldsymbol X}_{\rm power},
\label{eq:detector_coupling_chain}
\end{equation}
with detector-protection systems acting to prevent the abnormal state from
developing into equipment damage.

These examples show that the subsystem decomposition in
Eq.~\eqref{eq:lhc_coupled_state_vector} is primarily a convenient
organizational representation rather than a decomposition into dynamically
independent systems.  The complete facility is more appropriately viewed
as a network of coupled physical and operational states, in which a
perturbation originating in one subsystem may modify the state, the
protection requirements, or the admissible operating conditions of several
others.  Safety analysis must therefore address not only the stability and
protection of each subsystem separately, but also the possible
\emph{propagation of state changes across subsystem boundaries}.  The
corresponding protection problem is to identify these coupling paths,
detect abnormal transitions sufficiently early, interrupt or redirect
their propagation when necessary, and prevent the system from reaching
combinations of beam, magnet, cryogenic, detector, environmental, and
access states associated with unacceptable consequences.

The physical system is consequently characterized not simply by individual
hazards associated with separate subsystems, but by possible
\emph{transitions of the complete coupled facility state}.  A perturbation
originating in one subsystem can propagate through the couplings discussed
above and modify several other components of
${\boldsymbol X}$.  It is therefore more appropriate to write
symbolically
\begin{equation}
{\boldsymbol X}(t_a)
\xrightarrow{\;\mathcal{F}\;}
{\boldsymbol X}(t_b),
\label{eq:state_transition}
\end{equation}
where $\mathcal{F}$ denotes the process responsible for the transition.
It may represent an internal physical failure, an operational action, or
an external perturbation.  Schematically,
\begin{equation}
\mathcal{F}
\in
\left\{
\mathcal{F}_{\rm int},
\mathcal{F}_{\rm op},
\mathcal{F}_{\rm ext}
\right\},
\label{eq:transition_classes}
\end{equation}
where $\mathcal{F}_{\rm int}$ denotes internally generated failures or
instabilities, $\mathcal{F}_{\rm op}$ denotes deliberately initiated
changes of machine or detector configuration, and
$\mathcal{F}_{\rm ext}$ denotes perturbations entering through the
boundary conditions or supporting infrastructure of the facility.

The latter category is important because the accelerator and experiments
are not isolated physical systems.  Disturbances of electrical power,
cooling-water supplies, ventilation, cryogenic utilities, or other
infrastructure services can modify the state of accelerator or detector
subsystems even when the initiating perturbation originates outside them.
Environmental events may similarly act indirectly through these
infrastructure systems.  The resulting propagation can therefore take the
schematic form
\begin{equation}
\mathcal{F}_{\rm ext}
\rightarrow
{\boldsymbol X}_{\rm infrastructure}
\rightarrow
{\boldsymbol X}_{\rm mag,\ cryo,\ det,\ldots}
\rightarrow
{\boldsymbol X}_{\rm env}
\rightarrow
{\boldsymbol X}_{\rm access},
\label{eq:external_perturbation_chain}
\end{equation}
although the actual sequence depends on the nature of the initiating
event.  An external perturbation need not itself constitute the hazardous
state; its importance lies in its ability to drive one or more coupled
subsystems away from their nominal operating conditions.

A central purpose of protection systems is therefore to prevent such
transitions from propagating toward regions of state space associated with
unacceptable consequences, or, when an abnormal transition has already
begun, to redirect the evolution toward a controlled state.  In the
general case, protection does not act only on the instantaneous state
${\boldsymbol X}$ but on its subsequent evolution,
\begin{equation}
{\boldsymbol X}(t)
\xrightarrow{\;\mathcal{P}\;}
{\boldsymbol X}_{\rm protected}(t+\Delta t),
\label{eq:protective_state_transition}
\end{equation}
where $\mathcal{P}$ represents the relevant combination of monitoring,
interlock, control, relief, shutdown, or access-protection functions.

This observation refines the concept of the safe operational phase space
introduced in Sec.~\ref{sec:introduction}.  In a complex facility it is
not sufficient to require only
\begin{equation}
{\boldsymbol X}(t)\in\Omega_{\rm safe}.
\label{eq:instantaneous_safe_state}
\end{equation}
Safety also depends on how the facility reaches a given state and on the
intermediate states traversed during that evolution.  One must therefore
constrain the allowed trajectories,
\begin{equation}
{\boldsymbol X}(t_1)
\longrightarrow
{\boldsymbol X}(t_2),
\label{eq:safe_trajectory}
\end{equation}
because two individually acceptable endpoint states need not necessarily
be connected by an acceptable transition.

Magnet ramping provides a simple example.  Both the initial low-current
state and the final operating-current state may belong to
$\Omega_{\rm safe}$, while the ramp between them is permitted only if the
current evolution, quench-protection state, powering interlocks, cryogenic
conditions, and relevant access constraints remain valid throughout the
transition.  Similar considerations apply to beam injection, the
beam-energy ramp, the squeeze, establishment of stable beams, detector
power-up, magnet ramping, detector opening or closing, cryogenic cool-down
and warm-up, and transitions between beam-operation and personnel-access
states.  In each case, the sequence, rate, and intermediate configurations
are themselves relevant safety variables.

This leads naturally to a more general description
of operational trajectories. Let
\begin{equation}
\Gamma_{\rm op}:
t\longmapsto\boldsymbol X(t),
\qquad t\in[t_1,t_2],
\label{eq:safe_operational_trajectory}
\end{equation}
denote a particular time-ordered trajectory through
the coupled facility state space.

We denote by $\mathcal G_{\rm adm}$ the set of
admissible trajectories, namely those for which
the physical state remains within the applicable
safe domain and the complete evolution satisfies
the required protection and operational constraints.
These constraints may include limits on rates of
change, valid interlock states, the prescribed
sequence of operations, and permissible transitions
between discrete configurations.

In this formulation, safe operation requires both
\begin{equation}
\boldsymbol X(t)\in\Omega_{\rm safe}
\qquad \text{for all }t\in[t_1,t_2],
\end{equation}
and
\begin{equation}
\Gamma_{\rm op}\in\mathcal G_{\rm adm}.
\label{eq:safe_state_and_trajectory}
\end{equation}
The distinction is important: the first condition
constrains \emph{where} the accelerator and experiments
may operate, whereas the second constrains \emph{how}
they may move between permitted states.
Safety is consequently a property not only of the instantaneous
configuration of the facility but also of its dynamical evolution through
the coupled beam, magnet, cryogenic, detector, environmental, and access
state space.

\subsection{Energy control as a unifying physical principle}
\label{subsec:energy_control_principle}

The examples discussed above reveal a common physical structure behind
otherwise very different accelerator and detector hazards.  High-energy
physics facilities operate by creating and maintaining highly organized
physical states: a tightly confined relativistic beam, persistent currents
in superconducting magnet circuits, cryogenic systems maintained far below
ambient temperature, detector components biased at precisely controlled
electrical potentials, and massive mechanical structures held in accurately
defined configurations.  The scientific function of the facility depends
on preserving these states, and on moving between them through controlled
operational trajectories.

Many relevant failure processes can, in contrast, be interpreted as an
uncontrolled redistribution of energy or as an unintended transition
between coupled physical states.  Schematically,
\begin{equation}
E_{\rm stored}
\longrightarrow
E_{\rm thermal}
+
E_{\rm mechanical}
+
E_{\rm radiation}
+
E_{\rm fluid}
+
E_{\rm electrical}
+\cdots .
\label{eq:energy_redistribution}
\end{equation}
The particular terms that dominate depend on the subsystem and on the
failure sequence.  A beam loss converts organized kinetic energy into
particle showers, heat, radiation, and possibly mechanical damage.  A
magnet quench converts stored magnetic energy into resistive heating and
can transfer part of that energy to the cryogenic system.  A helium release
transforms a confined cryogenic state into pressure evolution, gas flow,
heat exchange, and potentially oxygen displacement.  An electrical or
cooling fault in a detector can similarly drive the detector away from its
nominal operating configuration and require an automatic transition to a
protected state.

Machine and detector protection can therefore be understood, at a
fundamental level, as the controlled management of both
\emph{energy pathways} and \emph{state transitions}.  Protection does not
remove the energetic or otherwise extreme physical conditions required for
the experiment.  Rather, it determines where energy may reside, how rapidly
it may be transferred, which transitions between states are permitted, and
under which conditions the system must be redirected toward a protected
configuration.

In the coupled-state description developed above, this can be represented
schematically as
\begin{equation}
{\boldsymbol X}(t)
\xrightarrow{\;\mathcal{F}\;}
{\boldsymbol X}'(t)
\xrightarrow{\;\mathcal{P}\;}
{\boldsymbol X}_{\rm protected},
\label{eq:failure_protection_state_chain}
\end{equation}
where $\mathcal{F}$ denotes an internal failure, an operational
perturbation, or an external disturbance, while $\mathcal{P}$ denotes the
corresponding protection response.  The role of the protection architecture
is to prevent the trajectory generated by $\mathcal{F}$ from entering
regions of state space associated with unacceptable consequences.

The same idea can be expressed in terms of an admissible operational
trajectory, presented by Eq.~\eqref{eq:safe_state_and_trajectory}. Protection is
therefore concerned not only
with limiting the magnitude of physical variables, but also with controlling
their rates of change, their mutual couplings, and the sequence in which
operational states are entered.

This viewpoint also clarifies why different protection systems can share a
common abstract logic despite acting on very different physical quantities.
A beam loss monitor (BLM) observes the consequences of particle loss; a quench
detector observes the onset of resistive behaviour in a superconducting
circuit; a pressure sensor observes the evolution of a cryogenic system;
an oxygen monitor observes a change in the underground atmosphere; and a
detector safety system (DSS) observes abnormal detector conditions.  In each
case, the measurement is interpreted in relation to the current operating
state and may trigger an action intended to redirect the subsequent
evolution.

One may therefore write the common protection logic as
\begin{equation}
\text{physical state}
\rightarrow
\text{deviation}
\rightarrow
\text{measurement}
\rightarrow
\text{decision}
\rightarrow
\text{protective action}
\rightarrow
\text{controlled state}.
\label{eq:general_protection_logic}
\end{equation}

The relevant question is consequently not simply whether the LHC contains
large stored energies or potentially hazardous physical conditions---any
facility operating at such beam energies, magnetic fields, cryogenic
temperatures, electrical powers, and radiation levels necessarily does.
The more meaningful question is how the corresponding physical quantities
and subsystem couplings are translated into identifiable failure modes,
how abnormal transitions propagate through the facility, how the resulting
risks are quantified, and how sufficiently independent protective barriers
are placed along those propagation paths.

The concepts introduced in this section therefore provide the bridge
between the description of the LHC as an energetic coupled physical system
and the detailed protection architectures discussed below.  In the next
section, we examine how physical states and failure trajectories are mapped
onto hazards, monitoring functions, interlocks, passive and active
protection, and operational constraints.

\section{From physical quantities to hazards and protective barriers}
\label{sec:hazards_barriers}

Section~\ref{sec:lhc_energetic_system} described the LHC and its
experiments as a set of coupled physical and operational subsystems and
introduced the corresponding state vector
${\boldsymbol X}$, together with the possibility of transitions and
trajectories through the combined facility state space.  In that
description, an internal failure, operational action, or external
perturbation can initiate a change in one subsystem and, through the
couplings between beam, magnet, cryogenic, detector, environmental, and
access states, propagate to several others.

The next step is to determine which of these possible transitions are
safety relevant and where protective functions must act.  Not every
departure from a nominal state constitutes a hazard, and not every
abnormal state leads to an unacceptable consequence.  A safety-relevant
sequence is instead characterized by the existence of a physically
credible propagation path connecting an initiating event to a state in
which personnel, equipment, the environment, or continued operation may
be adversely affected.

It is therefore useful to refine the generic state transition introduced
in Sec.~\ref{sec:lhc_energetic_system} into a \emph{hazard chain}. For a generic subsystem, such a chain can be represented schematically as
\begin{equation}
{\boldsymbol X}_{0}
\xrightarrow{\;F_i\;}
{\boldsymbol X}_{1}
\xrightarrow{\;{\cal P}_i\;}
{\boldsymbol X}_{2}
\xrightarrow{\;{\cal E}_i\;}
{\boldsymbol{\cal C}}_i ,
\label{eq:generic_hazard_chain}
\end{equation}
where ${\boldsymbol X}_{0}$ denotes a nominal physical state, $F_i$ an
initiating failure or perturbation, ${\cal P}_i$ the subsequent physical
propagation of the abnormal state, ${\cal E}_i$ an energy-transfer,
exposure, or loss mechanism, and ${\boldsymbol{\cal C}}_i$ the resulting
consequence vector.  A protective system acts by interrupting, redirecting,
or sufficiently slowing one or more links of this chain.
Equation~\eqref{eq:generic_hazard_chain} should be understood as a
coarse-grained representation of a trajectory through the coupled state
space rather than as a sequence of dynamically independent states.  In
general, the propagation represented by ${\cal P}_i$ may involve several
subsystem couplings.  For example,
\begin{equation}
\text{beam perturbation}
\rightarrow
\text{beam loss}
\rightarrow
\text{local energy deposition}
\rightarrow
\text{magnet quench}
\rightarrow
\text{cryogenic response},
\label{eq:multisystem_hazard_example}
\end{equation}
or
\begin{equation}
\text{cryogenic loss of containment}
\rightarrow
\text{helium discharge}
\rightarrow
\text{environmental change}
\rightarrow
\text{access or personnel consequence}.
\label{eq:cryo_hazard_example}
\end{equation}
The hazard chain therefore identifies not only the initiating event and
final consequence, but also the physical pathway by which the facility
can evolve from one to the other.

A protective barrier acts by modifying this pathway.  Depending on its
position in the sequence, it may prevent the initiating disturbance from
developing, detect an abnormal evolution, interrupt propagation between
subsystems, redirect energy toward a controlled destination, reduce the
severity of the consequence, or prevent personnel from being exposed to
the resulting state.  Schematically,
\begin{equation}
{\boldsymbol X}_{0}
\xrightarrow{\;F_i\;}
{\boldsymbol X}_{1}
\xrightarrow{\;B_k\;}
{\boldsymbol X}_{\rm protected},
\label{eq:barrier_interrupts_trajectory}
\end{equation}
where $B_k$ denotes the relevant preventive or mitigative barrier.

This representation also clarifies the relation between the concepts of
\emph{hazard} and \emph{risk}.  A hazard is associated with the potential
of a physical state, stored-energy reservoir, substance, or process to
produce an adverse consequence.  Risk additionally depends on the
likelihood that a physically credible sequence reaches that consequence
and on its severity.  Thus, a large stored energy or extreme physical
condition can constitute a major hazard while its residual operational
risk remains controlled if sufficiently effective and reliable barriers
prevent the corresponding trajectory from reaching an unacceptable
region of state space.

The purpose of this section is therefore to move from the dynamical
description developed in Sec.~\ref{sec:lhc_energetic_system} to a
protection-oriented description.  We examine how failure trajectories are
translated into hazard and consequence classes, how preventive and
mitigative barriers are positioned along those trajectories, how their
independence and response time determine their effectiveness, and how
measurements and interlocks convert physical information into operational
constraints.  The same framework will subsequently be applied to the
specific cases of beam and magnet protection, cryogenic safety, and
detector protection.

\subsection{Hazard chains and consequence spaces}
\label{subsec:hazard_chains}

Once a physically credible initiating event has been identified, the next
question is how its possible propagation paths and consequences should be
represented and how their likelihood enters the protection problem.  The
coupled-state description developed above implies that an initiating event
does not, in general, determine a unique outcome.  The same perturbation
may generate several possible trajectories depending on the instantaneous
facility state, the response of protection systems, and the availability
of alternative barriers. This representation follows the general logic of hazard-sequence and
barrier-based risk analysis, in which initiating events, propagation
mechanisms, protective functions, and consequences are treated as distinct
elements of an accident scenario
\cite{Sklet:2006SafetyBarriers,DeDianousFievez:2006ARAMIS}.

A generic initiating event $F_i$ may therefore be associated with a set of
possible trajectories,
\begin{equation}
F_i
\longrightarrow
\left\{
\Gamma_{1}^{(i)},
\Gamma_{2}^{(i)},
\ldots,
\Gamma_{n_i}^{(i)}
\right\},
\label{eq:hazard_trajectory_set}
\end{equation}
where the superscript $(i)$ identifies the initiating event and the
subscript $j$ labels the different physically possible propagation paths
that may follow from it.  The number of such paths, $n_i$, may depend on
the initiating event.  The corresponding trajectories may terminate in
different consequences,
\begin{equation}
\Gamma_{j}^{(i)}
\longrightarrow
{\cal C}_{j}^{(i)}.
\label{eq:trajectory_to_consequence}
\end{equation}

This branching structure is important because an initiating failure need
not be intrinsically damaging.  Its safety significance is determined by
the subsequent propagation path.  A magnet-powering fault, for example,
may be detected sufficiently early for the beam to be extracted without
further consequence, or it may generate a beam-orbit perturbation followed
by localized losses.  Similarly, a cryogenic abnormality may remain
confined within the designed pressure-relief system or, under a different
failure sequence, propagate toward an environmental or personnel-safety
condition.

The relevant object is therefore not only the probability or frequency of
the initiating event, but also the conditional probability that the event
follows a particular trajectory.  For trajectory $\Gamma_{j}^{(i)}$ one
may write schematically
\begin{equation}
P_{j}^{(i)}
=
P(F_i)\,
P\!\left(
\Gamma_{j}^{(i)}
\mid
F_i
\right),
\label{eq:trajectory_probability}
\end{equation}
where $P(F_i)$ characterizes the occurrence of the initiating event and
$P(\Gamma_{j}^{(i)}\mid F_i)$ denotes the conditional probability that,
once $F_i$ has occurred, the system evolves along trajectory
$\Gamma_{j}^{(i)}$.

If a trajectory consists of a sequence of successive transitions,
\begin{equation}
T_{0}^{(i,j)}
\rightarrow
T_{1}^{(i,j)}
\rightarrow
\cdots
\rightarrow
T_{m}^{(i,j)},
\label{eq:trajectory_transition_sequence}
\end{equation}
one may write, schematically,
\begin{equation}
P\!\left(
\Gamma_{j}^{(i)}
\mid
F_i
\right)
=
\prod_{k=0}^{m-1}
P
\left(
T_{k+1}^{(i,j)}
\mid
T_{k}^{(i,j)}
\right),
\label{eq:conditional_transition_product}
\end{equation}
provided that the conditional structure is specified appropriately.
The factorized form in Eq.~\eqref{eq:conditional_transition_product} is
only a conceptual representation.  Real failure sequences may contain
dependent events, common-cause failures, feedback between subsystems, and
protection actions that make a simple product representation inadequate.
Its purpose is to emphasize that risk depends on the probability of
\emph{propagation} through the hazard chain, not solely on the frequency
of the initiating event.

The consequence associated with a given trajectory is generally
multidimensional, as stated in Eq.~\eqref{eq:generic_hazard_chain}.  It is therefore useful to introduce a consequence
vector
\begin{equation}
{\boldsymbol {\cal C}}_{j}^{(i)}
=
\left(
{\cal C}_{j,{\rm pers}}^{(i)},
{\cal C}_{j,{\rm mach}}^{(i)},
{\cal C}_{j,{\rm det}}^{(i)},
{\cal C}_{j,{\rm rad}}^{(i)},
{\cal C}_{j,{\rm env}}^{(i)},
{\cal C}_{j,{\rm op}}^{(i)}
\right),
\label{eq:consequence_vector}
\end{equation}
whose components may refer, respectively, to consequences for personnel,
accelerator equipment, detector equipment, radiation exposure, the
environment, and scientific operation. These components are not generally reducible to a single physical unit.
An event may have negligible personnel consequence but cause prolonged
accelerator downtime; another may produce little equipment damage but
require immediate evacuation or generate an unacceptable radiation
condition.  Consequently, the notion of ``severity'' should be understood
as depending on the class of consequence under consideration rather than
as a universal scalar observable.

At the most elementary level, if a single dominant consequence pathway is
assumed, risk may be represented schematically as
\begin{equation}
R_i
\sim
P_i\,{\cal C}_i,
\label{eq:simple_risk_definition}
\end{equation}
where $P_i$ is the occurrence probability of the undesired event within a
specified reference interval and ${\cal C}_i$ is a scalar measure of its
consequence.  Equation~\eqref{eq:simple_risk_definition} is only a
conceptual approximation.  In a complex facility, an initiating event can
develop through several possible pathways and its consequences are
generally multidimensional.

The same probability--consequence logic underlies risk-matrix approaches.
In such approaches, occurrence probability and consequence severity are
assigned to discrete categories, and their combination is used to classify
the corresponding level of risk
\cite{Schmidt:2016MPS,CartierMichaud:2021RiskMatrices}.  A risk matrix
should not, however, be identified with the numerical product in
Eq.~\eqref{eq:simple_risk_definition}; rather, it provides a discrete
mapping from probability--consequence categories to predefined risk
classes.

For an initiating event $F_i$ with several possible propagation branches,
along which the system may evolve through pathways
$\Gamma_{j}^{(i)}$, the corresponding risk can be represented more
generally as
\begin{equation}
{\cal R}^{(i)}
=
{\cal R}
\left[
P(F_i),
\left\{
P\!\left(
\Gamma_{j}^{(i)}
\mid
F_i
\right)
\right\}_{j=1}^{n_i},
\left\{
{\boldsymbol{\cal C}}_{j}^{(i)}
\right\}_{j=1}^{n_i}
\right].
\label{eq:general_risk_function}
\end{equation}
Here the symbol ${\cal R}$ denotes the risk-evaluation mapping that combines
the probability $P(F_i)$ of the initiating event, the conditional
probabilities $P(\Gamma_{j}^{(i)}\mid F_i)$ of the subsequent evolution
pathways, and their associated multidimensional consequences
${\boldsymbol{\cal C}}_{j}^{(i)}$ into a risk measure or risk class.
The precise form of ${\cal R}$ depends on the adopted risk-assessment
methodology and need not correspond to a simple analytic product~\cite{KaplanGarrick:1981,Aven:2016Risk}.

Equation~\eqref{eq:general_risk_function} makes explicit three logically
distinct ingredients:
\begin{equation}
\text{initiating-event likelihood}
\;+\;
\text{propagation likelihood}
\;+\;
\text{consequence severity}.
\label{eq:risk_three_components}
\end{equation}
Protection can act on any of these components.  Preventive design can
reduce the probability of the initiating event; interlocks can reduce the
probability that an abnormal state propagates along a dangerous branch;
and mitigation systems can reduce the final consequence even if the
preceding transitions have already occurred.

To evaluate the risk reduction provided by a particular
protection measure or combination of barriers, it is
useful to distinguish between a reference configuration
and the configuration in which the protection measures
under consideration have been implemented.

We denote by $P_0$ the probability assessment for the
reference configuration and by $P_{\rm res}$ the
corresponding assessment for the protected configuration.
The subscripts $0$ and ${\rm res}$ therefore identify
two specified protection configurations, rather than
different types of probability. The reference
configuration may already include other existing
safety barriers.

For the same initiating-event class $F_i$ and a common
reference interval, the initial risk is represented as
\begin{equation}
{\cal R}_{0}^{(i)}
=
{\cal R}
\left[
P_{0}(F_i),
\left\{
P_{0}\!\left(
\Gamma_{j}^{(i)}
\mid F_i
\right)
\right\}_{j=1}^{n_i},
\left\{
{\boldsymbol{\cal C}}_{0,j}^{(i)}
\right\}_{j=1}^{n_i}
\right],
\label{eq:initial_risk_function}
\end{equation}
where $P_0(F_i)$ is the probability of occurrence
of the initiating event in the reference configuration,
and $P_0(\Gamma_j^{(i)}\mid F_i)$ is the corresponding
conditional probability of its propagation along
branch $\Gamma_j^{(i)}$.

The residual risk after implementation of the
considered protection measures is
\begin{equation}
{\cal R}_{\rm res}^{(i)}
=
{\cal R}
\left[
P_{\rm res}(F_i),
\left\{
P_{\rm res}\!\left(
\Gamma_{j}^{(i)}
\mid F_i
\right)
\right\}_{j=1}^{n_i},
\left\{
{\boldsymbol{\cal C}}_{{\rm res},j}^{(i)}
\right\}_{j=1}^{n_i}
\right].
\label{eq:residual_risk}
\end{equation}
Here, $P_{\rm res}(F_i)$ denotes the initiating-event
probability in the protected configuration, while
$P_{\rm res}(\Gamma_j^{(i)}\mid F_i)$ describes
the corresponding conditional propagation probability.
The vectors
${\boldsymbol{\cal C}}_{0,j}^{(i)}$ and
${\boldsymbol{\cal C}}_{{\rm res},j}^{(i)}$
represent the consequences associated with the
corresponding branches in the two configurations.

The comparison assumes a consistent definition of
the initiating event, a common reference interval,
and comparable propagation outcomes. When the
introduction of protection measures changes the
possible failure pathways, the branch classification
must be adapted accordingly.

The intended effect of the protection measures is
to reduce the evaluated risk, schematically expressed
as
\begin{equation}
{\cal R}_{\rm res}^{(i)}
<
{\cal R}_{0}^{(i)}.
\label{eq:risk_reduction_condition}
\end{equation}
Here, the comparison is understood within a common
risk-evaluation methodology that provides an ordering
of the resulting risk measures or classes.

Risk reduction does not necessarily require that the occurrence
probability, propagation probability, and consequence severity all decrease
simultaneously.  A preventive barrier may primarily reduce
\begin{equation}
P_{\rm res}(F_i)
<
P_{0}(F_i),
\label{eq:initiating_event_reduction}
\end{equation}
whereas an interlock may act mainly through
\begin{equation}
P_{\rm res}
\left(
\Gamma_{j}^{(i)}
\mid
F_i
\right)
<
P_{0}
\left(
\Gamma_{j}^{(i)}
\mid
F_i
\right),
\label{eq:propagation_probability_reduction}
\end{equation}
and a mitigative barrier modifies the consequence vector associated with a
given propagation pathway,
\begin{equation}
{\boldsymbol{\cal C}}_{j,0}^{(i)}
\longrightarrow
{\boldsymbol{\cal C}}_{j,\rm mit}^{(i)},
\label{eq:consequence_reduction}
\end{equation}
with the intended effect of reducing the relevant consequence components
and, correspondingly, the evaluated residual risk. A complete protection
architecture normally combines several of these mechanisms.

The objective is not, in general, to establish \({\cal R}_{\rm res}^{(i)}=0\),
which would be unrealistic for a complex scientific installation.  The
purpose is to reduce the residual risk to the level required for the
corresponding activity while maintaining the scientific and operational
function of the facility.

This formulation also clarifies the relation between the safe operational
phase space and risk.  A state may belong to $\Omega_{\rm safe}$ not
because every conceivable failure has been eliminated, but because the
physically credible trajectories originating from that state are either
sufficiently improbable, intercepted by effective barriers, or constrained
to consequences compatible with the applicable protection requirements.

Schematically,
\begin{equation}
{\boldsymbol X}
\in
\Omega_{\rm safe}
\quad\Longrightarrow\quad
{\cal R}_{\rm res}^{(i)}
<
{\cal R}_{\rm acc}^{(i)}
\qquad
\text{for all relevant initiating-event classes }i,
\label{eq:safe_space_residual_risk}
\end{equation}
where ${\cal R}_{\rm acc}^{(i)}$ denotes the applicable acceptance
criterion or operational risk boundary associated with event class $i$.
The notation is conceptual rather than regulatory; different safety
domains employ different methods for defining acceptable or tolerable
conditions.

The trajectory-based formulation is especially useful for the LHC because
the same energetic reservoir can participate in very different hazard
chains.  Stored beam energy may be harmless while confined to the nominal
orbit, deliberately absorbed in the beam dump, intercepted in a controlled
manner by collimators, or damaging if deposited in a vulnerable component.
Likewise, cryogenic helium is an essential working fluid while contained
within its intended thermodynamic system but can become relevant to
pressure or oxygen-deficiency hazards after loss of containment.

The hazard therefore cannot be identified solely with the presence of an
energy reservoir, substance, or extreme physical condition.  Safety
analysis must specify the \emph{trajectory by which the corresponding
physical state can be converted into an adverse consequence}.  The
branching notation introduced above makes this point explicit:
\begin{equation}
F_i
\rightarrow
\Gamma_{j}^{(i)}
\rightarrow
{\boldsymbol {\cal C}}_{j}^{(i)}.
\label{eq:event_trajectory_consequence_summary}
\end{equation}
The next question is where along such trajectories protective functions
should act and whether they prevent the hazardous evolution or mitigate
its consequences.  This provides the basis for the distinction between
preventive and mitigative barriers considered in the following subsection.

\subsection{Preventive and mitigative barriers}
\label{subsec:preventive_mitigative}

Having introduced the general hazard-chain description in
Sec.~\ref{subsec:hazard_chains}, it is useful to distinguish barriers
according to the stage of the sequence at which they act.  In particular,
a separation can be made between \emph{preventive} barriers, which reduce
the likelihood that a hazardous energy transfer or exposure occurs, and
\emph{mitigative} barriers, which limit the consequences once such a
transfer or exposure has already begun.  The distinction between barriers
acting upstream to prevent an undesired
event and barriers acting downstream to limit its consequences is widely
used in barrier-based safety analysis
\cite{Sklet:2006SafetyBarriers,DeDianousFievez:2006ARAMIS}.

Consider the generic sequence introduced in
Eq.~\eqref{eq:generic_hazard_chain}.
A preventive barrier acts upstream of the undesired transfer or exposure.
Its function is to interrupt the sequence before the transition associated
with ${\cal E}_i$ occurs, or to reduce the probability that the system
reaches the corresponding hazardous state.  Schematically,
\begin{equation}
{\boldsymbol X}_{1}
\xrightarrow{\;{\cal P}_i\;}
{\boldsymbol X}_{2}
\xrightarrow{\;B_{\rm prev}\;}
{\boldsymbol X}_{\rm ctrl},
\label{eq:preventive_barrier}
\end{equation}
where ${\boldsymbol X}_{\rm ctrl}$ denotes a controlled state in which the
subsequent harmful energy-transfer or exposure process is avoided.

At the level of event occurrence, the action of a preventive barrier can
be represented schematically as a reduction of the effective event
frequency or likelihood,
\begin{equation}
f_i^{\rm prev}
<
f_i^{(0)},
\label{eq:preventive_frequency_reduction}
\end{equation}
where $f_i^{(0)}$ is the frequency associated with the unprotected or
less-protected sequence and $f_i^{\rm prev}$ the corresponding frequency
after introduction of the preventive barrier.  Equation~
\eqref{eq:preventive_frequency_reduction} is schematic: in a detailed
analysis the barrier may act on a conditional transition probability,
failure rate, or event frequency rather than on a single universal scalar
quantity.

A mitigative barrier acts differently. It does not
necessarily prevent the initiating event or the
hazardous transfer from occurring, but modifies
the subsequent evolution so that the resulting
consequences are reduced or otherwise altered.

For an initiating event $F_i$ and a specified
propagation branch $\Gamma_j^{(i)}$, its action
may be represented as a mapping in consequence
space,
\begin{equation}
\boldsymbol{\mathcal C}_{j,0}^{(i)}
\xrightarrow{\;B_{\rm mit}\;}
\boldsymbol{\mathcal C}_{j,\rm mit}^{(i)},
\label{eq:mitigated_consequence}
\end{equation}
where the subscripts $0$ and $\rm mit$ denote
the consequences evaluated in the reference
and mitigated configurations, respectively.
The comparison refers to the same initiating-event
class and to corresponding consequence scenarios.

Because the consequence vector is multidimensional,
its components cannot generally be compared through
a single scalar inequality. In particular, a
protection measure may reduce equipment damage
without producing an equivalent reduction in
operational disruption or other consequences.

When a scalar assessment is appropriate, one may
introduce a severity functional
\begin{equation}
\mathcal S_j^{(i)}
=
\mathcal S\left(
\boldsymbol{\mathcal C}_j^{(i)}
\right),
\label{eq:consequence_severity_functional}
\end{equation}
which maps the consequence vector associated with
branch $\Gamma_j^{(i)}$ onto a scalar severity
measure or an ordered severity category.  Such a representation is consistent
with established approaches to quantitative
and semi-quantitative risk assessment
\cite{KaplanGarrick:1981,Aven:2016Risk}.

The functional $\mathcal S$ is not a universal
physical observable. Its definition depends on
the consequence criteria and risk-assessment
methodology adopted for the hazard under
consideration. It may, for example, characterize
personnel injury, equipment damage, radiological
consequences, operational disruption, or an
explicitly defined combination of such effects.
Where several consequence dimensions are retained
independently, a scalar severity functional need
not be introduced.

If the same severity criteria are applied to
the reference and mitigated configurations,
the intended consequence reduction can be
expressed as
\begin{equation}
\mathcal S\left(
\boldsymbol{\mathcal C}_{j,\rm mit}^{(i)}
\right)
<
\mathcal S\left(
\boldsymbol{\mathcal C}_{j,0}^{(i)}
\right).
\label{eq:mitigation_severity_reduction}
\end{equation}
This relation states that the consequence
associated with the considered branch has
a lower assessed severity after mitigation.
It is meaningful only when the severity
measure or category admits the corresponding
ordering and remains consistently defined
between the two configurations.

The connection with risk assessment follows
from the generalized risk function introduced
in Eq.~\eqref{eq:general_risk_function}.
The severity functional provides one possible
evaluation of its consequence component,
whereas the initiating-event and propagation
probabilities determine how the corresponding
scenarios contribute to the assessed risk.
The effectiveness of preventive and mitigative
barriers must consequently be evaluated
through their influence on the complete
risk assessment, including the possible
changes in propagation pathways and
residual consequences
\cite{Sklet:2006SafetyBarriers,
DeDianousFievez:2006ARAMIS}.

For example, if $\mathcal S$ is a quantitatively
meaningful, real-valued severity measure and
the propagation branches form a mutually exclusive
and collectively exhaustive classification of
the outcomes following $F_i$, an expected-severity
risk measure may be constructed as
\begin{equation}
\mathcal R_{\mathcal S}^{(i)}
=
P(F_i)
\sum_{j=1}^{n_i}
P\left(
\Gamma_j^{(i)}\mid F_i
\right)
\mathcal S\left(
\boldsymbol{\mathcal C}_j^{(i)}
\right).
\label{eq:risk_expected_severity}
\end{equation}
Here, $P(F_i)$ refers to a specified reference
interval, and the propagation probabilities
are conditional on the occurrence of $F_i$.
The resulting quantity represents the expected
severity contribution of the considered
initiating-event class over that interval,
according to the chosen severity measure.

Equation~\eqref{eq:risk_expected_severity}
is an illustrative quantitative realization
of the more general mapping
$\mathcal R$; it is not proposed as a
universal definition of risk.
In particular, it is not applicable
as a numerical expectation when
$\mathcal S$ represents only ordinal
severity categories. In that case,
a risk matrix or another explicitly
defined mapping may combine occurrence
and severity categories into risk classes,
as discussed in
Sec.~\ref{subsec:hazard_chains}.

Furthermore, a scalar expected-severity
measure need not adequately represent
all safety-relevant information.
Consequences involving personnel,
equipment, the environment, and scientific
operation may be subject to different
assessment criteria or acceptance limits.
Low-probability events with particularly
severe consequences may also require
separate consideration rather than
aggregation into a single expected value.

The distinction between prevention and
mitigation can now be expressed directly
through Eq.~\eqref{eq:general_risk_function}.
A preventive measure may reduce the probability
of the initiating event or the conditional
probability of propagation along a hazardous
branch. A mitigative measure primarily
changes the consequence vector associated
with one or more branches and, where a
suitable severity functional is employed,
the corresponding assessed severity.

Consequently, the initial and residual
risk evaluations introduced in
Eqs.~\eqref{eq:initial_risk_function}
and~\eqref{eq:residual_risk}
may differ through the initiating-event
probabilities, propagation probabilities,
consequence vectors, or several of
these quantities simultaneously.

A reduction in the severity of a particular
branch does not necessarily imply the same
reduction in the total assessed risk.
The latter also depends on the probabilities
and consequences of the other propagation
branches and on the risk-evaluation methodology.
In practice, a protection measure may alter
both the possible propagation pathways
and their associated consequences.
Accordingly, the effectiveness of mitigation
must ultimately be evaluated through the
corresponding residual-risk assessment,
rather than through the severity reduction
of an isolated scenario alone.

For beam-induced hazards, a beam abort initiated in response to an
incipient equipment failure is predominantly preventive with respect to a
subsequent damaging beam loss.  Collimators and absorbers, by contrast,
control where unavoidable or rapidly developing losses are deposited and
therefore also have an important mitigative function.  In cryogenic
systems, containment and early isolation act mainly on the probability or
extent of a release, whereas pressure-relief systems limit the
consequences of a release that has already entered a protected enclosure.
For radiation hazards, shielding mitigates the exposure produced by an
existing radiation field, while access interlocks prevent personnel from
being present under conditions in which such a field can be generated.

The same protection element may therefore play different roles depending
on the hazard sequence being considered.  A collimator, for example, can
be viewed as a preventive barrier with respect to losses in a vulnerable
superconducting magnet, while simultaneously acting as a mitigative barrier
with respect to the unavoidable interception of beam halo.  The
classification is thus functional rather than intrinsic to the hardware.

This distinction also clarifies the role of defence in depth (DiD).  A robust
protection architecture generally combines barriers acting at different
positions along the same hazard chain:
\begin{equation}
B_{\rm prev}^{(1)}
\oplus
B_{\rm prev}^{(2)}
\oplus
B_{\rm mit}^{(1)}
\oplus
B_{\rm mit}^{(2)}
\oplus\cdots ,
\label{eq:defence_in_depth}
\end{equation}
where the symbol $\oplus$ is used schematically to indicate complementary
protection functions rather than a mathematical addition or an assumption
of statistical independence.

The usefulness of such layering is that it avoids relying on a single
physical mechanism.  If an initiating failure cannot be eliminated
completely, a downstream preventive barrier may still interrupt its
propagation.  If that barrier also fails or is bypassed by a sufficiently
rapid event, a mitigative barrier may still reduce the resulting
consequence.  In this sense, prevention and mitigation are not competing
strategies but complementary ways of modifying different parts of the same
failure trajectory.

The distinction is especially important in complex accelerator systems,
where protection must remain effective over failure modes with very
different characteristic timescales.  Fast failures may leave insufficient
time for an active preventive response and therefore require passive
mitigation, while slower failures may permit detection, diagnosis, and
controlled transition to a protected state.  This dynamical aspect is
considered explicitly in Sec.~\ref{subsec:protection_time}.

\subsection{Barrier effectiveness, independence, and common-cause failure}
\label{subsec:barrier_independence}

The effectiveness of a layered protection architecture depends not only on
the number of barriers present, but also on the extent to which their
protection functions are genuinely independent.  This distinction is
particularly important in the DiD picture introduced in
Sec.~\ref{subsec:preventive_mitigative}.  Several barriers may act at
different stages of the same hazard chain, but their combined effectiveness
cannot in general be inferred simply by counting them.
Reliability, independence, response time, and robustness are distinct
properties of a protection barrier, while common-cause and dependent
failures can substantially reduce the protection gained from nominal
redundancy
\cite{Sklet:2006SafetyBarriers,Filippini:2005Reliability}.

Consider a particular initiating event $F_i$ and a specified undesirable
consequence or consequence class.  Let $P_i^{(0)}$ denote the probability
that the corresponding hazardous sequence would progress to that
consequence in the absence of the protection barriers under consideration.
For barrier $B_k$, define
\begin{equation}
q_k
=
P
\left(
B_k\ {\rm fails\ on\ demand}
\,\big|\,
\text{barrier }B_k\text{ is required}
\right).
\label{eq:barrier_failure_probability}
\end{equation}
If $N$ barriers must all fail for the specified sequence to reach the
consequence, and if their failures are statistically independent under the
conditions being considered, the residual probability may be written
schematically as
\begin{equation}
P_i^{\rm res}
\simeq
P_i^{(0)}
\prod_{k=1}^{N} q_k .
\label{eq:independent_barriers}
\end{equation}

Equation~\eqref{eq:independent_barriers} is deliberately restricted to a
specific failure sequence and consequence.  It should not be interpreted
as a complete expression for the multidimensional risk
${\cal R}_i(f_i,{\boldsymbol{\cal C}}_i)$ introduced in
Sec.~\ref{subsec:hazard_chains}.  In particular, different barriers may
modify different components of the consequence vector, may act on
different branches of the failure sequence, or may influence both the
event frequency and the final consequences.

The independence assumption in
Eq.~\eqref{eq:independent_barriers} is strong.  Two nominally redundant
barriers may depend on a common sensor, power supply, communication
network, software process, environmental condition, mechanical support, or
maintenance procedure.  A single disturbance can then impair both barriers
simultaneously.  Their combined failure probability cannot be represented
by the simple product $q_1q_2$.

A schematic representation of a common-cause contribution is
\begin{equation}
P_{\rm fail}^{(1,2)}
\simeq
q_{\rm cc}
+
\left(1-q_{\rm cc}\right)
q_1 q_2 ,
\label{eq:common_cause}
\end{equation}
where $q_{\rm cc}$ denotes the probability that
a common-cause mechanism defeats both barriers
when protection is required, while $q_1$ and $q_2$
represent their remaining independent failure-on-demand
probabilities, conditional on the absence of such
a common-cause failure. The quantity $P_{\rm fail}^{(1,2)}$ is therefore
the combined probability that both barriers fail
to provide the required protection. It is distinct
from the residual probability $P_i^{\rm res}$
of the specified hazardous consequence introduced
in Eq.~\eqref{eq:independent_barriers}.
For the simplified failure sequence considered here,
the latter may be estimated as
$P_i^{\rm res}\simeq
P_i^{(0)}P_{\rm fail}^{(1,2)}$. Equation~\eqref{eq:common_cause} is illustrative
rather than a general reliability formula. It
demonstrates that, when the independent failure
probabilities are small, even a comparatively
small common-cause contribution can dominate
the probability of failure of the combined
protection arrangement.

The relevant concept is therefore not merely redundancy but
\emph{functional independence}.  Redundancy means that more than one
element can provide a required function.  Independence requires, in
addition, that failure of one element does not substantially increase the
probability that another will fail when required.  These concepts may be
represented schematically as
\begin{equation}
\text{redundancy}
\not\Longrightarrow
\text{independence}.
\label{eq:redundancy_not_independence}
\end{equation}

A related concept is \emph{diversity}.  Two protection channels can be made
less vulnerable to common-cause failure if they rely on different physical
observables, technologies, signal paths, actuation principles, or power
sources.  Diversity can therefore support independence, although it does
not by itself guarantee it:
\begin{equation}
\text{diversity}
\;\Longrightarrow\;
\text{reduced common-cause susceptibility in suitable cases}.
\label{eq:diversity_common_cause}
\end{equation}

This distinction is particularly relevant in accelerator protection.
Several physically different observables may indicate the development of
the same hazardous sequence.  A magnet powering fault may be detected
through electrical quantities before significant beam motion occurs,
while a subsequent orbit excursion or beam loss may be detected by other
systems.  The corresponding protection chain can therefore contain
barriers based on different physical information,
\begin{equation}
\text{powering state}
\rightarrow
\text{beam state}
\rightarrow
\text{loss state},
\label{eq:diverse_observables_chain}
\end{equation}
with protection possible at more than one stage.

The same principle appears in cryogenic and detector protection.  Pressure
relief, oxygen monitoring, ventilation, and access control do not measure
or control the same physical quantity, yet they can act sequentially on
the same accident progression.  Similarly, a detector cooling failure may
be detected through flow, pressure, or temperature information, while a
dedicated safety system may subsequently remove electrical power.  Such
barriers are complementary because they act on different physical links of
the hazard chain.

Barrier independence should also be distinguished from barrier
\emph{effectiveness}.  A barrier may be fully independent of another and
still provide insufficient risk reduction if its probability of failure on
demand is too large or if it acts too late.  Conversely, a highly reliable
barrier may not provide the expected additional protection if it shares a
dominant common-cause failure with another layer.  For a specified protection function $B_k$, it is useful
to characterize its performance schematically through
a collection of attributes,
\begin{equation}
{\cal B}_k
=
\left\{
q_k,
\tau_k,
{\cal I}_k,
{\cal D}_k,
{\cal T}_k
\right\},
\label{eq:barrier_attributes}
\end{equation}
where $q_k$ denotes the probability of failure on
demand and $\tau_k$ the time required to perform
the specified protective action. The remaining
attributes describe properties that influence
the effectiveness of the barrier within a
layered protection architecture: its independence
from other protection functions, the diversity
of the principles on which those functions rely,
and the possibility of verifying their continued
availability. These concepts are established in
the general literature on safety barriers and
reliability engineering
\cite{Sklet:2006SafetyBarriers,
DeDianousFievez:2006ARAMIS,
Yuan:2022SafetyBarriers}.

The attribute ${\cal I}_k$ represents the
\emph{independence} of barrier $B_k$ from the
other barriers credited against the same
hazardous sequence. In practice, independence
is assessed by examining whether failures
of one protection function can impair
another through shared components,
services, or operating conditions.
Relevant dependencies include common
sensors, power supplies, communication
paths, logic systems, actuators,
environmental conditions, software,
maintenance activities, and
configuration errors. Independence
is therefore not an intrinsic property
of an isolated component; it must
be assessed relative to the other
barriers and the failure scenario
under consideration.

The distinction is important when
evaluating redundancy. Two channels
may separately detect a hazardous
condition but remain dependent on
the same downstream protection
equipment. Conversely, physically
separate channels may remain
vulnerable to common disturbances
or identical systematic errors.
Evidence for independence may
therefore include analysis of
functional dependencies,
electrical and mechanical
segregation, shared services,
common-cause failure modes,
and the operational and
maintenance arrangements.
Such evidence determines
whether the independent failure
assumption used in
Eq.~\eqref{eq:independent_barriers}
can reasonably be applied
\cite{Filippini:2005Reliability,
Rausand:2014SafetyCritical}.

The attribute ${\cal D}_k$ represents
\emph{functional or technological
diversity} relative to other barriers.
Diversity may arise from the use
of different physical observables,
measurement principles, processing
technologies, actuation mechanisms,
or protection strategies.
For example, a developing
accelerator fault may be
recognized through electrical
measurements, beam-orbit
information, or beam-loss
signals. These observables
provide different information
about the physical evolution
of the same hazardous sequence,
allowing protective actions
at different stages.

However, diversity should not
be confused with independence.
Protection functions based
on different sensors or
physical principles may still
depend on a common interlock
path, final actuator, power
source, or configuration
procedure. Likewise, two
technologically different
barriers may be exposed
to the same environmental
or operational disturbance.
Diversity can reduce
susceptibility to particular
common-cause failures, but
its contribution must be
established for the relevant
failure mechanisms rather
than assumed from the
difference in technologies
alone.

The attribute ${\cal T}_k$
characterizes the
\emph{testability and
verifiability} of the
protection function.
It concerns the extent
to which failures
that could prevent
the required protective
action can be detected
through diagnostics,
inspection, functional
testing, or dedicated
proof tests.
The assessment must
consider which elements
of the protection
chain can be exercised,
which failure modes
can be revealed,
and whether the
complete response,
including the final
protective action,
can be verified
under representative
conditions.

In practice, a test
may verify only
the operation of
a sensor or an
interlock logic
channel, whereas
an end-to-end
test may also
exercise the
communication path
and final actuator.
These tests provide
different levels
of evidence.
A successful
partial test
does not establish
that the complete
protection function
will operate correctly
on demand.
Furthermore,
some failures
may remain latent
between successive
tests, while others
can be detected
continuously by
internal diagnostics.
The test interval,
failure-mode coverage,
and conditions under
which testing is
performed therefore
influence the
confidence that
the protection
function remains
available
\cite{Sklet:2006SafetyBarriers,
Rausand:2014SafetyCritical}.

The attributes
${\cal I}_k$,
${\cal D}_k$,
and ${\cal T}_k$
are introduced
here as
\emph{qualitative
or methodology-dependent
descriptors},
not as universally
defined numerical
quantities.
Their evaluation
requires a specified
protection function,
hazardous sequence,
technical configuration,
and assessment
procedure.
Where quantitative
models are available,
the results may
be incorporated
into estimates
of common-cause
failure,
barrier unavailability,
or failure-on-demand
probabilities.
Otherwise, the
attributes provide
structured evidence
for assessing
the robustness
of the layered
protection architecture.

An illustrative
example is
provided by
the LHC machine-protection
system. Powering
surveillance,
beam-loss monitoring,
interlock logic,
beam extraction,
and passive
absorbers perform
different functions
along possible
beam-loss sequences.
Their complementary
physical roles
provide diversity,
but their
independence
must be assessed
at the level
of complete
protection paths.
For example,
different
monitoring systems
may ultimately
rely on the
same beam-dumping
equipment.
Testing must
therefore establish
not only the
correct operation
of individual
monitoring channels
but also the
availability
of the relevant
interlock and
protective
response paths
\cite{Puccio:2010BIS,
Filippini:2005Reliability}.

Thus, the collection
in
Eq.~\eqref{eq:barrier_attributes}
should not be
interpreted as
a vector of
independent
numerical
performance
parameters.
It is a
schematic
description
of complementary
properties that
must be assessed
for each credited
protection function.
Their relevance
lies in determining
whether the
assumptions used
to estimate
barrier effectiveness
and residual risk
are justified
for the corresponding
hazardous sequence.

The LHC protection architecture reflects these requirements through
distributed monitoring, dedicated interlock systems, redundant signal
paths, autonomous protection functions, and testing of critical systems
\cite{Bordry:2001MPS,Puccio:2010BIS,Filippini:2005Reliability}.
Reliability studies of the machine-protection system explicitly distinguish
between failures that could prevent a required protection action and
failures that generate unnecessary protection requests.

It is therefore useful to distinguish
\begin{equation}
\begin{aligned}
&\text{dangerous failure:}
&&
\text{a required protection function is unavailable},
\\
&\text{spurious action:}
&&
\text{a protection function acts when its intervention is not required}.
\end{aligned}
\label{eq:dangerous_spurious}
\end{equation}
The first can increase the probability that a hazard sequence reaches an
unacceptable consequence; the second primarily affects operational
availability and scientific efficiency.

These two failure classes need not be assigned equal significance.
Protection design is generally driven first by the requirement that the
probability of a dangerous failure remain sufficiently small for the
hazard being controlled.  At the same time, excessive spurious actions can
make the accelerator or detector operationally inefficient and may obscure
genuine faults among frequent unnecessary interventions.  The design
objective is therefore to achieve sufficiently high protection integrity
while maintaining acceptable availability
\cite{Filippini:2005Reliability,Wenninger:2016MPO}.

In the framework developed in the previous subsections, independence,
diversity, and redundancy should therefore be regarded as properties of the
\emph{barrier architecture}, not as direct measures of risk themselves.
Their importance lies in determining how effectively several protection
functions modify the propagation of a specific hazard sequence and how
robustly the resulting residual risk remains controlled when individual
components fail.

The temporal effectiveness of those barriers introduces a further
constraint: even an independent and reliable protection function is useful
only if it acts before the corresponding hazardous evolution reaches an
irreversible consequence.  This dynamical requirement is considered in
Sec.~\ref{subsec:protection_time}.

\subsection{Time as a protection variable}
\label{subsec:protection_time}

The effectiveness of a protection barrier depends not only on its
reliability and independence, but also on whether it can act within the
time available before the corresponding hazard sequence reaches an
unacceptable state.  Protection is therefore intrinsically dynamical.
This temporal aspect complements the barrier properties discussed in
Sec.~\ref{subsec:barrier_independence}.

Consider a particular hazard sequence initiated at time $t_0$.  Let
$t_{\rm crit}$ denote the earliest time at which the evolution reaches a
specified critical state beyond which the consequence under consideration
can no longer be prevented by the protection function being analysed.  The
available protection time is then
\begin{equation}
\tau_{\rm avail}
=
t_{\rm crit}-t_0 .
\label{eq:available_protection_time}
\end{equation}
The meaning of $t_{\rm crit}$ is necessarily hazard dependent.  It may
correspond, for example, to the onset of irreversible material damage, a
magnet quench, an unacceptable pressure, a hazardous oxygen concentration,
or another explicitly defined limit relevant to the failure sequence.

For an active protection chain, the total response time may be decomposed
schematically as
\begin{equation}
\tau_{\rm prot}
=
\tau_{\rm det}
+
\tau_{\rm logic}
+
\tau_{\rm comm}
+
\tau_{\rm act}
+
\tau_{\rm phys},
\label{eq:protection_latency}
\end{equation}
where $\tau_{\rm det}$ is the time required to detect the abnormal
condition, $\tau_{\rm logic}$ the time required to evaluate the protection
condition, $\tau_{\rm comm}$ the signal-transmission time,
$\tau_{\rm act}$ the actuator response time, and $\tau_{\rm phys}$ the
additional physical time required for the system to reach the intended
protected state.

A necessary timing condition for that active protection function is
therefore
\begin{equation}
\tau_{\rm prot}
<
\tau_{\rm avail}.
\label{eq:protection_time_condition}
\end{equation}
In practice, operation close to equality would provide little robustness
against modelling uncertainty, signal jitter, degraded equipment
performance, or variation of the underlying physical process.  It is
therefore useful to introduce a timing margin
\begin{equation}
\Delta\tau_{\rm prot}
=
\tau_{\rm avail}
-
\tau_{\rm prot},
\label{eq:protection_time_margin}
\end{equation}
for which a robust protection design requires
\begin{equation}
\Delta\tau_{\rm prot}>0
\label{eq:positive_protection_margin}
\end{equation}
with an appropriate additional engineering margin.

A dimensionless measure of the protection response
relative to the available time is
\begin{equation}
\kappa_{\rm prot}
=
\frac{\tau_{\rm prot}}
     {\tau_{\rm avail}},
\label{eq:protection_time_ratio}
\end{equation}
where $\kappa_{\rm prot}$ denotes the protection
timing ratio. For $\tau_{\rm avail}>0$, the
necessary timing condition becomes
\begin{equation}
\kappa_{\rm prot}<1.
\label{eq:protection_ratio_condition}
\end{equation}
Values approaching unity indicate a decreasing
timing margin and increasing sensitivity to
uncertainties in both the hazard evolution
and the protection response.

The quantities in Eqs.~\eqref{eq:available_protection_time}--%
\eqref{eq:protection_ratio_condition} should be associated with a
\emph{specified hazard path}.  A single initiating event may have several
possible consequences characterized by different critical times,
\begin{equation}
\tau_{\rm avail}^{(1)},
\;
\tau_{\rm avail}^{(2)},
\ldots,
\tau_{\rm avail}^{(n)}.
\label{eq:multiple_critical_times}
\end{equation}
For example, beam loss may reach a superconducting-magnet quench threshold
before it reaches a material-damage threshold.  Similarly, a cryogenic
release may challenge a pressure limit on a timescale different from that
associated with the development of an oxygen-deficient atmosphere.
Protection timing must therefore be evaluated against the relevant
consequence or transition rather than against a universal ``damage time.''

This observation is also consistent with the multidimensional consequence
description introduced in Sec.~\ref{subsec:hazard_chains}.  Different
components of ${\boldsymbol{\cal C}}_i$ may become relevant on different
timescales, and a barrier that is sufficiently fast to prevent one
consequence need not necessarily prevent another.  Timing is thus a
property of the combination
\begin{equation}
\left[
\text{hazard sequence},\
\text{barrier},\
\text{specified consequence}
\right],
\label{eq:timing_triplet}
\end{equation}
rather than of the barrier alone.

The distinction between preventive and mitigative barriers introduced in
Sec.~\ref{subsec:preventive_mitigative} is also important here.  For a
preventive barrier, the relevant timing condition is that the protection
act before the undesired energy-transfer or exposure mechanism becomes
unavoidable.  For a mitigative barrier, the hazardous transfer may already
have begun, and the relevant condition is instead that the barrier act
before the consequence exceeds the level for which the mitigation was
designed.  Schematically,
\begin{equation}
\tau_{\rm prev}
<
\tau_{\rm transfer},
\qquad
\tau_{\rm mit}
<
\tau_{\rm consequence},
\label{eq:preventive_mitigative_times}
\end{equation}
where the two characteristic times refer to different positions along the
same hazard chain.

At the LHC, relevant failure times span many orders of magnitude.  Some
beam-related failures can drive the machine toward an unacceptable state
within one or a few turns and therefore require extremely fast
hardware-based detection and beam extraction.  Other failures evolve over
milliseconds, seconds, or substantially longer periods and allow more
complex diagnostic or control responses
\cite{Schmidt:2016MPS,Wenninger:2016MPO}.  Cryogenic pressure evolution,
thermal excursions, detector cooling failures, smoke propagation, and
oxygen depletion similarly possess characteristic timescales determined
by their respective physical processes.

This hierarchy implies that the architecture of a protection system must
follow the dynamics of the hazard rather than a uniform technological
prescription.  A supervisory software process cannot replace a dedicated
hardware interlock when
\begin{equation}
\tau_{\rm software}
>
\tau_{\rm avail},
\label{eq:software_too_slow}
\end{equation}
whereas a slowly evolving condition does not necessarily require a
microsecond-scale hardware response.  The required response technology is
therefore determined in part by the temporal structure of the underlying
physics.

There is, moreover, a class of failures for which the available time is
too short for any active detection--decision--actuation chain to complete.
If
\begin{equation}
\tau_{\rm avail}
\lesssim
\tau_{\rm prot}^{\rm min},
\label{eq:active_protection_impossible}
\end{equation}
where $\tau_{\rm prot}^{\rm min}$ is the shortest realistically achievable
active response time, protection must rely on passive or intrinsically
protective features.  At the LHC this principle is particularly important
for very fast injection or extraction failures, for which dedicated
absorbers and collimators must protect vulnerable equipment even when
active intervention cannot occur sufficiently rapidly
\cite{Schmidt:2016MPS,Bartmann:2010InjectionDump}.

Passive protection should not, however, be described using the same
latency decomposition as Eq.~\eqref{eq:protection_latency}.  A passive
absorber, mechanical containment boundary, radiation shield, or pressure
relief path may act through the immediate physical response of its design
rather than through a sequence of detection and logical decision.  Its
effectiveness is therefore determined by quantities such as geometry,
material properties, response capacity, and the timescale of the physical
interaction.

The temporal structure of protection can consequently be summarized as
\begin{equation}
\begin{array}{ll}
\tau_{\rm avail}>\tau_{\rm prot}
&
\Rightarrow
\text{active protection can in principle interrupt the sequence},
\\[1ex]
\tau_{\rm avail}\lesssim\tau_{\rm prot}^{\rm min}
&
\Rightarrow
\text{passive or intrinsic protection is required}.
\end{array}
\label{eq:active_passive_timing}
\end{equation}

For layered protection, several barriers may also act at different times
along the same sequence.  If $B_k$ becomes effective at time
$t_k^{\rm act}$, one may schematically write
\begin{equation}
t_1^{\rm act}
<
t_2^{\rm act}
<
\cdots
<
t_{\rm crit},
\label{eq:sequential_barrier_times}
\end{equation}
where upstream barriers attempt to prevent progression of the event and
downstream barriers remain available to mitigate its consequences if the
earlier layers do not succeed.  This temporal ordering is another aspect
of DiD.

The response time $\tau_k$ introduced among the barrier attributes in
Eq.~\eqref{eq:barrier_attributes} should therefore be understood in this
context.  Reliability, independence, and diversity determine whether a
barrier is likely to remain available when demanded; timing determines
whether its successful action is still physically capable of altering the
hazard trajectory.  An independent and highly reliable barrier that acts
after $t_{\rm crit}$ cannot provide the intended protection.

Protection timing is consequently another manifestation of the
trajectory-based description developed in
Sec.~\ref{subsec:underground_environment_state}.  Safety is not determined
only by whether the instantaneous facility state belongs to
$\Omega_{\rm safe}$, but also by whether abnormal trajectories can be
detected and redirected before they cross the relevant boundary of that
region.  The subsequent subsection considers how such physical protection
conditions are translated into logical permits and interlocks that
constrain accelerator operation.

\subsection{Interlocks, permits, and logical representation of operational conditions}
\label{subsec:interlocks_permits}

The protection concepts developed above must ultimately be translated into
conditions that determine whether a particular accelerator or detector
state may be entered or maintained.  In many cases this translation is
implemented through interlocks and permits.  Their role is to convert
information about physical and technical subsystem states into logical
conditions that constrain the subsequent evolution of the facility.

Consider a set of protection-relevant conditions indexed
by $j$. For a specified operational mode $m$, one may
define a binary permit indicator
\begin{equation}
\chi_j^{\rm permit}(\boldsymbol X,m)
=
\begin{cases}
1, & \text{condition $j$ permits the specified operation},\\
0, & \text{otherwise}.
\end{cases}
\label{eq:permit_variable}
\end{equation}
The variable $\chi_j^{\rm permit}$ is a binary
indicator specifying whether condition $j$
permits the corresponding operation.\footnote{
Throughout this work, the symbol $\chi$, with
appropriate indices, is used to denote a binary
logical, validity, or admissibility indicator,
$\chi\in\{0,1\}$. Its precise interpretation
depends on the physical or operational condition
being represented.}
The explicit dependence on $\boldsymbol X$
emphasizes that the permit is derived from
the physical and technical state of the
facility, while the dependence on $m$
reflects its operational-mode dependence.
The dependence on $\boldsymbol X$ emphasizes that
the permit is derived from the physical and technical
state of the facility, while the dependence on $m$
reflects the mode-dependent applicability of
protection conditions.

For a set of conditions that must all be satisfied,
the corresponding global permit indicator is
\begin{equation}
\chi_{\rm global}^{\rm permit}
\left(\boldsymbol X,m\right)
=
\bigwedge_{j=1}^{N_m}
\chi_j^{\rm permit}
\left(\boldsymbol X,m\right),
\label{eq:global_permit}
\end{equation}
where $\wedge$ denotes logical AND and $N_m$
is the number of permit conditions required
for operational mode $m$.

Operation requiring this permit is allowed only when
\begin{equation}
\chi_{\rm global}^{\rm permit}
\left(\boldsymbol X,m\right)=1.
\label{eq:permit_condition}
\end{equation}

Equation~\eqref{eq:global_permit} is deliberately schematic.  Real
protection architectures may include redundant channels, voting logic,
mode-dependent masks, hierarchical permits, diagnostic states, and
conditions that are valid only during particular phases of operation.
Nevertheless, the expression captures the essential idea: many
heterogeneous physical measurements are reduced to a logical statement
about whether a defined operation may continue.

The LHC Beam Interlock System provides a particularly clear realization
of this principle.  Accelerator subsystems generate beam-permit
information from quantities such as beam losses, powering conditions,
magnet states, radio-frequency systems, beam instrumentation, and the
readiness of the beam dumping system.  The Beam Interlock System combines
these distributed inputs and, when a required permit is lost, requests
beam extraction
\cite{Bordry:2001MPS,Puccio:2010BIS,Schmidt:2016MPS,Wenninger:2016MPO}.

The corresponding information flow may be represented as
\begin{equation}
\left\{
{\boldsymbol X}_{1},
{\boldsymbol X}_{2},
\ldots,
{\boldsymbol X}_{N}
\right\}
\longrightarrow
\left\{
\chi_1,\chi_2,\ldots,\chi_N
\right\}
\longrightarrow
\chi_{\rm global}^{\rm permit},
\label{eq:state_to_permit}
\end{equation}
where the ${\boldsymbol X}_j$ may correspond to physically different
subsystems.  The permit is therefore not itself a physical observable; it
is a logical representation derived from the measured and validated state
of the installation.

It is useful to distinguish this concept from an alarm.  An alarm
communicates that a condition requires attention, whereas an interlock
changes the set of operational transitions available to the system.  In
schematic form,
\begin{equation}
\text{alarm}
\;:\;
{\boldsymbol X}
\rightarrow
\text{information},
\label{eq:alarm_logic}
\end{equation}
while
\begin{equation}
\text{interlock}
\;:\;
{\boldsymbol X}
\rightarrow
\text{constraint on subsequent evolution}.
\label{eq:interlock_logic}
\end{equation}
An alarm may lead to human intervention, while an interlock can
automatically inhibit a transition or initiate a predefined protective
action.

The permit concept can be connected directly to the state-space
description developed in the preceding subsections.  For a specified
operational mode $m$, define
\begin{equation}
\Omega_{\rm permit}^{(m)}
=
\left\{
{\boldsymbol X}\in\Omega_{\rm phys}
\;\bigg|\;
\chi_{\rm global}^{\rm permit}({\boldsymbol X},m)=1
\right\}.
\label{eq:permitted_region}
\end{equation}
This is the subset of physically realizable states for which the logical
conditions required for operation in mode $m$ are satisfied.

Importantly, $\Omega_{\rm permit}^{(m)}$ should not be identified with the
complete safe operational phase space.  A state for which
$\chi_{\rm global}^{\rm permit}=0$ may be entirely safe---for example, a machine with the
beam already extracted and a protection system intentionally withholding
the beam permit.  Conversely, satisfaction of a particular permit does not
by itself establish that every safety condition relevant to the facility
has been fulfilled.

For the specified operation, the admissible region is therefore more
appropriately written as
\begin{equation}
\Omega_{\rm op}^{(m)}
=
\Omega_{\rm safe}
\cap
\Omega_{\rm permit}^{(m)}
\cap
\Omega_{\rm auth}^{(m)},
\label{eq:operational_region_with_permit}
\end{equation}
where $\Omega_{\rm auth}^{(m)}$ schematically represents any additional
authorization, access, or procedural conditions required for that
operational mode.  Equation~\eqref{eq:operational_region_with_permit}
makes explicit that an interlock permit is one component of the admissible
operating state, rather than a complete definition of safety.

The logical state also constrains transitions.  Consider a machine state
${\boldsymbol X}_a$ from which a transition to
${\boldsymbol X}_b$ is requested.  The transition may be written
schematically as
\begin{equation}
{\boldsymbol X}_a
\xrightarrow{
\;\chi_{\rm global}^{\rm permit}=1\;
}
{\boldsymbol X}_b,
\label{eq:permit_conditioned_transition}
\end{equation}
where the permit condition must remain valid according to the requirements
of the corresponding operational mode.

If a required permit is withdrawn while the system is operating, the
interlock initiates or enables a transition toward a protected state,
\begin{equation}
\chi_{\rm global}^{\rm permit}:1\rightarrow0
\quad\Longrightarrow\quad
{\boldsymbol X}
\xrightarrow{\;\mathcal{P}_{\rm int}\;}
{\boldsymbol X}_{\rm protected},
\label{eq:interlock_state_transition}
\end{equation}
where $\mathcal{P}_{\rm int}$ denotes the associated interlock-driven
protection action.  Depending on the subsystem, this may correspond to beam
extraction, inhibition of injection, power removal, magnet discharge,
closure of a valve, detector shutdown, or another predefined response.

The effectiveness of the transition in
Eq.~\eqref{eq:interlock_state_transition} remains subject to the timing
condition discussed in Sec.~\ref{subsec:protection_time}.  For an active
interlock to prevent the specified consequence, its complete
detection--decision--actuation sequence must satisfy Eq.~\eqref{eq:protection_time_condition}.
An interlock that correctly identifies an abnormal condition but acts only
after the relevant critical state has been reached does not provide the
intended protection.

The state dependence of permits is also important
during commissioning. A protection condition that
may be maskable or otherwise tolerated with a
low-intensity setup beam can become mandatory
when the beam parameters enter a regime in which
uncontrolled losses may cause equipment damage.
Schematically,
\begin{equation}
\chi_j^{\rm permit}
=
\chi_j^{\rm permit}
\left(
\boldsymbol X,m
\right),
\label{eq:state_dependent_permit}
\end{equation}
where the dependence on $\boldsymbol X$ includes
the relevant beam parameters, in particular
the beam energy and circulating particle
population, as well as other physical and
technical conditions affecting the required
protection. The operational mode $m$ identifies
the corresponding operating or commissioning
conditions. Thus, the permit logic may depend
on the beam regime without introducing an
additional independent state variable.

This does not imply that protection requirements become arbitrary.  On the
contrary, mode dependence is acceptable only when the corresponding change
in hazard has been explicitly established.  A protection function that may
be bypassed under a non-damaging commissioning condition must become
effective before the machine enters the region in which the associated
failure can produce unacceptable consequences.

The permit structure therefore provides a logical realization of the
trajectory constraints introduced in
Eq.~\eqref{eq:safe_operational_trajectory}.  The facility is not merely
required to occupy an admissible state; transitions into more demanding
states are conditional on the continued validity of the relevant
protection functions.  Schematically,
\begin{equation}
\Gamma_{\rm op}
=
\left\{
{\boldsymbol X}(t)
\;\bigg|\;
{\boldsymbol X}(t)\in\Omega_{\rm safe},
\;
\chi_{\rm global}^{\rm permit}
\left(
{\boldsymbol X}(t),m(t)
\right)
=1
\ \text{whenever required}
\right\}.
\label{eq:permit_constrained_trajectory}
\end{equation}

Interlocks and permits therefore provide the interface between continuous
physical variables and discrete operational decisions.  Measurements of
beam loss, current, voltage, temperature, pressure, position, or other
quantities are converted into logical conditions that determine whether a
physical trajectory may continue.  In this sense, the interlock
architecture is a practical implementation of the safe-operational-phase-
space concept: it does not define all aspects of safety, but it prevents
selected trajectories from entering regions for which specified protection
conditions are no longer satisfied.

For this logic to remain effective, the protection architecture must also
respond safely to failures of its own sensors, communication paths,
controllers, or actuators.  This requirement leads naturally to the
fail-safe and autonomous-protection principles considered in
Sec.~\ref{subsec:failsafe_protection}.

\subsection{Fail-safe behaviour and autonomous protection}
\label{subsec:failsafe_protection}

The preceding discussion has treated protection barriers primarily in terms
of their response to failures occurring in the accelerator, detector, or
supporting infrastructure.  A complete protection architecture must also
consider failures occurring within the protection system itself.  Sensors,
signal-transmission paths, logic units, power supplies, communication
links, and actuators are physical components and may themselves fail,
degrade, or become unavailable.

Let $F_{B_k}$ denote an internal failure affecting
protection barrier $B_k$. For the specified class
of credible failures, the required protection
function should either remain effective or
initiate an appropriate fail-safe response.
Schematically,
\begin{equation}
F_{B_k}
\quad\Longrightarrow\quad
\left\{
\begin{array}{l}
B_k\ \text{remains effective},
\\[1ex]
\text{or}
\\[1ex]
\chi_{\rm req}^{\rm permit}\rightarrow 0
\quad\text{and/or}\quad
\boldsymbol X
\xrightarrow{\;\mathcal P_{B_k}\;}
\boldsymbol X_{\rm protected}.
\end{array}
\right.
\label{eq:failsafe_condition}
\end{equation}
Here, $\chi_{\rm req}^{\rm permit}\in\{0,1\}$
denotes the binary permit associated with
the operation requiring protection barrier
$B_k$, while $\mathcal P_{B_k}$ denotes
the corresponding protective action.
The value $\chi_{\rm req}^{\rm permit}=1$
indicates that the specified operation is
permitted, whereas
$\chi_{\rm req}^{\rm permit}=0$
indicates that the permission is withdrawn. The subscript \(\mathrm{req}\)
indicates that the relevant permit need not be the global operational permit.
It identifies the permission associated with the operation that depends on barrier \(B_k\).

Equation~\eqref{eq:failsafe_condition} expresses the fail-safe principle
in a form consistent with the permit logic of
Sec.~\ref{subsec:interlocks_permits}.  The essential requirement is not
that every internal fault should force the complete facility into a unique
``safe state,'' but that a credible failure of the protection architecture
should not silently preserve permission for an operation whose required
protection can no longer be assured.

The undesired alternative is a latent loss
of protection,
\begin{equation}
F_{B_k}
\quad\Longrightarrow\quad
B_k\ \text{unavailable},
\qquad
\chi_{\rm req}^{\rm permit}=1,
\label{eq:latent_protection_loss}
\end{equation}
for which the facility may continue operating
without recognizing that a required protection
function has become unavailable.  Such failures are particularly important
because they can remain hidden until a second event demands the missing
protection function.

This distinction connects directly to the dangerous-failure concept
introduced in Sec.~\ref{subsec:barrier_independence}.  A protection-system
failure is dangerous when it increases the probability that the relevant
hazard sequence can propagate while the loss of protection remains
undetected or insufficiently acted upon.  By contrast, a failure that
withdraws a permit unnecessarily or initiates an unwarranted protective
action is predominantly a spurious failure and affects availability rather
than directly removing the required protection.

Fail-safe design therefore tends to favour, for critical functions,
\begin{equation}
\text{uncertain or invalid protection state}
\quad\Longrightarrow\quad
\text{operation not permitted},
\label{eq:invalid_state_no_permit}
\end{equation}
rather than
\begin{equation}
\text{uncertain or invalid protection state}
\quad\Longrightarrow\quad
\text{continued permission by default}.
\label{eq:invalid_state_continue}
\end{equation}

This principle is particularly evident in accelerator interlock systems,
where interruption or invalidation of a required permit path is designed,
for the relevant failure assumptions, to remove the corresponding beam
permit rather than to be interpreted as continued authorization for beam
operation
\cite{Schmidt:2016MPS,Puccio:2010BIS}.  The same general philosophy is
used in detector-protection systems, where dedicated safety logic can
initiate predefined equipment-protection actions without relying on the
continued availability of the normal supervisory-control layer
\cite{Luders:2003DSS}.

Fail-safe behaviour should nevertheless be distinguished from
\emph{fault tolerance}.  A fault-tolerant system attempts to maintain its
required function despite one or more internal failures,
\begin{equation}
F_{B_k}
\quad\Longrightarrow\quad
B_k^{\rm function}
\ {\rm maintained},
\label{eq:fault_tolerant_condition}
\end{equation}
whereas a fail-safe system may terminate
or inhibit the hazardous operation by
withdrawing the corresponding permit,
\begin{equation}
F_{B_k}
\quad\Longrightarrow\quad
\chi_{\rm req}^{\rm permit}=0.
\label{eq:failsafe_vs_fault_tolerant}
\end{equation}
The two strategies are complementary.  Redundancy and diversity may allow
a protection function to remain available after a component failure,
while fail-safe logic defines the response when continued protection can
no longer be guaranteed.
Fail-safe behaviour, fault tolerance, diagnostic coverage, and the response
of a safety-related system to detected internal faults are central concepts
of functional-safety engineering~\cite{Willeke:2016Reliability}.

Autonomy introduces a related requirement.  A protection function should,
where appropriate, remain capable of responding even if the higher-level
control or supervisory system is unavailable.  If the same computing,
communication, or operator-interface infrastructure is required both to
operate the equipment and to protect it, a common failure may remove the
normal function and the corresponding protection simultaneously.

During normal operation, changes of detector or accelerator configuration
are produced by intentional control actions.  Let
${\boldsymbol X}_{a}$ denote the initial operational state and
${\boldsymbol X}_{b}$ the requested subsequent state.  A normal control
transition may then be represented schematically as
\begin{equation}
{\boldsymbol X}_{a}
\xrightarrow{\;{\cal U}_{\rm ctrl}\;}
{\boldsymbol X}_{b},
\qquad
{\boldsymbol X}_{a},
{\boldsymbol X}_{b}
\in
\Omega_{\rm op},
\label{eq:normal_control_path}
\end{equation}
where ${\cal U}_{\rm ctrl}$ denotes an intentional control action generated
by the normal operational or supervisory control system, and
$\Omega_{\rm op}\subseteq\Omega_{\rm safe}$ denotes the region of state
space corresponding to normal permitted operation.  Examples include
controlled detector power-up or power-down, modification of voltage
settings, magnet ramping, or a commanded change of detector configuration.

The situation is different when an abnormal condition is detected.  Let
${\boldsymbol X}_{\rm abn}$ denote a state in which one or more monitored
variables have departed from their permitted operational range, while
${\boldsymbol X}_{\rm prot}$ denotes a predefined protected state.  An
autonomous protection path can then be written as
\begin{equation}
{\boldsymbol X}_{\rm abn}
\xrightarrow{\;{\cal P}_{\rm aut}\;}
{\boldsymbol X}_{\rm prot},
\qquad
{\boldsymbol X}_{\rm prot}
\in
\Omega_{\rm safe},
\label{eq:autonomous_protection_path}
\end{equation}
where ${\cal P}_{\rm aut}$ denotes an autonomous protection action
initiated by a dedicated protection system rather than by the normal
operational control sequence.

The distinction between
Eqs.~\eqref{eq:normal_control_path} and
\eqref{eq:autonomous_protection_path} is therefore both functional and
dynamical.  In the normal control path, the system is deliberately moved
between two permitted operational states,
\begin{equation}
{\boldsymbol X}_{a},
{\boldsymbol X}_{b}
\in
\Omega_{\rm op},
\end{equation}
whereas in the autonomous protection path the objective is to redirect an
abnormal evolution toward a state in which the relevant hazard is
controlled,
\begin{equation}
{\boldsymbol X}_{\rm abn}
\rightarrow
{\boldsymbol X}_{\rm prot}.
\end{equation}

The state ${\boldsymbol X}_{\rm abn}$ need not already correspond to an
unacceptable consequence.  In many cases it belongs to the intermediate
protection region introduced in Sec.~\ref{subsec:barriers_phase_space},
\begin{equation}
{\boldsymbol X}_{\rm abn}
\in
\Omega_{\rm protect},
\end{equation}
where abnormal evolution has been detected but irreversible damage can
still be avoided.  The role of ${\cal P}_{\rm aut}$ is precisely to
prevent the subsequent trajectory from entering
$\Omega_{\rm unacceptable}$.

Terefore, the distinction between control and an autonomous protection paths can be
summarized schematically as
\begin{equation}
\begin{array}{lll}
\text{normal control:}
&
{\boldsymbol X}_{a}
\xrightarrow{{\cal U}_{\rm ctrl}}
{\boldsymbol X}_{b},
&
{\boldsymbol X}_{a},{\boldsymbol X}_{b}
\in\Omega_{\rm op},
\\[1ex]
\text{autonomous protection:}
&
{\boldsymbol X}_{\rm abn}
\xrightarrow{{\cal P}_{\rm aut}}
{\boldsymbol X}_{\rm prot},
&
{\boldsymbol X}_{\rm abn}\in\Omega_{\rm protect},
\quad
{\boldsymbol X}_{\rm prot}\in\Omega_{\rm safe}.
\end{array}
\label{eq:control_protection_comparison}
\end{equation}
Functional independence between these paths reduces the probability that a
single failure disables both normal control and the protection action.
Autonomy does not imply complete isolation.  Protection systems generally
exchange status information with control, diagnostic, and logging systems.
The essential requirement is instead that the execution of the critical
protection action should not depend unnecessarily on components whose
failure could plausibly be part of the initiating event.

The protection architecture must also distinguish between
\emph{detected} and \emph{undetected} internal failures via \emph{diagnostic coverage}.
For protection barrier $B_k$, the diagnostic coverage
$\eta_{{\rm diag},k}$ quantifies the capability of the associated
diagnostic and test functions to reveal relevant internal failures of the
barrier before those failures can remain latent until a protection demand
occurs.
Let $t_{F,k}$ denote the time at which a relevant internal failure of
barrier $B_k$ occurs, $t_{D,k}$ the time at which that failure is detected,
and $t_{{\rm dem},k}$ the time of the next demand on the barrier.  Diagnostic
coverage can then be understood as
\begin{equation}
\eta_{{\rm diag},k}
=
P\left(
t_{D,k}<t_{{\rm dem},k}
\,\big|\,
F_k
\right).
\label{eq:diagnostic_coverage}
\end{equation}
Thus, $\eta_{{\rm diag},k}=1$ would correspond, within the set of failure
modes included in the reliability model, to complete diagnostic coverage:
every relevant internal failure is expected to be detected before it can
remain hidden until the next protection demand.  Conversely,
$\eta_{{\rm diag},k}=0$ corresponds to the absence of diagnostic capability
for those failures.
The complementary quantity \(1-\eta_{{\rm diag},k}\)
represents the probability that a relevant internal failure remains
undetected, conditional on such a failure having
occurred~\footnote{
Testability ${\cal T}_k$, discussed in Sec.~\ref{subsec:barrier_independence},
should be distinguished from the diagnostic coverage $\eta_{{\rm diag},k}$. The former is a broader
design and verification property, whereas the latter characterizes
the ability
to reveal
specified internal
failures under
the adopted
diagnostic model.
High testability
does not imply
complete diagnostic
coverage, nor
does either
property guarantee
an adequate
response time
when protection
is demanded.
}.  These
undetected or \emph{latent} failures are particularly important for
protection systems because the loss of the protection function may become
apparent only when the barrier is subsequently demanded.
This distinction provides a direct connection with the barrier
unavailability discussed later in
Sec.~\ref{subsec:testing_observability}.  Continuous diagnostics and
periodic functional tests reduce the time during which a latent protection
failure can remain present without being recognized.

The temporal requirement developed in
Sec.~\ref{subsec:protection_time} also remains essential.  Detection of an
internal protection-system failure is useful only if the resulting
fail-safe response occurs before the corresponding hazardous operation can
reach the relevant critical state.  For an active fail-safe transition,
one therefore requires schematically
\begin{equation}
\tau_{\rm diag}
+
\tau_{\rm logic}
+
\tau_{\rm act}
<
\tau_{\rm avail},
\label{eq:failsafe_timing_condition}
\end{equation}
where $\tau_{\rm diag}$ is the time required to recognize the internal
failure and the remaining terms characterize the associated protective
response.

A fail-safe architecture may consequently
act on both the logical authorization and
the physical state of the facility.
For a failure requiring permit withdrawal
and an active protective response,
one may write schematically
\begin{equation}
F_{B_k}
\quad\Longrightarrow\quad
\left(
\chi_{\rm req}^{\rm permit},
\boldsymbol X
\right)
\longrightarrow
\left(
0,
\boldsymbol X_{\rm protected}
\right).
\label{eq:failsafe_state_permit_mapping}
\end{equation}
The relation represents a particular
fail-safe response rather than a universal
requirement for every internal barrier
failure. Depending on the operation,
permit withdrawal may prevent a future
transition, while an ongoing hazardous
operation may additionally require
an autonomous protective action.

The designation ${\boldsymbol X}_{\rm protected}$ should be interpreted
carefully.  It does not necessarily denote a state free of all hazards.
A beam dump, magnet discharge, detector power removal, or cryogenic
isolation can itself involve substantial physical processes.  Rather,
${\boldsymbol X}_{\rm protected}$ denotes a state or trajectory in which
the particular hazardous evolution under consideration has been
interrupted, contained, or redirected according to the intended protection
strategy.

Fail-safe behaviour is therefore a property of a specified protection
function relative to a specified hazard sequence, not an absolute property
of a component.  The relevant question is whether credible failures of the
protection architecture preserve, restore, or deliberately terminate the
conditions required to keep that hazard sequence controlled.

This perspective also explains the connection between fail-safe design and
DiD.  No single barrier should be assumed infallible.
Redundancy can maintain a protection function after some failures;
diversity can reduce susceptibility to common-cause mechanisms;
diagnostics can reveal latent degradation; and fail-safe logic can remove
permission for continued operation when the required protection state
cannot be demonstrated.

The resulting architecture may be summarized schematically as
\begin{equation}
\text{failure}
\rightarrow
\text{diagnosis}
\rightarrow
\text{loss of permit or maintained redundancy}
\rightarrow
\text{protected trajectory}.
\label{eq:failsafe_protection_chain}
\end{equation}

The next question is how these technical and logical protection functions
fit into a broader hierarchy of controls, ranging from reduction of the
hazard by design to automatic protection, collective mitigation,
procedural controls, and emergency response.  This hierarchy is considered
in Sec.~\ref{subsec:protection_layers}.

\subsection{Protection layers and the hierarchy of controls}
\label{subsec:protection_layers}

The foregoing discussion shows that protection in a complex accelerator
facility is rarely provided by a single barrier.  Instead, risk reduction
is achieved through a set of complementary measures acting at different
stages of the hazard chain and, in many cases, on different physical
variables.  These measures may reduce the probability that an initiating
event develops into a hazardous state, interrupt the subsequent
propagation, limit the transfer of energy or exposure, or reduce the
severity of the resulting consequences.

It is therefore useful to distinguish a broad sequence of protection
strategies,
\begin{equation}
\begin{split}
\text{hazard elimination or reduction}
&\rightarrow
\text{intrinsic design measures}
\\
&\rightarrow
\text{engineered preventive barriers}
\rightarrow
\text{automatic protection}
\\
&\rightarrow
\text{mitigative barriers}
\rightarrow
\text{procedural and organisational controls}
\\
&\rightarrow
\text{personal protection and emergency response}.
\end{split}
\label{eq:protection_hierarchy}
\end{equation}
Equation~\eqref{eq:protection_hierarchy} is schematic and should not be
interpreted as a universal temporal sequence or as a strict ranking that
applies identically to every hazard.  Rather, it summarizes a general
principle: where practicable, protection should be introduced as far
upstream in the hazard chain as possible, while additional downstream
barriers are retained to control the consequences of failures that cannot
be completely excluded.

Using the generic hazard-chain representation introduced in
Eq.~\eqref{eq:generic_hazard_chain}, different protection layers may be
associated with different links of the sequence.  An intrinsic design
measure may reduce the probability of $F_i$; an engineered preventive
barrier may interrupt ${\cal P}_i$; an automatic protection function may
redirect the system before ${\cal E}_i$ occurs; and a mitigative barrier
may modify the resulting consequence vector
${\boldsymbol{\cal C}}_i$.

The resulting architecture may be represented schematically as
\begin{equation}
\left\{
B_{\rm design},
B_{\rm prev}^{(1)},
B_{\rm prev}^{(2)},
B_{\rm mit}^{(1)},
B_{\rm mit}^{(2)},
B_{\rm proc},
B_{\rm emerg}
\right\},
\label{eq:layered_barrier_set}
\end{equation}
where the individual barriers need not act on the same physical quantity
or at the same stage of the sequence.  Their combined purpose is to
prevent the facility trajectory from entering regions of state space
associated with unacceptable consequences.

This formulation is more appropriate than treating all protection
measures as equivalent.  For example, in accelerator machine protection,
robust component design, collimation, equipment surveillance, beam-loss
monitoring, interlocks, beam dumping, absorbers, and operational
procedures act on different parts of the beam-loss hazard chain.  In a
cryogenic installation, pressure-vessel design, containment, relief
capacity, discharge routing, ventilation, oxygen monitoring, access
restrictions, and emergency response address different stages of a
possible helium-release sequence.  For detector systems, design margins,
cooling and power monitoring, subsystem interlocks, Detector Safety System
actions, fire detection, and procedural controls likewise provide
complementary functions.

The distinction between preventive and mitigative functions introduced in
Sec.~\ref{subsec:preventive_mitigative} remains important within this
layered picture.  Different barriers may act at different stages of the
same hazard sequence: some primarily reduce the probability that the
sequence progresses toward a hazardous transfer or exposure, whereas
others limit the consequences after such a transfer has occurred.
Individual protection elements may also perform both functions, depending
on the particular failure trajectory being considered.  The classification
is therefore functional rather than intrinsic to the hardware and should
always be defined with respect to a specified hazard chain.

A layered architecture also provides protection against uncertainty in the
failure~\footnote{The use of multiple protection layers acting at different stages of a
hazard sequence is a standard principle of functional-safety engineering
\cite{Sklet:2006SafetyBarriers}.
At CERN, an explicit protection-layer methodology has,
for example, been applied to the HL-LHC Full Remote Alignment System
\cite{FernandezAdiego:2023ProtectionLayers}.}.  If one barrier is ineffective because an event develops
along an unexpected path, a downstream barrier may still interrupt or
mitigate the sequence.  This is the practical meaning of DiD:
\begin{equation}
B_1
\oplus
B_2
\oplus
\cdots
\oplus
B_N ,
\label{eq:layered_defence_depth}
\end{equation}
where, as before, $\oplus$ denotes complementary protection functions and does not
imply mathematical addition, equal effectiveness, or statistical
independence.

The latter point is essential.  The presence of several barriers does not
automatically imply a corresponding multiplication of risk reduction.
As discussed in Sec.~\ref{subsec:barrier_independence}, barriers that
share a common sensor, power supply, communication channel, software
process, maintenance procedure, or environmental dependency may be
susceptible to common-cause failure.  A robust layered architecture should
therefore consider not only the number of barriers but also the properties
already summarized in Eq.~\eqref{eq:barrier_attributes}, including their
failure-on-demand probability, response time, degree of independence,
functional or technological diversity, diagnostic coverage, and
testability.

These properties characterize the effectiveness of individual barriers
within the wider protection architecture.  Their relevance is
configuration and hazard dependent: a barrier may be highly reliable but
too slow for a particular failure trajectory, or sufficiently fast but
strongly correlated with another barrier through a common-cause failure.
Similarly, a barrier with high diagnostic coverage may still provide
limited protection if its actuation mechanism is not independent of the
failure being controlled.

Diagnostic coverage is particularly relevant in a layered architecture
because a barrier can appear available while containing a latent internal
failure.  As defined in Eq.~\eqref{eq:diagnostic_coverage}, the quantity
$\eta_{{\rm diag},k}$ characterizes the probability that a relevant
internal failure of barrier $B_k$ is detected before the next demand on
that protection function.  Periodic testing, continuous diagnostics, and
condition monitoring therefore contribute directly to maintaining the
assumed effectiveness of the protection layer.

The hierarchy of controls also contains an important distinction between
collective and individual protection.  Measures such as shielding,
ventilation, automatic beam inhibition, pressure relief, or power removal
act on the source or environment and can protect several persons
simultaneously without requiring each individual to recognize the hazard
and react correctly.  Personal protective equipment and individual
actions remain necessary for many activities, but they generally form only
one part of the overall protection architecture.

Schematically,
\begin{equation}
{\cal P}_{\rm collective}
\oplus
{\cal P}_{\rm procedural}
\oplus
{\cal P}_{\rm individual}
\label{eq:collective_individual_protection}
\end{equation}
represents complementary protection layers rather than interchangeable
alternatives.  Where a recurring hazard can reasonably be reduced through
intrinsic design or collective engineered protection, it is generally more
robust to do so than to rely exclusively on repeated human intervention.

The layered-barrier picture can also be expressed in the state-space
language introduced earlier.  Let an abnormal trajectory generated by an
initiating event $F_i$ be denoted by
\begin{equation}
\Gamma_i^{(0)}
:
{\boldsymbol X}_0
\rightarrow
\Omega_{\rm unacceptable}.
\label{eq:unprotected_hazard_trajectory}
\end{equation}
A protection barrier $B_k$ acts by modifying this trajectory,
\begin{equation}
\Gamma_i^{(0)}
\xrightarrow{\;B_k\;}
\Gamma_i^{(k)},
\label{eq:barrier_trajectory_mapping}
\end{equation}
such that the modified trajectory either remains within the admissible
region or terminates in a protected state,
\begin{equation}
\Gamma_i^{(k)}
\subset
\Gamma_{\rm allowed}
\qquad
\text{or}
\qquad
{\boldsymbol X}
\rightarrow
{\boldsymbol X}_{\rm prot}.
\label{eq:barrier_safe_trajectory}
\end{equation}

This representation emphasizes that protection layers do not merely
``reduce risk'' in an abstract sense.  They alter the physical evolution
of the facility by preventing, interrupting, redirecting, or mitigating
specific transitions through the coupled state space.

The effectiveness of a protection architecture therefore depends on how
well its layers cover the relevant failure trajectories, how independent
their failure modes are, whether latent failures can be detected, and
whether the barriers act sufficiently rapidly.  The last requirement is
dynamical: even a reliable and independent barrier is ineffective if its
response occurs after the hazardous process has already reached an
irreversible consequence.  This temporal aspect is considered in
Sec.~\ref{subsec:protection_time}.

\subsection{Protection integrity and residual risk}
\label{subsec:protection_integrity}

The effectiveness of a protection architecture cannot be characterized
solely by the presence of individual barriers.  What ultimately matters is
whether the required protection function remains available and performs
correctly when demanded, under the physical and operational conditions for
which it was designed.  We use the term \emph{protection integrity} in
this general engineering sense.

Protection integrity is therefore a property of a specified protection
function rather than an intrinsic scalar property of a piece of hardware.
Its assessment requires the corresponding hazard sequence, protection
demand, facility configuration, and required response to be identified.
For example, the relevant protection claim may be that a beam must be
extracted before an evolving failure produces damaging losses, that
electrical power must be removed before a detector reaches an unacceptable
thermal state, or that a cryogenic release must remain within the capacity
of the designed containment and relief paths.

The quantities introduced in the preceding subsections describe different
aspects of this integrity.  In particular, the failure-on-demand
probability $q_k$, response time $\tau_k$,  as well as and testability ${\cal T}_k$, summarized in
Eq.~\eqref{eq:barrier_attributes}, as well as diagnostic coverage $\eta_{{\rm diag},k}$, introduced in
Eq.~\eqref{eq:diagnostic_coverage}, should not be interpreted as independent
figures of merit.  Their significance arises from their combined effect on
a particular protection function.

A small value of $q_k$, for example, is useful only if the corresponding
barrier responds sufficiently rapidly to the relevant failure trajectory.
Likewise, nominal redundancy does not provide the expected increase in
integrity when redundant channels remain vulnerable to a common-cause
failure, as discussed in Sec.~\ref{subsec:barrier_independence}.  High
diagnostic coverage improves confidence that latent failures will be
revealed before demand, but does not compensate for an inadequate
actuation mechanism or for a protection response that occurs too late.
The temporal requirement discussed in Sec.~\ref{subsec:protection_time}
is therefore an integral part of protection integrity rather than an
independent consideration.

The same reasoning applies at the level of a layered architecture.
Several protection layers may contribute to the control of one hazard
sequence, but the resulting integrity cannot in general be obtained by
simply multiplying nominal barrier efficiencies.  Such a treatment is
justified only for a specified sequence under the independence assumptions
discussed in Sec.~\ref{subsec:barrier_independence}.  In a realistic
facility, different layers may act on different branches of the sequence,
modify different physical variables, or share common dependencies.

Protection integrity should consequently be related to the
\emph{residual risk} associated with a specified hazard.  The risk
function introduced in Sec.~\ref{subsec:hazard_chains} already provides
the required framework: protection measures modify the frequency of the
relevant sequence, its possible consequences, or both.  The residual risk
is the risk that remains after the applicable preventive and mitigative
barriers have been taken into account.  No additional universal risk
formula is required, because the mapping from barrier performance to
residual risk depends on the hazard under consideration and on the
structure of its possible propagation paths.

This distinction is important when discussing ``risk reduction.''  A
quoted reduction has meaning only relative to a defined reference
condition.  It must specify what initiating event is considered, which
consequence or consequence class is being controlled, which barriers are
credited, and under what facility configuration their protection functions
are assumed to be available.  A protection measure that provides strong
risk reduction for one failure trajectory may contribute little to another
trajectory involving the same equipment.

The facility configuration is particularly important at the LHC and its
experiments.  A protection function that is valid during stable-beam
operation need not provide the same protection during injection,
commissioning, access, maintenance, detector movement, magnet ramping, or
cryogenic intervention.  The corresponding protection claim must therefore
be associated with the operational state in which the barrier is expected
to act.  In the state-space language used throughout this work, protection
integrity concerns the capability of the barrier architecture to prevent
an abnormal trajectory from progressing toward an unacceptable region, or
to redirect it toward an admissible protected state.

This viewpoint also clarifies the distinction between dangerous failures
and spurious protection actions introduced in
Sec.~\ref{subsec:barrier_independence}.  A dangerous failure reduces
protection integrity because a required protection function may be
unavailable when demanded.  A spurious action does not normally represent
loss of the protection function itself, but reduces operational
availability and may interrupt scientific operation unnecessarily.
Protection-system design must therefore maintain sufficiently strong
protection against dangerous failures while avoiding an excessive rate of
unnecessary interventions.

The appropriate level of protection integrity is necessarily
hazard-dependent.  Failure trajectories capable of producing rapid and
large energy deposition, significant cryogenic releases, unacceptable
radiation exposure, or serious damage to unique accelerator or detector
equipment require correspondingly robust protection architectures.  Less
severe or slowly developing deviations may permit greater reliance on
diagnostics, operator intervention, or controlled operational recovery.
Thus, protection requirements should follow from the physical consequences
and characteristic timescales of the hazard rather than from a universal
integrity criterion applied identically to all systems.

Finally, protection integrity is not established once at the design stage.
It must be maintained throughout operation.  Changes in hardware,
software, configuration, operating parameters, maintenance state, or
diagnostic capability can alter the validity of the assumptions on which
the original protection argument was based.  Periodic testing,
configuration control, diagnostics, and analysis of real protection
actions are therefore necessary to demonstrate that the credited barriers
remain effective.

Protection integrity is consequently best regarded as an experimentally
and operationally maintained property of the protection architecture:
\emph{the required protection function must remain available, sufficiently
independent, diagnosable, and fast enough to control the specified hazard
under the facility configuration in which the demand may occur}.  How
these properties are verified and maintained through testing,
observability, and post-event analysis is discussed in the following
subsection.

\subsection{Testing, observability, and post-event validation}
\label{subsec:testing_observability}

The protection integrity discussed in
Sec.~\ref{subsec:protection_integrity} cannot be established solely from
the design of the protection architecture.  It must also be possible to
verify that the credited protection functions remain available and behave
as intended under the configurations in which they may be demanded.
Testing, continuous diagnostics, state monitoring, and analysis of actual
protection actions therefore form an essential part of the protection
strategy.

A distinction should first be made between the \emph{existence} of a
protection function and its \emph{demonstrated availability}.  A barrier
may be correctly designed but unavailable because of a latent hardware
fault, a communication failure, an incorrect configuration, a disabled
channel, an incomplete maintenance intervention, or a software state that
differs from that assumed in the protection logic.  For this reason,
critical protection functions must be accompanied by sufficient
observability to determine whether the conditions required for their
operation are actually satisfied.

Here, observability is used in an operational sense.  It denotes the
ability to infer the relevant state of the protected system and of its
protection functions from available measurements, status information,
diagnostics, and recorded events.  It should not be confused with the more
restrictive mathematical definition of observability used in control
theory.  The relevant information may include physical measurements,
interlock states, permit states, equipment modes, communication status,
power-supply conditions, diagnostic flags, timestamps, and configuration
information.

This information serves two related purposes.  First, it enables abnormal
conditions to be detected early enough for the protection system to act.
Second, it provides evidence that the protection system itself remains
capable of performing the required action.  The latter role is directly
connected with the diagnostic coverage introduced previously: continuous
diagnostics can reveal some internal failures while the system is
operating, whereas failures not detectable online may require periodic
functional testing.

Different testing strategies therefore address different classes of
failure.  Continuous self-diagnostics can identify faults that produce an
observable change in the internal state of a protection channel.  Periodic
tests can reveal failures that remain latent during normal operation.
Functional tests verify that an input condition produces the expected
logical response, while end-to-end tests additionally check the complete
chain from sensing and decision logic to the final protective action.
These approaches are complementary rather than interchangeable.

For a protection function involving several subsystems, end-to-end testing
is particularly important.  Correct operation of the individual sensor,
logic, communication, and actuator elements does not by itself guarantee
that the complete protection chain will perform correctly when demanded.
Configuration errors, timing mismatches, interface failures, or unexpected
dependencies between subsystems may become visible only when the complete
sequence is exercised.

At the LHC, this principle is especially relevant because machine
protection relies on distributed systems whose actions must remain
coherent across accelerator equipment, interlock logic, beam permits,
beam-dump requests, and the associated actuators
\cite{Bordry:2001MPS,Puccio:2010BIS,Wenninger:2016MPO}.  Comparable
considerations apply to experiment safety systems, where detector
conditions, cooling, electrical power, environmental monitoring, and
dedicated safety actions may be distributed across several control and
protection domains \cite{Ninin:2005LACS,Luders:2003DSS}.

Testing must also be configuration aware.  A protection function verified
in one operational state cannot automatically be assumed to provide the
same protection after a change of hardware configuration, firmware,
software, cabling, operating parameters, detector geometry, access state,
or maintenance condition.  The validity of the protection claim therefore
depends not only on whether a test has been performed, but also on whether
the tested configuration corresponds to the configuration in which the
protection function is subsequently credited.

This requirement connects testing directly to configuration control.
Following an intervention, the return to operation should establish that
the relevant protection functions have been restored, that temporary
bypasses or inhibited channels have been removed or formally accounted
for, and that the resulting configuration is consistent with the
conditions assumed by the interlock and safety logic.  A successful
component repair is therefore not necessarily equivalent to restoration
of the complete protection function.

Tests themselves must also be interpreted carefully.  A successful test
provides evidence that the tested function operated correctly under the
conditions of that test; it does not prove that every possible failure
mode has been excluded.  The significance of the result depends on the
scope of the test, the interval since the previous test, the failure modes
that can develop between tests, and the extent to which the test represents
a realistic protection demand.  Protection integrity is therefore built
from a combination of design evidence, diagnostic information, periodic
verification, and operational experience rather than from a single test
result.

Actual protection actions provide an additional and particularly valuable
source of information.  When an interlock, beam dump, equipment trip, or
detector safety action occurs, the recorded data can be used to reconstruct
the evolution that preceded the event and the subsequent response of the
protection architecture.  Such post-event analysis can establish whether
the initiating condition was correctly identified, whether the expected
barriers acted, whether their timing was adequate, and whether additional
or unexpected propagation paths were present.

For this purpose, synchronized and sufficiently detailed event records are
essential.  The reconstruction of a complex event may require correlation
of measurements and state information originating from several
subsystems.  Accurate timestamps, archived interlock transitions,
equipment-status histories, physical measurements, and configuration
records permit the causal sequence to be distinguished from secondary
effects that occurred after the initial protection action.

Post-event analysis is therefore not restricted to identifying the
component that first reported a fault.  Its more important role is to
reconstruct the physical trajectory of the facility and determine whether
the observed behaviour is consistent with the hazard and protection models
used in the safety argument.  An event may reveal, for example, an
unanticipated coupling between subsystems, a common dependency between
nominally separate barriers, an insufficient diagnostic capability, or a
response time different from that assumed in the protection analysis.

This creates an important feedback from operation to protection design.
Observed anomalies, protection actions, near misses, and test results can
lead to revised thresholds, additional diagnostics, modified interlock
logic, improved procedures, changes in hardware, or refinement of the
underlying failure model.  Operational experience therefore acts as an
empirical test of the assumptions used when the protection architecture
was designed.

The same reasoning applies to events that do not produce damage.
Successful protection actions are valuable because they demonstrate how
the architecture behaves under real disturbances, while near misses may
identify trajectories for which the available safety margin was smaller
than expected.  Both types of event can provide information that is not
available from nominal-operation data alone.

Testing and observability consequently have complementary roles.
Observability determines what can be known about the instantaneous state
of the facility and its protection functions; testing deliberately probes
whether credited functions remain available; and post-event analysis
compares the observed evolution with the behavior expected from the
protection model.  Together they provide the evidence required to maintain
confidence that the barriers credited in the safety analysis remain
effective throughout operation.

Protection should therefore be regarded as a continuously validated
physical and operational capability rather than as a static property
established once during design.  This continual validation is also central
to the balance between protection and availability: insufficient testing
or observability can leave dangerous latent failures undetected, whereas
poorly designed testing or excessive spurious protection actions can
unnecessarily reduce operational efficiency.  The resulting relation
between safety, availability, and scientific performance is considered in
the following subsection.

\subsection{Safety, availability, and scientific performance}
\label{subsec:safety_availability}

Protection systems necessarily influence accelerator and detector
availability.  Interlocks, automatic shutdowns, beam dumps, power removal,
access restrictions, and other protective actions can interrupt data
taking or delay the return to operation.  This does not imply, however,
that safety and scientific performance should be regarded as competing
objectives of equal status.  Protection integrity defines the conditions
under which operation is admissible, whereas availability and scientific
performance are optimized within those conditions.

This distinction is important because reducing the number of protection
actions does not necessarily improve the overall performance of the
facility.  A protection system that intervenes too frequently because of
poorly chosen thresholds, inadequate diagnostics, or spurious signals can
produce unnecessary downtime.  Conversely, weakening a required
protection function in order to increase short-term availability can
increase the probability of equipment damage, prolonged recovery, or loss
of experimental capability.  The relevant objective is therefore not
maximum instantaneous availability, but sustainable operation within the
validated protection envelope.

Recall that $\Omega_{\rm op}$ denote the region of normal permitted operation
introduced above and $\Omega_{\rm safe}$ the corresponding safe region.
Scientific operation is meaningful only within the intersection of the
states allowed by the relevant physical, protection, and operational
constraints.  In this sense, the accessible domain for data taking may be
written schematically as
\begin{equation}
\Omega_{\rm science}
\subseteq
\Omega_{\rm op}
\subseteq
\Omega_{\rm safe}.
\label{eq:science_operational_safe_domains}
\end{equation}
The inclusion in Eq.~\eqref{eq:science_operational_safe_domains} is
conceptual rather than universal in detail: different accelerator and
detector modes may have distinct operational subspaces, and some safe
states may correspond to maintenance, recovery, access, or other
non-physics conditions.

Availability should therefore be interpreted conditionally on the
protection functions that must remain active.  If $T_{\rm phys}$ denotes
the time during which the facility is in a configuration suitable for
physics operation and $T_{\rm sched}$ the corresponding scheduled
operational interval, a simple operational availability may be written as
\begin{equation}
A_{\rm phys}
=
\frac{T_{\rm phys}}{T_{\rm sched}}.
\label{eq:physics_availability}
\end{equation}
This quantity is useful operationally, but it does not by itself measure
scientific performance or protection quality.  Two periods with the same
$A_{\rm phys}$ may differ substantially in delivered luminosity, detector
live time, data quality, beam conditions, or experimental efficiency.

For this reason, scientific output is more naturally regarded as a
functional of the operational history,
\begin{equation}
{\cal Y}_{\rm sci}
=
{\cal Y}
\left[
{\boldsymbol X}(t),
{\cal Q}(t),
t\in T_{\rm phys}
\right],
\label{eq:scientific_yield_functional}
\end{equation}
where ${\boldsymbol X}(t)$ represents the evolving facility state and
${\cal Q}(t)$ denotes the relevant measure of data-taking quality or
scientific usefulness.  The precise form of ${\cal Y}_{\rm sci}$ depends
on the experiment and may involve integrated luminosity, detector live
fraction, data quality, trigger performance, or other physics-specific
quantities.

The influence of protection on scientific performance is therefore
two-sided.  On short timescales, a beam dump, detector trip, or access
restriction may reduce $A_{\rm phys}$.  On longer timescales, however,
effective protection preserves the accelerator, detector, and technical
infrastructure from failures that could otherwise produce much larger
losses of operating time.  Protection can therefore increase the
\emph{sustainable} scientific yield even when individual interventions
temporarily interrupt operation.

This point is particularly relevant to the distinction between dangerous
failures and spurious protection actions discussed in
Sec.~\ref{subsec:barrier_independence}.  Dangerous failures threaten the
availability of a required protection function and can increase residual
risk.  Spurious actions, by contrast, primarily reduce availability by
initiating a protective response when no corresponding hazardous
condition is present.  Their mitigation is therefore important for
operational efficiency, but it must not compromise the capability of the
system to respond to genuine demands.

The preferred response to excessive spurious interventions is consequently
not to weaken the protection requirement, but to improve the quality of
the protection architecture.  Possible measures include better
diagnostics, improved signal validation, removal of common technical
failure sources, refinement of thresholds, increased robustness of
communication paths, and more effective testing of the complete
protection chain.  In this way, protection integrity and availability can
often be improved simultaneously.

A similar consideration applies to recovery after a genuine protection
action.  The duration of the interruption depends not only on the initial
trip or beam dump, but also on the ability to identify the initiating
condition, verify the state of the affected systems, restore the required
protection functions, and demonstrate that the configuration is again
compatible with permitted operation.  The observability and post-event
validation discussed in Sec.~\ref{subsec:testing_observability} therefore
contribute directly to availability by reducing unnecessary uncertainty
during recovery.

The relation between safety and performance can thus be summarized as a
constrained optimization problem.  Scientific operation seeks to maximize
a suitable measure of useful physics output,
\begin{equation}
\max\;{\cal Y}_{\rm sci},
\qquad
\text{subject to}
\qquad
\Gamma_{\rm op}\in\Gamma_{\rm allowed},
\label{eq:constrained_scientific_operation}
\end{equation}
where the allowed operational trajectories are determined by the physical,
protection, access, and procedural constraints discussed in the preceding
subsections.  Equation~\eqref{eq:constrained_scientific_operation} should
not be interpreted as a literal real-time optimization algorithm; it
expresses the conceptual relation between physics performance and the
conditions required for safe and protected operation.

This viewpoint avoids a false opposition between safety and physics.
Robust protection may impose operational constraints, but those
constraints preserve the facility state space within which reproducible
and sustainable scientific operation is possible.  Conversely, excessive
or poorly designed protection actions can unnecessarily contract the
practically accessible part of that state space.  The engineering task is
therefore to maintain the required protection integrity while minimizing
unnecessary restrictions on valid operation.

For a large accelerator facility, the relevant performance objective is
thus not simply the largest fraction of time with beam or detector systems
enabled.  It is the largest sustainable scientific output compatible with
the validated protection architecture, the current facility
configuration, and the admissible operational trajectories~\cite{Filippini:2005Reliability,Willeke:2016Reliability,
Wenninger:2016MPO}.

\subsection{Protection barriers and the admissible operational phase space}
\label{subsec:barriers_phase_space}

The state-space formulation introduced in
Sec.~\ref{sec:lhc_energetic_system} provides a useful way to interpret the
collective role of protection barriers.  In this context, the term
``phase space'' is used in a generalized operational sense: the facility
state ${\boldsymbol X}$ contains both continuous physical variables and
discrete configuration variables, and therefore need not constitute a
Hamiltonian phase space in the strict dynamical-systems sense.

Protection barriers do not simply assign a scalar reduction factor to
risk.  They constrain which states may be entered, which transitions are
permitted, and how the facility is redirected when an abnormal evolution
is detected.  The protection architecture therefore helps define the
portion of the physically accessible state space that can be used for
operation.

It is useful to distinguish the physically safe region
$\Omega_{\rm safe}$ from the region that is actually admissible for a
given operational configuration.  Let ${\cal B}_{\rm av}$ denote the set
of protection functions that are currently available and credited for the
operation under consideration.  We may then define schematically
\begin{equation}
\Omega_{\rm adm}
\left(
{\cal B}_{\rm av}
\right)
\subseteq
\Omega_{\rm safe},
\label{eq:barrier_dependent_admissible_space}
\end{equation}
where $\Omega_{\rm adm}({\cal B}_{\rm av})$ denotes the states that may be
admitted while the specified set of protection functions is available.

The distinction is important.  A state may be physically benign under
nominal conditions but nevertheless be excluded from operation if the
barrier required to control a credible subsequent failure is unavailable.
Conversely, restoration of the corresponding protection function may
permit operation in a larger region of the safe state space.

In this sense, the admissible operational region is configuration
dependent.  If a required protection function becomes unavailable and no
equivalent compensating measure is introduced, the permitted domain must
generally contract.  Schematically, if
\begin{equation}
{\cal B}_{\rm av}'
\subset
{\cal B}_{\rm av},
\end{equation}
then one expects
\begin{equation}
\Omega_{\rm adm}
\left(
{\cal B}_{\rm av}'
\right)
\subseteq
\Omega_{\rm adm}
\left(
{\cal B}_{\rm av}
\right),
\label{eq:protection_loss_contracts_space}
\end{equation}
provided that all other conditions remain unchanged.  Equation~
\eqref{eq:protection_loss_contracts_space} is schematic rather than a
universal mathematical theorem, because compensating operational
restrictions or alternative barriers may preserve part of the previously
available domain.

This provides a natural interpretation of degraded operational modes.
If one protection channel, diagnostic function, cooling circuit, or
interlock path is unavailable, continued operation need not be described
simply as either ``safe'' or ``unsafe.''  Instead, the admissible region
may be reduced by imposing additional constraints on beam intensity,
magnet current, detector voltage, access state, operating mode, or other
components of ${\boldsymbol X}$.

The same reasoning applies to transitions.  As emphasized previously,
safe endpoints are not sufficient: the trajectory connecting them must
also remain compatible with the applicable protection constraints.
Barrier availability therefore affects not only the set of admissible
states but also the set of admissible trajectories.  In particular, a
transition that is normally permitted may become unavailable if the
protection function required during that transition cannot be guaranteed.

This distinction is especially relevant for operational phases such as
beam injection, energy ramping, squeezing, detector power-up, magnet
ramping, cryogenic transitions, detector movement, and access changes.
Different barriers may be required in different phases, so the admissible
domain is not fixed once for the entire facility but evolves with the
operational configuration.

Protection barriers also define boundaries between qualitatively
different responses to system evolution.  Within normal permitted
operation, intentional control actions move the facility between allowed
states.  When monitored quantities approach a protection boundary, the
control logic may inhibit a requested transition, remove a permit, or
initiate an autonomous protective action.  The role of the barrier is
therefore not necessarily to wait until an unacceptable state is reached,
but to prevent the corresponding trajectory from developing.

A useful conceptual decomposition is consequently
\begin{equation}
\Omega_{\rm phys}
\supset
\Omega_{\rm safe}
\supseteq
\Omega_{\rm adm}
\supseteq
\Omega_{\rm op}.
\label{eq:state_space_hierarchy}
\end{equation}
Recall, that $\Omega_{\rm phys}$ is the physically accessible state space,
$\Omega_{\rm safe}$ the region compatible with the relevant safety
conditions, $\Omega_{\rm adm}$ the part admissible under the currently
available protection architecture, and $\Omega_{\rm op}$ the subset
associated with normal permitted operation.

Equation~\eqref{eq:state_space_hierarchy} should be interpreted
schematically.  The precise relations depend on the definitions adopted
for a particular subsystem and operational mode, and the corresponding
sets may contain both continuous and discrete conditions.  Its purpose is
to distinguish physical possibility from safe operation and, in turn,
safe operation from operation that is actually authorized under the
available protection configuration.

This formulation also clarifies the role of interlocks and permits.
They provide an operational realization of boundaries in the admissible
state space: when the conditions required for a transition are not
satisfied, the corresponding permit is withheld and the trajectory is
prevented from entering a region for which the necessary protection
assumptions do not hold.

The protection architecture therefore acts both dynamically and
geometrically.  Dynamically, it detects and redirects abnormal
trajectories.  Geometrically, it restricts the portion of the generalized
state space in which operation is permitted.  These two roles are
complementary: the first controls the evolution of the facility after a
deviation is detected, while the second prevents operation from beginning
in configurations for which the required protection cannot be assured.

From this perspective, safety does not merely impose an external
constraint on accelerator and detector operation.  It contributes to the
definition of the operational domain itself.  The available physics
programme is realized within the subset of facility states and trajectories
for which the required barriers, diagnostics, interlocks, access
conditions, and recovery mechanisms are demonstrably available.

The safe operational phase space is therefore not immutable.  It evolves
with hardware configuration, maintenance status, protection availability,
operating mode, and accumulated operational knowledge.  Improvements in
diagnostics, protection reliability, modelling, and validation can enlarge
the practically accessible domain without relaxing the underlying safety
requirements, whereas unavailable or degraded protection functions may
temporarily contract it.

\section{Stored energy and machine protection}
\label{sec:machine_protection}

The general protection framework developed in
Sec.~\ref{sec:hazards_barriers} acquires a particularly stringent form in
the LHC because of the large amount of energy stored in the circulating
beams and in the superconducting magnet system.  Machine protection is
therefore concerned not merely with detecting abnormal conditions, but
with preventing the coupled accelerator state from evolving along
trajectories that can lead to equipment damage, loss of superconducting
state, uncontrolled energy deposition, or other unacceptable machine
conditions
\cite{Schmidt:2016MPS,Wenninger:2016MPO}.

In the state-space language introduced in
Sec.~\ref{subsec:underground_environment_state}, the relevant problem may
be written schematically as
\begin{equation}
{\boldsymbol X}_{\rm beam}
\oplus
{\boldsymbol X}_{\rm mag}
\oplus
{\boldsymbol X}_{\rm infrastructure}
\;\longrightarrow\;
{\boldsymbol X}(t),
\label{eq:mps_coupled_state}
\end{equation}
where the symbol $\oplus$  denotes the coupled participation of the
corresponding subsystems rather than a direct mathematical sum.
A perturbation originating in one block can propagate rapidly into the
others.  Beam-orbit errors, powering failures, extraction faults, or
injection failures can produce localized beam losses; beam losses can in
turn generate intense energy deposition, heating, secondary radiation,
and superconducting-magnet quenches.  Conversely, magnet or powering
failures can perturb the beam sufficiently rapidly to create a damaging
loss pattern.

The central machine-protection problem is therefore naturally expressed
as the control of a hazard trajectory,
\begin{equation}
{\boldsymbol X}(t_0)
\xrightarrow{\;\mathcal{F}_i\;}
{\boldsymbol X}(t)
\xrightarrow{\;{\cal E}_i\;}
{\boldsymbol{\cal C}}_i ,
\label{eq:mps_hazard_trajectory}
\end{equation}
where $\mathcal{F}_i$ denotes an initiating failure or perturbation,
${\cal E}_i$ the corresponding undesired energy-transfer process, and
${\boldsymbol{\cal C}}_i$ the resulting consequence vector, following the
notation of Sec.~\ref{subsec:hazard_chains}.  The purpose of the machine-
protection system is to prevent or redirect this evolution before the
relevant critical state is reached.

This requirement immediately connects machine protection to the
preventive--mitigative distinction introduced in
Sec.~\ref{subsec:preventive_mitigative}.  Beam interlocks and beam
extraction can act preventively by interrupting the evolution before a
damaging loss develops, while collimators, absorbers, dilution systems,
and other passive elements can limit the consequences of losses that
cannot be avoided completely.  The same hardware element may contribute
to both functions depending on the particular hazard sequence under
consideration
\cite{Redaelli:2016Collimation,Schmidt:2016MPS}.

The temporal structure of the failure is equally important.  Some
failure modes evolve sufficiently slowly that an active chain of
detection, logic, communication, and beam extraction can act before the
critical state is reached.  Other failures develop on timescales
comparable to, or shorter than, the minimum achievable active response.
The machine-protection architecture must therefore combine active
interlocks with passive protection according to the condition developed
in Sec.~\ref{subsec:protection_time},
\begin{equation}
\tau_{\rm prot}
<
\tau_{\rm avail},
\label{eq:mps_timing_condition}
\end{equation}
whenever active intervention is expected to interrupt the hazard
trajectory.

Machine protection system (MPS) is consequently not represented by a single device or
interlock.  It is a layered architecture comprising beam-loss monitoring,
beam and powering permits, collimation, injection and extraction
protection, the beam interlock system (BIS), the LHC beam dumping system (LBDS), quench and
powering protection, and the associated passive absorbers and energy-
dissipation structures
\cite{Puccio:2010BIS,Bartmann:2010InjectionDump,
RodriguezMateos:2004QPS,Kramer:2008LBDS}.
These elements act at different positions along a failure sequence and on
different characteristic timescales.

The beam permit is determined by the machine-protection
conditions applicable to the current physical state
and operational mode. Schematically,
\begin{equation}
\chi_{\rm beam}^{\rm permit}(\boldsymbol X,m)
=
\bigwedge_{j\in\mathcal J_{\rm MPS}(m)}
\chi_j^{\rm MPS}(\boldsymbol X,m),
\label{eq:mps_beam_permit}
\end{equation}
where $\chi_j^{\rm MPS}(\boldsymbol X,m)$
denotes the validity of the $j$th machine-protection
condition, and $\mathcal J_{\rm MPS}(m)$ denotes
the set of protection conditions required for
operational mode $m$. The dependence on $\boldsymbol X$ accounts for
the physical conditions relevant to machine
protection, including beam parameters and
equipment states, while the dependence on $m$
reflects the applicability of particular
protection conditions in different operational
modes. The expression represents the logical
combination of the required protection conditions;
each individual indicator may itself incorporate
more complex interlock logic. Loss of a required permit initiates, or prevents continuation
of, the corresponding operational trajectory.  The BIS
provides the central implementation of this principle by combining
distributed permit information and transmitting a beam-dump request when
the required conditions are no longer satisfied~\cite{Puccio:2010BIS}.

It is important, however, not to identify the beam permit with the complete
safe state of the machine.  A machine with the beam already extracted may
have no beam permit and nevertheless occupy a protected state.  The permit
instead defines whether a particular beam-related operational trajectory
may be entered or maintained.  In the notation of
Sec.~\ref{subsec:interlocks_permits},
\begin{equation}
\Omega_{\rm beam}^{(m)}
=
\Omega_{\rm safe}
\cap
\Omega_{\rm permit}^{(m)}
\cap
\Omega_{\rm auth}^{(m)}
\end{equation}
is the relevant admissible region for the corresponding beam operation.

The LHC MPS can therefore be viewed as a physical
implementation of the trajectory-based safety concept introduced earlier:
measurements determine the evolving accelerator state, protection logic
tests whether the corresponding trajectory remains admissible, and active
or passive barriers prevent unacceptable energy transfer when the
trajectory approaches a critical boundary.  The following subsections
develop this principle for beam losses, fast and slow failure modes,
collimation, beam interlocks and extraction, injection and dump
protection, and the coupling between beam and superconducting-magnet
protection.

A particularly direct illustration is provided by the stored energy of the
circulating beams.  A failure does not make the stored beam energy
intrinsically hazardous; rather, it can redirect a fraction of that energy
into an unintended beam-loss process and subsequently into localized
particle--matter interactions.  The essential physical sequence may be
represented schematically as
\begin{equation}
E_{\rm beam}
\quad
\xrightarrow{\;\text{failure}\;}
\quad
\Delta E_{\rm loss}
\quad
\xrightarrow{\;\text{particle--matter interaction}\;}
\quad
\left\{
\begin{array}{l}
\text{magnet quench},\\
\text{local heating},\\
\text{material damage},\\
\text{activation},\\
\text{secondary radiation}.
\end{array}
\right.
\label{eq:beam_energy_hazard_chain}
\end{equation}
Here $\Delta E_{\rm loss}$ denotes the fraction of the stored beam energy
transferred out of the nominal circulating-beam state during the loss
process.  The resulting consequences depend not only on the total energy
lost, but also on its spatial distribution, temporal structure, and the
properties of the material in which the energy is deposited.

The corresponding protection problem is therefore not to eliminate beam
losses completely---which is neither physically possible nor required for
normal accelerator operation---but to control their location, amplitude,
duration, and consequences.  This simple energy-flow picture will be
generalized below to coupled subsystem states, hazard trajectories, and
protective barriers, and will later provide the physical basis for the
discussion of LHC machine protection in
Sec.~\ref{sec:machine_protection}.

\subsection{Beam loss, energy deposition, and damage mechanisms}
\label{subsec:beam_loss_damage}

The circulating LHC beam constitutes a highly concentrated reservoir of
stored kinetic energy.  Under nominal high-energy operation, the total
beam energy is sufficiently large that only a small fraction of the beam
needs to be lost locally to quench a superconducting magnet or, at higher
deposited-energy densities, to damage accelerator components
\cite{Tahir:2005LHCBeam,Tahir:2009Graphite,Schmidt:2016MPS}.
Machine protection must therefore control not only the total amount of beam
lost, but also the spatial distribution, duration, and material-specific
energy deposition associated with that loss.

In the state-space description introduced previously, a beam-loss event
may be regarded as a transition from the nominal beam state toward a
different coupled accelerator state,
\begin{equation}
{\boldsymbol X}_{\rm beam}^{(0)}
\xrightarrow{\;\mathcal{F}_i\;}
{\boldsymbol X}_{\rm beam}^{\rm loss},
\label{eq:beam_loss_state_transition}
\end{equation}
where $\mathcal{F}_i$ may represent an orbit perturbation, optics error,
powering fault, injection or extraction failure, radio-frequency fault,
beam instability, or another disturbance capable of moving particles out
of the nominal machine acceptance.

The corresponding beam loss is not itself the final consequence.  It
initiates an energy-transfer process from the circulating beam to matter,
\begin{equation}
E_{\rm beam}
\rightarrow
E_{\rm dep}
\rightarrow
\left\{
E_{\rm thermal},
E_{\rm mechanical},
E_{\rm rad},
\ldots
\right\},
\label{eq:beam_energy_deposition_chain}
\end{equation}
where $E_{\rm dep}$ denotes the energy deposited in accelerator material.
This redistribution may produce heating, thermo-mechanical stress,
particle showers, superconducting quenches, material deformation, or
irreversible damage, depending on the magnitude and localization of the
deposition.

The hazard sequence can therefore be written schematically as
\begin{equation}
{\boldsymbol X}_{\rm beam}
\xrightarrow{\;\mathcal{F}_i\;}
{\boldsymbol X}_{\rm loss}
\xrightarrow{\;{\cal E}_{\rm dep}\;}
{\boldsymbol{\cal C}}_i ,
\label{eq:beam_loss_hazard_chain}
\end{equation}
where ${\cal E}_{\rm dep}$ denotes the beam--matter energy-transfer process
and ${\boldsymbol{\cal C}}_i$ is the corresponding consequence vector in
the notation of Sec.~\ref{subsec:hazard_chains}.

The scale separation involved is substantial.  At the nominal LHC design
energy of $7~\mathrm{TeV}$, a full proton beam stores approximately
0.3~GJ (see Eq.~\eqref{eq:lhc_nominal_beam_energy},
corresponding to more than $3\times10^{14}$ circulating protons
\cite{Schmidt:2016MPS,Holzer:2012BLM}.  Uncontrolled deposition of the
full stored beam energy would therefore be highly destructive, but machine
protection must operate far below this extreme limit.

The relevant thresholds are strongly dependent on the spatial and temporal
structure of the loss.  Representative estimates for LHC superconducting
magnets at $7~\mathrm{TeV}$ illustrate the hierarchy: local deposited
energy densities of order
\begin{equation}
\varepsilon_{\rm q}
\sim
10^{-3}~\mathrm{J\,cm^{-3}}
\end{equation}
can be sufficient to induce a quench, whereas local energy densities of
order
\begin{equation}
\varepsilon_{\rm dam}
\sim
1~\mathrm{J\,cm^{-3}}
\end{equation}
enter the regime in which magnet damage becomes possible
\cite{Holzer:2012BLM}.  These values should be regarded as representative
scales rather than universal thresholds: the quench and damage limits
depend on loss duration, beam energy, shower geometry, magnet location,
material properties, and the local superconducting temperature margin.

The smallness of the beam fraction required to produce a relevant
consequence is illustrated experimentally by transient dust-induced losses
observed in the LHC cryogenic arcs during Run~II.  Magnet quenches were
observed for events involving approximately
$6\times10^{7}$ inelastic proton--dust collisions, corresponding to only
about
\begin{equation}
f_{\rm loss}
\sim
2\times10^{-7}
\label{eq:dust_quench_fraction}
\end{equation}
fraction of the maximum stored beam intensity in the data set
\cite{Lechner:2022Dust}.  This number is specific to the highly localized
loss geometry of dust events and should not be interpreted as a universal
quench fraction.  It nevertheless demonstrates directly why protection of
a high-energy hadron collider must be sensitive to losses involving an
extremely small fraction of the circulating beam.

At the opposite end of the scale, losses in the megajoule range are
unambiguously capable of producing severe macroscopic damage.  Historical
high-intensity proton-beam incidents relevant to LHC machine-protection
studies showed substantial damage to vacuum equipment for beam energies of
a few megajoules
\cite{Holzer:2012BLM}.  Relative to the nominal $362~\mathrm{MJ}$ stored
beam energy, a $1~\mathrm{MJ}$ uncontrolled loss corresponds to
\begin{equation}
\frac{\Delta E_{\rm loss}}{E_{\rm beam}^{\rm nom}}
\simeq
2.8\times10^{-3},
\label{eq:one_MJ_beam_fraction}
\end{equation}
or only about $0.3\%$ of the full nominal beam energy.  Thus, depending on
the loss geometry and duration, relevant machine-protection consequences
span a very broad range---from extremely small localized fractions capable
of inducing a superconducting quench to sub-percent beam-energy fractions
capable of causing major material damage.

These examples also show why a beam-loss fraction by itself is not an
adequate measure of hazard.  For beam-induced damage, the relevant physical
quantity is often not the total lost energy alone but the local
energy-deposition density and its characteristic timescale.  One may
write schematically the local deposited energy per unit mass as
\begin{equation}
\varepsilon_{\rm dep}
=
\frac{dE_{\rm dep}}{dM},
\label{eq:specific_energy_deposition}
\end{equation}
where $M$ denotes the mass of the material receiving the deposited energy.
When spatial localization is important, the energy deposit may be better characterized
by local volumetric power density
\begin{equation}
u_{\rm dep}({\boldsymbol r},t)
=
\frac{dE_{\rm dep}}
     {dV\,dt}.
\label{eq:local_energy_deposition}
\end{equation}
The resulting temperature increase may be estimated locally as
\begin{equation}
\Delta T
\simeq
\frac{\varepsilon_{\rm dep}}
     {c_p},
\label{eq:beam_loss_temperature_rise}
\end{equation}
for a sufficiently simple short-time approximation, where $c_p$ is the
specific heat of the irradiated material.  In realistic accelerator
components, phase transitions, temperature-dependent material properties,
particle-shower development, heat transport, and mechanical response must
generally be included explicitly.

For superconducting magnets, the most relevant threshold may be the
quench threshold rather than permanent material damage.  If local energy
deposition raises part of the superconducting cable above the stability
margin, the conductor can undergo a transition to the normal-conducting
state,
\begin{equation}
{\boldsymbol X}_{\rm mag}^{\rm SC}
\xrightarrow{\;E_{\rm dep}\;}
{\boldsymbol X}_{\rm mag}^{\rm quench}.
\label{eq:beam_induced_quench}
\end{equation}
Thus, a beam-loss sequence can couple directly the beam and magnet
subsystems,
\begin{equation}
{\boldsymbol X}_{\rm beam}
\rightarrow
{\boldsymbol X}_{\rm loss}
\rightarrow
{\boldsymbol X}_{\rm mag}
\rightarrow
{\boldsymbol X}_{\rm cryo},
\label{eq:beam_loss_coupling_chain}
\end{equation}
illustrating the inter-subsystem propagation discussed in
Sec.~\ref{subsec:underground_environment_state}.

A quench and irreversible material damage are physically distinct
consequences and generally correspond to different thresholds and
timescales.  Schematically,
\begin{equation}
E_{\rm dep}^{\rm quench}
<
E_{\rm dep}^{\rm damage}
\label{eq:quench_damage_thresholds}
\end{equation}
for many superconducting-magnet loss scenarios, although the precise
thresholds depend strongly on loss duration, spatial distribution,
material, cooling conditions, and beam energy.  The relevant protection
criterion must therefore always be associated with a specified consequence
rather than with a universal beam-loss threshold.

This distinction is naturally expressed through the consequence vector,
\begin{equation}
{\boldsymbol{\cal C}}_{\rm loss}
=
\left(
{\cal C}_{\rm quench},
{\cal C}_{\rm damage},
{\cal C}_{\rm rad},
{\cal C}_{\rm downtime},
\ldots
\right),
\label{eq:beam_loss_consequence_vector}
\end{equation}
whose components need not become significant at the same deposited energy
or on the same timescale.  A loss that is sufficient to quench a magnet
may remain below the threshold for material damage, while another loss
pattern may generate intense localized heating or mechanical stress.

The response time of the loss process is therefore an essential machine
protection variable.  Different failure mechanisms produce losses with
widely different characteristic timescales.  For a given failure mode,
one may define a consequence-specific available time
\begin{equation}
\tau_{\rm avail}^{(k)}
=
t_{\rm crit}^{(k)}-t_0,
\label{eq:beam_loss_available_time}
\end{equation}
where $t_{\rm crit}^{(k)}$ is the earliest time at which consequence $k$
can no longer be prevented by the protection function under consideration.
The active protection condition introduced in
Sec.~\ref{subsec:protection_time} then becomes
\begin{equation}
\tau_{\rm prot}
<
\tau_{\rm avail}^{(k)}.
\label{eq:beam_loss_protection_condition}
\end{equation}

This formulation is especially important for beam losses because the same
stored beam can generate qualitatively different hazards depending on the
loss mechanism.  Slowly developing losses may permit detection by beam
instrumentation followed by beam extraction.  Very fast failures, by
contrast, may evolve too rapidly for a complete active
detection--decision--actuation chain, making passive absorbers and
collimation elements indispensable
\cite{Bartmann:2010InjectionDump,Burkhardt:2003TransferProtection,
Kain:2005InjectionProtection}.

The preventive--mitigative distinction of
Sec.~\ref{subsec:preventive_mitigative} is therefore particularly useful
for beam-loss protection.  An interlock-triggered beam dump acts
predominantly preventively when it removes the circulating beam before an
incipient fault develops into damaging losses,
\begin{equation}
{\boldsymbol X}_{\rm beam}^{\rm abnormal}
\xrightarrow{\;B_{\rm prev}\;}
{\boldsymbol X}_{\rm beam}^{\rm extracted}.
\label{eq:beam_dump_preventive}
\end{equation}
Collimators and absorbers, on the other hand, provide a mitigative function
when they intercept particles that have already left the nominal beam
phase space and direct the resulting energy deposition toward components
designed to withstand it,
\begin{equation}
{\cal E}_{\rm loss}^{\rm unprotected}
\xrightarrow{\;B_{\rm mit}\;}
{\cal E}_{\rm loss}^{\rm controlled}.
\label{eq:beam_loss_mitigation}
\end{equation}

The distinction is not absolute.  A collimator can be preventive with
respect to a downstream superconducting-magnet quench while simultaneously
being mitigative with respect to the unavoidable loss of halo particles.
Its classification therefore depends on the particular hazard sequence and
consequence being considered.

The machine-protection problem is consequently not to eliminate all beam
losses.  Some losses are intrinsic to high-energy accelerator operation.
Rather, the objective is to ensure that the trajectories generated by
normal and abnormal losses remain within the domain for which the
resulting energy deposition is acceptable by the MPS,
\begin{equation}
\Gamma_{\rm loss}
\subset
\Gamma_{\rm allowed}^{\rm MPS},
\label{eq:allowed_beam_loss_trajectory}
\end{equation}
or, when this condition can no longer be guaranteed, to redirect the
machine toward a protected state before the relevant critical boundary is
crossed.

This viewpoint also clarifies why beam-loss monitoring and collimation are
complementary rather than competing protection functions.  Beam-loss
monitors provide information about the developing loss trajectory and can
trigger active protection, while the collimation system determines where
a substantial fraction of the unavoidable beam halo and abnormal losses
is intercepted.  Their combined role is therefore both diagnostic and
physical: one system measures the approach to an unacceptable state,
while the other shapes the energy-deposition pathway itself
\cite{Holzer:2005BLM,Redaelli:2016Collimation}.

The following subsections consider how the characteristic timescale of the
initiating failure determines whether active or passive protection is
required, and how the LHC collimation and beam-interlock systems implement
these principles in practice.

\subsection{Fast and slow failures: the hierarchy of protection timescales}
\label{subsec:fast_slow_failures}

The effectiveness of machine protection depends critically on the
timescale over which an initiating failure drives the beam toward an
unacceptable state.  The distinction between ``fast'' and ``slow''
failures should therefore not be regarded as an absolute classification
of accelerator faults.  Rather, it expresses the relation between the
characteristic evolution of a particular hazard sequence and the time
available for the corresponding protection function to act.

A natural reference timescale is the LHC revolution period, Eq.~\eqref{eq:lhc_revolution_time}.
Thus, a failure that significantly changes the beam trajectory within only
a few turns can evolve on a timescale of a few hundred microseconds,
whereas other failures may develop over milliseconds, seconds, or much
longer periods
\cite{Schmidt:2016MPS,Wenninger:2016MPO}.

For machine-protection purposes, it is useful to distinguish schematically
three temporal regimes:
\begin{equation}
\begin{array}{lll}
\text{ultra-fast failures:}
&
\tau_{\rm fail}\lesssim 3T_{\rm rev}

\sim 270~\mu{\rm s},
&
\\[1ex]
\text{fast failures:}
&
3T_{\rm rev}\lesssim\tau_{\rm fail}
\lesssim \text{a few ms},
&
\\[1ex]
\text{slow failures:}
&
\tau_{\rm fail}\gtrsim \text{a few ms}.
&
\end{array}
\label{eq:failure_time_classes}
\end{equation}
These boundaries are intentionally approximate.  The physically relevant
quantity is not the nominal failure time alone but the consequence-specific
available time \(\tau_{\rm avail}^{(k)}\) (see Eq.~\eqref{eq:beam_loss_available_time})
introduced in Sec.~\ref{subsec:protection_time}.

The finite response time of an active protection chain provides the
physical origin of this classification.  Representative fast LHC
protection systems operate on timescales well below one millisecond.
Fast magnet current change monitors can detect certain powering failures
on a timescale of approximately
\begin{equation}
\tau_{\rm FMCCM}
\simeq
20~\mu{\rm s},
\label{eq:fmcm_detection_time}
\end{equation}
while the overal characteristic beam loss monitor (BLM) response is of order
\begin{equation}
\tau_{\rm BLM}^{\rm overal}
\simeq
80~\mu{\rm s}.
\label{eq:blm_fast_detection_time}
\end{equation}
Once a dump request has reached the interlock system, the BIS and the LBDS require less than approximately
\begin{equation}
\tau_{\rm BIS+LBDS}
\lesssim
280~\mu{\rm s}
\simeq
3T_{\rm rev}
\label{eq:bis_lbds_response_time}
\end{equation}
to complete removal of the affected beam from the ring
\cite{Schmidt:2016MPS}.

The complete protection time is therefore not simply the detection time.
As discussed in Sec.~\ref{subsec:protection_time}, it contains the
successive contributions described in Eq.~\eqref{eq:protection_latency}.
For beam extraction, these terms include detection of the developing
failure, generation and transmission of the dump request, synchronization
with the abort gap, firing of the extraction system, and removal of the
beam.

A useful practical scale follows from this latency budget.  Failures for
which critical beam losses do not develop in less than approximately
\begin{equation}
\tau_{\rm avail}
\sim
1~{\rm ms}
\sim
10T_{\rm rev}
\label{eq:one_ms_active_protection}
\end{equation}
can, in principle, be addressed by sufficiently fast active interlocking
and beam extraction.  This does not mean that every failure with a
timescale longer than $1~{\rm ms}$ is automatically safe: the required
condition remains
\begin{equation}
\tau_{\rm prot}
<
\tau_{\rm avail}^{(k)}
\label{eq:fast_failure_protection_condition}
\end{equation}
for the particular hazard path and consequence under consideration.

For slower failures, the available margin is generally larger.  Failures
evolving over tens of turns or more---for example,
\begin{equation}
20T_{\rm rev}
\simeq
1.8~{\rm ms},
\label{eq:twenty_turn_timescale}
\end{equation}
or longer---allow beam surveillance systems such as the BLM system sufficient time to detect the developing abnormal
condition and request a beam dump before the corresponding critical
boundary is reached
\cite{Schmidt:2016MPS}.  Many powering failures, magnet quenches,
radio-frequency failures, and slowly developing orbit or beam-instability
conditions belong to this general class, although their detailed
protection times vary substantially.

The situation changes qualitatively for failures developing on the
timescale of only one to three turns,
\begin{equation}
T_{\rm rev}
\lesssim
\tau_{\rm fail}
\lesssim
3T_{\rm rev}
\simeq
270~\mu{\rm s}.
\label{eq:ultrafast_failure_timescale}
\end{equation}
In this regime, observation of the resulting beam motion or beam loss may
occur too late for an ordinary detection--interlock--dump sequence to
prevent the first potentially damaging impact.  Active equipment
surveillance can still be useful if it detects the initiating equipment
fault before the beam itself reaches the critical trajectory, but active
beam observation alone cannot in general provide sufficient protection.

This distinction can be expressed directly in the framework of
Sec.~\ref{subsec:protection_time}:
\begin{equation}
\begin{array}{ll}
\tau_{\rm avail}^{(k)}
>
\tau_{\rm prot}
&
\Rightarrow
\text{active protection can in principle interrupt the sequence},
\\[1ex]
\tau_{\rm avail}^{(k)}
\lesssim
\tau_{\rm prot}^{\rm min}
&
\Rightarrow
\text{passive or intrinsic protection is required}.
\end{array}
\label{eq:fast_slow_active_passive}
\end{equation}

Ultra-fast failures therefore represent a particularly demanding class.
Examples include abnormal injection or extraction kicker behaviour and
other failures capable of changing the beam trajectory within one or a few
turns.  Because the beam may reach vulnerable equipment before an active
protection sequence can be completed, passive absorbers, collimators, and
dedicated injection- and extraction-protection devices must be positioned
so that the resulting energy deposition occurs preferentially in
components designed to tolerate it
\cite{Bartmann:2010InjectionDump,
Burkhardt:2003TransferProtection,Kain:2005InjectionProtection}.

This provides a direct example of the preventive--mitigative distinction
introduced in Sec.~\ref{subsec:preventive_mitigative}.  For a relatively
slow failure, early detection followed by beam extraction can act
preventively:
\begin{equation}
{\boldsymbol X}_{\rm beam}^{\rm abnormal}
\xrightarrow{\;\text{detection/interlock}\;}
{\boldsymbol X}_{\rm beam}^{\rm extracted},
\label{eq:slow_failure_prevention}
\end{equation}
before significant uncontrolled energy deposition occurs.  For an
ultra-fast failure, the first abnormal beam impact may be unavoidable;
passive protection must then control the location and consequences of that
impact,
\begin{equation}
{\cal E}_{\rm loss}^{\rm uncontrolled}
\xrightarrow{\;B_{\rm passive}\;}
{\cal E}_{\rm loss}^{\rm controlled}.
\label{eq:ultrafast_passive_mitigation}
\end{equation}

Fast failures also illustrate why the time associated with the initiating
equipment failure should not automatically be identified with
$\tau_{\rm avail}$.  A magnet-current change, for example, may begin at
$t_0$, while the resulting orbit displacement requires several additional
turns to drive the beam into a limiting aperture.  Conversely, a kicker
misfire can alter the beam trajectory essentially immediately.  The
relevant protection time is therefore the interval between the earliest
detectable precursor and the particular critical beam state,
\begin{equation}
\tau_{\rm avail}^{(k)}
=
t_{\rm crit}^{(k)}
-
t_{\rm detectable},
\label{eq:detectable_to_critical_time}
\end{equation}
rather than a universal characteristic time assigned to the hardware
failure itself.

This distinction is important because different observables can provide
different amounts of warning.  Detection of a powering anomaly at the
equipment level may precede measurable beam losses by several turns,
whereas a BLM necessarily responds after particles have
already begun to leave the nominal circulating-beam state.  The diverse
observables discussed in Sec.~\ref{subsec:barrier_independence} can
therefore increase the available protection margin by detecting different
stages of the same hazard trajectory.

The temporal hierarchy of machine protection can consequently be
summarized schematically as
\begin{equation}
\begin{array}{c}
\text{equipment precursor}
\\
\downarrow
\\
\text{beam perturbation}
\\
\downarrow
\\
\text{beam loss}
\\
\downarrow
\\
\text{critical energy deposition}
\end{array}
\label{eq:failure_temporal_hierarchy}
\end{equation}
with different protection systems acting at different positions along this
sequence.  The earlier a reliable precursor can be detected, the larger
the effective $\tau_{\rm avail}$ for active intervention.

The distinction among ultra-fast, fast, and slow failures is therefore
not merely a convenient classification of hardware faults.  It determines
which combinations of active surveillance, interlocks, beam extraction,
and passive protection can physically modify the resulting trajectory.
This temporal structure provides the basis for the complementary roles of
the LHC collimation system, BLM, BIS, and
LBDS discussed in the following subsections.

\subsection{Collimation as a machine-protection system}
\label{subsec:collimation_machine_protection}

The LHC collimation system provides a direct physical realization of the
principle introduced in Sec.~\ref{subsec:beam_loss_damage}: unavoidable
beam losses should, as far as possible, be transferred from vulnerable
machine elements to locations specifically designed to intercept and
withstand them.  Collimation therefore does not eliminate the loss process.
Instead, it modifies its spatial development and the resulting
energy-deposition pathway
\cite{Redaelli:2016Collimation,Schmidt:2016MPS}.

The corresponding protection function may be represented schematically as
\begin{equation}
{\boldsymbol X}_{\rm halo}
\longrightarrow
{\boldsymbol X}_{\rm loss}
\xrightarrow{\;{\cal P}_{\rm coll}\;}
{\boldsymbol X}_{\rm loss}^{\rm controlled},
\label{eq:collimation_state_mapping}
\end{equation}
where ${\cal P}_{\rm coll}$ denotes the action of the collimation system in
redirecting beam losses toward designated absorbing structures.  The
resulting consequence vector is then changed from that associated with an
uncontrolled loss pattern,
\begin{equation}
{\boldsymbol{\cal C}}_{\rm loss}^{\rm uncoll}
\xrightarrow{\;{\cal P}_{\rm coll}\;}
{\boldsymbol{\cal C}}_{\rm loss}^{\rm coll}.
\label{eq:collimation_consequence_mapping}
\end{equation}

In this sense, collimation is predominantly a mitigative protection
function with respect to unavoidable beam losses: particles have already
left the nominal beam core, but their subsequent trajectories and energy
deposition are controlled.  At the same time, the same system acts
preventively with respect to downstream consequences such as
superconducting-magnet quenches or material damage.  This illustrates the
point made in Sec.~\ref{subsec:preventive_mitigative} that the classification
of a barrier as preventive or mitigative depends on the particular hazard
sequence and consequence under consideration.

The LHC uses a multistage collimation hierarchy
\cite{Redaelli:2016Collimation}.  Primary collimators define the first
controlled aperture encountered by beam-halo particles.  Their interaction
with the halo increases particle amplitudes through scattering and energy
loss, allowing secondary collimators to intercept a much larger fraction
of the scattered population.  Downstream absorbers capture secondary
particle showers and further reduce leakage toward superconducting
magnets and other vulnerable equipment.  Additional collimators near the
experimental insertions protect the low-$\beta$ triplet magnets and help
control collision-related backgrounds.

The hierarchy may be represented schematically as
\begin{equation}
\text{beam halo}
\rightarrow
\text{primary collimator}
\rightarrow
\text{secondary collimators}
\rightarrow
\text{absorbers}
\rightarrow
\text{residual leakage}.
\label{eq:collimation_hierarchy}
\end{equation}
The successive stages are arranged so that the effective machine aperture
is deliberately defined by robust collimating elements rather than by
sensitive superconducting components.
Two dedicated cleaning insertions provide the principal halo-cleaning
functions.  Interaction Region~7 (IR7) performs primarily
\emph{betatron cleaning}, while Interaction Region~3 (IR3) performs
primarily \emph{momentum cleaning}.\footnote{The LHC ring is conventionally
divided into eight long straight sections, referred to as Interaction
Regions (IRs), associated with the eight insertion points around the
27-km circumference.  IR7 is the main betatron-cleaning insertion.  It is
optimized to intercept particles whose transverse oscillation amplitudes
have grown sufficiently large that they approach the machine aperture,
even when their momenta remain close to the nominal beam momentum.  IR3,
by contrast, is the main momentum-cleaning insertion.  Its optics provides
significant horizontal dispersion, so that particles with sufficiently
large relative momentum deviations are displaced transversely from the
reference orbit and can be intercepted by the collimation system.  Thus,
in simplified terms, IR7 primarily removes particles that are too far from
the beam core in transverse phase space, whereas IR3 primarily removes
particles that are too far from the nominal momentum.}
These two functions address physically different mechanisms by which
particles can leave the stable circulating-beam phase space
\cite{Redaelli:2016Collimation}.

For transverse halo cleaning, the physical sequence is approximately
\begin{equation}
{\boldsymbol X}_{\rm halo}
\xrightarrow{\rm TCP}
{\boldsymbol X}_{\rm scattered}
\xrightarrow{\rm TCS}
{\boldsymbol X}_{\rm intercepted}
\xrightarrow{\rm TCLA}
{\boldsymbol X}_{\rm absorbed},
\label{eq:multistage_collimation_sequence}
\end{equation}
where TCP, TCS, and TCLA denote the primary, secondary, and absorber
collimator families, respectively.  The notation in
Eq.~\eqref{eq:multistage_collimation_sequence} is schematic: particle
showers and repeated turns make the actual dynamics more complex than a
strictly sequential single-pass process.

A useful measure of the overall cleaning performance is the fraction of
halo losses prevented from reaching sensitive downstream regions.  In a
schematic global form one may write
\begin{equation}
\epsilon_{\rm clean}
=
1-
\frac{N_{\rm leak}}
     {N_{\rm halo}},
\label{eq:collimation_cleaning_efficiency}
\end{equation}
where $N_{\rm halo}$ represents particles entering the relevant cleaning
process and $N_{\rm leak}$ those escaping into regions that the
collimation system is intended to protect.  In accelerator operation,
performance is generally characterized more locally through cleaning
inefficiencies and BLM signals rather than through a single
global number.

Representative LHC measurements have demonstrated cleaning efficiencies
exceeding
\begin{equation}
\epsilon_{\rm clean}
>
99.99\% ,
\label{eq:collimation_efficiency_scale}
\end{equation}
with a measured value of approximately $99.993\%$ reported for a
representative IR7 loss configuration
\cite{Redaelli:2016Collimation}.  Equivalently, only a few parts in
$10^{5}$ of the relevant losses leaked toward the limiting cold regions
in that particular measurement.  Such values should not be interpreted
as universal constants of the collimation system: the local cleaning
performance depends on beam energy, optics, collimator settings, orbit,
loss mechanism, and the location at which leakage is evaluated.

The need for such high cleaning efficiency follows directly from the
extreme hierarchy of energy scales discussed in
Sec.~\ref{subsec:beam_loss_damage}.  A superconducting magnet can quench
after receiving only a minute fraction of the energy stored in the
circulating beam.  Consequently, even a cleaning system intercepting more
than $99.99\%$ of the relevant halo losses must control carefully where the
remaining fraction is deposited.

The collimator hierarchy is therefore also an aperture
hierarchy. Each collimator consists of movable jaws
positioned at specified transverse distances from
the circulating beam. These distances are conventionally
expressed in units of the local rms transverse beam size
$\sigma$, rather than directly in millimetres.

For a collimator family $k$, its normalized setting
$n_k$ represents the distance between the reference
beam orbit and the relevant jaw surface, divided by
the local beam size in the corresponding transverse
plane. Thus, $n_k=6$, for example, means that the
jaw surface is positioned six local rms beam sizes
from the beam orbit. The physical distance depends
on the local accelerator optics and beam emittance:
in the horizontal plane,
$\sigma_x=\sqrt{\beta_x\varepsilon_x}$,
where $\beta_x$ is the horizontal betatron function
and $\varepsilon_x$ the geometric beam emittance.
An analogous relation applies in the vertical plane.

The successive collimator families are positioned
at increasing normalized distances from the beam
core. Schematically, the required hierarchy is
\begin{equation}
n_{\rm primary}
<
n_{\rm secondary}
<
n_{\rm absorber}
<
n_{\rm protected\ aperture},
\label{eq:collimation_aperture_hierarchy}
\end{equation}
where $n_{\rm primary}$ denotes the normalized
setting of the primary collimators (TCP), which
define the smallest controlled transverse aperture
and intercept particles diffusing out of the beam
core. The secondary collimators (TCS), positioned
farther from the beam, intercept particles whose
transverse amplitudes have increased following
scattering at the primary collimators.
The absorber collimators (TCLA), with larger
normalized openings, intercept additional
scattered particles and secondary showers.

The final quantity,
$n_{\rm protected\ aperture}$, does not denote
another collimator setting. It represents
schematically the effective normalized aperture
of vulnerable accelerator equipment, particularly
superconducting magnets, that must remain
protected from uncontrolled beam losses.
The hierarchy is arranged so that the
collimators intercept the relevant halo
population before it reaches these sensitive
apertures.

A smaller value of $n_k$ consequently means
that the corresponding collimator jaw is
positioned closer to the beam in units of
the local beam size. The hierarchy is
expressed in normalized coordinates because
the physical beam size varies around the ring
with the optical functions. Two collimators
having different openings in millimetres
can therefore occupy the same normalized
position relative to the circulating beam.

Equation~\eqref{eq:collimation_aperture_hierarchy}
is a schematic representation of the principal
multistage cleaning hierarchy, rather than a
complete ordering of every collimator in the
LHC. Additional protection collimators,
including tertiary collimators (TCT) near
the experimental insertions, are positioned
according to their specific functions and
the apertures they must protect.
The actual settings and required separations
between successive stages depend on beam
energy, emittance, optics, orbit, and the
operational configuration. Sufficient margins
must also be maintained to accommodate
relevant uncertainties and orbit or optics
variations
\cite{Redaelli:2016Collimation}.

The hierarchy Eq.~~\eqref{eq:collimation_aperture_hierarchy} converts an otherwise uncontrolled aperture limitation into
a deliberately engineered one:
\begin{equation}
\text{machine aperture}
\quad\longrightarrow\quad
\text{controlled collimator-defined aperture}.
\label{eq:controlled_aperture}
\end{equation}
In the trajectory-based language developed earlier, the collimators
therefore constrain which loss trajectories remain physically accessible
to halo particles.

The protection function of collimation is not limited to normal halo
cleaning.  Collimators also contribute to passive machine protection by
shielding sensitive apertures against certain abnormal beam trajectories.
This function becomes especially important for the fast and ultra-fast
failure regimes discussed in Sec.~\ref{subsec:fast_slow_failures}, where
the first abnormal beam impact may occur before an active
detection--interlock--dump sequence can be completed.

It is nevertheless important not to interpret the standard collimation
system as a universal absorber capable of safely intercepting every
possible full-beam accident.  Different collimator materials and designs
have different robustness limits, and specific injection and extraction
failure scenarios require dedicated passive protection devices.  These
specialized systems are discussed separately in the context of injection
and dump protection.

Collimation is also intimately coupled to active machine protection.
Beam-loss monitors located around the collimation regions measure the
secondary-particle showers produced by intercepted beam and provide
information on both normal cleaning and abnormal loss patterns.  If the
measured losses exceed predefined thresholds, the beam loss monitoring
system can request beam extraction.  The combined protection chain may
therefore be written schematically as
\begin{equation}
\text{halo or abnormal loss}
\rightarrow
\text{collimator interception}
\rightarrow
\text{BLM measurement}
\rightarrow
\text{interlock decision}
\rightarrow
\text{beam extraction},
\label{eq:collimation_blm_chain}
\end{equation}
where the first stages are primarily physical and passive, while the later
stages constitute active protection.

The collimation system must consequently satisfy two complementary
requirements.  During normal operation it must intercept unavoidable halo
losses with sufficiently high efficiency to prevent quenches and excessive
backgrounds, while during abnormal operation it must contribute to the
protection of the machine aperture against trajectories that cannot be
corrected or extracted rapidly enough.

This dual function can be summarized as
\begin{equation}
\begin{array}{c}
\text{unavoidable beam loss}
\\
\downarrow
\\
\text{controlled interception}
\\
\downarrow
\\
\text{localized energy deposition in robust elements}
\\
\downarrow
\\
\text{suppressed leakage toward vulnerable equipment}
\end{array}
\label{eq:collimation_protection_logic}
\end{equation}

In terms of the allowed-trajectory formulation introduced in
Sec.~\ref{subsec:beam_loss_damage}, the collimation system acts to enforce
\begin{equation}
\Gamma_{\rm loss}^{\rm coll}
\subset
\Gamma_{\rm allowed}^{\rm MPS},
\label{eq:collimation_allowed_trajectory}
\end{equation}
where $\Gamma_{\rm allowed}^{\rm MPS}$ denotes the set of beam-loss
trajectories compatible with the constraints imposed by the MPS.
If the measured loss pattern indicates that this
condition can no longer be maintained, active protection must terminate
the circulating-beam trajectory through beam extraction.

Collimation therefore provides the physical link between beam dynamics and
machine protection: it transforms an uncontrolled loss distribution into a
designed energy-deposition pattern, while beam-loss monitoring determines
whether the residual losses remain compatible with continued operation.
This connection motivates the discussion of beam-loss monitoring and
interlock thresholds in the following subsection.

\subsection{Beam-loss monitoring and threshold-based protection}
\label{subsec:beam_loss_monitoring}

The collimation system discussed in
Sec.~\ref{subsec:collimation_machine_protection} determines, to a large
extent, where unavoidable beam losses are intercepted.  Machine protection
also requires continuous information on whether the resulting loss pattern
remains compatible with the tolerable energy deposition in accelerator
components.  This function is provided primarily by the LHC BLM system
\cite{Holzer:2012BLM,Dehning:2016BLM}.

A beam-loss monitor does not in general measure the number of lost
circulating protons directly.  When high-energy particles strike accelerator
material they initiate hadronic and electromagnetic showers.  The BLM
detectors sample the secondary-particle field produced by these
interactions, providing a signal correlated with the underlying beam loss
and with the corresponding energy deposition in nearby equipment.  The
measurement chain may therefore be represented schematically as
\begin{equation}
{\boldsymbol X}_{\rm beam}
\rightarrow
{\boldsymbol X}_{\rm loss}
\rightarrow
{\cal E}_{\rm dep}
\rightarrow
\text{secondary-particle shower}
\rightarrow
S_{\rm BLM},
\label{eq:blm_measurement_chain}
\end{equation}
where $S_{\rm BLM}$ denotes the detector signal associated with the local
loss process.

This distinction is important.  The protection quantity is ultimately the
energy deposition and its possible consequences, whereas the BLM system
observes a measurable proxy for that process.  The relation
\begin{equation}
S_{\rm BLM}
\longleftrightarrow
E_{\rm dep}
\label{eq:blm_energy_deposition_relation}
\end{equation}
therefore depends on the monitor position, accelerator geometry, beam
energy, loss location, shower development, and the temporal structure of
the loss.  Establishing appropriate protection thresholds consequently
requires simulations, dedicated loss measurements, quench tests where
appropriate, and operational validation
\cite{Holzer:2012BLM,Dehning:2016BLM}.

A central feature of the LHC BLM architecture is that losses are evaluated
simultaneously over a broad hierarchy of timescales.  The detector signals
are integrated over 12 characteristic time windows extending approximately
from
\begin{equation}
\tau_{\rm BLM}^{\rm min}
\simeq
40~\mu{\rm s}
\label{eq:blm_shortest_window}
\end{equation}
to
\begin{equation}
\tau_{\rm BLM}^{\rm max}
\simeq
84~{\rm s}.
\label{eq:blm_longest_window}
\end{equation}
The shortest interval corresponds to approximately one half of the LHC
revolution period,
whereas the longest intervals allow slowly accumulating losses to be
identified
\cite{Holzer:2012BLM}.

The broad range in
Eqs.~\eqref{eq:blm_shortest_window}  and \eqref{eq:blm_longest_window}
reflects the physical diversity of beam-loss processes.  A very intense
short loss can threaten equipment within a fraction of a millisecond,
whereas a much smaller but persistent loss can generate excessive
integrated heating or eventually quench a superconducting magnet.  No
single integration time can therefore provide adequate protection against
all relevant loss mechanisms.

For a monitor $i$, integration window $w$, and beam-energy regime $E$, the
protection decision may be written schematically as
\begin{equation}
S_{i,w}
=
\int_{t-\Delta t_w}^{t}
s_i(t')\,dt',
\label{eq:blm_integrated_signal}
\end{equation}
followed by the comparison
\begin{equation}
S_{i,w}
>
S_{i,w}^{\rm th}(E)
\quad
\Longrightarrow
\quad
D_{i,w}=1,
\label{eq:blm_threshold_condition}
\end{equation}
where $s_i(t)$ is the instantaneous detector signal,
$\Delta t_w$ the corresponding integration interval,
$S_{i,w}^{\rm th}$ the predefined protection threshold, and
$D_{i,w}=1$ denotes a dump condition generated by that monitor and
integration window.

The threshold should therefore not be regarded as a single universal
``maximum beam loss.''  Rather, it belongs to a multidimensional
protection configuration,
\begin{equation}
S^{\rm th}
=
S^{\rm th}
\left(
i,
\Delta t_w,
E_{\rm beam},
{\boldsymbol\Xi}_{\rm mach},
\ldots
\right),
\label{eq:blm_threshold_dependence}
\end{equation}
where $\boldsymbol\Xi_{\rm mach}$ denotes the
relevant accelerator technical configuration,
including the applicable machine optics,
collimator settings, and equipment configuration.
The dependence of the BLM thresholds on this
configuration reflects the fact that the
relationship between measured beam losses
and the resulting energy deposition in
vulnerable equipment varies with the
machine configuration. In the original LHC implementation, threshold values were
organized for the different integration windows and for 32 beam-energy
ranges~\cite{Holzer:2012BLM}.

This dependence connects directly to the consequence-specific description
developed in Sec.~\ref{subsec:beam_loss_damage}.  A BLM threshold is not a
fundamental boundary of the beam state itself.  It is an operational
representation of the requirement that the measured loss remain below a
level expected to produce an unacceptable consequence in the protected
equipment.  Schematically,
\begin{equation}
S_{i,w}
<
S_{i,w}^{\rm th}
\quad\Longrightarrow\quad
{\cal E}_{\rm dep}
\ \text{remains within the validated protection domain},
\label{eq:blm_validated_domain}
\end{equation}
subject to the assumptions and uncertainties entering the loss-to-signal
and signal-to-consequence mappings.

If a threshold is exceeded, the BLM system generates a beam-dump request
that is transmitted to the BIS.  The corresponding
active protection chain can be represented as
\begin{equation}
\text{beam loss}
\rightarrow
S_{\rm BLM}
\rightarrow
\text{integration}
\rightarrow
\text{threshold comparison}
\rightarrow
\text{dump request}
\rightarrow
\text{BIS}
\rightarrow
\text{beam extraction}.
\label{eq:blm_interlock_chain}
\end{equation}
Thus, the BLM system provides the measurement and decision interface
between the physical beam-loss process and the logical permit architecture
introduced in Sec.~\ref{subsec:interlocks_permits}.

The timing requirement of Sec.~\ref{subsec:protection_time} applies
directly.  For a loss process that can be actively intercepted before the
critical consequence develops, the complete response must satisfy
\begin{equation}
\tau_{\rm BLM}
+
\tau_{\rm BIS}
+
\tau_{\rm dump}
<
\tau_{\rm avail}^{(k)},
\label{eq:blm_timing_condition}
\end{equation}
where $\tau_{\rm BLM}$ includes detection, integration, and threshold
evaluation, $\tau_{\rm BIS}$ denotes communication through the interlock
chain, and $\tau_{\rm dump}$ represents the time required to synchronize
and extract the beam.

The shortest BLM integration window of approximately $40~\mu{\rm s}$ does
not therefore imply that a hazardous beam can always be removed within
$40~\mu{\rm s}$.  As discussed in
Sec.~\ref{subsec:fast_slow_failures}, additional communication,
synchronization, kicker, and extraction times are required.  The complete
machine-protection response after failure detection is consequently of
order several LHC turns.  This is why ultra-fast failures remain dependent
on passive protection even though the BLM system itself can detect losses
on a sub-turn timescale.

The relationship between collimation and beam-loss monitoring is
particularly close.  Collimators deliberately intercept halo particles
and thereby generate secondary-particle showers that are observed by
nearby BLMs.  BLM signals are consequently used not only for emergency
protection but also for collimator setup, loss maps, and verification of
cleaning performance
\cite{Holzer:2012BLM,Redaelli:2016Collimation}.  The combined system forms
a feedback between the physical and diagnostic protection layers,
\begin{equation}
\text{halo}
\rightarrow
\text{collimator}
\rightarrow
\text{controlled loss}
\rightarrow
\text{BLM}
\rightarrow
\left\{
\begin{array}{l}
\text{validated continued operation},\\
\text{or}\\
\text{beam extraction}.
\end{array}
\right.
\label{eq:collimation_blm_feedback}
\end{equation}

The BLM architecture also illustrates the fail-safe principles discussed
in Sec.~\ref{subsec:failsafe_protection}.  Protection does not depend only
on detecting excessive beam losses; the monitoring system must also
establish that its own protection channels remain functional.  Internal
validation and integrity tests are therefore part of the protection
architecture, and certain detected failures of the BLM system themselves
result in withdrawal of the conditions required for continued beam
operation
\cite{Holzer:2012BLM,Wenninger:2016MPO}.

Schematically,
\begin{equation}
\left[
S_{i,w}>S_{i,w}^{\rm th}
\right]
\;\lor\;
\left[
\chi_{\rm BLM}^{\rm valid}=0
\right]
\quad\Longrightarrow\quad
\chi_{\rm beam}^{\rm permit}\rightarrow 0,
\label{eq:blm_loss_or_invalid}
\end{equation}
where $\chi_{\rm BLM}^{\rm valid}$
is a binary indicator representing the validity
of the BLM protection function required for
the current operating conditions, while $\chi_{\rm beam}^{\rm permit}=0$ indicates
withdrawal of the beam permit.

Equation~\eqref{eq:blm_loss_or_invalid}
expresses the schematic machine-protection
logic: either a beam-loss signal exceeding
its applicable threshold or the loss of
a required BLM protection function can
lead to withdrawal of the beam permit.
It does not imply that every diagnostic
warning necessarily initiates beam extraction;
the response depends on the corresponding
protection logic and operational configuration.

This self-monitoring requirement is important because a BLM channel that
silently ceases to detect losses could constitute a latent dangerous
failure in the sense of Sec.~\ref{subsec:barrier_independence}.  Diagnostic
coverage, periodic integrity tests, and redundant signal-processing paths
therefore contribute directly to the reliability of the protection
function.

At the same time, protection thresholds cannot be made arbitrarily low.
Excessively conservative settings could generate unnecessary beam dumps
from losses that do not threaten the machine, thereby reducing accelerator
availability.  Threshold design therefore embodies the safety--availability
balance discussed in Sec.~\ref{subsec:safety_availability}:
\begin{equation}
\text{sufficient sensitivity to dangerous losses}
\quad\leftrightarrow\quad
\text{acceptable rate of unnecessary dumps}.
\label{eq:blm_safety_availability}
\end{equation}
The appropriate operating point is constrained primarily by protection of
the machine, but it must also permit stable and efficient physics
operation.

From the trajectory perspective developed throughout
this manuscript, the BLM system can be interpreted
as an observer of the beam-loss component of the
facility state. Its function is to determine, from
measured secondary-particle signals and their
temporal evolution, whether the developing beam-loss
conditions remain compatible with the machine-protection
requirements defining $\mathcal G_{\rm adm}^{\rm MPS}$.

When a monitored loss exceeds the applicable
protection threshold, the BLM system generates
a beam-dump request signal. Schematically,
\begin{equation}
S_{\rm BLM}(t)
\rightarrow
\chi_{\rm dump}^{\rm req}=1
\rightarrow
\chi_{\rm beam}^{\rm permit}=0
\rightarrow
\boldsymbol X_{\rm beam}^{\rm extracted},
\label{eq:blm_trajectory_to_dump}
\end{equation}
where $\chi_{\rm dump}^{\rm req}$
denotes the binary beam-dump request indicator. The final term
represents the beam state following successful
extraction. The relation is schematic: beam
extraction is achieved through the subsequent
response of the beam-dumping system, rather
than instantaneously upon permit withdrawal.

Beam-loss monitoring therefore provides a particularly clear example of
the general protection logic introduced in
Sec.~\ref{subsec:energy_control_principle}:
measurement of the physical state is converted into a protection decision,
and that decision constrains the subsequent trajectory of a high-energy
system.  The Beam Interlock System considered next provides the logical
infrastructure through which this and other distributed protection
requests are combined into the global beam-permit architecture.

\subsection{Beam Interlock System and the beam-permit architecture}
\label{subsec:bis_machine_protection}

The BLM system discussed in
Sec.~\ref{subsec:beam_loss_monitoring} provides one of the principal
measurements from which a machine-protection decision can originate.
The subsequent transmission and combination of such protection decisions
is the function of the LHC BIS.  The BIS therefore
does not normally diagnose the underlying physical failure itself.
Rather, it provides the fast and reliable logical infrastructure linking
many distributed protection systems to the conditions required for beam
operation and, when necessary, to the LBDS
\cite{Puccio:2010BIS,Schmidt:2016MPS}.

In the terminology introduced in
Sec.~\ref{subsec:interlocks_permits}, each connected
protection system provides one or more binary
permit conditions,
\begin{equation}
\chi_j^{\rm permit}(\boldsymbol X,m)
\in\{0,1\},
\label{eq:bis_user_permit}
\end{equation}
In the terminology introduced in
Sec.~\ref{subsec:interlocks_permits}, each connected protection system
provides one or more binary permit conditions, \(\chi_{\rm beam}^{\rm permit}\left(\boldsymbol X,m\right)\),
where $\boldsymbol X$ is the relevant facility state
and $m$ denotes the operational mode.
The value $\chi_j^{\rm permit}=1$ indicates that
the corresponding protection condition permits
the beam-related operation under consideration,
whereas $\chi_j^{\rm permit}=0$ indicates that
this permission is withheld.

The LHC BIS is distributed around the accelerator.
Beam Interlock Controllers (BICs) act as local
concentrators for permit signals generated by
protection-relevant subsystems in their vicinity.
For BIC $k$, the corresponding local permit may
be represented schematically as
\begin{equation}
\chi_k^{\rm loc,permit}(\boldsymbol X,m)
=
\bigwedge_{j\in\mathcal J_k}
\chi_j^{\rm permit}(\boldsymbol X,m),
\label{eq:bic_local_permit}
\end{equation}
where $\mathcal J_k$ denotes the set of permit
inputs associated with BIC $k$, and
$\chi_k^{\rm loc,permit}\in\{0,1\}$ represents
their combined logical status.

For circulating-beam protection, the local
permit conditions are combined through the
beam-permit architecture to establish the
global beam permit
$\chi_{\rm beam}^{\rm permit}(\boldsymbol X,m)$,
introduced in Eq.~\eqref{eq:mps_beam_permit}.
Equation~\eqref{eq:bic_local_permit} represents
the logical principle rather than every detail
of the implemented BIS architecture.

The existing LHC architecture uses 16 BICs
distributed around the ring, approximately one on either side of each of
the eight interaction regions.  Protection-relevant information is
provided by accelerator systems including beam-loss monitoring, powering
and quench protection, collimation, radio-frequency systems, beam
instrumentation, vacuum and access systems, the experiments, and the beam
dumping system itself
\cite{Puccio:2010BIS}.

The distinction between the physical measurement and the BIS logic can
therefore be represented as
\begin{equation}
\begin{array}{c}
\text{physical subsystem state}
\\
\downarrow
\\
\text{subsystem-specific protection decision}
\\
\downarrow
\\
\chi_j^{\rm permit}
\\
\downarrow
\\
\text{BIC / BIS}
\\
\downarrow
\\
\chi_k^{\rm loc,permit}
\\
\downarrow
\\
\text{continued operation or beam extraction}.
\end{array}
\label{eq:bis_protection_logic}
\end{equation}
Thus, the BIS is primarily an aggregation and transmission layer between
distributed protection functions and the final beam-related action.

For circulating beams, the beam permit is transmitted continuously around
the machine through dedicated optical beam-permit loops.  The implemented
system uses a carrier signal of approximately
\begin{equation}
f_{\rm permit}
\simeq
10~{\rm MHz},
\label{eq:beam_permit_frequency}
\end{equation}
whose presence represents a valid beam-permit state.  Withdrawal of a
required user permit causes the corresponding BIC to interrupt the
beam-permit transmission.  The absence of the valid permit signal is then
interpreted as a request for beam extraction
\cite{Puccio:2010BIS}.

This architecture provides a direct physical implementation of the
fail-safe principle discussed in
Sec.~\ref{subsec:failsafe_protection}.  Schematically,
\begin{equation}
\text{valid continuous signal}
\quad\Longleftrightarrow\quad
\chi_k^{\rm loc,permit}=1,
\label{eq:bis_signal_present}
\end{equation}
whereas
\begin{equation}
\text{required signal absent}
\quad\Longrightarrow\quad
\chi_k^{\rm loc,permit}=0.
\label{eq:bis_signal_absent}
\end{equation}
Consequently, interruption of a protection-relevant communication path is
not, under the corresponding failure assumptions, interpreted as
continued authorization for beam operation.

For a circulating beam, loss of the global permit produces the logical
transition
\begin{equation}
\chi_k^{\rm loc,permit}:1\rightarrow0
\quad\Longrightarrow\quad
\chi_{\rm dump}^{\rm req}=1,
\label{eq:bis_dump_request}
\end{equation}
where $\chi_{\rm dump}^{\rm req}$ denotes the beam-dump request transmitted to the
LBDS.  The physical accelerator state then follows the separate extraction
process,
\begin{equation}
\chi_{\rm dump}^{\rm req}=1
\quad\Longrightarrow\quad
{\boldsymbol X}_{\rm beam}^{\rm circulating}
\xrightarrow{\;{\cal P}_{\rm LBDS}\;}
{\boldsymbol X}_{\rm beam}^{\rm extracted}.
\label{eq:bis_to_beam_extraction}
\end{equation}
This separation is useful: the BIS generates and transmits the logical
protection request, whereas the LBDS performs the physical removal of the
stored beam.

The same logical infrastructure also prevents injection when the required
machine conditions are not satisfied.  If
\begin{equation}
\chi_k^{\rm loc,permit}({\boldsymbol X},m)=0,
\label{eq:no_permit_no_injection}
\end{equation}
the corresponding beam-related transition is not permitted.  The meaning
of permit withdrawal therefore depends on the instantaneous operational
state: with circulating beam it can require beam extraction, whereas
before injection it can inhibit the transition into a beam-filled state.

This distinction reinforces the point made in
Sec.~\ref{subsec:interlocks_permits}: $\chi_k^{\rm loc,permit}=0$ does not imply that
the machine is unsafe.  Indeed, an empty machine with the beam permit
withdrawn may occupy a perfectly controlled state.  The permit determines
whether a particular beam-related operational trajectory may be entered or
maintained; it is not itself a complete definition of
$\Omega_{\rm safe}$.

The BIS response must also satisfy the timing requirements established in
Sec.~\ref{subsec:protection_time}.  Because dump requests can originate
from systems distributed around the approximately $27$-km LHC ring, their
propagation to the dumping system must be much faster than the evolution
of the failure modes for which active protection is intended.  The BIS
architecture transmits beam-dump requests on a timescale below
approximately
\begin{equation}
\tau_{\rm BIS}
\lesssim
100~\mu{\rm s},
\label{eq:bis_transmission_time}
\end{equation}
depending on the position at which the request originates
\cite{Puccio:2010BIS,Schmidt:2016MPS}.

This value should not be confused with the complete machine-protection
response time.  For a loss-triggered beam dump, the relevant sequence is
approximately
\begin{equation}
\tau_{\rm prot}
=
\tau_{\rm det}
+
\tau_{\rm BIS}
+
\tau_{\rm LBDS}
+
\tau_{\rm extr},
\label{eq:bis_complete_response}
\end{equation}
where $\tau_{\rm det}$ is associated with the system detecting the failure,
$\tau_{\rm BIS}$ with propagation of the interlock decision, and the
remaining terms describe synchronization and physical extraction of the
beam.  As discussed in
Sec.~\ref{subsec:fast_slow_failures}, the complete response occupies
several LHC turns even though the BIS transmission itself is substantially
faster.

The resulting timing condition remains defined in Eq.~\eqref{eq:beam_loss_protection_condition}.
Thus, fast transmission by the BIS is necessary but
not sufficient for protection against ultra-fast failures.  When the
critical beam trajectory develops before the complete active response can
be completed, passive protection remains indispensable.

Reliability is equally important because a failure of the interlock
infrastructure could otherwise defeat several upstream protection
functions simultaneously.  The BIS therefore uses redundant signal paths
and continuously transmitted permit information.  The protection chain is
also subject to systematic functional testing
\cite{Puccio:2010BIS,Filippini:2005Reliability}.  In particular, critical
paths extending from connected protection systems through the permit loops
to beam extraction are tested before beam operation and checked following
protection events.

This illustrates the distinction among redundancy, independence, and
diagnostic coverage introduced in
Secs.~\ref{subsec:barrier_independence} and
\ref{subsec:failsafe_protection}.  Redundant permit paths reduce the
probability that a single internal fault prevents transmission of a
required dump request, while functional tests reduce the probability that
a latent failure remains undetected until the protection function is
demanded.

The BIS must simultaneously limit dangerous failures and unnecessary beam
dumps.  The former correspond to situations in which a valid protection
request is not transmitted correctly; the latter arise when the permit is
withdrawn although continued beam operation would have remained
acceptable.  These two failure classes connect directly to the
safety--availability discussion of
Sec.~\ref{subsec:safety_availability},
\begin{equation}
\begin{aligned}
&\text{dangerous BIS failure}
&&\rightarrow
\text{loss of required protection},
\\
&\text{spurious BIS action}
&&\rightarrow
\text{unnecessary loss of beam availability}.
\end{aligned}
\label{eq:bis_failure_classes}
\end{equation}
The first is constrained primarily by machine-protection requirements,
while excessive occurrence of the second reduces operational efficiency
and integrated physics performance
\cite{Filippini:2005Reliability}.

The beam-permit logic is also configuration dependent.  The set of
conditions required for high-intensity operation need not be identical to
that applicable to low-intensity commissioning beams.
Information on beam intensity, beam energy, and other protection-relevant
machine parameters is distributed by the Safe Machine Parameter (SMP)
system and used to derive flags defining whether particular operational
conditions correspond to a sufficiently low-hazard beam regime.  Selected
maskable interlocks may then be treated differently during commissioning
with such setup beams, while the full protection configuration is restored
automatically once the beam parameters leave the established safe regime
\cite{Blaszkiewicz:2023SMP,Bolton:2024SMP}. The mode dependence of
\(\chi_{\rm beam}^{\rm permit}\left(\boldsymbol X,m\right)\) does not weaken the protection
principle: before the beam enters a regime in which a given failure can
produce unacceptable consequences, all protection functions required for
that regime must be active.

From the trajectory perspective of this manuscript, the BIS therefore
provides the discrete logical boundary between an admissible and a
non-admissible continuation of a beam-related trajectory:
\begin{equation}
\chi^{\rm loc,permit}=1
\quad\Longrightarrow\quad
\Gamma_{\rm beam}
\ \text{may continue subject to the other protection constraints},
\label{eq:bis_trajectory_permitted}
\end{equation}
whereas
\begin{equation}
\chi^{\rm loc,permit}=0
\quad\Longrightarrow\quad
\Gamma_{\rm beam}
\ \text{must not enter or remain in the corresponding beam-operating
state}.
\label{eq:bis_trajectory_forbidden}
\end{equation}

The BIS is therefore a particularly direct realization of the general
logic developed in Sec.~\ref{sec:hazards_barriers}.  Continuous physical
measurements and subsystem states are converted into local permit
decisions; the BIS combines and transmits these decisions; and loss of the
required global permit constrains the subsequent accelerator trajectory.
The final physical removal of the stored beam is performed by the LHC Beam
Dumping System, which is considered in the following subsection.

\subsection{Safe Machine Parameters and configuration-dependent protection}
\label{subsec:safe_machine_parameters}

The protection requirements of the LHC depend strongly on the physical
state of the beam.  A machine configuration that is acceptable during
commissioning with a low-intensity beam may be unacceptable once the
stored beam energy becomes capable of damaging accelerator equipment.
Machine protection therefore requires not only information on whether
individual subsystems are ready, but also reliable knowledge of parameters
that characterize the hazard associated with the beam itself.

This function is provided by the Safe Machine Parameter (SMP) system,
which distributes protection-relevant information such as beam energy,
beam intensity, optical parameters, and logical flags describing the
corresponding beam-safety regime to other elements of the machine protection architecture
\cite{Blaszkiewicz:2023SMP,Bolton:2024SMP}.  Its information outputs may be represented
schematically by
\begin{equation}
\boldsymbol Y_{\rm SMP}
=
\left(
E_{\rm beam},
N_{\rm beam},
\boldsymbol\beta,
\boldsymbol\chi_{\rm SMP}^{\rm flags},
\ldots
\right),
\label{eq:smp_parameter_vector}
\end{equation}
where $E_{\rm beam}$ denotes the beam energy
parameter distributed by the SMP system,
$N_{\rm beam}$ the circulating beam population,
and $\boldsymbol\beta$ represents
protection-relevant optical parameters.
The quantity
$\boldsymbol\chi_{\rm SMP}^{\rm flags}$
collects the binary logical flags derived
or distributed by the SMP system.
The representation is schematic and
combines physical machine parameters
with discrete protection-related information.

The SMP should therefore not be interpreted as a detector of a particular
hazard or as a protection barrier acting directly on the beam.  Its role
is instead to provide trusted state information from which other machine-
protection systems determine which protection conditions are required in
the current configuration.  Schematically,
\begin{equation}
\boldsymbol X_{\rm beam}
\longrightarrow
\boldsymbol Y_{\rm SMP}
\longrightarrow
\text{protection configuration}
\longrightarrow
\left\{
\chi_j^{\rm permit}
\right\},
\label{eq:smp_configuration_chain}
\end{equation}
where the $\chi_j^{\rm permit}$ are the permit conditions introduced in
Sec.~\ref{subsec:interlocks_permits}.

A central concept is the setup beam flag (SBF),
which indicates whether the beam belongs to a predefined
low-hazard regime for which selected machine-protection
interlocks may be masked during commissioning or setup
activities \cite{Wenninger:2016MPO,Blaszkiewicz:2023SMP}.
It may be represented as
\begin{equation}
\chi_{\rm SBF}
=
\begin{cases}
1, & \text{beam satisfies the defined setup-beam conditions},\\
0, & \text{beam lies outside the setup-beam regime}.
\end{cases}
\label{eq:setup_beam_flag}
\end{equation}
Its value reflects the applicable setup-beam
classification and does not by itself constitute
a general statement that the beam is free
of hazardous consequences.

The terminology ``safe beam'' requires some care.  The relevant criterion
was developed primarily with respect to the capability of the beam to
produce material damage in representative accelerator components.  A beam
classified as a setup beam is therefore not free of all machine-protection
consequences.  In particular, the energy required to quench a
superconducting magnet can be substantially smaller than that required to
produce permanent material damage.
Thus, classification as a setup beam does not imply the absence of all
possible machine-protection consequences,
\begin{equation}
\chi_{\rm SBF}=1
\quad\not\Longrightarrow\quad
{\boldsymbol{\cal C}}_{\rm beam}
=
{\boldsymbol 0},
\label{eq:setup_beam_not_hazard_free}
\end{equation}
where ${\boldsymbol 0}$ denotes the zero consequence vector, corresponding
to the simultaneous absence of all consequence components considered in
the analysis.
Rather, the flag indicates that the beam satisfies the specified
low-damage-potential criterion under the assumptions used to define that
operational regime
\cite{Wenninger:2016MPO}.

The underlying physical reason for the strong energy dependence of the
setup-beam limit is straightforward.  Increasing the proton energy
increases the energy deposited per lost particle, while the transverse
beam size generally decreases with increasing energy at fixed normalized
emittance.  Shower development provides a partial counteracting dilution,
but the net result obtained from simulations can be represented
approximately by
\begin{equation}
N_{\rm SB}(E)
\propto
E^{-1.7},
\label{eq:safe_beam_scaling}
\end{equation}
where $N_{\rm SB}$ denotes the beam population corresponding to the
reference setup-beam limit
\cite{Wenninger:2016MPO}.

A convenient historical representation of the normal setup-beam criterion
is
\begin{equation}
N_{\rm SB}(E)
\simeq
N_{0}
\left(
\frac{E_{0}}{E}
\right)^{1.7},
\label{eq:safe_beam_intensity}
\end{equation}
with $E_{0}=450~\mathrm{GeV}$ and a reference beam population $N_{0}$ defined
for the corresponding injection-energy configuration.  Copper was used as
a reference material in establishing the damage-related criterion
\cite{Wenninger:2016MPO}.  The numerical normalization and the detailed
operational implementation have evolved with experience, beam emittance,
commissioning requirements, and machine configuration; consequently,
Eq.~\eqref{eq:safe_beam_intensity} should be regarded as a physical scaling
relation rather than a universal operational limit.

This qualification is important because beam-damage potential cannot in
general be represented by energy and total intensity alone.  It also
depends on transverse emittance, loss geometry, impact parameters,
material properties, and the spatial and temporal distribution of the
deposited energy.  The setup-beam criterion therefore provides a
conservative operational classification rather than a complete
calculation of the consequence vector introduced in
Sec.~\ref{subsec:hazard_chains}.

The purpose of the SBF is primarily to introduce controlled flexibility
during low-intensity commissioning.  Some BIS inputs are designated as
\emph{maskable}, whereas protection functions whose availability remains
necessary even for setup beams are \emph{unmaskable}.  Schematically,
\begin{equation}
\chi_{\rm SBF}=1
\quad\Longrightarrow\quad
\text{selected maskable interlocks may be masked},
\label{eq:sbf_masking_allowed}
\end{equation}
whereas
\begin{equation}
\chi_{\rm SBF}=0
\quad\Longrightarrow\quad
\text{the corresponding masks are not permitted}.
\label{eq:sbf_masking_forbidden}
\end{equation}
The distinction allows commissioning activities that would otherwise be
prevented by protection conditions designed for high-intensity operation,
without removing those barriers that remain mandatory in the setup-beam
regime
\cite{Wenninger:2016MPO}.

Here ``masking'' should not be understood as disabling machine protection
as a whole.  It means that a specifically designated interlock input is
temporarily excluded from the beam-permit decision under an explicitly
validated low-hazard condition.  The protection architecture therefore
remains configuration dependent,
\begin{equation}
\chi_j^{\rm permit}
=
\chi_j^{\rm permit}
\left(
{\boldsymbol X},
m,
\chi_{\rm SBF}
\right),
\label{eq:smp_permit_dependence}
\end{equation}
where $m$ denotes the operational mode.

This gives a concrete realization of the mode-dependent permit concept
introduced in Sec.~\ref{subsec:interlocks_permits}.  The beam permit can
be written schematically as
\begin{equation}
\chi_{\rm beam}^{\rm permit}
=
\chi_{\rm beam}^{\rm permit}
\left(
\boldsymbol X,
m,
\boldsymbol Y_{\rm SMP}
\right),
\label{eq:smp_beam_permit}
\end{equation}
so that the set of active permit conditions changes only according to
predefined and validated rules associated with the physical beam state.

An essential requirement is that this flexibility disappears when the
beam leaves the setup-beam regime.  Once the intensity--energy combination
no longer satisfies the relevant low-hazard criterion,
\begin{equation}
\chi_{\rm SBF}:1\rightarrow0,
\label{eq:sbf_transition_unsafe}
\end{equation}
the corresponding maskable protection inputs must again participate in
the beam-permit logic.  Thus, increasing the beam hazard contracts the
set of admissible protection configurations,
\begin{equation}
\Omega_{\rm permit}^{\rm high\;hazard}
\subset
\Omega_{\rm permit}^{\rm setup},
\label{eq:smp_permitted_regions}
\end{equation}
where the relation refers specifically to the additional operational
freedom provided by maskable interlocks in the setup-beam regime.

This relation should not be interpreted as implying that
$\Omega_{\rm permit}^{\rm setup}$ is intrinsically ``safer.''  Rather, a
larger set of machine configurations is tolerable because the physical
consequences of the corresponding beam are constrained by its lower
damage potential.  As the available stored energy increases, additional
protection conditions become mandatory before the associated trajectory
is allowed to continue.

The SMP therefore provides an important connection between
the physical beam parameters and the logical machine-protection
architecture:
\begin{equation}
\begin{array}{c}
\text{beam energy and intensity}
\\
\downarrow
\\
\text{damage-potential classification}
\\
\downarrow
\\
\boldsymbol\chi_{\rm SMP}^{\rm flags}
\\
\downarrow
\\
\text{allowed interlock configuration}
\\
\downarrow
\\
\chi_{\rm beam}^{\rm permit}(\boldsymbol X,m).
\end{array}
\label{eq:smp_protection_logic}
\end{equation}
Here, $\boldsymbol\chi_{\rm SMP}^{\rm flags}$ denotes
the collection of binary flags distributed by the SMP
system, including  $\chi_{\rm SBF}$,
while $\chi_{\rm beam}^{\rm permit}(\boldsymbol X,m)$
denotes the resulting beam-permit status, determined
by the applicable machine-protection logic.
The arrows represent functional dependencies,
rather than a sequence of distinct physical operations.

Reliability of this information is itself a machine-protection
requirement.  An incorrect beam-energy or beam-intensity classification
could affect several downstream protection systems simultaneously and
could therefore constitute a common input failure rather than an isolated
loss of a single barrier.  Modern reliability studies of the SMP
accordingly identify hazard chains associated with its end users and use
risk-based requirements to determine acceptable failure frequencies
\cite{Blaszkiewicz:2023SMP}.

This point connects directly to the discussion
of barrier independence in
Sec.~\ref{subsec:barrier_independence}.
Because the same SMP information is distributed
to several protection users, a common failure
in the parameter-generation or distribution
chain may affect several downstream protection
decisions simultaneously. Schematically,
\begin{equation}
\chi_{\rm SMP}^{\rm fault}=1
\quad\Longrightarrow\quad
\left\{
\chi_1^{\rm permit},
\chi_2^{\rm permit},
\ldots,
\chi_N^{\rm permit}
\right\}
\ \text{may be affected},
\label{eq:smp_common_input_failure}
\end{equation}
where $\chi_{\rm SMP}^{\rm fault}\in\{0,1\}$
denotes the occurrence of a fault affecting
the relevant SMP information chain, and
$\chi_j^{\rm permit}\in\{0,1\}$ denotes a
downstream protection-permit condition
depending on the distributed information.

The relation indicates a potential common
dependency; it does not imply that all
downstream permits necessarily become zero.
Depending on the failure mechanism and its
detection, affected protection functions
may withdraw their permits or, in the
undesired case of an undetected failure,
retain an incorrect permit state.
Redundancy, validation, diagnostics,
and reliable distribution are therefore
integral parts of the SMP architecture.

These requirements are also driving the development of the
second-generation SMP for the High-Luminosity LHC era.  Recent designs
retain the hardware-based character of the protection system while
expanding monitoring and diagnostic capabilities and adapting the
distribution architecture to the evolving requirements of connected
machine-protection systems
\cite{Blaszkiewicz:2023SMP,Bolton:2024SMP}.

In the trajectory-based formulation of this manuscript, the SMP can
therefore be regarded as a mechanism for making the boundary of the
admissible operational region explicitly dependent on the physical beam
state:
\begin{equation}
\Gamma_{\rm beam}
\in
\mathcal G_{\rm adm}^{\rm MPS}
\left(
\boldsymbol Y_{\rm SMP}
\right).
\label{eq:smp_allowed_trajectory}
\end{equation}
A trajectory acceptable for a low-intensity setup beam need not remain
admissible after the beam energy or intensity is increased.  The SMP
ensures that this change in physical hazard is propagated to the logical
protection configuration before the machine is allowed to enter the more
demanding operational regime.

Safe Machine Parameters therefore provide the interface between
quantitative beam properties and configuration-dependent machine
protection.  Rather than defining safety independently, they ensure that
the appropriate set of interlocks, permits, and protection functions is
associated with the actual damage potential of the beam.

\subsection{The LHC Beam Dumping System}
\label{subsec:lhc_beam_dumping}

The BIS discussed in
Sec.~\ref{subsec:bis_machine_protection} determines whether the logical
conditions required for continued beam operation remain satisfied.  Once
a valid beam-dump request has been generated, however, the stored beam
energy must be removed physically from the accelerator.  This function is
performed by the LBDS, which fast-extracts the
circulating beam from the machine and transports it to a dedicated external
beam dump
\cite{Kramer:2008LBDS,Schmidt:2016MPS}.

It is useful to distinguish explicitly the logical
and physical stages of beam extraction. For a
circulating beam, withdrawal of the beam permit
initiates the corresponding dump request,
schematically represented as
\begin{equation}
\chi_{\rm beam}^{\rm permit}:1\rightarrow0
\quad\Longrightarrow\quad
\chi_{\rm dump}^{\rm req}=1
\quad\Longrightarrow\quad
\boldsymbol X_{\rm beam}^{\rm circulating}
\xrightarrow{\;\mathcal P_{\rm LBDS}\;}
\boldsymbol X_{\rm beam}^{\rm extracted},
\label{eq:lbds_logical_physical_chain}
\end{equation}
where
$\chi_{\rm dump}^{\rm req}\in\{0,1\}$
the beam-dump request indicator, and
$\mathcal P_{\rm LBDS}$ the physical extraction
process performed by the LBDS.
The BIS propagates the protection decision,
whereas the LBDS physically removes the
circulating beam from the accelerator.

The dumping system is installed around Interaction Region~6 (IR6).  For
each beam, the principal extraction chain comprises 15 fast extraction kicker magnets
(MKD), followed by 15 extraction septum magnets (MSD), which direct the
beam into the extraction line.  Ten dilution kicker magnets (MKB) then
sweep the beam over the front face of the external beam-dump absorber,
the Target Dump External (TDE)~\cite{Kramer:2008LBDS,Barlow:2005BETS}. The corresponding sequence may be
written schematically as
\begin{equation}
\text{circulating beam}
\rightarrow
\text{MKD extraction kickers}
\rightarrow
\text{MSD septa}
\rightarrow
\text{MKB dilution kickers}
\rightarrow
\text{TDE dump block}.
\label{eq:lbds_extraction_chain}
\end{equation}

The extraction kickers cannot reach their nominal deflecting field
instantaneously.  Their field rises over approximately
\begin{equation}
\tau_{\rm MKD}^{\rm rise}
\simeq
3~\mu{\rm s}.
\label{eq:mkd_rise_time}
\end{equation}
Particles traversing the kickers during this rise would receive an
intermediate deflection and could be directed toward sensitive machine
apertures rather than into the nominal extraction channel.  The circulating
beam therefore contains a dedicated particle-free interval, the
\emph{abort gap}, whose duration is approximately
\begin{equation}
\tau_{\rm abort}
\simeq
3~\mu{\rm s}.
\label{eq:abort_gap_duration}
\end{equation}
The MKD firing must be synchronized with this gap
\cite{Kramer:2008LBDS,Schmidt:2016MPS}.

The abort gap is therefore itself part of the machine-protection
configuration.  Schematically,
\begin{equation}
\chi_{\rm abort\,gap}
=
\begin{cases}
1,
&
\text{particle population compatible with safe extraction},
\\[1ex]
0,
&
\text{abort-gap population exceeds the permitted level}.
\end{cases}
\label{eq:abort_gap_state}
\end{equation}
Here, $\chi_{\rm abort\,gap}=1$ indicates an
abort-gap population compatible with safe beam
extraction, while $\chi_{\rm abort\,gap}=0$
indicates that the permitted population has
been exceeded.

Since a dump request can arrive at essentially any phase of the LHC
revolution period $T_{\rm rev}$, the extraction kickers cannot in general
fire immediately.  Their firing must first be synchronized with the
arrival of the particle-free abort gap, which introduces a waiting time of
up to approximately one revolution period.  Once the abort gap reaches
the extraction region, the MKD field rises during the gap and the beam is
extracted from the ring within the following turn.  Consequently, the
relevant LBDS response time is determined by the combination of abort-gap
synchronization, kicker response, and single-turn beam extraction, rather
than by the approximately $3~\mu{\rm s}$ MKD field-rise time alone.
The physical extraction latency is therefore naturally of order
one to a few LHC turns.

For a failure detected by an upstream protection system, the complete
active response contains several contributions,
\begin{equation}
\tau_{\rm prot}
=
\tau_{\rm det}
+
\tau_{\rm BIS}
+
\tau_{\rm sync}
+
\tau_{\rm MKD}
+
\tau_{\rm extr},
\label{eq:lbds_total_response_time}
\end{equation}
where $\tau_{\rm sync}$ includes synchronization with the abort gap,
$\tau_{\rm MKD}$ represents the relevant kicker response, and
$\tau_{\rm extr}$ describes completion of the physical extraction.

As discussed in Sec.~\ref{subsec:fast_slow_failures} (Eq.~\eqref{eq:bis_lbds_response_time}) following detection
of a critical failure the combined BIS and LBDS sequence requires less
than approximately \(3T_{\rm rev}\)
to complete removal of the affected beam from the LHC ring
\cite{Schmidt:2016MPS}.  This numerical scale should be distinguished from
both the sub-$100~\mu{\rm s}$ BIS transmission time discussed in
Sec.~\ref{subsec:bis_machine_protection} and the MKD field
rise time.

The timing condition derived in
Sec.~\ref{subsec:protection_time} consequently becomes
\begin{equation}
\tau_{\rm det}
+
\tau_{\rm BIS}
+
\tau_{\rm LBDS}
<
\tau_{\rm avail}^{(k)}
\label{eq:lbds_timing_condition}
\end{equation}
for the consequence $k$ that the extraction is intended to prevent.  The
LBDS therefore provides effective active protection only when a reliable
precursor can be detected sufficiently early.  Ultra-fast failures for
which the critical trajectory develops before completion of this sequence
require additional passive protection.

Extraction alone is not sufficient, because the stored beam energy cannot
be deposited safely in a small spot on the dump block.  At the nominal
design energy Eq.~\eqref{eq:lhc_nominal_beam_energy}, the circulating proton beam can contain about
\(0.3~{\rm GJ}\)
and the extracted beam must therefore be spatially diluted before reaching
the absorber.  Horizontal and vertical MKB dilution kickers generate a
time-dependent deflection that sweeps successive bunches over an extended
pattern on the front face of the TDE
\cite{Kramer:2008LBDS,Schmidt:2016MPS}.

The dilution system provides a particularly transparent example of how
beam physics enters directly into machine protection.  The extracted beam
cannot be allowed to strike the dump block repeatedly at essentially the
same transverse position.  Instead, the horizontal and vertical dilution
kickers generate time-dependent magnetic fields that sweep successive
bunches across the dump face, distributing the beam energy over a much
larger area.

For an ultra-relativistic proton beam, the transverse angular deflection
produced by a kicker magnet may be written as
\begin{equation}
\theta(t)
=
\frac{q}{p}\int B_{\perp}(s,t)\,ds
=
\frac{1}{B\rho}{\int B_{\perp}(s,t)\,ds},
\label{eq:kicker_deflection}
\end{equation}
where $q$ and $p$ are the particle charge and momentum, $s$ denotes the longitudinal coordinate along the beam trajectory
through the kicker,
$B_{\perp}$ is the transverse magnetic field, and
$B\rho=p/q$ is the magnetic rigidity of the beam.
The integral in Eq.~\eqref{eq:kicker_deflection} is taken over the effective magnetic length of the magnet.
The inverse dependence
on $B\rho$ makes explicit why the required kicker strengths must track the
actual beam energy.

For the horizontal and vertical dilution systems one may therefore write
\begin{equation}
\theta_x(t)
=
\frac{1}{B\rho}{\displaystyle\int B_y(s,t)\,ds},
\qquad
\theta_y(t)
=
\frac{1}{B\rho}{\displaystyle\int B_x(s,t)\,ds},
\label{eq:dump_dilution}
\end{equation}
with the orthogonal magnetic fields producing transverse deflections in
the two planes.

After transport through the extraction line, these angular kicks are
converted into transverse displacements on the dump face.  In linear
beam optics,
\begin{equation}
\begin{pmatrix}
x_{\rm dump}(t)\\[1ex]
y_{\rm dump}(t)
\end{pmatrix}
\simeq
\begin{pmatrix}
R_{12}^{(x)}\,\theta_x(t)\\[1ex]
R_{12}^{(y)}\,\theta_y(t)
\end{pmatrix},
\label{eq:dilution_transport}
\end{equation}
where $R_{12}^{(x,y)}$ denote the relevant transport-matrix elements
between the dilution kickers and the dump block.  The precise trajectory
also contains contributions from the extracted-beam coordinates and the
full extraction-line optics, but Eq.~\eqref{eq:dilution_transport}
captures the essential protection principle.

The horizontal and vertical kicker waveforms are chosen so that successive
bunches arrive at different positions,
\begin{equation}
{\boldsymbol r}_n
=
\left(
x_{\rm dump}(t_n),
y_{\rm dump}(t_n)
\right),
\label{eq:dilution_bunch_positions}
\end{equation}
where $t_n$ is the arrival time of bunch $n$.  Their superposition produces
the characteristic dilution pattern on the absorber surface rather than a
single concentrated beam spot.

The protection effect may be expressed schematically through the deposited
energy density.  Without dilution, the energy of many bunches would be
concentrated within an area comparable to the transverse beam size,
whereas dilution distributes the bunch energies over positions
${\boldsymbol r}_n$:
\begin{equation}
u_{\rm dep}({\boldsymbol r})
\propto
\sum_{n}
E_n\,
G\!\left(
{\boldsymbol r}-{\boldsymbol r}_n
\right),
\label{eq:dilution_energy_density}
\end{equation}
where $E_n$ is the energy carried by bunch $n$ and $G$ represents the transverse energy-deposition profile generated by that
bunch and its particle shower.

The protection condition is therefore the essential reduction of the local deposited-energy density $u_{\rm dep}$
\begin{equation}
u_{\rm dep}^{\rm diluted,max}
\ll
u_{\rm dep}^{\rm undiluted,max},
\label{eq:dilution_peak_reduction}
\end{equation}
while the total deposited beam energy remains essentially unchanged:
\begin{equation}
\int dV\,
u_{\rm dep}^{\rm diluted}
\simeq
\int dV\,
u_{\rm dep}^{\rm undiluted}
\simeq
E_{\rm beam},
\label{eq:dilution_energy_conservation}
\end{equation}
apart from energy deposited elsewhere along the extraction and shower
development.

Thus, the dilution system does not reduce the stored beam energy that must
ultimately be absorbed.  Its function is to reshape the spatial
distribution of that energy so that the peak temperature rise,
thermo-mechanical stress, and local material loading remain within the
validated limits of the dump system.

The corresponding energy-transfer process may be represented as
\begin{equation}
E_{\rm beam}
\xrightarrow{\;\text{extraction}\;}
E_{\rm ext}
\xrightarrow{\;\text{dilution}\;}
E_{\rm dep}({\boldsymbol r})
\xrightarrow{\;\text{TDE}\;}
E_{\rm thermal}
+
E_{\rm shower}
+
\cdots .
\label{eq:lbds_energy_path}
\end{equation}

This distinction is important in the barrier language developed in
Sec.~\ref{subsec:preventive_mitigative}.  Beam extraction is predominantly
preventive with respect to uncontrolled losses in the LHC ring:
\begin{equation}
{\boldsymbol X}_{\rm beam}^{\rm abnormal}
\xrightarrow{\;{\cal P}_{\rm LBDS}\;}
{\boldsymbol X}_{\rm beam}^{\rm extracted}.
\label{eq:lbds_preventive_function}
\end{equation}
The dilution system and dump block are predominantly mitigative with
respect to the unavoidable deposition of the extracted beam energy:
\begin{equation}
{\boldsymbol{\cal C}}_{\rm dump}^{\rm concentrated}
\xrightarrow{\;\text{dilution and absorption}\;}
{\boldsymbol{\cal C}}_{\rm dump}^{\rm controlled}.
\label{eq:lbds_mitigative_function}
\end{equation}
Thus, a single dumping sequence combines preventive and mitigative
protection functions.

The TDE is located in a dedicated cavern several hundred metres downstream
of the extraction region.  Its carbon- and graphite-based absorbing
structure is surrounded by substantial shielding and is designed to
receive the extracted beam after dilution.  The distance between the
machine and the dump provides space for beam transport, dilution, and
shower development, while physically separating the principal beam-energy
absorption from the superconducting accelerator ring
\cite{Kramer:2008LBDS}.

Correct extraction also requires the strength of the active LBDS elements
to track the actual beam energy.  A dump request may occur at any point
during operation, from injection energy to the highest machine energy.
An extraction kick appropriate at one energy would not provide the correct
trajectory at another. The beam energy tracking system (BETS) verifies that the
energy-dependent settings of the LBDS remain compatible with the actual beam energy.
Schematically, this validation can be represented as
\begin{equation}
E_{\rm beam}(t)
\longrightarrow
\left\{
U_{\rm MKD}^{\rm ref},
U_{\rm MKB}^{\rm ref},
I_{\rm MSD}^{\rm ref},
\ldots
\right\}
\xrightarrow{\rm BETS}
\chi_{\rm BETS}^{\rm valid}(t)
\in\{0,1\},
\label{eq:bets_energy_tracking}
\end{equation}
where $U_{\rm MKD}^{\rm ref}$ and $U_{\rm MKB}^{\rm ref}$
denote the beam-energy-dependent reference charging
voltages of the extraction and dilution kicker systems,
respectively, and $I_{\rm MSD}^{\rm ref}$ denotes the
reference current of the extraction septum magnets.
All these reference settings depend on $E_{\rm beam}(t)$.
The binary indicator $\chi_{\rm BETS}^{\rm valid}(t)=1$
denotes successful validation of the monitored settings
against their energy-dependent reference values within
the prescribed tolerances, while
$\chi_{\rm BETS}^{\rm valid}(t)=0$ indicates that the
validation conditions are not satisfied.

The BETS continuously verifies that the relevant kicker charging voltages
and magnet currents remain within predefined tolerances compatible with
the beam energy and extraction aperture.  A developing tracking fault can
therefore itself generate a dump request before the system becomes unable
to extract the beam correctly
\cite{Barlow:2005BETS}.  This is an example of detecting a protection-
system precursor before it develops into a failure of the final actuator.

The LBDS must also satisfy unusually demanding reliability requirements.
It is required both for routine termination of fills and for emergency
machine protection, and it must remain available whenever circulating beam
is present.  A failure to execute a required dump could expose the machine
to a large stored-energy hazard, while an asynchronous or incorrectly
executed dump can itself generate dangerous beam losses.

The protection problem is therefore not simply
\begin{equation}
\begin{aligned}
&\text{valid dump request} \\
&\rightarrow \text{synchronized extraction} \\
&\rightarrow \text{correct extraction trajectory} \\
&\rightarrow \text{beam dilution} \\
&\rightarrow \text{controlled absorption}.
\end{aligned}
\label{eq:lbds_complete_protection_chain}
\end{equation}
Failure at any of these stages can produce a different beam-loss
trajectory and must therefore be included in the machine-protection
analysis.

This requirement provides another concrete realization of the fail-safe
and DiD concepts discussed in
Secs.~\ref{subsec:failsafe_protection} and
\ref{subsec:protection_layers}.  Redundant extraction channels, continuous
surveillance of critical parameters, energy tracking, synchronization,
passive absorbers, and dedicated protection devices are combined so that
credible individual failures do not directly expose vulnerable machine
components to the full circulating beam
\cite{Schmidt:2016MPS,Kramer:2008LBDS}.

The trajectory-based interpretation is particularly
transparent. Under normal operation, the circulating
beam follows an individual trajectory
$\Gamma_{\rm circ}$ that must belong to the set
$\mathcal G_{\rm adm}^{\rm MPS}$ of trajectories
compatible with the applicable machine-protection
requirements:
\begin{equation}
\Gamma_{\rm circ}
\in
\mathcal G_{\rm adm}^{\rm MPS}.
\label{eq:mps_trajectory_allowed}
\end{equation}
Here, $\mathcal G_{\rm adm}^{\rm MPS}$ denotes
the set of trajectories satisfying the relevant
machine-protection constraints, including
the applicable equipment limits, protection
conditions, and timing requirements.
Its definition depends on the physical
and technical configuration of the
accelerator and the corresponding
operational mode.

When continued circulation is no longer admissible, the LBDS deliberately
replaces that trajectory by an extraction trajectory,
\begin{equation}
\Gamma_{\rm circ}
\xrightarrow{\;D_{\rm dump}\;}
\Gamma_{\rm ext}
\xrightarrow{\;\text{dilution}\;}
\Gamma_{\rm dump},
\label{eq:lbds_trajectory_replacement}
\end{equation}
whose endpoint is the controlled absorption of the beam energy in the TDE.

The LBDS is therefore the final active element in the
beam-related machine-protection chain.  Measurements and equipment states
identify an unacceptable trajectory, the BIS converts the corresponding
protection decisions into a dump request, and the LBDS transforms that
logical decision into controlled physical removal, dilution, and
absorption of the stored beam energy.

\subsection{Injection and dump protection}
\label{subsec:injection_dump_protection}

Injection and beam extraction represent particularly demanding
machine-protection situations because the beam trajectory is deliberately
changed by fast pulsed magnets.  Failures of these systems can therefore
produce large trajectory errors within a single passage of the beam,
leaving insufficient time for a conventional
detection--interlock--actuation sequence to prevent the first impact.
Injection and dump protection consequently provide important examples of
the complementary roles of active and passive barriers discussed in
Secs.~\ref{subsec:preventive_mitigative} and
\ref{subsec:fast_slow_failures}.

Injection into the LHC takes place at an energy of \(E_{\rm inj}=450~\mathrm{GeV}\)
through the transfer lines TI2 and TI8 into Interaction Regions~2 and~8,
respectively.  Septum magnets provide the first separation from the
transfer line, while the LHC injection kicker magnets (MKI) provide the
fast vertical deflection required to place the incoming beam onto the
circulating orbit
\cite{Burkhardt:2003TransferProtection,Kain:2005InjectionProtection,
Bracco:2020InjectionDumping}.

The MKI systems operate with a field-rise time below approximately
\begin{equation}
\tau_{\rm MKI}^{\rm rise}
\simeq
0.9~\mu{\rm s},
\label{eq:mki_rise_time}
\end{equation}
illustrating immediately why injection failures belong to the ultra-fast
class.  A mistimed pulse, missing module, erratic firing, or incorrect
kicker strength can place the injected beam on a strongly perturbed
trajectory before an active protection system could detect the resulting
beam loss and respond.

The relevant hazard sequence may be represented schematically as
\begin{equation}
{\boldsymbol X}_{\rm inj}^{\rm nominal}
\xrightarrow{\;{\cal F}_{\rm MKI}\;}
{\boldsymbol X}_{\rm inj}^{\rm mis-steered}
\xrightarrow{\;{\cal E}_{\rm impact}\;}
{\boldsymbol{\cal C}}_{\rm inj},
\label{eq:injection_failure_chain}
\end{equation}
where ${\cal F}_{\rm MKI}$ denotes an injection-kicker failure or timing error and
${\cal E}_{\rm impact}$ the resulting beam--matter interaction.

Active protection acts principally \emph{before} the beam is transferred.
Equipment states, timing, magnet currents, beam parameters, and machine
permits are verified before extraction from the SPS and injection into the
LHC.  If the required injection conditions are not satisfied, the
corresponding transition is inhibited,
\begin{equation}
\chi_{\rm inj}^{\rm permit}=0
\quad\Longrightarrow\quad
\Gamma_{\rm SPS\rightarrow LHC}
\ \text{not permitted}.
\label{eq:injection_permit_condition}
\end{equation}
Here, $\chi_{\rm inj}^{\rm permit}(t)\in\{0,1\}$ denotes
the binary injection-permit indicator, with the value $1$
allowing injection and $0$ inhibiting it.
Once the transfer has started, however, a fault of the fast injection
kickers can develop too rapidly for an active response to prevent the
first abnormal impact.  Passive protection must therefore remain capable
of intercepting the mis-steered beam.

The principal LHC injection-protection absorber is the segmented Target
Dump Injection Segmented (TDIS), which replaced the earlier TDI during
Long Shutdown~2.  TDIS units are installed in both injection regions,
Points~2 and~8, and consist of approximately $5.8~\mathrm{m}$ long,
three-module segmented assemblies containing graphite and higher-$Z$
absorbing materials
\cite{CarbajoPerez:2021TDIS,Bracco:2020InjectionDumping}.

The placement of the injection absorber is itself determined by beam
optics.  For a localized angular error $\theta_k$ produced at the kicker,
the resulting transverse displacement at a downstream position $s$ is,
in linear optics,
\begin{equation}
y(s)
\simeq
\sqrt{
\beta_y(s)\,
\beta_y(s_k)
}
\,
\sin\Delta\mu_y
\,
\theta_k ,
\label{eq:injection_kick_transport}
\end{equation}
where $\beta_y$ is the vertical betatron function and
$\Delta\mu_y$ the betatron phase advance between the kicker location
$s_k$ and the observation point.  An absorber located near
\begin{equation}
\Delta\mu_y
\simeq
\frac{\pi}{2}
\label{eq:tdis_phase_advance}
\end{equation}
therefore experiences close to the maximum transverse displacement
generated by the kicker error and can intercept the mis-steered beam
before it reaches more vulnerable downstream elements
\cite{Kain:2005InjectionProtection,Bracco:2020InjectionDumping}.

The injection-protection principle can consequently be represented as
\begin{equation}
{\boldsymbol X}_{\rm inj}^{\rm missteered}
\xrightarrow{\;{\cal P}_{\rm TDIS}\;}
{\boldsymbol X}_{\rm loss}^{\rm controlled},
\label{eq:tdis_protection_mapping}
\end{equation}
where ${\cal P}_{\rm TDIS}$ redirects the inevitable energy deposition
toward an absorber specifically designed to withstand it.  Additional
masks and collimating elements suppress the secondary-particle showers and
protect downstream superconducting magnets.

This is predominantly a mitigative function with respect to the kicker
failure itself: the abnormal trajectory has already been generated.
Relative to downstream magnet damage, however, the same absorber acts
preventively by stopping the trajectory before it reaches the protected
aperture.  The example therefore again demonstrates that the
preventive--mitigative classification depends on the consequence being
considered.

Correct positioning of the TDIS jaws is itself a protection
requirement. If an absorber is too far from the nominal
trajectory, a dangerous mis-steered beam may escape; if it
is too close, normal beam operation may be affected.
Its configuration must therefore satisfy the conditions
required for injection,
\begin{equation}
\chi_{\rm inj}^{\rm permit}(t)=1
\quad\Longrightarrow\quad
{\boldsymbol X}_{\rm TDIS}(t)
\in
\Omega_{\rm TDIS}^{\rm allowed},
\label{eq:tdis_allowed_state}
\end{equation}
where ${\boldsymbol X}_{\rm TDIS}(t)$ denotes the
actual TDIS configuration, including the relevant jaw
positions, and $\Omega_{\rm TDIS}^{\rm allowed}$ is
the set of configurations compatible with safe injection.
The binary indicator $\chi_{\rm inj}^{\rm permit}(t)$
is defined in Eq.~\eqref{eq:injection_permit_condition}.
Redundant position measurements and interlock checks
consequently form part of the protection architecture
\cite{CarbajoPerez:2021TDIS}.

A closely related problem occurs during beam extraction.  Under nominal
conditions, the MKD extraction kickers described in
Sec.~\ref{subsec:lhc_beam_dumping} rise during the approximately
$3~\mu{\rm s}$ particle-free abort gap.  If an extraction kicker fires
asynchronously, or if synchronization with the abort gap is lost, some
bunches encounter the rising kicker field and therefore receive an
intermediate deflection
\cite{Wiesner:2018AsynchronousDump,Bracco:2020InjectionDumping}.

The abnormal extraction trajectory may be written as
\begin{equation}
t_{\rm MKD}
\notin
{t}_{\rm gap-abort}
\quad\Longrightarrow\quad
0<\theta_{\rm MKD}(t)<\theta_{\rm MKD}^{\rm nom},
\label{eq:asynchronous_dump_kick}
\end{equation}
where ${t}_{\rm gap-abort}$ denotes the time interval occupied by the
abort gap.  During the kicker rise, successive bunches consequently
receive different angular deflections and are swept across the downstream
machine aperture.

This is one of the clearest examples of an ultra-fast failure for which
active protection cannot prevent the initial abnormal beam trajectory.
The corresponding sequence is
\begin{equation}
\text{asynchronous MKD firing}
\rightarrow
\text{partial kicks}
\rightarrow
\text{swept beam}
\rightarrow
\text{potential aperture impact}.
\label{eq:asynchronous_dump_chain}
\end{equation}
Dedicated passive absorbers are therefore required downstream of the
extraction kickers.

The principal dump-protection elements are the TCDS and TCDQ absorbers.
The fixed TCDS protects the extraction septum magnets (MSD) against
particles swept across the aperture during an asynchronous dump, whereas
the movable TCDQ protects the downstream superconducting quadrupole Q4 and
more distant machine elements.  A fixed TCDQM mask immediately upstream
of Q4 further reduces the secondary-particle shower reaching the
superconducting coils
\cite{Wiesner:2018AsynchronousDump,Bracco:2020InjectionDumping}.

The upgraded TCDQ has an absorber length of approximately
\begin{equation}
L_{\rm TCDQ}
\simeq
9~{\rm m},
\label{eq:tcdq_length}
\end{equation}
and uses carbon-fibre-composite materials chosen to withstand the intense
thermo-mechanical loading associated with asynchronous extraction
failures
\cite{Bracco:2020InjectionDumping}.

The passive dump-protection chain may therefore be represented as
\begin{equation}
{\boldsymbol X}_{\rm dump}^{\rm asynchronous}
\xrightarrow{\;{\cal P}_{\rm TCDS/TCDQ}\;}
{\boldsymbol X}_{\rm loss}^{\rm controlled}
\rightarrow
{\boldsymbol{\cal C}}_{\rm dump}^{\rm reduced}.
\label{eq:dump_passive_protection}
\end{equation}
The function of these devices is not to restore the nominal extraction
trajectory, which is no longer possible on the relevant timescale, but to
intercept the resulting bunch trajectories before they reach components
with substantially lower damage tolerance.

The distinction developed in
Sec.~\ref{subsec:protection_time} is especially transparent here:
\begin{equation}
\tau_{\rm fail}
\lesssim
\tau_{\rm prot}^{\rm min}
\quad\Longrightarrow\quad
\text{passive protection required}.
\label{eq:injection_dump_passive_condition}
\end{equation}
Injection-kicker and asynchronous-dump failures can generate dangerous
trajectories in a single passage or within only a few turns.  No
measurement made after the first abnormal impact can retroactively prevent
that impact; protection must therefore already be present in the physical
beam path.

This does not reduce the importance of active protection.  Active and
passive functions act at different stages of the same hazard sequence:
\begin{equation}
\begin{array}{c}
\text{equipment surveillance and permits}
\\
\downarrow
\\
\text{prevention of invalid injection or extraction states}
\\
\downarrow
\\
\text{fast pulsed-magnet failure}
\\
\downarrow
\\
\text{passive interception by dedicated absorbers}
\\
\downarrow
\\
\text{controlled energy deposition}.
\end{array}
\label{eq:injection_dump_layered_protection}
\end{equation}

Injection and dump protection thus provide a particularly clear example
of DiD.  Interlocks and permits reduce the probability that
an invalid beam transfer is initiated; surveillance can detect precursors
before the pulsed system is triggered; and passive absorbers remain
available for failure modes whose timescale is shorter than that of any
active response.

In the trajectory-based language used throughout this
manuscript, the objective is to ensure that even when
the nominal transfer trajectory is lost,
\begin{equation}
\Gamma_{\rm transfer}^{\rm nominal}
\rightarrow
\Gamma_{\rm transfer}^{\rm fault},
\end{equation}
the resulting faulty trajectory is intercepted within
a region specifically designed to tolerate its energy
deposition,
\begin{equation}
\Gamma_{\rm transfer}^{\rm fault}
\xrightarrow{\;\text{passive protection}\;}
\Gamma_{\rm loss}^{\rm controlled}
\in
{\cal G}_{\rm adm}^{\rm MPS}.
\label{eq:injection_dump_allowed_trajectory}
\end{equation}
Here, $\Gamma_{\rm loss}^{\rm controlled}$ represents a
trajectory in which the beam energy is deposited
within the designed protection system.

The injection and dump systems therefore illustrate the physical limit of
purely active protection: when a high-energy beam can reach a damaging
trajectory faster than a protection decision can be executed, safety must
be embedded directly in the accelerator optics, aperture hierarchy, and
material placement.

\subsection{Interface between machine protection and magnet-quench protection}
\label{subsec:machine_quench_interface}

Superconducting-magnet protection provides an important example of the
coupled-state character of LHC machine protection.  A magnet quench may be
initiated by beam-induced energy deposition, but it may also originate from
electrical, powering, mechanical, or cryogenic disturbances that are not
themselves caused by the circulating beam.  Conversely, once a quench has
developed, the resulting change in magnet current and field can perturb
the beam and thereby generate a secondary beam-loss hazard.  Magnet
protection and beam protection must therefore be treated as interacting
rather than independent functions
\cite{RodriguezMateos:2004QPS,Schmidt:2016MPS,Wenninger:2016MPO}.

In the coupled-state notation introduced in
Sec.~\ref{subsec:underground_environment_state}, one important propagation
path is
\begin{equation}
{\boldsymbol X}_{\rm beam}
\rightarrow
{\boldsymbol X}_{\rm loss}
\rightarrow
{\boldsymbol X}_{\rm mag}^{\rm quench}
\rightarrow
{\boldsymbol X}_{\rm cryo},
\label{eq:beam_quench_coupling}
\end{equation}
where local beam energy deposition drives a superconducting magnet across
its stability boundary.  A second path begins within the magnet or powering
system,
\begin{equation}
{\boldsymbol X}_{\rm power}
\rightarrow
{\boldsymbol X}_{\rm mag}^{\rm abnormal}
\rightarrow
{\boldsymbol X}_{\rm beam}^{\rm perturbed}
\rightarrow
{\boldsymbol X}_{\rm loss},
\label{eq:powering_beam_coupling}
\end{equation}
because a sufficiently rapid change of magnet current or field can modify
the closed orbit, optics, or beam trajectory.

The superconducting state may be represented schematically by a stability
condition
\begin{equation}
{\boldsymbol X}_{\rm mag}
\in
\Omega_{\rm SC},
\label{eq:superconducting_domain}
\end{equation}
where $\Omega_{\rm SC}$ denotes the region of magnet-state space compatible
with superconducting operation.  It depends on quantities such as current,
temperature, magnetic field, cooling conditions, and deposited energy.
A quench corresponds to a transition
\begin{equation}
{\boldsymbol X}_{\rm mag}^{\rm SC}
\xrightarrow{\;\mathcal{F}_{\rm q}\;}
{\boldsymbol X}_{\rm mag}^{\rm N},
\label{eq:quench_state_transition}
\end{equation}
where ${\boldsymbol X}_{\rm mag}^{\rm N}$ denotes a state containing a
normal-conducting region and $\mathcal{F}_{\rm q}$ represents the
initiating disturbance.

For beam-induced quenches, the relevant disturbance is the local energy
deposition discussed in Sec.~\ref{subsec:beam_loss_damage}.  The quench
condition may be expressed schematically as
\begin{equation}
\varepsilon_{\rm dep}
>
\varepsilon_{\rm q}
\left(
\tau_{\rm loss},
E_{\rm beam},
{\boldsymbol X}_{\rm mag}
\right),
\label{eq:quench_threshold_general}
\end{equation}
where $\varepsilon_{\rm q}$ is the quench threshold appropriate to the
loss duration, beam energy, magnet state, and deposition geometry.
Equation~\eqref{eq:quench_threshold_general} emphasizes that no single
universal deposited-energy threshold characterizes all beam-induced
quenches.

Once a normal-conducting zone forms, Joule heating develops locally,
\begin{equation}
P_{\rm J}(t)
=
I_{\rm mag}^{2}(t)\,
R_{\rm q}(t),
\label{eq:quench_joule_heating}
\end{equation}
where $R_{\rm q}(t)$ is the resistance of the growing normal zone.
Without rapid protection, this positive feedback can concentrate a large
fraction of the magnet's stored energy in a limited conductor volume.

The purpose of the Quench Protection System (QPS) is
therefore to detect the resistive transition and initiate
a controlled redistribution of the stored magnetic energy
\cite{RodriguezMateos:2004QPS}. Schematically,
\begin{equation}
{\boldsymbol X}_{\rm mag}^{\rm quench}
\xrightarrow{\;\text{QPS detection}\;}
\chi_{\rm QPS}^{\rm det}(t)=1
\xrightarrow{\;{\cal A}_{\rm QPS}\;}
{\boldsymbol X}_{\rm mag}^{\rm protected},
\label{eq:qps_protection_chain}
\end{equation}
where $\chi_{\rm QPS}^{\rm det}(t)\in\{0,1\}$ is the
binary quench-detection indicator, taking the value
$1$ when a quench has been detected and validated
by the QPS, and ${\cal A}_{\rm QPS}$ denotes the
subsequent magnet-protection actions.

These actions can include current interruption, energy extraction,
activation of quench heaters or other quench-propagation devices, and
controlled discharge of the affected electrical circuit.  Their common
physical purpose is to prevent excessive localization of the stored
magnetic energy.  The energy-flow problem can be represented as
\begin{equation}
E_{\rm mag}
\rightarrow
E_{\rm ext}
+
E_{\rm coil}
+
E_{\rm leads}
+
E_{\rm cryo}
+\cdots ,
\label{eq:quench_energy_redistribution}
\end{equation}
where $E_{\rm ext}$ denotes energy transferred to dedicated extraction
elements and the remaining terms represent the fractions dissipated in
the magnet circuit and its surroundings, like coils, leads, cryo-cooling equipment etc.

Magnet protection and beam protection become directly coupled whenever
the quench or powering disturbance can modify the beam trajectory before
the beam has been extracted.  A protection-relevant magnet or powering
fault can therefore lead to withdrawal of the corresponding beam permit,
\begin{equation}
\chi_{\rm mag}^{\rm valid}=0
\quad\Longrightarrow\quad
\chi_{\rm beam}^{\rm permit}=0,
\label{eq:magnet_state_beam_permit}
\end{equation}
where
$\chi_{\rm mag}^{\rm valid}\in\{0,1\}$ is the binary validity indicator
for the magnet and associated powering-protection state:
$\chi_{\rm mag}^{\rm valid}=1$ denotes a state compatible with the
corresponding beam-permit condition, whereas
$\chi_{\rm mag}^{\rm valid}=0$ denotes an invalid state and therefore
withdraws the beam permit and $\chi_{\rm beam}^{\rm permit}$ is the beam permit defined in
Eq.~\eqref{eq:mps_beam_permit}.

The complete protection sequence may then take the form
\begin{equation}
\text{magnet or powering fault}
\rightarrow
\left\{
\begin{array}{l}
\text{QPS / powering protection},\\
\text{beam-permit withdrawal}
\end{array}
\right.
\rightarrow
\left\{
\begin{array}{l}
\text{controlled magnet discharge},\\
\text{beam extraction}.
\end{array}
\right.
\label{eq:magnet_beam_parallel_protection}
\end{equation}
The two responses address different stored-energy reservoirs: the QPS
controls the magnetic energy, whereas the MPS and LBDS control the
circulating beam energy.

This parallel response is important because removing the beam does not
protect the magnet against the energy already stored in its electrical
circuit, while discharging the magnet does not by itself remove the beam
that may be perturbed by the developing field change.  The two protection
functions are therefore complementary,
\begin{equation}
{\cal P}_{\rm QPS}
\oplus
{\cal P}_{\rm beam},
\label{eq:qps_mps_complementarity}
\end{equation}
where $\oplus$ again denotes complementary protection functions rather
than mathematical addition.

The temporal relationship between the two responses depends on the
failure mechanism.  For a beam-induced quench, the BLM
system may detect the loss and request beam extraction before or close to
the onset of the resistive transition.  For a powering failure, the
electrical disturbance may be detected directly by the powering or magnet
protection systems before substantial beam loss occurs.  The available
protection time is therefore determined by the earliest reliably
detectable precursor,
\begin{equation}
\tau_{\rm avail}^{(k)}
=
t_{\rm crit}^{(k)}
-
t_{\rm detectable},
\label{eq:magnet_interface_available_time}
\end{equation}
consistent with the definition introduced in
Sec.~\ref{subsec:fast_slow_failures}.

This illustrates the advantage of diverse observables.  The same developing
hazard may be visible through magnet voltage, current change, quench
signals, orbit motion, or beam-loss measurements.  Detection at the
earliest stage can increase the available time for beam extraction:
\begin{equation}
\text{electrical precursor}
\rightarrow
\text{field deviation}
\rightarrow
\text{beam perturbation}
\rightarrow
\text{beam loss}.
\label{eq:powering_failure_precursor_chain}
\end{equation}
A beam-loss monitor acts relatively late in this sequence, whereas a
dedicated powering or magnet-protection channel can in some cases respond
before significant beam displacement develops.

The Fast Magnet Current change Monitor (FMCM) provides an example of this
principle for selected fast powering failures.  Rather than waiting for the
resulting orbit excursion to produce measurable beam losses, the system
detects sufficiently rapid changes of magnet current and can initiate a
beam dump on a timescale of order
\begin{equation}
\tau_{\rm FMCM}
\sim
20~\mu{\rm s},
\label{eq:fmcm_machine_quench_interface}
\end{equation}
thereby increasing the effective protection margin for the corresponding
failure modes
\cite{Schmidt:2016MPS}.

The coupling also extends in the opposite direction.  When a beam-induced
quench occurs, the quench is a machine-protection consequence even if no
permanent material damage results.  As emphasized in
Sec.~\ref{subsec:beam_loss_damage},
\begin{equation}
{\cal C}_{\rm quench}
\neq
{\cal C}_{\rm damage},
\label{eq:quench_not_damage}
\end{equation}
and the relevant thresholds can differ by orders of magnitude.  Beam-loss
thresholds must therefore protect not only against irreversible material
damage but also, where required, against quenches that would terminate the
fill and disturb accelerator operation.

A quench also couples the magnet system to the cryogenic state.  Resistive
heating and energy extraction produce local thermal loads, helium warming,
and, depending on the event, pressure transients,
\begin{equation}
{\boldsymbol X}_{\rm mag}^{\rm quench}
\rightarrow
{\boldsymbol X}_{\rm thermal}
\rightarrow
{\boldsymbol X}_{\rm cryo}.
\label{eq:quench_cryo_coupling}
\end{equation}
The magnitude of this coupling depends strongly on the magnet circuit and
on the effectiveness of the protection response.  Large-scale loss of
containment is therefore not an intrinsic consequence of an ordinary
protected quench; rather, it belongs to more severe failure sequences in
which cryogenic containment or pressure-control functions are also
challenged.  These issues are discussed further in
Sec.~\ref{sec:cryogenic_safety}.

The interface between machine protection and magnet protection is therefore
not a simple serial chain in which one system hands control to another.
It is better represented as a coupled protection network:
\begin{equation}
\begin{array}{ccc}
{\boldsymbol X}_{\rm beam}
&
\longleftrightarrow
&
{\boldsymbol X}_{\rm mag}
\\[1ex]
\downarrow
&&
\downarrow
\\
\text{BLM / MPS}
&&
\text{QPS / powering protection}
\\[1ex]
\downarrow
&&
\downarrow
\\
\text{beam extraction}
&&
\text{magnet-energy control}.
\end{array}
\label{eq:mps_qps_coupled_network}
\end{equation}

The logical interface between these systems is provided through interlocks
and permits.  Magnet and powering systems communicate protection-relevant
states to the machine-protection architecture, while beam-loss systems can
initiate beam extraction before beam-induced energy deposition develops
into a quench.  Protection of the complete accelerator therefore requires
that both stored-energy reservoirs be controlled consistently.

In the trajectory-based formulation used throughout this
manuscript, the combined objective is
\begin{equation}
\left(
\Gamma_{\rm beam},
\Gamma_{\rm mag}
\right)
\in
{\cal G}_{\rm allowed}^{\rm coupled},
\label{eq:coupled_beam_magnet_trajectory}
\end{equation}
where $\Gamma_{\rm beam}$ and $\Gamma_{\rm mag}$ denote
the beam and magnet subsystem trajectories, respectively,
and ${\cal G}_{\rm allowed}^{\rm coupled}$ is the set
of admissible coupled trajectories for which both
beam-energy deposition and magnet-energy evolution
remain within their validated protection domains.
If either subsystem leaves this domain, the corresponding
protection actions redirect the accelerator toward a
state in which the two stored-energy reservoirs can
be dissipated or removed in a controlled manner.

The machine--quench interface thus provides one of the clearest examples
of the central principle developed in this manuscript: safety at a
high-energy accelerator is determined not by isolated subsystem states,
but by the controlled evolution of strongly coupled physical systems.

\subsection{Machine protection in state space}
\label{subsec:mps_state_space}

The preceding examples show that LHC machine protection cannot be reduced
to a collection of independent thresholds.  Beam losses, magnet and
powering faults, injection and extraction failures, collimator states,
beam-dump readiness, and protection permits are coupled through the
physical evolution of the accelerator.  The state-space formulation
introduced in Sec.~\ref{subsec:underground_environment_state} therefore
provides a useful common representation of the machine-protection problem.

For the present purpose, one may introduce a machine-protection state
vector containing the variables most directly relevant to beam and magnet
protection,
\begin{equation}
{\boldsymbol X}_{\rm MPS}
=
\left(
{\boldsymbol X}_{\rm beam},
{\boldsymbol X}_{\rm mag},
{\boldsymbol X}_{\rm coll},
{\boldsymbol X}_{\rm inj},
{\boldsymbol X}_{\rm dump},
{\boldsymbol X}_{\rm prot}
\right),
\label{eq:mps_state_vector}
\end{equation}
where the individual blocks represent, respectively, the circulating-beam
state, magnet and powering state, collimation configuration, injection
configuration, beam-dumping configuration, and the state of the associated
protection systems.  The latter includes quantities such as beam-loss
monitoring status, interlock permits, SMP information,
and protection-system readiness.

Equation~\eqref{eq:mps_state_vector} is a reduced representation of the
full facility state ${\boldsymbol X}$ introduced earlier.  It is not meant
to imply dynamical independence of its components.  On the contrary, the
principal machine-protection hazards arise precisely because changes in
one block can drive rapid evolution in another:
\begin{equation}
{\boldsymbol X}_{i}
\longrightarrow
{\boldsymbol X}_{j}
\longrightarrow
{\boldsymbol X}_{k}
\longrightarrow\cdots .
\label{eq:mps_state_couplings}
\end{equation}
Examples include beam loss driving a magnet quench, a powering failure
perturbing the beam trajectory, an asynchronous dump generating abnormal
aperture impacts, or an injection-kicker failure transferring beam energy
to a dedicated passive absorber.

For a specified operational mode $m$, one may define a
machine-protection-admissible region
\begin{equation}
\Omega_{\rm MPS}^{(m)}
\subset
\Omega_{\rm phys},
\label{eq:mps_admissible_region}
\end{equation}
containing those physical states for which the beam- and magnet-related
machine-protection requirements associated with that mode are satisfied.
This region is determined by quantities such as beam energy and intensity,
collimator and absorber positions, magnet and powering conditions, beam
dump readiness, injection configuration, and the validity of the required
protection functions.

It is important not to identify
$\Omega_{\rm MPS}^{(m)}$ with the complete safe operational phase space
$\Omega_{\rm safe}$.  Machine protection addresses primarily the control
of accelerator and equipment hazards associated with stored beam and
magnetic energy.  Personnel safety, radiation protection, cryogenic
conditions, access authorization, and other facility-level constraints
introduce additional requirements.  Thus,
\begin{equation}
\Omega_{\rm MPS}^{(m)}
\neq
\Omega_{\rm safe}
\label{eq:mps_not_complete_safe_space}
\end{equation}
in general.

Likewise, the machine-protection region should be
distinguished from the logical permit region introduced
in Sec.~\ref{subsec:interlocks_permits}.
For beam operation in mode $m$,
\begin{equation}
\Omega_{\rm permit}^{(m)}
=
\left\{
\boldsymbol X
\;\bigg|\;
\chi_{\rm beam}^{\rm permit}(\boldsymbol X,m)=1
\right\},
\label{eq:mps_permit_region}
\end{equation}
where $\chi_{\rm beam}^{\rm permit}(\boldsymbol X,m)
\in\{0,1\}$ denotes the binary beam-permit indicator
defined in Eq.~\eqref{eq:mps_beam_permit}.
The machine-protection-admissible region for
continued beam operation is therefore constrained
by both the physical protection conditions and
the corresponding permit logic:
\begin{equation}
\Omega_{\rm beam,op}^{(m)}
=
\Omega_{\rm MPS}^{(m)}
\cap
\Omega_{\rm permit}^{(m)}.
\label{eq:mps_operational_intersection}
\end{equation}
Additional facility-level safety and authorization
conditions must be included when the full
operational state is considered.

The
SMP system makes
$\Omega_{\rm MPS}^{(m)}$ explicitly dependent on the physical hazard
associated with the beam.  In schematic form,
\begin{equation}
\Omega_{\rm MPS}^{(m)}
=
\Omega_{\rm MPS}^{(m)}
\left(
{\boldsymbol Y}_{\rm SMP}
\right),
\label{eq:mps_region_smp_dependence}
\end{equation}
where ${\boldsymbol Y}_{\rm SMP}$ contains protection-relevant quantities
such as beam energy, intensity, optical parameters, and safety flags.
A configuration admissible for a low-intensity setup beam need not remain
admissible after the beam energy or intensity is increased.

The boundary of the admissible region is therefore not fixed.  It depends
on the machine configuration and on the physical consequences that become
possible in that configuration.  Increasing the stored beam energy, for
example, can require additional interlocks, tighter collimation conditions,
validated dump-system readiness, and removal of masks permitted during
low-intensity commissioning.  In this sense, the physical hazard and the
logical protection configuration evolve together.

Machine protection is, however, fundamentally a problem of trajectories
rather than isolated states.  A state may satisfy all instantaneous
protection conditions and yet evolve toward an unacceptable configuration
if the transition proceeds too rapidly for the available protection to
act.

Define the set of dynamic machine-protection constraints that must be
satisfied for the corresponding beam and operational state to remain
admissible,
\begin{equation}
{\cal K}_{\rm MPS}
=
\left\{
\begin{array}{l}
\tau_{\rm prot}^{(k)}<\tau_{\rm avail}^{(k)},\\
\chi_{\rm beam}^{\rm permit}=1\quad\text{when required},\\
\sigma_{\rm coll}\in\Sigma_{\rm coll}^{\rm allowed},\\
\chi_{\rm LBDS}^{\rm valid}=1,\\
\ldots
\end{array}
\right\}.
\label{eq:mps_dynamic_constraints}
\end{equation}
Here $\sigma_{\rm coll}$ denotes the actual collimation configuration,
while $\Sigma_{\rm coll}^{\rm allowed}$ is the set of configurations
compatible with the relevant beam and operational conditions.
The binary indicator
$\chi_{\rm LBDS}^{\rm valid}\in\{0,1\}$ specifies whether the state of
the LHC Beam Dumping System satisfies the validity conditions required
for the corresponding operational mode.  The index $k$ labels the
applicable failure or protection case for which the available reaction
time is evaluated.

The simultaneous satisfaction of the applicable constraints in
${\cal K}_{\rm MPS}$ is represented by the binary admissibility
indicator
\begin{equation}
\chi_{\rm MPS}^{\rm adm}(t)
=
\begin{cases}
1,
&
\text{all applicable constraints in ${\cal K}_{\rm MPS}$ are satisfied
at time $t$},
\\[1ex]
0,
&
\text{otherwise}.
\end{cases}
\label{eq:mps_constraint_indicator}
\end{equation}

In the framework adopted here, the relevant object is therefore the set
of allowed machine-protection trajectories,
\begin{equation}
\Gamma_{\rm allowed}^{\rm MPS}
=
\left\{
\gamma(t)={\boldsymbol X}(t)
\;\middle|\;
{\boldsymbol X}(t)\in\Omega_{\rm MPS}^{(m(t))},
\;
\chi_{\rm MPS}^{\rm adm}(t)=1,
\;
\forall t\in[t_0,t_f]
\right\}.
\label{eq:mps_allowed_trajectory_definition}
\end{equation}

Equation~\eqref{eq:mps_dynamic_constraints} is schematic rather than a
complete specification of machine conditions.  Its purpose is to
emphasize that machine protection combines continuous physical variables,
discrete logical states, configuration constraints, and timing
requirements.

Normal beam operation requires
\begin{equation}
\Gamma_{\rm beam}
\in
{\cal G}_{\rm adm}^{\rm MPS},
\label{eq:mps_normal_trajectory_condition}
\end{equation}
where the notation \({\cal G}_{\rm adm}^{\rm MPS}\) for the resulting admissible trajectory
was introduced in Eq.~\eqref{eq:mps_trajectory_allowed}.
When this condition can no longer be maintained, the protection
architecture must either inhibit the corresponding transition or redirect
the accelerator toward a protected trajectory.

The latter situation may be represented as
\begin{equation}
\Gamma_{\rm beam}
\xrightarrow{\;\mathcal{F}\;}
\Gamma_{\rm fault}
\xrightarrow{\;\mathcal{P}_{\rm MPS}\;}
\Gamma_{\rm protected},
\label{eq:mps_fault_trajectory_redirection}
\end{equation}
where $\mathcal{F}$ denotes the initiating failure and
$\mathcal{P}_{\rm MPS}$ the combined machine-protection response.
Depending on the failure, the protected trajectory may involve beam
extraction, interception by passive absorbers, controlled magnet discharge,
inhibition of injection, or another predefined transition.

This formulation naturally incorporates the distinction between active
and passive protection.  For a failure satisfying
\begin{equation}
\tau_{\rm prot}
<
\tau_{\rm avail}^{(k)},
\end{equation}
an active protection system can, in principle, redirect the trajectory
before consequence $k$ becomes unavoidable.  If instead
\begin{equation}
\tau_{\rm avail}^{(k)}
\lesssim
\tau_{\rm prot}^{\rm min},
\end{equation}
the geometry of the accessible trajectory itself must be constrained by
passive protection.  Injection absorbers, dump-protection devices, and
collimators therefore act directly on the structure of
${\cal G}_{\rm adm}^{\rm MPS}$.

The collimation system provides a particularly transparent example.
Without collimation, halo particles~\footnote{Beam halo particles are particles in the
outer transverse regions of the circulating beam
distribution, at relatively large amplitudes with
respect to the beam core. They may arise from
beam diffusion and other dynamical processes
and are intercepted by the collimation system
to limit uncontrolled losses in sensitive
accelerator components.} can explore loss trajectories leading
to vulnerable superconducting apertures.  With the collimator hierarchy
properly configured, those trajectories are redirected toward robust
absorbing elements:
\begin{equation}
\Gamma_{\rm halo}^{\rm uncoll}
\xrightarrow{\;{\cal P}_{\rm coll}\;}
\Gamma_{\rm loss}^{\rm coll}
\in
{\cal G}_{\rm adm}^{\rm MPS}.
\label{eq:mps_collimation_trajectory}
\end{equation}

Similarly, the BIS acts primarily on the logical
accessibility of trajectories. If a required beam permit is withdrawn,
\begin{equation}
\chi_{\rm beam}^{\rm permit}:1\rightarrow0,
\end{equation}
continued beam circulation is no longer an
admissible continuation, and beam extraction
is requested. Schematically,
\begin{equation}
\Gamma_{\rm circ}
\xrightarrow{\;\chi_{\rm dump}^{\rm req}=1\;}
\Gamma_{\rm ext}.
\label{eq:mps_bis_trajectory_transition}
\end{equation}
The LBDS then performs the corresponding physical
transition from circulating to extracted beam,
provided that the beam-dumping system operates
successfully.

The magnet-protection interface requires an extension of the same picture
to coupled trajectories.  A powering or quench event can require
simultaneous control of the beam and magnetic-energy reservoirs:
\begin{equation}
\left(
\Gamma_{\rm beam},
\Gamma_{\rm mag}
\right)
\xrightarrow{\;\mathcal{P}_{\rm coupled}\;}
\left(
\Gamma_{\rm beam}^{\rm extracted},
\Gamma_{\rm mag}^{\rm discharged}
\right).
\label{eq:mps_coupled_protected_trajectory}
\end{equation}
Machine protection is therefore not associated with a unique endpoint but
with a set of controlled trajectories appropriate to different failure
classes.

This viewpoint also clarifies the role of DiD.  Different
protection layers constrain different parts of state space and different
segments of a possible failure trajectory:
\begin{equation}
\Gamma_{\rm fault}
\cap
\left(
{\cal B}_1
\oplus
{\cal B}_2
\oplus
\cdots
\oplus
{\cal B}_N
\right)
\longrightarrow
\Gamma_{\rm protected},
\label{eq:mps_layered_trajectory_control}
\end{equation}
where ${\cal B}_k$ denotes schematically the action of protection layer
$k$ and $\oplus$ again indicates complementary functions rather than
mathematical addition.  A BLM may detect the loss trajectory, the BIS may
remove the logical permission for continued circulation, the LBDS may
extract the beam, and a passive absorber may control the first unavoidable
impact of an ultra-fast failure.

The machine-protection architecture can consequently be interpreted as a
distributed mechanism for restricting both the states and the trajectories
accessible to the accelerator:
\begin{equation}
\begin{array}{c}
{\boldsymbol X}_{\rm MPS}
\\
\downarrow
\\
\Omega_{\rm MPS}^{(m)}
\\
\downarrow
\\
\Gamma_{\rm allowed}^{\rm MPS}
\\
\downarrow
\\
\text{measurement, permits, active protection, passive protection}
\\
\downarrow
\\
\Gamma_{\rm protected}.
\end{array}
\label{eq:mps_state_space_summary}
\end{equation}

This representation should not be understood as a replacement for the
detailed engineering description of the MPS.
Rather, it provides a common physical language for the systems discussed
in this section.  Beam-loss monitoring, collimation, Safe Machine
Parameters, interlocks, beam dumping, injection and extraction protection,
and magnet protection all act by restricting which coupled states and
trajectories remain accessible as the stored beam and magnetic energies
increase.

In this sense, the LHC MPS  provides a concrete
realization of the safe-operational-phase-space concept introduced earlier:
it continuously maps measured physical quantities and validated technical
states into constraints on the future evolution of the accelerator.

The state-space description also provides a compact way of summarizing how
the different protection systems discussed above act on representative
failure trajectories.  Some systems identify a developing departure from
the admissible region, some remove the logical permission for the
trajectory to continue, and others physically redirect an unavoidable
abnormal trajectory toward components designed to tolerate the resulting
energy deposition.  Table~\ref{tab:machine_protection_layers} summarizes
these complementary functions.

The classification in the table is necessarily schematic.  A given
protection system may act preventively with respect to one consequence and
mitigatively with respect to another, and several systems may participate
in the response to the same initiating event.  The table should therefore
be read as an illustration of how representative machine-protection layers
constrain or redirect trajectories within
${\cal G}_{\rm adm}^{\rm MPS}$, rather than as a one-to-one mapping
between failures and protection devices.

\begin{table*}[t]
\caption{\label{tab:machine_protection_layers}
Representative failure conditions and corresponding functions within the
LHC machine-protection architecture.  The entries illustrate complementary
protection layers and are not intended as an exhaustive classification.}
\begin{ruledtabular}
\begin{tabular}{lll}
\textbf{Failure or condition}
&
\textbf{Relevant observable or state}
&
\textbf{Protection response}
\\
\hline

Beam halo or diffusion
&
Beam-loss pattern
&
\shortstack[l]{Collimation intercepts losses in robust elements;\\
leakage toward vulnerable equipment is suppressed.}
\\
\hline

Excessive local beam loss
&
\shortstack[l]{BLM signal and\\integration threshold}
&
\shortstack[l]{BLM generates a dump request; BIS withdraws\\
the beam permit and initiates beam extraction.}
\\
\hline

Magnet or powering fault
&
\shortstack[l]{Current, voltage,\\equipment state}
&
\shortstack[l]{Powering protection can request beam extraction\\
before field changes produce damaging losses.}
\\
\hline

Superconducting transition
&
Resistive-voltage signature
&
\shortstack[l]{QPS initiates controlled magnet-energy protection;\\
beam extraction is requested when required.}
\\
\hline

Invalid machine configuration
&
\shortstack[l]{Subsystem permits\\and SMP information}
&
\shortstack[l]{Interlock logic inhibits the transition\\
or terminates continued beam operation.}
\\
\hline

Injection-system failure
&
\shortstack[l]{Kicker state, timing,\\transfer trajectory}
&
\shortstack[l]{Active checks inhibit invalid injection; TDIS\\
intercepts a mis-steered injected beam.}
\\
\hline

Abnormal or asynchronous dump
&
\shortstack[l]{Kicker timing and\\extraction state}
&
\shortstack[l]{TCDS, TCDQ, and associated elements intercept\\
abnormal extraction trajectories.}
\\
\hline

Normal or requested beam disposal
&
\shortstack[l]{Dump request and\\LBDS readiness}
&
\shortstack[l]{LBDS extracts the beam; MKB kickers dilute it,\\
and the TDE absorbs the stored beam energy.}
\\

\end{tabular}
\end{ruledtabular}
\end{table*}

Table~\ref{tab:machine_protection_layers} also emphasizes that machine
protection is not a simple serial chain.  Measurement, logical
interlocking, active beam removal, passive interception, and magnet-energy
control act at different stages and on different timescales.  Collectively,
they restrict the physical and logical trajectories accessible to the
machine and redirect abnormal evolution toward validated protected states.

\subsection{Machine protection, availability, and physics output}
\label{subsec:mps_physics_output}

Machine protection is sometimes perceived primarily as a constraint on
accelerator operation because interlocks, beam dumps, setup-beam limits,
collimator settings, and protection validation can prevent or terminate a
physics fill.  At a machine such as the LHC, however, the opposite
interpretation is more fundamental: reliable machine protection is one of
the conditions that makes sustained operation with large stored beam and
magnetic energies physically possible.

The relevant objective is therefore not to maximize beam availability
independently of protection, nor to minimize all beam interruptions at any
cost.  It is to maximize the amount of scientifically useful operation
that can be achieved while the accelerator remains inside the validated
machine-protection domain.

In the state-space notation developed in
Sec.~\ref{subsec:mps_state_space}, useful beam operation requires
\begin{equation}
\Gamma_{\rm phys}
\in
{\cal G}_{\rm adm}^{\rm MPS},
\label{eq:mps_physics_allowed_trajectory}
\end{equation}
where $\Gamma_{\rm phys}$ denotes a trajectory associated with physics
production.

A protection action that terminates a fill therefore need not represent a
failure of physics operation in the broader sense.  If the alternative
would be progression toward equipment damage, an uncontrolled beam loss,
or a prolonged machine outage, beam extraction preserves the possibility
of subsequent physics operation.  Schematically,
\begin{equation}
\text{short interruption}
\quad\longrightarrow\quad
\text{avoidance of severe damage}
\quad\longrightarrow\quad
\text{preservation of long-term physics availability}.
\label{eq:mps_short_long_availability}
\end{equation}

This distinction is particularly important for protection systems with
high actuation rates.  Excessively sensitive thresholds or unreliable
interlock channels can generate unnecessary beam dumps and reduce
short-term availability, whereas insufficiently sensitive protection can
increase the probability of equipment damage and substantially longer
interruptions.  The relevant optimization therefore involves both
dangerous failures and spurious actions,
\begin{equation}
\begin{aligned}
&\text{dangerous failure}
&&\rightarrow
\text{insufficient protection},
\\
&\text{spurious protection action}
&&\rightarrow
\text{unnecessary loss of availability}.
\end{aligned}
\label{eq:mps_physics_failure_classes}
\end{equation}
This is the accelerator-specific realization of the more general
safety--availability balance discussed in
Sec.~\ref{subsec:safety_availability}
\cite{Filippini:2005Reliability,Willeke:2016Reliability}.

For collider operation, scientific performance is ultimately determined
not by instantaneous luminosity alone but by the luminosity accumulated
over an extended period.  A useful schematic relation is therefore
\begin{equation}
{\cal L}_{\rm int}
=
\int_{\rm physics}
{\cal L}(t)\,dt ,
\label{eq:integrated_luminosity}
\end{equation}
where ${\cal L}(t)$ is the instantaneous luminosity and
${\cal L}_{\rm int}$ the integrated luminosity delivered during periods of
physics operation.

The contribution of machine protection to this quantity is indirect but
fundamental.  Protection influences the fraction of time during which the
machine can remain in stable-beam operation, the frequency of unnecessary
beam dumps, the duration of recovery following protection events, and,
most importantly, the probability of major equipment damage that would
remove the machine from operation for a much longer period.

One may therefore write schematically
\begin{equation}
{\cal L}_{\rm int}
\sim
{\cal L}_{\rm peak}
\,
T_{\rm op}
\,
A_{\rm mach}
\,
\eta_{\rm phys},
\label{eq:physics_output_schematic}
\end{equation}
where ${\cal L}_{\rm peak}$ denotes a characteristic luminosity scale,
$T_{\rm op}$ the available operational time,
$A_{\rm mach}$ the machine availability, and
$\eta_{\rm phys}$ a schematic efficiency factor describing the fraction of
available time converted into useful physics production.  Equation~
\eqref{eq:physics_output_schematic} is not intended as an exact luminosity
model; its purpose is to make explicit that protection and availability
enter scientific output through the time structure of accelerator
operation.

The role of the MPS is therefore twofold.  First, it
allows operation with beam parameters that would otherwise present an
unacceptable risk to accelerator equipment.  Second, it limits the
duration and severity of abnormal events when they do occur.  In this
sense, protection enlarges the practically accessible physics domain,
\begin{equation}
\Omega_{\rm physics}^{\rm accessible}
\subset
\Omega_{\rm MPS}^{\rm validated},
\label{eq:mps_accessible_physics_domain}
\end{equation}
where $\Omega_{\rm MPS}^{\rm validated}$ denotes the region of machine
states for which the required protection functions, settings, and
operational procedures have been established.

The qualification ``validated'' is essential.  Increasing beam energy,
intensity, brightness, or luminosity cannot simply be interpreted as a
motion toward a more productive point in accelerator state space.
Additional machine-protection conditions may become necessary as the
consequences associated with abnormal trajectories increase.  Safe
Machine Parameters, collimator settings, beam-loss thresholds, dump-system
readiness, injection protection, and magnet-protection conditions must
therefore evolve together with the beam parameters.

Commissioning can consequently be viewed as a controlled enlargement of
the validated machine-protection domain.  Starting from low-intensity
setup beams, progressively more demanding beam conditions are introduced
only after the corresponding protection functions have been tested and
validated.  Schematically,
\begin{equation}
\Omega_{\rm MPS}^{(1)}
\subset
\Omega_{\rm MPS}^{(2)}
\subset
\cdots
\subset
\Omega_{\rm MPS}^{(\rm physics)},
\label{eq:mps_commissioning_expansion}
\end{equation}
where the successive regions represent increasing experimentally validated
operational capability rather than an unconditional relaxation of
protection requirements.

The same principle applies during a physics fill.  Changes in beam energy,
optics, intensity, collimator hierarchy, or detector conditions alter the
set of admissible trajectories and therefore the corresponding protection
configuration.  Scientific operation proceeds only while the combined
beam, magnet, collimation, injection, extraction, and protection states
remain compatible with the relevant machine mode.

Machine protection thus contributes to physics output not by removing
operational constraints but by defining a controlled region in which high-
performance operation can be sustained.  This relation can be summarized
schematically as
\begin{equation}
\begin{aligned}
&\text{higher stored energy} \\
&\rightarrow
\text{larger potential consequences} \\
&\rightarrow
\text{stronger validated protection} \\
&\rightarrow
\text{sustainable high-performance operation} \\
&\rightarrow
\text{integrated physics output}.
\end{aligned}
\label{eq:mps_physics_output_chain}
\end{equation}

This viewpoint also clarifies why protection integrity and availability
cannot be optimized independently.  A machine-protection system that
permits operation but cannot reliably interrupt hazardous trajectories
does not provide sustainable physics.  Conversely, a system that
generates excessive unnecessary dumps may protect the machine but degrade
scientific efficiency.  The operational objective is therefore a
protection architecture that keeps Eq.~\eqref{eq:mps_physics_allowed_trajectory}
with sufficiently high reliability while minimizing unnecessary
departures from the physics trajectory.

The LHC MPS  consequently illustrates the broader
``safety in physics'' principle developed in this manuscript.  Protection
is not external to scientific operation: it determines which beam states
can be reached, how transitions between them may occur, and how long
high-energy operation can be sustained without unacceptable risk to the
accelerator.

The discussion so far has concentrated primarily on the control of stored
beam and magnetic energy.  A major magnet fault or quench can, however,
transfer part of the stored energy into the cryogenic system and ultimately
into the underground environment.  The next section therefore extends the
coupled-state analysis to helium release, pressure evolution, oxygen
deficiency, ventilation, and personnel protection in the LHC cryogenic
environment.

\section{Cryogenics, helium release, and underground personnel safety}
\label{sec:cryogenic_safety}

The machine-protection discussion of
Sec.~\ref{sec:machine_protection} concentrated primarily on controlling
the large energies stored in the circulating beams and superconducting
magnet circuits.  The cryogenic system introduces a further coupling:
energy deposited in, or released from, superconducting equipment can
modify the thermodynamic state of the helium inventory and, in sufficiently
severe failure sequences, the pressure, gas composition, and habitability
of the underground environment.  Cryogenic safety therefore provides a
particularly direct example of how an equipment-level disturbance can
propagate into a personnel-safety problem.

The LHC superconducting magnet system operates at cryogenic temperatures,
with the main dipole and quadrupole circuits cooled by superfluid helium
at approximately $1.9~\mathrm{K}$.  The accelerator consequently contains
a large cryogenic inventory distributed over the underground machine
\cite{Lebrun:1999Cryogenics,Claudet:2010LHCcryo}.  Under normal conditions
this inventory is confined within the cryogenic system and forms part of
the controlled operating state.  A safety-relevant sequence arises when
thermal, mechanical, electrical, or containment failures transfer helium
from its nominal confined state toward regions in which pressure,
temperature, or atmospheric composition may become hazardous.

In the coupled-state notation introduced in
Sec.~\ref{subsec:underground_environment_state}, the corresponding
propagation can be represented schematically as
\begin{equation}
{\boldsymbol X}_{\rm mag}
\rightarrow
{\boldsymbol X}_{\rm cryo}
\rightarrow
{\boldsymbol X}_{\rm env}
\rightarrow
{\boldsymbol X}_{\rm access},
\label{eq:cryo_environment_coupling}
\end{equation}
where ${\boldsymbol X}_{\rm cryo}$ contains quantities such as helium
temperature, pressure, inventory, and mass-flow rate,
${\boldsymbol X}_{\rm env}$ describes the underground atmosphere and
ventilation state, and ${\boldsymbol X}_{\rm access}$ characterizes
occupancy and access conditions.

The physical meaning of this chain is important.  A disturbance beginning
in a superconducting magnet or cryogenic circuit can generate local
heating and helium expansion; if the resulting pressure transient exceeds
the capacity of the normal containment and relief paths, helium may enter
the tunnel or experimental cavern.  The released gas can then alter the
local atmospheric composition and, through displacement of air, reduce
the oxygen concentration available to personnel.

A generic cryogenic hazard sequence may therefore be written as
\begin{equation}
{\boldsymbol X}_{\rm cryo}^{(0)}
\xrightarrow{\;\mathcal{F}_{i}\;}
{\boldsymbol X}_{\rm cryo}^{\rm abnormal}
\xrightarrow{\;{\cal E}_{\rm rel}\;}
{\boldsymbol X}_{\rm env}^{\rm perturbed}
\xrightarrow{\;{\cal E}_{\rm exp}\;}
{\boldsymbol{\cal C}}_{i},
\label{eq:cryogenic_hazard_chain}
\end{equation}
where $\mathcal{F}_{i}$ denotes the initiating failure,
${\cal E}_{\rm rel}$ the helium-release or pressure-propagation process,
${\cal E}_{\rm exp}$ the subsequent exposure mechanism, and
${\boldsymbol{\cal C}}_{i}$ the corresponding consequence vector.

The relevant consequences are multidimensional.  They can include
mechanical loading from overpressure, damage to cryogenic or accelerator
equipment, degradation of ventilation conditions, oxygen deficiency,
restricted access, operational interruption, and, in sufficiently severe
cases, direct risk to personnel.  It is therefore useful to write
schematically
\begin{equation}
{\boldsymbol{\cal C}}_{\rm cryo}
=
\left(
{\cal C}_{\rm pressure},
{\cal C}_{\rm equipment},
{\cal C}_{\rm ODH},
{\cal C}_{\rm access},
{\cal C}_{\rm pers},
{\cal C}_{\rm op},
\ldots
\right),
\label{eq:cryogenic_consequence_vector}
\end{equation}
where ODH denotes oxygen-deficiency hazard.

A magnet quench is one possible initiator of cryogenic loading, but it
should not be identified automatically with a major helium-release
accident.  In normal protected operation, the QPS
redistributes the stored magnetic energy and the cryogenic system absorbs
the associated thermal load within its designed operating envelope.
A severe cryogenic-safety sequence requires additional conditions, such as
large energy release, loss of containment, inadequate relief, or abnormal
propagation into the surrounding environment.  Thus,
\begin{equation}
{\boldsymbol X}_{\rm mag}^{\rm quench}
\not\Longrightarrow
{\boldsymbol X}_{\rm env}^{\rm hazardous}
\label{eq:quench_not_environmental_accident}
\end{equation}
in general.

The distinction is essential for the barrier analysis developed in
Sec.~\ref{sec:hazards_barriers}.  Different protection layers act at
different stages of the cryogenic hazard trajectory.  Energy extraction
and quench protection limit the initial thermal loading; cryogenic
containment and relief systems control pressure and mass flow; ventilation
limits the persistence and spatial extent of released helium; oxygen
monitoring detects deterioration of the atmosphere; and access control,
alarms, and evacuation procedures reduce personnel exposure.

The corresponding DiD structure can be represented as
\begin{equation}
\begin{array}{c}
\text{thermal or mechanical disturbance}
\\
\downarrow
\\
\text{cryogenic containment and pressure control}
\\
\downarrow
\\
\text{helium release, if containment is challenged}
\\
\downarrow
\\
\text{ventilation and atmospheric monitoring}
\\
\downarrow
\\
\text{access restriction, warning, and evacuation}
\end{array}
\label{eq:cryogenic_protection_layers}
\end{equation}
with several barriers acting preventively and others mitigatively,
depending on the consequence being considered.

The temporal structure is also different from that of beam protection.
Some cryogenic transients can evolve rapidly through pressure-wave
propagation or flashing and expansion of helium, while atmospheric
dispersion, ventilation, oxygen depletion, and recovery can occur over
longer timescales.  The available protection time is therefore still
obeying Eq.~\eqref{eq:magnet_interface_available_time},
but the relevant observables now include pressure, temperature, mass-flow
rate, gas composition, oxygen concentration, and ventilation state rather
than beam-loss and orbit variables.

Cryogenic safety also makes explicit why the complete safe operational
phase space cannot be defined from machine protection alone.  A magnet or
cryogenic state may be technically acceptable for equipment while the
corresponding underground environment is unsuitable for personnel
occupancy.  Conversely, an unoccupied tunnel may tolerate transient
atmospheric conditions that would be unacceptable during access.  Thus,
the admissible region depends on occupancy,
\begin{equation}
\Omega_{\rm safe}^{\rm occupied}
\neq
\Omega_{\rm safe}^{\rm unoccupied},
\label{eq:occupied_unoccupied_safe_spaces}
\end{equation}
and the access state becomes a genuine component of the physical safety
problem rather than an external administrative variable.

The cryogenic system therefore extends the central argument of this
manuscript from machine protection to personnel safety.  Stored magnetic
energy, helium thermodynamics, pressure propagation, gas transport,
ventilation, atmospheric composition, and access conditions form a single
coupled protection problem.  The following subsections examine these
links in more detail, beginning with the thermodynamic consequences of
helium expansion and release.

\subsection{Superconducting operation and the origin of cryogenic hazards}
\label{subsec:superconducting_cryo_hazard}

The cryogenic hazards of the LHC originate from the physical conditions
required for superconducting accelerator operation.  The main dipole and
quadrupole magnets operate with superfluid helium at approximately
$1.9~\mathrm{K}$, allowing the superconducting coils to carry the large
currents required to generate the magnetic fields of the accelerator
\cite{Lebrun:1999Cryogenics,Claudet:2010LHCcryo}.  This operating state
combines substantial stored magnetic energy with a large cryogenic helium
inventory confined at very low temperature.

The stored magnetic energy of a magnet circuit carrying current is defined
in Eq.~\eqref{eq:inductive_energy}, while the corresponding magnetic-field
energy density is given in Eq.~\eqref{eq:magnetic_energy_density}.

Under nominal conditions, the magnet, cryogenic, and electrical states
occupy a coupled superconducting operating domain,
\begin{equation}
\left(
{\boldsymbol X}_{\rm mag},
{\boldsymbol X}_{\rm cryo},
{\boldsymbol X}_{\rm power}
\right)
\in
\Omega_{\rm SC},
\label{eq:cryo_superconducting_domain}
\end{equation}
where $\Omega_{\rm SC}$ denotes the region compatible with stable
superconducting operation.  The boundary of this region depends on
current, magnetic field, temperature, cooling conditions, local energy
deposition, and conductor properties.

A sufficiently large disturbance can drive part of a superconducting
coil into the normal-conducting state,
\begin{equation}
{\boldsymbol X}_{\rm mag}^{\rm SC}
\xrightarrow{\;\mathcal{F}_{\rm q}\;}
{\boldsymbol X}_{\rm mag}^{\rm N},
\label{eq:cryo_quench_transition}
\end{equation}
where $\mathcal{F}_{\rm q}$ denotes the initiating perturbation.  Possible
initiators include beam-induced energy deposition, powering faults,
mechanical disturbances, conductor motion, local thermal disturbances, or
other events capable of exceeding the stability margin of the
superconductor.

The appearance of a normal-conducting region generates resistive heating Eq.~\eqref{eq:quench_joule_heating},
\begin{equation}
P_{\rm J}(t)
=
I^{2}(t)\,
R_{\rm q}(t),
\label{eq:cryo_quench_joule_power}
\end{equation}
where $R_{\rm q}(t)$ is the resistance of the growing normal zone.  The
resulting thermal load must be controlled rapidly to prevent excessive
temperature rise or localization of the stored magnetic energy.

As discussed in
Sec.~\ref{subsec:machine_quench_interface}, the QPS
detects the resistive transition and initiates controlled redistribution
of the magnetic energy.  The corresponding energy flow may be represented
schematically as
\begin{equation}
E_{\rm mag}
\rightarrow
E_{\rm ext}
+
E_{\rm coil}
+
E_{\rm structure}
+
E_{\rm cryo}
+\cdots ,
\label{eq:magnet_energy_to_cryo}
\end{equation}
where $E_{\rm ext}$ denotes energy removed through dedicated extraction
paths and $E_{\rm cryo}$ the part transferred ultimately to the
cryogenic system.

The cryogenic consequence of a quench is therefore determined not simply
by whether superconductivity is lost, but by how the magnetic energy is
redistributed and how efficiently the resulting heat is accommodated.
The helium absorbs part of the thermal load and undergoes changes of
temperature, pressure, density, and phase.  One may write schematically
\begin{equation}
{\boldsymbol X}_{\rm cryo}
=
\left(
T_{\rm He},
P_{\rm He},
\rho_{\rm He},
m_{\rm He},
\dot m_{\rm He},
\ldots
\right),
\label{eq:cryogenic_state_variables}
\end{equation}
so that a thermal disturbance produces
\begin{equation}
{\boldsymbol X}_{\rm cryo}^{(0)}
\rightarrow
{\boldsymbol X}_{\rm cryo}^{\rm transient}.
\label{eq:cryogenic_state_transition}
\end{equation}

For a helium mass $m_{\rm He}$ undergoing a thermodynamic transition from
state 1 to state 2, the corresponding heat input may be represented in
terms of the specific enthalpy difference as
\begin{equation}
Q
=
m_{\rm He}
\left(
h_{2}-h_{1}
\right),
\label{eq:helium_enthalpy_load}
\end{equation}
where $h_{1}$ and $h_{2}$ denote the specific enthalpies of the initial and
final states.  In realistic cryogenic transients, the relation among
energy deposition, pressure rise, phase change, and mass flow requires the
appropriate helium equation of state and the geometry of the cryogenic
circuit.

The large thermodynamic expansion associated with warming cryogenic helium
is central to the safety problem.  Helium initially confined as a dense
cryogenic fluid occupies a much larger volume when warmed toward ambient
temperature and reduced toward atmospheric pressure.  For scale, one
kilogram of helium gas at approximately room temperature and atmospheric
pressure occupies several cubic metres, whereas the same mass occupies a
much smaller volume in its cryogenic state Eq.~\eqref{eq:helium_expansion_factor}.  Depending on the initial
thermodynamic state, the corresponding volume expansion can therefore be
of order~\cite{Chorowski:2000Helium,Chorowski:2001ODH}
\begin{equation}
\frac{V_{\rm warm}}{V_{\rm cryo}}
\sim
10^{3}.
\label{eq:helium_expansion_scale}
\end{equation}

This large expansion converts a thermal disturbance into a potential
pressure and mass-flow problem.  If the helium remains within the
cryogenic circuit, the principal quantities of interest are the pressure
rise and the capacity of the relief system.  If containment is lost, the
same expanding helium can enter the tunnel or cavern atmosphere and
generate a different hazard chain,
\begin{equation}
\begin{array}{c}
\text{cryogenic heating}
\\
\downarrow
\\
\text{helium expansion}
\\
\downarrow
\\
\text{pressure and mass flow}
\\
\downarrow
\\
\text{release to environment}
\\
\downarrow
\\
\text{oxygen displacement}.
\end{array}
\label{eq:cryo_hazard_sequence}
\end{equation}

The mass-flow rate becomes a particularly important variable,
\begin{equation}
\dot m_{\rm He}
=
\frac{d m_{\rm He}}{dt},
\label{eq:helium_mass_flow}
\end{equation}
because pressure relief, gas transport, ventilation loading, and the
development of an oxygen-deficient atmosphere depend not only on the
total helium mass released but also on the rate at which the release
occurs.

It is therefore useful to distinguish three successive physical stages:
\begin{equation}
\begin{array}{c}
\text{loss of superconducting or cryogenic equilibrium}
\\
\downarrow
\\
\text{thermal and pressure transient within the cryogenic system}
\\
\downarrow
\\
\text{possible transfer of helium to the underground environment}.
\end{array}
\label{eq:cryo_three_stage_hazard}
\end{equation}
The first two stages concern primarily equipment and cryogenic integrity;
the third introduces direct coupling to environmental and personnel-safety
variables.

This distinction also clarifies why an ordinary protected quench should
not be equated with a helium-release accident.  A protected quench can
produce significant thermal and pressure transients while remaining
within the design envelope of the cryogenic containment and relief
systems.  A hazardous environmental release requires an additional
failure or exceedance of the corresponding containment or pressure-control
functions.  In the notation of the preceding subsection,
\begin{equation}
{\boldsymbol X}_{\rm mag}^{\rm quench}
\not\Longrightarrow
{\boldsymbol X}_{\rm env}^{\rm hazardous}.
\end{equation}

The relevant protection objective is consequently to prevent the
cryogenic trajectory from progressing from a controlled internal transient
to an uncontrolled release.  Schematically,
\begin{equation}
{\boldsymbol X}_{\rm cryo}^{\rm transient}
\xrightarrow{\;\mathcal{P}_{\rm cryo}\;}
{\boldsymbol X}_{\rm cryo}^{\rm controlled},
\label{eq:cryo_protected_transition}
\end{equation}
where $\mathcal{P}_{\rm cryo}$ represents containment, pressure relief,
energy extraction, isolation, and related cryogenic protection functions.

If these functions are insufficient for the developing event, the
protection problem moves from the cryogenic circuit to the underground
environment,
\begin{equation}
{\boldsymbol X}_{\rm cryo}
\rightarrow
{\boldsymbol X}_{\rm env},
\label{eq:cryo_to_environment_transition}
\end{equation}
and ventilation, oxygen monitoring, access control, alarms, and evacuation
become the relevant downstream barriers.

The superconducting cryogenic system therefore illustrates a central
feature of the coupled-state approach: the same initiating disturbance can
move successively through electromagnetic, thermal, fluid-dynamic, and
environmental regimes.  The next subsection examines the thermodynamic
expansion and pressure evolution of released helium in greater detail.

\subsection{Helium expansion and cryogenic release}
\label{subsec:helium_expansion_cryo}

A distinctive feature of the cryogenic hazard is the very large change in
specific volume that accompanies the transition of helium from its
cryogenic state to warm gas.  In the LHC magnet system, helium is maintained
at cryogenic temperatures, including operation in the superfluid regime
near \(1.9~\mathrm{K}\), and therefore occupies a comparatively small volume
for a given mass.  The thermodynamic and transport properties of helium,
including those of liquid and superfluid helium, are reviewed in
Ref.~\cite{VanSciver:2012Helium}.  If helium is released from the cold system
and receives heat from the surrounding structures and environment,
vaporization and warming produce a large volumetric expansion
\cite{Lebrun:1999Cryogenics,Claudet:2010LHCcryo,Chorowski:2000Helium}.

The corresponding expansion scale has already been quantified in
Eqs.~\eqref{eq:helium_expansion} and
\eqref{eq:helium_expansion_numeric}.  In particular, \(1~\mathrm{kg}\) of
helium warmed to approximately room temperature at atmospheric pressure
occupies about \(6.1~\mathrm{m^3}\).  Relative to its cryogenic volume, this
corresponds to an expansion by approximately three orders of magnitude,
with the precise ratio depending on the initial and final thermodynamic
states.  This large expansion is one of the principal reasons why even a
localized cryogenic release can become relevant on the scale of an
underground accelerator enclosure.

The expansion factor should not, however, be interpreted as an
instantaneous pressure multiplication.  A real release is a transient
thermodynamic process involving heat transfer, phase change, potentially
two-phase flow, and transport through a finite hydraulic geometry.  Its
evolution therefore depends not only on the helium inventory, but also on
the initial thermodynamic state, the deposited heat, the available
expansion volume, flow restrictions, and the discharge path.  The helium
enthalpy load and mass-flow rate introduced in
Eqs.~\eqref{eq:helium_enthalpy_load} and
\eqref{eq:helium_mass_flow} provide the corresponding energetic and
dynamic measures of this process.

From the safety perspective, the important quantity is consequently not
the expansion factor alone, but the amount of helium transferred out of
the cryogenic system, the rate at which this transfer occurs, and the
environment into which it is discharged.  These quantities determine how
a disturbance of the cryogenic state can propagate into the environmental
part of the facility state introduced above.  A sufficiently rapid release
can challenge the capacity of the surrounding volume and discharge paths,
while the resulting helium transport can modify the local atmospheric
conditions.

The large thermodynamic expansion of helium thus provides the physical
link between a cryogenic initiating event and two distinct downstream
hazards: pressure development in confined volumes and displacement of
breathable air.  The corresponding protection functions are considered
separately below, beginning with pressure limitation and relief.

\subsection{Pressure evolution and relief capacity}
\label{subsec:pressure_relief}

The large volumetric expansion discussed in
Sec.~\ref{subsec:helium_expansion_cryo} becomes a pressure hazard when
helium is released into a volume from which it cannot be discharged at a
sufficient rate.  Pressure protection must therefore be considered as a
dynamic problem: helium enters the protected volume, continues to receive
heat, and is simultaneously discharged through the available relief paths.
For a control volume containing helium mass \(m\), the corresponding mass
balance may be written as
\begin{equation}
\frac{dm}{dt}
=
\dot m_{\rm in}
-
\dot m_{\rm out},
\label{eq:cryogenic_mass_balance}
\end{equation}
where \(\dot m_{\rm in}\) denotes the helium flow into the protected volume
and \(\dot m_{\rm out}\) the flow removed through the relief system.

For illustration, if the helium in a fixed volume \(V\) is represented by
the same warm-gas approximation used in
Sec.~\ref{subsec:helium_expansion_cryo}, its pressure evolution can be
written schematically as
\begin{equation}
\frac{dP}{dt}
=
\frac{R_{\rm He}}{V}
\left[
T\left(\dot m_{\rm in}-\dot m_{\rm out}\right)
+
m\frac{dT}{dt}
\right],
\label{eq:pressure_evolution}
\end{equation}
where \(R_{\rm He}\) is the specific gas constant of helium.  This relation
is not intended as a detailed model of an actual cryogenic discharge,
which may involve two-phase flow, strong spatial gradients, rapidly
changing thermodynamic properties, and pressure-dependent flow through
the relief devices.  It instead makes explicit the basic competition
between helium accumulation, warming, available volume, and discharge.

The safety function of the pressure-relief system is consequently to
prevent the transient pressure from exceeding the allowable pressure of
the protected enclosure,
\begin{equation}
P_{\rm max} \leq P_{\rm allow},
\label{eq:pressure_allowable}
\end{equation}
for the set of credible accidental conditions considered in the design.
Safety valves, rupture disks, and other pressure-relief devices therefore
constitute passive mitigating barriers whose required capacity follows
from the thermodynamic state and mass-flow rate generated by the relevant
initiating event \cite{ParmaLeclercq:2017CryogenicSafety}.

For the LHC, this requirement is particularly demanding because a severe
failure can transfer helium from the magnet cold mass into the insulation
vacuum enclosure at a very high rate.  For a standard cryogenic
subsector, the maximum credible incident (MCI) adopted for the revised
protection scheme corresponds to a maximum breach area of approximately
\(120~\mathrm{cm^2}\) and a helium flow to be discharged of approximately
\(30~\mathrm{kg\,s^{-1}}\) at \(90~\mathrm{K}\)
\cite{CryogenicRisk:2010,Cruikshank:2010Relief}.  The relief capacity must
therefore be established from the credible failure scenario and the
resulting transient flow, rather than from the static helium inventory
alone.

In the terminology of the risk framework introduced in
Sec.~\ref{sec:hazards_barriers}, the pressure-relief design sequence can
be represented schematically as
\begin{equation}
F_i
\;\longrightarrow\;
\left\{
\dot m_{\rm design}^{(i)},
T_{\rm design}^{(i)}
\right\}
\;\longrightarrow\;
A_{\rm relief}
\;\longrightarrow\;
P_{\rm max}^{(i)}
\leq
P_{\rm allow},
\label{eq:relief_design_chain}
\end{equation}
where \(F_i\) is a credible initiating event,
\(\dot m_{\rm design}^{(i)}\) and \(T_{\rm design}^{(i)}\) characterize
the corresponding design discharge, and \(A_{\rm relief}\) is the
effective relief area.  In practice, the required relief configuration
also depends on the hydraulic impedance upstream and downstream of the
relief devices and on the thermodynamic evolution of the discharged
helium \cite{Cruikshank:2010Relief,ParmaLeclercq:2017CryogenicSafety}.

Pressure limitation does not terminate the hazard chain.  Helium safely
discharged from the cryogenic enclosure must still be transported to an
appropriate destination.  In an underground environment, its release can
alter the local atmosphere and thereby couple the cryogenic state to the
environmental and access states.  The resulting oxygen-displacement
hazard is considered in the following subsection.

\subsection{Oxygen displacement and atmospheric hazard}
\label{subsec:oxygen_displacement}

Pressure relief limits the pressure consequences of a cryogenic transient,
but it does not necessarily eliminate the downstream consequences of the
released helium.  Helium is chemically inert and non-toxic, but its release
into an occupied underground volume can displace air and thereby reduce the
local oxygen concentration.  In the confined environment of the LHC this
constitutes an oxygen deficiency hazard (ODH) and provides a direct coupling
between the cryogenic, environmental, and access states of the facility
\cite{Chorowski:2000Helium,Chorowski:2001ODH}.

A useful first estimate can be obtained by assuming homogeneous mixing of
helium with initially normal air.  Let $C_{\rm He}$ denote the helium
fraction in the resulting gas mixture and $C_{O_2}^{(0)}$ the initial
oxygen fraction.  Neglecting chemical reactions and treating the gas
fractions as molar fractions, the oxygen fraction after dilution is
\begin{equation}
C_{O_2}
=
C_{O_2}^{(0)}
\left(
1-C_{\rm He}
\right).
\label{eq:oxygen_dilution}
\end{equation}
For a specified minimum oxygen fraction $C_{O_2}^{\rm min}$, the
corresponding helium fraction is therefore
\begin{equation}
C_{\rm He}^{\rm crit}
=
1-
\frac{C_{O_2}^{\rm min}}
     {C_{O_2}^{(0)}}.
\label{eq:critical_helium_fraction}
\end{equation}
Taking $C_{O_2}^{(0)}\simeq0.21$ and, for illustration, an oxygen level
of $C_{O_2}^{\rm min}=0.18$ gives
\begin{equation}
C_{\rm He}^{\rm crit}
\simeq
0.14.
\label{eq:critical_helium_fraction_example}
\end{equation}
Thus, a helium fraction of order ten percent can already produce a
substantial reduction of the oxygen concentration.  The numerical value
in Eq.~\eqref{eq:critical_helium_fraction_example} should be regarded as
an illustrative scale associated with the adopted oxygen criterion rather
than as a universal boundary for all ODH classifications.

The homogeneous-mixing model is useful for establishing this scale but is
not, in general, an adequate description of an actual cryogenic release.
Following discharge, the helium distribution is governed by the initial
temperature and momentum of the released flow, turbulent entrainment,
buoyancy, heat exchange with surrounding structures, underground geometry,
and forced ventilation.  The resulting atmosphere can therefore be
strongly nonuniform and time dependent
\cite{Chorowski:2000Helium,Chorowski:2001ODH}.

The safety-relevant oxygen concentration should consequently be treated as
a local field,
\begin{equation}
C_{O_2}
=
C_{O_2}({\boldsymbol r},t),
\qquad
C_{\rm He}
=
C_{\rm He}({\boldsymbol r},t),
\label{eq:oxygen_concentration_field}
\end{equation}
rather than solely through an enclosure-averaged concentration.  This
description is consistent with the environmental-state block introduced
in Eq~\eqref{eq:environment_state_vector} of Sec.~\ref{subsec:underground_environment_state}.

The spatial dependence is particularly important for personnel protection.
A total released helium inventory, or even an acceptable volume-averaged
oxygen concentration, does not guarantee that the atmosphere is acceptable
at every location where personnel may be present.  A necessary local
occupancy condition may be written schematically as
\begin{equation}
C_{O_2}({\boldsymbol r},t)
\geq
C_{O_2}^{\rm min},
\qquad
{\boldsymbol r}
\in
{\cal V}_{\rm occ}(t),
\label{eq:oxygen_safe_access_condition}
\end{equation}
where ${\cal V}_{\rm occ}(t)$ denotes the spatial region occupied by, or
accessible to, personnel at time $t$.  Equation~
\eqref{eq:oxygen_safe_access_condition} represents only the oxygen-related
condition; other environmental, technical, and procedural requirements
may also be necessary for safe occupancy.

This distinction avoids identifying a spatial region with the state-space
domains introduced earlier.  In terms of those domains, the corresponding
requirement is that the environmental state remain within the region
compatible with occupancy,
\begin{equation}
{\boldsymbol X}_{\rm env}(t)
\in
\Omega_{\rm env}^{\rm occupied}.
\label{eq:oxygen_occupied_state_condition}
\end{equation}
A cryogenic release can therefore make an area incompatible with personnel
occupancy even when the pressure transient within the cryogenic system has
been successfully controlled.

The physical sequence can be represented schematically as
\begin{equation}
{\boldsymbol X}_{\rm cryo}
\xrightarrow{\;{\cal E}_{\rm rel}\;}
{\boldsymbol X}_{\rm env}
\rightarrow
C_{O_2}({\boldsymbol r},t)\downarrow
\rightarrow
{\boldsymbol X}_{\rm access},
\label{eq:cryo_odh_hazard_chain}
\end{equation}
where ${\cal E}_{\rm rel}$ denotes the helium-release process.  The final
coupling reflects the fact that sufficiently degraded atmospheric
conditions require warning, evacuation, access restriction, or inhibition
of re-entry.

Several protection functions act at different stages of this sequence.
Containment and appropriately routed relief paths act upstream by reducing
the probability or extent of helium entering occupied volumes.
Ventilation acts downstream on the environmental state by diluting and
removing released helium.  Oxygen-deficiency monitoring does not modify
the atmosphere itself, but provides the measurement required to determine
whether the evolving state remains compatible with occupancy.

The corresponding functional distinction may be summarized as
\begin{equation}
\text{helium release}
\rightarrow
\text{transport and mixing}
\rightarrow
C_{O_2}({\boldsymbol r},t)
\rightarrow
\left\{
\begin{array}{l}
\text{ventilation and atmospheric recovery},\\
\text{ODH monitoring and personnel protection}.
\end{array}
\right.
\label{eq:oxygen_displacement_protection_chain}
\end{equation}

Thus, oxygen displacement provides the physical link between loss of
cryogenic containment and personnel exposure.  The detailed evolution of
the atmospheric state depends on helium transport and ventilation, while
the translation of that state into alarms and access decisions depends on
ODH monitoring.  These two complementary functions are considered in
Secs.~\ref{subsec:cryo_ventilation} and \ref{subsec:odh_monitoring},
respectively.

\subsection{Helium transport, ventilation, and atmospheric recovery}
\label{subsec:cryo_ventilation}

Once helium has escaped from the cryogenic system, the protection problem
changes character.  The relevant dynamics are no longer governed primarily
by the thermodynamic state inside the cryogenic circuit, but by the
transport, dilution, and removal of helium within the underground
environment.  Ventilation therefore provides an important downstream
mitigative barrier between a cryogenic release and the development or
persistence of an oxygen-deficient atmosphere.

In the coupled-state description introduced in
Sec.~\ref{subsec:underground_environment_state}, this stage corresponds to
the transition
\begin{equation}
{\boldsymbol X}_{\rm cryo}
\xrightarrow{\;{\cal E}_{\rm rel}\;}
{\boldsymbol X}_{\rm env},
\label{eq:cryo_release_environment_state}
\end{equation}
where the environmental state includes, among other quantities, the local
helium and oxygen concentrations, air temperature, pressure, and
ventilation conditions.

The personnel-relevant quantity is consequently not the total released
helium mass alone, but the resulting spatially and temporally varying
atmospheric composition.  One may write
\begin{equation}
C_{O_2}
=
C_{O_2}({\boldsymbol r},t),
\qquad
C_{\rm He}
=
C_{\rm He}({\boldsymbol r},t),
\label{eq:local_gas_concentrations}
\end{equation}
where ${\boldsymbol r}$ denotes position within the tunnel, cavern, shaft,
or other underground volume.  A single volume-averaged oxygen
concentration can therefore be insufficient to characterize the local
hazard.

The transport of released helium is determined by several competing
physical processes: the initial momentum and temperature of the release,
buoyancy, turbulent mixing, confinement by the tunnel or cavern geometry,
forced ventilation, heat exchange with surrounding structures, and
communication with adjacent volumes.  The relative importance of these
effects changes as the initially cold helium warms and mixes with the
surrounding air.

For an idealized well-mixed volume $V$, ventilation introduces a
characteristic exchange timescale
\begin{equation}
\tau_{\rm vent}
=
\frac{V}{\dot V_{\rm vent}},
\label{eq:ventilation_timescale}
\end{equation}
where $\dot V_{\rm vent}$ is the effective volumetric ventilation flow
through the considered region.  Equation~\eqref{eq:ventilation_timescale}
provides a useful scale for atmospheric renewal but should not be
interpreted as the actual recovery time of an arbitrary underground
release.

Indeed, the assumption of complete mixing can fail significantly during
the transient following a cryogenic release.  Local oxygen concentration
may then differ substantially from the value inferred from a simple
volume-averaged dilution model,
\begin{equation}
C_{O_2}({\boldsymbol r},t)
\neq
\left\langle C_{O_2}(t)\right\rangle_V .
\label{eq:local_vs_average_oxygen}
\end{equation}
This distinction is particularly relevant in large underground volumes,
regions with restricted air exchange, local recesses, shafts, or areas in
which buoyancy and directed ventilation establish preferential flow
paths.

Ventilation should therefore be regarded primarily as a physical
mitigation mechanism,
\begin{equation}
{\cal P}_{\rm vent}:
{\boldsymbol X}_{\rm env}^{\rm perturbed}
\rightarrow
{\boldsymbol X}_{\rm env}^{\rm controlled},
\label{eq:ventilation_environment_control}
\end{equation}
whose objective is to dilute and remove released helium and to restore an
atmospheric state compatible with the intended occupancy conditions.

For a continuing release, however, ventilation must compete with the
helium source.  The relevant balance depends on the helium mass-flow rate
$\dot m_{\rm He}$ introduced in
Eq.~\eqref{eq:helium_mass_flow}, the effective ventilation flow, and the
geometry of the affected volume.  A sufficiently large release can
temporarily exceed the capacity of normal ventilation to maintain an
acceptable atmosphere:
\begin{equation}
\dot m_{\rm He}
\ \text{large compared with the effective removal capacity}
\quad\Longrightarrow\quad
C_{O_2}({\boldsymbol r},t)
\ \text{may decrease locally}.
\label{eq:release_ventilation_competition}
\end{equation}
The statement is deliberately qualitative because the relation between
mass-flow rate and local oxygen concentration depends strongly on the
thermodynamic state of the released helium and on the flow geometry.

Ventilation also plays different roles during the successive phases of an
event.  During the release it can reduce the rate at which helium
accumulates in occupied regions.  After the source has been isolated or
exhausted, it determines the atmospheric recovery toward normal
conditions.  Schematically,
\begin{equation}
\text{release}
\rightarrow
\text{dispersion and dilution}
\rightarrow
\text{source termination}
\rightarrow
\text{ventilated recovery}.
\label{eq:ventilation_event_phases}
\end{equation}

The corresponding recovery condition cannot be defined solely by the
elapsed time after the initiating event.  Re-entry or restoration of
normal access requires demonstration that the relevant environmental
variables have returned to an acceptable region,
\begin{equation}
{\boldsymbol X}_{\rm env}(t)
\in
\Omega_{\rm env}^{\rm admissible},
\label{eq:environment_admissible_region}
\end{equation}
where $\Omega_{\rm env}^{\rm admissible}$ includes the applicable
conditions on oxygen concentration, gas composition, temperature,
pressure, ventilation state, and any other variables relevant to the
affected area.

This point connects ventilation directly to the occupancy-dependent safe
phase space introduced in
Eq.~\eqref{eq:occupied_unoccupied_safe_spaces}.  Atmospheric conditions
that can be tolerated temporarily in an evacuated tunnel need not be
compatible with personnel access:
\begin{equation}
\Omega_{\rm env}^{\rm occupied}
\subseteq
\Omega_{\rm env}^{\rm unoccupied}.
\label{eq:environment_occupancy_regions}
\end{equation}
The precise relation is configuration dependent, but the general
principle is that evacuation changes the consequence structure without
removing the underlying atmospheric disturbance.

Ventilation is therefore only one layer of the protection architecture.
It modifies the physical atmospheric state, but it does not by itself
establish whether that state is acceptable for personnel occupancy.
That determination requires measurement of the atmosphere and translation
of those measurements into alarms, access restrictions, or evacuation
actions.

The protection chain may consequently be represented as
\begin{equation}
{\boldsymbol X}_{\rm cryo}
\rightarrow
{\boldsymbol X}_{\rm env}^{\rm perturbed}
\xrightarrow{\;{\cal P}_{\rm vent}\;}
{\boldsymbol X}_{\rm env}^{\rm controlled}
\rightarrow
\text{atmospheric verification}
\rightarrow
{\boldsymbol X}_{\rm access}.
\label{eq:ventilation_access_chain}
\end{equation}

In the terminology of Sec.~\ref{subsec:preventive_mitigative},
ventilation is predominantly mitigative with respect to a helium release:
it does not normally prevent the initial loss of containment, but reduces
the spatial extent, duration, and potentially the severity of the
resulting atmospheric disturbance.  With respect to personnel exposure,
however, sufficiently effective ventilation can also have a preventive
role by maintaining occupied regions outside the oxygen-deficiency
domain.

The effectiveness of ventilation must consequently be assessed together
with the release scenario, underground geometry, occupancy state, and
atmospheric monitoring.  This coupling motivates the following discussion
of oxygen-deficiency monitoring, which provides the observational link
between the evolving environmental state and personnel-protection
decisions.

\subsection{Oxygen-deficiency monitoring and personnel warning}
\label{subsec:odh_monitoring}

The ventilation system discussed in
Sec.~\ref{subsec:cryo_ventilation} acts directly on the physical state of
the underground atmosphere by diluting and removing released helium.
Personnel protection requires an additional and conceptually distinct
function: determining whether the resulting atmospheric state remains
compatible with occupancy.  This observational role is provided by
ODH monitoring and the associated alarm and
access-protection systems.

ODH monitoring should therefore not be regarded as a mechanism that
reduces the helium concentration itself.  Rather, it observes the evolving
environmental state and converts the measured information, together with
the validity status of the monitoring channels, into protection-relevant
alarm and access decisions.  In the notation used above,
\begin{equation}
{\cal M}_{\rm ODH}:
{\boldsymbol X}_{\rm env}
\longrightarrow
\left(
C_{O_2,j}^{\rm meas},
\chi_{{\rm ODH},j}^{\rm valid}
\right)
\longrightarrow
\left(
\chi_{\rm alarm}^{\rm ODH},
\sigma_{\rm access}
\right),
\label{eq:odh_monitoring_chain}
\end{equation}
where ${\cal M}_{\rm ODH}$ denotes the ODH monitoring and decision
function, $C_{O_2,j}^{\rm meas}$ is the oxygen concentration measured by
channel $j$, and
$\chi_{{\rm ODH},j}^{\rm valid}\in\{0,1\}$ is the corresponding binary
channel-validity indicator.  The quantity
$\chi_{\rm alarm}^{\rm ODH}\in\{0,1\}$ denotes the binary ODH-alarm
indicator, while
$\sigma_{\rm access}\in\Sigma_{\rm access}$ denotes the discrete
multivalued access status of the relevant area.
Here the parentheses group quantities that jointly characterize the
corresponding stage of the monitoring and protection chain.  Thus,
$\left(C_{O_2,j}^{\rm meas},
\chi_{{\rm ODH},j}^{\rm valid}\right)$ combines the measured oxygen
concentration with the validity status of monitoring channel $j$, while
$\left(\chi_{\rm alarm}^{\rm ODH},\sigma_{\rm access}\right)$ combines
the resulting alarm indicator with the access status of the relevant
area.

The alarm decision and the access decision should be distinguished.
The former is determined by the measured environmental condition,
whereas the latter can also depend on the validity of the monitoring
channels required for safe occupancy.  Accordingly,
$\sigma_{\rm access}^{\rm restricted}\in\Sigma_{\rm access}$ denotes
an access state in which personnel entry is restricted either because an
ODH alarm condition is present or because a required monitoring channel
is invalid.

This distinction complements the ventilation mapping
introduced in Eq.~\eqref{eq:ventilation_environment_control}.
Ventilation modifies the atmosphere, whereas ODH monitoring determines
whether the modified atmosphere is acceptable for personnel occupancy.

Because the oxygen concentration is generally a local field, \(C_{O_2}({\boldsymbol r},t)\),
the measurement obtained from  of channel $j$ is more appropriately written as
\begin{equation}
C_{O_2,j}^{\rm meas}(t)
=
{\cal M}_{j}
\left[
C_{O_2}({\boldsymbol r},t)
\right]
+
\delta C_j(t),
\label{eq:oxygen_sensor_measurement}
\end{equation}
where ${\cal M}_{j}$ represents the spatial and instrumental response of
the sensor and $\delta C_j$ denotes the corresponding measurement
uncertainty or instrumental deviation.  A sensor therefore samples the
atmospheric state only in the region to which it is sensitive.

This observation is important because, as emphasized in
Eq.~\eqref{eq:local_vs_average_oxygen}, local atmospheric conditions can
differ substantially from a volume average.  Sensor placement must
therefore reflect the credible transport paths of released helium, the
geometry of the tunnel or cavern, ventilation flows, possible accumulation
regions, and the locations in which personnel may be present~\cite{Chorowski:2001ODH}.  The
monitoring problem is consequently spatial as well as temporal.

For a sensor $j$, the simplest protection logic may be represented as
\begin{equation}
C_{O_2,j}^{\rm meas}
<
C_{O_2,j}^{\rm th}
\quad\Longrightarrow\quad
A_j=1,
\label{eq:odh_threshold_condition}
\end{equation}
where $C_{O_2,j}^{\rm th}$ denotes the applicable alarm threshold and
$A_j=1$ represents the corresponding ODH alarm condition.  In practice,
several alarm levels, delays, voting conditions, or mode-dependent
responses may be used; Eq.~\eqref{eq:odh_threshold_condition} is intended
only to represent the underlying physical logic.

The relevant consequence of an ODH alarm depends on the operational and
occupancy state.  A detected atmospheric deterioration may initiate one
or more actions,
\begin{equation}
{\mathcal A}_j=1
\quad\Longrightarrow\quad
\left\{
\begin{array}{l}
\text{local warning},\\
\text{evacuation request},\\
\text{access inhibition},\\
\text{restriction of re-entry},\\
\text{notification of the relevant safety systems}.
\end{array}
\right.
\label{eq:odh_alarm_actions}
\end{equation}
The precise response is determined by the protection configuration of the
area and by the severity and persistence of the measured atmospheric
deviation.

ODH monitoring therefore couples the environmental and access blocks of
the facility state,
\begin{equation}
{\boldsymbol X}_{\rm env}
\xrightarrow{\;{\cal M}_{\rm ODH}\;}
{\boldsymbol X}_{\rm access},
\label{eq:environment_to_access_monitoring}
\end{equation}
and provides a direct physical basis for changing occupancy conditions.
Access restriction is not introduced merely because an abnormal cryogenic
event has occurred, but because the measured or inferred environmental
state is no longer compatible with the conditions required for personnel
presence.

The occupancy dependence introduced in
Eq.~\eqref{eq:occupied_unoccupied_safe_spaces} is therefore especially
important.  For personnel access, the relevant condition can be written
schematically as
\begin{equation}
{\boldsymbol X}_{\rm env}
\in
\Omega_{\rm env}^{\rm occupied},
\label{eq:odh_occupied_environment_condition}
\end{equation}
whereas after evacuation the facility may temporarily occupy states in
\(
\Omega_{\rm env}^{\rm unoccupied}
\setminus
\Omega_{\rm env}^{\rm occupied}
\)
without implying that the environmental disturbance itself has already
been removed.

This distinction is particularly relevant during atmospheric recovery.
The end of a helium release does not by itself justify re-entry.  The
environment must first return to a validated state,
\begin{equation}
{\boldsymbol X}_{\rm env}(t)
\in
\Omega_{\rm env}^{\rm occupied},
\label{eq:odh_reentry_condition}
\end{equation}
and the corresponding monitoring channels must indicate that the relevant
conditions have been restored.  ODH monitoring therefore contributes not
only to the initial detection of an event but also to verification of the
subsequent recovery.

The protection function is also subject to the fail-safe and diagnostic
principles developed in
Sec.~\ref{subsec:failsafe_protection}.  A protection-relevant failure of an
oxygen-monitoring channel should not be interpreted automatically as
evidence of a normal atmosphere.  Let
${\chi}_{{\rm ODH},j}^{\rm valid}$ denote the validated state of sensor
channel $j$.  The logic can then be represented as
\begin{equation}
\left[
C_{O_2,j}^{\rm meas}<C_{O_2,j}^{\rm th}
\right]
\;\lor\;
\left[
\chi_{{\rm ODH},j}^{\rm valid}=0
\right]
\quad\Longrightarrow\quad
\sigma_{\rm access}
=
\sigma_{\rm access}^{\rm restricted},
\label{eq:odh_invalid_or_alarm}
\end{equation}
where the  value $\sigma_{\rm access}^{\rm restricted}\in\Sigma_{\rm access}$
denotes the access configuration in which personnel entry is restricted
as a consequence of an ODH alarm or an invalid monitoring channel.
Equation~\eqref{eq:odh_invalid_or_alarm} is schematic; the actual response
depends on the redundancy, voting logic, and operational configuration of
the monitoring system.

This requirement illustrates why monitoring availability and diagnostic
coverage are themselves safety quantities.  If $F_{{\rm ODH},j}$ denotes
a failure of monitoring channel $j$, one may define
\begin{equation}
{\rm DC}_{j}^{\rm ODH}
=
P\!\left(
D_{{\rm ODH},j}
\mid
F_{{\rm ODH},j}
\right),
\label{eq:odh_diagnostic_coverage}
\end{equation}
where $F_{{\rm ODH},j}$ denotes a failure of ODH monitoring channel $j$
and $D_{{\rm ODH},j}$ the event that this failure is detected by the
associated diagnostic functions.  The quantity
${\rm DC}_{j}^{\rm ODH}\in[0,1]$ denotes the diagnostic coverage of
channel $j$, i.e. the conditional probability that a channel failure is
detected given that the failure has occurred.

Timing is also relevant, although the characteristic timescales differ
from those encountered in beam protection.  For an atmospheric hazard
developing after a helium release, personnel warning must occur before the
local environment reaches the critical occupancy condition.  A
corresponding requirement is
\begin{equation}
\tau_{\rm det}^{\rm ODH}
+
\tau_{\rm alarm}
+
\tau_{\rm response}
<
\tau_{\rm avail}^{\rm pers},
\label{eq:odh_timing_condition}
\end{equation}
where $\tau_{\rm avail}^{\rm pers}$ denotes the time available before the
specified personnel-exposure condition becomes unacceptable.  This
quantity depends on the release rate, transport path, ventilation,
geometry, sensor location, and personnel location.

The effectiveness of ODH monitoring therefore cannot be evaluated
independently of ventilation and helium transport.  A sensor positioned
far from the developing low-oxygen region may respond only after a
substantial delay, while effective forced ventilation can modify both the
spatial evolution of the oxygen field and the time available for warning.
The relevant protection system is consequently the coupled combination
\begin{equation}
\left(
{\cal P}_{\rm vent},
{\cal M}_{\rm ODH},
{\cal P}_{\rm access}
\right),
\label{eq:ventilation_monitoring_access_system}
\end{equation}
where the three functions respectively modify the atmosphere, observe its
state, and control personnel exposure.

The combined protection chain may be summarized as
\begin{equation}
{\boldsymbol X}_{\rm env}
\xrightarrow{{\cal P}_{\rm vent}}
{\boldsymbol X}_{\rm env}^{\rm controlled}
\xrightarrow{{\cal M}_{\rm ODH}}
\left(
C_{O_2,j}^{\rm meas},
\chi_{{\rm ODH},j}^{\rm valid}
\right)
\longrightarrow
\left(
\chi_{\rm alarm}^{\rm ODH},
\sigma_{\rm access}
\right).
\label{eq:ventilation_odh_access_chain}
\end{equation}
Equation~\eqref{eq:ventilation_odh_access_chain} summarizes the
complementary roles of ventilation, ODH monitoring, and access control.
The ventilation function ${\cal P}_{\rm vent}$ acts on the physical
environmental state, whereas ${\cal M}_{\rm ODH}$ observes that state
through the measured oxygen concentration and the validity of the
monitoring channels.  The resulting information contributes to the ODH
alarm state and to the access status of the relevant area.

Ventilation and oxygen-deficiency monitoring are thus complementary rather
than interchangeable barriers.  Ventilation acts on the physical
atmospheric state, while ODH monitoring provides the information required
to determine whether occupancy can continue, must be terminated, or may
later be restored.  Their combination connects helium transport directly
to personnel-protection decisions and provides the physical basis for the
access restrictions discussed in the following subsection.

\subsection{Access as a cryogenic-protection variable}
\label{subsec:cryo_access}

The atmospheric-protection mechanisms discussed in
Secs.~\ref{subsec:cryo_ventilation} and \ref{subsec:odh_monitoring}
provide the physical and monitoring conditions needed to determine
whether personnel may remain in, enter, or re-enter an underground area.
Access should therefore not be treated as an administrative condition
external to the physical safety problem.  In particular, personnel
occupancy changes the consequence associated with a given cryogenic or
atmospheric state and must be included explicitly in the coupled
description.

In the state-space framework introduced in
Eq.~\eqref{eq:lhc_subsystem_state_vectors} of
Sec.~\ref{subsec:underground_environment_state}, the access component may
be written as
\begin{equation}
{\boldsymbol X}_{\rm access}
=
\left(
\sigma_{\rm access},
\chi_{\rm occupancy},
\sigma_{\rm evacuation},
\ldots
\right),
\label{eq:cryo_access_state}
\end{equation}
where $\sigma_{\rm access}$ is the multivalued access-status variable
introduced above,
$\chi_{\rm occupancy}\in\{0,1\}$ indicates whether the relevant area is
occupied, and
$\sigma_{\rm evacuation}\in\Sigma_{\rm evacuation}$ denotes its discrete
evacuation status.  In particular,
$\chi_{\rm occupancy}=1$ denotes personnel presence in the area, whereas
$\chi_{\rm occupancy}=0$ denotes an unoccupied area.

The safety significance of a cryogenic or atmospheric condition depends
explicitly on these access variables.  An environmental state that is
incompatible with personnel occupancy can, for example, remain within an
admissible unoccupied domain while ventilation acts to restore conditions
compatible with re-entry.  As already expressed by
Eq.~\eqref{eq:occupied_unoccupied_safe_spaces}, the admissible domains
therefore depend on occupancy.

More specifically, the environmental states compatible with personnel
presence form a restricted subset,
\begin{equation}
\Omega_{\rm env}^{\rm occupied}
\subseteq
\Omega_{\rm env}^{\rm unoccupied},
\label{eq:cryo_environment_access_relation}
\end{equation}
because occupancy imposes additional constraints on quantities such as
oxygen concentration, gas composition, temperature, pressure, and their
spatial distributions.  Satisfaction of these physical environmental
conditions is necessary but not, by itself, sufficient for access:
the relevant monitoring and protection functions must also be available
and validated.

At the subsystem level, the coupling may therefore be represented
schematically as
\begin{equation}
{\boldsymbol X}_{\rm cryo}
\longrightarrow
{\boldsymbol X}_{\rm env}
\longrightarrow
{\boldsymbol X}_{\rm access}.
\label{eq:cryo_environment_access_chain}
\end{equation}
The first transition represents the influence of a cryogenic disturbance
on the underground environment, while the second represents the resulting
change in personnel-access conditions.  This second coupling is mediated
by ventilation, ODH monitoring, alarm logic, and access-protection
functions, whose detailed roles were described in
Secs.~\ref{subsec:cryo_ventilation} and \ref{subsec:odh_monitoring}.

A deterioration of the atmospheric state can require several distinct
personnel-protection responses:
\begin{equation}
{\boldsymbol X}_{\rm env}
\notin
\Omega_{\rm env}^{\rm occupied}
\quad\Longrightarrow\quad
\left\{
\begin{array}{l}
\text{warning of personnel already present},\\
\text{evacuation of the affected area},\\
\text{inhibition of further access},\\
\text{restriction of re-entry}.
\end{array}
\right.
\label{eq:cryo_access_responses}
\end{equation}
These actions primarily reduce personnel exposure rather than modifying
the helium release itself and are therefore mitigative with respect to
the cryogenic initiating event.

The distinction between evacuation and access inhibition is important.
Evacuation acts on personnel already present in the affected region,
whereas access inhibition prevents additional personnel from entering a
region whose state is incompatible with occupancy.  Successful evacuation
corresponds schematically to
\begin{equation}
\chi_{\rm occupancy}:1\rightarrow0,
\label{eq:cryo_occupancy_transition}
\end{equation}
while restriction of further entry is represented by
\begin{equation}
\sigma_{\rm access}
\rightarrow
\sigma_{\rm access}^{\rm restricted}.
\label{eq:cryo_access_restriction}
\end{equation}
The variable $\sigma_{\rm evacuation}$ in
Eq.~\eqref{eq:cryo_access_state} may be used to describe the intermediate
status of the evacuation process itself, whereas
Eq.~\eqref{eq:cryo_occupancy_transition} represents its intended outcome
for the affected area.

Access protection can therefore change the personnel-consequence
structure without immediately changing the underlying physical
disturbance.  For example,
\begin{equation}
{\boldsymbol X}_{\rm env}^{\rm perturbed},
\qquad
\chi_{\rm occupancy}=0,
\label{eq:cryo_perturbed_unoccupied}
\end{equation}
can correspond to a substantially smaller personnel-exposure consequence
than the same environmental state with personnel present.  The
environment nevertheless remains perturbed and must not be interpreted as
having returned to its normal state.

Access restriction should therefore be distinguished from environmental
recovery.  Following a helium release,
\begin{equation}
\sigma_{\rm access}
=
\sigma_{\rm access}^{\rm restricted}
\label{eq:cryo_access_closed}
\end{equation}
may remain appropriate while ventilation, mixing, extraction, and
atmospheric recovery continue.  Re-entry requires an independent
validation that both the environmental conditions and the protection
functions required for occupancy have been restored.

A schematic set of re-entry conditions may be written as
\begin{equation}
{\boldsymbol X}_{\rm env}
\in
\Omega_{\rm env}^{\rm occupied},
\qquad
\chi_{\rm alarm}^{\rm ODH}=0,
\qquad
\chi_{{\rm ODH},j}^{\rm valid}=1
\quad
\text{for all required }j,
\qquad
\chi_{\rm vent}^{\rm valid}=1,
\label{eq:cryo_reentry_conditions}
\end{equation}
together with any additional technical or procedural requirements
applicable to the affected area.  Here
$\chi_{\rm vent}^{\rm valid}\in\{0,1\}$ denotes the binary indicator that
the ventilation condition required for occupancy has been validated.
When ventilation validity is not an independent requirement for a
particular area or mode, the corresponding condition is omitted.

Let
$\sigma_{\rm access}^{\rm admissible}\in\Sigma_{\rm access}$ denote an
access status compatible with personnel entry under the validated
conditions.  Re-opening can then be represented schematically as
\begin{equation}
\sigma_{\rm access}^{\rm restricted}
\rightarrow
\sigma_{\rm access}^{\rm admissible}.
\label{eq:cryo_access_reopen}
\end{equation}
This transition does not itself constitute an individual work
authorization; it indicates that the cryogenic and environmental
conditions no longer require the area to remain in the restricted-access
state.

Re-entry is therefore a state-validation problem rather than a purely
time-based decision.  The end of a release, completion of a nominal
ventilation period, or disappearance of an alarm does not by itself
demonstrate that the area has returned to a state compatible with
occupancy.

The timing of personnel protection is also important.  If atmospheric
conditions deteriorate after a release, the warning and evacuation chain
must complete before the specified exposure condition becomes
unacceptable:
\begin{equation}
\tau_{\rm det}^{\rm ODH}
+
\tau_{\rm alarm}
+
\tau_{\rm evac}
<
\tau_{\rm avail}^{\rm pers},
\label{eq:cryo_access_timing}
\end{equation}
where $\tau_{\rm avail}^{\rm pers}$ denotes the time available before the
relevant personnel-exposure condition is reached.  This interval depends
on the release rate, helium transport, ventilation, sensor location,
geometry, and the location of personnel.

Equation~\eqref{eq:cryo_access_timing} also shows why access control can
act preventively before an intervention begins.  If a credible release
scenario can develop more rapidly than safe evacuation can be ensured,
the corresponding occupied configuration should not be admitted in the
first place.  Schematically,
\begin{equation}
\text{cryogenic configuration incompatible with safe occupancy}
\quad\Longrightarrow\quad
\sigma_{\rm access}
=
\sigma_{\rm access}^{\rm restricted}.
\label{eq:cryo_preventive_access}
\end{equation}
In this case, access restriction acts upstream of personnel exposure.

Access control can therefore play either preventive or mitigative roles,
depending on the hazard trajectory being considered.  It is preventive
when entry into a configuration incompatible with safe occupancy is
restricted before exposure occurs, and mitigative when personnel are
warned or evacuated after an abnormal atmospheric condition has already
developed.  This is consistent with the consequence-dependent
classification of barriers introduced in
Sec.~\ref{subsec:preventive_mitigative}.

The same reasoning applies to maintenance and intervention states.
Opening cryogenic equipment, changing containment boundaries, disabling
or reconfiguring ventilation elements, modifying relief paths, or
performing work close to cryogenic installations can alter the relevant
hazard trajectories.  The access statuses that can be admitted therefore
depend on the cryogenic and environmental states and on the operational
or intervention mode:
\begin{equation}
\sigma_{\rm access}
\in
\Sigma_{\rm access}^{\rm allowed}
\left(
{\boldsymbol X}_{\rm cryo},
{\boldsymbol X}_{\rm env},
m
\right)
\subseteq
\Sigma_{\rm access},
\label{eq:cryo_access_configuration_dependence}
\end{equation}
where $m$ denotes the applicable operational or intervention mode.
Relevant monitoring, protection, and procedural requirements impose
additional conditions on membership in
$\Sigma_{\rm access}^{\rm allowed}$.

This configuration dependence is a direct realization
of the safe operational phase-space concept.
Let $\mathcal G_{\rm access}^{\rm occupied}$
denote the set of operational trajectories involving
personnel occupancy in the considered underground
region, and let $\mathcal G_{\rm adm}^{\rm occupied}$
denote the set of trajectories satisfying the
corresponding cryogenic, environmental, and
personnel-access protection requirements.
The admissibility of occupied operation requires
\begin{equation}
\mathcal G_{\rm access}^{\rm occupied}
\subseteq
\mathcal G_{\rm adm}^{\rm occupied}.
\label{eq:cryo_access_allowed_trajectory}
\end{equation}
If these conditions cease to be satisfied, the
relevant activities must be interrupted and
appropriate protective measures initiated,
including evacuation when required.

The coupled protection sequence can therefore be summarized at the
subsystem level as
\begin{equation}
{\boldsymbol X}_{\rm cryo}
\longrightarrow
{\boldsymbol X}_{\rm env}
\longrightarrow
{\boldsymbol X}_{\rm access}
\longrightarrow
\left\{
\begin{array}{l}
\text{continued occupancy},\\
\text{evacuation},\\
\text{access inhibition},\\
\text{validated re-entry}.
\end{array}
\right.
\label{eq:cryo_access_protection_chain}
\end{equation}
The detailed measurement and decision logic connecting
${\boldsymbol X}_{\rm env}$ to the access response is given by
Eqs.~\eqref{eq:odh_monitoring_chain} and
\eqref{eq:odh_invalid_or_alarm}.

Cryogenic access protection therefore closes the chain connecting the
physical state of the cryogenic system to potential personnel exposure.
The relevant safety boundary is determined not by the cryogenic state
alone, but by its coupling to the underground environment, the
availability and validity of the required protection functions, and the
occupancy and access states of the affected region.  This coupling becomes
particularly important in severe cryogenic incidents, for which the
protection architecture must control not only equipment and pressure
transients but also the conditions under which the affected underground
region may remain occupied or subsequently be re-entered.

\subsection{The 19 September 2008 sector 3--4 incident}
\label{subsec:2008_cryo_incident}

The incident that occurred in LHC sector~3--4 on 19 September 2008
provides an important example of the coupled failure propagation discussed
throughout this section.  The event occurred during powering tests of the
main dipole circuit, with no circulating beam present.  A defective
high-current electrical connection between superconducting magnets
developed excessive resistance, leading to local heating and an electrical
arc.  The arc damaged the helium enclosure and initiated a rapid transfer
of helium from the magnet cold mass into the insulation vacuum
\cite{Bajko:2009Incident,CryogenicRisk:2010}.

The initiating sequence may be represented schematically as
\begin{equation}
{\boldsymbol X}_{\rm power}
\xrightarrow{\;\text{resistive fault}\;}
{\boldsymbol X}_{\rm electrical}^{\rm arc}
\xrightarrow{\;\text{containment breach}\;}
{\boldsymbol X}_{\rm cryo}^{\rm release}.
\label{eq:2008_initiating_sequence}
\end{equation}
The event therefore did not begin as an oxygen-deficiency hazard or even
primarily as a cryogenic failure.  It originated in the electrical
interconnection and propagated across subsystem boundaries.

Once the helium enclosure had been breached, helium entered the insulation
vacuum at a rate substantially larger than had been assumed in the
pre-incident cryogenic risk analysis.  The resulting pressure rise in the
vacuum enclosure exceeded the capacity of the then-existing relief
configuration in the affected region.  Large pressure forces consequently
developed on the vacuum barriers and associated cryostat structures,
producing substantial mechanical displacement and damage
\cite{Bajko:2009Incident,CryogenicRisk:2010}.

The corresponding propagation can be represented as
\begin{equation}
{\boldsymbol X}_{\rm cryo}^{\rm release}
\rightarrow
{\boldsymbol X}_{\rm vacuum}^{\rm overpressure}
\rightarrow
{\boldsymbol X}_{\rm mech}^{\rm displaced}
\rightarrow
{\boldsymbol X}_{\rm cryo}^{\rm further\;release}.
\label{eq:2008_cryo_mechanical_feedback}
\end{equation}
This sequence illustrates that the cryogenic and mechanical states were
not dynamically independent.  Pressure loading generated mechanical
displacement, while the resulting damage to interconnected equipment
created additional opportunities for helium release.

Approximately $6~\mathrm{t}$ of helium were ultimately released into
sector~3--4, with about $2~\mathrm{t}$ discharged into the machine tunnel
during the first approximately two minutes of the event
\cite{LHCSafetyTaskForce:2009,CryogenicRisk:2010}.  The total released mass
alone, however, does not characterize the severity of the transient.
For pressure-relief and personnel-safety analysis, the mass-flow rate,
release location, thermodynamic state, tunnel geometry, and resulting
atmospheric transport are equally important.

The event therefore followed a much longer coupled chain than the initial
electrical fault alone would suggest:
\begin{equation}
\begin{array}{c}
\text{electrical interconnection fault}
\\
\downarrow
\\
\text{electrical arc}
\\
\downarrow
\\
\text{helium-containment breach}
\\
\downarrow
\\
\text{insulation-vacuum overpressure}
\\
\downarrow
\\
\text{mechanical displacement and damage}
\\
\downarrow
\\
\text{helium release to the tunnel}
\\
\downarrow
\\
\text{perturbation of the underground atmosphere}.
\end{array}
\label{eq:2008_complete_failure_chain}
\end{equation}

In the state-space notation used in this manuscript, the same sequence can
be written more compactly as
\begin{equation}
{\boldsymbol X}_{\rm power}
\rightarrow
{\boldsymbol X}_{\rm mag}
\rightarrow
{\boldsymbol X}_{\rm cryo}
\rightarrow
{\boldsymbol X}_{\rm mech}
\rightarrow
{\boldsymbol X}_{\rm env}
\rightarrow
{\boldsymbol X}_{\rm access}.
\label{eq:2008_state_space_chain}
\end{equation}
The incident is therefore a particularly clear example of why safety
analysis based on independent subsystem boundaries can be insufficient.

It is also important to distinguish this event from an ordinary protected
magnet quench.  During a normal quench, the QPS acts
to redistribute the stored magnetic energy while the helium and pressure
transients remain within the intended protection envelope.  The 2008 event
instead involved an electrical arc, physical rupture of the helium
enclosure, rapid transfer of helium into the insulation vacuum, and
subsequent overpressure of that volume.  Thus,
\begin{equation}
{\boldsymbol X}_{\rm mag}^{\rm quench}
\not\equiv
{\boldsymbol X}_{\rm 2008}^{\rm incident},
\label{eq:2008_not_ordinary_quench}
\end{equation}
and the event should not be used to imply that a normal protected quench
generically leads to a major helium release.

The personnel-safety implications were also determined by the coupled
environmental and access states.  The release altered atmospheric
conditions in the tunnel and generated pressure effects that affected
tunnel infrastructure, including ventilation doors.  At the time of the
incident, however, the corresponding access and operational conditions
prevented personnel exposure to the developing hazard; no personnel were
injured
\cite{Bajko:2009Incident,LHCSafetyTaskForce:2009}.

This distinction is important in the framework developed in
Sec.~\ref{subsec:cryo_access}.  A severe environmental disturbance does
not by itself determine the personnel consequence; the latter also depends
on occupancy, access control, warning, and evacuation conditions.  In
schematic form,
\begin{equation}
{\boldsymbol{\cal C}}_{\rm pers}
=
{\boldsymbol{\cal C}}_{\rm pers}
\left(
{\boldsymbol X}_{\rm env},
{\boldsymbol X}_{\rm access}
\right).
\label{eq:2008_personnel_consequence}
\end{equation}
The 2008 event therefore illustrates simultaneously the severity of the
physical disturbance and the importance of access as an independent
protection variable.

From the barrier perspective, the incident also demonstrates why the
performance of a protection layer must be evaluated against the actual
failure trajectory.  Pressure-relief devices were present and operated,
but the helium discharge into the insulation vacuum substantially exceeded
the release scenario assumed in the earlier analysis.  The relevant issue
was therefore not simply whether a relief device existed, but whether its
capacity was sufficient for the realized mass-flow and pressure transient.

Schematically,
\begin{equation}
\dot m_{\rm He}^{\rm incident}
>
\dot m_{\rm He}^{\rm design\;scenario}
\quad\Longrightarrow\quad
{\cal P}_{\rm relief}^{\rm existing}
\ \text{insufficient for the realized transient}.
\label{eq:2008_relief_capacity}
\end{equation}
Equation~\eqref{eq:2008_relief_capacity} is intended conceptually rather
than as a quantitative reconstruction of the event.  Its significance is
that a protection function must be validated against the envelope of
credible physical trajectories rather than only against the nominal
presence of the corresponding hardware.

Post-incident analysis showed that the helium discharge from the cold mass
into the insulation vacuum exceeded by approximately an order of magnitude
the maximum credible incident flow used in the earlier cryogenic risk
analysis
\cite{CryogenicRisk:2010}.  The event therefore provided direct empirical
evidence that the assumed boundary of the credible cryogenic accident
space had been too restrictive.

In the language of the safe-operational-phase-space
framework, one may represent the lesson schematically as
\begin{equation}
\Gamma_{\rm 2008}
\notin
\mathcal G_{\rm accident}^{\rm assumed},
\qquad
\Gamma_{\rm 2008}
\in
\mathcal G_{\rm accident}^{\rm physically\;accessible},
\label{eq:2008_model_boundary}
\end{equation}
where $\Gamma_{\rm 2008}$ represents the physical
accident trajectory reconstructed from the 2008
incident, $\mathcal G_{\rm accident}^{\rm assumed}$
denotes the set of accident trajectories encompassed
by the pre-incident risk-model assumptions, and
$\mathcal G_{\rm accident}^{\rm physically\;accessible}$
denotes the broader set of physically realizable
accident trajectories under the relevant facility
conditions.
The incident demonstrated that the assumed accident
envelope did not encompass the full range of
physically possible helium-release and pressure
transients. Consequently, the protection arrangements
had to be reassessed against an expanded set of
credible failure scenarios.

This observation led to substantial revision of the cryogenic risk
analysis and of the corresponding protection architecture.  Measures
included reinforcement and extension of pressure-relief capacity on the
insulation vacuum enclosures, reconsideration of credible helium-release
rates, improvements to electrical-joint monitoring and protection, and
review of tunnel infrastructure and personnel-safety conditions
\cite{Bajko:2009Incident,CryogenicRisk:2010,LHCSafetyTaskForce:2009}.

The event therefore exemplifies an important feedback process,
\begin{equation}
\begin{aligned}
&\text{incident} \\
&\rightarrow \text{physical reconstruction} \\
&\rightarrow \text{revised failure model} \\
&\rightarrow \text{redefined design scenario} \\
&\rightarrow \text{modified protection architecture}.
\end{aligned}
\label{eq:2008_learning_cycle}
\end{equation}
Operational experience did not merely lead to replacement of damaged
components; it changed the physical assumptions used to define the
required protection envelope.

This feedback is central to the relation between physics and safety
developed in this manuscript.  The 2008 sector~3--4 incident demonstrated
that the validity of a protection architecture depends on the completeness
of the underlying physical model of credible failure propagation.  Once a
previously underestimated trajectory had been observed, the corresponding
risk model, pressure-relief assumptions, and protection requirements had
to be enlarged accordingly.

The reformulation of the maximum credible incident following the 2008
event provides a concrete example of this process and is considered in the
next subsection.

\subsection{Redefinition of the maximum credible incident}
\label{subsec:cryo_mci_redefinition}

The post-incident analysis of the 19 September 2008 event showed that the
cryogenic failure envelope used before the incident did not encompass the
actual helium-transfer rate from the magnet cold mass into the insulation
vacuum.  The resulting revision of the maximum credible incident (MCI)
therefore represented more than a numerical adjustment of a relief-system
parameter: it enlarged the class of cryogenic trajectories against which
the protection architecture had to remain effective
\cite{CryogenicRisk:2010,LHCSafetyTaskForce:2009,Cruikshank:2010Relief}.

In the trajectory-based language introduced above, the
pre-incident assumption may be represented schematically as
\begin{equation}
\Gamma_{\rm cryo}
\in
\mathcal G_{\rm MCI}^{\rm pre},
\label{eq:preincident_mci_domain}
\end{equation}
where $\Gamma_{\rm cryo}$ represents a cryogenic accident
trajectory considered in the pre-incident assessment, and
$\mathcal G_{\rm MCI}^{\rm pre}$ denotes the set of such
trajectories covered by the earlier maximum credible
incident (MCI) definition.

The 2008 event demonstrated that the actual accident
trajectory lay outside this previously considered envelope:
\begin{equation}
\Gamma_{\rm 2008}
\notin
\mathcal G_{\rm MCI}^{\rm pre}.
\label{eq:2008_outside_preincident_mci}
\end{equation}
The event nevertheless represented a physically realizable
failure sequence, demonstrating that the earlier design
assumptions did not encompass all relevant accident
conditions.

The protection basis therefore had to be revised to
include the newly identified failure conditions.
Schematically,
\begin{equation}
\Gamma_{\rm 2008}
\in
\mathcal G_{\rm MCI}^{\rm post},
\qquad
\mathcal G_{\rm MCI}^{\rm pre}
\subset
\mathcal G_{\rm MCI}^{\rm post},
\label{eq:mci_domain_expansion}
\end{equation}
where $\mathcal G_{\rm MCI}^{\rm post}$ denotes the
enlarged set of cryogenic accident trajectories
considered in the revised protection design basis.
The inclusion represents the extension of the
design-basis accident envelope, rather than
an enlargement of the domain of permitted
normal operation.

A central quantity in this redefinition is the helium mass-flow rate.
For the revised severe-release scenario, representative post-incident
analyses considered a discharge of order
\begin{equation}
\dot m_{\rm He}^{\rm MCI}
\sim
30~{\rm kg\,s^{-1}}
\label{eq:revised_mci_mass_flow}
\end{equation}
at a characteristic helium temperature of order
\begin{equation}
T_{\rm He}^{\rm MCI}
\sim
90~{\rm K},
\label{eq:revised_mci_temperature}
\end{equation}
through an effective breach area of order
\begin{equation}
A_{\rm breach}^{\rm MCI}
\sim
120~{\rm cm^{2}}.
\label{eq:revised_mci_breach_area}
\end{equation}
These values should be interpreted as representative parameters of the
redefined severe-release scenario rather than as universal constants for
all cryogenic failures
\cite{CryogenicRisk:2010}.

The importance of Eq.~\eqref{eq:revised_mci_mass_flow} follows from the
fact that the consequences of a helium release are controlled not only by
the total released mass but also by the rate at which the cryogenic
inventory is transferred into the insulation vacuum or surrounding
environment.  A large $\dot m_{\rm He}$ can produce a pressure transient
faster than the available relief paths can accommodate it,
\begin{equation}
\dot m_{\rm He}^{\rm rel}
>
\dot m_{\rm He}^{\rm relief}
\quad\Longrightarrow\quad
\frac{dP}{dt}>0,
\label{eq:mci_pressure_imbalance}
\end{equation}
where $\dot m_{\rm He}^{\rm rel}$ denotes the effective helium source term
and $\dot m_{\rm He}^{\rm relief}$ the mass flow that can be discharged
through the available relief paths under the corresponding conditions.

Equation~\eqref{eq:mci_pressure_imbalance} is schematic: the actual
pressure evolution depends on the thermodynamic state of helium, the
effective volume, heat transfer, two-phase flow, pressure-dependent relief
capacity, and the geometry of the cryogenic enclosure.  Its purpose is to
make explicit why the mass-flow rate is a primary design quantity for the
revised MCI.

The redefinition therefore changed the relevant protection problem from
one of controlling a relatively limited helium release to one of
demonstrating sufficient capacity for a substantially more severe
transient.  The required protection functions include containment,
pressure relief, mechanical integrity of vacuum barriers and cryostat
elements, controlled routing of released helium, and downstream
atmospheric protection.

The corresponding DiD sequence may be represented as
\begin{equation}
\begin{aligned}
&\text{credible breach} \\
&\rightarrow \dot m_{\mathrm{He}}^{\mathrm{MCI}}
\rightarrow \text{pressure transient} \\
&\rightarrow \text{relief and containment response} \\
&\rightarrow \text{controlled environmental consequence}.
\end{aligned}
\label{eq:mci_protection_sequence}
\end{equation}

The post-incident MCI should therefore not be interpreted simply as a
prediction of what will occur in every severe cryogenic accident.  It is a
design-basis envelope used to define the performance that relevant
protection layers must be capable of accommodating.  In this sense, the
MCI plays a role analogous to the consequence-specific protection
boundaries introduced earlier: it defines a class of physically credible
trajectories against which the system must remain robust.

This distinction is important because different initiating events can
generate different combinations of breach area, helium temperature,
pressure, mass-flow rate, and duration while still producing comparable
demands on the protection system.  The appropriate safety requirement is
therefore not tied to a single microscopic failure mechanism but to the
resulting thermodynamic and mechanical load envelope.

The relevant cryogenic evolution is described by the
facility-state framework introduced earlier, while
the particular MCI scenario is characterized by
the physical conditions governing helium release,
including the breach geometry, helium temperature
and pressure, mass-flow rate, and release duration.
These quantities determine the transient loads
to which the protection system must respond.

The revised MCI envelope
$\mathcal G_{\rm MCI}^{\rm post}$
therefore represents the set of credible accident
scenarios included in the post-incident design basis.
Its distinction from the resulting protected
trajectories is essential: the occurrence of an
accident is not itself an admissible operational
transition, whereas its consequences must be
controlled by the available protection functions.

Denoting by $\mathcal G_{\rm MCI}^{\rm resp}$
the set of resulting trajectories, including
the credited protection responses, the corresponding
design requirement may be expressed schematically as
\begin{equation}
\mathcal G_{\rm MCI}^{\rm resp}
\subseteq
\mathcal G_{\rm adm}^{\rm cryo},
\label{eq:mci_allowed_trajectory}
\end{equation}
where $\mathcal G_{\rm adm}^{\rm cryo}$ denotes
the set of trajectories satisfying the applicable
cryogenic protection constraints, including
those associated with pressure, mechanical integrity,
helium discharge, and the resulting environmental
conditions.

This relation expresses a protection requirement
that must be demonstrated for the design-basis
accident scenarios, rather than an assertion
that every possible cryogenic failure can be
controlled. Its validity depends on the physical
models, the assumed accident envelope, and the
successful performance of the credited
protection functions.

Adequate pressure relief and containment can
limit the mechanical consequences of a severe
cryogenic transient, while ventilation, oxygen
deficiency monitoring, access restriction,
and evacuation provide additional protection
against the resulting environmental hazards.
These functions address different stages of
the hazardous sequence and are not interchangeable.
In particular, personnel-protection measures
cannot compensate for insufficient mechanical
integrity or pressure-relief capacity.

If this condition Eq.~\eqref{eq:mci_allowed_trajectory} is satisfied, the severe cryogenic transient can be
redirected through the intended relief and containment paths without
progressing toward uncontrolled mechanical failure or an unacceptable
environmental state.  If it is not, downstream barriers---ventilation,
ODH monitoring, access restriction, and evacuation---must still limit the
personnel consequences, but they cannot substitute for insufficient
pressure-relief or containment capacity.

The post-2008 redesign therefore modified both the physical protection
hardware and the assumptions used to define its required performance.
Pressure-relief capacity was increased, the mechanical response of the
vacuum enclosure was reconsidered, and the credible helium-release
scenarios used in risk analysis were enlarged
\cite{CryogenicRisk:2010,LHCSafetyTaskForce:2009}.

This sequence illustrates an important general principle:
\begin{equation}
\begin{aligned}
&\text{observed failure trajectory} \\
&\rightarrow \text{revised physical model} \\
&\rightarrow \text{enlarged design envelope} \\
&\rightarrow \text{modified protection requirements}.
\end{aligned}
\label{eq:mci_model_update_cycle}
\end{equation}
The MCI is therefore not a fixed property of the installation independent
of experience; it is part of the engineering representation of the
physically credible failure space and must be revised when operational
evidence demonstrates that the previous envelope was incomplete.

The redefined maximum credible incident consequently provides a concrete
example of the feedback between physics and safety that motivates this
manuscript.  Measurements and reconstruction of the 2008 event changed
the assumed cryogenic state space, and that revised physical understanding
was translated directly into new design and protection requirements.

\subsection{Post-incident strengthening of cryogenic protection barriers}
\label{subsec:cryo_postincident_barriers}

The redefinition of the MCI incident discussed in
Sec.~\ref{subsec:cryo_mci_redefinition} required a corresponding revision
of the protection architecture.  The principal lesson of the 2008
sector~3--4 incident was not simply that an individual component had
failed, but that the realized failure trajectory exceeded the envelope for
which several downstream protection functions had originally been
dimensioned.  The post-incident response therefore addressed multiple
stages of the hazard sequence
\cite{Bajko:2009Incident,Cruikshank:2010Relief,
Chorowski:2011RiskUpdate}.

In the barrier terminology introduced in
Sec.~\ref{subsec:preventive_mitigative}, the resulting measures can be
separated schematically into upstream preventive functions and downstream
mitigative functions,
\begin{equation}
\left\{
\begin{array}{l}
{\cal B}_{\rm prev}:
\text{detect or suppress the initiating electrical fault},\\[1ex]
{\cal B}_{\rm mit}:
\text{limit pressure, mechanical, environmental, and personnel consequences}.
\end{array}
\right.
\label{eq:postincident_barrier_classes}
\end{equation}
The distinction is functional rather than absolute, since some measures
influence more than one stage of the failure trajectory.

A first class of improvements addressed the electrical origin of the
incident.  The scope of the QPS was enlarged to
include more sensitive supervision of superconducting busbar and splice
resistances and to provide rapid intervention when abnormal resistive
behaviour is detected.  The upgraded protection architecture introduced
distributed monitoring capable of detecting small anomalous resistances
and initiating rapid circuit discharge before the local heating develops
into a destructive electrical arc
\cite{Bajko:2009Incident,Cruikshank:2010Relief}.

The corresponding preventive chain may be
represented schematically as
\begin{equation}
R_{\rm splice}(t)>R_{\rm splice}^{\rm th}
\;\longrightarrow\;
\text{electrical fault detection}
\;\longrightarrow\;
\text{interlock}
\;\longrightarrow\;
\text{controlled circuit discharge},
\label{eq:postincident_splice_protection}
\end{equation}
where $R_{\rm splice}(t)$ denotes the effective
electrical resistance of the monitored superconducting
busbar joint, and $R_{\rm splice}^{\rm th}$ is the
corresponding protection threshold. An excessive
splice resistance can produce localized Joule
heating proportional to $I_{\rm mag}^{2}R_{\rm splice}$,
potentially initiating damage to the electrical
connection. The resistance anomaly is inferred
from electrical measurements, with appropriate
account taken of circuit conditions. Its early
detection allows the protection system to initiate
circuit discharge before the fault develops
into a more severe electrical and cryogenic event.

This development illustrates the precursor-detection principle introduced
in Sec.~\ref{subsec:fast_slow_failures}.  Detection of the initiating
electrical anomaly provides a larger protection margin than waiting for
the later stages of the sequence,
\begin{equation}
\text{resistive heating}
\rightarrow
\text{electrical arc}
\rightarrow
\text{helium breach}
\rightarrow
\text{overpressure}.
\label{eq:postincident_precursor_chain}
\end{equation}
The protection objective is therefore to act as far upstream in this chain
as the available observables permit.

A second and particularly important class of modifications addressed the
pressure response of the insulation vacuum enclosure.  The relief capacity
was redesigned to accommodate the substantially larger helium mass-flow
rates associated with the redefined MCI.  Additional and enlarged relief
devices were introduced so that rapid helium transfer from the cold mass
could be discharged without reproducing the excessive vacuum-enclosure
pressures observed in 2008
\cite{Cruikshank:2010Relief,Chorowski:2011RiskUpdate}.

The corresponding requirement may be written schematically as
\begin{equation}
\dot m_{\rm relief}^{\rm cap}
\gtrsim
\dot m_{\rm He}^{\rm MCI},
\label{eq:postincident_relief_capacity}
\end{equation}
where $\dot m_{\rm relief}^{\rm cap}$ denotes the effective mass-flow
capacity of the pressure-relief system under the relevant thermodynamic
conditions.  Equation~\eqref{eq:postincident_relief_capacity} is not
intended as a complete sizing relation; actual relief performance depends
on pressure, temperature, two-phase flow, discharge coefficients, geometry,
and the distribution of relief devices.

The important conceptual change is that the relief system was dimensioned
against the revised physical accident envelope rather than the
pre-incident release assumption:
\begin{equation}
{\cal P}_{\rm relief}:
\Gamma_{\rm MCI}^{\rm post}
\rightarrow
\Gamma_{\rm pressure}^{\rm controlled}.
\label{eq:postincident_relief_mapping}
\end{equation}
Thus, the barrier requirement was tied directly to the enlarged set of
credible cryogenic trajectories.

Mechanical protection was strengthened in parallel.  The 2008 event had
shown that overpressure of the insulation vacuum could generate forces
large enough to displace cryostated components and vacuum barriers,
thereby producing secondary mechanical and electrical damage.  The
post-incident consolidation therefore included reinforcement of selected
supports and anchoring arrangements in addition to the increased relief
capacity
\cite{Bajko:2009Incident,LHCSafetyTaskForce:2009,Cruikshank:2010Relief,Chorowski:2011RiskUpdate}.

The physical purpose of these measures can be
represented schematically as
\begin{equation}
\Delta P_{\rm vac}
\rightarrow
\boldsymbol F_{\rm mech}
\rightarrow
\Delta\boldsymbol r_{\rm cryostat},
\label{eq:pressure_force_displacement}
\end{equation}
where $\Delta P_{\rm vac}$ denotes the pressure
difference acting across the relevant mechanical
structure following pressurization of the insulation
vacuum enclosure, $\boldsymbol F_{\rm mech}$ is
the resulting pressure-induced mechanical force,
and $\Delta\boldsymbol r_{\rm cryostat}$ represents
the displacement of the cryostat or associated
mechanical structure.

The magnitude of the applied force depends on
the pressure distribution and the geometry of
the affected surfaces. For an approximately
uniform pressure difference acting on an
effective area $A_{\rm eff}$, its magnitude
may be estimated schematically as
$F_{\rm mech}\sim\Delta P_{\rm vac}A_{\rm eff}$.
The resulting displacement also depends on
the stiffness, supports, and anchoring
arrangements of the structure.

The corresponding mechanical protection
objective may be expressed as
\begin{equation}
\left\lVert
\Delta\boldsymbol r_{\rm cryostat}
\right\rVert
<
\Delta r_{\rm allowable},
\label{eq:postincident_mechanical_limit}
\end{equation}
where $\Delta r_{\rm allowable}$ denotes the
allowable displacement magnitude for the
considered component or structural arrangement.
Pressure relief and mechanical reinforcement
are therefore complementary barriers:
the former limits the pressure-induced loads,
while the latter increases the ability of
the structure to withstand the resulting forces
without unacceptable displacement or damage.

The post-incident protection architecture consequently acts on several
successive stages of the failure trajectory,
\begin{equation}
\begin{array}{c}
\text{abnormal electrical resistance}
\\
\downarrow
\\
\text{enhanced splice and busbar monitoring}
\\
\downarrow
\\
\text{rapid circuit protection}
\\
\downarrow
\\
\text{enhanced cryogenic pressure relief}
\\
\downarrow
\\
\text{reinforced mechanical containment}
\\
\downarrow
\\
\text{ventilation, ODH monitoring, and access protection}.
\end{array}
\label{eq:postincident_layered_protection}
\end{equation}

The downstream atmospheric and personnel-protection layers remain
important because even a correctly operating pressure-relief system may
intentionally discharge helium from a protected cryogenic volume.
Pressure protection therefore does not imply absence of an environmental
disturbance.  Instead, the relief system changes the trajectory from an
uncontrolled overpressure sequence to a controlled discharge sequence,
which must then be managed by ventilation, atmospheric monitoring, and
access protection.

This distinction may be written as
\begin{equation}
{\boldsymbol X}_{\rm cryo}^{\rm overpressure}
\xrightarrow{\;{\cal P}_{\rm relief}\;}
{\boldsymbol X}_{\rm cryo}^{\rm relieved}
\rightarrow
{\boldsymbol X}_{\rm env}^{\rm perturbed},
\label{eq:relief_to_environment}
\end{equation}
where the second transition need not imply an unacceptable personnel
consequence if the downstream protection barriers perform as intended.

The access and occupancy state therefore remains part
of the post-incident protection architecture. A severe
helium-release scenario may require personnel access
to the affected underground region to be restricted,
even when the cryogenic and mechanical systems remain
within their revised design limits. Schematically,
\begin{equation}
\boldsymbol X_{\rm env}
\notin
\Omega_{\rm env}^{\rm occupied}
\quad\Longrightarrow\quad
\sigma_{\rm access}
\rightarrow
\sigma_{\rm access}^{\rm restricted}.
\label{eq:postincident_access_condition}
\end{equation}
Here, $\Omega_{\rm env}^{\rm occupied}$ denotes
the environmental-state domain compatible with
personnel occupancy, while
$\sigma_{\rm access}^{\rm restricted}$ represents
the access status in which entry into the affected
region is restricted. This formulation is consistent
with the occupancy-dependent safe regions and
access-status definitions introduced in
Sec.~\ref{subsec:cryo_access}.
Access restriction does not by itself restore
acceptable environmental conditions or establish
that an already occupied region has been evacuated.
These require the corresponding protective and
operational responses.

The revisions also strengthened the importance of diagnostics and
validation.  A protection barrier whose nominal capacity is sufficient
for the redefined MCI is useful only if it remains available and its state
is known.  Periodic testing, monitoring of electrical joints, validation
of relief devices, surveillance of ventilation conditions, and
verification of atmospheric-protection systems therefore form part of the
protection architecture rather than being external maintenance activities.

In the terminology of Sec.~\ref{subsec:barrier_independence}, the
post-incident measures also increased the functional diversity of the
protection system.  Electrical supervision, cryogenic pressure relief,
mechanical reinforcement, ventilation, oxygen monitoring, and access
control rely on different physical observables and act on different
segments of the same accident trajectory.  Their complementarity reduces
the dependence of the overall protection function on any single mechanism.

The resulting DiD structure can be represented schematically
as
\begin{equation}
{\cal B}_{\rm electrical}
\oplus
{\cal B}_{\rm QPS}
\oplus
{\cal B}_{\rm relief}
\oplus
{\cal B}_{\rm mechanical}
\oplus
{\cal B}_{\rm vent}
\oplus
{\cal B}_{\rm ODH}
\oplus
{\cal B}_{\rm access},
\label{eq:postincident_defence_in_depth}
\end{equation}
where $\oplus$ denotes complementary protection functions rather than
statistical independence or mathematical addition.

From the trajectory perspective, the post-incident objective is therefore
not to guarantee that every initiating fault is eliminated, but to ensure
that a credible severe trajectory is intercepted or redirected at several
successive stages:
\begin{equation}
\Gamma_{\rm fault}
\xrightarrow{\;{\cal B}_{1}\;}
\Gamma_{1}^{\rm controlled}
\xrightarrow{\;{\cal B}_{2}\;}
\Gamma_{2}^{\rm controlled}
\rightarrow\cdots
\rightarrow
\Gamma_{\rm protected}.
\label{eq:postincident_trajectory_control}
\end{equation}
Failure of one layer need not then imply immediate progression to the
largest possible consequence.

The post-2008 consolidation thus provides a concrete example of how an
observed event can reshape the protection architecture at several physical
levels.  The initiating electrical failure motivated improved precursor
detection; the unexpectedly large helium-transfer rate required a new
pressure-relief basis; the observed mechanical displacement motivated
structural reinforcement; and the environmental consequences reinforced
the role of ventilation, ODH monitoring, and access control.

The resulting protection architecture is therefore a direct physical
response to the enlarged accident space established in
Sec.~\ref{subsec:cryo_mci_redefinition}.  Rather than treating electrical,
cryogenic, mechanical, environmental, and personnel protection as
independent systems, it constrains the propagation of a severe event across
all of these coupled states.

\subsection{Empirical feedback and validation of cryogenic safety models}
\label{subsec:cryo_empirical_feedback}

The discussion of the 2008 sector~3--4 incident and the subsequent
redefinition of the MCI illustrates a broader
principle: cryogenic safety cannot rely exclusively on design-stage models.
The relevant physical processes involve superconducting transitions,
two-phase helium flow, pressure propagation, structural response, gas
transport, ventilation, and personnel occupancy, and the resulting coupled
system contains uncertainties that can only partly be constrained before
operation.

Operational experience therefore provides an additional source of
information about the facility state and its possible failure trajectories.
Measurements obtained during commissioning, controlled tests, protection
events, abnormal transients, near misses, and incidents can all be used to
compare the predicted response with the observed one and to revise the
underlying physical and engineering models.

The corresponding feedback process may be written schematically as
\begin{equation}
\text{model}
\rightarrow
\text{predicted trajectory}
\rightarrow
\text{measurement or event}
\rightarrow
\text{comparison}
\rightarrow
\text{model update}.
\label{eq:cryo_empirical_model_cycle}
\end{equation}

In the trajectory-based formulation developed in this manuscript, a model
may predict a family of cryogenic evolutions
\begin{equation}
\Gamma_{\rm cryo}^{\rm pred}
=
\Gamma_{\rm cryo}
\left(
{\boldsymbol \theta},
{\boldsymbol X}_{0}
\right),
\label{eq:cryo_predicted_trajectory}
\end{equation}
where ${\boldsymbol X}_{0}$ denotes the initial facility state and
${\boldsymbol \theta}$ represents the physical and engineering parameters
entering the model.  These may include, for example, effective breach
areas, heat-transfer coefficients, relief characteristics, helium
thermodynamic properties, structural response parameters, ventilation
rates, or detector-response parameters.

Measurements provide constraints on these quantities through observables
such as
\begin{equation}
{\boldsymbol Y}_{\rm cryo}
=
\left(
P(t),
T(t),
\dot m_{\rm He}(t),
C_{O_2}({\boldsymbol r},t),
\dot V_{\rm vent}(t),
\ldots
\right),
\label{eq:cryo_empirical_observables}
\end{equation}
which can be compared with the corresponding model predictions.

The comparison
\begin{equation}
{\boldsymbol Y}_{\rm cryo}^{\rm meas}
\quad\leftrightarrow\quad
{\boldsymbol Y}_{\rm cryo}^{\rm pred}
\label{eq:cryo_prediction_measurement_comparison}
\end{equation}
tests not only individual parameters but also the assumed structure of the
failure model.  Agreement supports the use of the corresponding model
within its validated domain, whereas systematic disagreement may indicate
that an important physical mechanism, coupling, or failure path has been
omitted or inadequately represented.

The 2008 incident provides an especially clear example.  The observed
helium-transfer rate and pressure evolution showed that the physically
accessible cryogenic trajectory extended beyond the envelope represented
in the pre-incident analysis.  The empirical information therefore did
not merely refine a numerical coefficient; it changed the assumed
boundary of the credible accident space,
\begin{equation}
\Gamma_{\rm credible}^{\rm old}
\rightarrow
\Gamma_{\rm credible}^{\rm revised}.
\label{eq:cryo_credible_space_revision}
\end{equation}

This distinction is important.  Empirical feedback can act at several
different levels:
\begin{equation}
\left\{
\begin{array}{l}
\text{parameter refinement},\\
\text{validation or modification of a physical model},\\
\text{identification of a previously neglected coupling},\\
\text{revision of a failure probability or frequency},\\
\text{revision of a consequence model},\\
\text{redefinition of the credible trajectory envelope}.
\end{array}
\right.
\label{eq:cryo_feedback_levels}
\end{equation}
The last of these is particularly important for rare, high-consequence
events because an incomplete failure model can invalidate the assumed
performance margin of several downstream barriers simultaneously.

Empirical information also constrains the performance of the protection
layers themselves.  For a barrier $B_k$, the relevant quantities include
not only whether the barrier is present, but whether its observed response
is consistent with the assumed performance,
using the barrier attributes introduced in
Eq.~\eqref{eq:barrier_attributes}.  Functional tests and operational
events can therefore provide information on response time, diagnostic
coverage, availability, and possible common-cause dependencies.

For example, a pressure-relief system can be validated by comparing its
measured opening and discharge behaviour with the expected pressure and
mass-flow trajectory.  Similarly, ventilation performance can be checked
against measured gas transport and atmospheric recovery, while ODH
monitoring can be tested through calibration, functional checks, and the
observed relation between sensor response and the environmental state.

The protection requirement expressed in
Eq.~\eqref{eq:mci_allowed_trajectory} must
be supported by validated physical models
and verified protection functions.
In particular, the models used to predict
the response to the accident scenarios
included in $\mathcal G_{\rm MCI}^{\rm post}$
must remain applicable over the relevant
range of physical conditions and uncertainties.
The credited barriers must likewise be
demonstrated to provide the required
protection performance, including the
response times and capacities assumed
in the assessment.

Consequently, the inclusion Eq.~\eqref{eq:mci_allowed_trajectory}
is not merely a formal design requirement:
its validity depends on the experimental
and operational evidence supporting the
physical models, the verification of the
protection functions, and the assumptions
used to construct the design-basis
accident scenarios.

This validation is necessarily configuration dependent.  Changes in
cryogenic hardware, ventilation, relief devices, magnet circuits, access
conditions, or operating parameters can modify the relation between the
initial disturbance and the resulting consequence.  A validation obtained
for one configuration cannot automatically be transferred to another
without demonstrating that the relevant physical assumptions remain
applicable.

One may therefore associate the validated cryogenic domain with the
available empirical evidence,
\begin{equation}
\Omega_{\rm cryo}^{\rm validated}
=
\Omega_{\rm cryo}^{\rm validated}
\left(
{\cal D}_{\rm op},
{\cal D}_{\rm test},
{\cal D}_{\rm incident}
\right),
\label{eq:cryo_validated_domain}
\end{equation}
where ${\cal D}_{\rm op}$, ${\cal D}_{\rm test}$, and
${\cal D}_{\rm incident}$ denote, schematically, operational, dedicated
test, and incident-derived information.

As new evidence becomes available, the validated domain may either expand
or contract.  Successful testing can support operation in a previously
unvalidated region, while an unexpected event can demonstrate that an
assumed boundary was incomplete or that a protection margin was smaller
than expected:
\begin{equation}
\Omega_{\rm cryo}^{\rm validated}
\rightarrow
\Omega_{\rm cryo}^{\rm validated\,\prime}.
\label{eq:cryo_validated_domain_update}
\end{equation}

The same feedback applies to the risk description introduced in
Sec.~\ref{subsec:hazard_chains}.  New empirical information can modify the
estimated event frequency $f_i$, the conditional propagation of the
failure sequence, the consequence vector ${\boldsymbol{\cal C}}_i$, or the
performance assigned to the relevant barriers.  Thus,
\({\cal R}_{i}={\cal R}\left(f_i,{\boldsymbol{\cal C}}_i\right)\)
should be understood as depending on quantities that are themselves subject
to experimental and operational revision.

This does not imply that safety analysis becomes purely empirical.  Rare
severe accidents cannot be characterized statistically from operating
experience alone, and many credible failure combinations may never be
observed directly.  Physical modelling, simulation, conservative design
assumptions, and engineering judgement therefore remain essential~\cite{Cruikshank:2010Relief,Chorowski:2011RiskUpdate}.
Empirical feedback instead provides a means of testing whether those
assumptions continue to represent the real installation.

The resulting methodology can be summarized as
\begin{equation}
\begin{aligned}
&\text{physical model} \\
&\rightarrow \text{design and protection} \\
&\rightarrow \text{operation and measurement} \\
&\rightarrow \text{validation or discrepancy} \\
&\rightarrow \text{updated model and protection}.
\end{aligned}
\label{eq:cryo_empirical_feedback_cycle}
\end{equation}

Cryogenic safety is therefore an iterative rather than static property of
the facility.  The safe operational domain is defined initially from
physical models and engineering requirements, but its credibility is
strengthened---and, when necessary, revised---through measurements and
operational experience.  This feedback between physical understanding and
protection design is one of the clearest examples of the reciprocal
relation between physics and safety developed throughout this manuscript.

\subsection{Coupled cryogenic, environmental, and access state}
\label{subsec:cryo_coupled_state}

The preceding subsections show that cryogenic safety cannot be described
by the state of the helium system alone.  A disturbance can originate in
the superconducting magnet or powering system, propagate through the
cryogenic inventory and pressure boundaries, modify the underground
atmosphere, and ultimately change the conditions under which personnel
occupancy is permitted.  The relevant safety problem is therefore one of
coupled subsystem evolution.

For the purposes of cryogenic safety, a reduced facility state may be
written as
\begin{equation}
{\boldsymbol X}_{\rm cryo\text{-}saf}
=
\left(
{\boldsymbol X}_{\rm mag},
{\boldsymbol X}_{\rm cryo},
{\boldsymbol X}_{\rm env},
{\boldsymbol X}_{\rm access}
\right),
\label{eq:cryo_coupled_state_vector}
\end{equation}
where the four blocks describe, respectively, the superconducting-magnet
state, the cryogenic state, the underground environmental state, and the
access or occupancy state.  Electrical, mechanical, and powering
variables enter these blocks or their couplings as required by the
particular failure sequence.

The significance of this representation is that
the subsystem states are not independent. Their
evolution involves both continuous physical
processes and discrete transitions associated
with equipment protection, operational modes,
and personnel-access conditions.

To distinguish these two types of evolution,
let $\boldsymbol X_{\rm cryo\text{-}saf}^{\rm cont}$
denote the continuous components of the coupled
state vector introduced in
Eq.~\eqref{eq:cryo_coupled_state_vector},
and let $\boldsymbol\sigma_{\rm cryo\text{-}saf}$
collect its discrete status variables.
Between discrete transitions, the evolution
of the continuous components may be represented
schematically as
\begin{equation}
\frac{d\boldsymbol X_{\rm cryo\text{-}saf}^{\rm cont}}{dt}
=
\boldsymbol{\mathcal F}_{\rm cryo}
\left(
\boldsymbol X_{\rm cryo\text{-}saf}^{\rm cont},
\boldsymbol\sigma_{\rm cryo\text{-}saf},
m;
\boldsymbol\Xi
\right),
\label{eq:cryo_coupled_dynamics}
\end{equation}
where ${\boldsymbol{\mathcal F}}_{\rm cryo}$ denotes the
coupled evolution law for the continuous
physical quantities, $m$ is the applicable
operational or intervention mode, and
$\boldsymbol\Xi$ represents the relevant
technical configuration. The time derivative on the left-hand side
describes the instantaneous rates of change
of the continuous state variables. These may
include the magnet current, helium temperature
and pressure, cryogenic inventory, helium
mass-flow conditions, and environmental
quantities such as oxygen concentration.
The corresponding components of
${\boldsymbol{\mathcal F}}_{\rm cryo}$ describe their
evolution according to the underlying
electromagnetic, thermodynamic, fluid-dynamic,
and transport processes. Importantly, the evolution of each continuous
quantity may depend on the states of several
subsystems. For example, the helium pressure
evolution can depend on heat deposition
following a magnet quench, the cryogenic
inventory, and the available pressure-relief
capacity. Similarly, the oxygen concentration
in an underground region can depend on the
helium release, ventilation conditions,
and the geometry of the affected volume. The discrete variables collected in
$\boldsymbol\sigma_{\rm cryo\text{-}saf}$
describe conditions such as the quench
status, equipment-protection states,
and personnel-access or occupancy
statuses. Their changes are represented
by discrete transitions rather than
ordinary time derivatives. Such transitions
may be triggered by threshold crossings,
interlocks, protection commands, or
authorized operational actions.
They can, in turn, modify the continuous
evolution by changing the applicable
protection functions or technical
configuration.

Equation~\eqref{eq:cryo_coupled_dynamics}
therefore represents the continuous part
of a coupled hybrid dynamical system.
It is not intended as a complete set
of differential equations for the
cryogenic installation. Its purpose
is to emphasize that the physical
evolution, protection responses,
and operational conditions must
be considered together when assessing
cryogenic safety trajectories.

A representative propagation sequence is
\begin{equation}
{\boldsymbol X}_{\rm mag}
\rightarrow
{\boldsymbol X}_{\rm cryo}
\rightarrow
{\boldsymbol X}_{\rm env}
\rightarrow
{\boldsymbol X}_{\rm access},
\label{eq:cryo_coupled_propagation}
\end{equation}
but the coupling need not be strictly unidirectional.  For example,
access conditions determine whether personnel consequences are possible,
while intervention or maintenance states can themselves modify the
cryogenic configuration and therefore the set of credible failure
trajectories.

The safe cryogenic domain is consequently not, in general, a Cartesian
product of independently acceptable subsystem regions:
\begin{equation}
\Omega_{\rm safe}^{\rm cryo}
\neq
\Omega_{\rm mag}^{\rm safe}
\times
\Omega_{\rm cryo}^{\rm safe}
\times
\Omega_{\rm env}^{\rm safe}
\times
\Omega_{\rm access}^{\rm safe}.
\label{eq:cryo_nonfactorizable_safe_space}
\end{equation}
A combination of subsystem states that would appear acceptable when each
is considered separately may become unacceptable because of the coupling
between them.

This non-factorizability is particularly clear for occupancy.  A
perturbed environmental state can remain manageable when the affected
region is evacuated while being unacceptable if personnel remain present.
Thus, the admissible cryogenic safety region depends explicitly on the
occupancy condition,
\begin{equation}
\Omega_{\rm safe}^{\rm cryo,occupied}
\neq
\Omega_{\rm safe}^{\rm cryo,unoccupied}.
\label{eq:cryo_coupled_occupancy_spaces}
\end{equation}
The difference does not imply that evacuation restores the physical
environment; rather, it changes the possible personnel consequence while
the atmospheric disturbance is being controlled.

The personnel consequence should therefore be regarded
as depending on both the environmental and access states.
Schematically,
\begin{equation}
{\cal C}_{\rm pers}
=
{\cal C}_{\rm pers}
\left(
\boldsymbol X_{\rm env},
\boldsymbol X_{\rm access}
\right).
\label{eq:cryo_personnel_coupled_consequence}
\end{equation}
Here, ${\cal C}_{\rm pers}$ denotes the
personnel-consequence component of the
multidimensional consequence vector
$\boldsymbol{\cal C}$ introduced in
Eq.~\eqref{eq:consequence_vector}.
The expression emphasizes that the personnel
consequences of an abnormal cryogenic event
depend not only on the resulting atmospheric
conditions, but also on the presence of personnel,
their locations, and the applicable access and
evacuation conditions.

The same atmospheric disturbance may therefore
produce different personnel consequences depending
on whether the affected region is occupied,
personnel are being evacuated, or access has
already been restricted. The expression is
schematic: a quantitative assessment must also
account for the temporal evolution of exposure
conditions and personnel occupancy.

The protection architecture acts on different parts of this coupled
evolution.  Quench and powering protection constrain the transfer of
magnetic energy into the cryogenic system; containment and pressure relief
constrain the cryogenic transient; ventilation modifies the environmental
state; ODH monitoring observes the atmospheric state; and access control
acts on personnel occupancy.  These functions should not be interpreted
as a single serial barrier chain, because several can act in parallel and
at different stages of the event.

A protected cryogenic trajectory may therefore contain a sequence such as
\begin{equation}
{\boldsymbol X}_{\rm cryo}^{\rm abnormal}
\xrightarrow{\;{\cal P}_{\rm relief}\;}
{\boldsymbol X}_{\rm cryo}^{\rm controlled}
\rightarrow
{\boldsymbol X}_{\rm env}^{\rm perturbed}
\xrightarrow{\;{\cal P}_{\rm vent}\;}
{\boldsymbol X}_{\rm env}^{\rm recovering}
\xrightarrow{\;{\cal M}_{\rm ODH}\;}
{\boldsymbol X}_{\rm access}^{\rm restricted},
\label{eq:cryo_protected_coupled_trajectory}
\end{equation}
where a protected state need not correspond to the immediate restoration
of all subsystem variables to their nominal values.

This point is important.  Safety does not require that every disturbance
be eliminated instantaneously.  A cryogenic transient can remain within a
protected trajectory while pressure is relieved, helium is removed by
ventilation, atmospheric conditions are monitored, and access remains
restricted.  What matters is that the evolution remains within a
validated domain in which the relevant consequences are controlled.

One may define the set of admissible cryogenic-safety
trajectories schematically as
\begin{equation}
\mathcal G_{\rm adm}^{\rm cryo}
=
\left\{
\Gamma:
t\mapsto\boldsymbol X_{\rm cryo\text{-}saf}(t)
\;\middle|\;
\begin{aligned}
&\boldsymbol X_{\rm cryo\text{-}saf}(t)
\in\Omega_{\rm safe}^{\rm cryo}[m(t)],
\\
&\chi_{\rm cryo}^{\rm prot}(t)=1,
\quad \forall\,t\in[t_0,t_f]
\end{aligned}
\right\}.
\label{eq:cryo_allowed_trajectory_set}
\end{equation}
Here, $\mathcal G_{\rm adm}^{\rm cryo}$ denotes the
set of time-ordered trajectories of the coupled
magnet, cryogenic, environmental, and access states
that satisfy the applicable cryogenic-safety
requirements. The first condition requires the
facility state to remain within the safe domain
corresponding to the operational or intervention
mode $m(t)$. The explicit time dependence of
$m$ allows for changes between operating,
protective, and intervention configurations.
The second condition,
$\chi_{\rm cryo}^{\rm prot}(t)=1$, expresses the
satisfaction of the applicable dynamic protection
constraints throughout the considered evolution.
These may include limits on pressure and temperature
transients, the required response of pressure-relief
and isolation systems, and environmental conditions
relevant to personnel protection. The indicator
does not represent the activation status of a
particular interlock or protective device; rather,
it expresses the combined validity of the
protection requirements associated with the
considered trajectory. The two conditions are complementary. Membership
in the instantaneous safe domain does not by itself
establish that the evolution between successive
states is admissible. For example, an abnormal
helium release may initiate a rapid pressure
transient requiring timely pressure relief,
while the resulting change in oxygen concentration
may require access restrictions or evacuation.
The admissibility of the complete trajectory
therefore depends on both the physical evolution
and the effectiveness and timing of the
corresponding protective responses. The formulation is schematic and assumes that
the relevant continuous evolution and discrete
protection transitions are included in the
trajectory description. The actual admissible
set depends on the technical configuration,
the credible failure scenarios, and the
protection functions available for the
considered operating conditions.

The requirement on an actual coupled cryogenic-safety
trajectory can then be expressed as
\begin{equation}
\Gamma_{\rm cryo}
\in
\mathcal G_{\rm adm}^{\rm cryo},
\label{eq:cryo_trajectory_requirement}
\end{equation}
where $\Gamma_{\rm cryo}$ denotes an individual
trajectory of the coupled magnet, cryogenic,
environmental, and access states, while
$\mathcal G_{\rm adm}^{\rm cryo}$ is the set
of trajectories satisfying the applicable
cryogenic protection constraints.

For the design-basis MCI scenarios discussed above,
the corresponding requirement extends to the set
$\mathcal G_{\rm MCI}^{\rm resp}$ of trajectories
resulting from the credited protective responses.
As expressed in Eq.~\eqref{eq:mci_allowed_trajectory},
these response trajectories must satisfy the
admissibility conditions defining
$\mathcal G_{\rm adm}^{\rm cryo}$.
The initiating accident need not be prevented,
but its consequences must remain controlled
within the applicable protection requirements,
assuming successful operation of the credited
barriers.

As discussed in Sec.~\ref{subsec:cryo_empirical_feedback},
the physically admissible domain is constrained not
only by analytical models and design calculations
but also by commissioning measurements, functional
tests, operational experience, and incident
reconstruction. It is therefore
useful to distinguish the nominal design domain from the empirically
validated one,
\begin{equation}
\Omega_{\rm op}^{\rm cryo}
\subseteq
\Omega_{\rm safe}^{\rm cryo}
\cap
\Omega_{\rm cryo}^{\rm validated},
\label{eq:cryo_safe_validated_intersection}
\end{equation}
where $\Omega_{\rm op}^{\rm cryo}$ denotes the region in which the
corresponding cryogenic operation or occupancy condition is permitted.

The 2008 sector~3--4 incident illustrates why this distinction is
necessary.  The observed trajectory was physically accessible even though
it extended beyond the severe-release envelope represented in the
pre-incident analysis.  Subsequent redefinition of the MCI and
strengthening of the protection barriers effectively modified both the
modelled accident domain and the validated response region.

The coupled-state formulation therefore provides a unified interpretation
of the cryogenic protection measures discussed above.  Pressure relief,
mechanical robustness, ventilation, ODH monitoring, and access restriction
do not protect independent systems; they constrain different parts of the
same evolving facility state.  Their combined purpose is to prevent an
initiating disturbance from propagating into an unacceptable consequence
trajectory.

The coupling of the relevant subsystems may be
summarized schematically as
\begin{equation}
\boldsymbol X_{\rm mag}
\rightarrow
\boldsymbol X_{\rm cryo}
\rightarrow
\boldsymbol X_{\rm env}
\rightarrow
\boldsymbol X_{\rm access}.
\label{eq:cryo_coupled_state_summary}
\end{equation}
The resulting evolution must satisfy the
admissible-trajectory requirement already
expressed in
Eq.~\eqref{eq:cryo_trajectory_requirement}.

Cryogenic safety is thus a property of the coupled trajectory of the
facility rather than of any single cryogenic variable.  This provides a
direct realization of the safe-operational-phase-space concept introduced
earlier in the manuscript: the relevant safety boundary is determined by
the simultaneous state of the magnet, cryogenic, environmental, and access
systems, together with the validity and response of the protection
barriers that couple them.

\subsection{Layered cryogenic protection architecture}
\label{subsec:layered_cryo_architecture}

The preceding discussion shows that cryogenic safety at the LHC is not
provided by a single protection system.  It results from a layered
architecture in which different barriers act on different stages of the
coupled evolution from an initiating electrical, magnetic, or cryogenic
disturbance to possible mechanical, environmental, and personnel
consequences.

For the reduced cryogenic-safety state introduced in
Eq.~\eqref{eq:cryo_coupled_state_vector}, the protection problem can be
represented schematically as
\begin{equation}
{\boldsymbol X}_{\rm cryo\text{-}saf}
\xrightarrow{\;{\cal F}_i\;}
{\boldsymbol X}_{\rm cryo\text{-}saf}^{\rm abnormal}
\xrightarrow{\;\{{\cal B}_k\}\;}
\Gamma_{\rm protected},
\label{eq:layered_cryo_general_mapping}
\end{equation}
where ${\cal F}_i$ denotes the initiating failure process and
$\{{\cal B}_k\}$ the set of protection barriers that can intercept,
redirect, or limit the subsequent trajectory.

The relevant barriers act at physically different stages.  A schematic
ordering is
\begin{equation}
\begin{array}{c}
\text{electrical and magnet protection}
\\
\downarrow
\\
\text{cryogenic containment and pressure relief}
\\
\downarrow
\\
\text{mechanical integrity}
\\
\downarrow
\\
\text{ventilation and atmospheric control}
\\
\downarrow
\\
\text{ODH monitoring}
\\
\downarrow
\\
\text{access restriction and evacuation}.
\end{array}
\label{eq:layered_cryo_architecture_chain}
\end{equation}
This ordering describes the propagation of a representative severe event,
not a strictly serial control system.  Several barriers can act in
parallel, and the relevance of a particular layer depends on the
initiating event and the consequence being considered.

The first layers act as far upstream as practicable.  Electrical
supervision, powering protection, and the QPS can
interrupt or limit the transfer of stored magnetic energy before a fault
develops into a severe cryogenic disturbance.  These functions are
predominantly preventive with respect to loss of cryogenic containment,
although their classification depends on the specified failure sequence
and consequence.

If a cryogenic transient nevertheless develops, containment and pressure
relief act directly on the helium and pressure state.  Their objective is
not necessarily to maintain all cryogenic variables at their nominal
values, but to prevent the transient from evolving into an uncontrolled
pressure or structural failure.  In trajectory language,
\begin{equation}
\Gamma_{\rm cryo}^{\rm abnormal}
\xrightarrow{\;{\cal B}_{\rm relief}\;}
\Gamma_{\rm cryo}^{\rm controlled},
\label{eq:layered_cryo_relief_trajectory}
\end{equation}
where the controlled trajectory can still involve helium discharge and
substantial departure from normal operating conditions.

Mechanical integrity provides an additional and partly passive protection
layer.  Relief capacity reduces the pressure load, while supports,
vacuum barriers, anchoring, and associated structural elements must
withstand the residual forces associated with the design-basis transient.
The two functions are complementary rather than interchangeable:
pressure relief reduces the applied load, whereas mechanical robustness
limits the consequence of the load that remains.

If helium reaches the underground environment, the relevant protection
problem moves to the environmental and access blocks of the facility
state.  Ventilation acts on the physical atmosphere,
\begin{equation}
{\boldsymbol X}_{\rm env}^{\rm perturbed}
\xrightarrow{\;{\cal P}_{\rm vent}\;}
{\boldsymbol X}_{\rm env}^{\rm recovering},
\label{eq:layered_cryo_environment_control}
\end{equation}
while ODH monitoring provides the observational link between the
atmospheric state and the personnel-protection response.  Access
restriction and evacuation then act on the occupancy state rather than on
the helium source itself.

These downstream layers therefore address different
physical and operational quantities:
\begin{equation}
\left\{
\begin{array}{ll}
{\cal P}_{\rm vent}:&
{\boldsymbol X}_{\rm env}
\rightarrow
{\boldsymbol X}_{\rm env}^{\rm controlled},
\\[1ex]
{\cal M}_{\rm ODH}:&
{\boldsymbol X}_{\rm env}
\rightarrow
\chi_{\rm alarm}^{\rm ODH},
\\[1ex]
{\cal P}_{\rm access}:&
{\boldsymbol X}_{\rm access}
\rightarrow
{\boldsymbol X}_{\rm access}^{\rm protected}.
\end{array}
\right.
\label{eq:layered_cryo_downstream_functions}
\end{equation}
Here, ${\cal P}_{\rm vent}$ denotes the ventilation
response acting on the atmospheric state, while
${\cal M}_{\rm ODH}$ represents the monitoring
function that evaluates the environmental conditions
and provides the corresponding ODH-alarm indication.
The binary variable $\chi_{\rm alarm}^{\rm ODH}$
was introduced in Eq.~\eqref{eq:odh_monitoring_chain}.
The function ${\cal P}_{\rm access}$ represents
the resulting personnel-protection actions,
which may include access restriction and evacuation.
These actions modify the access and occupancy
conditions described by
${\boldsymbol X}_{\rm access}$,
including the discrete access status
$\sigma_{\rm access}$. The relations are schematic: the ODH-alarm
indication also depends on the applicable
monitoring logic, while the access-protection
response may be required when a monitoring
channel becomes invalid, even in the absence
of a confirmed oxygen-deficiency alarm. The three functions should therefore not
be conflated. Ventilation modifies the
atmospheric conditions, ODH monitoring
evaluates them, and personnel-protection
measures control access and exposure.
Their actions may occur simultaneously
and are subject to the applicable
protection and operational conditions.

The barrier classification is also consequence dependent.  For example,
pressure relief is mitigative with respect to a helium-containment breach
that has already occurred, but preventive with respect to subsequent
mechanical rupture of the vacuum enclosure.  Similarly, access inhibition
is preventive with respect to personnel exposure when imposed before
entry, but evacuation is mitigative once personnel are already present in
an affected region.

Consequently, no single label such as ``preventive'' or ``mitigative''
should be regarded as an intrinsic property of a cryogenic protection
element.  The appropriate classification refers to a specified segment of
the hazard trajectory and to a specified consequence.

The same observation applies to active and passive protection.  Electrical
interlocks, relief devices, ventilation controls, ODH sensors, and access
systems involve active detection or response to varying degrees, whereas
structural containment and mechanical robustness can provide passive
resistance to failure propagation.  A credible protection architecture
combines these different mechanisms because they have different
dependencies, response times, and failure modes.

The timing requirements are correspondingly heterogeneous.  Electrical
precursor detection may need to act before local heating develops into an
arc; pressure relief must respond on the timescale of the cryogenic
pressure transient; ODH detection and warning must precede unacceptable
personnel exposure; and atmospheric recovery may occur on a much longer
timescale.  There is therefore no unique cryogenic protection time
$\tau_{\rm prot}$ applicable to all layers.

Instead, each relevant failure--barrier--consequence combination has its
own requirement,
\begin{equation}
\tau_{{\rm prot},k}
<
\tau_{{\rm avail},k},
\label{eq:layered_cryo_timing}
\end{equation}
where the index $k$ identifies the corresponding protection function and
consequence.  This is the cryogenic realization of the
consequence-specific timing framework introduced in
Sec.~\ref{subsec:protection_time}.

The layered architecture should also not be interpreted as implying
statistical independence among all barriers.  Electrical supervision,
cryogenic instrumentation, ventilation, atmospheric monitoring, and
access control can share electrical power, communications,
instrumentation infrastructure, or environmental dependencies.  The
existence of several layers therefore provides DiD,
but does
not by itself establish barrier independence.

In the notation of Sec.~\ref{subsec:barrier_independence}, the relevant
protection set may be written schematically as
\begin{equation}
\left\{
{\cal B}_{\rm electrical},
{\cal B}_{\rm QPS},
{\cal B}_{\rm containment},
{\cal B}_{\rm relief},
{\cal B}_{\rm mechanical},
{\cal B}_{\rm vent},
{\cal B}_{\rm ODH},
{\cal B}_{\rm access}
\right\},
\label{eq:layered_cryo_barrier_set}
\end{equation}
with each barrier characterized by the applicable failure probability,
response time, independence, diagnostic, and testability attributes
introduced earlier.

Table~\ref{tab:cryogenic_protection_layers} summarizes representative
functions within this layered architecture.  The classification is
necessarily schematic: several systems can participate in the response
to the same failure, and a given barrier can be preventive with respect
to one consequence while remaining mitigative with respect to another.
The table should therefore be interpreted as a map of complementary
protection functions rather than as a strictly serial sequence.
\begin{table*}[t]
\caption{\label{tab:cryogenic_protection_layers}
Schematic relation between representative cryogenic failure conditions,
physical observables, protection functions, and principal objectives in
an underground superconducting accelerator.  The entries illustrate
complementary layers of the cryogenic protection architecture and are not
intended as an exhaustive classification of LHC safety systems.}
\begin{ruledtabular}
\begin{tabular}{llll}
\textbf{Condition or failure}
&
\textbf{Relevant observable}
&
\textbf{Protection function}
&
\textbf{Principal objective}
\\
\hline
Abnormal electrical or splice state
&
Resistance, voltage, current
&
\shortstack[l]{Electrical supervision /\\circuit protection}
&
\shortstack[l]{Interrupt the trajectory before\\destructive heating or arcing}
\\
\hline
Superconducting transition
&
Voltage, current, temperature
&
Quench protection
&
\shortstack[l]{Control redistribution and\\dissipation of magnetic energy}
\\
\hline
Abnormal cryogenic state
&
\shortstack[l]{Temperature, pressure, level,\\mass-flow rate}
&
\shortstack[l]{Cryogenic monitoring and\\control}
&
\shortstack[l]{Detect and constrain departure\\from the validated state}
\\
\hline
Loss into insulation vacuum
&
Pressure, mass-flow rate
&
\shortstack[l]{Vacuum-enclosure\\pressure relief}
&
\shortstack[l]{Limit overpressure and prevent\\structural failure}
\\
\hline
Pressure-induced mechanical loading
&
\shortstack[l]{Pressure, force,\\displacement}
&
\shortstack[l]{Mechanical integrity and\\structural reinforcement}
&
\shortstack[l]{Limit displacement and\\secondary damage}
\\
\hline
Helium release to underground volume
&
\shortstack[l]{Gas flow, pressure,\\environmental state}
&
\shortstack[l]{Discharge routing and\\ventilation}
&
\shortstack[l]{Limit accumulation and promote\\atmospheric recovery}
\\
\hline
Oxygen dilution
&
$C_{O_2}({\boldsymbol r},t)$
&
ODH monitoring
&
\shortstack[l]{Detect an atmosphere incompatible\\with personnel occupancy}
\\
\hline
Hazardous or unvalidated atmosphere
&
\shortstack[l]{ODH alarm, environmental\\and access state}
&
\shortstack[l]{Access restriction /\\evacuation}
&
\shortstack[l]{Prevent or terminate\\personnel exposure}
\\
\end{tabular}
\end{ruledtabular}
\end{table*}

The objective of the architecture is not to guarantee that every
initiating disturbance is suppressed.  Instead, it is to prevent a
physically credible disturbance from progressing through all barriers to
an unacceptable consequence.  A failure trajectory can therefore be
redirected at several possible stages,
\begin{equation}
\Gamma_{\rm fault}
\xrightarrow{\;{\cal B}_{1}\;}
\Gamma_{1}^{\rm protected}
\quad\text{or}\quad
\xrightarrow{\;{\cal B}_{2}\;}
\Gamma_{2}^{\rm protected}
\quad\text{or}\quad
\cdots ,
\label{eq:layered_cryo_trajectory_redirection}
\end{equation}
where the protected trajectories need not coincide with the nominal
operating trajectory.

This distinction is essential for cryogenic safety.  A state involving a
magnet quench, controlled helium discharge, temporary oxygen alarm, or
restricted access can be abnormal while remaining protected.  Safety
therefore does not require
\begin{equation}
{\boldsymbol X}_{\rm cryo\text{-}saf}
\equiv
{\boldsymbol X}_{\rm nominal},
\end{equation}
but rather that the evolving stateremain within the validated protected
trajectory set  Eq.~\eqref{eq:cryo_trajectory_requirement}, defined in Eq.~\eqref{eq:cryo_allowed_trajectory_set}.

For the redefined MCI, the same principle requires
that the corresponding severe trajectory remain within the validated
response envelope of the combined barriers.  Pressure relief and
mechanical protection must control the primary cryogenic transient, while
environmental and access barriers must limit the consequences of any
helium that is nevertheless discharged into underground volumes.

Barrier availability and validation therefore form a cross-cutting part of
the architecture.  A nominally adequate protection function cannot be
credited if its state is unknown, if required diagnostics are unavailable,
or if its performance has not been demonstrated for the relevant
configuration.  Functional testing, instrumentation validation,
maintenance, commissioning measurements, and operational feedback are
therefore integral to the protection architecture rather than external
administrative activities.

This observation also connects the layered architecture to the empirical
feedback discussed in Sec.~\ref{subsec:cryo_empirical_feedback}.  The
validated protection envelope is not immutable: measurements and
operational experience can reveal previously underestimated couplings or
failure trajectories and consequently require modification of individual
barriers or of the design basis itself.

The overall cryogenic protection concept can therefore be summarized as
\begin{equation}
{\cal F}_i
\rightarrow
\left\{
\begin{array}{c}
\text{energy and electrical protection}\\
\text{containment and pressure relief}\\
\text{mechanical protection}\\
\text{environmental mitigation}\\
\text{atmospheric monitoring}\\
\text{access and occupancy protection}
\end{array}
\right\}
\rightarrow
\Gamma_{\rm protected}
\in
\mathcal G_{\rm adm}^{\rm cryo}.
\label{eq:layered_cryo_summary}
\end{equation}

The LHC cryogenic safety architecture is thus a distributed system of
complementary barriers acting on different physical variables, timescales,
and consequence pathways.  Its effectiveness lies not in any single
protection element, but in the ability of the combined architecture to
prevent or redirect propagation across the coupled magnet, cryogenic,
mechanical, environmental, and access states.  This layered structure is a
direct realization of the broader principle developed in this manuscript:
safety is obtained by constraining the physically accessible trajectories
of the facility rather than by assigning safety independently to each
subsystem.

The discussion so far has concentrated primarily on the LHC machine.
Large experimental detectors introduce another set of coupled physical
states---high and low voltages, superconducting detector magnets, cooling
and gas systems, movable structures, environmental conditions, and
radiation fields---which must be protected while preserving detector
availability.  In Sec.~\ref{sec:atlas_cms_safety}, we therefore turn to
ATLAS and CMS as complementary examples of how safety is embedded directly
within the architecture and operation of large particle detectors.

\section{Safety embedded in ATLAS and CMS}
\label{sec:atlas_cms_safety}

The preceding sections considered safety first at the level of the LHC
machine and subsequently through the coupling of superconducting magnets,
cryogenics, the underground environment, and personnel access.  The same
general framework applies to the large experiments installed at the LHC
interaction points, but the relevant state space becomes more strongly
detector dependent.  ATLAS and CMS operate in the same accelerator
complex and are exposed to many common facility-level conditions, yet
their different detector architectures, services, stored-energy systems,
cryogenic installations, mechanical configurations, and access
arrangements generate different combinations of hazards and protection
requirements
\cite{ATLAS:2008Detector,CMS:2008Detector}.

The safety problem is therefore not obtained by simply adding a generic
``detector'' block to the machine state.
For an experiment
$\alpha\in\{\mathrm{ATLAS},\mathrm{CMS}\}$, the safety-relevant evolution
can be regarded schematically as a coupled trajectory
\begin{equation}
{\boldsymbol X}_{\alpha}(t)
=
\left(
{\boldsymbol X}_{\rm det},
{\boldsymbol X}_{\rm env},
{\boldsymbol X}_{\rm access}
\right)_{\alpha},
\label{eq:experiment_coupled_state}
\end{equation}
where ${\boldsymbol X}_{\rm det}$ represents the detector and its
technical systems, ${\boldsymbol X}_{\rm env}$ the local cavern or service
environment, and ${\boldsymbol X}_{\rm access}$ the corresponding
occupancy and access state.  Additional machine or infrastructure
variables enter whenever they couple materially to the experimental
configuration.

The components of ${\boldsymbol X}_{\rm det}$ are experiment specific.
They can include high-voltage and low-voltage states, magnet conditions,
cryogenic variables, gas-system states, cooling and ventilation
conditions, movable detector elements, stored mechanical or electrical
energy, and the status of detector-control and protection systems.  The
important point is not the precise decomposition, but that these variables
are coupled to one another and to the environment in which personnel may
be present.

A representative experimental hazard trajectory may therefore be written
as
\begin{equation}
{\boldsymbol X}_{\alpha}^{(0)}
\xrightarrow{\;{\cal F}_i\;}
{\boldsymbol X}_{\alpha}^{\rm abnormal}
\xrightarrow{\;\{{\cal B}_k\}\;}
{\boldsymbol X}_{\alpha}^{\rm protected},
\label{eq:experiment_hazard_trajectory}
\end{equation}
where ${\cal F}_i$ denotes an initiating failure or abnormal configuration
and $\{{\cal B}_k\}$ the set of technical, environmental, access, or
procedural barriers acting on the subsequent evolution.  As in the
machine and cryogenic cases, a protected experimental state need not be
identical to the nominal operating state: detector shutdown, inhibited
high voltage, restricted access, stopped motion, or a temporarily
unavailable subsystem can all represent intended protected states.

The corresponding safe domain is therefore intrinsically
configuration dependent,
\begin{equation}
\Omega_{\rm safe}^{\alpha}
=
\Omega_{\rm safe}^{\alpha}
\left(
m,
{\boldsymbol \Xi}_{\alpha}
\right)
\label{eq:experiment_safe_domain}
\end{equation}
where $m$ denotes the operational or intervention mode and
${\boldsymbol \Xi}_{\alpha}$ is the experiment-specific configuration
vector.
It specifies,
\begin{equation}
{\boldsymbol \Xi}_{\alpha}
=
\left(
{\boldsymbol \xi}_{\rm mech},
{\boldsymbol \xi}_{\rm elec},
{\boldsymbol \xi}_{\rm cryo},
{\boldsymbol \xi}_{\rm access},
{\boldsymbol \xi}_{\rm prot}
\right)_{\alpha},
\label{eq:experiment_configuration_vector}
\end{equation}
the technical and mechanical configuration relevant
to the safety analysis, including, for example, the positions and states
of movable detector subsystems or mechanical structures, the status of
energized systems, cryogenic configurations, access conditions, and the
availability of the protection functions required for that configuration.

Consequently, the safe operational domains of ATLAS and CMS
are generally different, reflecting their distinct physical
architectures, technical configurations, and protection
requirements. This difference does not imply that one
experiment is intrinsically safer than the other.
A more explicit comparison of the corresponding
experiment-specific safe domains is given in
Sec.~\ref{subsec:atlas_cms_risk_comparison}.

This distinction is particularly important for personnel access.  A
detector state that is admissible during closed-cavern physics operation
may be incompatible with maintenance access, while a state acceptable
during an intervention may require the removal of beam, magnet excitation,
high voltage, motion or reconfiguration of movable detector subsystems or
mechanical structures, or another stored-energy condition.  Accordingly,
\begin{equation}
\Omega_{\rm safe}^{{\alpha},{\rm occupied}}
\neq
\Omega_{\rm safe}^{{\alpha},{\rm unoccupied}},
\label{eq:experiment_occupancy_safe_spaces}
\end{equation}
in direct analogy with the occupancy-dependent cryogenic safety domains
introduced in Sec.~\ref{subsec:cryo_coupled_state}.

The protection architecture of ATLAS and CMS is therefore layered and
heterogeneous.  Detector control systems (DCS) provide continuous information
and operational control, dedicated detector safety systems (DSS) can act
autonomously on selected technical hazards, and personnel-safety systems
protect occupancy and access.  These functions overlap in the physical
variables they observe but should not be regarded as interchangeable.
In particular, detector control, equipment protection, and personnel
safety correspond to different protection objectives and consequence
classes.

The following discussion therefore does not attempt to compare ATLAS and
CMS through a single scalar measure of safety.  Instead, it examines how
their different constructions lead to different couplings among detector,
environmental, and access states, and how the corresponding protection
layers constrain the trajectories that remain compatible with detector
integrity, personnel safety, and sustained physics operation.

\subsection{Common protection principles in ATLAS and CMS}
\label{subsec:atlas_cms_common_protection}

Despite their different detector architectures, ATLAS and CMS share a
number of general protection principles.
Both experiments combine continuous detector supervision, autonomous
equipment-protection functions, infrastructure monitoring, and
personnel-protection systems---including access control, fire and smoke
detection, gas and oxygen-deficiency monitoring, alarms, and evacuation
functions---in order to constrain the evolution of the detector and its
environment during physics operation, technical stops, maintenance, and
interventions~\cite{ATLAS:2008Detector,CMS:2008Detector,Luders:2003DSS,CMS:2009DCS}.

The common structure can be represented schematically as
\begin{equation}
{\boldsymbol X}_{\alpha}
\rightarrow
{\boldsymbol Y}_{\alpha}
\rightarrow
\left\{
\begin{array}{l}
\mathrm{DCS}_{\rm ctrl},\\
{\cal P}_{\rm DSS},\\
{\cal P}_{\rm pers},
\end{array}
\right.
\label{eq:experiment_common_protection_structure}
\end{equation}
where ${\boldsymbol X}_{\alpha}$ is the physical state of experiment
${\alpha}$, ${\boldsymbol Y}_{\alpha}$ denotes the corresponding measured
or derived observables, $\mathrm{DCS}_{\rm ctrl}$ represents the
supervisory and operational-control functions of  the DCS,
${\cal P}_{\rm DSS}$ the autonomous DSS
protectiv-action function, and ${\cal P}_{\rm pers}$ the
personnel-protection function associated with environmental and access
hazards.

These functions are complementary but should not be conflated.  In both
ATLAS and CMS, the DCS provides continuous
supervision and operational control of detector subsystems and associated
technical infrastructure, including the monitoring and control of
quantities such as voltages, currents, temperatures, cooling conditions,
gas-system parameters, and other detector states~\cite{Boterenbrood:2004ATLASDCS,CMS:2009DCS}.
The DSS, developed within the CERN Joint Controls Project for the LHC
experiments, provides a distinct and highly reliable autonomous
equipment-protection layer.  It monitors selected safety-critical
conditions and can execute predefined protective actions independently of
normal DCS operation, for example by removing power or otherwise driving
affected detector equipment toward a protected state~\cite{Luders:2003DSS,Schmeling:2004DSS}.
The DSS is therefore intended to
complement rather than duplicate the DCS.  Personnel-protection functions
have a different objective again: through systems such as access control,
fire and smoke detection, gas and oxygen-deficiency monitoring, alarms,
and evacuation functions, they act primarily to prevent or limit
personnel exposure to hazardous environmental or operational conditions
\cite{Luders:2003DSS}.

A generic protection action may therefore be
written as
\begin{equation}
{\boldsymbol X}_{\rm det}^{\rm abnormal}
\xrightarrow{\;{\cal P}_{\rm DSS}\;}
{\boldsymbol X}_{\rm det}^{\rm protected},
\label{eq:experiment_detector_protection}
\end{equation}
where the protected state can involve, for example, removal of high or
low voltage, inhibition of a subsystem, interruption of a technical
process, or another predefined transition intended to prevent equipment
damage.

The distinction between the DCS and the DSS is also reflected in the way
measured information is translated into action.  In the DCS, measured or
derived detector quantities are used primarily for supervision,
operational control, alarm generation, and operator-supported or
software-mediated responses.  The DSS, by contrast, implements dedicated
autonomous protection logic for selected safety-critical detector
conditions and can initiate predefined equipment-protection actions
independently of normal DCS operation
\cite{Boterenbrood:2004ATLASDCS,CMS:2009DCS,
Luders:2003DSS,Schmeling:2004DSS}.

At the most general level, the protection sequence may still be written as
\begin{equation}
\text{measurement}
\rightarrow
\text{decision}
\rightarrow
\text{protective action},
\label{eq:experiment_measurement_decision_action}
\end{equation}
but the implementation of the intermediate decision stage depends on the
protection function.  For normal detector control, the decision can belong
to DCS logic, higher-level control software, or operational intervention,
whereas for DSS-protected conditions it is implemented by dedicated
autonomous safety logic.  Schematically,
\begin{equation}
{\boldsymbol Y}_{\alpha}
\rightarrow
\left\{
\begin{array}{ll}
\mathrm{DCS}_{\rm ctrl}
&\rightarrow
\text{supervision or operational control},
\\[1ex]
{\cal L}_{\rm DSS}
&\rightarrow
{\cal P}_{\rm DSS},
\end{array}
\right.
\label{eq:dcs_dss_action_paths}
\end{equation}
where $\mathrm{DCS}_{\rm ctrl}$ denotes the supervisory and
operational-control branch of the DCS introdused in Eq.~\eqref{eq:experiment_common_protection_structure}, ${\cal L}_{\rm DSS}$ denotes the dedicated DSS decision logic and
${\cal P}_{\rm DSS}$ denotes the same DSS protective-action function
introduced in Eq.~\eqref{eq:experiment_common_protection_structure}.
In Eq.~\eqref{eq:dcs_dss_action_paths}, it appears as the output of the
dedicated DSS decision logic ${\cal L}_{\rm DSS}$, while in
Eq.~\eqref{eq:experiment_detector_protection} it denotes the resulting
state transition from an abnormal detector condition to a protected
detector state.

The distinction is important because observation of an abnormal condition
by the DCS does not by itself imply execution of an autonomous safety
action, while a DSS protection channel is specifically designed to
translate selected abnormal inputs into predefined protective responses.
The two systems can therefore observe related physical quantities while
having different responsibilities, response paths, and degrees of
independence.

Personnel-safety systems form a further protection layer.  Their primary
objective is not detector integrity but the prevention or limitation of
personnel exposure to hazardous states.  Depending on the configuration,
these functions can include access control, fire or smoke detection,
gas or oxygen-deficiency monitoring, alarms, evacuation functions, and
other infrastructure-level protections.  Their relevant state transition
is therefore more naturally associated with the environmental and access
blocks,
\begin{equation}
{\boldsymbol X}_{\rm env}
\rightarrow
{\cal P}_{\rm pers}
\rightarrow
{\boldsymbol X}_{\rm access}^{\rm protected}.
\label{eq:experiment_personnel_protection}
\end{equation}

The separation of these functions is conceptually important because the
same measured quantity can be relevant to different consequence classes.
For example, an abnormal temperature may indicate a risk of detector
damage, a cooling failure, or the onset of a condition relevant to fire
protection.  The corresponding protective response depends on which
consequence is being controlled.

The consequence vector introduced earlier can therefore be specialized
to an experimental installation as
\begin{equation}
{\boldsymbol{\cal C}}_{\alpha}
=
\left(
{\cal C}_{\rm pers},
{\cal C}_{\rm det},
{\cal C}_{\rm rad},
{\cal C}_{\rm env},
{\cal C}_{\rm op}
\right)_{\alpha},
\label{eq:experiment_consequence_vector}
\end{equation}
where a given initiating event can contribute to more than one component.
Hence, the appropriate protection classification must remain
consequence dependent.

Several broad hazard classes are common to both experiments.  These
include abnormal electrical states, loss of cooling, overheating,
failures of gas or cryogenic services, water leaks, smoke or fire,
environmental deterioration, and unsafe access or intervention
configurations.  Their relative importance, however, differs between
ATLAS and CMS because the physical construction and technical services of
the two detectors are different.

The corresponding common protection logic can be expressed schematically
as
\begin{equation}
{\boldsymbol X}_{\alpha}^{(0)}
\xrightarrow{\;{\cal F}_i\;}
{\boldsymbol X}_{\cal E}^{\rm abnormal}
\xrightarrow{\;\{{\cal B}_k\}_{\alpha}\;}
{\boldsymbol X}_{\alpha}^{\rm protected},
\label{eq:experiment_common_barrier_logic}
\end{equation}
where the barrier set $\{{\cal B}_k\}_{\alpha}$ can include operational
control, autonomous detector protection, infrastructure protection, and
personnel-safety functions.

As in the machine and cryogenic systems, these barriers do not necessarily
form a strictly serial chain.  Some act in parallel, some operate on
different consequence pathways, and some are relevant only for particular
operational or intervention modes.  Their effectiveness therefore depends
on the complete detector configuration and not merely on the presence of
the individual protection elements.

A protected detector state should likewise not be confused with the
normal physics state.  Conditions such as
\begin{equation}
\text{high voltage off},\qquad
\text{subsystem inhibited},\qquad
\text{access restricted},\qquad
\text{movable subsystem secured}
\end{equation}
can represent intended protected configurations even though they are
incompatible with normal data taking.

The common objective of the protection architecture
is consequently to ensure that the evolution of
the experiment follows an admissible trajectory,
\begin{equation}
\Gamma_{\alpha}
\in
\mathcal G_{\rm adm}^{\alpha},
\label{eq:experiment_allowed_trajectory}
\end{equation}
where $\Gamma_{\alpha}$ denotes an individual
trajectory of experiment
$\alpha$,
and $\mathcal G_{\rm adm}^{\alpha}$ represents
the set of trajectories satisfying the applicable
detector, environmental, and personnel-protection
requirements.
The admissible trajectory set depends on the
experiment-specific technical configuration
$\boldsymbol\Xi_\alpha$, operational mode
$m_\alpha$, personnel-occupancy conditions,
and availability of the required protection
functions. Its definition accounts not only
for the admissibility of the instantaneous
physical states but also for the permitted
transitions between them, including the
timing and effectiveness of protective
responses.

Consequently, the admissible trajectory sets
of ATLAS and CMS need not be identical,
even when the experiments perform broadly
analogous scientific or operational activities.

The commonality between ATLAS and CMS therefore lies primarily in this
layered protection philosophy rather than in an identical realization of
individual systems.  Both experiments constrain abnormal trajectories
through combinations of detector control, autonomous equipment
protection, environmental monitoring, and personnel-safety functions,
while the detailed hazard pathways and required barriers remain specific
to their respective detector architectures.

\subsection{ATLAS-specific safety configuration}
\label{subsec:atlas_safety_configuration}

The generic protection framework introduced in
Sec.~\ref{subsec:atlas_cms_common_protection} applies directly to ATLAS and
will not be repeated here.  What distinguishes the ATLAS safety
configuration is the particular combination of detector technologies,
stored-energy systems, cryogenic installations, mechanical structures,
services, and access conditions that determine the experiment-specific
configuration vector ${\boldsymbol\Xi}_{\rm ATLAS}$.

ATLAS combines a central superconducting solenoid with large barrel and
end-cap toroidal magnet systems, liquid-argon calorimetry, extensive
electrical and cooling services, massive detector structures, and several
underground working levels within the experimental cavern
\cite{ATLAS:2008Detector,tenKate:2005ATLASMagnets}.  These systems do not
constitute independent safety domains: abnormal conditions in one system
can modify the state or accessibility of others.

A particularly important ATLAS-specific feature is the extensive use of
liquid argon in the electromagnetic, hadronic end-cap, and forward
calorimeter systems~\cite{Bremer:2000ATLASLAr,Fabre:2009ATLASLArCryo}.
  The liquid-argon cryogenic installations introduce hazards associated with low
temperature, pressure, loss of containment, and displacement of air by
released argon.  The relevant propagation is therefore analogous in
structure, although not in thermodynamic detail, to the helium-release
problem discussed for the LHC cryogenic system:
\begin{equation}
{\boldsymbol X}_{\rm LAr}
\rightarrow
{\boldsymbol X}_{\rm env}
\rightarrow
{\boldsymbol X}_{\rm access}.
\label{eq:atlas_lar_environment_coupling}
\end{equation}
Here ${\boldsymbol X}_{\rm LAr}$ denotes the state of the ATLAS
liquid-argon cryogenic system.  A loss of containment can modify the local
atmospheric state and can consequently require ventilation, atmospheric
monitoring, alarms, and access restrictions.

The significance of a liquid-argon release cannot be inferred from the
cryogenic inventory alone.  As in the helium case, the personnel hazard
depends on the release rate and location, the geometry of the affected
volume, ventilation, gas transport, and occupancy.  The atmospheric and
access states must therefore be considered together when defining whether
a given detector configuration remains compatible with personnel
presence.

The ATLAS magnet system introduces a different class of stored-energy
hazards.  The central solenoid and the barrel and end-cap toroids contain
large magnetic-energy reservoirs and generate substantial electromagnetic
forces during normal operation and during abnormal or discharge
conditions
\cite{ATLAS:2008Detector,tenKate:2005ATLASMagnets}.  Their protection
therefore involves the coupled electrical, superconducting, cryogenic,
and mechanical response of the magnet system rather than any one of these
quantities separately.

For safety analysis, it is particularly important that a protected magnet
state need not coincide with the nominal energized state.  Controlled
discharge, inhibited powering, or another predefined magnet configuration
can constitute the intended protected condition if it prevents the
trajectory from progressing toward equipment damage or an unsafe
intervention state.

The mechanical configuration of ATLAS is likewise relevant during
technical stops and maintenance.  The experiment as a whole remains fixed
at its interaction point, but particular detector subsystems, shielding
elements, service structures, platforms, or other mechanical components
can be moved, opened, secured, or placed in intervention configurations.
Such changes modify clearances, access routes, possible stored mechanical
energy, working-at-height conditions, and the spatial relationship between
personnel and detector services.

These changes are represented through
${\boldsymbol\Xi}_{\rm ATLAS}$ rather than by introducing a separate
mechanical safety variable.  Consequently, two ATLAS intervention
configurations can require different admissible access conditions and
different combinations of protection barriers even when the accelerator
and the main detector remain otherwise unchanged.

Electrical and cooling services provide another important coupling.
ATLAS contains extensive high- and low-voltage distributions,
front-end electronics, cooling circuits, and associated technical
infrastructure.  Abnormal currents, voltages, temperatures, cooling
conditions, or water leaks can threaten detector equipment and can also
modify the admissibility of an intervention.

The distinction between DCS and DSS introduced in
Sec.~\ref{subsec:atlas_cms_common_protection} is particularly important in
this context.  The ATLAS DCS provides supervision and operational control
of detector and infrastructure states, whereas the DSS provides an
independent autonomous equipment-protection layer for selected abnormal
conditions
\cite{Boterenbrood:2004ATLASDCS,Luders:2003DSS,Schmeling:2004DSS}.
Personnel-protection functions remain distinct from both: a condition can
require access restriction or evacuation even when the detector equipment
has already been placed in a protected state.

This separation is especially useful for coupled hazards.  For example,
loss of cooling can initially be an equipment-protection problem, while
the resulting overheating or associated secondary effects can lead to a
different consequence class.  Similarly, a cryogenic or gas-system
abnormality can require detector protection, environmental monitoring,
and personnel-access actions at different stages of the same event.

The ATLAS-specific safety configuration should therefore be understood
not as a list of independent hazards but as the particular pattern of
couplings created by its detector architecture.  Liquid-argon cryogenics,
large superconducting magnets, distributed electrical and cooling
services, massive mechanical structures, and changing intervention
configurations define the relevant components of
${\boldsymbol\Xi}_{\rm ATLAS}$ and determine which protection layers are
required for a given operational state.

Thus, the principal ATLAS-specific question is not whether the generic
protection framework differs from that of CMS, but how the ATLAS detector
architecture populates that framework with different physical states,
failure pathways, and intervention constraints.  These differences are
what make
$\Omega_{\rm safe}^{\rm ATLAS}({\boldsymbol\Xi}_{\rm ATLAS})$
experiment specific, without implying any ranking of the overall safety
performance of ATLAS relative to CMS.

\subsection{Liquid argon as an ATLAS-specific underground hazard}
\label{subsec:atlas_lar_safety}

One of the most important differences between the ATLAS and CMS safety
configurations is the presence of the ATLAS liquid-argon calorimeter
cryogenic system in the underground experimental cavern.

The ATLAS electromagnetic and hadronic end-cap calorimeters use liquid
argon as the active ionization medium.  The three major calorimeter
cryostats contain a total liquid-argon volume of approximately
$80~{\rm m^3}$, with values of about $78$--$83~{\rm m^3}$ quoted for
the installed and design configurations
\cite{Bremer:2000ATLASLAr,Fabre:2009ATLASLArCryo}.
The liquid is maintained near
\begin{equation}
T_{\rm LAr}
\simeq
88~{\rm K}.
\label{eq:atlas_lar_temperature}
\end{equation}

Liquid argon is neither combustible nor chemically toxic under normal
conditions.  Its principal large-release personnel hazard in an enclosed
underground volume is displacement of air and therefore oxygen deficiency.
In contrast to helium, gaseous argon is denser than air at comparable
temperature, while freshly released cryogenic argon is additionally cold.
Accumulation in low-lying or poorly ventilated regions must therefore be
included in the analysis of a significant release.

The scale of the volumetric transformation can be illustrated by considering
a liquid volume $V_{\rm LAr}$ with density $\rho_{\rm LAr}$.  The contained
argon mass is
\begin{equation}
m_{\rm Ar}
=
\rho_{\rm LAr}V_{\rm LAr},
\label{eq:atlas_argon_mass}
\end{equation}
and, after vaporization and warming sufficiently for an ideal-gas
approximation to be useful,
\begin{equation}
V_{\rm Ar}^{\rm gas}
\simeq
\frac{m_{\rm Ar}R_{\rm Ar}T}{P}.
\label{eq:atlas_argon_gas_volume}
\end{equation}
The resulting gas volume is many hundreds of times greater than the
original liquid volume at atmospheric conditions.  As in the helium case,
Eq.~\eqref{eq:atlas_argon_gas_volume} is a scale estimate rather than an
accident model; actual release behaviour depends on spill rate,
evaporation, heat transfer, confinement, ventilation, and geometry.

The safety architecture of the liquid-argon system therefore contains
several layers.  Under normal conditions the liquid remains confined in
the calorimeter cryostats.  The design includes underground storage
capacity into which the calorimeter argon can be transferred in an
emergency.  Large argon- and nitrogen-containing volumes are protected by
relief systems connected to dedicated discharge infrastructure leading
toward the surface, while retention, ventilation, insulation-vacuum
monitoring, and oxygen monitoring provide additional protection
\cite{Bremer:2000ATLASLAr,Fabre:2009ATLASLArCryo}.

The protection sequence can be represented schematically as
\begin{equation}
\begin{split}
\text{cryostat integrity}
&\rightarrow
\text{controlled emergency transfer}
\\
&\rightarrow
\text{pressure relief/discharge}
\rightarrow
\text{retention and ventilation}
\\
&\rightarrow
O_2\text{ monitoring}
\rightarrow
\text{alarm and evacuation}.
\end{split}
\label{eq:atlas_lar_protection_chain}
\end{equation}

This sequence is conceptually similar to the helium-protection layers
discussed in Sec.~\ref{sec:cryogenic_safety}, but the fluid properties,
inventory geometry, detector integration, and release scenarios are
different.  The liquid-argon system is therefore not simply another
instance of the LHC helium problem.

An important consequence is that the ATLAS
underground risk configuration contains a
persistent cryogenic contribution even when
no beam is circulating:
\begin{equation}
N_{\rm beam}=0
\;\not\Rightarrow\;
\boldsymbol X_{\rm LAr}
\in
\Omega_{\rm safe}^{\rm LAr}.
\label{eq:atlas_beamoff_cryo}
\end{equation}
Here, $N_{\rm beam}=0$ denotes the absence
of circulating beam particles, while
$\boldsymbol X_{\rm LAr}$ represents the
physical state of the ATLAS liquid-argon
calorimeter cryogenic system and
$\Omega_{\rm safe}^{\rm LAr}$ its corresponding
safe-state domain.

Thus, the absence of circulating beams does
not by itself establish a safe cryogenic
state. The liquid-argon inventory, containment,
pressure, transfer and relief systems,
environmental conditions, and associated
protection functions remain relevant
independently of the beam state.

\subsection{CMS modularity and configuration-dependent safety}
\label{subsec:cms_modular_safety}

A characteristic feature of CMS is its highly modular mechanical
construction.  The detector was designed as a set of large barrel and
end-cap structures that could be assembled and tested at the surface,
lowered separately into the experimental cavern, and subsequently
separated when access to internal detector systems is required
\cite{CMS:2008Detector}.  This architecture provides important advantages
for installation, maintenance, and detector upgrades, but it also makes
the mechanical configuration of the experiment an explicit safety
variable.

The CMS superconducting solenoid is surrounded by a massive steel
flux-return yoke that also supports substantial parts of the muon system
and associated services.  The barrel yoke is divided into large wheels,
while the end-cap region consists of massive disk-like structures.  During
normal physics operation these elements form the compact closed detector
configuration.  During technical stops and interventions, selected
structures can be separated or displaced to create access to otherwise
inaccessible detector regions
\cite{CMS:2008Detector}.

The relevant distinction is therefore not between a fixed and a moving
CMS detector as a whole.  CMS remains installed at its interaction point;
rather, its internal large-scale mechanical configuration can change
substantially between closed physics operation and open intervention
states.
In the notation introduced above, an opening operation therefore
corresponds to a sequence of configuration changes,
\begin{equation}
{\boldsymbol\Xi}_{\rm CMS}^{\rm closed}
\rightarrow
{\boldsymbol\Xi}_{\rm CMS}^{\rm moving}
\rightarrow
{\boldsymbol\Xi}_{\rm CMS}^{\rm open},
\label{eq:cms_configuration_transition}
\end{equation}
which consequently modifies the set of conditions required for safe
access and work. Here ``open'' denotes the mechanically secured intervention configuration
reached after completion of the detector movement, rather than a detector
structure that remains in motion.

This configuration dependence is particularly important because the
moving structures are extremely massive.  Their displacement requires
dedicated mechanical systems and controlled procedures for lifting,
support, translation, positioning, and securing.  The safety problem
therefore involves not only the final position of a detector element, but
also the trajectory followed during its movement and the state of the
associated supports, restraints, services, clearances, and exclusion
zones.

The scale of these operations is unusual for an experimental detector.
Individual CMS yoke structures have masses ranging from several hundred
tonnes to well above $10^{3}~{\rm t}$, and the detector-moving system was
designed for a maximum load of approximately \(M_{\rm move}^{\rm max} \simeq 2600~{\rm t},\),
corresponding, in the design specification, to three end-cap sections
considered together \cite{CMS:2008Detector}.  The importance of the
movement system therefore follows not only from the dimensions of the
structures but also from the forces required to control such large masses
with high positional accuracy.

For long displacements, CMS employs heavy-duty air pads that strongly
reduce the effective friction between the detector structure and its
support surface.  Controlled traction and retention are provided by a
hydraulic strand-jack system.  The design must also account for the
approximately $1.2\%$ slope of the experimental cavern, as well as
friction, cable-chain forces, and the requirement to control both uphill
and downhill movement \cite{CMS:2008Detector}.

These features make the movement process itself a safety-relevant physical
trajectory rather than merely a transition between two static detector
configurations.
The deliberately low movement velocity of detector sections limits this kinetic energy, but kinetic
energy alone does not characterize the mechanical risk.  Because the
masses are very large, the relevant protection problem also depends on
controlled forces, geometry, support conditions, clearances, and the
possibility of unintended motion.

A reduced movement-state vector may therefore be introduced as
\begin{equation}
{\boldsymbol X}_{\rm move}^{\rm CMS}
=
\left(
M_{\rm sec},
x,
v,
a,
\theta_{\rm cav},
F_{\rm tr},
F_{\rm ret},
d_{\rm clr},
\sigma_{\rm supp}
\right),
\label{eq:cms_movement_state}
\end{equation}
where $x$, $v$, and $a$ denote the position, velocity, and acceleration
of the moving section, $\theta_{\rm cav}$ represents the cavern slope,
$F_{\rm tr}$ and $F_{\rm ret}$ characterize the traction and retaining
forces, $d_{\rm clr}$ denotes a relevant mechanical clearance, and
$\sigma_{\rm supp}\in\Sigma_{\rm supp}$ denotes the discrete status of
the supporting and guiding structures.

The vector in Eq.~\eqref{eq:cms_movement_state} should be regarded as a
reduced mechanical subset of the complete CMS physical state
${\boldsymbol X}_{\rm CMS}$ rather than as an independent safety system.
Its purpose is to make explicit why the safety of a detector-opening
operation cannot be inferred from the mass or final position alone.

In particular, the intermediate configurations occurring during movement
cannot simply be treated as equivalent to either the closed or open
endpoint.  A mechanically admissible final configuration does not imply
that every trajectory leading to it is admissible.  During motion,
personnel exclusion zones, mechanical clearances, supported loads,
cable and service connections, hydraulic and traction systems, braking
or retaining capability, nearby fixed equipment, and emergency stopping
conditions become part of the protection problem.  The central beam line
and detector components that remain fixed impose additional geometrical
constraints during opening and closing.  The protection requirements
therefore apply to the complete movement trajectory rather than only to
its endpoints.

This point may be expressed through the configuration-dependent safe
domains,
\begin{equation}
\Omega_{\rm safe}^{\rm CMS}
\left(
m_{\rm phys},
{\boldsymbol\Xi}_{\rm CMS}^{\rm closed}
\right)
\neq\Omega_{\rm safe}^{\rm CMS}
\left(
m_{\rm move},
{\boldsymbol\Xi}_{\rm CMS}^{\rm moving}
\right)
\neq\Omega_{\rm safe}^{\rm CMS}
\left(
m_{\rm int},
{\boldsymbol\Xi}_{\rm CMS}^{\rm open}
\right),
\label{eq:cms_configuration_safe_spaces}
\end{equation}
where $m_{\rm phys}$, $m_{\rm move}$, and $m_{\rm int}$ denote,
schematically, the physics, detector-movement, and intervention modes.
The inequalities indicate different admissible domains rather than a
ranking of their safety.

Once the detector has reached an open and mechanically secured
configuration, the risk structure changes again.  Internal detector
regions become accessible, but opening can expose personnel to new
working conditions involving work at height, falling-object hazards,
electrical and cooling services, confined working volumes, and
simultaneous activities.  Components and services that are inaccessible
during normal physics operation can also become directly involved in an
intervention.

A mechanically admissible final configuration does not by itself guarantee
that the transition leading to it is safe.  The movement trajectory must
remain compatible with the required clearances, supported loads, connected
services, nearby fixed equipment, and personnel-exclusion conditions.
Consequently, the protection requirements apply to the complete opening or
closing operation rather than only to its endpoints.

The modularity of CMS consequently creates a class of hazards that is
strongly configuration dependent.  During an intervention, separation of
large detector structures changes available working volumes and access
routes, exposes surfaces and services that are inaccessible during physics
operation, and can introduce temporary interfaces among mechanical,
electrical, cooling, gas, and detector-support systems.  The intervention
configuration must therefore be treated as a different technical state,
not simply as the normal detector with beam removed.

This distinction is especially relevant to the configuration vector
${\boldsymbol\Xi}_{\rm CMS}$.  The latter characterizes the technical
configuration in which the physical state ${\boldsymbol X}_{\rm CMS}$ is
to be interpreted: for example, whether major detector structures are
closed, moving, parked, mechanically secured, or opened for intervention,
and which electrical, cooling, gas, magnet, and protection systems are
configured or enabled for the corresponding mode.  The instantaneous
physical quantities associated with those systems---such as currents,
temperatures, pressures, positions, velocities, and forces---remain
components of ${\boldsymbol X}_{\rm CMS}$.  The corresponding access
conditions can differ substantially between configurations.

CMS modularity nevertheless provides an important safety advantage:
opening the detector can create controlled access to components that would
otherwise be difficult or impossible to reach.  Maintenance and upgrade
activities can therefore be performed in deliberately prepared
configurations rather than requiring personnel to work within the compact
closed geometry.  Modularity is thus not intrinsically a hazard; it
changes the set of hazards and protection requirements associated with
different operational phases.

The same feature illustrates the consequence-dependent character of
barrier classification developed earlier in this manuscript.  Mechanical
restraints, movement controls, position verification, and exclusion of
personnel from appropriate regions can act preventively with respect to
uncontrolled movement or personnel exposure.  Once a detector element has
reached its intervention position, mechanical securing and verification
of the configuration provide the conditions under which subsequent work
can proceed.

The transition from physics operation to an intervention state also
requires coordination with non-mechanical systems.  Depending on the work
to be performed, magnet state, high and low voltage, cooling, gas systems,
detector services, radiation conditions, and access permissions may all
have to satisfy specified conditions before movement or personnel access
is authorized.  Thus, mechanical opening is only one component of a
larger configuration transition.

The DCS and DSS functions introduced in
Sec.~\ref{subsec:atlas_cms_common_protection} remain relevant throughout
these changes, but their generic roles need not be redefined here.  The
important CMS-specific point is that the observables and protection
conditions relevant during an open intervention configuration can differ
from those required during closed physics operation.  Personnel-protection
functions similarly depend on the actual spatial and mechanical
configuration of the detector.

CMS therefore provides a particularly transparent example of why the safe
operational domain of a large experiment cannot be defined independently
of its mechanical configuration.  Its modular construction creates
distinct closed, moving, and mechanically secured open configurations,
each with its own combination of mechanical, technical, and access
constraints.  The same modularity that enables controlled access for
maintenance consequently makes both the final configuration and the
trajectory by which it is reached relevant to safety.

The emphasis on CMS should not be interpreted as implying that large-scale
mechanical reconfiguration is unique to this experiment.  ATLAS also
requires lifting, displacement, opening, and repositioning of massive
detector components and associated structures during shutdowns,
maintenance, and upgrades.  CMS is used here as a particularly explicit
conceptual example because its modular detector-opening scheme makes the
sequence from closed through moving to open configurations especially
clear.  More generally, for experimental installations of this scale,
movement or reconfiguration of massive structures changes the admissible
domain through the combined mechanical configuration, movement
trajectory, connected services, and personnel occupancy.  The CMS example
therefore complements the ATLAS liquid-argon discussion while remaining
within the common state-space and layered-protection framework introduced
in Sec.~\ref{subsec:atlas_cms_common_protection}.

\subsection{Magnet configuration and personnel access in ATLAS and CMS}
\label{subsec:atlas_cms_magnet_access}

The superconducting magnet systems of ATLAS and CMS provide another
example of how detector architecture influences the relation between
technical configuration and personnel access.  In both experiments, the
magnet state is relevant not only to detector operation but also to the
conditions under which interventions can be performed safely
\cite{ATLAS:2008Detector,CMS:2008Detector,tenKate:2005ATLASMagnets}.

The scale of the magnet systems makes this coupling quantitatively
important.  The ATLAS magnet system distributes its stored magnetic
energy among the central solenoid, the barrel toroid, and the two end-cap
toroids.  The central solenoid operates at a nominal current of
$7.73~{\rm kA}$ and provides a field of approximately $2~{\rm T}$ in the
inner-detector volume, with a stored energy of about $40~{\rm MJ}$.  It is
approximately $5.8~{\rm m}$ long and $2.5~{\rm m}$ in diameter.  The much
larger barrel toroid operates at $20.4~{\rm kA}$, stores approximately
$1.08~{\rm GJ}$, and extends over about $25.3~{\rm m}$ in length and
$20.1~{\rm m}$ in outer diameter.  Each end-cap toroid also operates at
$20.4~{\rm kA}$ and stores approximately $0.25~{\rm GJ}$
\cite{ATLAS:2008Detector,tenKate:2005ATLASMagnets}.

The magnetic field of the ATLAS toroidal system is intrinsically
nonuniform.  Representative field ranges in the useful bore are about
$0.2$--$2.5~{\rm T}$ for the barrel toroid and
$0.2$--$3.5~{\rm T}$ for the end-cap toroids, while the peak fields in
the windings reach approximately $3.9~{\rm T}$ and $4.1~{\rm T}$,
respectively \cite{ATLAS:2008Detector}.  The corresponding mechanical
loads therefore cannot be characterized by a single uniform magnetic
pressure.

CMS concentrates the dominant magnetic-energy reservoir in a single
large superconducting solenoid.  The solenoid has a free bore of
approximately $6~{\rm m}$ diameter and a magnetic length of
$12.5~{\rm m}$, with a cold mass of about $220~{\rm t}$.  It is operated
nominally at $3.8~{\rm T}$, corresponding to a current of approximately
$18.16~{\rm kA}$, while commissioning at $4~{\rm T}$ demonstrated a
stored magnetic energy of approximately $2.6~{\rm GJ}$
\cite{CMS:2008Detector,Cure:2024CMSMagnet}.

The magnetic-field energy density introduced in
Eq.~\eqref{eq:magnetic_energy_density} also provides a useful scale for
the associated electromagnetic loading.  Interpreted as an equivalent
magnetic-pressure scale,
\begin{equation}
p_B
\equiv
u_B,
\label{eq:detector_magnetic_pressure_scale}
\end{equation}
a field of $2~{\rm T}$ corresponds to approximately
$1.6~{\rm MPa}$, while $3.8~{\rm T}$ and $4~{\rm T}$ correspond to
approximately $5.7~{\rm MPa}$ and $6.4~{\rm MPa}$, respectively.
Using the peak winding fields, the corresponding local scales are about
$6.1~{\rm MPa}$ for the ATLAS barrel toroid and
$6.7~{\rm MPa}$ for the end-cap toroids.
\begin{table*}[t]
\caption{\label{tab:atlas_cms_magnet_scales}
Representative parameters of the principal ATLAS and CMS superconducting
magnet systems.  The magnetic-pressure values are equivalent local scales
derived from the magnetic-field energy density and should not be
interpreted as uniform structural pressures.}
\begin{ruledtabular}
\begin{tabular}{llllll}
\textbf{Magnet}
&
\textbf{Field}
&
\textbf{Current}
&
\textbf{Stored energy}
&
\textbf{Characteristic size}
&
\textbf{$p_B$ scale}
\\
\hline
ATLAS central solenoid
&
$2.0~{\rm T}$
&
$7.73~{\rm kA}$
&
$0.04~{\rm GJ}$
&
$\sim2.5~{\rm m}\times5.8~{\rm m}$
&
$\sim1.6~{\rm MPa}$
\\
\hline
ATLAS barrel toroid
&
\shortstack[l]{$0.2$--$2.5~{\rm T}$ bore;\\
$3.9~{\rm T}$ peak}
&
$20.4~{\rm kA}$
&
$1.08~{\rm GJ}$
&
$\sim20.1~{\rm m}\times25.3~{\rm m}$
&
$\sim6.1~{\rm MPa}$ peak
\\
\hline
ATLAS end-cap toroid
&
\shortstack[l]{$0.2$--$3.5~{\rm T}$ bore;\\
$4.1~{\rm T}$ peak}
&
$20.4~{\rm kA}$
&
$0.25~{\rm GJ}$ each
&
$\sim10.7~{\rm m}\times5.0~{\rm m}$
&
$\sim6.7~{\rm MPa}$ peak
\\
\hline
CMS solenoid
&
\shortstack[l]{$3.8~{\rm T}$ nominal;\\
$4.0~{\rm T}$ maximum}
&
\shortstack[l]{$18.16~{\rm kA}$ nominal;\\
$\sim19.2~{\rm kA}$ at $4~{\rm T}$}
&
$2.6~{\rm GJ}$ at $4~{\rm T}$
&
$\sim6~{\rm m}\times12.5~{\rm m}$
&
\shortstack[l]{$\sim5.7~{\rm MPa}$ at $3.8~{\rm T}$;\\
$\sim6.4~{\rm MPa}$ at $4~{\rm T}$}
\\
\end{tabular}
\end{ruledtabular}
\end{table*}
These values should not be interpreted as uniform mechanical pressures
acting on the complete detector structures.  Actual electromagnetic
forces and stresses depend on the coil geometry, field distribution,
current distribution, supports, and magnetic return structure.  The
quantity $p_B$ is used here only as a physically transparent scale for
the magnitude of the electromagnetic loading associated with fields of
several tesla.

The corresponding coupling can be represented schematically as
\begin{equation}
{\boldsymbol X}_{\rm mag}^{\alpha}
\longrightarrow
\Omega_{\rm access}^{\alpha}
\left(
m,
{\boldsymbol\Xi}_{\alpha}
\right),
\label{eq:magnet_access_coupling}
\end{equation}
where ${\boldsymbol X}_{\rm mag}^{\alpha}$ denotes the physical state of
the experiment magnet system and
$\Omega_{\rm access}^{\alpha}$ the access domain compatible with the
corresponding operational or intervention configuration.  The relation is
not intended to imply that the magnet state alone determines access, but
that it forms one of the conditions entering the complete access decision.

For ATLAS, the magnet system consists of a central solenoid together with
the large barrel and end-cap toroidal magnet systems.  The toroidal
geometry produces an extended magnetic-field configuration throughout a
large volume surrounding the detector and is associated with substantial
stored magnetic energy and electromagnetic forces
\cite{ATLAS:2008Detector,tenKate:2005ATLASMagnets}.  The safety relevance
of the magnet state therefore extends beyond the immediate superconducting
coils.

During interventions, the admissible configuration depends on whether the
relevant magnet circuits are energized, discharged, or otherwise placed
in a validated state.  Magnetic-field conditions can affect personnel
access, the use of tools and equipment, the admissibility of work near
magnetic components, and the mechanical state of structures subject to
electromagnetic forces.  The appropriate intervention conditions must
therefore be defined together with the actual detector configuration and
the work to be performed.

CMS realizes the same general coupling through a different magnet
architecture.  Its high-field superconducting solenoid is surrounded by
the massive steel return yoke, which strongly shapes the magnetic flux
distribution and forms an integral part of the modular detector structure
\cite{CMS:2008Detector}.  The magnet state is consequently coupled not
only to the electromagnetic environment but also to the mechanical and
intervention configuration discussed in
Sec.~\ref{subsec:cms_modular_safety}.

In particular, major detector-opening and movement operations require a
magnet configuration compatible with the mechanical transition and with
personnel access.  Thus, the mechanically defined sequence of Eq.~\eqref{eq:cms_configuration_transition}
cannot be considered independently of the corresponding magnet state.
The mechanical and magnetic configurations together determine which
movement and intervention trajectories are admissible.

This coupling is not unique to CMS.  ATLAS also requires major detector
opening, lifting, and repositioning activities during shutdowns and
upgrades, and these operations likewise have to be compatible with the
magnet and powering states.  The difference is therefore not that one
experiment has mechanically relevant magnet conditions while the other
does not, but that the spatial distribution of the magnetic field and the
mechanical realization of the detector are different.

The magnet-related access problem can be summarized more generally as
\begin{equation}
\left(
{\boldsymbol X}_{\rm mag}^{\alpha},
{\boldsymbol\Xi}_{\alpha},
m
\right)
\in
\Omega_{\rm int}^{\alpha}
\quad\Longrightarrow\quad
\text{intervention permitted},
\label{eq:magnet_intervention_condition}
\end{equation}
where $\Omega_{\rm int}^{\alpha}$ denotes the set of combined magnetic,
technical, and operational conditions under which the specified
intervention is admissible.  This notation emphasizes that an intervention
cannot be authorized on the basis of the detector geometry or magnet
current alone.

A further distinction is important.  De-energization of a superconducting
magnet does not by itself imply that the complete experimental
installation is ready for unrestricted access.  Other hazards associated
with cryogenics, high voltage, cooling, detector services, mechanical
configuration, radiation conditions, or simultaneous work can remain
present.  Conversely, some forms of controlled access may be compatible
with particular magnet states if the corresponding intervention has been
explicitly validated.

Accordingly,
\begin{equation}
{\boldsymbol X}_{\rm mag}^{\alpha}
=
{\boldsymbol X}_{\rm mag}^{\rm off}
\quad\not\Longrightarrow\quad
\Omega_{\rm access}^{\alpha}
=
\Omega_{\rm access}^{\rm unrestricted}.
\label{eq:magnet_off_not_unrestricted_access}
\end{equation}
Equation~\eqref{eq:magnet_off_not_unrestricted_access} expresses the same
non-factorizable character of the safe domain encountered previously for
cryogenic and mechanical configurations: removal of one hazard source does
not eliminate the constraints associated with the remaining coupled
subsystems.

The magnet systems therefore provide a further example of the general
principle developed in this section.  \mbox{ATLAS} and CMS share the requirement
that magnetic, mechanical, environmental, and access states remain
mutually compatible, but their different detector architectures produce
different realizations of that requirement.  The magnet state is thus one
component of the complete experiment-specific configuration that
determines whether a particular intervention trajectory belongs to the
admissible operational domain.

\subsection{Infrastructure services: electrical power, cooling, gases, and
fire protection}
\label{subsec:atlas_cms_services}

The safety configuration of ATLAS and CMS depends not only on the detector
elements themselves but also on an extensive infrastructure of electrical
power distribution, cooling, ventilation, gas supply, environmental
control, fire detection, and emergency protection.  These services are
distributed between surface installations, underground service caverns,
and the experimental caverns, and they connect large numbers of otherwise
independent detector subsystems.  A failure of a common infrastructure
service can therefore constitute a common-cause disturbance affecting
several detector systems simultaneously
\cite{ATLAS:2008Detector,CMS:2008Detector}.

In the notation introduced above, the instantaneous physical variables
associated with these services form part of
${\boldsymbol X}_{\alpha}$, whereas their required availability and
configuration form part of
${\boldsymbol\Xi}_{\alpha}$.
Thus, for example, cooling-water temperature and flow are physical-state
variables, while the selection and availability of the cooling circuits
required for a particular intervention belong to the corresponding
technical configuration.  This distinction is useful because an
infrastructure service can be physically healthy but unavailable for a
specific configuration, or conversely can be nominally configured but
develop an abnormal physical state during operation.

The scale of the electrical infrastructure illustrates the magnitude of
these common services.  The original ATLAS Point~1 infrastructure was
dimensioned for an electrical demand of order $13~{\rm MW}$ and included
21 transformers with a combined installed rating of about $42~{\rm MW}$.
Different power classes were provided according to the required
availability, including normal, assured, secured, and uninterruptible
supplies.  The assured supply included two approximately $1~{\rm MW}$
diesel generators, while the installed uninterruptible-power capacity was
of order $700~{\rm kW}$
\cite{ATLAS:2008Detector}.  The corresponding original CMS infrastructure
was designed for a steady-state electrical consumption of order
$9~{\rm MW}$, distributed among detector electronics, front-end
low-voltage systems, magnet and cryogenic services, ventilation, cooling,
and general site infrastructure
\cite{CMS:2008Detector}.

The magnitude of these electrical loads reflects the extraordinary
granularity of the detector electronics.  ATLAS reads out more than
$10^{8}$ electronic channels distributed among silicon tracking
detectors, calorimeters, muon chambers, trigger systems, and associated
monitoring and control electronics
\cite{ATLAS:2008Detector,ATLAS:2020DataQuality}.  The electrical
infrastructure therefore powers not a small number of large loads but a
highly distributed population of front-end modules, power converters,
off-detector crates, network and trigger electronics, and computing
equipment.

The liquid-argon calorimeter provides a representative ATLAS example.
Its approximately $1.9\times10^{5}$ readout channels are serviced by
1524 front-end boards installed in 58 front-end crates mounted on the
detector.  The original off-detector readout system occupied approximately
60 additional crates.  The front-end boards dissipate heat in a confined
and radiation-exposed region and are thermally coupled to water-circulating
cooling plates
\cite{ATLAS:2008Detector,ATLAS:LArElectronics}.  During the Run-3
electronics upgrade, each of the 1524 original front-end boards had to be
removed from its water-cooled plate and subsequently reassembled and
tested for cooling-system integrity.

The scale extends well beyond the calorimeter electronics.  The original
ATLAS infrastructure comprised approximately 500 racks when electronics,
gas, and water-service racks are counted together.  About 70 rack
locations in the underground US15 service cavern were provided with a
dedicated cooling capacity of approximately $500~{\rm kW}$, while the
surface SDX1 data-acquisition area provided space for approximately
100 racks with a further $500~{\rm kW}$ of water-cooling capacity
\cite{ATLAS:2008Detector}.  These values characterize the installed
infrastructure rather than the exact present rack population, which
evolves as detector and data-acquisition electronics are upgraded.

The Run-3 New Small Wheel provides a more recent illustration of the
growth of this distributed electronics infrastructure.  Its Micromegas
and sTGC systems together contain approximately $2.4\times10^{6}$ trigger
and readout channels and require of order $8\times10^{3}$ front-end
boards, with an electronics power consumption of order $75~{\rm kW}$.
Such detector-mounted electronics require local thermal management in
addition to the larger rack- and plant-level cooling infrastructure.

CMS exhibits a comparable scale with a different distribution of loads.
The original infrastructure allocated approximately $2.3~{\rm MW}$ to
electronics racks and approximately $1.0~{\rm MW}$ to low-voltage
front-end electronics.  The detector-mounted front-end electronics
dissipated approximately $800~{\rm kW}$ in the experimental cavern,
while the rack infrastructure represented a thermal load of approximately
$1.6~{\rm MW}$
\cite{CMS:2008Detector}.  These figures again represent design-scale
values; subsystem upgrades have subsequently changed the detailed
distribution of electrical and thermal loads.

The CMS tracker alone illustrates the granularity behind these power
figures.  In its Run-3 configuration it contains 1856 pixel modules with
approximately $1.24\times10^{8}$ pixel channels and 15148 silicon-strip
modules with approximately $10^{7}$ additional readout channels
\cite{CMS:Phase1Pixel,CMS:TrackerRun3}.  These detector-mounted electronic
systems require dedicated low-temperature cooling rather than direct rack
water cooling: the Phase-1 pixel detector uses two-phase CO$_2$, while
the strip tracker uses its dedicated fluorocarbon cooling system.
Consequently, water cooling at the infrastructure level should not be
identified directly with the coolant circulating through every front-end
detector module.

Other CMS subsystems use water much more directly.  The ECAL, HCAL, and
muon front-end electronics are served by approximately $18^\circ{\rm C}$
water circuits, while the original infrastructure assigned cooling powers
of approximately $300~{\rm kW}$ to ECAL, $60~{\rm kW}$ to HCAL and the
barrel yoke, and about $150~{\rm kW}$ to the combined Tracker, Pixel, and
Preshower services
\cite{CMS:2008Detector}.  The ECAL alone contains 75848 crystal readout
channels, and its thermal-stability requirements led to a water-cooling
plant capable of circulating approximately $62~{\rm L\,s^{-1}}$ with
temperature stability at the level of a few hundredths of a kelvin.

The off-detector data-acquisition infrastructure introduces another
concentrated electrical and cooling load.  As an illustrative Run-3-era
CMS configuration, the surface counting room contained approximately
140 water-cooled DAQ racks of several generations, with nominal
per-rack cooling capacities ranging from about $10$ to $30~{\rm kW}$.
The highest-density generation used water flows of order
$4~{\rm m^3\,h^{-1}}$ per rack
\cite{CMS:DAQPhase2TDR}.  The rack population and power density continue
to evolve during preparation for the HL-LHC and should therefore be
interpreted as configuration-dependent infrastructure quantities rather
than immutable detector parameters.

These values characterize the installed design scale rather than the
instantaneous electrical consumption of the present detector
configurations; both experiments have undergone substantial upgrades since
their original construction.  They nevertheless demonstrate that ATLAS
and CMS should be regarded as large industrial electrical installations as
well as particle detectors.  Electrical power is distributed among very
large populations of front-end modules, power converters, off-detector
crates, electronics racks, cooling plants, gas installations, ventilation,
controls, and protection systems.  A disturbance of a common electrical
supply can therefore propagate simultaneously into several otherwise
distinct detector and infrastructure subsystems.

The large channel, crate, and rack populations also explain why electrical
power and cooling cannot be treated as independent infrastructure
functions.  The electronic heat load is distributed over many thousands
of modules and boards, several of which are inaccessible during physics
operation.  Loss of a common cooling circuit can consequently affect a
large population of powered electronics, while local temperatures can rise
on timescales shorter than those available for manual intervention.
Conversely, loss of electrical power can stop cooling pumps or associated
control equipment and thereby generate a secondary thermal disturbance.
Electrical and thermal states are therefore dynamically coupled,
schematically,
\begin{equation}
\text{power disturbance}
\rightarrow
\left\{
\begin{array}{l}
\text{loss of electronic function},\\
\text{loss or degradation of cooling},\\
\text{loss of auxiliary services}
\end{array}
\right.
\rightarrow
{\boldsymbol X}_{\rm det}^{\rm abnormal}.
\label{eq:experiment_power_service_coupling}
\end{equation}

Critical infrastructure consequently uses combinations of independent
power classes, uninterruptible power supplies, and diesel-backed supplies
so that loss of the normal mains supply does not immediately remove the
functions required to monitor the detector or bring affected equipment
toward a controlled state.  At the subsystem level, cooling supervision
and temperature monitoring provide the corresponding observational link.
An abnormal thermal condition can first generate DCS information and
alarms and, for selected conditions requiring autonomous equipment
protection, subsequently activate dedicated DSS logic and protective
actions.  The scale and inaccessibility of the distributed electronics
therefore provide a direct physical motivation for the coupling among
electrical-power availability, cooling supervision,
$\mathrm{DCS}_{\rm ctrl}$, and
${\cal L}_{\rm DSS}\rightarrow{\cal P}_{\rm DSS}$.

Cooling constitutes an equally important distributed service.  The heat
loads originate from front-end electronics, power converters, electronics
racks, detector sensors, magnets, cryogenic equipment, and computing
systems, and are removed through several physically distinct cooling
networks.  Cooling failure can therefore evolve from an infrastructure
fault into an equipment-protection problem,
\begin{equation}
\text{loss of cooling}
\rightarrow
T_{\rm det}\uparrow
\rightarrow
\text{equipment-protection action},
\label{eq:experiment_cooling_failure_chain}
\end{equation}
where the appropriate response can include power reduction or shutdown
through DCS-controlled operation or autonomous DSS action.

The current ATLAS Run-3 Inner Detector provides a useful illustration of
the coexistence of different cooling technologies.  The pixel detector
and SCT are cooled by an evaporative C$_3$F$_8$ system whose primary
thermosiphon circuit removes a heat load of approximately
$60~{\rm kW}$, whereas the Insertable B-Layer (IBL) employs a separate
two-phase CO$_2$ cooling system
\cite{ATLAS:2024Run3}.  The Run-3 New Small Wheels introduced further
water-cooling requirements; the associated chilled-water infrastructure
provides approximately $175~{\rm kW}$ on each side of ATLAS
\cite{ATLAS:2024Run3}.

CMS also employs two-phase CO$_2$ cooling, notably for the Phase-1 pixel
detector.  Two parallel cooling plants were installed, each designed for
a cooling capacity of approximately $15~{\rm kW}$ and capable of
supporting the full pixel detector independently.  Representative design
heat loads are about $6~{\rm kW}$ for the barrel pixel detector,
$3~{\rm kW}$ for the forward pixel detector, together with an additional
heat leak of order $2~{\rm kW}$
\cite{CMS:2020PixelUpgrade}.  The CMS silicon-strip tracker, however,
continues to use a fluorocarbon-based cooling infrastructure and has been
operated at progressively lower temperatures in Run~3 to control
radiation-induced leakage current.  The distinction again shows that
``detector cooling'' is not one homogeneous service but a set of
subsystem-specific circuits whose simultaneous availability enters the
experiment configuration.

Dry-gas circulation provides a related but physically distinct function.
It should not be confused with detector cooling.  In ATLAS, the Inner
Detector volume is maintained dry by a combination of dry N$_2$ and
CO$_2$ flows.  In the original service architecture, dedicated CO$_2$
ventilation channels flush the Inner Detector envelope and gaps, while
the Pixel and SCT detector volumes use dry nitrogen
\cite{ATLAS:2008Detector}.  The purpose is to control humidity and gas
composition around cold detector structures rather than to remove their
thermal load.  It is therefore more precise to state that parts of the
ATLAS Inner Detector volume are \emph{flushed with dry CO$_2$} than that
the Inner Detector is simply ``filled with CO$_2$''.

CMS uses the same general principle with a different implementation.
The Tracker and Pixel volumes are maintained at very low humidity by a
dedicated dry-gas infrastructure.  The CMS tracker volume is of order
$25~{\rm m^3}$ and has been operated with continuous dry-air or
dry-nitrogen flushing, while larger flows are supplied around service
bulkheads and cold pipework.  A dedicated Point~5 plant was developed
with a capacity up to approximately $400~{\rm m^3\,h^{-1}}$ of dry
nitrogen, or dry air during appropriate shutdown configurations
\cite{CMS:DryGasPlant,CMS:TrackerThermal}.  These dry-gas systems are
particularly important when detector structures operate substantially
below the cavern dew point, because loss of dry conditions can lead to
condensation or icing even when the cooling plant itself remains
functional.

Detector gases constitute a separate and extensive infrastructure.  ATLAS
contains several large gaseous detector systems with very different gas
compositions.  In the original detector configuration, the MDT chambers
alone represented an active gas volume of approximately
$710~{\rm m^3}$, while the RPC and TGC systems had active volumes of order
$14~{\rm m^3}$ and $16~{\rm m^3}$, respectively
\cite{ATLAS:2008Detector}.  The gas systems include mixing, distribution,
recirculation where applicable, purification, pressure regulation,
analysis, exhaust, and remote monitoring and control.

The Run-3 ATLAS muon system provides an especially important example of
the safety implications of detector gases.  The small-strip Thin Gap
Chambers (sTGCs), like the existing TGCs, operate with a mixture of
approximately \(55\%\,{\rm CO_2}+45\%\,n\text{-}{\rm C_5H_{12}}\),
where $n$-pentane is flammable.  The sTGC gas volumes are therefore
surrounded by a CO$_2$ safety envelope, and the exhaust from this envelope
is monitored by flammable-gas sniffers.  A leak of the chamber mixture can
thus be detected in the surrounding CO$_2$ stream before it is released
directly into the cavern.  For Run~3, the complete TGC/sTGC gas system
retains a maximum flow of approximately
$5000~{\rm L\,h^{-1}}$
\cite{ATLAS:2024Run3}.

The New Small Wheel (NSW) Micromegas use a different mixture \(93\%\,{\rm Ar}+5\%\,{\rm CO_2}+2\%\,i{\rm C_4H_{10}}\), with a nominal system flow of approximately
$600~{\rm L\,h^{-1}}$
\cite{ATLAS:2024Run3}.  Isobutane is itself flammable, illustrating why
gas composition, dilution, leak rate, ventilation, and detection must be
considered together rather than classifying a detector gas only by its
dominant component.  The ATLAS RPC mixture likewise contains an
isobutane component, although its complete mixture and operating
conditions differ from those of the TGC system.

Flammable-gas considerations are not unique to ATLAS.  The CMS RPC system
also uses a gas mixture containing isobutane.  The standard CMS RPC
mixture is approximately \(95.2\%\,{\rm C_2H_2F_4}+4.5\%\,i{\rm C_4H_{10}}+0.3\%\,{\rm SF_6}\)
with current reasearch and development (R\&D) directed toward lower global warming potential
alternatives
\cite{CMS:2008Detector,Ramos:2026CMSRPC}.  The installed RPC gas volume
is of order $14~{\rm m^3}$; the original closed-loop system circulated
approximately $10~{\rm m^3\,h^{-1}}$ through several hundred distribution
channels while requiring only a small fraction of fresh-gas replacement
\cite{CMS:2008Detector}.  The closed-loop architecture extends from the
surface gas building through the underground service cavern to the
experimental cavern.

Other CMS muon technologies use different, predominantly
argon--CO$_2$-based mixtures.  The cathode-strip chambers presently use
an Ar/CO$_2$/CF$_4$ mixture, while the Run-3 GEM chambers use an
Ar/CO$_2$ mixture without CF$_4$
\cite{CMS:2024Run3}.  Thus, as in ATLAS, the CMS gas infrastructure is a
network of detector-specific services rather than a single experiment-wide
gas system.

The safety significance of these gases is not limited to intrinsic
flammability.  Large inert-gas releases can alter the oxygen concentration,
greenhouse gases require containment and recirculation for environmental
reasons, and loss of pressure or mixture composition can damage detector
performance or create abnormal high-voltage conditions.  Gas-system
monitoring therefore includes combinations of flow, pressure, mixture
composition, leak detection, and environmental sampling.

Fire and gas detection provide a further common protection layer.  ATLAS
is equipped with smoke detection in the main and service caverns and in
electronics-rack regions; selected racks also have local CO$_2$
extinguishing systems.  The large liquid-argon inventory is complemented
by oxygen-deficiency monitoring and dedicated low-level extraction
arrangements, while detector regions containing flammable gases are
equipped with corresponding detection systems
\cite{ATLAS:2008Detector}.  The ATLAS DSS receives information from
distributed infrastructure sensors, including cooling status, smoke, and
flammable-gas detection, and can autonomously initiate detector-protection
actions such as removal of high or low voltage
\cite{Luders:2003DSS,Schmeling:2004DSS}.

CMS likewise employs distributed smoke detection and protection of
electronics racks.  A distinctive ultimate fire-mitigation layer at Point~5
is the high-expansion foam system for the experimental cavern.  The
published CERN description gives a full-cavern design in which
approximately $20~{\rm m^3}$ of water mixed with foaming agent can produce
about $1.65\times10^{4}~{\rm m^3}$ of high-expansion foam within several
minutes.  The system is passive in the sense that it uses the hydrostatic
head of the approximately $100~{\rm m}$ vertical water column rather than
requiring an electrically powered pump
\cite{CMS:FireFoam,CMS:FireFoamTest}. 

These examples also clarify the separation among the protection functions
introduced in Sec.~\ref{subsec:atlas_cms_common_protection}.  The DCS
supervises service variables and supports normal operational control;
dedicated DSS logic can autonomously protect detector equipment when
selected infrastructure conditions become unacceptable; fire, gas,
oxygen-deficiency, alarm, evacuation, and access systems protect personnel
and the underground installation.  Observation of the same infrastructure
variable can therefore contribute to different protection paths.

A cooling-water failure, for example, can first be detected as a service
deviation, cause the DCS to issue alarms, trigger a DSS action that removes
power from sensitive electronics, and, if accompanied by another
environmental consequence, also modify the access state.  As in Eq.~\eqref{eq:experiment_common_protection_structure},
the protection logic is consequently better represented as a coupled response than as a
single serial chain:
\begin{equation}
\text{infrastructure disturbance}
\rightarrow
\left\{
\begin{array}{l}
\mathrm{DCS}_{\rm ctrl},\\
{\cal L}_{\rm DSS}\rightarrow{\cal P}_{\rm DSS},\\
{\cal P}_{\rm pers},
\end{array}
\right.
\label{eq:experiment_infrastructure_protection_paths}
\end{equation}
with the applicable branch determined by the physical disturbance and the
consequence being controlled.

The infrastructure of ATLAS and CMS therefore provides another direct
example of the non-factorizable safe-domain concept developed throughout
this manuscript.  Electrical power, cooling, dry-gas systems, detector
gases, ventilation, fire detection, and emergency protection are shared
resources whose states can couple otherwise distinct detector systems.
Their availability and configuration must consequently be considered
together with ${\boldsymbol X}_{\alpha}$ and
${\boldsymbol\Xi}_{\alpha}$ when determining whether a particular physics,
movement, maintenance, or intervention state is admissible.

\subsection{Radiation environment, activation, and personnel access}
\label{subsec:atlas_cms_radiation_access}

Radiation protection provides another important coupling between the
physical state of the accelerator--experiment system and personnel
access.  In both ATLAS and CMS, the radiation environment depends on the
beam state, collision and loss conditions, accumulated luminosity,
detector and shielding configuration, material activation, and the time
elapsed since beam operation
\cite{CERN:RadiationProtectionCode,Ninin:2005LACS,
ATLAS:2008Detector,CMS:2008Detector}.

Two physically distinct radiation conditions must be separated.  During
beam operation, prompt radiation is generated by proton--proton
collisions, beam losses, secondary-particle transport, and interactions
with accelerator and detector material.  In beam-exposed underground
zones, personnel access is therefore prevented by the accelerator access
and safety systems.  After the beam has been removed, the prompt component
rapidly disappears, but radionuclides produced in detector, shielding,
beam-line, and accelerator components can continue to generate a residual
radiation field.

Consequently,
\begin{equation}
N_{\rm beam}=0
\quad\not\Longrightarrow\quad
\dot H^{*}(10)=0,
\label{eq:beamoff_not_zero_radiation}
\end{equation}
where $\dot H^{*}(10)$ denotes the ambient dose-equivalent rate, with the
value in parentheses specifying the reference depth of $10~{\rm mm}$ in
the ICRU sphere used for this operational radiation-protection quantity
\cite{ICRP:2021Publication147,Otto:2024ICRU95}.
Equation~\eqref{eq:beamoff_not_zero_radiation} expresses an important
operational distinction: removal of beam eliminates the dominant prompt
radiation source but does not remove the consequences of accumulated
activation.

The residual radiation field is spatially and temporally dependent,
\begin{equation}
\dot H_{\rm res}^{*}
=
\dot H_{\rm res}^{*}
\left(
10;
{\boldsymbol r},
t_{\rm cool}
\right),
\label{eq:residual_dose_rate_field}
\end{equation}
where ${\boldsymbol r}$ denotes position and $t_{\rm cool}$ the time
elapsed since the relevant irradiation period.  Its magnitude also
depends on the preceding luminosity and loss history, material
composition, shielding, and the experiment-specific configuration
${\boldsymbol\Xi}_{\alpha}$.  Thus, a radiological condition measured or
calculated for one shutdown, cooling time, or detector configuration
cannot automatically be transferred to another.

This dependence is especially important close to the interaction region.
Components near the beam line and in the forward regions can accumulate
substantial activation because they are exposed to intense secondary
particle fields.  Beam pipes, shielding elements, forward absorbers,
inner tracking structures, calorimeter components, and nearby services
can therefore become important sources of residual dose during shutdown
interventions.  Radiation-protection studies for the LHC experiments
accordingly rely on calculated and measured dose maps that are specific
to the detector configuration and the intervention being planned.

The radiation environment of ATLAS and CMS is not identical.  Differences
in detector geometry, material distribution, forward shielding, magnet
structures, and service layouts modify both the production and transport
of secondary radiation and the resulting residual fields.  These
differences should not be interpreted as a ranking of radiological
safety; rather, they imply different spatial dose distributions and
different intervention constraints for the two experiments.

The radiological state is therefore another component of the environment
that constrains the experiment-specific access domain.  Schematically,
\begin{equation}
\dot H^{*}(10,{\boldsymbol r})
\rightarrow
\text{radiological classification and work conditions}
\rightarrow
\Omega_{\rm access}^{\alpha},
\label{eq:radiation_access_coupling}
\end{equation}
where the access conditions can include restrictions on occupancy time,
required dosimetry, specific authorization, local shielding, or additional
radiation-protection measures.

The relation in Eq.~\eqref{eq:radiation_access_coupling} should not be
interpreted as a simple threshold decision based on one dose-rate value.
CERN radiation-area classification also considers the expected exposure
of personnel, occupancy, possible surface or airborne contamination, and
the type of work to be performed
\cite{CERN:RadiationProtectionCode}.  The radiological significance of a
location is therefore determined by the combination of the radiation
field and the planned activity.

For interventions in activated regions, the relevant protection principle
is optimization according to the ALARA principle~\cite{FassoRokni:2009ALARA}.  The work is prepared
so that individual and collective exposures are kept as low as reasonably
achievable while the intervention remains technically feasible.  Measures
can include delaying work to benefit from radioactive decay, minimizing
the duration of exposure, increasing distance from activated components,
installing temporary shielding, optimizing the sequence of tasks,
preparing tools and procedures in advance, and using remote handling where
appropriate.

This can be represented schematically as
\begin{equation}
\text{radiation map}
\rightarrow
\text{intervention planning}
\rightarrow
\left\{
\begin{array}{l}
t_{\rm cool},\\
\text{work duration},\\
\text{distance and shielding},\\
\text{work sequence},\\
\text{dosimetry and access conditions}
\end{array}
\right.
\rightarrow
\text{optimized exposure}.
\label{eq:alara_intervention_chain}
\end{equation}

The configuration dependence introduced earlier is especially relevant
here.  Opening or moving detector structures can expose components that
were previously shielded or inaccessible and can change the spatial
relation between activated material and personnel.  Thus, the transition
from a closed detector to an intervention configuration changes not only
the mechanical access geometry but potentially the radiation field
experienced at the workplace.

For CMS, the modular opening sequence discussed in
Sec.~\ref{subsec:cms_modular_safety} can therefore produce working regions
with radiological conditions different from those of the closed detector.
Similarly, major ATLAS openings and interventions can expose personnel to
activated beam-line, inner-detector, forward, shielding, or calorimeter
components that are inaccessible during normal operation.  Radiation
protection must therefore be evaluated for the actual intervention
configuration rather than for an abstract ``beam-off detector'' state.

Radiation monitoring and radiological surveys provide the observational
basis for these decisions.  Fixed monitors characterize selected
radiation fields during operation and shutdown, while dedicated
radiation-protection measurements are used to establish local dose-rate
conditions before and during interventions.  Individual or operational
dosimetry provides the corresponding measurement of personnel exposure
where required.

These functions should be distinguished from the DCS and DSS roles
described in Sec.~\ref{subsec:atlas_cms_common_protection}.  DCS and DSS
primarily supervise and protect detector equipment, whereas radiation
monitoring, radiological classification, dosimetry, and access
authorization belong principally to personnel and radiation protection.
Information can be exchanged among these systems, but their protection
objectives are different.

The access decision is consequently based on several simultaneous
conditions.  A beam-free state can be necessary, but the detector and
environment must also be compatible with the planned intervention, and
the applicable radiation-protection requirements must be satisfied.
Schematically,
\begin{equation}
{\boldsymbol X}_{\alpha}
\in
\Omega_{\rm int}^{\alpha}
\left(
z,w;
m,
{\boldsymbol\Xi}_{\alpha}
\right)
\quad\Longrightarrow\quad
\chi_{\rm access}^{\alpha}(z,w)=1,
\label{eq:radiation_access_conditions}
\end{equation}
where $\chi_{\rm access}^{\alpha}(z,w)$ is a binary access indicator:
$\chi_{\rm access}^{\alpha}=1$ denotes authorization for the specified
zone $z$ and activity $w$, whereas
$\chi_{\rm access}^{\alpha}=0$ denotes that the corresponding access is
not authorized.  The value $\chi_{\rm access}^{\alpha}=1$ should not be
interpreted as unrestricted permission to enter other regions or to
perform activities different from the one considered.

This formulation also makes clear why access authorization is
activity-dependent.  A region can be admissible for a short inspection
while requiring additional preparation for prolonged work close to an
activated component.  Conversely, a location with a relatively low
ambient dose rate can still require restrictions if the planned activity
involves handling activated material or a potential contamination source.

The relation between radiation and access can therefore be summarized as
\begin{equation}
\begin{aligned}
&\text{beam and collision history} \\
&\rightarrow \text{activation} \\
&\rightarrow \dot H_{\mathrm{res}}^{*}(10,\boldsymbol{r},t_{\mathrm{cool}}) \\
&\rightarrow \text{RP assessment and ALARA optimization} \\
&\rightarrow \boldsymbol{X}_{\mathrm{access}}.
\end{aligned}
\label{eq:radiation_access_summary}
\end{equation}

ATLAS and CMS thus provide another realization of the general
configuration-dependent safety framework.  Beam removal suppresses the
prompt-radiation hazard but does not erase the radiological history of the
installation.  Safe intervention consequently depends on the coupled beam,
detector, environmental, and access states, together with the accumulated
activation and the radiation-protection measures appropriate to the
actual detector configuration and work being performed.

\subsection{Personnel access as a detector-configuration-dependent variable}
\label{subsec:detector_personnel_access}

The preceding discussion shows that personnel access to ATLAS and CMS
cannot be determined from a single machine or detector condition.
Access to a given underground region depends on the simultaneous state
of the accelerator, detector, magnet systems, cryogenic and gas
installations, electrical and cooling services, radiation environment,
mechanical configuration, and the protection functions required for the
intended activity.  Personnel access is therefore a
configuration-dependent property of the coupled experimental system,
rather than an administrative condition imposed only after the physical
state has been established
\cite{Ninin:2005LACS,CERN:RadiationProtectionCode,
ATLAS:2008Detector,CMS:2008Detector}.

This conclusion follows directly from several examples considered above.
Removal of beam suppresses the prompt collision- and beam-related
radiation environment, but does not eliminate residual activation,
cryogenic inventories, energized detector systems, or mechanical
hazards.  Similarly, de-energization of an experiment magnet does not by
itself establish that a detector is ready for arbitrary access, while
opening a detector can remove geometrical barriers and simultaneously
introduce new mechanical, electrical, work-at-height, service, and
radiological conditions.

As established in Eq.~\eqref{eq:radiation_access_conditions}, access is
associated with both a specified zone $z$ and a specified activity $w$.
The actual experiment state must belong to the corresponding intervention
domain
$\Omega_{\rm int}^{\alpha}
(z,w;m,{\boldsymbol\Xi}_{\alpha})$
before the activity-specific access indicator
$\chi_{\rm access}^{\alpha}(z,w)$ can take the authorized value.
The intervention domain is therefore generally a restricted part of the
safe state space,
\begin{equation}
\Omega_{\rm int}^{\alpha}
\left(
z,w;
m,
{\boldsymbol\Xi}_{\alpha}
\right)
\subseteq
\Omega_{\rm safe}^{\alpha}
\left(
m,
{\boldsymbol\Xi}_{\alpha}
\right).
\end{equation}
The inclusion emphasizes that a state can be safe in a broader
operational sense without necessarily being compatible with a particular
intervention in a particular location.

The conditions defining
$\Omega_{\rm int}^{\alpha}$ are necessarily heterogeneous.  Depending on
the intervention, they can include
\begin{equation}
\left\{
\begin{array}{l}
\text{beam state compatible with access},\\
\text{magnet and powering state compatible with the work},\\
\text{movable structures stopped, positioned, and secured},\\
\text{required high- and low-voltage states established},\\
\text{cooling and gas systems in the prescribed configuration},\\
\text{cryogenic and atmospheric conditions compatible with occupancy},\\
\text{radiological conditions assessed and validated},\\
\text{required fire, gas, ODH, and other protection functions available}.
\end{array}
\right.
\label{eq:detector_access_condition_set}
\end{equation}
The precise subset of applicable conditions depends on the zone and the
work activity; Eq.~\eqref{eq:detector_access_condition_set} is therefore
a conceptual summary rather than a universal access checklist.

This formulation makes explicit why the access problem is
non-factorizable.  Satisfaction of one condition cannot compensate for
the failure of another.  In particular,
\begin{equation}
N_{\rm beam}=0
\quad\not\Longrightarrow\quad
\chi_{\rm access}^{\alpha}(z,w)=1.
\label{eq:beamoff_not_detector_access}
\end{equation}
Analogous statements apply to magnet, mechanical, radiological,
cryogenic, electrical, or service conditions considered individually.
Access is authorized only when the relevant combination of conditions is
compatible with the specified intervention.

The distinction between the physical state
${\boldsymbol X}_{\alpha}$ and the configuration vector
${\boldsymbol\Xi}_{\alpha}$ is particularly useful in this context.
The former contains instantaneous physical quantities---for example
currents, temperatures, pressures, gas concentrations, dose rates,
positions, and cooling flows---whereas the latter describes the technical
configuration in which those quantities are interpreted.  A measured
cooling flow can therefore be acceptable for one configuration but
insufficient for another, while a mechanical position admissible for
maintenance can be incompatible with detector movement or physics
operation.

The activity-specific indicator
$\chi_{\rm access}^{\alpha}(z,w)$ should also be distinguished from the
multivalued access status $\sigma_{\rm access}$ introduced in the
cryogenic discussion.  The latter describes the status of an area or
access system, whereas
$\chi_{\rm access}^{\alpha}(z,w)$ answers the more specific question of
whether access to zone $z$ for activity $w$ is authorized under the
prevailing conditions.  A general area status compatible with access does
not therefore imply authorization for every location or activity.

For CMS, the transition from the closed detector through a movement state
to the mechanically secured open configuration provides a particularly
clear example of this configuration dependence.  Access conditions during
detector motion differ from those applicable after the structures have
reached and been verified in their intervention positions.  Personnel
exclusion during motion is itself a protection barrier, while successful
completion of the movement does not authorize work until the resulting
mechanical configuration, connected services, and environmental
conditions have also been validated.

The same principle applies to ATLAS.  Although its large-scale mechanical
architecture differs from that of CMS, shutdown and upgrade activities can
involve opening, lifting, displacement, or repositioning of substantial
detector components and associated structures.  In addition, the
persistent liquid-argon inventory, large magnet systems, detector gases,
dry-CO$_2$ flushing, electrical services, cooling circuits, and radiation
environment continue to constrain access independently of whether
circulating beam is present.

The service infrastructure discussed in
Sec.~\ref{subsec:atlas_cms_services} further illustrates why access
authorization must refer to the actual configuration.  A detector
subsystem can be electrically isolated while nearby cooling, gas,
ventilation, or power-distribution equipment remains active.
Conversely, an intervention on a common service can modify the protection
conditions of detector systems that are not themselves being worked on.
The relevant intervention domain must therefore account for both the
local worksite and the coupled infrastructure affected by the work.

Radiation protection introduces an additional activity-dependent
constraint.  As discussed in
Sec.~\ref{subsec:atlas_cms_radiation_access}, residual dose rates after
beam operation depend on position, cooling time, preceding irradiation
history, shielding, and detector configuration.  A zone can therefore be
compatible with a short inspection while requiring additional ALARA
preparation, dosimetry, shielding, or work sequencing for a longer or more
intrusive intervention
\cite{CERN:RadiationProtectionCode,FassoRokni:2009ALARA}.

Access authorization should also be distinguished from DCS and DSS
functions.  The DCS supervises and controls detector and infrastructure
states, while the DSS can autonomously protect detector equipment against
selected abnormal conditions.  Their information and status can provide
essential prerequisites for an intervention, but neither system by itself
constitutes personnel-access authorization.  Personnel protection
additionally requires the applicable access, radiation-protection, fire,
gas, ODH, and work-control conditions to be satisfied.

The corresponding logical structure can therefore be represented
schematically as
\begin{equation}
\left(
{\boldsymbol X}_{\alpha},
{\boldsymbol\Xi}_{\alpha},
m,
z,
w
\right)
\longrightarrow
\text{validation of intervention conditions}
\longrightarrow
\chi_{\rm access}^{\alpha}(z,w).
\label{eq:detector_access_validation_chain}
\end{equation}
The intermediate validation step is essential: the access decision is not
inferred from one measured quantity, but from the compatibility of the
complete physical state and technical configuration with the intended
activity.

Access restriction can act either preventively or mitigatively, depending
on the stage of the hazard trajectory.  Preventing entry into a region
whose configuration is incompatible with the intended activity is
preventive with respect to personnel exposure.  Evacuating or restricting
an already occupied region after an abnormal condition has developed is
mitigative.  As elsewhere in the manuscript, this classification refers
to a specified consequence and hazard trajectory rather than being an
intrinsic property of the access system.

The transition back from an intervention state to physics operation
requires the inverse form of configuration validation.  Completion of the
work does not itself establish readiness for detector closure or beam
operation.  Temporary equipment, open services, modified protection
states, mechanical restraints, access equipment, or deliberately disabled
functions must be returned to the required configuration and verified
before the experiment can leave the intervention domain.

Personnel access is therefore embedded in a bidirectional configuration
cycle,
\begin{align}
&\text{physics configuration} \notag \\
&\longrightarrow \text{prepared intervention configuration} \notag \\
&\longrightarrow \text{authorized access and work} \notag \\
&\longrightarrow \text{validated restoration} \notag \\
&\longrightarrow \text{physics configuration}.
\label{eq:detector_access_configuration_cycle}
\end{align}
Each transition changes the set of admissible detector states and requires
the corresponding technical and protection conditions to be established.

Personnel access in ATLAS and CMS should therefore not be regarded as an
external constraint on detector operation.  Rather, the specific access
decision is derived from the coupled physical state, technical
configuration, operational mode, location, and intended activity.  The
same physical quantities that determine whether equipment can operate can
also determine whether personnel may enter, remain in, or work within a
given region.  The experiment-specific safe domain consequently includes
not only states compatible with data taking, but also deliberately
prepared configurations in which installation, maintenance, repair, and
upgrade activities can be performed safely.

\subsection{ATLAS and CMS: comparison of risk configurations}
\label{subsec:atlas_cms_risk_comparison}

The preceding examples show that ATLAS and CMS share many broad classes
of hazard while realizing them through substantially different detector
architectures.  A meaningful comparison should therefore not ask which
experiment is generically ``safer.''  Rather, it should identify how
common physical hazard families are transformed, by experiment-specific
states and configurations, into different initiating events, hazard
trajectories, consequence structures, protection requirements, and
intervention conditions.

At the level of broad hazard classes, one may schematically write
\begin{equation}
{\cal H}_{\rm common}
=
\left\{
\begin{array}{l}
\text{magnetic and stored-energy hazards},\\
\text{cryogenic and oxygen-displacement hazards},\\
\text{electrical and powering hazards},\\
\text{cooling and service failures},\\
\text{detector-gas and fire hazards},\\
\text{mechanical and movement hazards},\\
\text{prompt and residual radiation},\\
\text{personnel-access hazards}
\end{array}
\right\}.
\label{eq:atlas_cms_common_hazards}
\end{equation}
The presence of a common hazard family does not, however, imply a common
risk.  For a hazard class $h\in{\cal H}_{\rm common}$,
its realization depends on the physical state
${\boldsymbol X}_{\alpha}$, the
experiment-specific configuration
${\boldsymbol\Xi}_{\alpha}$, and the
corresponding operational or intervention
mode $m$.

This distinction is important in the risk framework introduced earlier.
The same broad hazard class can correspond in the two experiments to
different initiating-event frequencies or probabilities, different
conditional branches, and different consequence vectors
${\boldsymbol{\cal C}}_{j,\alpha}^{(h)}$.  Accordingly,
\begin{equation}
h\in{\cal H}_{\rm common}
\quad\not\Longrightarrow\quad
{\cal R}_{\rm ATLAS}^{(h)}
=
{\cal R}_{\rm CMS}^{(h)}.
\label{eq:common_hazard_not_common_risk}
\end{equation}
Equation~\eqref{eq:common_hazard_not_common_risk} should not be read as
asserting that the numerical risks must always differ.  It states instead
that equality cannot be inferred from the common physical origin of the
hazard alone.  The risk must be evaluated using the experiment-specific
hazard trajectories, protection functions, and consequences.

The comparison of consequence vectors requires the same care.  Since
${\boldsymbol{\cal C}}$ is multidimensional, the vectors
${\boldsymbol{\cal C}}_{j,\rm ATLAS}^{(h)}$ and
${\boldsymbol{\cal C}}_{j,\rm CMS}^{(h)}$ should not in general be ordered
by a scalar inequality.  A ranking is meaningful only after specifying
the relevant consequence components, a partial ordering, or the scalar
severity functional ${\cal S}({\boldsymbol{\cal C}})$ introduced in
Sec.~\ref{subsec:preventive_mitigative}.  This prevents an apparent
reduction of personnel, equipment, environmental, availability, and
scientific consequences to a single implicit scale.

ATLAS illustrates particularly clearly how detector architecture can
introduce a persistent cryogenic contribution to the risk configuration.
Its liquid-argon calorimeters retain a substantial cryogenic inventory
independently of the beam state.  Consequently, removal of beam does not
establish a safe liquid-argon state, as expressed by
Eq.~\eqref{eq:atlas_beamoff_cryo}.  Containment, relief, transfer,
ventilation, atmospheric monitoring, and access conditions remain relevant
during shutdown and intervention states.

CMS provides a complementary example through its highly modular
mechanical architecture.  Opening the detector deliberately changes
${\boldsymbol\Xi}_{\rm CMS}$ and introduces movement trajectories,
support conditions, mechanical clearances, connected-service constraints,
and personnel-exclusion requirements that are absent, or have different
importance, in the closed physics configuration.  The relevant distinction
is therefore not simply between an operating and a stopped detector, but
among different mechanically and technically prepared configurations, as
described in Sec.~\ref{subsec:cms_modular_safety}.

These examples should not be interpreted as assigning cryogenic hazards
to ATLAS and mechanical hazards to CMS exclusively.  Both experiments
contain superconducting magnets, electrical and cooling infrastructure,
gas systems, massive mechanical structures, radiation fields, and
configuration-dependent access conditions.  ATLAS also requires movement,
lifting, opening, and repositioning of substantial detector components,
while CMS also contains cryogenic systems associated with its
superconducting solenoid.  The comparison concerns the relative prominence
and coupling of hazard classes within different detector architectures,
not the exclusive ownership of particular hazards.

The different detector architectures are reflected
directly in their corresponding safe-state domains.
For broadly analogous operational or intervention
modes $m_{\rm ATLAS}$ and $m_{\rm CMS}$, one may write
schematically
\begin{equation}
\Omega_{\rm safe}^{\rm ATLAS}
\left(
m_{\rm ATLAS},
{\boldsymbol\Xi}_{\rm ATLAS}
\right)
\not\equiv
\Omega_{\rm safe}^{\rm CMS}
\left(
m_{\rm CMS},
{\boldsymbol\Xi}_{\rm CMS}
\right).
\label{eq:atlas_cms_different_safe_spaces}
\end{equation}
Here, $m_{\rm ATLAS}$ and $m_{\rm CMS}$ denote
experiment-specific realizations of broadly
comparable operational activities, rather
than identical physical or technical states.
The relation expresses the different
definitions and constraints of the
corresponding safe domains; it does not
constitute a ranking of the safety of
the two experiments.

The same architecture dependence carries over to intervention states.
As developed in Sec.~\ref{subsec:detector_personnel_access}, access to a
zone $z$ for an activity $w$ is authorized only when the actual detector
state belongs to the corresponding intervention domain
\begin{equation}
{\boldsymbol X}_{\alpha}
\in
\Omega_{\rm int}^{\alpha}
\left(
z,w;
m,
{\boldsymbol\Xi}_{\alpha}
\right),
\label{eq:atlas_cms_intervention_comparison}
\end{equation}
with the resulting decision represented by the previously defined
indicator $\chi_{\rm access}^{\alpha}(z,w)$~\footnote{
Throughout this work, the indices identifying a particular
experiment or accelerator facility are omitted from the
variables $z$, $w$, and $m$ for notational simplicity,
unless explicitly required. In general, these variables
are facility-dependent: they denote the zone, work activity,
and operational mode applicable to the experiment or
accelerator under consideration.}.  Because the detector
configurations, services, radiation fields, mechanical arrangements, and
persistent hazards differ, the intervention domains of ATLAS and CMS need
not coincide even for broadly analogous activities.

Common protection principles nevertheless remain visible beneath these
architectural differences.  Both experiments rely on monitoring,
interlocks, independent protection functions, validated technical states,
controlled access, radiation protection, and configuration management.
The DCS, DSS, and personnel-protection functions discussed above therefore
play analogous functional roles, even though the observables, thresholds,
actuators, and protected equipment to which they are applied are
experiment specific.

The comparison can consequently be summarized schematically as
\begin{equation}
\left(
{\cal H}_{\rm common},
m
\right)
+
\left(
{\boldsymbol X}_{\alpha},
{\boldsymbol\Xi}_{\alpha}
\right)
\longrightarrow
\left\{
\begin{array}{l}
\text{experiment-specific hazard trajectories},\\
\text{experiment-specific protection requirements},\\
\text{experiment-specific consequence vectors},\\
\text{experiment-specific safe and intervention domains}.
\end{array}
\right.
\label{eq:atlas_cms_architecture_risk_mapping}
\end{equation}
Here the symbol $+$ is not intended as an algebraic operation, but only as
a visual indication that the common hazard classes must be considered
together with the actual experiment state and configuration.

The principal lesson is therefore not that ATLAS and CMS possess
fundamentally different concepts of safety.  They implement the same
general principles---physical-state monitoring, DiD,
fail-safe protection, configuration validation, and controlled
access---within detector architectures that generate different couplings
among hazards and barriers.  Their safe operational phase spaces are
consequently different realizations of the same underlying framework.

This distinction is important for the broader argument of this
manuscript.  Safety cannot be inferred from a catalogue of hazards alone.
It emerges from the relation between physical processes, technical
configuration, protection functions, personnel occupancy, and the
trajectory followed by the system.  ATLAS and CMS provide complementary
examples of how the same general safety principles acquire different
physical realizations in large underground experiments.

\subsection{From detector safety to safety in physics}
\label{subsec:detector_safety_physics_transition}

The ATLAS and CMS examples complete the first direction of the argument
developed in this manuscript.  Physical understanding of stored energy,
cryogenic inventories, radiation fields, gas transport, electrical power,
cooling, mechanical motion, and detector response provides the basis for
identifying hazards and constructing protection functions.  At the same
time, the preceding discussion shows that the relation also operates in
the opposite direction: the conditions required for safe operation define
which experimental states, configurations, and transitions can actually
be used for physics.

This point follows naturally from the state-space description.  For an
experiment
$\alpha\in\{\mathrm{ATLAS},\mathrm{CMS}\}$, the instantaneous physical
state ${\boldsymbol X}_{\alpha}$ is interpreted together with the
technical configuration ${\boldsymbol\Xi}_{\alpha}$ and the applicable
operational mode $m$.  Safety therefore does not correspond to a single
fixed detector condition, but to membership in a configuration- and
mode-dependent admissible domain,
\begin{equation}
{\boldsymbol X}_{\alpha}(t)
\in
\Omega_{\rm safe}^{\alpha}
\left(
m,
{\boldsymbol\Xi}_{\alpha}
\right).
\label{eq:detector_safe_state_condition}
\end{equation}

Physics operation imposes additional requirements beyond generic safety.
Using the operational domain introduced in the general framework, one may
write
\begin{equation}
\Omega_{\rm operational}^{\alpha}
\left(
m,
{\boldsymbol\Xi}_{\alpha}
\right)
\subseteq
\Omega_{\rm safe}^{\alpha}
\left(
m,
{\boldsymbol\Xi}_{\alpha}
\right).
\label{eq:detector_operational_safe_relation}
\end{equation}
Thus, a detector state compatible with physics operation must also lie
within the corresponding safe domain, whereas a safe state need not be a
physics-ready state.  A detector can, for example, be safely parked,
de-energized, opened for intervention, maintained in a controlled
cryogenic configuration, or placed in another protected state without
being ready for collision data taking.

This distinction is important because safety and physics readiness are
not interchangeable concepts.  Safe shutdown, safe intervention, safe
commissioning, and safe physics operation correspond to different
regions of the accessible state space and can require different technical
configurations.  The transition between them is consequently part of the
physical and operational problem rather than an administrative change of
label.

The preceding discussion of personnel access provides a direct example.
For a zone $z$ and activity $w$, the intervention domain
$\Omega_{\rm int}^{\alpha}
(z,w;m,{\boldsymbol\Xi}_{\alpha})$
defines states compatible with that specific work, while
$\chi_{\rm access}^{\alpha}(z,w)$ represents the corresponding binary
authorization decision.  These intervention states need not coincide
with those required for physics operation.  Conversely, a detector
configuration optimized for data taking can deliberately exclude
personnel from regions that are accessible only after the experiment has
been transferred to an intervention configuration.

The distinction also applies to detector protection.  The DCS, DSS, and
personnel-protection functions discussed above do not merely respond after
a hazardous condition develops.  By determining which combinations of
powering, cooling, gas, cryogenic, mechanical, radiation, and access
conditions are admissible, they participate in defining the operational
boundary within which detector physics can proceed.  Protection therefore
acts not only on abnormal trajectories but also on the set of trajectories
that may be entered intentionally.

This may be expressed schematically as
\begin{equation}
\left(
{\boldsymbol X}_{\alpha},
{\boldsymbol\Xi}_{\alpha},
m
\right)
\longrightarrow
\text{validation of applicable constraints}
\longrightarrow
\left\{
\begin{array}{l}
\text{physics operation},\\
\text{commissioning or technical operation},\\
\text{prepared intervention},\\
\text{protected or recovery state}.
\end{array}
\right.
\label{eq:detector_mode_validation}
\end{equation}
The alternatives in
Eq.~\eqref{eq:detector_mode_validation} represent classes of admissible
operation rather than mutually exclusive universal detector states.
Their precise realization is experiment- and configuration-dependent.

The ATLAS--CMS comparison makes this especially clear.  The different
detector architectures discussed in
Sec.~\ref{subsec:atlas_cms_risk_comparison} lead to different safe and
intervention domains, but the underlying principle is the same.  In
ATLAS, persistent cryogenic inventories, the magnet systems, detector
services, radiation conditions, and mechanical configurations constrain
the transition toward both access and physics operation.  In CMS, the
large-scale modular mechanical configuration adds particularly explicit
closed, moving, and open intervention states.  Neither example can be
reduced to a single ``detector ready'' variable.

The same argument applies to transitions back toward physics operation.
Completion of maintenance or disappearance of an alarm does not by itself
establish physics readiness.  Mechanical configurations must be restored
and verified, temporary intervention conditions removed, required services
returned to their prescribed states, protection functions made available,
and the resulting detector state validated against the conditions
applicable to the intended operational mode.  Physics readiness is
therefore a validated state of the coupled system rather than the absence
of a known fault.

In this sense, safety constraints do not simply remove undesirable states
from an otherwise independently defined physics programme.  They help
define the operational phase space in which that programme is physically
realizable.  The safe domain determines which states and trajectories may
be occupied, while the scientific objective selects, within that domain,
the subset appropriate for data taking, commissioning, calibration,
maintenance, or other activities.

This provides the transition to the second direction of the central theme
of this manuscript.  The preceding sections have shown how physics enters
safety through quantitative models, measurements, monitoring, and
protection systems.  The discussion that follows considers the converse:
how safety requirements influence detector design, commissioning,
operational modes, access planning, maintenance, availability, and
ultimately the practical organization of experimental physics.

\section{Safety in physics}
\label{sec:safety_in_physics}

The preceding sections have developed the first direction of the relation
between physics and safety: physical laws, quantitative models,
measurements, and engineering knowledge make it possible to identify
hazards, construct protection systems, and define admissible operating
conditions.  The discussion of the LHC, ATLAS, and CMS also reveals the
converse relation.  Safety requirements do not merely act after the
scientific system has been designed; they influence which detector
configurations can be constructed, commissioned, operated, accessed,
maintained, and ultimately used for physics.

This second direction is what we refer to here as \emph{safety in
physics}.  The expression does not mean that safety replaces scientific
criteria or determines the scientific objectives of an experiment.
Rather, it means that scientific activity is carried out within a
physically and technically admissible domain whose boundaries are partly
defined by protection requirements.  Within this domain, experimental
goals determine which operating points are scientifically useful; outside
it, a state or trajectory may be physically accessible but cannot be
accepted for routine operation or a specified intervention.

The distinction developed in the preceding sections between the physical
state ${\boldsymbol X}$, the technical configuration
${\boldsymbol\Xi}$, the operational mode $m$, and the corresponding
admissible domains is particularly useful in this context.  Safety does
not generally impose a single numerical limit on experimental activity.
Instead, it constrains combinations of beam conditions, magnet states,
cryogenic inventories, detector powering, cooling and gas systems,
mechanical configurations, radiation conditions, personnel occupancy, and
the availability of the required protection functions.  Consequently,
the admissible region can change when the operational mode or technical
configuration changes.

This configuration dependence has an immediate consequence for the
practice of experimental physics.  Physics operation, commissioning,
calibration, detector opening, maintenance, upgrade work, and recovery
from abnormal conditions correspond to different regions of the
accessible state space and therefore require different combinations of
technical and protection conditions.  A state that is appropriate for
one activity need not be admissible for another.  In particular, the
absence of beam, the de-energization of a magnet, or the completion of a
mechanical movement does not by itself establish readiness for either
personnel intervention or physics operation.

Safety therefore enters scientific work through the control of both
states and transitions.  Commissioning is typically staged because
stored energy, beam intensity, detector voltage, cryogenic inventory, or
other relevant parameters cannot always be increased independently of the
validation of the corresponding protection functions.  Maintenance and
upgrade activities require deliberately prepared intervention
configurations, while restoration to physics operation requires
verification that temporary conditions introduced for the intervention
have been removed and that the required technical and protection states
have been re-established.  The trajectory between configurations is thus
as important as the endpoint.

This perspective also clarifies the relation between safety and
performance.  Protection requirements can restrict the immediately
accessible operating space, but this should not be interpreted simply as
a loss of scientific freedom.  Reliable operation at high beam intensity,
high stored energy, strong magnetic field, or demanding detector
conditions is possible precisely because these regimes are embedded in a
validated protection architecture.  Safety constraints can therefore
limit particular trajectories while enabling sustained access to much
more demanding regions of experimental performance.

The relevant interaction is consequently bidirectional.  Scientific
objectives motivate new machine and detector configurations, higher
performance, new operational modes, and more ambitious interventions.
These changes modify the accessible state space and can introduce new
hazard trajectories or alter existing ones.  Safety analysis and
operational experience then feed back into the design of protection
functions, procedures, commissioning sequences, and admissible
configurations.  The resulting framework is not static: it evolves with
the facility and with the scientific programme.

The discussion in this section therefore moves from the physical
description of hazards and barriers to the way in which safety shapes the
practice of physics itself.  The emphasis is on design choices,
commissioning strategy, operational modes, access and maintenance,
configuration control, availability, and organizational decision making.
The central point is that safety should not be regarded as a boundary
external to scientific activity.  In a complex facility such as the LHC
and its experiments, it is one of the conditions that make sustained,
repeatable, and progressively more demanding physics operation possible.

\subsection{Commissioning as staged validation of the operational domain}
\label{subsec:safety_commissioning}

Commissioning provides one of the clearest examples of how safety
requirements shape the practice of physics.  A complex accelerator or
detector is not taken directly from an installed state to its full
performance configuration.  Instead, progressively more demanding
operating conditions are admitted only after the relevant physical
response, monitoring, protection functions, and recovery mechanisms have
been demonstrated to behave as required.  Commissioning is therefore not
merely preparation for physics operation; it is the process by which
parts of the physically accessible state space are validated for
intentional use.

In the state-space language developed above, a commissioning programme
can be represented schematically as a sequence
\begin{equation}
\left(
{\boldsymbol X}_{n},
{\boldsymbol\Xi}_{n},
m_{n}
\right)
\xrightarrow{\;\text{validated commissioning step}\;}
\left(
{\boldsymbol X}_{n+1},
{\boldsymbol\Xi}_{n+1},
m_{n+1}
\right),
\label{eq:commissioning_stage_transition}
\end{equation}
where $n$ labels a commissioning stage,
${\boldsymbol X}_{n}$ denotes the corresponding physical state,
${\boldsymbol\Xi}_{n}$ its technical configuration, and $m_{n}$ the
applicable operational mode.  Equation~\eqref{eq:commissioning_stage_transition}
is schematic: a commissioning step can involve a change in physical
parameters, technical configuration, operational mode, or several of
these simultaneously.

The important requirement is not that a later
commissioning stage be ``larger'' than an earlier
one in any scalar sense. Rather, each transition
must remain compatible with the applicable safe
domain and with the protection constraints
relevant to the new configuration.
In terms of the trajectory formulation introduced
earlier, the commissioning trajectory can be
represented as
\begin{equation}
\Gamma_{\rm comm}:
t\mapsto\boldsymbol X(t),
\qquad
\Gamma_{\rm comm}\in\mathcal G_{\rm adm}.
\label{eq:commissioning_allowed_trajectory}
\end{equation}
Here, $\Gamma_{\rm comm}$ denotes the individual
time-ordered trajectory followed by the coupled
facility state during the commissioning interval,
while $\mathcal G_{\rm adm}$ is the set of
trajectories satisfying the applicable physical,
protection, and operational constraints.

The membership condition expresses that not only
the intended commissioning endpoint, but also
the complete sequence of intermediate states and
transitions through which it is reached, must
remain admissible. In particular, the applicable
safe domain and protection requirements may change
as equipment is commissioned, operating parameters
are increased, or new technical functions are
progressively enabled.

This distinction is important because commissioning often changes the
hazard potential of the system.  Increasing beam intensity or energy,
raising magnet current, powering additional detector channels, introducing
high voltage, changing a cryogenic inventory, enabling gas systems, or
moving large mechanical structures can open new physical trajectories or
increase the consequences associated with already accessible ones.
Progression to a new stage therefore requires validation of the protection
functions appropriate to that stage rather than extrapolation from the
previous one.

For the LHC, this principle is particularly visible in beam commissioning.
Low-intensity or setup-beam conditions allow parts of the machine,
collimation, beam-loss monitoring, interlock, and beam-dump protection
chain to be established and tested before substantially larger stored beam
energies are admitted
\cite{Bordry:2001MPS,Schmidt:2016MPS}.  Such conditions reduce the
potential consequences of selected failure scenarios but should not be
interpreted as hazard-free states.  In particular, classification as a
setup or low-intensity beam does not imply that the corresponding
consequence vector ${\boldsymbol{\cal C}}$ vanishes.

The same logic applies to superconducting and cryogenic systems.
Commissioning at progressively different currents, temperatures,
pressures, or cryogenic configurations tests not only whether the intended
technical performance can be reached, but also whether quench detection,
energy extraction, pressure relief, monitoring, and associated protection
functions remain compatible with the relevant state.  A successful test
at one operating point therefore validates that condition and the
associated transition; it does not automatically validate every
physically accessible higher-energy state.

Detector commissioning follows an analogous pattern.  ATLAS and CMS
subsystems are progressively brought into operating configurations through
controlled changes of powering, high and low voltage, cooling, gas,
cryogenic, magnet, readout, and protection states
\cite{ATLAS:2008Detector,CMS:2008Detector}.  The required sequence depends
on the subsystem and activity.  The essential point is that a detector
element can be technically functional while the coupled experiment is not
yet ready for the intended physics mode.  Cooling, environmental
conditions, connected services, protection functions, and configuration
compatibility must be established together.

Commissioning therefore provides an operational realization of the
distinction between the safe and operational domains discussed above.
A state can be safe without yet being commissioned for physics, while a
physics configuration can be admitted only after the relevant part of the
operational domain has been validated.  Commissioning progressively
provides the evidence required to justify intentional occupation of such
states.

The validation concerns more than nominal functionality.  Protection
timing, interlock paths, diagnostic coverage, fail-safe responses,
communication among coupled systems, and observability of relevant
failures can all determine whether progression is justified.  The
principles developed in
Secs.~\ref{subsec:protection_time}--\ref{subsec:testing_observability}
therefore apply directly to commissioning: a protection function that is
present in the design but has not been shown to respond appropriately in
the applicable configuration cannot simply be assumed to provide the
required protection.

Commissioning is also intrinsically a learning process.  Measurements made
during staged operation can reveal couplings, transients, tolerances, or
failure responses that were imperfectly represented in the design model.
Such observations can lead to changes in thresholds, interlock logic,
operating procedures, technical configuration, or the definition of the
admissible domain.  The resulting feedback can be summarized
schematically as
\begin{align}
&\text{commissioning observations}  \notag \\
&\longrightarrow \text{updated physical and protection models}  \\
&\longrightarrow \text{revised operational constraints}.  \notag
\label{eq:commissioning_feedback}
\end{align}
This feedback is an important example of the two-way relation between
physics and safety: physical measurements validate the protection model,
while the resulting safety constraints determine which subsequent
physical states may be explored.

The staged character of commissioning also provides a natural response to
uncertainty.  When the behaviour of a coupled system has not yet been
demonstrated under a particular condition, operation can initially be
restricted to a domain in which the credible consequences are smaller or
recovery is more readily controlled.  Additional states are admitted only
after the relevant assumptions and protection functions have been tested.
This does not require the commissioning domains to form a mathematically
nested sequence; changes of mode and configuration can alter the
applicable constraints as well as the accessible performance.

Commissioning should therefore not be regarded as a temporary relaxation
of the requirements applicable to normal operation.  In many cases the
opposite is true: because system behaviour is still being established,
commissioning requires additional diagnostics, restricted configurations,
lower stored energy, increased observation, or dedicated recovery
conditions.  The allowable trajectory during commissioning can thus be
more constrained than the trajectory permitted after the system has been
fully validated.

At the same time, these restrictions are enabling rather than merely
limiting.  They provide a controlled path from an installed or modified
system toward increasingly demanding scientific operation.  High beam
intensity, large stored magnetic energy, complex detector powering, and
tightly coupled experimental configurations become usable not because
their hazards disappear, but because commissioning establishes evidence
that the corresponding states and transitions can be controlled within
the validated protection architecture.

Commissioning therefore lies directly at the interface between
\emph{physics in safety} and \emph{safety in physics}.  Physical models
and measurements are used to test the behaviour of the machine and its
protection systems, while the outcome of those tests determines which
states and trajectories become available for subsequent scientific
operation.  The operational phase space is not assumed in advance; it is
progressively demonstrated, validated, and, where necessary, revised.

\subsection{Access constraints and operational planning}
\label{subsec:safety_access_operational}

Personnel access provides a second direct example of how safety
requirements shape the practice of physics.  As developed in
Sec.~\ref{subsec:detector_personnel_access}, access is not determined by
the geographical location alone, nor by a single condition such as the
absence of beam.  Authorization is associated with a specified zone $z$
and activity $w$, and is represented by the previously defined indicator
$\chi_{\rm access}^{\alpha}(z,w)$.  The corresponding intervention domain
$\Omega_{\rm int}^{\alpha}
(z,w;m,{\boldsymbol\Xi}_{\alpha})$
therefore depends on the physical state, technical configuration,
operational mode, and intended work.

This formulation has an immediate operational consequence.  A state that
is compatible with personnel intervention need not be a state appropriate
for physics operation, and conversely a physics-ready configuration need
not permit the required intervention.  In general,
\begin{equation}
\Omega_{\rm int}^{\alpha}
\left(
z,w;
m,
{\boldsymbol\Xi}_{\alpha}
\right)
\not\equiv
\Omega_{\rm operational}^{\alpha}
\left(
m,
{\boldsymbol\Xi}_{\alpha}
\right).
\label{eq:access_operational_domains}
\end{equation}
The symbol $\not\equiv$ emphasizes that the two domains represent
different admissibility conditions; it does not imply that they must be
disjoint.  Some technical activities can coexist with selected forms of
operation, whereas others require the experiment or accelerator to be
placed in a dedicated intervention configuration.

Access therefore often requires a deliberate transition away from the
configuration optimized for physics.  Depending on the location and the
work to be performed, preparation can involve beam inhibition, changes of
magnet or detector powering, isolation of electrical or fluid services,
changes of gas or cooling configuration, mechanical opening or securing,
radiation assessment, establishment of ventilation or atmospheric
conditions, and validation of the required protection functions.  The
particular combination is activity dependent and should not be interpreted
as a universal sequence.

The distinction between the multivalued area status
$\sigma_{\rm access}$ and the activity-specific Boolean indicator
$\chi_{\rm access}^{\alpha}(z,w)$ remains important here.
The former describes the status of the relevant access system or area,
whereas the latter expresses whether the specified work in zone $z$ is
authorized under the prevailing conditions.  Consequently,
\begin{equation}
\sigma_{\rm access}
=
\sigma_{\rm access}^{\rm admissible}
\quad\not\Longrightarrow\quad
\chi_{\rm access}^{\alpha}(z,w)=1
\quad
\text{for arbitrary }z,w .
\label{eq:access_status_not_universal_authorization}
\end{equation}
An area can therefore be generally accessible while a particular
intervention remains incompatible with the detector, radiation,
electrical, cryogenic, or mechanical conditions applicable to that work.

The operational importance of this distinction becomes particularly clear
during technical stops and shutdowns.  Access to one region can require
changes in services that affect other detector subsystems, while an
intervention on a common cooling, gas, power-distribution, ventilation, or
cryogenic system can modify the admissible conditions in regions remote
from the immediate worksite.  Access planning must therefore take account
of coupled infrastructure rather than treating each intervention as a
strictly local activity.

The same coupling is relevant to simultaneous work.  Two interventions
that are individually admissible need not remain admissible when performed
together.  Their coexistence can modify personnel occupancy, evacuation
conditions, electrical isolation boundaries, mechanical exclusion zones,
service availability, fire load, gas configuration, or the set of
protection functions that must remain active.  Operational planning must
therefore consider the compatibility of concurrent activities as part of
the overall configuration rather than assuming independent access
permissions.

Access restrictions consequently influence the temporal organization of
experimental work.  Maintenance, repair, inspection, calibration, and
upgrade activities are assigned to periods in which the corresponding
intervention configurations can be established without conflicting with
higher-priority operational requirements.  Some work can be performed
during short technical stops, while other interventions require extended
shutdown conditions, detector opening, radiation cool-down, service
reconfiguration, or large-scale mechanical preparation.  The duration of
an intervention is therefore not determined only by the time needed to
perform the technical task itself.

In particular, the operational cost of an intervention includes the
preparation and restoration of the surrounding system.  The
configuration cycle described in
Eq.~\eqref{eq:detector_access_configuration_cycle} makes this explicit:
physics operation is followed by preparation of an intervention
configuration, authorization and execution of the work, and validated
restoration before the experiment can return to the required physics
state.  Completion of the local task is only one stage of this cycle.

The restoration step is safety-relevant for the same reason as the
preparation step.  Temporary equipment must be removed or placed in its
prescribed state, electrical and fluid connections restored as required,
mechanical structures secured, temporary access conditions closed,
disabled or bypassed functions returned to their required configuration,
and the relevant detector and protection systems revalidated.  A return
to physics operation is therefore not inferred simply from the end of
personnel presence in the work area.

This point also connects access planning with commissioning.  After a
major intervention or configuration change, restoration can require
partial recommissioning before the corresponding operational state is
again admitted.  The commissioning principles developed in
Sec.~\ref{subsec:safety_commissioning} then apply: the relevant physical
response, observability, interlocks, protection functions, and coupled
services must be demonstrated for the restored configuration before the
associated operational domain is treated as validated.

Access requirements can therefore affect accelerator and detector
availability, but safety and availability should not be identified.
Restricting access, delaying restoration, or requiring additional
validation can reduce the immediately available time for physics, while
at the same time preventing an uncontrolled transition that could produce
a much longer interruption or more serious consequences.  Conversely,
well-designed access systems, modular detector configurations, reliable
isolation procedures, and efficient validation can reduce intervention
time without weakening the required protection conditions.

The relation is consequently not a simple opposition between safety and
physics time.  Operational planning seeks trajectories through the
available configurations that satisfy both the scientific programme and
the applicable safety constraints.  In schematic form,
\begin{align}
&\text{scientific and technical objectives} \notag \\
&\longrightarrow
\text{planned configuration transitions} \notag \\
&\longrightarrow
\text{validated operation and intervention} \\
&\longrightarrow
\text{restored physics configuration}. \notag
\label{eq:access_operational_planning_chain}
\end{align}
Each transition must remain compatible with the corresponding admissible
state and trajectory conditions established elsewhere in this manuscript.

Personnel access thus illustrates the broader meaning of
\emph{safety in physics}.  Safety requirements influence when an
experiment can be opened, which activities can be performed
simultaneously, how services must be configured, when the system can be
restored, and when physics operation can resume.  These constraints do
not merely interrupt scientific operation; they define the controlled
transitions by which maintenance and modification can be performed while
preserving the capability for sustained future physics.

\subsection{Maintenance as preservation and restoration of validated states}
\label{subsec:safety_maintenance}

Maintenance provides another direct connection between safety and the
long-term practice of experimental physics.  Accelerator, detector, and
infrastructure systems do not remain indefinitely in the state in which
they were originally commissioned.  Components age, calibrations drift,
mechanical interfaces change, electrical contacts degrade, cooling and
ventilation performance can evolve, sensors can lose accuracy, and
protection channels themselves require periodic verification.  Sustained
physics operation therefore depends not only on the initial validation of
a safe operational domain, but also on maintaining the physical and
technical conditions on which that validation was based.

In this sense, maintenance is not external to the state-space framework
developed above.  An intervention deliberately transfers the system from
an operational configuration to a configuration in which inspection,
repair, replacement, adjustment, or testing can be performed.  For an
experiment $\alpha$, this transition may be represented schematically as
\begin{equation}
\left(
{\boldsymbol X}_{\alpha}^{\rm pre},
{\boldsymbol\Xi}_{\alpha}^{\rm pre}
\right)
\longrightarrow
\left(
{\boldsymbol X}_{\alpha}^{\rm maint},
{\boldsymbol\Xi}_{\alpha}^{\rm maint}
\right)
\longrightarrow
\left(
{\boldsymbol X}_{\alpha}^{\rm post},
{\boldsymbol\Xi}_{\alpha}^{\rm post}
\right),
\label{eq:maintenance_state_transition}
\end{equation}
where the superscripts denote the pre-intervention, maintenance, and
post-intervention conditions.  The intermediate maintenance state is not
in general compatible with physics operation; it is instead a deliberately
prepared intervention configuration subject to the access and work
conditions discussed in
Sec.~\ref{subsec:safety_access_operational}.

The transition in
Eq.~\eqref{eq:maintenance_state_transition} should not be interpreted as
requiring
${\boldsymbol\Xi}_{\alpha}^{\rm post}
=
{\boldsymbol\Xi}_{\alpha}^{\rm pre}$.
Maintenance can include replacement, modification, upgrade, or
reconfiguration of equipment, so that the validated post-maintenance
configuration can differ from the one that existed before the
intervention.  What is required is that the resulting physical state and
technical configuration satisfy the conditions applicable to the
intended subsequent mode of operation.

For a return to physics operation, this requirement can be expressed as
\begin{equation}
{\boldsymbol X}_{\alpha}^{\rm post}
\in
\Omega_{\rm operational}^{\alpha}
\left(
m_{\rm phys},
{\boldsymbol\Xi}_{\alpha}^{\rm post}
\right),
\label{eq:maintenance_return_operational}
\end{equation}
where $m_{\rm phys}$ denotes the relevant physics operating mode.
Completion of the maintenance task itself is therefore not sufficient.
The post-intervention state must be validated against the operational
conditions appropriate to the restored or modified configuration.

This distinction is particularly important when maintenance affects a
protection function.  Replacement of a sensor, repair of an interlock
channel, modification of a cooling circuit, reconnection of a power
system, or intervention on a mechanical restraint can change the
conditions on which previous validation was based.  Functional testing,
diagnostic checks, calibration, and, where necessary, partial
recommissioning are therefore part of the transition back toward
operation rather than optional activities performed after the repair.

The barrier framework developed in
Sec.~\ref{sec:hazards_barriers} makes this role explicit.  Barrier
properties such as failure-on-demand probability, response time,
independence, diagnostic capability, and testability can themselves
degrade with time.  Maintenance is one of the mechanisms by which these
properties are kept within their validated range.  Periodic testing also
provides evidence that latent failures have not accumulated in functions
that may remain inactive during normal operation but are required on
demand.

The same argument applies to diagnostic coverage.  A protection channel
can remain apparently available while its ability to reveal internal
failure has degraded.  Maintenance of sensors, diagnostics, communication
paths, actuators, and associated test functions therefore contributes not
only to equipment reliability but also to preservation of the
observability assumptions represented by quantities such as
${\rm DC}_k$.  A nominally functioning component and a validated
protection function are not necessarily equivalent concepts.

Maintenance can consequently play different roles with respect to
different hazard trajectories.  Preventive maintenance can reduce the
probability that degradation develops into an initiating event or barrier
failure.  Corrective maintenance restores functionality after a detected
fault or loss of capability.  At the same time, the maintenance activity
itself can introduce temporary hazards through electrical isolation,
opened containment, modified gas or cooling configurations, mechanical
movement, temporary equipment, altered fire load, or the deliberate
inhibition of selected functions.  The preventive or mitigative
classification therefore remains consequence- and trajectory-dependent,
as emphasized earlier in the manuscript.

The temporary configuration created for maintenance deserves particular
attention.  A system placed in a maintenance state is not simply the
normal system with one component switched off.  Isolation boundaries,
temporary services, removed shielding, opened detector structures,
mechanical supports, access equipment, ventilation conditions, and
disabled or bypassed protection functions can produce a configuration
with its own admissible domain.  This is why the activity-specific
intervention domain
$\Omega_{\rm int}^{\alpha}
(z,w;m,{\boldsymbol\Xi}_{\alpha})$
introduced above is more appropriate than a generic statement that the
facility is ``safe for maintenance.''

Large maintenance campaigns also illustrate the importance of coupled
systems.  An intervention on one subsystem can modify the conditions of
another through shared electrical distribution, cooling, ventilation,
cryogenic infrastructure, detector gases, mechanical supports, or common
protection systems.  Maintenance planning must therefore consider the
state of the surrounding installation and the compatibility of concurrent
activities, not only the component on which the work is performed.

For ATLAS and CMS, this coupling becomes especially visible during
shutdown periods.  Detector opening, access to internal components,
changes of cooling or gas configuration, work on magnet or cryogenic
systems, electrical interventions, and radiation-protection requirements
can coexist within the same period.  The resulting configuration differs
substantially from that used for data taking and can evolve repeatedly as
successive activities are prepared, performed, and completed.  Safe
maintenance therefore depends on configuration control throughout the
intervention rather than only on the initial establishment of access.

The return from maintenance deserves the same level of attention as the
preparation for the work.  Temporary equipment must be removed or
accounted for, mechanical structures placed in their required
configuration, electrical and fluid connections restored as appropriate,
temporary isolations and bypasses reviewed, and the required monitoring
and protection functions returned to service.  Depending on the scope of
the intervention, the subsequent validation can range from a local
functional test to a broader recommissioning sequence of the type
described in Sec.~\ref{subsec:safety_commissioning}.

This creates a natural maintenance cycle,
\begin{align}
&\text{validated operation}\notag \\
&\longrightarrow
\text{prepared intervention}\notag \\
&\longrightarrow
\text{maintenance activity} \\
&\longrightarrow
\text{post-maintenance validation} \notag \\
&\longrightarrow
\text{validated operation}. \notag
\label{eq:maintenance_validation_cycle}
\end{align}
The final state need not reproduce the initial configuration exactly, but
it must be demonstrated to satisfy the conditions required for the next
intended mode.

Maintenance also generates information.  Repeated failures, unexpected
wear, calibration drift, inspection findings, or recurrent faults can
reveal deficiencies in assumptions made during design or commissioning.
Such observations can lead to changes in inspection intervals,
replacement strategies, monitoring thresholds, spare-component policies,
protection logic, technical configuration, or the definition of the
admissible operating domain.  Maintenance history therefore provides a
feedback path from operational experience to the physical and engineering
models used to manage the facility.

This feedback can be summarized schematically as
\begin{align}
&\text{maintenance observations}\notag \\
&\longrightarrow
\text{updated knowledge of degradation and failure modes} \\
&\longrightarrow
\text{revised maintenance and protection strategy}.\notag
\label{eq:maintenance_feedback}
\end{align}
Maintenance is thus not only a means of restoring equipment after loss of
function.  It is also a continuing source of empirical information about
how the real system evolves with time.

The relation between maintenance and physics availability is therefore
more subtle than a direct loss of operating time.  Maintenance consumes
time that could otherwise be available for data taking, but insufficient
maintenance can increase the probability of failures, prolonged
unplanned interventions, or degradation of protection functions.
Conversely, well-planned preventive maintenance can reduce future
interruptions and preserve the validated conditions required for
high-performance operation.

Maintenance consequently represents another form of \emph{safety in
physics}.  Safety requirements determine how equipment can be accessed,
isolated, modified, tested, and returned to service, while maintenance
preserves the physical and technical conditions that make future physics
operation possible.  The long-term scientific capability of a facility
therefore depends not only on reaching demanding operating states, but on
retaining the ability to validate and restore those states throughout the
lifetime of the machine and its experiments.

\subsection{Simultaneous activities and coordination of safety}
\label{subsec:simultaneous_activities}

The operation of a large experimental facility involves
activities that may take place simultaneously in different
areas or within the same technical infrastructure.
At ATLAS and CMS, detector maintenance, infrastructure
interventions, equipment testing, calibration,
commissioning, and preparation for physics operation
may overlap in time. Although each activity can be
individually authorized and performed under appropriate
safety conditions, their coexistence may introduce
additional hazards or invalidate assumptions on which
the individual authorizations were based.
Consequently, the safety of simultaneous activities
cannot generally be established by considering each
intervention independently~\cite{Leveson:2004AccidentModel,Perrow:1999NormalAccidents}.

This problem is particularly relevant in underground
experimental areas, where personnel access, electrical
distribution, cooling and ventilation systems, gas
installations, detector services, and protection
infrastructure are strongly interconnected.
An intervention affecting a shared technical system
may modify the conditions under which another activity
is performed, even when the corresponding work areas
are physically separated. For example, a modification
of ventilation conditions may affect an intervention
involving gas installations, while electrical isolation
or restoration may influence equipment being tested
elsewhere. Similarly, detector movements, lifting
operations, or changes in access restrictions may
alter the safety conditions of personnel working
in neighboring areas.

Within the state-space description introduced above,
simultaneous activities are represented through their
combined influence on the facility state
$\boldsymbol X_\alpha$, the configuration
$\boldsymbol\Xi_\alpha$, and the applicable operational
mode $m$. The relevant safety question is therefore
not only whether each activity is individually
permitted, but whether their combined execution
preserves the conditions defining the admissible
facility state and its evolution.

In particular, two activities that are separately
compatible with the applicable safety requirements
need not remain compatible when performed together.
Their interaction may introduce new initiating events,
modify the possible propagation of hazardous
conditions, or reduce the effectiveness of preventive
and mitigative barriers~\cite{Leveson:2012SaferWorld}. This illustrates the
importance of the coupled-system perspective developed
in Sec.~\ref{sec:hazards_barriers}: the consequences
of an intervention depend not only on its immediate
technical objective, but also on the configuration
and operating conditions of the surrounding facility.

The coordination of simultaneous activities therefore
requires a common understanding of the current
facility configuration, the technical dependencies
between interventions, and the protection functions
that must remain available. Work authorization,
access control, operational supervision, and
communication between responsible teams provide
complementary means of maintaining this consistency.
The authorization of an individual activity should
not be interpreted as an unconditional permission
independent of subsequent changes in the facility
state or in other ongoing activities.

An important distinction must also be made between
the authorization of personnel access and the
authorization of the work itself. The satisfaction
of the access conditions represented by
$\chi_{\rm access}^\alpha(z,w)$ establishes whether
entry into a particular area is permitted under
the applicable conditions. It does not, by itself,
establish that every planned intervention is
compatible with other activities or with the
configuration of the facility. The latter requires
verification of the relevant technical and
organizational constraints, including the conditions
associated with the intervention domain
$\Omega_{\rm int}^\alpha(z,w;m,\boldsymbol\Xi_\alpha)$.

The dynamic character of simultaneous activities
is equally important. An intervention that is
compatible with the initial facility configuration
may cease to be admissible following a change in
equipment status, the beginning of another activity,
or the restoration of a previously
isolated system~\cite{Rasmussen:1997RiskManagement}.
Coordination must therefore extend beyond initial
work authorization to include the management of
configuration changes, communication of modified
conditions, and, when necessary, suspension or
rescheduling of activities whose continued execution
would no longer satisfy the applicable safety
requirements~\cite{Hollnagel:2006Resilience}.

This approach is particularly significant during
technical stops and shutdown periods, when the
density and diversity of interventions increase
substantially. Activities associated with different
detector subsystems and infrastructure services
may be organized according to separate technical
objectives, but their safety implications must be
evaluated within the common facility environment.
The resulting coordination is not simply an
administrative requirement; it constitutes an
operational mechanism for preserving the integrity
of the protection system while allowing multiple
technical tasks to proceed.

Simultaneous-activity management thus provides
another example of safety influencing the practical
organization of experimental physics. By determining
which activities may coexist, which must be
separated in time or space, and which require
additional protective conditions, safety constraints
shape the scheduling and execution of detector
interventions. Conversely, the physical understanding
of technical dependencies, energy sources, and
hazard propagation provides the basis for identifying
potentially incompatible activities.

In this sense, coordination contributes directly
to maintaining the facility within its admissible
operational domain. It enables maintenance,
commissioning, and detector preparation to proceed
without compromising the safety conditions required
for subsequent restoration of normal operation
and physics data taking.

\subsection{Technical and procedural safety barriers}
\label{subsec:technical_procedural_barriers}

The safety of accelerator experiments relies on the combined
application of technical protection systems and procedural
controls. Although these measures differ in their physical
implementation and mechanisms of enforcement, they contribute
to the same objective: preventing hazardous conditions,
limiting their consequences, and maintaining the facility
within the admissible safety domain. Their complementarity
becomes particularly important during commissioning,
maintenance, and simultaneous interventions, when the
configuration of the experimental infrastructure may differ
substantially from that required for routine physics operation.

Technical barriers act through engineered functions whose
operation is determined by the state of equipment, measured
physical quantities, or predefined protection conditions.
Examples include electrical protection and isolation devices,
gas detection and alarm systems, ventilation-related safety
functions, emergency shutdown arrangements, and access
interlocks. Depending on their design, such systems may
prevent the initiation of a hazardous event, interrupt its
development, or mitigate its consequences. Their effectiveness
depends on the availability of the required equipment, the
validity of the monitored conditions, and the integrity of
the associated protection functions.

Procedural barriers operate through prescribed activities,
organizational responsibilities, and controlled sequences
of actions. They include intervention planning, work
authorization, verification of isolation conditions,
communication between technical teams, operational
handover, and confirmation that equipment has been
restored to its authorized configuration. Unlike
automatically enforced technical barriers, procedural
measures generally depend on correct human execution,
the availability of reliable information, and effective
coordination between the persons responsible for an
intervention.

The distinction between these categories does not imply
that technical and procedural barriers are independent.
A technical protection system may require procedural
verification before an intervention, while the validity
of a procedural authorization may depend on the status
of specific technical protections. For example, an
electrical intervention may require physical isolation
of the relevant equipment, verification of the absence
of hazardous voltage, and protection against unintended
re-energization. The isolation arrangement provides
a technical means of controlling the energy source,
whereas the corresponding procedures establish how
the isolation is implemented, verified, maintained,
and eventually removed.

Within the facility-state description developed in
the preceding sections, these complementary measures
contribute to controlling the evolution of the physical
state $\boldsymbol X_\alpha$ under the applicable
configuration $\boldsymbol\Xi_\alpha$ and operational
mode $m$. Technical barriers constrain particular
physical processes or equipment states, while procedural
controls govern the activities and configuration changes
through which those processes may be initiated or
modified. Neither category alone necessarily guarantees
that the conditions defining an admissible intervention
domain remain satisfied.

This distinction is especially relevant when personnel
perform interventions within the domain
$\Omega_{\rm int}^{\alpha}
(z,w;m,\boldsymbol\Xi_\alpha)$.
The authorization of access and the authorization of
a specific intervention remain separate decisions.
Even when the applicable access conditions are
satisfied, the intervention may require additional
technical isolations, operational restrictions,
or verification steps. Conversely, the completion
of an intervention does not automatically establish
that the equipment can be returned to normal operation;
the relevant protection functions and configuration
conditions must first be restored and, where required,
validated.

At ATLAS and CMS, the relationship between technical
and procedural barriers is particularly important
because interventions frequently involve detector
subsystems connected to shared infrastructure.
Electrical distribution, cooling circuits, gas
installations, ventilation, and detector services
may be affected by activities performed in different
areas or under different technical responsibilities.
A protection measure established for one intervention
may therefore depend on conditions modified by another
activity. The coordination requirements discussed
in the preceding subsection are consequently part
of maintaining the effectiveness of the overall
barrier arrangement, rather than merely an
administrative addition to technical protection.

The reliability of procedural barriers also requires
particular attention. Their effectiveness may be
compromised by incomplete information, ambiguous
responsibilities, incorrect assumptions about
equipment status, or changes in operating conditions
that have not been communicated. Such limitations
are especially consequential when a procedure is
expected to compensate for an unavailable technical
protection. The general safety-engineering principles
discussed in Refs.~\cite{Rasmussen:1997RiskManagement,
Leveson:2004AccidentModel} emphasize that accidents
may emerge from interactions between technical
systems, human actions, and organizational controls,
even when individual components or activities appear
to satisfy their respective requirements.

A practical illustration is provided by the protection
of detector electronics against cooling failures in
ATLAS and CMS. In ATLAS, the Detector Safety System
DSS independently monitors conditions relevant to
equipment safety and can initiate automatic protective
actions when abnormal conditions are detected
\cite{Beltramello:2009DSS}. Similarly, the CMS
electromagnetic calorimeter employs a dedicated
safety system with hardwired interlocks capable of
disconnecting the relevant power supplies in response
to cooling failures or excessive temperatures
\cite{DiCalafiori:2020CMS_ECAL}.
These arrangements illustrate the importance of
engineered protection functions that do not depend
on the timely intervention of an operator.

Consider, for example, a hypothetical maintenance
intervention during which the cooling supply to a
detector subsystem is interrupted while its electronics
remain powered. Even if the intervention has been
formally authorized, incomplete communication between
the maintenance team and detector operators could
leave the equipment in a configuration incompatible
with the assumed cooling conditions. A procedural
instruction requiring an operator to recognize the
problem and switch off the affected equipment would
then introduce additional dependencies on information
availability, communication, and human response time.
By contrast, an appropriately configured automatic
interlock can initiate the required protective action
when the monitored conditions reach predefined limits.
Procedural measures remain essential for coordinating
the intervention, establishing the required initial
conditions, and verifying that the cooling and
protection functions have been restored before
re-energization. The example therefore illustrates
that technical and procedural barriers are complementary
but cannot be regarded as interchangeable.

Accordingly, procedural measures should not be
considered interchangeable with engineered
protection functions. In particular, where a hazard
requires rapid detection or intervention beyond
practical human response capabilities, a procedural
instruction cannot substitute for an appropriately
designed automatic protection system. Procedural
controls remain essential, however, for establishing
the conditions under which such systems operate,
managing authorized exceptions, and ensuring
controlled transitions between equipment states.
Any temporary modification or unavailability of
a protection function must therefore be assessed
against the applicable safety requirements rather
than treated as acceptable solely because additional
instructions have been issued.

The same consideration applies to the independence
of protection measures. The presence of several
technical or procedural barriers does not necessarily
provide independent protection if their effectiveness
depends on a common power supply, communication
channel, configuration assumption, or organizational
decision. The identification of such dependencies
is essential when determining whether the combined
barrier arrangement provides the intended protection,
as discussed in Sec.~\ref{sec:hazards_barriers}.

The restoration of normal operating conditions
provides a further illustration of this
complementarity. Following maintenance or
commissioning, procedural checks establish
whether the required interventions have been
completed, isolations removed where appropriate,
temporary arrangements resolved, and operational
responsibilities transferred. Technical verification
then provides evidence that the relevant equipment
and protection functions have returned to their
required states. The transition to a subsequent
operational mode is admissible only when the
corresponding safety conditions have been established.

Technical and procedural barriers thus perform
distinct but interconnected functions within the
safety architecture of large experimental facilities.
Their combined implementation connects physical
protection mechanisms with the organizational
processes required to configure, operate, and
maintain complex detectors. From the perspective
of the present work, this relationship illustrates
how physical understanding determines the
requirements of engineered protection, while
safety constraints shape the procedures through
which experimental activities can be performed.

\subsection{Competence, training, and authorization}
\label{subsec:competence_training}

The effectiveness of procedural safety barriers depends not only on
the existence of appropriate instructions and authorization
mechanisms, but also on the competence of the personnel responsible
for implementing them. In a complex experimental facility,
interventions may require an understanding of the physical hazards,
the applicable protection functions, the current technical
configuration, and the possible consequences of modifying that
configuration. Competence, training, qualification, and authorization
therefore constitute essential elements of the organizational
conditions under which experimental activities can be performed.

These concepts have distinct roles. Training provides the knowledge
and practical preparation required for particular activities,
whereas competence concerns the demonstrated ability to apply
that knowledge under relevant operating conditions. Qualification
establishes that specified competence requirements have been met,
while authorization grants permission to perform a defined activity
under prescribed conditions. Authorization consequently depends not
only on the qualifications of the individual but also on the
characteristics of the planned work, the responsibilities assigned
to the personnel involved, and the current state of the facility.
The management of interventions at CERN illustrates the importance
of integrating these requirements with operational preparation
and work coordination~\cite{Garino:2013IMPACT}.

At CERN, these requirements are supported by an extensive
programme of safety training, combining introductory
courses with specialized instruction adapted to particular
hazards, activities, and operational responsibilities.
The programme includes classroom and e-learning training,
with dedicated instruction in areas such as radiation
protection, electrical safety, and the management of
occupational hazards. The development of specific courses
for supervisors and safety officers further illustrates
the importance attributed to organizational responsibilities
and competence in the management of complex scientific
facilities. The scope and continuing development of these
activities are documented in CERN's annual reports
\cite{CERN:AnnualReport2023,CERN:ProgressReport2024}.
Such training provides a common foundation for personnel
working in the accelerator complex and its experiments,
while task-specific qualifications and authorizations
remain necessary for interventions involving specialized
equipment or protection systems.

Within the state-space framework developed in the preceding sections,
personnel competence and authorization do not constitute additional
physical coordinates of the facility state
$\boldsymbol X_\alpha$. Instead, they enter the organizational
conditions under which particular interventions and configuration
transitions are permitted. For a specified zone $z$, activity $w$,
operational mode $m$, and technical configuration
$\boldsymbol\Xi_\alpha$, the intervention domain
$\Omega_{\rm int}^{\alpha}
(z,w;m,\boldsymbol\Xi_\alpha)$
describes the admissible physical and technical conditions.
Satisfaction of these conditions is necessary, but does not
independently establish that a particular individual is qualified
and authorized to perform the corresponding task.

The distinction is important because permission to enter an
experimental area, represented by the access indicator
$\chi_{\rm access}^{\alpha}(z,w)$ introduced above, does not
necessarily confer authorization to manipulate the equipment
located there. An individual may be permitted to enter a detector
cavern without being qualified to perform electrical isolation,
operate lifting equipment, modify gas or cooling installations,
or change the configuration of a protection system. Such activities
may require additional technical qualifications, specific
authorization, supervision, or the participation of designated
specialists. The admissibility of an intervention therefore depends
on both the state of the installation and the personnel-related
conditions governing its execution.

This distinction can be illustrated by maintenance activities
at ATLAS and CMS. During an intervention involving detector
electronics, establishing the required electrical and cooling
conditions may involve personnel responsible for different
technical systems. The operator authorized to control detector
power need not be authorized to modify the cooling installation,
while personnel performing mechanical work may not be qualified
to alter electrical isolation or protection arrangements.
Similarly, operations involving large detector structures,
cryogenic systems, or radiation-controlled equipment require
competence appropriate to the particular physical hazards and
technical configuration. The different detector architectures
discussed in Sec.~\ref{sec:atlas_cms_safety} therefore imply
different distributions of technical responsibilities, even
though the underlying principles of competence and authorization
are common.

An additional consideration concerns the composition of the
teams responsible for safety assessment and operational
supervision. In large scientific facilities such as the LHC
and its underground experiments, it is particularly
advantageous for these teams to include personnel with
direct experience in accelerator or experimental
high-energy physics, alongside professionals specializing
in occupational safety, radiation protection, and other
relevant safety disciplines. Familiarity with detector
operation, beam conditions, superconducting magnets,
cryogenic infrastructure, and the practical organization
of experimental activities can substantially improve
the identification of system-specific hazards and the
interpretation of abnormal technical conditions.
Such expertise is especially valuable when evaluating
interactions between subsystems, assessing unconventional
interventions, or determining whether a proposed change
of configuration remains compatible with the established
protection requirements. Conversely, specialized safety
professionals contribute essential knowledge of hazard
assessment, protection methodology, regulatory requirements,
and organizational controls. The combination of these
competencies provides a more comprehensive basis for
safety decisions than either background alone, consistent
with the systems-oriented treatment of safety in complex
technical installations
\cite{Leveson:2004AccidentModel,
Rasmussen:1997RiskManagement}.

The importance of competence becomes particularly evident during
transitions between configurations. An intervention may require
equipment isolation, temporary modification of services,
verification of mechanical stability, or restoration of
protection functions before operation can resume. The personnel
involved must understand not only the actions assigned to them,
but also the boundaries of their responsibility and the conditions
under which the intervention must be suspended or referred to
another competent authority. Effective communication and
handover are consequently part of maintaining continuity between
the assumptions used to authorize the work and the actual state
of the facility.

The relationship to the preceding discussion of technical and
procedural barriers is direct. Training and competence contribute
to the reliability of procedural measures by reducing the
probability of incorrect actions, incomplete verification, or
misinterpretation of equipment status. Their effectiveness,
however, remains dependent on the available information,
organizational arrangements, and working conditions.
Consequently, neither competence nor training can guarantee
error-free execution or compensate unconditionally for an
unavailable technical protection function. This interpretation
is consistent with systems-oriented approaches to safety, in
which human actions, technical equipment, and organizational
conditions are treated as interacting elements rather than
independent sources of protection
\cite{Rasmussen:1997RiskManagement,
Leveson:2004AccidentModel}.

Competence requirements must also evolve with the facility.
Detector upgrades, modified protection systems, new operating
modes, changes in access conditions, and the introduction of
new technical procedures can invalidate assumptions underlying
previously acquired experience. Training and authorization
therefore require appropriate review when the technical
configuration or the responsibilities associated with an
activity change. In particular, successful experience with
an earlier configuration does not automatically establish
competence to perform an intervention involving substantially
modified equipment or protection functions.

This consideration is especially relevant during major
shutdowns, when extensive maintenance, consolidation,
installation, and commissioning activities may involve
personnel with different specializations and organizational
responsibilities.
The ongoing third long shutdown (LS3) of the CERN
accelerator complex, together with the continued
development of CERN personnel-safety systems,
provides a particularly relevant context in which
intervention management, technical coordination,
personnel competence, and protection arrangements
play essential roles
\cite{Tock:2025LS3,Evrard:2025CERNPersonnelSafety}.
These activities require not only the availability of
appropriately trained personnel, but also a consistent
definition of responsibilities and authorization conditions
throughout successive phases of work.

Competence, training, and authorization consequently influence
which operational trajectories can be realized in practice.
A technically feasible intervention may remain inadmissible
until the necessary personnel qualifications, responsibilities,
supervision, and work authorization have been established.
Conversely, the availability of competent personnel allows
the facility to undertake interventions, commissioning
activities, and configuration changes whose execution would
otherwise introduce unacceptable uncertainty or uncontrolled
risk.

In this sense, personnel competence is not an external
administrative attribute added to an otherwise complete
technical safety system. It contributes to maintaining the
organizational conditions under which technical protection
and procedural barriers remain effective. Within the
perspective of \emph{safety in physics}, training and
authorization therefore help determine not only who may
perform an activity, but also which experimental
configurations and transitions can be safely established,
modified, and validated.

\subsection{Radiation protection and optimisation of interventions}
\label{subsec:radiation_work_planning}

Radiation protection provides a particularly instructive example
of how safety requirements influence not only whether an experimental
activity may be performed, but also its preparation, organization,
and execution. As discussed in
Sec.~\ref{subsec:atlas_cms_radiation_access}, residual activation
following accelerator operation produces a spatially and temporally
dependent radiation environment. An intervention may therefore be
technically feasible while its immediate execution, duration,
location, or sequence would result in avoidable occupational
exposure. Radiation protection consequently introduces additional
constraints on the admissible intervention configurations and
trajectories.

The relevant radiological quantity is not simply the ambient
dose-equivalent rate at a particular location, but the exposure
accumulated by personnel throughout the activity. For a worker
$i$, the estimated external occupational dose associated with
a planned intervention ${\cal A}$ may be represented schematically
as
\begin{equation}
D_i^{\rm plan}({\cal A})
=
\int_{t_0}^{t_1}
\dot D_i^{\rm est}
\left[
{\boldsymbol r}_i(t),t;
{\boldsymbol\Xi}_{\alpha}(t)
\right]
dt,
\label{eq:individual_work_dose}
\end{equation}
where ${\boldsymbol r}_i(t)$ denotes the position of the worker
during the intervention, ${\boldsymbol\Xi}_{\alpha}(t)$ the
relevant detector and infrastructure configuration, and
$\dot D_i^{\rm est}$ the estimated occupational dose rate
appropriate to the planned exposure conditions. Here
$D_i^{\rm plan}$ denotes a dose estimate expressed in an
appropriate radiation-protection quantity, rather than
absorbed dose in the general physical sense.

The distinction from the ambient dose-equivalent rate
$\dot H^{*}(10)$ introduced previously is important.
Ambient dose-equivalent measurements provide information
about the radiation environment and contribute to the
assessment of occupational exposure, but the dose
received by a particular worker also depends on the
individual trajectory, exposure geometry, radiation
field, and applicable dosimetric quantities. The
interpretation of occupational dose therefore requires
the corresponding radiation-protection framework
\cite{CERN:RadiationProtectionCode,
ICRP:2021Publication147}.

For an intervention involving several workers, the
corresponding planned collective dose may be written as
\begin{equation}
D_{\rm coll}^{\rm plan}({\cal A})
=
\sum_{i=1}^{N_{\rm w}}
D_i^{\rm plan}({\cal A}),
\label{eq:collective_work_dose}
\end{equation}
where $N_{\rm w}$ is the number of exposed workers
participating in the activity. This quantity characterizes
the sum of the estimated individual doses associated with
the intervention, but does not by itself describe the
distribution of exposure among personnel. In particular,
a reduction of collective dose must not be achieved at
the expense of unacceptable individual exposure.

Equations~\eqref{eq:individual_work_dose} and
\eqref{eq:collective_work_dose} illustrate why radiation
protection is inherently a problem of intervention
planning. The occupational exposure depends not only
on the residual radiation field, but also on the time
spent in different locations, the number of workers,
the sequence of operations, the detector configuration,
and the protective measures implemented during the work.
Depending on the activity, possible optimization measures
include allowing additional radioactive decay before
access, reducing the duration of work in higher-dose
regions, increasing distance from activated components,
installing temporary shielding, preparing tools and
procedures in advance, or employing remote handling
where technically appropriate.

These considerations form the basis of the ALARA
principle, under which occupational exposures are
maintained as low as reasonably achievable, taking
account of relevant technical, economic, and
organizational factors
\cite{FassoRokni:2009ALARA,
CERN:RadiationProtectionCode}.
The objective is not an unconstrained mathematical
minimization of collective dose. Rather, intervention
planning seeks a configuration and sequence of
activities that satisfy the required radiological
protection conditions while achieving the intended
technical objective.

Within the state-space framework of the present work,
radiation-protection optimization can therefore be
understood as the selection of an admissible intervention
trajectory. For a specified zone $z$, activity $w$,
operational mode $m$, and technical configuration
${\boldsymbol\Xi}_{\alpha}$, the intervention must remain
compatible with the corresponding domain
$\Omega_{\rm int}^{\alpha}
(z,w;m,{\boldsymbol\Xi}_{\alpha})$.
Radiological constraints contribute to defining this
domain, while the planned duration, personnel movements,
and sequence of operations determine the exposure
accumulated along the trajectory. A configuration that
permits a short inspection may consequently be
incompatible with a prolonged intervention unless
additional protection measures are established.

An illustrative situation arises during maintenance
of detector components close to the beam line or
within the forward regions of ATLAS and CMS. These
components may have accumulated activation during
preceding accelerator operation, while their
accessibility depends on the particular detector
geometry and shutdown configuration. Consider a
planned intervention requiring the replacement or
inspection of an activated component. Performing the
work immediately after beam operation may result in
higher occupational exposure than performing it after
an appropriate cooling interval. The intervention may
also be reorganized so that preparatory operations,
tool assembly, and functional checks are completed
outside the higher-dose area, reducing the time
required for direct access. Depending on the
configuration, temporary shielding or remote
handling may provide further reductions in exposure.

This example illustrates a direct connection between
physical knowledge and the organization of scientific
work. The radiation field results from particle
interactions, material activation, and radioactive
decay, whereas the intervention strategy depends on
the predicted or measured evolution of that field.
The same technical objective can therefore be
achieved through different operational trajectories
with different radiological consequences. The
preferred trajectory is determined by the applicable
radiation-protection requirements and the feasibility
of the corresponding protective measures.

Radiation-protection planning is also closely connected
to the coordination of simultaneous activities discussed
in Sec.~\ref{subsec:simultaneous_activities}. Changes
in detector geometry, movement of shielding elements,
or interventions involving activated material may modify
the radiological conditions of neighboring work areas.
Consequently, the dose assessment prepared for one
activity may require reconsideration if another
intervention changes the configuration or exposure
conditions on which the original assessment was based.
The management of such dependencies is part of the
broader intervention-coordination problem
\cite{Garino:2013IMPACT}.

The competence and authorization requirements discussed
in Sec.~\ref{subsec:competence_training} are equally
relevant. Personnel must understand the radiological
conditions associated with their tasks, the applicable
work restrictions, the significance of dosimetric
information, and the circumstances requiring
reassessment or suspension of an intervention.
Radiation-protection specialists provide the necessary
expertise in exposure assessment and optimization,
while detector and accelerator experts contribute
knowledge of the installation, its activated components,
and the technical requirements of the planned work.
Their collaboration is particularly important when
radiological constraints must be reconciled with
complex detector configurations or specialized
maintenance procedures.

Finally, planned exposure and actual exposure must
be distinguished. Radiological surveys, operational
dosimetry, and work supervision provide information
that can be compared with the assumptions used during
intervention planning. Significant changes in measured
conditions or deviations from the authorized work
sequence may require reassessment before the activity
continues. The resulting operational experience can
also improve subsequent dose estimates, work procedures,
and intervention strategies.

Radiation protection thus provides a quantitative
illustration of the central relationship developed
in this manuscript. Physical models and measurements
determine the radiation environment and its evolution,
while the resulting safety requirements constrain the
timing, geometry, personnel organization, and
configuration of experimental interventions. In this
sense, ALARA optimization does not merely limit
scientific activity; it helps determine how the
required work can be performed within an admissible
operational trajectory, while minimizing avoidable
occupational exposure.

\subsection{Configuration control and the return to physics}
\label{subsec:return_to_physics}

The completion of an intervention does not, by itself,
establish that an accelerator or experimental detector
is ready to resume physics operation. Maintenance,
repair, installation, and upgrade activities may alter
the physical state of equipment, its technical
configuration, the availability of protection functions,
and the conditions governing personnel access.
Consequently, the return to operation requires a
controlled transition from the intervention configuration
to a validated state compatible with the intended
operational mode.

This requirement is particularly important following
major technical stops  (TSs) and LSs, during which
temporary installations, electrical isolations,
modified cooling or gas connections, opened detector
structures, displaced shielding, newly installed
electronics, and changes to control or interlock
configurations may have been introduced. Although
individual interventions can be completed and verified
locally, their combined effects must also be considered
before the corresponding accelerator or detector
systems are returned to operation. The management of
such dependencies forms an essential part of
intervention coordination
\cite{Garino:2013IMPACT,Tock:2025LS3}.

The maintenance-state transition introduced in
Eq.~\eqref{eq:maintenance_state_transition} provides
the appropriate description of this process.
For an experiment $\alpha$, the post-intervention
physical state ${\boldsymbol X}_{\alpha}^{\rm post}$
and technical configuration
${\boldsymbol\Xi}_{\alpha}^{\rm post}$ need not
coincide with their pre-intervention values.
Replacement of components, modification of services,
detector upgrades, or changes to protection
arrangements may intentionally establish a new
configuration. The relevant requirement is therefore
not the reproduction of the previous state, but
validation of the resulting configuration against
the conditions applicable to the next operational
mode.

For physics operation, this requirement has already
been expressed in
Eq.~\eqref{eq:maintenance_return_operational}
through the domain
$\Omega_{\rm operational}^{\alpha}
(m_{\rm phys},{\boldsymbol\Xi}_{\alpha}^{\rm post})$.
Membership in this domain must be supported by
appropriate evidence that the required technical,
physical, and protection conditions have been
established. Moreover, the transition toward this
state must itself remain admissible. Two individually
acceptable configurations do not necessarily imply
that every intermediate configuration, or every
sequence of operations connecting them, is permitted.

Configuration control consequently requires
identification of the changes introduced during
the intervention and verification of their effects
on the surrounding installation. This includes
not only the equipment directly modified, but also
its interfaces with electrical distribution,
cooling, gas and cryogenic systems, ventilation,
mechanical structures, monitoring, and protection
functions. An intervention that has no apparent
effect on the nominal performance of one subsystem
may nevertheless modify assumptions underlying
the protection of another.

The restoration of protection functions deserves
particular attention. A protection system verified
before an intervention cannot automatically be
assumed to remain valid after modifications to
sensors, cabling, power distribution, communication
paths, software, interlock logic, or associated
equipment. Depending on the scope of the changes,
restoration may require functional tests,
calibration, diagnostic verification, confirmation
of fail-safe responses, or tests involving several
interconnected subsystems. The required evidence
must correspond to the actual post-intervention
configuration rather than relying exclusively on
the successful operation of the previous one.

For ATLAS and CMS, the distinction between normal
detector control and autonomous equipment protection
is particularly relevant. DCS provide the functions required to monitor
and establish operational configurations, while
dedicated detector safety functions can initiate
protective actions under abnormal conditions.
Restoring the control system to a nominal operating
state does not, by itself, demonstrate that the
associated protection functions have been verified
in the modified configuration
\cite{Boterenbrood:2004ATLASDCS,
Schmeling:2004DSS,CMS:2009DCS,
DiCalafiori:2020CMS_ECAL}.

Consider, for example, the replacement of detector
electronics or an intervention involving their
cooling infrastructure. After completion of the
mechanical and electrical work, the affected
equipment may appear operational from the
perspective of its control system. Nevertheless,
the cooling conditions, temperature monitoring,
power-removal functions, and relevant protection
interlocks must remain compatible with the restored
configuration. Where modifications have affected
these functions or their interfaces, appropriate
verification is required before the equipment can
be returned to its intended operating state.
This illustrates why the completion of a technical
task and validation of the corresponding protected
configuration are distinct stages.

The same principle applies at the level of the
complete experiment. In ATLAS, the restoration
of operational conditions must account for the
relevant magnet, cryogenic, liquid-argon,
mechanical, detector-service, and protection
configurations. In CMS, the return from an opened
or otherwise modified detector configuration
may additionally involve large-scale mechanical
closure, verification of the positions and
restraints of movable structures, and restoration
of the associated services. Although the specific
sequences differ because of the detector
architectures, the underlying requirement is
identical: the resulting coupled configuration
must be validated for the intended operational
mode
\cite{ATLAS:2008Detector,CMS:2008Detector,
ATLAS:2024Run3,CMS:2024Run3}.

Personnel access constitutes another important
part of this transition. During maintenance,
access may be authorized under conditions that
are incompatible with normal physics operation.
The subsequent restoration of operational
conditions therefore requires the appropriate
termination or reassessment of intervention
authorizations, verification of personnel
occupancy where applicable, and establishment
of the access and protection states required
for the next mode. Closing an intervention
or restoring an access system to its normal
configuration does not independently establish
readiness for beam operation. The relevant
machine, detector, environmental, and personnel
protection conditions must remain mutually
compatible
\cite{Ladzinski:2009LHCAccess,
Evrard:2025CERNPersonnelSafety}.

The transition from a shutdown configuration
to physics operation also need not occur in
a single step. Depending on the modifications
performed, it may be necessary to restore
infrastructure services, verify safety-critical
systems, commission individual detector
subsystems, establish the required magnet and
cryogenic conditions, and progressively enable
the operating configurations needed for data
taking. At the accelerator level, the restoration
of beam operation similarly depends on the
validation of the relevant machine-protection
functions and the authorized progression through
successive operating states
\cite{Schmidt:2016MPS,Wenninger:2016MPO}.

This staged restoration is closely related to
the commissioning trajectory discussed in
Sec.~\ref{subsec:safety_commissioning}. The
difference is primarily the starting point:
commissioning may establish the operational
properties of newly installed equipment,
whereas post-intervention validation must
demonstrate that an existing or modified
installation remains compatible with the
required protection conditions. In both cases,
the progression is conditional on the successful
verification of the relevant functions.
The fact that a configuration was previously
validated does not eliminate the need to
reassess those parts of the system affected
by subsequent modifications.

An important consequence is that readiness
for physics operation cannot generally be
represented by a single equipment status or
by the absence of active alarms. Equipment
may be functioning normally while its
associated protection conditions remain
incomplete, or individual subsystems may
be ready while their combined operation
has not yet been validated. The relevant
decision therefore concerns the compatibility
of the complete physical state
${\boldsymbol X}_{\alpha}^{\rm post}$,
technical configuration
${\boldsymbol\Xi}_{\alpha}^{\rm post}$,
and intended operational mode $m_{\rm phys}$
with the applicable safety and operational
requirements.

The return to physics consequently constitutes
the final stage of a controlled intervention
cycle rather than an automatic consequence
of completed maintenance. Configuration
records, functional tests, protection-system
verification, operational handover, and,
where necessary, recommissioning provide
the evidence required to authorize this
transition. They also establish the
reference conditions against which subsequent
operation, anomalies, and future interventions
can be evaluated.

From the perspective developed in this work,
configuration control provides another
illustration of the reciprocal relationship
between physics and safety. Physical
understanding and experimental measurements
are required to demonstrate that the restored
or modified system behaves as intended,
while safety requirements determine the
sequence and conditions under which the
facility may return to scientific operation.
The ability to restore a validated operational
configuration is therefore an essential
component of the sustainable physics
capability of large accelerator experiments.

\subsection{Near misses, anomalies, and operational learning}
\label{subsec:near_miss_learning}

The development of safety knowledge in a complex experimental
facility does not depend exclusively on major accidents or
events involving significant consequences. Operational
anomalies, unexpected equipment behaviour, protection-system
interventions, near misses, and difficulties encountered
during routine activities may provide valuable information
about the physical and organizational conditions under
which the installation operates. Such observations are
particularly important in highly coupled systems, where
a deviation originating in one subsystem may reveal
dependencies that were not fully represented in the
original hazard analysis.

It is useful to distinguish several categories of
operational observations. An anomaly is a departure
from expected behaviour that does not necessarily
constitute a hazardous event. A near miss is an event
in which an adverse consequence was narrowly avoided,
whether through the action of a protection barrier,
human intervention, or other circumstances. An actual
protection demand provides information about the
response of a barrier to the conditions that activated
it, whereas a spurious protection action may reveal
limitations in sensing, diagnostic functions, signal
transmission, or the definition of protection
thresholds. These categories may overlap, but they
should not be treated as equivalent because their
implications for the assessment of safety are different.

The value of such observations follows directly from
the hazard-chain framework introduced in
Sec.~\ref{sec:hazards_barriers}. An initiating event
$F_i$ may give rise to several propagation trajectories
$\Gamma_j^{(i)}$, depending on the facility state,
the technical configuration, and the response of
preventive or mitigative barriers. An operational event
that does not produce an adverse consequence may
nevertheless reveal information about the initiating
mechanism, an intermediate state transition, or the
performance of a barrier along one of these trajectories.
In particular, the absence of damage does not
necessarily imply that the initiating event was
insignificant or that the existing protection
arrangement provides a sufficient safety margin.

This distinction is especially important when
interpreting successful protection actions. The
observation that an interlock, shutdown function,
or other barrier has interrupted an abnormal
sequence provides evidence of its operation under
the conditions encountered. It does not, however,
establish that the same barrier will respond correctly
under every possible failure scenario or that all
dependencies affecting its performance have been
identified. Similarly, a protection action that
appears unnecessary should not automatically be
classified as a false alarm without reconstructing
the conditions responsible for its activation.
The evaluation of such events therefore requires
attention to both the physical response of the
installation and the assumptions underlying the
protection logic
\cite{Sklet:2006SafetyBarriers,
Leveson:2004AccidentModel}.

For the LHC, post-event analysis is supported by
the extensive monitoring, diagnostics, and recording
capabilities associated with accelerator operation
and machine protection. Beam losses, equipment
faults, interlock conditions, and beam-dump events
can provide information about the sequence of
physical and technical processes preceding a
protective response. Such information is useful
not only for identifying the immediate cause of
an interruption, but also for testing the
assumptions concerning failure propagation,
protection timing, and subsystem interactions
\cite{Wenninger:2016MPO,
Schmidt:2016MPS,Puccio:2010BIS}.

At the detector level, ATLAS and CMS provide
complementary examples of the same principle.
DCS and dedicated
equipment-protection functions monitor different
aspects of the experimental infrastructure,
including electrical conditions, temperatures,
cooling availability, and subsystem readiness.
The corresponding monitoring and protection
architectures provide a basis for identifying
abnormal behaviour and reconstructing its possible
consequences for detector operation
\cite{Boterenbrood:2004ATLASDCS,
Schmeling:2004DSS,CMS:2009DCS,
DiCalafiori:2020CMS_ECAL}.

Consider, for example, a recurrent temperature
warning affecting detector electronics during
otherwise stable operating conditions. Such an
observation may initially appear to be a local
instrumentation problem. A detailed investigation,
however, could reveal a drifting sensor, an
intermittent cooling disturbance, a configuration
inconsistency, or an unrecognized dependency on
another technical service. These possible
explanations have different implications for
the corresponding hazard model and protection
functions. The event is therefore instructive
even if no equipment damage occurs and normal
operation can subsequently be restored. This
hypothetical example illustrates why operational
anomalies should be examined in their broader
physical and technical context rather than
classified solely according to their immediate
consequences.

The reconstruction of an operational event
requires consideration of the facility state
before, during, and after the observed deviation.
Within the notation developed in this work,
the relevant information includes the evolution
of the physical state ${\boldsymbol X}_{\alpha}$,
the applicable technical configuration
${\boldsymbol\Xi}_{\alpha}$, the operational
mode $m$, and the response of the monitoring
and protection functions. Information from
equipment logs, diagnostic records, measurements,
intervention histories, and personnel observations
may therefore be complementary. Their combined
interpretation can reveal inconsistencies between
the assumed and actual system behaviour that
would remain difficult to identify from an
individual alarm or equipment status alone.

Operational learning must also distinguish
between a deficiency of the physical model,
a limitation of the protection architecture,
and a problem associated with organizational
or procedural controls. An unexpected event
may indicate that a failure mechanism was
incorrectly represented, that a coupling
between subsystems was neglected, that a
protection function did not perform as
assumed, or that the applicable work
conditions were not adequately communicated.
These possibilities require different
corrective measures and should not be
reduced automatically to an isolated
equipment failure or an individual
procedural error. This systems-oriented
interpretation is consistent with the
approaches developed by Rasmussen and
Leveson
\cite{Rasmussen:1997RiskManagement,
Leveson:2004AccidentModel}.

Within the risk framework of
Sec.~\ref{subsec:hazard_chains}, such
observations may contribute to revising
the estimated occurrence probability
$P(F_i)$, the conditional probabilities
$P(\Gamma_j^{(i)}\mid F_i)$, or the
consequences associated with particular
propagation paths. In other cases, the
principal result may be the identification
of a previously neglected trajectory
or a common dependency affecting several
barriers. Individual events do not
necessarily provide sufficient statistical
information to determine these quantities
quantitatively, but they can reveal
limitations of the assumptions used in
the existing safety assessment.

The interpretation of operational experience
must also account for the distinction between
safety and availability discussed in
Sec.~\ref{subsec:safety_availability}.
Repeated spurious interlocks may reduce
scientific availability and justify
improvements to monitoring or diagnostic
functions. However, reducing the frequency
of interruptions cannot be pursued by
weakening a protection function without
demonstrating that the required safety
conditions remain satisfied. Conversely,
an apparently successful operational
history cannot by itself establish the
absence of latent failures or the
adequacy of protection against rare,
high-consequence events.

The lessons extracted from anomalies and
near misses acquire practical significance
only when they are translated into
appropriate corrective or preventive
measures. Depending on the findings, these
may include refinement of hazard models,
modification of monitoring or interlock
logic, adjustment of maintenance and
testing strategies, improvement of
procedural controls, additional personnel
training, or revision of operational
restrictions. Where such modifications
affect the technical configuration or
protection assumptions, their implementation
must be followed by appropriate verification
before the revised arrangement is relied
upon during subsequent operation.

This requirement connects operational
learning to the configuration-control
and validation principles discussed in
Sec.~\ref{subsec:return_to_physics}.
Identifying the cause of an anomaly and
implementing a corrective measure do not
alone demonstrate that the resulting
facility configuration is admissible.
The modified system must be assessed
against the applicable physical,
technical, and protection requirements,
and the relevant functions must be
validated before the corresponding
operational mode is restored.

The broader significance of this feedback
was illustrated by the reassessment of
LHC cryogenic safety following the
2008 sector~3--4 incident, discussed
in Sec.~\ref{subsec:cryo_empirical_feedback}.
That experience demonstrated that
observations can require revision not
merely of numerical model parameters,
but of the assumed envelope of credible
failure trajectories and their
consequences
\cite{Chorowski:2011RiskUpdate}.
Near misses and lower-consequence
anomalies provide opportunities to
identify analogous limitations before
they develop into events with serious
consequences.

The effectiveness of this learning
process depends on the ability of
personnel to recognize, document, and
communicate relevant deviations.
This is particularly important in
large collaborations such as ATLAS
and CMS, where knowledge of detector
subsystems, technical infrastructure,
and operational conditions is
distributed among different teams.
Reliable reporting and subsequent
analysis allow local observations
to contribute to a more complete
understanding of the coupled
facility state.

Operational learning consequently
represents an essential connection
between the empirical character of
physics and the continuing development
of safety. Measurements, observations,
and reconstructed event sequences test
the validity of the physical and
engineering assumptions underlying
the protection architecture.
Their interpretation may lead to
revised models, improved barriers,
and better-defined operational
constraints. In this sense, the
scientific method contributes directly
to maintaining and refining the
admissible operational domain, while
the resulting safety knowledge
supports the long-term reliability
and sustainability of experimental
physics.

\subsection{Safety culture and distributed scientific responsibility}
\label{subsec:safety_culture}

The technical complexity of the LHC and its underground
experiments has an important organizational consequence:
no single individual or operational team can possess
complete real-time knowledge of the accelerator,
detector, infrastructure, environmental, and personnel
states. The relevant information is distributed among
machine operators, detector shifters, subsystem experts,
engineers, technicians, radiation-protection specialists,
safety professionals, intervention supervisors, and
emergency-response personnel. Each contributes knowledge
of particular components of the coupled facility state,
while monitoring systems, operational records, and
configuration databases provide complementary information.
Safe operation therefore depends not only on the
availability of this knowledge, but also on its timely
exchange, interpretation, and integration into operational
decisions.

This distribution of expertise is especially pronounced
in ATLAS and CMS, which are operated by large international
collaborations involving many research institutions and
specialized technical teams. Responsibility for detector
subsystems, control systems, infrastructure, maintenance,
and scientific operation is consequently distributed
across different organizational and technical structures
\cite{ATLAS:2024Run3,CMS:2024Run3}.
Although such specialization is indispensable for
developing and operating complex detectors, it also
creates interfaces at which incomplete communication
or inconsistent assumptions may affect safety.

Within the state-space framework developed in this work,
the complete physical state $\boldsymbol X_\alpha$ and
the applicable technical configuration
$\boldsymbol\Xi_\alpha$ cannot generally be reconstructed
from the information available to a single operator.
Different teams observe different physical quantities,
equipment states, and operational conditions, often
through separate control and monitoring systems.
Consequently, an operational decision concerning one
subsystem may depend on information about another
subsystem that is not directly available to the
person making that decision. The reliability of
the decision therefore depends partly on the
organizational mechanisms through which relevant
information is communicated and verified.

Safety culture can, in this context, be understood
as the combination of shared expectations, professional
practices, organizational responsibilities, and
communication mechanisms that support the recognition
and control of hazardous conditions. It includes
the willingness to report anomalies, question
inconsistent equipment states, communicate changes
of configuration, and suspend an activity when
its underlying safety assumptions are no longer
satisfied. These practices complement technical
protection functions by addressing situations
that may not be fully represented in automatic
monitoring or interlock logic. Their importance
is consistent with systems-oriented approaches
to safety, which emphasize interactions between
technical, human, and organizational elements
\cite{Rasmussen:1997RiskManagement,
Leveson:2012SaferWorld}.

The coordination of simultaneous activities provides
a particularly relevant illustration. During a
technical stop or shutdown, different teams may
perform interventions involving electrical
distribution, cooling, detector movement, gas
installations, or protection systems. An action
authorized within one technical domain may modify
conditions assumed by another team. For example,
the restoration of an electrical service may be
technically appropriate for one subsystem but
incompatible with an ongoing intervention that
requires the same service to remain isolated.
Avoiding such inconsistencies requires not only
well-defined procedures and technical isolation,
but also effective communication between the
responsible teams and explicit verification of
the conditions under which the activities may
continue. The management of interventions at
CERN provides an established operational context
for such coordination~\cite{Garino:2013IMPACT}.

Distributed responsibility must nevertheless be
distinguished from the absence of clearly assigned
authority. Although many specialists may contribute
to the assessment of a configuration or intervention,
the responsibilities for authorization, supervision,
execution, and restoration of operational conditions
must remain identifiable. Collective expertise
improves the completeness of the assessment, but
does not remove the need for defined decision-making
responsibilities. This distinction becomes
particularly important when an intervention affects
several subsystems or requires a transition between
operational modes.

The composition of safety and operational teams
also contributes to this organizational framework.
As discussed in
Sec.~\ref{subsec:competence_training}, the integration
of personnel experienced in accelerator or experimental
high-energy physics with professionals specializing
in safety engineering, radiation protection, and
other relevant disciplines is particularly valuable.
Direct experience with detector operation, beam
conditions, superconducting magnets, cryogenic
systems, and experimental infrastructure can
facilitate the recognition of system-specific
hazards and technical dependencies. Conversely,
professional safety expertise provides essential
methods for risk assessment, protection design,
organizational control, and the systematic
evaluation of operational requirements. A mature
safety culture encourages cooperation between
these complementary forms of expertise rather
than treating them as competing sources of authority.

An important element of this cooperation is the
recognition that operational knowledge is not
distributed uniformly and cannot always be
reduced to formal documentation. Experienced
personnel may recognize unusual equipment
behaviour, configuration inconsistencies, or
unexpected interactions on the basis of detailed
familiarity with the installation. Such experience
can contribute significantly to the assessment
of abnormal conditions, but its effectiveness
depends on the possibility of communicating
observations, questioning assumptions, and
escalating unresolved concerns. Formal procedures,
training, supervision, and technical documentation
remain necessary to ensure that protection does
not depend exclusively on individual experience
or informal knowledge.

The relationship between safety culture and
operational learning is equally important.
As discussed in
Sec.~\ref{subsec:near_miss_learning},
anomalies, near misses, and unexpected protection
actions can reveal limitations of existing
hazard models or barriers. Their scientific
and safety value depends, however, on whether
relevant observations are reported, reconstructed,
and examined without prematurely attributing
their causes to individual errors. A systems-oriented
investigation seeks to identify the technical,
physical, and organizational conditions that
allowed an abnormal sequence to develop.
The resulting findings can then inform
modifications of equipment, procedures,
training, or operational constraints
\cite{Leveson:2004AccidentModel,
Rasmussen:1997RiskManagement}.

The continued development of personnel-safety
systems at CERN and the ongoing LS3 provide a particularly relevant
context for these considerations. Extensive
installation, maintenance, consolidation, and
commissioning activities involve multiple
technical domains, different organizational
responsibilities, and frequent changes of
facility configuration
\cite{Tock:2025LS3,Evrard:2025CERNPersonnelSafety}.
Under such conditions, effective safety culture
supports the consistent application of protection
requirements throughout the successive phases
of an intervention, from preparation and
authorization to execution, verification,
and restoration of operational readiness.

From the perspective of the present work,
safety culture contributes to maintaining
the conditions under which the physical
and technical state of the facility remains
within the admissible operational domain.
It does not replace engineered barriers
or provide an independent guarantee of
safety. Rather, it supports the organizational
processes required to interpret physical
information, recognize deviations from
assumed configurations, and ensure that
operational decisions remain compatible
with the established protection requirements.

Safety culture therefore represents an
important connection between scientific
knowledge and operational responsibility.
The distributed expertise required to
construct and operate a complex experimental
facility must also be available for
understanding its hazards, evaluating
configuration changes, and responding
to abnormal conditions. In this sense,
safety is not exclusively the responsibility
of dedicated safety personnel, nor can
it be delegated entirely to individual
detector or accelerator specialists.
It emerges from the coordinated application
of scientific understanding, engineering
protection, professional safety expertise,
and clearly defined operational
responsibilities.

This organizational dimension reinforces
the central concept of \emph{safety in physics}:
safety requirements influence not only
the technical configurations accessible
to an experiment, but also the collective
processes through which those configurations
are established, validated, and maintained.
A strong safety culture thus contributes
to the ability of large scientific facilities
to sustain demanding experimental programmes
without compromising the protection conditions
on which their continued operation depends.

\subsection{Safety constraints and scientific freedom}
\label{subsec:safety_scientific_freedom}

Safety requirements may initially appear to restrict the freedom
with which experimental physics can be performed. Interlocks,
access restrictions, operating limits, work authorizations, and
mandatory verification procedures can delay interventions,
prevent particular configurations, or require additional
commissioning before more demanding operating conditions
are permitted. At the level of an individual activity, these
constraints may indeed reduce the range of immediately
available choices. However, the relationship between safety
and scientific freedom becomes more subtle when considered
at the level of the complete facility and its long-term
scientific programme.

The distinction follows naturally from the state-space
framework developed in this work. A physical state may be
technically achievable without being admissible for routine
operation. For example, an accelerator may be capable of
producing a particular beam intensity, or a detector may
be capable of operating at a specified voltage or magnetic
field, while the corresponding protection conditions have
not yet been demonstrated to be adequate. Physical
achievability therefore does not imply operational
admissibility, and neither necessarily establishes
readiness for scientific data taking.

As discussed in
Sec.~\ref{subsec:barriers_phase_space}, the admissible
domain depends on the protection functions available
for the operating conditions under consideration.
The distinction between the safe and operational
domains was further developed in
Sec.~\ref{subsec:detector_safety_physics_transition}.
In particular, the operational domain
$\Omega_{\rm operational}^{\alpha}
(m,\boldsymbol\Xi_{\alpha})$
represents states compatible with the requirements
of the specified experimental configuration and
operational mode. A state outside this domain may
remain physically realizable, but its intentional
use for the corresponding activity is not justified
until the necessary operating and protection
conditions have been established.

The central point is that these domains are not
necessarily fixed throughout the lifetime of a
facility. Improvements in monitoring, protection
reliability, failure modelling, diagnostic coverage,
and operational validation can make previously
excluded states accessible under controlled
conditions. Schematically, for an unchanged physical
installation and comparable operating requirements,
a validated improvement in the protection
arrangement may permit
\begin{equation}
\Omega_{\rm adm}
\left({\cal B}_{\rm av}^{(0)}\right)
\subseteq
\Omega_{\rm adm}
\left({\cal B}_{\rm av}^{(1)}\right),
\label{eq:sustainable_space_expansion}
\end{equation}
where ${\cal B}_{\rm av}^{(0)}$ and
${\cal B}_{\rm av}^{(1)}$ denote the initially
available and subsequently improved sets of
credited protection functions, respectively.
The inclusion expresses a possible enlargement
of the admissible domain, not a universal consequence
of adding protection measures. It applies when
the modified arrangement has been demonstrated
to preserve the required safety conditions while
permitting additional operating states or
transitions.

The LHC provides a particularly clear illustration.
Higher stored beam energies and intensities increase
the consequences of uncontrolled beam losses and
therefore impose increasingly demanding requirements
on collimation, beam-loss monitoring, interlocks,
and beam extraction. Improvements in these protection
systems, together with their commissioning and
validation, contribute to making more demanding
beam conditions compatible with controlled operation.
The additional scientific capability is not obtained
by relaxing the underlying damage limits, but by
improving the ability to detect and interrupt
hazardous trajectories before unacceptable
consequences develop
\cite{Schmidt:2016MPS,Redaelli:2016Collimation,
Wenninger:2016MPO}.

A related argument applies to detector operation.
At ATLAS and CMS, increasing detector performance
may require more demanding configurations of
electrical power, cooling, magnetic systems,
readout electronics, or other technical services.
Reliable monitoring and equipment-protection
functions help establish the conditions under
which such configurations can be operated
reproducibly. For example, an adequately validated
response to loss of cooling can protect detector
electronics against overheating and thereby
support sustained operation under the specified
powering and thermal conditions. The protection
function does not eliminate the physical hazard;
rather, it reduces the probability that an
abnormal transition develops into equipment damage
\cite{Schmeling:2004DSS,
DiCalafiori:2020CMS_ECAL}.

The same principle extends beyond automatic
protection systems to maintenance and intervention
planning. Radiation-protection constraints may
exclude a particular intervention immediately
after beam operation because of residual activation.
However, an appropriate cooling interval, temporary
shielding, improved tooling, or remote-handling
arrangements may make the same technical activity
admissible at a later time or in a modified
configuration. In this case, scientific and
technical freedom is increased not by removing
the exposure constraints, but by identifying
an alternative intervention trajectory compatible
with the required protection conditions
\cite{FassoRokni:2009ALARA,
CERN:RadiationProtectionCode}.

These examples also clarify the importance of
the distinction between an admissible state
and an admissible trajectory. Improved protection
may allow a more demanding final operating state,
but the transition toward that state must still
satisfy the applicable commissioning, interlock,
and configuration requirements. Similarly,
improved intervention techniques may make a
previously impractical activity possible only
through a particular sequence of preparation,
isolation, access, execution, and restoration.
The enlargement of scientific possibilities
therefore concerns not only the accessible
states, but also the set of validated
trajectories connecting them.

The enlargement represented by
Eq.~\eqref{eq:sustainable_space_expansion}
should not be confused with an unconditional
increase in performance. A newly identified
failure mechanism may require additional
restrictions, and an unavailable protection
function may temporarily reduce the admissible
domain. Conversely, improvements that reduce
spurious protection actions or facilitate
recovery may increase operational availability
without changing the boundaries of the
admissible state space. These distinctions
are important for interpreting the relationship
between protection integrity and scientific
performance discussed in
Sec.~\ref{subsec:safety_availability}.

Scientific freedom in a complex accelerator
facility is consequently better understood
as the ability to explore, establish, and
reproduce physically meaningful experimental
conditions within a validated protection
framework, rather than as the unrestricted
possibility of selecting any technically
achievable configuration. Safety constraints
can prohibit particular activities while
creating the conditions under which more
demanding experimental states become
reliably accessible.

The argument also has an organizational
dimension. The competence, communication,
configuration control, and operational
learning discussed in the preceding
subsections contribute to maintaining
confidence in the validity of the protection
framework. Even technically well-designed
systems cannot provide their intended
operational capability if their configuration
is uncertain, their limitations are not
understood, or their required conditions
are not maintained. Scientific freedom
therefore depends on both the technical
availability of protection and the ability
of the organization to operate within its
validated constraints.

From the perspective of \emph{safety in physics},
the essential conclusion is that safety
requirements and scientific capability are
not necessarily competing objectives.
Safety imposes boundaries on individual
activities and operating configurations,
but effective protection can also make
additional regions of the physically
accessible state space available for
sustained scientific use. The resulting
scientific benefit is not merely the
absence of accidents, but the ability
to reach, maintain, and reproduce
experimental conditions whose operation
would otherwise involve unacceptable
or insufficiently controlled risks.

\subsection{From safe operation to scientific continuity}
\label{subsec:safety_scientific_continuity}

The long-term scientific objective of a large research
facility is not simply to maximize its instantaneous
performance, but to sustain the production of reliable
scientific results over many years or decades. At the
LHC and its underground experiments, the scientific
value of high beam intensity, luminosity, detector
efficiency, or measurement precision depends on the
ability to establish the corresponding operating
conditions repeatedly, maintain them for sufficiently
long periods, and recover them following interruptions.
Personnel protection, equipment preservation,
maintainability, recoverability, and the continuity
of technical knowledge therefore contribute to the
long-term scientific capability of the facility.

This observation extends the relationship between
safety and availability discussed in
Sec.~\ref{subsec:safety_availability}. The operational
availability defined in Eq.~\eqref{eq:physics_availability}
characterizes the fraction of scheduled time during
which physics operation is possible, but does not
fully describe the corresponding scientific output.
As expressed more generally in
Eq.~\eqref{eq:scientific_yield_functional}, the latter
depends on the evolving facility state, the conditions
under which measurements are performed, and the
scientific usefulness of the resulting data.
For collider experiments, integrated luminosity
provides an important measure of accumulated
exposure to collisions, while detector live time,
data quality, trigger performance, and reconstruction
conditions determine how effectively those collisions
can be used for physics analysis.

Protection systems influence this cumulative
performance through several distinct mechanisms.
On short timescales, an interlock, beam dump,
detector shutdown, or access restriction can
interrupt data taking and reduce operational
availability. On longer timescales, however,
effective protection can prevent damage to
accelerator or detector components, reduce the
duration of recovery, and preserve the possibility
of subsequent operation. The scientific cost of
a controlled interruption must therefore be
distinguished from the potentially much greater
consequences of allowing an abnormal physical
state to propagate without adequate protection
\cite{Filippini:2005Reliability,
Willeke:2016Reliability,Wenninger:2016MPO}.

The importance of this distinction becomes
particularly clear for highly energetic systems.
An appropriate beam dump may terminate a period
of physics operation, but can prevent uncontrolled
energy deposition and damage whose consequences
would extend far beyond the immediate interruption.
Similarly, an automatic detector shutdown following
a loss of cooling or another abnormal condition
may temporarily reduce data-taking efficiency
while preserving sensitive equipment and its
subsequent operational capability. In both cases,
the protection function contributes to scientific
continuity by preventing a short-duration
disturbance from developing into a prolonged
loss of experimental capability
\cite{Schmidt:2016MPS,
Schmeling:2004DSS,DiCalafiori:2020CMS_ECAL}.

Scientific continuity also depends on the
recoverability of the facility following an
interruption. Recovery is not equivalent to
the simple restoration of power, beam, or
detector readiness signals. The initiating
condition must be understood, the relevant
physical and technical states assessed, and
the required protection functions verified
before operation resumes. Depending on the
event, recovery may involve diagnosis,
repair, configuration restoration, functional
testing, or partial recommissioning. The
principles developed in
Sec.~\ref{subsec:return_to_physics}
therefore apply not only after planned
maintenance, but also to the restoration
of operation following abnormal events.

Within the state-space framework of the
present work, scientific continuity requires
more than repeated access to individual
states belonging to the admissible
operational domain. The facility must also
retain the ability to follow admissible
trajectories between physics operation,
controlled interruption, maintenance,
recovery, and subsequent restoration.
For an experiment $\alpha$, these
transitions involve the evolution of
the physical state $\boldsymbol X_\alpha$
under the applicable technical
configuration $\boldsymbol\Xi_\alpha$
and operational mode $m$. The relevant
protection and authorization conditions
may differ between successive phases,
but their consistency must be preserved
throughout the complete operational cycle.

The corresponding operational cycle can be  schematically  represented as follows
\begin{equation}
\begin{array}{ccccc}
&&
\boxed{\substack{
\text{Validated physics}\\
\text{operation}
}}
&&
\\[6pt]
&
\nearrow
&&
\searrow
&
\\[6pt]
\boxed{\substack{
\text{Recovery and}\\
\text{revalidation}
}}
&&&&
\boxed{\substack{
\text{Controlled}\\
\text{interruption}
}}
\\[6pt]
&
\nwarrow
&&
\swarrow
&
\\[6pt]
&&
\boxed{\substack{
\text{Maintenance and}\\
\text{intervention}
}}
&&
\end{array}
\label{eq:safety_scientific_cycle}
\end{equation}
Each transition in Eq.~\eqref{eq:safety_scientific_cycle} is subject to the physical, technical,
organizational, and protection conditions appropriate
to the current facility configuration and operational
mode. The diagram does not imply that every interruption
necessarily requires maintenance, or that all stages
must be traversed in a fixed sequence. Rather, it
illustrates the recurrent character of scientific
operation, in which controlled interruptions,
interventions, and subsequent validation contribute
to the continued availability of admissible physics
configurations.

An important consequence is that
maintainability and recoverability must
be considered alongside nominal
performance when evaluating the design
of experimental systems. Accessibility
of components, diagnostic coverage,
segmentation of technical services,
availability of replacement equipment,
and the possibility of testing protection
functions without unnecessary disturbance
to other subsystems can substantially
influence the duration and complexity
of interventions. A system optimized
exclusively for its nominal operating
state may consequently offer poorer
long-term scientific performance if
maintenance and recovery require
disproportionately complex operations.

The same consideration applies to the
evolution of the experimental apparatus.
ATLAS and CMS have undergone successive
technical interventions and detector
upgrades while continuing their long-term
physics programmes. Such modifications
may improve detector performance,
extend experimental capabilities, or
address limitations identified during
previous operation. However, the
scientific benefits of an upgrade
depend on successful integration,
commissioning, and validation of the
modified system, including its
interfaces with the existing
infrastructure and protection
architecture
\cite{ATLAS:2024Run3,CMS:2024Run3}.
Scientific continuity therefore does
not imply that the facility remains
technically unchanged. Rather, it
requires that controlled evolution
of the installation remain compatible
with the preservation of its scientific
and safety functions.

Continuity has an organizational
dimension as well. The long operational
lifetime of a major experimental
facility inevitably involves changes
in personnel, technical responsibilities,
equipment generations, and operational
procedures. The preservation of
configuration records, intervention
histories, diagnostic information,
protection-system documentation,
and experience acquired during
commissioning and operation is
therefore essential. As discussed
in Secs.~\ref{subsec:competence_training}
and~\ref{subsec:safety_culture},
training, knowledge transfer, and
cooperation between scientific,
technical, and safety specialists
contribute to maintaining the
competence required for reliable
operation across successive
generations of personnel.

The operational learning discussed
in Sec.~\ref{subsec:near_miss_learning}
provides an additional mechanism
through which scientific continuity
is strengthened. Anomalies, near
misses, protection actions, and
maintenance experience can reveal
limitations of existing technical
assumptions or operational procedures.
When the resulting information is
incorporated into validated changes
of equipment, protection functions,
or working methods, the facility
can improve its ability to prevent
recurrence, reduce unnecessary
interruptions, and recover from
future disturbances. The protection
architecture therefore evolves
together with the experimental
installation and the knowledge
accumulated through its operation.

The contribution of safety to
scientific continuity should
nevertheless not be interpreted
as a requirement to avoid every
interruption. Planned shutdowns,
preventive maintenance, protection
testing, and systematic verification
may reduce immediate data-taking
time while preserving the conditions
necessary for reliable future
operation. Conversely, postponing
essential interventions or relaxing
protection requirements to improve
short-term availability can undermine
the long-term scientific capability
of the installation. The appropriate
objective is therefore sustained
scientific performance within the
validated protection framework,
rather than uninterrupted operation
as an end in itself.

This distinction brings together
the two complementary directions
developed throughout the manuscript.
\emph{Physics in safety} provides
the models, measurements, and
technical understanding required
to control hazardous processes,
diagnose abnormal behaviour,
and validate protection functions.
\emph{Safety in physics} ensures
that these capabilities are
translated into conditions under
which experimental systems can
be operated, maintained, recovered,
and progressively improved.
The resulting relationship is
not confined to individual
protection actions: it extends
over the complete scientific
lifetime of the facility.

The LHC, ATLAS, and CMS thus
illustrate that safety is not
an activity performed only before
physics operation begins or after
an incident has occurred. It is
an integral part of the continuing
process through which scientific
capability is established,
exercised, preserved, and extended.
This perspective also has direct
implications for future accelerator
and underground experimental
facilities, where protection,
maintainability, recoverability,
and the preservation of operational
knowledge should be incorporated
from the earliest stages of design,
as discussed in
Sec.~\ref{sec:future_facilities}.

\section{Lessons for the HL-LHC and future large scientific facilities}
\label{sec:future_facilities}

The experience accumulated at the LHC and its underground experiments
provides a particularly instructive basis for considering the design
and operation of future large-scale scientific facilities. As
discussed in the preceding sections, accelerator and detector safety
cannot be reduced to the independent protection of individual
components or to the enforcement of fixed operational limits.
It depends on the interactions between energetic systems,
experimental equipment, technical infrastructure, environmental
conditions, personnel access, and organizational controls.
The corresponding protection architecture determines not only
which physical states can be safely established, but also which
transitions between configurations can be performed and validated.

These considerations acquire particular importance for the
High-Luminosity LHC (HL-LHC) and future accelerator concepts,
including the Future Circular Collider (FCC) and possible muon
colliders. Their scientific objectives involve different combinations
of stored beam and magnetic energy, superconducting technologies,
radiation environments, cryogenic requirements, detector systems,
and underground infrastructure
\cite{Apollinari:2017HLLHC,
Adorisio:2015HLLHCSafety,
Benedikt:2025FCCFeasibility,
Accettura:2024MuonCollider}.
Although the relevant physical scales and hazard configurations
depend on the accelerator concept, increasing technical complexity
and subsystem coupling place additional demands on monitoring,
protection, intervention planning, and operational coordination.
The experience gained at the LHC is therefore valuable not as
a universal protection architecture to be reproduced, but as
a source of physical understanding, engineering principles,
and methods for controlling complex operational states.

A central lesson is that safety requirements should enter
the design process from its earliest stages, rather than
being addressed primarily after the scientific and engineering
configuration has been established. The geometry of underground
installations, distribution of stored energy, segmentation
of technical systems, positioning of protection barriers,
accessibility of equipment, radiation shielding, ventilation,
evacuation arrangements, and provisions for maintenance
contribute directly to determining the admissible operational
domain. Once civil engineering and major technical systems
have been defined, some of these characteristics may be
difficult or prohibitively expensive to modify. Safety
considerations must therefore participate in the selection
and optimization of the facility configuration itself.

The safe operational phase-space framework developed in
this work provides a common conceptual basis for such
integration. For a prospective facility, the admissible
domain depends on its physical and technical configuration,
the intended operational modes, and the protection functions
available under those conditions. The design problem is
consequently not limited to identifying acceptable values
of individual physical parameters. It must also establish
whether the combinations of subsystem states required for
scientific operation are mutually compatible and whether
the trajectories connecting them can be executed within
the applicable protection constraints. This includes both
normal operational transitions and credible abnormal
evolutions that must be prevented, interrupted, or redirected
toward controlled states.

The preceding discussion of scientific continuity introduces
a further requirement. Future facilities should be designed
not only for achieving their nominal scientific performance,
but also for maintaining, recovering, and progressively
extending that performance throughout their operational
lifetime. Accessibility of critical equipment, diagnostic
capabilities, replaceability of components, remote-handling
provisions, isolation of technical services, and the
possibility of validating protection functions after
interventions can substantially influence both safety
and long-term scientific availability. In this respect,
maintainability and recoverability are not secondary
operational considerations, but characteristics that
should influence the physical and engineering design
from the outset.

The same reasoning applies to the organization of future
experimental facilities. Increasing complexity requires
the integration of accelerator and detector expertise,
specialized safety knowledge, operational experience,
and clearly defined responsibilities. Provisions for
personnel access, simultaneous interventions, training,
configuration control, and communication between technical
teams should therefore be considered together with the
corresponding engineered protection functions. This
approach recognizes that the effectiveness of a protection
architecture depends not only on its technical design,
but also on the conditions under which it is commissioned,
operated, maintained, and restored.

The resulting objective is not simply to maximize
scientific performance subject to a collection of
safety restrictions imposed afterwards. Rather, it
is to develop the scientific installation, its
protection architecture, and its operational
arrangements as mutually dependent elements of
a common design. Such an approach can improve the
ability of a facility to establish demanding
experimental configurations while preserving the
conditions required for their safe and sustainable
operation. The following subsections examine the
implications of this principle for the scaling of
the admissible operational domain, the HL-LHC,
radiation protection and maintainability, the
management of abnormal transitions, and the
specific technological challenges of future
circular and muon colliders.

\subsection{The safe operational phase space as a scaling principle}
\label{subsec:safe_phase_space_scaling}

The safe operational phase-space concept developed in this work
becomes particularly relevant when considering the transition
from the LHC to the High-Luminosity LHC (HL-LHC) and future
large accelerator and underground experimental facilities.
The significance of this approach does not arise solely from
possible increases in stored energy, radiation exposure,
cryogenic inventories, or the physical dimensions of the
installations. Equally important is the increasing complexity
of the interactions between technical subsystems and the
corresponding difficulty of establishing safe operating
conditions through independent limits on individual
physical quantities.

The state-space formulation introduced in
Sec.~\ref{sec:lhc_energetic_system} represents the facility
through a coupled physical and operational state
${\boldsymbol X}$, whose components describe beam, magnet,
cryogenic, detector, environmental, and access conditions.
The block decomposition in
Eq.~\eqref{eq:lhc_coupled_state_vector} is particularly
useful because it identifies the principal subsystems
without assuming their dynamical independence.
The corresponding protection problem concerns not only
the admissibility of individual subsystem states, but
also the compatibility of their simultaneous configurations
and the possible propagation of abnormal transitions
between them.

This distinction becomes increasingly important as the
number of interconnected technical systems grows.
In general, the safe domain of the complete installation
cannot be constructed as a Cartesian product of
independently determined safe regions for its subsystems.
Using the decomposition introduced above, one may write
schematically
\begin{equation}
\Omega_{\rm safe}
\neq
\Omega_{\rm beam}
\times
\Omega_{\rm mag}
\times
\Omega_{\rm cryo}
\times
\Omega_{\rm det}
\times
\Omega_{\rm env}
\times
\Omega_{\rm access},
\label{eq:nonfactorizable_safe_space}
\end{equation}
where the factors on the right-hand side represent
hypothetical subsystem domains defined without accounting
for all their mutual dependencies. The relation is intended
to express the general presence of cross-system constraints;
it does not exclude factorization in particular simplified
or effectively decoupled configurations.

The physical origin of this nonfactorizability is the
coupling between different components of the facility
state. A beam configuration that is admissible under
one magnet or collimation state may become inadmissible
under another. A cryogenic configuration that satisfies
the equipment-protection requirements may impose
additional restrictions on underground occupancy or
ventilation. Detector movement can modify mechanical
clearances, service connections, and personnel-access
conditions, while maintenance activities may temporarily
invalidate assumptions underlying the protection of
neighboring systems. The admissibility of the complete
configuration therefore depends on relationships between
subsystems rather than exclusively on their individual
operating limits.

These considerations connect directly to the barrier-dependent
admissible domain introduced in
Sec.~\ref{subsec:barriers_phase_space}. The set of states
available for intentional operation depends on the
protection functions ${\cal B}_{\rm av}$ that are available
and credited for the corresponding configuration.
Consequently, the effective operational domain can
change when protection systems are modified, additional
failure mechanisms are identified, or particular
protection functions become unavailable. The central
design requirement is therefore not merely to establish
safe limits for each component, but to demonstrate
that the complete system remains compatible with the
required protection conditions.

The second essential aspect of this scaling problem
is dynamical. As emphasized in the preceding sections,
the admissibility of an initial and final state does
not guarantee the admissibility of the trajectory
connecting them. The transition may involve intermediate
configurations, transient energy redistribution,
changes in protection availability, or temporary
incompatibilities between coupled systems. This
distinction is relevant to beam injection and extraction,
magnet powering, cryogenic transitions, detector movement,
access-mode changes, and restoration following
maintenance. In increasingly complex installations,
the number of possible interactions during such
transitions can make their validation particularly
demanding.

The HL-LHC provides a concrete illustration.
Its design foresees stored proton-beam energies
approaching $700~{\rm MJ}$ per beam, together with
new superconducting technologies, upgraded collimation,
and additional accelerator systems
\cite{Apollinari:2017HLLHC,
Adorisio:2015HLLHCSafety}.
These developments modify both the potential
consequences of selected failure scenarios and
the technical dependencies involved in their
detection and mitigation. The relevant safety
question is therefore not only whether each
new component possesses adequate local protection,
but also whether its normal operation and credible
failure trajectories remain compatible with
the protection requirements of the complete
accelerator. The corresponding beam-energy and
protection implications are examined in greater
detail in the following subsection.

The same principle applies to large underground
experimental facilities. Detector upgrades may
introduce additional electrical, cooling, gas,
mechanical, or readout dependencies without
necessarily increasing every individual stored-energy
reservoir. A modification that appears acceptable
when assessed locally may nevertheless change the
conditions under which another subsystem operates.
For future facilities, the design and validation
of interfaces between subsystems may therefore be
as important as the verification of their individual
protection functions.

Commissioning provides a practical means of addressing
these coupled constraints. As discussed in
Sec.~\ref{subsec:safety_commissioning},
progressively more demanding operating configurations
are admitted only after the required physical response
and protection functions have been validated.
This process does not necessarily enlarge the intrinsic
safe domain: rather, it establishes evidence that
particular states and trajectories within the
applicable admissible domain can be used intentionally.
Improvements in protection, monitoring, and modelling
may subsequently permit additional configurations,
whereas newly identified hazards or degraded protection
may require restrictions. Thus, the practically
accessible operational domain can expand, contract,
or change its shape as the facility and the knowledge
of its behaviour evolve.

For future circular colliders, the same reasoning
extends beyond the accelerator and detector systems.
The extensive underground infrastructure envisaged
in FCC studies introduces spatially distributed
dependencies involving electrical distribution,
cryogenic services, ventilation, radiation protection,
personnel access, maintenance, and emergency response
\cite{Benedikt:2025FCCFeasibility}.
The physical state of this infrastructure and its
interactions with accelerator operation must therefore
be included in the global safety assessment.
A larger geographical extent does not by itself
imply a proportionally greater risk, but it can
introduce additional interfaces, propagation
pathways, and constraints on operational coordination
and recovery.

An important extension of the state-space approach
concerns uncertainty. The boundaries defining
admissible operation depend on physical models,
instrumentation, material properties, failure
assumptions, and the characteristics of protection
systems. These quantities may not be known exactly,
particularly during the design and early commissioning
of new facilities. If the safe domain depends on
a set of uncertain parameters ${\boldsymbol\theta}$,
a conservative robust domain may be defined
schematically as
\begin{equation}
\Omega_{\rm safe}^{\rm robust}
=
\bigcap_{{\boldsymbol\theta}\in\Theta}
\Omega_{\rm safe}({\boldsymbol\theta}),
\label{eq:robust_safe_phase_space}
\end{equation}
where $\Theta$ denotes the set of parameter values
considered credible within the adopted uncertainty
model. The construction identifies states satisfying
the specified safety conditions for every parameter
combination included in $\Theta$, assuming that the
state coordinates and requirements have a common
definition across those combinations.

Equation~\eqref{eq:robust_safe_phase_space} is a
conceptual representation rather than a prescription
for constructing the complete operational domain
of a real facility. In practice, uncertainties may
involve correlated parameters, transient behaviour,
measurement errors, incomplete failure models, and
discrete configuration changes. Moreover, robust
admissibility of an instantaneous state does not
by itself establish that every trajectory originating
from that state remains admissible. The uncertainties
affecting the evolution and the response of protective
functions must also be considered.

This distinction is particularly important for
future systems whose nominal physical performance
may be well characterized while some abnormal
transitions remain insufficiently validated.
Conservative operating restrictions, additional
diagnostics, staged commissioning, and dedicated
protection tests can then provide a controlled
means of extending the experimentally supported
operational domain. As knowledge of the system
improves, the uncertainty assumptions and the
corresponding restrictions may be reassessed,
without implying that physical safety limits
can be relaxed arbitrarily.

The resulting scaling principle is therefore
qualitative rather than a universal mathematical
law relating facility size to risk. Increasing
complexity and subsystem coupling generally
strengthen the need for integrated state monitoring,
configuration management, protection coordination,
and validation of operational trajectories.
The required degree of integration depends on
the actual physical dependencies, failure
mechanisms, and protection architecture rather
than on geometric size or stored energy alone.

This perspective also connects the design of
future facilities with the scientific continuity
discussed in
Sec.~\ref{subsec:safety_scientific_continuity}.
The ability to reach a demanding experimental
configuration is scientifically valuable only
if that configuration can be maintained,
interrupted safely, recovered, and subsequently
re-established. Protection-system accessibility,
diagnostic capability, technical segmentation,
maintainability, and controlled restoration
therefore influence not only the admissible
operating domain but also the long-term ability
to exploit it.

The safe operational phase-space approach thus
provides a common conceptual framework for
integrating physical constraints, protection
functions, uncertainty, commissioning, and
operational evolution into the design of future
facilities. Its principal implication is not
that scientific performance and safety can
always be improved simultaneously, but that
they should be designed and assessed together.
The objective is to establish a protection
architecture that supports the required
scientific configurations and their admissible
transitions while maintaining the applicable
safety conditions throughout the operational
lifetime of the facility.

\subsection{HL-LHC: increasing performance within an established
protection framework}
\label{subsec:hllhc_safety_scaling}

The  HL-LHC represents the most immediate
extension of the physical and operational principles developed
in this work. Its principal objective is not to substantially
increase the proton collision energy, but to achieve significantly
higher luminosity through increased beam brightness, larger
bunch populations, stronger focusing at the interaction points,
superconducting crab cavities, upgraded collimation, and new
high-field superconducting magnets. These developments extend
the performance of an existing accelerator while modifying
its energetic characteristics, technical configuration,
failure modes, and protection requirements
\cite{Apollinari:2017HLLHC,
Adorisio:2015HLLHCSafety,Wollmann:2020HLLHCMPS}.

An important quantitative illustration is provided by the
stored beam energy. Using Eq.~\eqref{eq:lhc_beam_energy},
the larger bunch populations foreseen for the HL-LHC
increase the total stored energy even when the beam energy
per proton remains close to the nominal LHC design value.
The HL-LHC design considers a stored beam energy of order
\begin{equation}
E_{\rm beam}^{\rm HL}
\sim
700~{\rm MJ}
\label{eq:hllhc_beam_energy}
\end{equation}
per beam, compared with approximately $362~{\rm MJ}$
for the nominal LHC design configuration
\cite{Apollinari:2017HLLHC,
Redaelli:2020HLLHCCollimation}.
The increase is substantial because the additional
stored energy is not accompanied by a corresponding
increase in the damage thresholds of accelerator
components.

The consequences of beam losses nevertheless depend
on more than the total stored energy. The relevant
quantities include the spatial and temporal distribution
of the losses, the beam size, the material properties,
and the fraction of the lost energy deposited in
sensitive components. If a fraction $f_{\rm loss}$
of the stored beam energy is lost, and a fraction
$\eta_{\rm dep}$ of that energy is deposited in a
particular vulnerable region, the corresponding
deposited energy may be estimated schematically as
$\eta_{\rm dep}f_{\rm loss}E_{\rm beam}$.
For a fixed deposition geometry and damage criterion,
the tolerable fractional loss therefore scales as
\begin{equation}
f_{\rm tol}
\sim
\frac{E_{\rm dep}^{\rm lim}}
{\eta_{\rm dep}E_{\rm beam}},
\label{eq:tolerable_loss_fraction_scaling}
\end{equation}
where $E_{\rm dep}^{\rm lim}$ denotes the limiting
deposited energy for the particular component and
failure scenario under consideration.

Equation~\eqref{eq:tolerable_loss_fraction_scaling}
illustrates why the protection of increasingly
energetic beams requires control of progressively
smaller fractional losses under otherwise comparable
conditions. It is not a universal damage criterion:
both the deposited-energy limit and the deposition
fraction depend on the beam parameters, impacted
material, geometry, and duration of the loss.
Detailed energy-deposition calculations and
beam-tracking simulations are consequently required
to establish the actual protection limits
\cite{Tahir:2005LHCBeam,
Redaelli:2020HLLHCCollimation,
Tygier:2019Collimation}.

For the HL-LHC, the resulting protection challenge
involves the coordinated performance of collimation,
beam-loss monitoring, beam interlocks, injection
and extraction protection, and the beam-dumping
system. Collimators must intercept beam losses
before they produce unacceptable energy deposition
in vulnerable accelerator components, while the
monitoring and interlock systems must identify
conditions requiring beam extraction within
the relevant protection timescale. Their
effectiveness depends on the complete machine
configuration, including beam energy, intensity,
optics, collimator positions, and the status of
the protection functions
\cite{Redaelli:2020HLLHCCollimation,
Wollmann:2020HLLHCMPS}.

The increased stored beam energy is therefore
only one aspect of the scaling problem. The
HL-LHC also introduces operating configurations
and technical systems whose failure dynamics
may differ from those of the original LHC.
A particularly instructive example is provided
by superconducting crab cavities, which compensate
for the geometrical luminosity reduction associated
with the crossing angle at the interaction points.
An abnormal change in cavity voltage or radiofrequency
phase can produce a transverse beam perturbation
and, under sufficiently unfavorable conditions,
rapid beam losses. The corresponding consequences
depend on the time evolution of the cavity fields,
the response of the beam, the available protection
time, and the effectiveness of the associated
mitigation measures
\cite{Apsimon:2019CrabQuench,
Wollmann:2020HLLHCMPS}.

This example is particularly relevant to the
state-and-trajectory framework of the present work.
The instantaneous values of the cavity voltage
and phase do not fully characterize the potential
consequences of a failure. The rate and sequence
of their evolution, the resulting beam trajectory,
and the time available for detection and extraction
are equally important. Moreover, different cavity
failure scenarios need not produce the same loss
patterns or protection requirements. Studies of
realistic crab-cavity quench dynamics have shown
that the resulting beam losses can be considerably
smaller than those obtained under more severe
assumptions concerning rapid phase excursions
\cite{Apsimon:2019CrabQuench}.
The relevant protection limits must therefore
follow from physically justified failure models
rather than from stored-energy considerations alone.

The superconducting magnet system introduces
additional coupled protection requirements.
The HL-LHC employs new high-field magnets,
including Nb$_3$Sn-based insertion quadrupoles,
together with modified powering circuits,
superconducting links, and associated cryogenic
infrastructure. The protection of these systems
requires consideration of quench detection,
energy extraction or redistribution, electrical
insulation, thermal response, and the possible
consequences of abnormal energy transfer to
cryogenic components
\cite{Apollinari:2017HLLHC,
Wollmann:2020HLLHCMPS}.
The protection problem is consequently not
limited to maintaining acceptable operating
conditions within each superconducting device.
It also concerns the compatibility of magnetic,
electrical, cryogenic, and environmental states
during both nominal operation and credible
failure transitions.

The cryogenic infrastructure illustrates
this dependence at a larger scale.
The HL-LHC upgrades include new cryogenic
installations and distribution arrangements
associated with the high-luminosity interaction
regions, superconducting magnets, and links
\cite{Claudet:2020HLLHCCryogenics}.
The corresponding protection requirements
extend from equipment integrity and pressure
relief to the consequences of possible helium
release and the resulting underground
environmental conditions. Changes in the
cryogenic configuration may therefore affect
not only local equipment protection, but
also ventilation, oxygen-deficiency hazards,
personnel access, and the conditions under
which technical interventions are permitted.

Within the coupled-state representation
introduced in
Eq.~\eqref{eq:lhc_coupled_state_vector},
these examples demonstrate that the
HL-LHC protection problem cannot be reduced
to independent improvements of beam, magnet,
and cryogenic systems. A perturbation originating
in one subsystem may change the conditions
required for another, while the admissibility
of a particular operating configuration
depends on the simultaneous availability
of the corresponding protective functions.

The implications for operational trajectories
are equally important. The transition from
injection to higher beam energies, changes
in beam optics during the squeeze, the
establishment of collisions, and the
restoration of operation following a fault
must remain compatible with the relevant
protection conditions. In particular,
collimation and machine-protection studies
must account for intermediate configurations,
rather than considering only the final
high-luminosity operating state
\cite{Tygier:2019Collimation,
Wollmann:2020HLLHCMPS}.
The safe operational phase-space framework
thus provides a common language for expressing
the relationship between the available
protection architecture, the intended
machine configuration, and its permitted
evolution.

The transition toward HL-LHC operation
also illustrates the importance of
commissioning and validation. New protection
functions and modified technical interfaces
must be assessed under the conditions for
which they are intended, while their
interaction with existing machine systems
must be verified before the corresponding
operating configurations are authorized.
Progressively increasing beam intensity
or introducing more demanding optics
therefore requires evidence that the
protection functions remain adequate for
the relevant failure scenarios.

Within the notation developed earlier,
the corresponding operating states must
remain compatible with the admissible
domain associated with the available
protection functions
${\cal B}_{\rm av}$, while the trajectories
connecting them must satisfy the applicable
operational constraints. Improved protection
may permit additional configurations
to become accessible, but this enlargement
cannot be inferred from the installation
of new equipment alone. It requires
validation of the combined physical
and technical system, including relevant
interfaces and common dependencies.

A further consideration is the relationship
between protection and scientific continuity.
Higher luminosity increases the scientific
value of sustained accelerator availability,
but the consequences of damaging failures
may also become more significant.
Protection-system reliability, diagnostics,
testing, maintainability, and controlled
recovery consequently remain important
elements of the overall machine design.
The objective is not simply to minimize
the number of protective interruptions,
but to preserve the ability to operate
reproducibly at demanding beam parameters
without exposing the machine to
unacceptable consequences
\cite{Filippini:2005Reliability,
Wollmann:2020HLLHCMPS}.

The HL-LHC therefore provides an important
example of the reciprocal relationship
between physics and safety. Its luminosity
goals motivate new accelerator technologies
and more demanding operating conditions,
while the physical characteristics of
these systems determine the corresponding
protection requirements. Conversely,
the ability to establish and validate
those protection conditions determines
which combinations of beam parameters,
magnet states, and technical configurations
can be used for scientific operation.

The central lesson is that increased
accelerator performance requires not
merely stronger individual protection
functions, but an integrated architecture
capable of controlling coupled states
and their evolution. The HL-LHC thus
illustrates how protection requirements
must evolve together with scientific
performance, providing a concrete
application of the safe operational
phase-space approach developed in
this manuscript.

\subsection{Radiation, integrated luminosity, and maintainability}
\label{subsec:future_radiation_maintainability}

One of the most consequential safety challenges associated
with the HL-LHC concerns the radiation environment of
ATLAS and CMS, together with the increasingly demanding
radiological conditions in selected regions of the
accelerator itself. The higher integrated luminosity
and the more intense circulating beams will affect
the accumulated irradiation of detector components,
beam-line equipment, collimators, and surrounding
technical infrastructure. In addition to radiation
damage and equipment lifetime, these effects have
important consequences for residual activation,
personnel access, maintenance, replacement of
components, and radioactive-waste management.
Radiation protection and maintainability must
therefore be considered together throughout the
operational lifetime of the HL-LHC.

For a fixed experimental configuration and a
location where the radiation field is dominated
by particles produced in proton--proton collisions,
the accumulated particle fluence approximately
scales with integrated luminosity,
\begin{equation}
\Phi_{\rm rad}
\propto
{\cal L}_{\rm int},
\label{eq:radiation_luminosity_scaling}
\end{equation}
where the proportionality factor depends on the
collision energy, particle spectra, detector
geometry, materials, and shielding. A corresponding
approximate scaling may apply to accumulated
ionizing dose under comparable irradiation
conditions. The relation must not, however,
be applied without qualification to radiation
originating from beam losses elsewhere in the
accelerator or to residual dose rates after
operation.

The HL-LHC programme envisages integrated
luminosities of order
$3000$--$4000~{\rm fb}^{-1}$
for ATLAS and CMS. The resulting radiation
environment has motivated major detector
upgrades, including replacement of silicon
tracking systems and the development of
radiation-tolerant sensors, electronics,
and associated services
\cite{Apollinari:2017HLLHC,
Buttar:2025ATLASITk,
Kuo:2025CMSInnerTracker}.
These developments illustrate the direct
influence of radiation physics on detector
design, material selection, cooling, and
the expected lifetime of experimental
components.

Radiation damage to equipment and radiological
exposure of personnel must nevertheless be
distinguished. Quantities relevant to detector
performance, such as particle fluence and
total ionizing dose, do not directly determine
the exposure received by personnel during
maintenance. The latter depends on residual
radiation fields produced by radionuclides
generated during irradiation and subsequently
modified by radioactive decay. Consequently,
residual dose rates depend on the irradiation
history, radionuclide inventory, material
composition, shielding, geometry, and elapsed
cooling time.

This distinction becomes particularly important
for components situated close to the interaction
points and beam lines. In ATLAS, activation
studies have identified the beam pipe, nearby
collimators and shielding elements, inner
detector components, and endcap calorimeters
as important objects for radiation-protection
assessment. The ATLAS Phase-II upgrade studies
explicitly recognize that interventions near
activated detector components require
configuration-dependent radiation calculations,
dose maps, and optimization according to the
ALARA principle
\cite{ATLAS:2015PhaseIIScoping}.

The relevance of detector geometry is
particularly clear in CMS. Its large
movable structures permit access to
internal components by opening the
detector, but these movements can
substantially change the spatial
distribution of residual radiation.
Published activation studies demonstrate
that radiological conditions in accessible
areas depend strongly on the positions
of the endcap structures and associated
shielding. For one simulated configuration
after a cooling interval of one month,
opening an endcap was found to increase
the residual ambient dose-equivalent
rate at floor level by approximately
an order of magnitude
\cite{Lorenzon:2024CMSActivation}.
This is a scenario-specific result,
rather than a universal factor
applicable to all CMS configurations
or future irradiation conditions.

The importance of detailed activation
calculations is further illustrated
by studies of the CMS High Granularity
Calorimeter. The large mass of activated
materials, their composition, and their
location within the experimental
installation require dedicated
assessments of residual radiation
fields and the radiological
classification of components.
Such studies provide information
needed for maintenance planning,
material handling, and the
management of activated equipment
\cite{Bozzato:2023LHCExperiments}.

The ATLAS and CMS examples demonstrate
that the radiological state of an
experimental cavern cannot be
characterized solely by the time
elapsed since the end of beam
operation. Detector geometry,
the positions of movable structures,
shielding arrangements, and the
location of an intervention can
be equally important. An area
accessible under one detector
configuration may require additional
restrictions or protection measures
when the configuration changes.

For this reason, radiation-protection
planning must consider the sequence
of configurations required to perform
an intervention. Access to an inner
detector component may involve removal
of shielding, movement of large
detector structures, disconnection
of services, temporary storage of
activated equipment, and subsequent
restoration of the original or
a modified configuration. Each stage
may produce different radiological
conditions, while the mechanical,
electrical, and other technical
hazards associated with the
intervention must also remain
controlled.

Radiological challenges are not confined
to the underground experiments.
An important complementary example
is provided by the LHC collimation
system. Collimators intercept particles
outside the required beam envelope,
preventing excessive energy deposition
in vulnerable accelerator components.
However, the interaction of lost
particles with collimator jaws and
surrounding materials generates
secondary radiation and induces
radioactivity in the affected
equipment.

The beam-cleaning insertions IR3 and IR7,
responsible primarily for momentum
and betatron cleaning, respectively,
are particularly relevant in this
context. Since significant beam
losses are deliberately concentrated
in these regions, their collimators
and neighboring components can
develop substantial residual
radiation fields. Radiation-transport
and activation studies of the LHC
cleaning insertions have explicitly
demonstrated the importance of
estimating residual dose rates
for maintenance and repair
interventions
\cite{BruggerRoesler:2005CollimatorDose}.

The HL-LHC collimation system must
accommodate the increased beam
brightness and stored energy while
maintaining effective protection
of superconducting magnets and
other sensitive accelerator
components. Its development
includes upgraded collimator
designs, new materials, and
additional protection arrangements
in selected accelerator regions
\cite{Redaelli:2020HLLHCCollimation}.
These modifications have a
radiological dimension in addition
to their machine-protection
function. The distribution of
beam losses, material composition,
irradiation history, and geometry
influence the activation of
collimators and nearby equipment
and thus the conditions under
which future maintenance
can be performed.

Unlike the collision-induced
radiation fields near ATLAS and
CMS, the accumulated irradiation
of the principal beam-cleaning
insertions is governed primarily
by the history and spatial
distribution of beam losses.
Their residual activation
therefore cannot generally
be inferred from integrated
luminosity using
Eq.~\eqref{eq:radiation_luminosity_scaling}.
Dedicated beam-loss assumptions,
particle-transport calculations,
activation models, and cooling-time
scenarios are required to predict
the relevant radiation fields.

The maintenance of activated
collimators presents specific
technical difficulties. Direct
interventions may require
prolonged access to regions
with elevated residual dose
rates, while component replacement
involves mechanical handling,
vacuum connections, cooling
services, electrical interfaces,
and precise alignment.
CERN experience has therefore
motivated the development of
remote inspection, radiation
measurement, and handling
systems, including dedicated
equipment for collimator exchange
\cite{Kershaw:2013RemoteHandling}.
These developments illustrate
how radiation protection can
influence the construction
and installation of accelerator
components before maintenance
becomes necessary.

The corresponding challenge extends
to other beam-intercepting devices,
including absorbers and beam-dump
components. Their irradiation
and activation may create
substantial constraints on
inspection, replacement,
dismantling, and radioactive
material management. Published
studies of LHC beam-dump
interventions have demonstrated
the use of detailed radiation
transport and activation
calculations to optimize work
sequences and estimate
occupational exposure
\cite{Infantino:2023BeamDumpRP}.

These examples demonstrate that
radiation protection at the HL-LHC
must address both normal beam
operation and subsequent
interventions. During beam
operation, shielding, access
restrictions, monitoring, and
the controlled disposal of
beam losses contribute to
personnel and environmental
protection. During technical
stops and shutdowns, induced
radioactivity becomes an
important determinant of
accessibility, maintenance
conditions, and the radiological
management of materials.
The relevant requirements
therefore change with the
operational mode and the
evolving physical state
of the installation.
A common principle governing
these activities is the
optimization of radiological
protection according to
ALARA~\cite{FassoRokni:2009ALARA,
ICRP:2021Publication147,
CERN:RadiationProtectionCode}.

For the HL-LHC, ALARA is relevant
not only to the execution of
individual interventions but
also to the design of equipment,
shielding, underground
infrastructure, and maintenance
arrangements. The HL-LHC
technical design report (TDR) explicitly
incorporates radiation-protection
optimization into the design
of upgraded equipment and
facilities, including the
assessment of anticipated
interventions in activated
regions
\cite{Otto:2020HLLHCSafety}.

In practical terms, ALARA
optimization may involve
selecting appropriate cooling
intervals, minimizing unnecessary
work in radiation areas,
preparing and testing intervention
procedures outside activated
zones, installing temporary
shielding, improving access
geometry, and employing
remote-handling techniques.
The objective is not simply
to minimize intervention
duration, since a shorter
operation may involve a
greater dose rate or other
hazards. The relevant
assessment must consider
the expected individual
and collective exposures,
the feasibility of the
proposed measures, and their
compatibility with the
other protection
requirements.

The radiation-protection
challenges associated with
the HL-LHC upgrades at
Points~1 and~5 have been
examined using numerical
tools for radiation
transport, induced activation,
and intervention assessment.
These studies address
dismantling and replacement
of activated equipment,
anticipated residual
radiation fields,
material characterization,
and the optimization
of occupational exposure
during complex technical
operations
\cite{Elie:2023HLRadiationProtection,
Otto:2020HLLHCSafety}.
The same general methodology
is applicable to work
planning in the beam-cleaning
insertions and other
activated accelerator
regions.

Within the state-space
framework developed in
the present work, the
radiological conditions
associated with an
intervention contribute
to determining whether
the relevant configurations
and transitions remain
admissible. For an
experiment $\alpha$,
the intervention domain
$\Omega_{\rm int}^{\alpha}
(z,w;m,\boldsymbol\Xi_{\alpha})$
depends on the applicable
technical configuration,
including shielding and
detector arrangements,
and on the physical
conditions relevant
to the work.

The corresponding principles
extend to accelerator
interventions. Although
the experiment-specific
notation need not be
transferred unchanged
to the accelerator,
the underlying requirement
is the same: the physical
state, technical
configuration, operational
mode, and credited
protection measures
must remain compatible
throughout the intervention
trajectory. A configuration
that is radiologically
acceptable before the
movement of shielding
or activated equipment
may require reassessment
after the movement.

This interpretation provides
a direct extension of the
radiation-work planning
discussed in
Sec.~\ref{subsec:radiation_work_planning}.
The expected exposure
associated with a particular
activity depends on the
worker's position and time
in the radiation field,
the sequence of operations,
and the applicable
protection measures.
Radiological optimization
may therefore involve
not only waiting for
radioactive decay,
but also modifying
the order of interventions,
changing detector or
shielding configurations,
preparing components
outside activated areas,
and using remote-handling
techniques.

The long-term significance
of these considerations
extends beyond occupational
exposure. If an activated
detector or accelerator
component cannot be
accessed or replaced
within a reasonable
intervention period,
the corresponding
technical limitation
may affect the
availability and
scientific performance
of the facility.
Maintainability therefore
contributes directly
to the scientific
continuity discussed
in Sec.~\ref{subsec:safety_scientific_continuity}.

The development of future
facilities should consequently
include representative
intervention scenarios
already at the design stage.
Equipment expected to
require maintenance after
substantial irradiation
should, where practicable,
be provided with accessible
connections, appropriate
lifting and alignment
interfaces, compatible
shielding, and provisions
for remote or partly
remote operations.
These measures can reduce
occupational exposure
while improving the
predictability of
maintenance and
restoration activities.

Such provisions must
nevertheless remain
compatible with other
protection requirements.
Movement of activated
components may involve
substantial mechanical
loads, restricted
clearances, electrical
isolation, cooling
disconnections, and
temporary modifications
of the protection
configuration.
A radiologically
preferable intervention
is not automatically
the safest overall;
the combined hazards
must be assessed
throughout the
operation.

The HL-LHC thus provides
two complementary examples
of the reciprocal
relationship between
physics and safety.
In ATLAS and CMS,
collision-induced
radiation and activation
influence detector
architecture, shielding,
maintenance access,
and component replacement.
In the accelerator,
beam-loss distributions
and the resulting
activation of collimators
and other beam-intercepting
devices influence
equipment design,
remote-handling
capabilities, and
intervention planning.

In both cases, particle
transport, activation
calculations, dosimetry,
and operational
measurements provide
the physical basis
for predicting
radiological conditions.
Conversely, ALARA
requirements and
maintainability
considerations influence
the physical design
of the installation,
the available
intervention techniques,
and the trajectories
through which equipment
can be accessed,
repaired, replaced,
and returned to operation.

The resulting lesson
for future facilities
is that radiation
protection must be
integrated with
accelerator and
detector design
from the outset.
Its objective is
not only to limit
exposure during
individual activities,
but also to preserve
the possibility of
safely maintaining,
upgrading, and
recovering increasingly
complex experimental
and accelerator systems
throughout their
scientific lifetime.

\subsection{Designing for failure, not only for nominal operation}
\label{subsec:future_design_failure}

The design of a large scientific facility is necessarily driven
by its intended operating conditions and scientific performance.
However, the adequacy of its safety architecture cannot be
established solely by demonstrating that the nominal
configuration satisfies the applicable technical limits.
Abnormal transitions may involve physical conditions,
combinations of subsystem states, and rates of change
that differ substantially from those encountered during
routine operation. Future facilities must therefore be
designed not only for the states in which scientific
measurements are performed, but also for the credible
failure trajectories through which those states may
become hazardous.

This requirement follows directly from the hazard-chain
framework introduced in
Sec.~\ref{subsec:hazard_chains}. An initiating event
$F_i$ may produce several trajectories
$\Gamma_j^{(i)}$, depending on the initial facility
state, the technical configuration, and the response
of preventive and mitigative barriers. The corresponding
consequences may differ even when the initiating
mechanism is identical. The essential design problem
is therefore to identify which abnormal trajectories
are physically credible, how they propagate through
coupled subsystems, and which protection functions
are required to prevent or limit their consequences
\cite{Sklet:2006SafetyBarriers,
DeDianousFievez:2006ARAMIS,
Leveson:2004AccidentModel}.

The physical properties of an energetic system
determine not only the magnitude of a possible
release, but also its characteristic timescales,
spatial distribution, and propagation mechanisms.
An accelerator beam may deposit its energy in
sensitive material over an extremely short interval,
while a superconducting magnet fault may involve
electrical, thermal, and mechanical processes
evolving over different timescales. A cryogenic
failure may produce pressure transients, substantial
helium mass flows, and changes in underground
environmental conditions. The relevant protection
requirements consequently depend on the entire
failure evolution rather than exclusively on
the nominal stored energy or material inventory.

The 2008 LHC sector~3--4 incident provides
a particularly instructive example. An electrical
fault in a superconducting magnet interconnection
developed into an event involving mechanical
damage, helium release, pressure loading of
the insulation-vacuum enclosure, and damage
to accelerator infrastructure. The subsequent
reassessment of the cryogenic protection
architecture demonstrated the importance of
considering coupled failure mechanisms,
pressure-relief capacity, and the physical
consequences of helium discharge
\cite{Chorowski:2011RiskUpdate}.
As discussed in
Sec.~\ref{subsec:cryo_empirical_feedback},
the resulting lessons extended beyond
the modification of individual components:
they required reconsideration of the
credible failure scenarios and the
protection arrangements associated with them.

For future facilities, such analyses should
be undertaken sufficiently early to influence
the physical architecture of the installation.
The location and segregation of energetic
equipment, the geometry of pressure-relief
paths, the distribution of cryogenic inventories,
the positioning of shielding, the arrangement
of electrical circuits, and the provision
of protected access or evacuation routes
can all affect the development and
consequences of abnormal transitions.
Once civil engineering and major technical
systems have been installed, modifications
to these features may be technically
difficult or incompatible with the
existing infrastructure.

An important design principle is therefore
to reduce the possibility that a local
failure propagates into several otherwise
independent systems. Physical segregation,
segmentation of technical services,
appropriate energy-extraction or relief
paths, and limitation of the inventories
involved in an individual failure scenario
can reduce the extent of such propagation.
These provisions complement active
detection and interlocks by modifying
the physical conditions under which
an abnormal event can develop.
Their effectiveness must nevertheless
be assessed for the relevant failure
mechanisms rather than assumed from
their presence alone.

The distinction between preventive and
mitigative barriers developed in
Sec.~\ref{subsec:preventive_mitigative}
is particularly relevant at the
design stage. Prevention seeks to
reduce the occurrence of initiating
events or interrupt hazardous
developments before they reach
critical conditions. Mitigation
limits the consequences of an
event that has already occurred.
These functions are complementary,
but they are not generally
interchangeable. A fast beam
interlock cannot replace adequate
passive protection against failure
scenarios developing faster than
the available detection and
extraction time. Similarly,
cryogenic pressure relief cannot
replace measures intended to
prevent electrical faults,
although it may limit the
consequences of the resulting
helium release.

The temporal dimension of
protection must also enter
the engineering design.
As established in
Sec.~\ref{subsec:protection_time},
an active protection function
is effective against a particular
hazardous trajectory only if
the required detection,
decision, and response can
be completed within the
available protection time.
For sufficiently rapid
failures, the protection
architecture may therefore
require passive energy
absorption, mechanical
containment, intrinsically
limited inventories, or
other measures that do not
depend on the completion
of a delayed control action.
The choice between these
measures depends on the
underlying physical
timescales and the
consequences to be prevented.

The HL-LHC illustrates the
importance of this approach
for new accelerator systems.
Its upgraded superconducting
magnets, crab cavities,
powering arrangements, and
collimation systems introduce
technical configurations
whose abnormal behaviour
must be assessed together
with the existing machine
protection infrastructure
\cite{Apollinari:2017HLLHC,
Wollmann:2020HLLHCMPS}.
For example, a crab-cavity
fault may modify the
transverse beam dynamics,
but the resulting loss
trajectory depends on the
actual time evolution of
the cavity fields and
the subsequent beam response.
Physically realistic
failure models are therefore
essential for determining
whether additional
protection measures are
required and for specifying
their necessary response
times
\cite{Apsimon:2019CrabQuench}.

A corresponding requirement
arises in the experimental
detectors. ATLAS and CMS
contain closely coupled
electrical, cooling,
mechanical, and protection
systems. A loss of cooling,
for example, can produce
different consequences
depending on the powering
state of the associated
electronics, the response
of temperature monitoring,
and the availability of
automatic equipment
protection. Detector
movements can additionally
alter service connections,
mechanical clearances,
shielding configurations,
and personnel-access
conditions. Protection
against such events
therefore requires
consideration of the
complete experimental
configuration rather
than an isolated
subsystem
\cite{Schmeling:2004DSS,
DiCalafiori:2020CMS_ECAL}.

These examples also
illustrate the importance
of barrier independence
and common-cause failures,
discussed in
Sec.~\ref{subsec:barrier_independence}.
Several protective functions
cannot be regarded as
independent merely because
they are implemented by
different devices.
They may share electrical
supplies, cooling systems,
communication networks,
sensors, environmental
conditions, or supporting
infrastructure. An initiating
event that compromises
these shared resources
may simultaneously
reduce the effectiveness
of several nominally
separate protection
layers.

Consequently, future
protection architectures
should be assessed against
credible combinations
of equipment failures,
loss of supporting services,
and abnormal environmental
conditions. The objective
is not to assume the
simultaneous occurrence
of every conceivable
failure, but to identify
physically plausible
dependencies and
common-cause mechanisms
that could invalidate
the intended protection
strategy. Physical
diversity and independent
protection paths may
provide important
advantages, provided
their claimed independence
can be demonstrated
under the relevant
conditions
\cite{Sklet:2006SafetyBarriers,
Leveson:2012SaferWorld}.

The same principle applies
to the interaction
between equipment
protection and personnel
safety. A protective
response that successfully
limits damage to an
accelerator or detector
does not necessarily
establish acceptable
conditions for personnel.
For example, the discharge
or venting of a cryogenic
inventory may protect
equipment against excessive
pressure while generating
a helium release that
requires appropriate
ventilation, oxygen-deficiency
protection, or access
restrictions. Likewise,
the isolation of electrical
equipment may prevent
further damage without
eliminating residual
mechanical, thermal,
or radiological hazards.

The design assessment must
therefore consider the
different consequences
associated with each
credible trajectory,
including personnel
injury, equipment damage,
environmental effects,
and loss of scientific
capability. This is
consistent with the
multidimensional
consequence representation
introduced in
Sec.~\ref{subsec:hazard_chains}.
An arrangement that
reduces one component
of the consequences
may introduce or modify
another, requiring
assessment of the
complete protection
configuration.

Uncertainty constitutes
a further design
consideration. The
physical evolution
of a failure may
depend on quantities
that cannot be
determined precisely
during the initial
design phase, including
material properties,
energy-deposition
distributions, failure
rates, transient flow
conditions, and the
performance of protection
functions under abnormal
loads. As discussed in
Sec.~\ref{subsec:safe_phase_space_scaling},
robust design should
account for credible
uncertainties rather
than relying exclusively
on nominal model
parameters.

Conservative assumptions
can provide important
protection margins,
but their selection
must remain physically
justified. Excessively
simplified worst-case
models may obscure
the actual governing
failure mechanisms,
while insufficiently
conservative models
may underestimate
critical propagation
paths. A combination
of analytical
calculations, numerical
simulations, targeted
experiments, operational
experience, and
subsequent commissioning
is therefore necessary
to establish confidence
in the resulting
protection design.

The validation of
failure protection
also deserves attention.
Some protective functions
can be tested repeatedly
under operational
conditions, whereas
others are intended
to respond to rare
events that cannot
be reproduced in
their full severity
without unacceptable
consequences. Verification
may then require a
combination of component
tests, subsystem
experiments, validated
simulation models,
diagnostic monitoring,
and periodic checks
of the protection
chain. The design
should provide for
such verification
from the outset,
including the
observability and
testability of
safety-critical
functions discussed
in Sec.~\ref{subsec:testing_observability}.

From the perspective
of the safe operational
phase-space framework,
designing for failure
means establishing
not only the domain
of states compatible
with the intended
scientific operation,
but also the physical
and technical
conditions that
prevent credible
abnormal trajectories
from reaching
unacceptable
consequences.
This requirement
includes the ability
to detect deviations,
interrupt hazardous
propagation, contain
or dissipate stored
energy, and establish
controlled conditions
for subsequent
intervention and
recovery.

The resulting design
principle extends
beyond the LHC and
HL-LHC. Future
circular colliders
and other underground
experimental facilities
will require protection
architectures adapted
to their particular
energy inventories,
technologies,
infrastructure
topology, and
failure dynamics.
The relevant lesson
from existing
facilities is not
that a specific
collection of
interlocks or
barriers can be
transferred unchanged,
but that credible
failure trajectories
must be incorporated
into the physical
and operational
design from its
earliest stages.

Designing for failure
is therefore an
essential component
of \emph{safety in
physics}. It
influences the
geometry and
architecture of
scientific
installations,
the organization
of protection
systems, and
the configurations
in which the
facility can
subsequently be
operated and
maintained.
Conversely,
\emph{physics in
safety} provides
the models and
measurements
required to
understand
abnormal energy
transfer, failure
propagation,
and the response
of protective
barriers.
Together, these
two perspectives
establish a basis
for designing
future facilities
whose scientific
performance remains
compatible with
credible failure
conditions and
the protection
requirements
associated with
them.

\subsection{Future circular colliders and the scaling of underground
infrastructure}
\label{subsec:fcc_safety}

The Future Circular Collider (FCC) provides a particularly
instructive example of how the physical scale and spatial
distribution of a scientific installation influence its
safety architecture. The FCC feasibility studies consider
an underground accelerator tunnel of approximately
$91~{\rm km}$ circumference, with eight surface sites,
twelve permanent access and service shafts, and eight
sectors of approximately $11.3~{\rm km}$ each. The
proposed infrastructure is intended to support a staged
scientific programme, beginning with an electron--positron
collider (FCC-ee) and potentially followed by a
high-energy hadron collider (FCC-hh)
\cite{Benedikt:2025FCCFeasibility,
Benedikt:2025FCCCivilEngineering}.

These dimensions introduce safety considerations that
cannot be obtained by simply scaling the corresponding
LHC arrangements according to tunnel circumference.
The distribution of access points, the topology of
underground passages, the positions of experimental
and technical areas, and the availability of protected
routes can influence personnel safety independently
of the stored energies and radiation fields of the
accelerator. The design problem therefore involves
both the physical hazards associated with accelerator
operation and the spatial organization of the
infrastructure required to control their consequences.

A particularly important example concerns personnel
evacuation and emergency intervention. In an extensive
underground installation, the time required to reach
a protected location depends not only on the distance
from the nearest surface shaft, but also on the
available routes, personnel transport, access
restrictions, environmental conditions, and possible
obstruction of passages. During a fire, helium release,
or other abnormal event, the shortest geometrical
route may not remain usable. Evacuation planning
must therefore consider credible changes in the
availability and environmental conditions of
underground routes, rather than relying exclusively
on nominal travel distances.

The distinction between evacuation to a protected
underground location and subsequent access to the
surface is particularly important. Depending on
the hazard scenario, reaching a protected compartment
or refuge may provide the immediate means of
establishing acceptable conditions for personnel,
while complete evacuation to the surface may require
additional time and transport arrangements.
The relevant design requirements consequently involve
the compatibility of detection and alarm times,
the available escape or refuge arrangements,
the evolution of hazardous conditions, and the
response capabilities of emergency personnel.

For the FCC, the underground geometry and the
distribution of access shafts should therefore
be considered together with compartmentalization,
ventilation, smoke management, emergency lighting,
communication systems, personnel transport, and
the provision of protected electrical supplies.
The objective is not merely to provide sufficient
nominal capacity for these services, but to ensure
that their required functions remain available
under the credible abnormal conditions for which
they are intended
\cite{Benedikt:2025FCCFeasibility,
Benedikt:2025FCCCivilEngineering}.

The spatial distribution of the infrastructure
also introduces possible common dependencies.
Several underground sectors may rely on shared
electrical distribution, communication networks,
cooling services, or ventilation infrastructure.
An abnormal event affecting one of these services
can modify operating or access conditions in
regions that are physically separated from the
initiating event. Consequently, geographical
separation does not necessarily imply functional
independence of the corresponding protection
arrangements.

This consideration connects directly with the
discussion of barrier independence in
Sec.~\ref{subsec:barrier_independence}.
Redundancy and segmentation can improve protection,
but their effectiveness depends on the absence,
or adequate control, of relevant common-cause
failure mechanisms. A distributed system may
contain several nominally independent technical
installations while retaining dependencies on
common power supplies, control networks,
transport routes, or environmental conditions.
The corresponding safety assessment must therefore
consider both physical separation and the actual
functional dependencies between subsystems.

Cryogenics provides a further illustration of
the interaction between infrastructure topology
and physical hazards. The FCC programme
envisages different cryogenic requirements
for the FCC-ee and FCC-hh stages. In FCC-ee,
superconducting radiofrequency installations
constitute important cryogenic users, alongside
the requirements associated with experimental
magnets and interaction-region systems.
A possible subsequent FCC-hh would introduce
a substantially different cryogenic configuration
associated with extensive superconducting
magnet systems
\cite{Delprat:2025FCCCryogenics,
Benedikt:2025FCCFeasibility}.

The design of these cryogenic installations
must account for helium inventories,
distribution and recovery systems, pressure
relief, isolation possibilities, and the
consequences of credible release scenarios.
In an underground environment, the relevant
consequences can extend beyond equipment
protection to include oxygen displacement,
ventilation conditions, personnel access,
and emergency response. The geographical
distribution of cryogenic equipment and
the possibility of isolating affected
sections consequently influence both
the propagation of abnormal conditions
and the available protective responses.

A design based on distributed or segmented
technical systems can limit the extent
of certain failures, provided that the
segmentation remains effective under
the relevant conditions. However, the
presence of multiple sectors does not
automatically establish independent
containment of cryogenic or other hazards.
Pressure-relief paths, ventilation
connections, service penetrations,
and shared infrastructure may permit
physical interactions between regions.
These interfaces must therefore be
included in the analysis of credible
failure trajectories.

Radiation protection also depends on
the operating stage and the corresponding
accelerator technology. During FCC-ee
operation, the relevant radiation
environment will be influenced by
high-intensity electron and positron
beams, synchrotron radiation, beam
losses, and particle interactions.
A possible future FCC-hh stage would
introduce different beam energies,
stored-energy inventories, radiation
fields, and activation conditions.
Consequently, the radiation-protection
requirements of the two stages cannot
be assumed to be identical merely
because they would occupy the same
underground infrastructure
\cite{Benedikt:2025FCCFeasibility}.

The implications extend to maintenance
and long-term accessibility. Areas
designed for routine access during
one operational stage may require
different restrictions, shielding,
or intervention arrangements during
another. The location of replaceable
components, transport routes, temporary
storage areas, and provisions for
handling activated equipment can
therefore influence the ability to
adapt the installation to subsequent
scientific requirements. As discussed
in Sec.~\ref{subsec:future_radiation_maintainability},
these considerations should be
addressed through radiation-protection
and ALARA assessments already at
the design stage.

The staged FCC concept introduces
a further dimension of configuration
control. Although the proposed
underground infrastructure is intended
to accommodate successive collider
technologies, the transition from
FCC-ee to FCC-hh would involve
substantial changes in accelerator
equipment, technical services,
protection systems, and experimental
installations. It should therefore
not be regarded simply as a change
of accelerator operating mode.
The relevant physical and technical
configurations, credible failure
scenarios, and protection requirements
would have to be reassessed for the
subsequent installation.

This aspect is particularly relevant
to the safe operational phase-space
framework developed in the present
work. The physical configuration
of the underground installation
influences the safe domain through
the available access arrangements,
environmental protection,
infrastructure dependencies,
and paths for the propagation
of abnormal events. The applicable
technical configuration
$\boldsymbol\Xi$ includes the
distribution and availability
of infrastructure services and
protection arrangements, while
the coupled state $\boldsymbol X$
describes the corresponding
physical and operational conditions.

The admissibility of a complete
facility configuration therefore
depends not only on the individual
states of the accelerator,
detectors, and technical
installations, but also on
their compatibility with
the underground infrastructure.
For example, a cryogenic
operating condition may be
acceptable for the equipment
while imposing additional
requirements on ventilation
and occupancy. Similarly,
a maintenance operation
may require temporary
changes in electrical
distribution or access
arrangements that are
incompatible with the
normal operating
configuration of a
neighboring subsystem.

The distinction between
admissible states and
admissible trajectories
is equally important.
A planned intervention
may involve transport
through several underground
sectors, temporary
isolation of technical
services, removal of
shielding, and subsequent
restoration of equipment.
Each intermediate
configuration must
remain compatible
with the applicable
personnel and equipment
protection requirements.
A safe initial
configuration and
a safe final
configuration do
not, by themselves,
establish that the
complete intervention
trajectory is admissible.

The same reasoning
applies to emergency
response. The response
to a localized event
may require transitions
in ventilation,
electrical distribution,
access control,
personnel transport,
or other technical
systems. These
transitions can
modify conditions
elsewhere in the
installation and
must therefore be
considered as part
of the protection
architecture rather
than as independent
operational actions.
The effectiveness
of emergency provisions
depends on the
compatibility of
their responses
under the credible
abnormal conditions
being addressed.

Another consequence
of the FCC scale
concerns the
coordination of
distributed technical
and operational teams.
Maintenance,
commissioning,
equipment testing,
and intervention
activities may take
place simultaneously
at widely separated
locations while
sharing infrastructure
services or protection
conditions. The
coordination principles
developed in
Sec.~\ref{subsec:simultaneous_activities}
therefore acquire
an additional
spatial dimension.
Distance between
worksites may reduce
some direct interactions
but does not eliminate
dependencies transmitted
through common technical
systems.

The FCC feasibility
studies also demonstrate
that the physical
configuration of the
underground infrastructure
is influenced by
geological, environmental,
territorial, and
civil-engineering
constraints
\cite{Benedikt:2025FCCCivilEngineering}.
These considerations
affect the possible
locations of surface
sites, shaft depths,
tunnel alignment,
and underground
technical areas.
Consequently, the
protection architecture
must be integrated
with civil-engineering
design rather than
developed independently
after the basic layout
has been selected.

This integration
provides a direct
connection to the
discussion of
scientific continuity
in
Sec.~\ref{subsec:safety_scientific_continuity}.
The long-term
scientific value
of an installation
intended to support
several generations
of accelerator
technology depends
on its adaptability,
maintainability,
and recoverability.
A configuration
optimized exclusively
for the initial
scientific stage
may impose substantial
constraints on later
upgrades or the
introduction of
new protection
requirements.

For future facilities,
the design objective
is therefore not
simply to enlarge
the physical
installation while
preserving familiar
operational practices.
It is to establish
an infrastructure
whose geometry,
technical segmentation,
protection systems,
access arrangements,
and intervention
capabilities are
compatible with
the intended
scientific programme
and its credible
evolution.

The FCC consequently
provides a particularly
clear illustration
of \emph{safety in
physics}. The physical
layout and topology
of the facility
determine important
conditions for
personnel protection,
hazard containment,
emergency response,
maintenance, and
operational recovery.
Conversely, these
safety requirements
influence the
selection and
organization of
the underground
infrastructure
itself.

The transferable
lesson is that
the safe operational
phase space of a
large scientific
facility is partly
determined by
its civil-engineering
and infrastructure
architecture.
The topology of
the installation
affects not only
the states in
which scientific
operation can be
permitted, but
also the trajectories
through which
personnel, equipment,
and technical
systems can move
safely between
different operational
conditions.

\subsection{Muon colliders: new physical hazards require new
protection configurations}
\label{subsec:muon_collider_safety}

A future high-energy muon collider provides a particularly
instructive example of how the choice of accelerator
technology can modify the physical mechanisms that
determine safety requirements. Unlike the HL-LHC,
which extends an established proton-collider
technology, or the staged FCC concept, whose
different accelerator configurations would
occupy a common underground infrastructure,
a muon collider introduces a distinctive
combination of high-power particle production,
ionization cooling, rapid acceleration,
superconducting magnets, and radiation
generated by the decay of circulating muons.
These features require protection arrangements
adapted to the corresponding physical
processes rather than direct reproduction
of those developed for existing facilities
\cite{Accettura:2024MuonCollider}.

An important characteristic is the finite
muon lifetime. The mean lifetime at rest,
approximately $2.2~\mu{\rm s}$, is increased
by relativistic time dilation at high energy,
but remains sufficiently short that a large
fraction of the circulating particles decay
during the operational cycle. The principal
decay channels,
$\mu^-\rightarrow e^-+\bar{\nu}_e+\nu_\mu$
and their charge-conjugate counterparts,
produce energetic electrons or positrons
and neutrinos. Their consequences for
equipment protection, detector operation,
and environmental radiation protection
constitute one of the defining features
of the muon-collider concept.

The production of intense muon beams
introduces a first group of hazards.
Current muon-collider concepts envisage
a high-power proton beam incident on
a production target, followed by
capture and transport of secondary
particles and formation of the
required muon beam. The target
region must withstand intense
energy deposition, thermal
and mechanical loading, and
substantial irradiation.
Its surrounding capture
solenoids must operate under
conditions involving large
magnetic fields and significant
secondary-particle radiation
\cite{Accettura:2024MuonCollider,
Manczak:2024MuonTargetRadiation}.

The protection of this region
requires consideration of the
coupled evolution of beam,
target, magnet, shielding,
and cooling conditions.
A failure of the incident
beam-delivery or target-cooling
system may produce abnormal
energy deposition, while
secondary radiation can
generate additional thermal
loads and radiation damage
in neighboring superconducting
components. Radiation shielding
must therefore protect the
magnets against both excessive
instantaneous energy deposition
and accumulated degradation
of superconductors and
electrical insulation.

These requirements have direct
consequences for the physical
design of the target station.
Shielding geometry, extraction
of the residual proton beam,
cooling arrangements, and
the accessibility of activated
components must be considered
together rather than as
independent design problems.
Detailed radiation-transport
studies of the proposed
production region have
examined the radiation
loads on capture solenoids
and the possible extraction
of protons passing through
the target
\cite{Manczak:2024MuonTargetRadiation}.

The substantial activation
expected in the target and
front-end regions also makes
maintainability a fundamental
design requirement.
Replacement of targets,
shielding elements, or
irradiated nearby equipment
may require dedicated
remote-handling systems,
protected transport routes,
cooling and temporary
storage areas, and facilities
for treating radioactive
materials. The International
Muon Collider Collaboration
has identified the need
to incorporate such
possibilities into the
development of the
infrastructure
\cite{Accettura:2024MuonCollider}.
This provides a direct
extension of the radiation
protection and ALARA
considerations discussed
in Sec.~\ref{subsec:future_radiation_maintainability}.

The muon cooling complex
introduces additional
technological dependencies.
Ionization cooling involves
energy loss in material
absorbers, followed by
reacceleration and
appropriate magnetic
focusing. Depending
on the selected design,
the cooling system may
employ liquid-hydrogen
or other absorber
materials and high-gradient
radiofrequency cavities
operating in strong
magnetic fields.
The feasibility of
ionization cooling has
been demonstrated
experimentally, while
the parameters and
engineering arrangements
of a complete collider
cooling complex remain
subjects of development
\cite{MICE:2024IonizationCooling,
Accettura:2024MuonCollider}.

Where hydrogen is employed,
its use introduces
additional requirements
concerning containment,
leak detection, ventilation,
fire and explosion
prevention, electrical
equipment, and compatibility
with nearby cryogenic,
radiofrequency, and magnetic
systems. The corresponding
protection architecture
must be defined for the
actual material inventory
and technical configuration.
These hazards should not
be attributed universally
to every possible cooling
design, since the relevant
materials and engineering
solutions remain
configuration-dependent.

A second distinctive
challenge arises from
muon decay in the
acceleration systems
and collider ring.
The energetic electrons
and positrons produced
by these decays can
initiate electromagnetic
showers in the beam
pipe, absorbers,
magnets, and surrounding
equipment. The resulting
energy deposition
can produce substantial
distributed heat loads,
radiation damage, and
potential quench conditions
in superconducting
components. The protection
problem therefore differs
from one dominated
exclusively by occasional
localized beam losses:
a significant radiation
and thermal load is
intrinsic to the
normal circulation
of muon beams.

For the multi-TeV
collider concepts
under study, published
calculations indicate
that the power carried
by decay electrons
and positrons can
reach several hundred
watts per metre
of accelerator,
depending on the
beam parameters
\cite{Lechner:2024MuonMagnetShielding}.
This scale illustrates
why shielding and
heat removal must
be incorporated directly
into the superconducting
magnet architecture.

Dedicated shielding
inside the magnet
aperture can intercept
a substantial fraction
of the decay-induced
particle showers
before they reach
the superconducting
coils. However,
the thickness and
geometry of this
shielding influence
the required magnet
aperture, field quality,
thermal design,
cryogenic load,
and mechanical
integration. The
shielding must therefore
be optimized together
with the magnetic,
vacuum, and cryogenic
systems rather than
added after the
magnet configuration
has been established
\cite{Lechner:2024MuonMagnetShielding}.

The machine--detector
interface presents
another important
example of coupled
design requirements.
Muon decays upstream
of the interaction
point produce secondary
particles that can
reach the experimental
detector and generate
substantial beam-induced
backgrounds. The
radiation environment
depends on the collider
optics, the positions
and apertures of
nearby accelerator
elements, the properties
of the secondary
showers, and the
shielding configuration.

Studies of multi-TeV
muon-collider detectors
have therefore considered
specialized shielding
arrangements, including
conical absorbers
near the interaction
region, to reduce
the background
entering the sensitive
detector volume
\cite{Lucchesi:2024MuonMDI}.
Such arrangements
influence detector
acceptance and
the available
space for accelerator
components and
experimental services.
The scientific
performance of
the detector and
its radiation-protection
configuration must
therefore be
optimized jointly.

This represents a
particularly direct
illustration of
\emph{safety in
physics}. A shielding
configuration selected
to reduce radiation
damage and improve
equipment protection
can also modify the
angular acceptance
and background
conditions of the
experiment. Conversely,
the scientific
requirements of the
detector constrain
the geometrical
solutions available
for the protection
of the interaction
region.

A third radiological
challenge, qualitatively
different from the
activation of accelerator
equipment, concerns
the high-energy
neutrinos produced
by muon decay.
Because the muons
are highly relativistic,
the resulting
neutrino emission
is strongly
forward-directed.
Neutrinos can
propagate through
substantial thicknesses
of matter and,
in sufficiently
intense beams,
their interactions
may generate
secondary-particle
showers outside
the accelerator
installation.

This effect introduces
a potential off-site
radiation-protection
constraint that
must be assessed
during collider
design and site
selection. Published
FLUKA studies have
examined effective
dose from neutrino
interactions in soil
for representative
muon-beam energies
and propagation
distances
\cite{Lerner:2024MuonNeutrinoDose}.
The relevant exposure
depends on the muon
energy, beam intensity,
collider geometry,
angular distribution
of the neutrinos,
and the location
of potentially
exposed areas.

Unlike conventional
secondary radiation
from beam losses,
the neutrino flux
cannot be effectively
contained through
ordinary local
shielding surrounding
the accelerator.
Its mitigation
therefore requires
attention to collider
depth, tunnel
alignment, lattice
geometry, the length
and orientation
of straight sections,
and possible methods
of redistributing
the neutrino flux.
These are design
variables that
directly influence
the radiological
conditions beyond
the underground
installation
\cite{Accettura:2024MuonCollider,
Lerner:2024MuonNeutrinoDose}.

The corresponding
radiation-protection
assessment must
distinguish between
conservative dose
estimates and
predictions for
a fully specified
facility. Actual
exposure depends
on the adopted
beam parameters,
accelerator lattice,
site geometry,
geological conditions,
and operational
scenario. The
existence of a
potential neutrino
radiation hazard
does not imply
that unacceptable
public exposure
is unavoidable.
Rather, it makes
the demonstration
of adequate
mitigation an
important requirement
of future
collider design.

These examples
illustrate that
the radiological
environment of
a muon collider
has several
distinct origins.
Target interactions
produce intense
localized radiation
and activated
equipment;
muon decays produce
distributed
radiation and
thermal loading
along the machine;
and decay neutrinos
can generate
radiological
constraints beyond
the accelerator
complex. Their
different physical
mechanisms require
correspondingly
different combinations
of shielding,
containment,
equipment protection,
operational restrictions,
and design-stage
mitigation.

Other hazards
remain relevant
as well.
High-field
superconducting
magnets, cryogenic
systems, high-power
radiofrequency
equipment,
electrical
distribution,
and potentially
large inventories
of specialized
materials require
protection against
credible electrical,
thermal, mechanical,
and environmental
failure scenarios.
The new radiation
conditions must
be considered
together with
these more familiar
hazards rather
than independently
of them.

Within the
safe operational
phase-space
framework developed
in this work,
the choice of
muon-collider
technology modifies
both the relevant
physical state
and the protection
constraints applicable
to its evolution.
The initiating
events $F_i$,
possible propagation
trajectories
$\Gamma_j^{(i)}$,
and corresponding
consequences depend
on the actual
configuration
of the production,
cooling, acceleration,
and collider systems.
Consequently, the
admissible domain
cannot be obtained
by simply transferring
the operating limits
and protection
assumptions of
the LHC.

The corresponding
protection arrangements
must account for
both abnormal
transitions and
hazards intrinsic
to nominal
operation.
The continuous
radiation and
thermal loading
associated with
muon decay, for
example, is not
an accidental
failure trajectory.
It is a normal
physical process
whose consequences
must be controlled
by the design
of the machine.
Abnormal cooling,
shielding, magnet,
or beam conditions
may subsequently
modify these loads
and produce
additional hazardous
trajectories.

This distinction
reinforces the
importance of
integrated design.
Radiation shielding,
magnet aperture,
cryogenic heat
removal, beam optics,
detector acceptance,
maintenance access,
and off-site
radiation protection
are coupled
requirements.
A modification
that improves
one aspect of
the installation
may alter the
conditions under
which another
system can
operate safely.
The resulting
trade-offs must
therefore be
evaluated at
the level of
the complete
scientific
facility.

The lessons
drawn from
the LHC and
its underground
experiments remain
directly relevant
to this process,
although the
specific protection
architecture will
differ.
Physics-based
hazard identification,
independent and
complementary
protection functions,
configuration control,
ALARA-based intervention
planning, operational
monitoring, and
validation of
admissible transitions
provide a methodological
basis for developing
the new systems.
The design must
also incorporate
the maintainability
and recovery
requirements needed
to sustain operation
under the anticipated
radiation conditions.

The muon collider
thus illustrates
a central implication
of the present work:
the safe operational
phase space is
not an invariant
property shared
by all accelerator
technologies.
Its boundaries,
relevant physical
variables, and
admissible
trajectories
depend on the
specific processes
through which
particles are
produced,
accelerated,
stored, and
brought into
collision.

The transferable
lesson is therefore
not a particular
set of LHC
protection devices,
but the method
through which
physical understanding
is translated
into validated
protection
requirements.
In the case of
a future muon
collider, this
method must
incorporate
the distinctive
consequences
of muon decay,
the radiological
and technical
conditions of
high-power
muon production,
and the interactions
between machine,
detector,
environment,
and personnel
protection.

\subsection{Automation and observability}
\label{subsec:future_automation}

The increasing complexity of future accelerator and
experimental facilities will strengthen the dependence
on automated monitoring, control, diagnostics, and
protection systems. The large number of monitored
quantities, technical configurations, and interactions
between subsystems can make it difficult for human
operators to recognize and interpret hazardous
developments within the available protection time.
Automation is therefore an essential component
of the safety architecture of modern accelerator
facilities, particularly when abnormal physical
processes evolve too rapidly for direct human
intervention.

The LHC and its underground experiments already
provide extensive examples of this principle.
Beam-loss monitoring, beam interlocks, beam
extraction, superconducting magnet protection,
cryogenic control, and detector safety systems
rely on automated detection and response.
Their functions differ according to the
physical hazards being controlled, but they
share the requirement that relevant abnormal
conditions be recognized and that appropriate
actions be initiated within the available
protection time
\cite{Schmidt:2016MPS,
Puccio:2010BIS,
RodriguezMateos:2004QPS,
Schmeling:2004DSS}.

An important distinction must be maintained
between normal operational control and
autonomous protection, as established in
Sec.~\ref{subsec:interlocks_permits}.
Operational control establishes and maintains
the conditions required for the intended
scientific activity. Protection functions,
by contrast, respond to conditions that
threaten to invalidate the assumptions
under which continued operation is permitted.
The corresponding protective actions must
remain effective even when the normal control
system is unavailable or contributes to
the initiating failure.

The increasing sophistication of automated
control should therefore not lead to the
unnecessary integration of independent
protection functions into a single control
architecture. Functional separation,
appropriate diversity, and the management
of common dependencies remain important
design principles, particularly for
safety-critical functions whose failure
could produce severe consequences.
These requirements follow from the
discussion of barrier independence
in Sec.~\ref{subsec:barrier_independence}.

The effectiveness of automation also
depends on the interpretation of
measurements within the applicable
technical configuration. A measured
quantity does not necessarily possess
the same significance under all
operational conditions. For example,
the beam-loss signal requiring an
immediate beam dump may depend on
beam energy, integration time,
the position of the monitor,
and the operating configuration
of the accelerator. The corresponding
protection thresholds must therefore
be established and validated for
the relevant machine conditions
\cite{Holzer:2012BLM,
Dehning:2016BLM}.

A related situation occurs in the
underground experiments.
The response to a temperature
increase or loss of cooling
may depend on whether the
affected equipment is powered,
which detector configuration
is active, and which protective
functions are available.
Consequently, the interpretation
of individual measurements
must be consistent with the
relevant operating mode
and technical configuration
\cite{Schmeling:2004DSS,
DiCalafiori:2020CMS_ECAL}.

Within the state-space framework
developed in this work, monitoring
provides information about the
physical state $\boldsymbol X$,
while the interpretation of
that information depends on
the technical configuration
$\boldsymbol\Xi$ and the
operational mode $m$.
The complete facility state
cannot generally be determined
from a single measurement
or equipment status.
Rather, several complementary
observables may be required
to distinguish between
different physical conditions
that produce similar
individual signals.

This distinction gives
observability a central
role in the protection
architecture. A system
may possess effective
automatic responses
while remaining difficult
to diagnose if the
measurements and internal
states responsible for
those responses cannot
be reconstructed.
Conversely, extensive
monitoring does not
necessarily imply
adequate protection
if the required
measurements are
unreliable, insufficiently
timely, or unrelated
to the relevant
failure mechanisms.

As discussed in
Sec.~\ref{subsec:testing_observability},
observability must therefore
be considered together
with the capability
to verify protection
functions and diagnose
their failures.
The corresponding
instrumentation should
provide sufficient
information to determine
whether a protection
demand occurred,
which conditions
initiated the response,
whether the required
protective action
was issued, and
whether its expected
effect was achieved.

For future facilities,
these requirements
should influence
the design of
monitoring and
protection systems
from the outset.
The selection
and placement
of sensors,
the availability
of independent
diagnostic signals,
the synchronization
of measurements,
and the recording
of relevant
equipment states
can substantially
affect the ability
to reconstruct
abnormal events.
Such provisions
are particularly
important where
several subsystems
can contribute
to the same
hazardous trajectory.

The reconstruction
of an event may
require measurements
obtained from
different technical
systems, sometimes
operating with
different acquisition
frequencies and
response times.
A rapid beam-loss
event, for example,
may involve information
from beam diagnostics,
interlock signals,
magnet and powering
systems, and the
beam-dumping system.
The corresponding
records must permit
a sufficiently
reliable reconstruction
of the sequence
of physical and
technical events
\cite{Wenninger:2016MPO,
Schmidt:2016MPS}.

An important requirement
is that diagnostic
information remain
available even when
the monitored system
experiences an
abnormal transition.
A protection architecture
whose event records
are lost whenever
the initiating
failure affects
the normal control
network may provide
insufficient information
for subsequent
investigation.
Appropriate independence
of diagnostic recording,
reliable timestamps,
and preservation
of relevant information
can therefore
contribute to both
operational learning
and verification
of protection
performance.

Automation must
also account for
invalid or missing
measurements.
A protection
function cannot
necessarily
distinguish a
genuinely acceptable
physical condition
from a failed
sensor or unavailable
communication channel
unless these failure
modes have been
explicitly considered.
Where loss of
measurement validity
would undermine
the credited
protection function,
the resulting
response must
remain compatible
with the required
fail-safe conditions
discussed in
Sec.~\ref{subsec:failsafe_protection}.

This requirement
does not imply
that every
instrumentation
fault must produce
the same protective
action. The
appropriate response
depends on the
physical hazard,
the available
redundancy, the
required integrity
of the protection
function, and
the operational
configuration.
Nevertheless,
the treatment
of invalid signals
must be established
during design
and verified
during commissioning.
A missing measurement
cannot automatically
be interpreted
as evidence of
a safe condition.

The relationship
between automated
protection and
operational availability
also deserves
attention. Spurious
protection actions
may interrupt
scientific operation
without the
presence of
the hazardous
condition that
the barrier is
intended to control.
However, reducing
the frequency
of such interruptions
must not compromise
the capability
to detect genuine
protection demands.
Improved diagnostics,
appropriate redundancy,
and better understanding
of equipment
failure modes
can contribute
to both protection
integrity and
availability
\cite{Filippini:2005Reliability,
Sklet:2006SafetyBarriers}.

For the HL-LHC,
this balance becomes
particularly important
because the increased
stored beam energy
and more demanding
operating configurations
place substantial
requirements on
machine protection.
The relevant
monitoring and
interlock systems
must remain
compatible with
the possible
failure trajectories
and their available
protection times.
The extension
of automated
monitoring must
therefore be
accompanied by
verification
of the complete
protection chain,
including sensors,
logic, communication
interfaces, and
final protective
actions
\cite{Wollmann:2020HLLHCMPS}.

The same principles
apply to future
underground experiments
and distributed
technical infrastructure.
Electrical distribution,
cooling, cryogenics,
ventilation, and
environmental monitoring
may be operated
through interconnected
control systems,
while individual
protection functions
must remain effective
under credible
failures of the
supporting services.
The increasing
geographical extent
of future facilities,
as illustrated
by the FCC,
makes reliable
communication
and coherent
configuration
information
particularly
important.

The design of
automation must
therefore address
the relationship
between distributed
monitoring,
local protective
autonomy, and
centralized
operational
supervision.
Central monitoring
can improve
the integration
of information
from different
subsystems,
but safety-critical
responses should
not depend
unnecessarily
on communication
paths or supervisory
services whose
failure could
prevent timely
protection.

Another important
consideration
concerns configuration
management.
A protection
function validated
for one operating
configuration
cannot automatically
be assumed to
remain adequate
after changes
to sensors,
powering systems,
communication
interfaces,
protection thresholds,
or associated
equipment.
Modifications
to automatic
control or
protection logic
may therefore
require renewed
functional testing
and verification
before the
corresponding
operational
configuration
is restored.

This requirement
connects automation
directly with
the principles
developed in
Sec.~\ref{subsec:return_to_physics}.
The successful
completion of
a technical
intervention
does not, by
itself, establish
that the modified
control and
protection
configuration
is ready for
operation.
The applicable
functions must
be verified,
their interfaces
assessed,
and the resulting
physical and
technical states
validated for
the intended
operational mode.

The organizational
dimension of
automation is
equally important.
Automated systems
can reduce the
need for immediate
human intervention,
but they do not
eliminate the
requirement for
competent personnel
capable of
interpreting
unexpected
conditions,
examining
diagnostic
information,
and evaluating
the consequences
of configuration
changes.
The distribution
of such expertise
among accelerator,
detector,
infrastructure,
and safety teams
reinforces the
importance of
communication
and clearly
assigned
responsibilities,
as discussed
in
Sec.~\ref{subsec:safety_culture}.

From the perspective
of the safe
operational
phase-space
framework,
automation provides
the means to
monitor selected
components of
the facility
state, identify
conditions requiring
protection, and
initiate controlled
transitions
before unacceptable
consequences
develop.
Its effectiveness
depends not only
on the number
of monitored
variables or
the complexity
of the control
system, but also
on the physical
relevance of
the observables,
the reliability
of the corresponding
protection functions,
and the compatibility
of their actions
with the current
technical
configuration.

Automation and
observability
therefore illustrate
another aspect
of the reciprocal
relationship
between physics
and safety.
Physical understanding
determines which
quantities must
be monitored,
which deviations
are hazardous,
and which response
times are necessary.
Conversely,
the availability
and reliability
of monitoring
and protection
functions determine
which operating
conditions can
be established
and maintained
within the
validated
protection
framework.

For future
facilities,
the principal
design objective
is consequently
not simply
increasing the
degree of automation,
but establishing
a monitoring
and protection
architecture
whose physical
assumptions,
operational
configuration,
responses, and
internal failures
remain sufficiently
observable and
verifiable.
Such an architecture
supports both
the admissibility
of demanding
scientific
configurations
and the long-term
ability to
diagnose, maintain,
and recover
the facility
following
abnormal events.

\subsection{Remote handling as a safety-by-design principle}
\label{subsec:future_remote_handling}

The increasing radiation exposure and technical complexity
anticipated in future accelerator and experimental facilities
will strengthen the importance of remote inspection,
measurement, maintenance, and component replacement.
Although many interventions at the LHC and its underground
experiments can be performed directly after appropriate
cooling periods, radiological assessment, and work
preparation, some activities already require specialized
remote-handling techniques. Future facilities may extend
the range of conditions under which direct personnel
intervention is restricted, making remote-maintenance
capability an increasingly important element of the
initial engineering design.

Remote handling should nevertheless be distinguished
from the automation of normal operation and protective
responses discussed in
Sec.~\ref{subsec:future_automation}.
Automatic protection systems detect abnormal conditions
and initiate actions required to prevent or mitigate
their consequences, often within timescales incompatible
with human intervention. Remote-handling systems, by
contrast, permit personnel to inspect, manipulate,
remove, or replace equipment from a location offering
more favorable protection conditions. Such systems
may employ teleoperation, deterministic control
sequences, or combinations of both, without requiring
autonomous decision-making.

The principal safety benefit of remote handling is
the possibility of reducing personnel exposure to
hazardous environments. In accelerator facilities,
the most evident application concerns residual
radioactivity induced in materials by particle
interactions and beam losses. However, remote
techniques can also be valuable where direct
access is limited by mechanical, thermal,
electrical, or environmental hazards.
Their suitability must be established for
the particular intervention and the complete
set of associated risks.

The existing CERN accelerator complex provides
important practical experience. Remotely operated
systems have been developed for visual inspection,
radiation measurements, alignment surveys,
component manipulation, and the replacement
of activated equipment. These include remote
inspection systems for the LHC tunnel and
specialized handling equipment intended
for collimator exchange
\cite{Kershaw:2013RemoteHandling}.
Such developments demonstrate that remote
intervention can reduce occupational exposure
while providing diagnostic and maintenance
capabilities that would otherwise require
personnel to enter radiation areas.

The LHC collimation system provides a particularly
instructive example. Collimators intercept beam
halo particles and may accumulate substantial
residual activation during operation, especially
in the principal cleaning insertions. Their
replacement involves not only the handling
of heavy activated components, but also
vacuum connections, cooling services,
electrical interfaces, and precise alignment.
Consequently, the development of remote
replacement techniques has required dedicated
mechanical interfaces, guiding arrangements,
handling equipment, and visualization systems
\cite{Kershaw:2013RemoteHandling,
Redaelli:2020HLLHCCollimation}.

This experience illustrates an essential
design principle: remote maintainability
cannot be provided solely by introducing
a robotic or remotely operated device.
The equipment to be maintained and
the surrounding installation must
also be compatible with the intended
handling method. Component dimensions,
lifting points, connection mechanisms,
transport clearances, shielding
arrangements, and alignment requirements
must be considered together.

The radiological considerations developed in
Sec.~\ref{subsec:future_radiation_maintainability}
provide a further motivation. At the HL-LHC,
increased accumulated irradiation will
affect equipment near the ATLAS and CMS
interaction points, as well as collimators
and other beam-intercepting devices in
the accelerator. Depending on the
resulting activation, component location,
and planned intervention, remote inspection
or handling may provide an effective
means of reducing occupational exposure.
The practical benefit must be assessed
against the radiation fields and
technical conditions of the specific
intervention rather than assumed
from the use of remote equipment alone.

This distinction is consistent with
the ALARA principle discussed in
Sec.~\ref{subsec:radiation_work_planning}.
Radiological optimization may favor
remote intervention when it reduces
individual or collective exposure
to a reasonably achievable level,
taking account of the required
resources, technical feasibility,
and the potential exposure
associated with preparation,
installation, operation, recovery,
and dismantling of the handling
equipment
\cite{FassoRokni:2009ALARA,
ICRP:2021Publication147}.

In particular, remote operation does
not necessarily eliminate the need
for personnel access. Some activities
may require direct preparation,
connection or disconnection of
services, installation of handling
devices, or final verification.
The corresponding occupational
exposure must therefore be assessed
for the complete intervention
rather than exclusively for
the period of remote manipulation.
A combined sequence of manual
and remote operations may provide
a more effective and reliable
solution than either method alone.

Remote handling also introduces
additional technical hazards.
Manipulation of heavy or delicate
equipment can involve uncontrolled
movement, collisions, dropped loads,
damage to surrounding components,
or loss of mechanical stability.
Failures of communication,
actuators, sensors, power supplies,
or visualization systems may
prevent completion of a planned
operation. Where these failures
can produce hazardous consequences,
the appropriate mechanical
restraints, limits, interlocks,
and recovery provisions must
be included in the handling
system design.

The reliability of remote
operation is particularly
important in confined
underground environments.
Restricted clearances,
limited visibility,
complex service connections,
and difficult transport
routes can increase
the consequences of
positioning errors or
equipment failures.
Remote-handling equipment
must therefore be
validated under
representative conditions,
including foreseeable
abnormal situations.
Full-scale mock-ups,
functional tests,
and rehearsals can
provide important
evidence that the
intended operations
are mechanically
feasible and can
be performed within
the required protection
constraints
\cite{Kershaw:2013RemoteHandling}.

An essential design consideration
is the recovery of the remote
system itself. A remotely
operated manipulator or
transport vehicle that
becomes immobilized in
a radiation area may
create an additional
intervention requiring
personnel exposure.
For this reason,
recovery arrangements
should be assessed
together with the
normal handling
procedure. Depending
on the equipment,
these may include
mechanical release
functions, alternative
powering or control
arrangements, emergency
load support, or
means of retrieving
the device without
direct access to
the most hazardous
location.

The preceding discussion
of automation and
observability is directly
relevant to this requirement.
Successful teleoperation
depends on reliable
information about the
position and condition
of the handling device,
the state of the
manipulated component,
and the surrounding
environment. Cameras,
position sensors,
load monitoring,
limit switches,
and equipment-status
information may
provide complementary
observations.
Their selection
must reflect the
physical consequences
of incorrect or
incomplete information,
rather than simply
the convenience
of remote operation.

The loss of a
visualization or
communication channel
does not necessarily
justify an identical
response for every
handling system.
Nevertheless,
the required response
must be determined
and validated in
advance. Where
continued movement
without reliable
feedback would
be hazardous,
the corresponding
control architecture
should inhibit
the movement or
establish an
appropriate controlled
condition. Such
responses must
remain compatible
with the mechanical
and other hazards
associated with
the load being
handled.

Future muon-collider
concepts provide
an additional
motivation for
integrating remote
maintainability
into the physical
design of the
facility. The
high-power proton
target and surrounding
capture systems
are expected
to operate under
intense irradiation
and may require
replacement of
activated targets
or nearby components.
The corresponding
design studies
therefore identify
remote-handling
capabilities,
shielded maintenance
areas, and provisions
for the cooling,
storage, and treatment
of activated
materials as
important infrastructure
considerations
\cite{Accettura:2024MuonCollider}.

Such requirements
can influence
the layout of
the target station,
the positions
of shielding
structures,
the available
transport routes,
and the location
of maintenance
facilities.
They must be
addressed before
the civil-engineering
and equipment
configurations
are finalized,
since retrofitting
sufficient space
for handling
and transport
may be difficult
or incompatible
with the completed
installation.

The same general
principle applies
to future circular
colliders and
large underground
detectors. The
length and geometry
of transport routes,
dimensions of
access shafts,
locations of
service areas,
and distribution
of activated
equipment can
influence the
feasibility of
remote maintenance.
These considerations
connect remote-handling
requirements to
the infrastructure
topology discussed
in Sec.~\ref{subsec:fcc_safety},
without implying
that a single
remote-maintenance
technology would
be suitable for
all accelerator
and detector
configurations.

Within the state-space
framework developed
in this work,
remote handling
changes the
technical and
organizational
conditions under
which an intervention
can be performed.
For an experiment
$\alpha$, the
intervention domain
$\Omega_{\rm int}^{\alpha}
(z,w;m,\boldsymbol\Xi_{\alpha})$
depends on the
available protection
arrangements,
equipment configuration,
and characteristics
of the proposed work.
The introduction
of a validated
remote-handling
method can make
some interventions
admissible under
conditions where
direct personnel
work would be
restricted or
unacceptable.

This potential
extension is not
unconditional.
Remote manipulation
introduces its
own equipment
states, possible
failures, and
requirements for
protection and
recovery. Consequently,
the admissibility
of the intervention
must be assessed
for the complete
sequence of
preparation,
access or equipment
deployment,
manipulation,
transport,
withdrawal,
and restoration.
A safe initial
and final
configuration
does not guarantee
that every
intermediate
handling state
or trajectory
is acceptable.

The configuration-control
principles discussed
in
Sec.~\ref{subsec:return_to_physics}
therefore remain
applicable. After
remote replacement
or repair,
the equipment
and its interfaces
must be verified
before the
corresponding
operational mode
is restored.
Successful completion
of the mechanical
handling task
does not by
itself demonstrate
correct alignment,
service connections,
protection functions,
or operational
readiness.

Remote maintainability
also contributes
to scientific
continuity. The
ability to inspect
a failed component
without immediate
personnel access,
prepare an intervention
before radiological
conditions permit
manual work,
or replace
an activated
component through
a validated
remote sequence
can reduce
recovery delays
and improve
the predictability
of maintenance.
These benefits,
however, depend
on the reliability
of the handling
equipment and
the availability
of effective
recovery procedures
if the remote
operation fails.

The resulting
design principle
is broader than
the replacement
of manual work
by remotely
operated machinery.
Future facilities
should be designed
with explicit
consideration
of how equipment
will be inspected,
isolated,
disconnected,
handled,
replaced,
tested,
and returned
to operation
after substantial
irradiation
or other conditions
restricting
direct access.

Remote handling
thus provides
another illustration
of the reciprocal
relationship
between physics
and safety.
Radiation-transport
calculations,
activation studies,
mechanical analysis,
and operational
measurements
determine the
conditions under
which remote
interventions
may be required
or advantageous.
Conversely,
the need to
perform those
interventions
safely influences
the physical
architecture
of accelerator
and detector
systems, including
their interfaces,
shielding,
accessibility,
and maintenance
infrastructure.

In this sense,
remote maintainability
is not merely
a supplementary
operational
capability.
For highly
irradiated or
otherwise difficult
to access
installations,
it can become
an important
design condition
for preserving
both personnel
protection
and the long-term
scientific
capability
of the facility.

\subsection{Safety margins under increasing complexity}
\label{subsec:future_safety_margin}

The increasing complexity of accelerator and experimental
facilities introduces additional challenges for the
definition, assessment, and preservation of safety margins.
The relevant uncertainties concern not only the physical
limits of individual components, but also the accuracy
of measurements, the validity of physical models,
the evolution of abnormal trajectories, the response
times of protection functions, and the technical
configuration in which the corresponding limits
are applied. The protection margins of future
facilities must therefore be considered at the
level of the coupled physical and technical
system rather than exclusively through
individual equipment parameters.

The concept of a protection margin was introduced
in Sec.~\ref{subsec:barriers_phase_space},
where the separation between an interlock threshold
and a physical damage limit was used to illustrate
the principle of intervention before unacceptable
consequences develop. Such a separation provides
a useful first approximation, but does not
necessarily represent the margin available during
an actual abnormal event. The protected quantity
may continue to evolve after the detection of
a hazardous condition, and the final response
depends on the complete protection chain.

This distinction is particularly important for
hazards characterized by rapid physical evolution.
As discussed in Sec.~\ref{subsec:protection_time},
the protection response includes detection,
logic evaluation, communication, actuation,
and the physical time required to establish
the intended protected state. The relevant
margin must therefore account for the evolution
of the hazard during these processes.
A protection threshold that appears sufficiently
conservative under static conditions may
provide inadequate protection if the
physical quantity continues to increase
rapidly before the protective action
becomes effective.

For a scalar quantity $x$ whose excessive
value is associated with a specified
unacceptable consequence, a robust
protection margin may be expressed
schematically as
\begin{equation}
\Delta x_{\rm eff}^{(i,j)}
=
\inf_{\boldsymbol\theta\in\Theta}
\left[
x_{\rm lim}(\boldsymbol\theta)
-
\sup_{t\in I_{\rm prot}^{(i,j)}}
x^{(i,j)}(t;\boldsymbol\theta)
\right],
\label{eq:future_effective_margin}
\end{equation}
where $x_{\rm lim}$ denotes the relevant
physical limit, $x^{(i,j)}$ the evolution
of the protected quantity along the
failure trajectory $\Gamma_j^{(i)}$,
and $I_{\rm prot}^{(i,j)}$ the interval
covering the hazardous evolution and
the corresponding protective response.
The parameter set $\Theta$ represents
the credible uncertainties affecting
the physical process and protection
performance. In applying the expression,
the interval is understood to cover
the complete response for each
considered parameter realization.

For the specified failure scenario,
a positive value of
$\Delta x_{\rm eff}^{(i,j)}$
indicates that the considered trajectories
remain below the selected limit,
including the uncertainties represented
by $\Theta$. An adequate design requires
not merely a positive nominal margin,
but sufficient additional separation
to accommodate the uncertainties and
variations not explicitly resolved
in the assessment. The expression
is illustrative: hazards involving
lower physical limits, several
simultaneous constraints, or
nonmonotonic failure dynamics
require corresponding formulations.

Equation~\eqref{eq:future_effective_margin}
also clarifies the distinction between
the nominal operating margin and
the effective protection margin.
The former describes the separation
between an operating condition
and a selected limit, whereas
the latter must include the
transient evolution of the
physical quantity and the
performance of the protective
response. The difference can
be significant when the
available protection time
is short or the failure
dynamics are uncertain.

The HL-LHC provides an important
illustration. Increasing stored
beam energy and beam brightness
places demanding requirements
on collimation, beam-loss monitoring,
interlocks, and beam extraction.
The relevant protection margins
depend not only on the nominal
beam parameters, but also on
the distribution and duration
of beam losses, the local
energy deposition, the
response of the protection
systems, and the machine
configuration
\cite{Schmidt:2016MPS,
Wollmann:2020HLLHCMPS,
Redaelli:2020HLLHCCollimation}.

A margin established for one
beam energy, optical configuration,
or collimator arrangement cannot
automatically be transferred to
another. Changes in the beam
distribution or machine optics
may modify the relationship
between a measured beam-loss
signal and the corresponding
energy deposition in a
vulnerable component.
Consequently, improvements
in instrumentation precision
alone do not necessarily
increase the actual margin
against damage. The physical
model connecting measurements
with the protected equipment
must also remain valid under
the relevant operating conditions.

The same considerations apply
to superconducting magnets
and their cryogenic systems.
Protection margins may involve
temperature, current, voltage,
pressure, or energy-deposition
limits, with different
characteristic response times.
The thermal and electrical
evolution of a magnet quench,
for example, cannot generally
be represented by a static
difference between an operating
current and a limiting value.
The effectiveness of quench
protection depends on the
distribution of stored energy,
the development of resistive
regions, detection delays,
and the subsequent redistribution
or extraction of energy.

For cryogenic installations,
the corresponding margins
must account for the credible
rates of helium release,
pressure transients,
thermal conditions, and
the capacity of the
associated protective
arrangements. A design
margin in pressure-relief
capacity does not by itself
demonstrate that the resulting
underground environmental
conditions remain acceptable.
As emphasized earlier in
the manuscript, protection
of equipment and protection
of personnel may require
different physical criteria
and complementary barriers.

The underground experiments
provide further examples.
For ATLAS and CMS, equipment
protection margins can depend
on the combinations of
electrical power, cooling
capacity, temperature,
gas conditions, and detector
configuration. A temperature
limit appropriate for
one electronics operating
mode may not provide
equivalent protection
when the power dissipation
or cooling conditions change.
The interpretation of the
corresponding monitoring
signals must therefore
remain consistent with
the actual technical
configuration
\cite{Schmeling:2004DSS,
DiCalafiori:2020CMS_ECAL}.

An additional complication
is the coupling between
different margins.
An apparently conservative
operating condition in one
subsystem may reduce the
margin available in another.
For example, changes in
cooling performance can
affect the thermal response
of powered equipment,
while modifications of
shielding or mechanical
configuration may alter
the environmental conditions
relevant to personnel
protection. Such dependencies
reinforce the general
nonfactorizability of the
safe domain discussed in
Sec.~\ref{subsec:safe_phase_space_scaling}.

The management of uncertainty
must therefore account for
correlations between physical
parameters and protection
functions. Measurement errors,
modelling uncertainties,
and configuration variations
cannot always be represented
as independent contributions
whose magnitudes can simply
be added or combined
statistically. A common
technical dependency may
simultaneously modify the
physical hazard, its
observability, and the
effectiveness of the
corresponding protection
response.

The robust-domain formulation
introduced in
Eq.~\eqref{eq:robust_safe_phase_space}
provides a more general
interpretation of this
requirement. The admissible
region must be evaluated
under the relevant
uncertainties, while
the associated trajectories
must remain compatible
with the applicable
protection constraints.
A positive margin for
one selected physical
quantity is therefore
not sufficient to
establish the safety
of the complete
facility state.

The reliability of
configuration information
becomes particularly
important in this context.
As discussed in
Sec.~\ref{subsec:future_automation},
automated monitoring
and protection depend
on the correct
interpretation of
measurements under
the current operating
conditions. An incorrect
sensor calibration,
outdated protection
threshold, unidentified
equipment modification,
or inconsistent
configuration record
can undermine an
otherwise adequate
protection margin.
The resulting uncertainty
may not be reduced
simply by increasing
the number or
precision of sensors.

The same principle
applies to remote
handling, discussed
in
Sec.~\ref{subsec:future_remote_handling}.
Mechanical clearances,
positioning accuracy,
load capacity,
visualization,
and the response
to equipment failure
determine the margins
available during
an intervention.
A remotely operated
system may reduce
personnel exposure
while introducing
additional uncertainty
associated with
positioning or
load manipulation.
The corresponding
design must therefore
consider both the
intended operation
and credible deviations
from its planned
trajectory.

Radiation protection
introduces another
important distinction.
In the planning
of interventions
near activated
accelerator and
detector components,
uncertainties in
residual radiation
fields, work duration,
shielding effectiveness,
and the sequence
of operations
influence predicted
occupational exposure.
These uncertainties
must be addressed
through appropriate
radiological assessments,
measurements,
work preparation,
and optimization
according to ALARA
\cite{FassoRokni:2009ALARA,
ICRP:2021Publication147}.
The objective is
not to define
a universal dose
margin analogous
to a mechanical
damage threshold,
but to ensure
that the intervention
satisfies the
applicable protection
requirements and
that exposure is
reasonably optimized.

An important consequence
for future facilities
is that margins
cannot be considered
permanent properties
established once
during initial
commissioning.
The physical
installation,
instrumentation,
protection functions,
and operational
conditions evolve
through maintenance,
consolidation,
and upgrades.
Changes in
these conditions
may modify the
available margins,
even when the
nominal operating
parameters remain
unchanged.

Configuration control,
periodic verification,
and operational
learning are
therefore essential
for preserving
the validity
of protection
margins throughout
the facility
lifetime.
As discussed in
Sec.~\ref{subsec:return_to_physics},
the restoration
of equipment
following an
intervention
must include
appropriate
verification
of the modified
technical and
protection
configuration.
Similarly,
the analysis
of anomalies
and near misses
may reveal
that an assumed
margin was
smaller than
predicted or
that a relevant
uncertainty
had not been
adequately
represented.

The long-term
design objective
is consequently
not simply to
maximize individual
engineering margins.
Excessive conservatism
may impose substantial
technical or
operational costs,
while margins
defined without
adequate physical
understanding may
provide only
apparent protection.
The relevant
objective is
to establish
physically justified
margins, demonstrate
their adequacy
for credible
failure trajectories,
and preserve
their validity
under changing
operational
conditions.

This approach
reinforces the
reciprocal relationship
between physics
and safety.
Physical models,
measurements,
and simulations
provide the basis
for determining
the limits and
uncertainties
associated with
hazardous processes.
Conversely,
the resulting
protection margins
constrain the
configurations
and transitions
that can be
authorized
for scientific
operation.

For future
facilities,
the essential
principle is
therefore that
the safety margin
belongs not only
to the individual
component, but
also to the
complete protected
trajectory.
Its adequacy
depends on
the physical
evolution of
the hazard,
the technical
configuration,
the reliability
of monitoring
and protection,
and the uncertainties
affecting their
combined response.

\subsection{Graceful degradation and recoverable states}
\label{subsec:future_graceful_degradation}

The preceding discussion of safety margins raises an
important question concerning the response of a complex
scientific facility to the loss or degradation of
technical functions. The distinction between fully
operational and completely failed states is not always
sufficient to characterize the available responses.
Depending on the nature of the failure, the affected
subsystems, and the remaining protection capabilities,
it may be possible to continue a restricted activity
or establish a controlled configuration from which
subsequent recovery can be performed.

This possibility is commonly associated with
\emph{graceful degradation}, a concept developed
in the broader literature on dependable and
resilient engineering systems. It describes
the ability of a system to preserve selected
functions at reduced performance following
a disturbance or the loss of some of its
capabilities, rather than undergoing an
immediate and uncontrolled loss of all
functionality
\cite{Wied:2020Resilience,
Mentges:2023ResilienceGlossary}.
In the present context, graceful degradation
is understood more restrictively as the
controlled reduction of operational capability
while preserving the protection conditions
required for the activities that remain
permitted. Its purpose is not to maintain
scientific operation under all circumstances,
but to avoid unnecessary loss of functionality
when a restricted operating configuration
can be demonstrated to remain admissible.

The distinction follows naturally from the
barrier-dependent admissible domain introduced
in Sec.~\ref{subsec:barriers_phase_space}.
The unavailability of a protection function
may require the restriction of operating
parameters, the inhibition of particular
transitions, or the suspension of activities
that depend on the affected protection.
As illustrated by
Eq.~\eqref{eq:protection_loss_contracts_space},
a reduction in the available protection
functions may lead to a contraction of
the admissible operational domain,
provided that other relevant conditions
remain unchanged.

However, the loss of a technical function
does not always imply that every activity
must be interrupted. If the remaining
protection architecture is adequate for
a more restricted configuration, operation
may continue within the corresponding
admissible domain. For example, the
unavailability of a nonessential
instrumentation channel may be compatible
with continued operation when the
required protection integrity is
preserved by other validated channels.
By contrast, the loss of a function
essential for protection against a
credible rapid failure may require
immediate withdrawal of the relevant
operational permit.

Graceful degradation must therefore
be distinguished from the fail-safe
behavior discussed in
Sec.~\ref{subsec:failsafe_protection}.
Fail-safe protection requires that
credible failures of a protection
system do not silently permit
continued operation without
the required protection.
Graceful degradation concerns
the possible preservation of
limited functionality after
the resulting restrictions
have been established.
A fail-safe protective action
may terminate physics operation
completely, whereas graceful
degradation is appropriate
only when a restricted
operating mode remains
demonstrably admissible.

This distinction is particularly
important for accelerator
machine protection.
Certain equipment faults
may justify restrictions
on beam parameters or
the suspension of particular
machine operations.
Other faults, especially
those affecting functions
required to control
rapid beam-loss scenarios,
necessitate beam extraction.
The successful execution
of a protective beam dump
establishes a controlled
machine condition, but
does not imply that
continued beam operation
at reduced performance
would have been admissible
\cite{Schmidt:2016MPS,
Wollmann:2020HLLHCMPS}.

The recovery following
such an interruption
depends on the initiating
condition and the
resulting machine state.
In some cases,
restoration of the
required protection
conditions may permit
relatively straightforward
recommissioning.
In others, equipment
inspection, repair,
or additional testing
may be necessary.
The protected state
reached after beam
extraction is therefore
not equivalent to
an immediately
recoverable operating
configuration.

A related distinction
applies to superconducting
magnet systems.
An abnormal condition
may require automatic
quench protection,
energy extraction,
or controlled discharge
of the affected circuit.
These responses can
prevent or limit
equipment damage,
but they may be
followed by thermal
recovery, verification
of electrical integrity,
re-establishment of
cryogenic conditions,
and validation of
the powering system.
The protection action
is consequently only
the first stage of
a potentially extended
recovery process.

At the detector level,
some technical limitations
may permit operation
with reduced functionality.
The isolation of an
affected detector
subsystem or the
restriction of its
powering configuration,
for example, may allow
other parts of the
experiment to continue
operating, provided
that their protection
conditions and technical
dependencies remain valid.
The actual possibility
depends on the detector
architecture, the
distribution of services,
and the protection
functions involved.

Cooling failures illustrate
the limitations of this
approach. Where loss of
cooling threatens powered
electronics, the required
response may be automatic
power removal rather than
continued operation at
reduced performance.
If the affected equipment
can be isolated without
compromising neighboring
subsystems, the remaining
detector functions may
continue within a
validated restricted
configuration. However,
such continued operation
cannot be assumed
solely because the
initial fault appears
localized
\cite{Schmeling:2004DSS,
DiCalafiori:2020CMS_ECAL}.

The cryogenic infrastructure
provides another example
in which the consequences
of a protective response
must be considered at
the level of the complete
facility. Isolation of
an affected cryogenic
section or activation
of pressure-relief
arrangements may limit
damage to equipment,
but the resulting
helium discharge and
environmental conditions
may introduce additional
restrictions on personnel
access and nearby
technical activities.
A configuration that
is controlled from
the perspective of
cryogenic equipment
protection is not
necessarily compatible
with immediate
underground occupancy.

This illustrates why
degraded operation
cannot be characterized
through an independent
status assigned to
each subsystem.
The relevant condition
is the compatibility
of the coupled physical
state $\boldsymbol X$
with the applicable
technical configuration
$\boldsymbol\Xi$,
operational mode $m$,
and available protection
functions ${\cal B}_{\rm av}$.
Restrictions imposed
on one subsystem
may modify the
admissible activities
of another, even
when the latter
has not experienced
a direct failure.

The design of future
facilities should
therefore distinguish
between different
types of response
to abnormal conditions.
These include
continued operation
within a restricted
admissible domain,
controlled termination
of the affected
activity, establishment
of a protected
nonoperational state,
and subsequent
maintenance or
recovery. The
appropriate response
depends on the
physical hazard,
its evolution,
the available
protection functions,
and the conditions
required for
restoration.

The transition
between these
conditions is
itself part
of the safety
problem. A restricted
operating state
may be individually
admissible, but
the transition
toward it must
also satisfy
the applicable
protection requirements.
Similarly, a
controlled shutdown
may involve
temporary changes
in powering,
cooling, pressure,
mechanical loading,
or environmental
conditions that
must remain
within the
corresponding
protection limits.

This requirement
connects directly
to the treatment
of dynamic safety
margins in
Sec.~\ref{subsec:future_safety_margin}.
A reduction of
operational capability
is useful only
if the remaining
margins are
adequate for
the credible
failure trajectories
associated with
the restricted
configuration.
It is not
sufficient to
demonstrate that
the final state
satisfies a set
of static
operating limits;
the intermediate
physical evolution
and the performance
of the protective
responses must
also be considered.

An additional
requirement concerns
the observability
of degraded states.
As discussed in
Sec.~\ref{subsec:future_automation},
automated monitoring
and diagnostic
systems should
provide sufficient
information to
identify which
functions have
become unavailable,
which protection
conditions remain
satisfied, and
which activities
are consequently
permitted.
An uncertain
or incorrectly
interpreted
equipment status
must not be
treated as
evidence that
a restricted
operating mode
is safe.

The possibility
of graceful
degradation is
also influenced
by the physical
architecture
of the installation.
Segmentation
of technical
services,
independent
protection
functions,
appropriate
isolation
capabilities,
and accessible
diagnostics
may limit the
extent to which
a local failure
affects the
complete facility.
However, these
features must
be assessed
for credible
common-cause
failures and
interactions
between subsystems.
Physical separation
alone does not
establish functional
independence
\cite{Sklet:2006SafetyBarriers,
Leveson:2004AccidentModel}.

This consideration
becomes especially
relevant for
future facilities
with extensive
distributed
infrastructure,
such as the
proposed FCC.
Appropriate
segmentation
of electrical,
cooling, cryogenic,
and other
technical services
may permit
localized isolation
or maintenance
without requiring
the interruption
of every activity
elsewhere in
the installation.
Nevertheless,
continued operation
in unaffected
regions must
remain compatible
with the status
of shared services,
environmental
conditions,
personnel access,
and the protection
functions that
remain available.

Recoverability
introduces a
further distinction.
A state may
be protected
against immediate
hazards without
permitting a
simple return
to normal
operation.
The restoration
process may
require diagnosis,
equipment replacement,
re-establishment
of technical
services,
functional
testing,
and progressive
recommissioning.
Accordingly,
recoverability
is not determined
solely by
membership in
a particular
region of the
physical state
space. It also
depends on
the existence
of a feasible
and admissible
recovery trajectory.

This trajectory
may involve
personnel interventions
and temporary
configurations
that are different
from those
permitted during
normal physics
operation.
The principles
of access
authorization,
intervention
planning,
radiation
protection,
and configuration
control developed
in the preceding
sections therefore
remain applicable.
Where activated
equipment or
restricted
accessibility
complicates
recovery,
remote-handling
capabilities
may provide
additional
options, as
discussed in
Sec.~\ref{subsec:future_remote_handling}.

The restoration
of scientific
operation must
also be distinguished
from the completion
of the immediate
corrective action.
As established
in
Sec.~\ref{subsec:return_to_physics},
the post-intervention
physical state
and technical
configuration
must be validated
against the
conditions required
for the intended
operational mode.
The disappearance
of the initiating
alarm or the
successful reset
of an interlock
does not by
itself demonstrate
that the required
protection functions
and their interfaces
have been restored.

The corresponding
design objective
extends beyond
reducing the
probability of
technical failures.
Future facilities
should, where
practicable,
provide controlled
responses to
credible failures,
limit unnecessary
propagation
between subsystems,
preserve diagnostic
information,
and establish
clear conditions
for subsequent
recovery.
These capabilities
contribute to
the scientific
continuity
discussed in
Sec.~\ref{subsec:safety_scientific_continuity}
by improving
the predictability
of interruptions,
maintenance,
and restoration.

Nevertheless,
graceful degradation
should not be
treated as an
unconditional
means of improving
availability.
The preservation
of reduced
scientific
functionality
is justified
only when the
remaining
protection
architecture
has been
demonstrated
to be adequate.
In some cases,
complete termination
of the affected
operation is
the only
appropriate
response.
The objective
is therefore
not to maximize
the number of
degraded modes,
but to identify
those that can
be established
and maintained
without compromising
the applicable
safety requirements.

From the perspective
of the present
work, graceful
degradation and
recoverability
provide complementary
extensions of
the safe operational
phase-space
framework.
Graceful
degradation
concerns the
possibility of
preserving limited
operation within
a restricted
admissible domain,
whereas
recoverability
concerns the
existence of
validated
trajectories
from a protected
or interrupted
state toward
a configuration
supporting
subsequent
scientific
operation.

These concepts
illustrate how
\emph{physics in
safety} and
\emph{safety in
physics} remain
interdependent
even after
an abnormal
event has
occurred.
Physical
understanding
of failure
propagation,
energy
redistribution,
and subsystem
interactions
determines
which protective
and recovery
responses are
feasible.
Conversely,
the available
protection and
recovery
capabilities
determine which
restricted
configurations
can be authorized
and how the
facility may
return to
scientific
operation.

\subsection{Safety over the full facility life cycle}
\label{subsec:future_lifecycle}

The safe operational phase space of a large scientific
facility is not fixed throughout its lifetime.
Construction, installation, commissioning, physics
operation, technical stops, long shutdowns, upgrades,
and eventual dismantling or decommissioning involve
different physical configurations, activities,
hazards, and protection requirements. The conditions
under which a facility can be operated or accessed
must therefore be reassessed as its technical
configuration and intended activities evolve.

Within the framework developed in this work,
these changes can be described through the
physical state $\boldsymbol X$, the technical
configuration $\boldsymbol\Xi$, the applicable
operational mode $m$, and the available protection
functions ${\cal B}_{\rm av}$. Their evolution
modifies the corresponding safe and admissible
domains and the trajectories through which
transitions between configurations can be
performed. The relevant change is not simply
a dependence on calendar time, but a dependence
on the successive physical and technical
conditions established during the facility's
lifetime.

This distinction is important because the
different life-cycle phases do not constitute
a uniform progression through increasingly
or decreasingly hazardous conditions.
During construction, the dominant hazards
may involve civil engineering, excavation,
heavy lifting, temporary electrical
installations, work at height, confined
spaces, and simultaneous activities.
Beam-related hazards may not yet exist,
while mechanical and organizational
risks can be substantial. The corresponding
protection requirements must therefore
reflect the actual construction
configuration rather than assumptions
associated with the future operational
installation.

Installation and commissioning introduce
a different set of challenges.
Individual subsystems may be completed
and tested while neighboring equipment
remains under construction or modification.
Electrical powering, pressure testing,
cryogenic cool-down, magnet commissioning,
detector movement, and activation of
protection functions can introduce
hazards before the installation reaches
its intended operational configuration.
The admissibility of these activities
depends on their sequence and on the
compatibility of the temporary technical
states through which the facility passes.

As discussed in
Sec.~\ref{subsec:safety_commissioning},
commissioning is therefore not simply
the demonstration of nominal performance.
It also establishes evidence that
particular physical states and
operational transitions are compatible
with the applicable protection
requirements. The transition from
installation to scientific operation
must include verification of the
relevant functions and interfaces,
rather than assuming that the
completion of individual technical
systems establishes readiness
of the complete facility.

During routine physics operation,
the principal energetic hazards
arise from the accelerator beam,
superconducting magnets, cryogenic
systems, electrical installations,
detector equipment, and the
associated radiation environment.
The corresponding protection
architecture must maintain the
admissibility of the intended
scientific configurations and
respond to credible abnormal
transitions. However, these
protection arrangements are
not necessarily sufficient
for activities performed
during subsequent shutdowns
or major technical interventions.

Technical stops and long shutdowns
can fundamentally change the
hazard configuration.
Beam operation may be suspended,
but residual radioactivity,
stored electrical or magnetic
energy, cryogenic inventories,
mechanical movements, temporary
services, and large numbers
of simultaneous interventions
remain relevant.
Some hazards are reduced,
whereas others emerge from
the temporary configurations
required for maintenance,
consolidation, and upgrades.
The protection arrangements
must therefore be adapted
to the actual intervention
conditions rather than
derived solely from the
normal physics-operating state.

The ongoing LS3 of the CERN accelerator
complex provides a relevant
example of this transition.
The extensive activities
associated with accelerator
and experimental upgrades
require the coordination of
different technical domains,
personnel access, temporary
configurations, and the
restoration of protection
functions before subsequent
operation
\cite{Tock:2025LS3,
Evrard:2025CERNPersonnelSafety}.
Such activities illustrate
why intervention planning
and configuration control
form part of the safety
architecture throughout
the operational lifetime.

The successive upgrades
of ATLAS and CMS provide
another important example.
The replacement or
modification of detector
subsystems can alter
electrical distribution,
cooling arrangements,
mechanical configurations,
radiation shielding,
monitoring, and equipment
protection. The resulting
installation need not
reproduce its previous
technical state.
As established in
Sec.~\ref{subsec:return_to_physics},
the relevant requirement
is validation of the
post-intervention
configuration against
the conditions applicable
to the subsequent
operational mode.

An important consequence
is that the life-cycle
evolution of a facility
may involve both an
extension and a restriction
of its admissible
operational domain.
New equipment and
validated protection
functions may permit
additional scientific
configurations, while
aging components,
increased activation,
newly identified
failure mechanisms,
or temporary loss of
protection functions
may require restrictions.
Such changes need
not be monotonic,
and their implications
must be assessed for
the actual configuration
and intended activity.

Long-term radiation
protection provides
a particularly clear
illustration.
Accumulated irradiation
can modify the
radiological conditions
under which accelerator
and detector equipment
can be accessed or
replaced, even when
the nominal scientific
operating parameters
remain unchanged.
The activation
of collimators,
beam-line components,
shielding, and detector
structures can introduce
additional restrictions
on maintenance and
dismantling activities.
Radiation-protection
planning must therefore
account for the
irradiation history,
cooling intervals,
component geometry,
and applicable
intervention methods,
in accordance with
the ALARA principle
\cite{Otto:2020HLLHCSafety,
FassoRokni:2009ALARA}.

These considerations
have consequences
that extend to the
end of the facility's
scientific programme.
The permanent cessation
of beam operation
does not eliminate
all relevant hazards.
Activated components
and materials may
require radiological
characterization,
shielded handling,
temporary storage,
appropriate waste
management, or
controlled transport.
At the same time,
dismantling can
introduce substantial
mechanical, electrical,
and structural hazards,
particularly when
large equipment
must be removed
from confined
underground spaces.

Decommissioning must
therefore be treated
as a distinct phase
requiring its own
safety assessment
and protection
arrangements.
This principle is
consistent with
published international
guidance on the
decommissioning
of research and
other facilities,
which emphasizes
early planning,
assessment of
radiological and
non-radiological
hazards, and the
safe management
of dismantling
activities and
resulting materials
\cite{IAEA:2014Decommissioning,
IAEA:2019ResearchDecommissioning}.
Such guidance provides
a useful general
framework, although
the specific legal
and institutional
requirements applicable
to CERN must be
considered separately.

Decommissioning also
differs fundamentally
from maintenance
intended to restore
scientific operation.
During maintenance,
equipment and
protection functions
are generally
restored or replaced
to support a
subsequent operational
configuration.
During permanent
dismantling, selected
equipment and
services are
progressively removed,
and some original
protection functions
may become unnecessary
for their former
purpose while
remaining essential
to other activities.
The sequence in
which such functions
are disconnected
must therefore
remain compatible
with the hazards
present at each
stage of dismantling.

For example,
the removal of
shielding or
ventilation equipment
may be permissible
only after the
associated radiological
or environmental
conditions have
been reassessed.
Similarly,
electrical isolation,
discharge of stored
energy, cryogenic
warm-up, and removal
of heavy components
must be coordinated
with the remaining
mechanical and
personnel-protection
requirements.
The initial and
final configurations
of a dismantling
operation may
each be acceptable,
while an inappropriate
sequence of
intermediate actions
could create
additional hazards.

This provides another
application of the
distinction between
admissible states
and admissible
trajectories.
During the final
life-cycle phases,
the objective is
no longer to return
the installation
to a validated
physics-operating
domain, but to
reach a defined
end configuration
through a sequence
of controlled
interventions.
The applicable
safety requirements
continue to
govern the complete
trajectory, even
though the scientific
purpose of the
installation has
been discontinued.

Life-cycle safety
also depends
on the preservation
of technical
knowledge.
A large scientific
facility may remain
in operation for
several decades,
during which
personnel,
responsibilities,
equipment generations,
and technical
documentation evolve.
Information concerning
the original design,
subsequent modifications,
material inventories,
irradiation history,
protection assumptions,
and previous
interventions may
be essential for
safe maintenance
or dismantling
many years later.

The management
of such information
is therefore not
merely an
administrative
requirement.
It contributes
to the ability
to reconstruct
the actual
technical
configuration,
identify hidden
dependencies,
assess accumulated
hazards, and
plan interventions
under conditions
that may differ
substantially
from those
originally
anticipated.
This connects
life-cycle safety
with the
organizational
learning and
distributed
expertise discussed
in
Secs.~\ref{subsec:near_miss_learning}
and~\ref{subsec:safety_culture}.

An additional
design consideration
concerns the
future accessibility
and removability
of equipment.
Components installed
during construction
may subsequently
be surrounded
by other systems,
modified shielding,
services, or
activated materials.
Their eventual
replacement or
removal may
require dedicated
transport paths,
lifting arrangements,
remote-handling
equipment, or
temporary
reconfiguration
of the
infrastructure.
As discussed
in
Sec.~\ref{subsec:future_remote_handling},
such capabilities
are most effectively
incorporated
before the
corresponding
installation
becomes inaccessible
or highly
activated.

The same reasoning
applies to future
accelerator facilities
whose scientific
programmes may
involve substantial
changes of
technology.
The staged FCC
concept provides
an example in
which major
modifications
of accelerator
equipment and
supporting
infrastructure
may be required
between successive
scientific stages.
The ability
to introduce
new equipment,
remove obsolete
systems, and
establish the
protection
conditions
appropriate
to a different
accelerator
technology
should therefore
be considered
as part of
the original
facility design
\cite{Benedikt:2025FCCFeasibility}.

The importance
of life-cycle
planning also
follows from
the scientific
continuity
discussed in
Sec.~\ref{subsec:safety_scientific_continuity}.
A facility
optimized
exclusively
for its initial
operational
configuration
may become
difficult to
maintain,
consolidate,
or upgrade
when its
technical
requirements
evolve.
Conversely,
appropriate
provisions for
accessibility,
segmentation,
diagnostics,
component
replacement,
and configuration
validation can
support the
long-term
scientific
programme
without
compromising
the applicable
protection
requirements.

The resulting
life-cycle
perspective
extends the
concept of
graceful degradation
and recoverability
developed in
Sec.~\ref{subsec:future_graceful_degradation}.
Recoverability
concerns the
existence of
admissible
trajectories
toward subsequent
operation after
an interruption.
Life-cycle safety
addresses the
broader evolution
of the facility,
including changes
in its scientific
purpose and,
eventually,
the controlled
termination
of its operation.
Both require
understanding
not only the
current physical
state, but also
the configurations
and transitions
that the facility
may need to
support in
the future.

From the perspective
of the present
work, the principal
conclusion is
that safety
requirements
must accompany
the facility
throughout its
complete lifetime.
Their physical
realization
changes as
the installation
is constructed,
commissioned,
operated,
modified,
maintained,
and ultimately
dismantled.
The relevant
hazards and
protection
conditions
are phase-dependent,
but the underlying
requirement remains
the same:
the physical
states and
transitions
associated
with each
activity must
be compatible
with the
applicable
protection
framework.

This life-cycle
perspective
provides a
direct connection
between
\emph{physics in
safety} and
\emph{safety in
physics}.
Physical models,
measurements,
and operational
experience
support the
assessment
of hazards
at successive
stages of
the installation's
evolution.
Conversely,
the need to
control those
hazards influences
the original
design,
subsequent
modifications,
the organization
of interventions,
and the means
by which the
facility can
eventually
be dismantled
or decommissioned.
Safety is therefore
not a set of
conditions established
only for the
nominal scientific
operating state,
but a requirement
integrated into
the entire
physical and
technical
life cycle
of the facility.

\subsection{From safety constraints to safety-by-design}
\label{subsec:safety_by_design}

The considerations developed in the preceding subsections
suggest a general design principle for future accelerators
and underground experimental facilities: safety requirements
should be integrated into scientific and engineering design
from the earliest stages, rather than introduced primarily
as constraints on an already established installation.
The physical architecture of a facility, its protection
systems, operational arrangements, and provisions for
maintenance and recovery should be developed as mutually
dependent elements of a common design.

This principle follows directly from the safe operational
phase-space framework introduced in this work. The
scientific objectives of an accelerator or experiment
determine the physical conditions that must be established,
including beam parameters, magnetic fields, cryogenic
temperatures, radiation environments, and detector
configurations. However, the ability to establish these
conditions intentionally depends on the protection
functions available, the compatibility of the coupled
subsystem states, and the admissibility of the trajectories
required to reach them.

Safety requirements therefore influence the physical
configuration of the installation itself. The distribution
of stored energy, the geometry of accelerator and detector
components, the segmentation of cryogenic and electrical
systems, the arrangement of shielding, and the accessibility
of technical infrastructure can all affect the development
and consequences of abnormal transitions. These design
choices may determine which protection strategies are
physically feasible and which operational configurations
can subsequently be validated.

The distinction between prevention and mitigation is
particularly important in this context. Some hazardous
developments can be interrupted through sufficiently
rapid detection and automatic protection, while others
require passive containment, energy absorption, physical
segregation, or limitation of the inventories involved.
The appropriate combination depends on the physical
mechanisms, characteristic timescales, and credible
failure trajectories. Consequently, the design of
protective barriers must proceed together with the
development of the scientific and technical systems
they are intended to protect.

The HL-LHC illustrates this relationship through
the integration of more demanding beam parameters,
new superconducting technologies, upgraded collimation,
and modified machine-protection arrangements. The
associated protection requirements influence not
only the operation of the accelerator, but also
the physical configuration and technical interfaces
of its upgraded systems
\cite{Apollinari:2017HLLHC,
Wollmann:2020HLLHCMPS}.
The relevant design objective is not simply to
increase accelerator performance and subsequently
verify its safety, but to establish the technical
conditions under which the intended performance
can be achieved with adequate protection.

The underground experiments provide complementary
examples. The architectures of ATLAS and CMS determine
different combinations of magnetic, cryogenic,
mechanical, electrical, and radiological constraints.
These differences influence access arrangements,
equipment protection, detector movements, and
maintenance procedures. The corresponding safe
operational domains are therefore specific to
the physical and technical configurations of
the experiments, even when broadly analogous
scientific or operational activities are considered.

Future detector design must also account for the
radiological conditions expected after substantial
irradiation. As discussed in
Sec.~\ref{subsec:future_radiation_maintainability},
the HL-LHC radiation environment will influence
accessibility and maintenance near ATLAS and CMS,
as well as interventions involving activated
collimators and other accelerator equipment.
Appropriate shielding, material selection,
component replaceability, intervention geometry,
and remote-handling provisions can reduce
occupational exposure and improve maintainability.
The ALARA principle should therefore influence
the engineering configuration during design,
rather than being applied exclusively when
individual interventions are prepared
\cite{Otto:2020HLLHCSafety,
FassoRokni:2009ALARA}.

The spatial organization of future underground
facilities introduces another dimension of
safety-by-design. The FCC feasibility studies
demonstrate the importance of considering
tunnel and shaft arrangements, infrastructure
distribution, ventilation, cryogenic services,
personnel access, and emergency-response
capabilities within the overall civil-engineering
design
\cite{Benedikt:2025FCCFeasibility,
Benedikt:2025FCCCivilEngineering}.
Such characteristics influence not only
the consequences of credible abnormal events,
but also the feasibility of interventions
and the ability to maintain protection
functions across a geographically
distributed installation.

An essential requirement is that the protection
architecture remain compatible with the
possible evolution of the facility. Scientific
installations are modified through successive
upgrades, equipment replacements, changes in
operational requirements, and the introduction
of new technologies. The original design should
therefore provide, where practicable, sufficient
flexibility for testing, isolation, replacement,
and revalidation of equipment and protection
functions. This flexibility must itself be
subject to configuration control, since
modifications can introduce new dependencies
or invalidate assumptions underlying the
original safety assessment.

The same reasoning applies to the observability
of the installation. As discussed in
Sec.~\ref{subsec:future_automation},
monitoring and diagnostic capabilities should
be designed together with the equipment
and protection functions whose behavior
they are intended to assess. The placement
of sensors, availability of independent
diagnostic signals, synchronization of
event records, and possibility of functional
testing influence the ability to detect
abnormal conditions and verify protective
responses. Protection functions that
cannot be adequately observed or tested
may remain difficult to validate, even
when their nominal operating principles
are well understood.

The preceding discussion of safety margins
introduces a further requirement.
Design-stage assessments must account
for credible uncertainties in physical
models, measurements, technical
configurations, and protection responses.
As emphasized in
Sec.~\ref{subsec:future_safety_margin},
adequate protection cannot be established
solely through the separation between
nominal operating parameters and
individual equipment limits.
The evolution of credible failure
trajectories and the available protection
time must also be considered.

Such assessments should influence the
selection of operating limits, detection
thresholds, protection mechanisms, and
engineering margins. However, increasing
individual safety factors without considering
their physical meaning does not necessarily
improve the protection of the complete
facility. The relevant requirement is
to demonstrate that the coupled system
remains adequately protected under the
credible uncertainties and configurations
included in the design assessment.

Safety-by-design also requires attention
to the possible loss of technical functions.
The segmentation of infrastructure,
independence of protection arrangements,
and availability of controlled shutdown
and isolation mechanisms may determine
whether a local failure remains confined
or propagates to other subsystems.
Where justified by the remaining
protection conditions, these provisions
may also permit restricted operation
or controlled recovery, as discussed
in Sec.~\ref{subsec:future_graceful_degradation}.
Their purpose is not to guarantee
continued scientific operation after
every failure, but to establish
well-understood responses appropriate
to the resulting physical conditions.

Maintainability and recoverability
should consequently be regarded as
design properties rather than
exclusively operational concerns.
A component that can be diagnosed,
isolated, replaced, and revalidated
through a controlled intervention
may contribute more effectively
to long-term scientific performance
than an otherwise equivalent
component whose maintenance
requires substantial modification
of the surrounding installation.
The technical feasibility and
safety of future interventions
must therefore be considered
alongside nominal scientific
performance.

The life-cycle perspective developed
in Sec.~\ref{subsec:future_lifecycle}
extends this requirement to construction,
installation, commissioning, upgrades,
long shutdowns, and eventual dismantling
or decommissioning. These phases involve
different combinations of physical
hazards, technical configurations,
and protection requirements.
Design decisions made before the
installation enters operation can
determine whether equipment remains
accessible, whether activated materials
can be handled safely, and whether
protection functions can be progressively
modified or withdrawn during
decommissioning.

The corresponding safety assessment
must therefore address more than the
nominal scientific operating state.
It should consider representative
configurations and credible transitions
associated with commissioning,
routine operation, abnormal events,
maintenance, recovery, upgrades,
and the final life-cycle phases.
The objective is not to predict
every future intervention in detail,
but to ensure that the physical
architecture retains the necessary
protection and operational capabilities
as the installation evolves.

An important consequence concerns
the interpretation of scientific
optimization. Maximum instantaneous
accelerator or detector performance
does not necessarily correspond to
maximum scientific output over the
lifetime of the facility. A design
with demanding nominal performance
but poor maintainability, inadequate
protection, or difficult recovery
may produce less useful scientific
data over time than a configuration
whose protection and operational
capabilities are more robust.
This distinction follows from the
relationship between safety,
availability, and scientific
continuity developed in
Secs.~\ref{subsec:safety_availability}
and~\ref{subsec:safety_scientific_continuity}.

Integrated design does not imply
that scientific performance,
engineering feasibility,
protection integrity, and
maintainability can always
be improved simultaneously.
These requirements may compete
for space, resources, technical
complexity, or operating flexibility.
The purpose of safety-by-design
is therefore not to eliminate
such trade-offs, but to identify
and evaluate them before
fundamental design decisions
become difficult to modify.

Similarly, the objective is
not the elimination of all
hazards. High-energy beams,
powerful superconducting
magnets, cryogenic systems,
radiation fields, and
high-power electrical
installations are integral
to the scientific purpose
of modern accelerator
facilities. Their physical
properties cannot simply
be removed without
compromising the intended
scientific capability.
The relevant objective
is to establish a design
in which these energetic
and potentially hazardous
systems can be operated,
maintained, and recovered
within an adequately
validated protection
framework.

This requires cooperation
between accelerator and
detector physicists,
engineers, technical
specialists, radiation-
protection professionals,
safety personnel, and
the teams responsible
for operation and
maintenance. Their
respective expertise
must contribute to
the development and
assessment of the
complete installation.
As emphasized in
Sec.~\ref{subsec:safety_culture},
the distribution of
technical knowledge
makes clearly defined
responsibilities and
effective communication
essential components
of the design process.

The resulting approach
also requires continuous
verification throughout
the facility's lifetime.
A protection architecture
that has been validated
during initial commissioning
may require reassessment
following equipment
modifications, newly
identified failure
mechanisms, changes
in operating parameters,
or operational experience
revealing limitations
of the original models.
Safety-by-design must
therefore be understood
as an iterative process
linking physical design,
protection assessment,
commissioning evidence,
operational learning,
and subsequent
configuration control
\cite{Leveson:2004AccidentModel,
Leveson:2012SaferWorld}.

The safe operational
phase-space framework
provides a common
conceptual basis for
this integration.
Scientific and engineering
design determine the
physical configurations
that can be realized,
while protection design
and operational validation
establish which states
and transitions can
be admitted under
the corresponding
conditions.
Conversely, the
requirements associated
with these admissible
states and trajectories
can influence the
selection and development
of the physical
installation itself.

The central lesson
for future facilities
is therefore that
safety should not
be regarded as a
final verification
performed after
scientific design
has been completed.
It is a continuing
design requirement
that influences
the architecture,
operational capabilities,
and long-term evolution
of the installation.

In this sense,
\emph{physics in safety}
provides the physical
models, measurements,
and engineering
understanding required
to identify and
control hazards,
while \emph{safety
in physics}
determines how
those requirements
shape the scientific
facility and the
conditions under
which its capabilities
can be used.
Their integration
from the earliest
design stages is
essential for
developing future
accelerators and
underground experiments
capable of sustaining
ambitious scientific
programmes within
adequately controlled
physical and
operational conditions.

\section{Conclusions}
\label{sec:conclusions}

The operation of large particle-physics facilities reveals a
fundamental interdependence between scientific performance and
safety. The physical conditions required for frontier
experiments---high-energy particle beams, strong magnetic
fields, superconductivity, cryogenic temperatures, intense
radiation environments, and complex detector configurations---
also generate the principal hazards that must be controlled.
The LHC and its underground experiments ATLAS and CMS
demonstrate that these hazards cannot be considered
independently of the scientific and technical systems
responsible for producing them. Their understanding,
protection, and operational management form an integral
part of the experimental infrastructure.

This observation motivates the two complementary directions
developed in the present work: \emph{physics in safety}
and \emph{safety in physics}. The first describes how
physical laws, measurements, quantitative models, and
engineering knowledge provide the basis for identifying
hazards, reconstructing failure mechanisms, evaluating
their consequences, and developing preventive and
mitigative protection. The second expresses the converse
influence: the resulting safety requirements and
protection architecture determine the physical and
technical configurations in which scientific activities
can be performed, the conditions governing personnel
access and interventions, and the transitions through
which the facility may be operated, maintained,
and recovered.

The principal conceptual contribution of this work
is to bring these two directions together through
a \emph{safe operational phase-space framework}
for the coupled scientific facility. Rather than
representing safety as a collection of independent
equipment limits or administrative restrictions,
the framework considers the combined physical state,
technical configuration, operational mode, and
available protection functions. It thereby provides
a common description of the constraints imposed
by machine protection, detector protection,
cryogenic and environmental conditions,
radiation protection, and personnel access.

An essential feature of this formulation is that
the safe domain of a complex installation is
generally nonfactorizable. Subsystems that satisfy
their individual operating or protection limits
do not necessarily constitute an admissible
global configuration. The safe state of the
facility depends on interactions between beam,
magnet, cryogenic, detector, environmental,
and access conditions, as well as on the
availability and compatibility of the
corresponding protection functions.
This dependence becomes particularly important
when an abnormal event originating in one
subsystem modifies the conditions required
for the safe operation of another.

A second essential feature is the distinction
between admissible states and admissible
trajectories. Two configurations may each
satisfy the applicable safety requirements
without every transition between them
being permissible. Beam injection and
extraction, magnet powering, cryogenic
transitions, detector movement, personnel
access, maintenance, and the return to
physics all involve sequences in which
intermediate conditions and the timing
of protective responses can be decisive.
Safe operation must therefore be
characterized not only by the states
that may be established, but also by
the physical and operational paths
through which they can be reached.

The LHC machine-protection system provides
a particularly direct physical realization
of this principle. The large stored beam
energy required for high-energy collisions
can be exploited only under conditions
in which uncontrolled losses are prevented
or their consequences are limited by
adequate protection. Collimation,
beam-loss monitoring, interlocks,
passive protection, and beam extraction
form a coordinated architecture whose
effectiveness depends on the actual
machine configuration and the evolution
of credible failure scenarios.
The protection problem is therefore
inherently dynamical and cannot be
reduced to the magnitude of stored
energy alone.

The superconducting magnet and cryogenic
systems illustrate the same principle
through different physical mechanisms.
Electrical faults, quenches, energy
redistribution, helium expansion,
pressure transients, gas transport,
and oxygen displacement can couple
equipment-protection requirements
with underground environmental and
personnel-safety conditions.
The 2008 LHC sector~3--4 incident
provides a particularly important
historical example of how unexpected
failure propagation can reveal
limitations of an established
protection model and lead to
substantial revisions of the
corresponding safety architecture
\cite{Chorowski:2011RiskUpdate}.
It also demonstrates that safety
knowledge must remain open to
revision through operational
evidence and improved physical
understanding.

The comparison of ATLAS and CMS
extends this perspective to large
experimental detectors. Although
the experiments share several
general hazard classes, their
different magnet systems,
cryogenic installations,
mechanical architectures,
detector services, and
access arrangements produce
distinct risk configurations.
The same broadly defined
operational or intervention
activity may therefore require
different physical conditions
and protection arrangements
in the two experiments.
This difference does not
imply a ranking of their
safety performance; it
reflects the experiment-specific
realization of the corresponding
safe and admissible domains.

The analysis of commissioning,
access, maintenance, and
simultaneous interventions
further demonstrates that
safety cannot be identified
exclusively with the
availability of engineered
protection systems.
Configuration control,
personnel competence,
radiation-work planning,
coordination between
technical teams, and
verification of restored
conditions are essential
for maintaining the
assumptions under which
those systems provide
protection. Technical
and procedural measures
have complementary
functions, but they
are not interchangeable,
particularly when
hazardous developments
require responses faster
than human intervention
can provide.

Operational experience
also contributes directly
to the continuing
development of safety.
Anomalies, near misses,
successful protection
actions, and unexpected
equipment behavior
provide information
about initiating
events, subsystem
dependencies, barrier
performance, and
the validity of
existing failure
models. Their significance
does not depend
exclusively on whether
an adverse consequence
has occurred.
The resulting knowledge
can support revisions
of hazard assessments,
protection functions,
operational procedures,
and admissible
configurations, provided
that the corresponding
changes are appropriately
verified and validated.

These considerations
lead to an important
interpretation of
\emph{safety in physics}.
Safety requirements
may restrict particular
operating conditions,
delay interventions,
or require additional
commissioning and
verification. However,
effective and validated
protection can also
make more demanding
scientific configurations
accessible by controlling
the hazards associated
with their operation.
Such an enlargement
of the admissible
domain is conditional
on the adequacy of
the complete protection
architecture; it does
not follow automatically
from adding protective
devices or relaxing
individual operating
limits.

The scientific benefit
of protection also
extends beyond the
range of accessible
operating states.
The ability to
maintain, interrupt,
recover, and
re-establish
experimental
configurations
contributes to
the continuity
of scientific
operation over
the lifetime of
a facility.
Protection actions
may reduce
short-term
availability,
but can prevent
damage and prolonged
interruptions
that would otherwise
compromise the
scientific programme.
Consequently,
maximum instantaneous
performance does
not necessarily
correspond to
maximum sustainable
scientific output.

The implications
become particularly
important for the
HL-LHC and future
accelerator facilities.
Higher beam brightness
and stored energy,
new superconducting
technologies,
increasingly demanding
radiation environments,
and more complex
technical infrastructure
require the protection
architecture to
evolve together
with the scientific
installation.
For ATLAS and CMS,
radiation damage
and residual
activation will
place additional
requirements on
detector design,
shielding,
intervention
planning, and
maintainability.
The radiological
conditions associated
with activated
collimators and
other accelerator
components provide
a complementary
example of the
importance of
ALARA-based design
and maintenance.

Future facilities
also illustrate
that different
physical technologies
create different
safety problems.
The FCC concept
emphasizes the
importance of
underground
infrastructure
topology, distributed
technical services,
accessibility,
and emergency
response over
large spatial
scales. Muon-collider
concepts introduce
distinctive
radiological and
thermal challenges
associated with
high-power particle
production, muon
decay, detector
backgrounds, and
neutrino-induced
radiation beyond
the accelerator
complex.
The protection
principles developed
at the LHC are
transferable, but
their technical
implementation
must follow from
the relevant
physical processes
and credible
failure scenarios.

These examples
support a transition
toward
\emph{safety-by-design},
in which scientific,
engineering,
protection, and
operational
requirements are
developed together
from the earliest
stages of a
facility project.
Such an approach
must account not
only for nominal
performance,
but also for
uncertainty,
credible failure
trajectories,
protection margins,
observability,
maintainability,
remote handling,
controlled
degradation,
recoverability,
and the complete
facility life cycle,
including eventual
decommissioning.
The purpose is
not to eliminate
the physical
hazards intrinsic
to accelerator
science, but
to ensure that
their possible
consequences
remain adequately
controlled as
the installation
and its scientific
requirements
evolve.

The safe operational
phase-space
formulation is
intended as a
unifying conceptual
framework rather
than a substitute
for the detailed
quantitative
analyses required
to establish
specific protection
limits or assess
individual failure
scenarios.
Its value lies
in making explicit
the relationships
between the
physical state
of the facility,
its protection
architecture,
the admissibility
of operational
and intervention
trajectories,
and the evidence
required to
validate them.
In this sense,
the framework
connects
otherwise distinct
safety disciplines
without removing
the need for
their specialized
methods and
requirements.

The resulting
conclusion is
that physics
and safety are
not merely
compatible
activities
performed
within the same
facility.
They are
mutually
dependent
elements of
its scientific
capability.
\emph{Physics
in safety}
provides the
knowledge
required to
understand
and control
the physical
processes
that generate
hazards.
\emph{Safety
in physics}
determines
how the
corresponding
protection
requirements
shape the
configurations,
transitions,
and long-term
evolution
of the
scientific
installation.

A large
accelerator
or underground
experiment
is therefore
not made safe
simply by
adding
protection
systems to
an otherwise
completed
design.
Its scientific
capability
depends on
the coherent
integration
of physical
understanding,
engineering
protection,
operational
control,
and human
organization.
Safety is
thus neither
external
to experimental
physics nor
merely a
restriction
upon it:
it is one
of the
conditions
through
which
extreme
physical
states
become
accessible,
reproducible,
and
sustainable
in scientific
practice.

\section*{Acknowledgements}

The author wishes to acknowledge that the central
concept developed in this manuscript was inspired
by years of personal experience in safety
activities within the ATLAS experiment at CERN.
The close interaction between experimental and theoretical physics,
technical systems, and operational safety encountered
in this work motivated the broader perspective
presented here, in which physics provides the
foundation for safety, while safety in turn shapes
the conditions and possibilities of scientific
research.


\input{main.bbl}

\end{document}

%% file: main.bbl
%

%% file: mainV1.bbl
\begin{thebibliography}{115}%
\makeatletter
\providecommand \@ifxundefined [1]{%
 \@ifx{#1\undefined}
}%
\providecommand \@ifnum [1]{%
 \ifnum #1\expandafter \@firstoftwo
 \else \expandafter \@secondoftwo
 \fi
}%
\providecommand \@ifx [1]{%
 \ifx #1\expandafter \@firstoftwo
 \else \expandafter \@secondoftwo
 \fi
}%
\providecommand \natexlab [1]{#1}%
\providecommand \enquote  [1]{``#1''}%
\providecommand \bibnamefont  [1]{#1}%
\providecommand \bibfnamefont [1]{#1}%
\providecommand \citenamefont [1]{#1}%
\providecommand \href@noop [0]{\@secondoftwo}%
\providecommand \href [0]{\begingroup \@sanitize@url \@href}%
\providecommand \@href[1]{\@@startlink{#1}\@@href}%
\providecommand \@@href[1]{\endgroup#1\@@endlink}%
\providecommand \@sanitize@url [0]{\catcode `\\12\catcode `\$12\catcode `\&12\catcode `\#12\catcode `\^12\catcode `\_12\catcode `\%12\relax}%
\providecommand \@@startlink[1]{}%
\providecommand \@@endlink[0]{}%
\providecommand \url  [0]{\begingroup\@sanitize@url \@url }%
\providecommand \@url [1]{\endgroup\@href {#1}{\urlprefix }}%
\providecommand \urlprefix  [0]{URL }%
\providecommand \Eprint [0]{\href }%
\providecommand \doibase [0]{https://doi.org/}%
\providecommand \selectlanguage [0]{\@gobble}%
\providecommand \bibinfo  [0]{\@secondoftwo}%
\providecommand \bibfield  [0]{\@secondoftwo}%
\providecommand \translation [1]{[#1]}%
\providecommand \BibitemOpen [0]{}%
\providecommand \bibitemStop [0]{}%
\providecommand \bibitemNoStop [0]{.\EOS\space}%
\providecommand \EOS [0]{\spacefactor3000\relax}%
\providecommand \BibitemShut  [1]{\csname bibitem#1\endcsname}%
\let\auto@bib@innerbib\@empty
\bibitem [{\citenamefont {Br{\"u}ning}\ \emph {et~al.}(2004)\citenamefont {Br{\"u}ning}, \citenamefont {Collier}, \citenamefont {Lebrun}, \citenamefont {Myers}, \citenamefont {Ostojic}, \citenamefont {Poole},\ and\ \citenamefont {Proudlock}}]{Bruning:2004LHC}%
  \BibitemOpen
  \bibfield  {author} {\bibinfo {author} {\bibfnamefont {O.~S.}\ \bibnamefont {Br{\"u}ning}}, \bibinfo {author} {\bibfnamefont {P.}~\bibnamefont {Collier}}, \bibinfo {author} {\bibfnamefont {P.}~\bibnamefont {Lebrun}}, \bibinfo {author} {\bibfnamefont {S.}~\bibnamefont {Myers}}, \bibinfo {author} {\bibfnamefont {R.}~\bibnamefont {Ostojic}}, \bibinfo {author} {\bibfnamefont {J.}~\bibnamefont {Poole}},\ and\ \bibinfo {author} {\bibfnamefont {P.}~\bibnamefont {Proudlock}},\ }\href {https://doi.org/10.5170/CERN-2004-003-V-1} {\emph {\bibinfo {title} {{LHC Design Report, Volume I: The LHC Main Ring}}}},\ \bibinfo {series} {CERN Yellow Reports: Monographs}\ No.\ \bibinfo {number} {CERN-2004-003-V-1}\ (\bibinfo  {publisher} {CERN},\ \bibinfo {address} {Geneva},\ \bibinfo {year} {2004})\BibitemShut {NoStop}%
\bibitem [{\citenamefont {Wenninger}(2016)}]{Wenninger:2016MPO}%
  \BibitemOpen
  \bibfield  {author} {\bibinfo {author} {\bibfnamefont {J.}~\bibnamefont {Wenninger}},\ }\bibfield  {title} {\bibinfo {title} {{Machine Protection and Operation for LHC}},\ }in\ \href {https://doi.org/10.5170/CERN-2016-002.377} {\emph {\bibinfo {booktitle} {Proceedings of the 2014 Joint International Accelerator School: Beam Loss and Accelerator Protection}}}\ (\bibinfo  {publisher} {CERN},\ \bibinfo {address} {Geneva},\ \bibinfo {year} {2016})\ pp.\ \bibinfo {pages} {377--401},\ \Eprint {https://arxiv.org/abs/1608.03113} {arXiv:1608.03113 [physics.acc-ph]} \BibitemShut {NoStop}%
\bibitem [{\citenamefont {{ATLAS Collaboration}}(2008)}]{ATLAS:2008Detector}%
  \BibitemOpen
  \bibfield  {author} {\bibinfo {author} {\bibnamefont {{ATLAS Collaboration}}},\ }\bibfield  {title} {\bibinfo {title} {{The ATLAS Experiment at the CERN Large Hadron Collider}},\ }\href {https://doi.org/10.1088/1748-0221/3/08/S08003} {\bibfield  {journal} {\bibinfo  {journal} {JINST}\ }\textbf {\bibinfo {volume} {3}},\ \bibinfo {pages} {S08003}}\BibitemShut {NoStop}%
\bibitem [{\citenamefont {{CMS Collaboration}}(2008)}]{CMS:2008Detector}%
  \BibitemOpen
  \bibfield  {author} {\bibinfo {author} {\bibnamefont {{CMS Collaboration}}},\ }\bibfield  {title} {\bibinfo {title} {{The CMS Experiment at the CERN LHC}},\ }\href {https://doi.org/10.1088/1748-0221/3/08/S08004} {\bibfield  {journal} {\bibinfo  {journal} {JINST}\ }\textbf {\bibinfo {volume} {3}},\ \bibinfo {pages} {S08004}}\BibitemShut {NoStop}%
\bibitem [{\citenamefont {Schmidt}\ \emph {et~al.}(2007)\citenamefont {Schmidt}, \citenamefont {Assmann}, \citenamefont {Carlier}, \citenamefont {Dehning}, \citenamefont {Denz}, \citenamefont {Goddard}, \citenamefont {Holzer}, \citenamefont {Kain}, \citenamefont {Puccio}, \citenamefont {Todd}, \citenamefont {Uythoven}, \citenamefont {Wenninger},\ and\ \citenamefont {Zerlauth}}]{Bordry:2001MPS}%
  \BibitemOpen
  \bibfield  {author} {\bibinfo {author} {\bibfnamefont {R.}~\bibnamefont {Schmidt}}, \bibinfo {author} {\bibfnamefont {R.}~\bibnamefont {Assmann}}, \bibinfo {author} {\bibfnamefont {E.}~\bibnamefont {Carlier}}, \bibinfo {author} {\bibfnamefont {B.}~\bibnamefont {Dehning}}, \bibinfo {author} {\bibfnamefont {R.}~\bibnamefont {Denz}}, \bibinfo {author} {\bibfnamefont {B.}~\bibnamefont {Goddard}}, \bibinfo {author} {\bibfnamefont {E.~B.}\ \bibnamefont {Holzer}}, \bibinfo {author} {\bibfnamefont {V.}~\bibnamefont {Kain}}, \bibinfo {author} {\bibfnamefont {B.}~\bibnamefont {Puccio}}, \bibinfo {author} {\bibfnamefont {B.}~\bibnamefont {Todd}}, \bibinfo {author} {\bibfnamefont {J.}~\bibnamefont {Uythoven}}, \bibinfo {author} {\bibfnamefont {J.}~\bibnamefont {Wenninger}},\ and\ \bibinfo {author} {\bibfnamefont {M.}~\bibnamefont {Zerlauth}},\ }\bibfield  {title} {\bibinfo {title} {{LHC Machine Protection}},\ }in\ \href {https://doi.org/10.1109/PAC.2007.4441111} {\emph {\bibinfo {booktitle} {Proceedings of the 22nd
  IEEE Particle Accelerator Conference (PAC'07)}}}\ (\bibinfo {address} {Albuquerque, NM, USA},\ \bibinfo {year} {2007})\ pp.\ \bibinfo {pages} {878--882},\ \bibinfo {note} {paper TUZAC03}\BibitemShut {NoStop}%
\bibitem [{\citenamefont {Ninin}\ \emph {et~al.}(2008)\citenamefont {Ninin}, \citenamefont {Delamare}, \citenamefont {Di~Luca}, \citenamefont {Grau}, \citenamefont {Hakulinen}, \citenamefont {Hammouti}, \citenamefont {Havart}, \citenamefont {Juget}, \citenamefont {Ladzinski}, \citenamefont {Munoz~Codoceo}, \citenamefont {Nunes}, \citenamefont {Pettersson}, \citenamefont {Riesco}, \citenamefont {Sanchez-Corral}, \citenamefont {Scibile}, \citenamefont {Smith}, \citenamefont {Schmitt},\ and\ \citenamefont {Valentini}}]{Ninin:2005LACS}%
  \BibitemOpen
  \bibfield  {author} {\bibinfo {author} {\bibfnamefont {P.}~\bibnamefont {Ninin}}, \bibinfo {author} {\bibfnamefont {C.}~\bibnamefont {Delamare}}, \bibinfo {author} {\bibfnamefont {S.}~\bibnamefont {Di~Luca}}, \bibinfo {author} {\bibfnamefont {S.}~\bibnamefont {Grau}}, \bibinfo {author} {\bibfnamefont {T.}~\bibnamefont {Hakulinen}}, \bibinfo {author} {\bibfnamefont {L.}~\bibnamefont {Hammouti}}, \bibinfo {author} {\bibfnamefont {F.}~\bibnamefont {Havart}}, \bibinfo {author} {\bibfnamefont {J.-F.}\ \bibnamefont {Juget}}, \bibinfo {author} {\bibfnamefont {T.}~\bibnamefont {Ladzinski}}, \bibinfo {author} {\bibfnamefont {M.}~\bibnamefont {Munoz~Codoceo}}, \bibinfo {author} {\bibfnamefont {R.}~\bibnamefont {Nunes}}, \bibinfo {author} {\bibfnamefont {T.}~\bibnamefont {Pettersson}}, \bibinfo {author} {\bibfnamefont {T.}~\bibnamefont {Riesco}}, \bibinfo {author} {\bibfnamefont {E.}~\bibnamefont {Sanchez-Corral}}, \bibinfo {author} {\bibfnamefont {L.}~\bibnamefont {Scibile}}, \bibinfo {author} {\bibfnamefont
  {G.}~\bibnamefont {Smith}}, \bibinfo {author} {\bibfnamefont {F.}~\bibnamefont {Schmitt}},\ and\ \bibinfo {author} {\bibfnamefont {F.}~\bibnamefont {Valentini}},\ }\bibfield  {title} {\bibinfo {title} {{LHC Access System: From Design to Operation}},\ }in\ \href {https://proceedings.jacow.org/e08/papers/tupc129.pdf} {\emph {\bibinfo {booktitle} {Proceedings of the 11th European Particle Accelerator Conference (EPAC'08)}}}\ (\bibinfo {address} {Genoa, Italy},\ \bibinfo {year} {2008})\ pp.\ \bibinfo {pages} {1371--1373},\ \bibinfo {note} {paper TUPC129}\BibitemShut {NoStop}%
\bibitem [{\citenamefont {Ladzinski}\ \emph {et~al.}(2009)\citenamefont {Ladzinski}, \citenamefont {Valentini}, \citenamefont {Delamare}, \citenamefont {Di~Luca}, \citenamefont {Grau}, \citenamefont {Hakulinen}, \citenamefont {Hammouti}, \citenamefont {Havart}, \citenamefont {Juget}, \citenamefont {Ninin}, \citenamefont {Nunes}, \citenamefont {Pettersson}, \citenamefont {Riesco}, \citenamefont {Sanchez-Corral~Mena},\ and\ \citenamefont {Smith}}]{Ladzinski:2009LHCAccess}%
  \BibitemOpen
  \bibfield  {author} {\bibinfo {author} {\bibfnamefont {T.}~\bibnamefont {Ladzinski}}, \bibinfo {author} {\bibfnamefont {F.}~\bibnamefont {Valentini}}, \bibinfo {author} {\bibfnamefont {C.}~\bibnamefont {Delamare}}, \bibinfo {author} {\bibfnamefont {S.}~\bibnamefont {Di~Luca}}, \bibinfo {author} {\bibfnamefont {S.}~\bibnamefont {Grau}}, \bibinfo {author} {\bibfnamefont {T.}~\bibnamefont {Hakulinen}}, \bibinfo {author} {\bibfnamefont {L.}~\bibnamefont {Hammouti}}, \bibinfo {author} {\bibfnamefont {F.}~\bibnamefont {Havart}}, \bibinfo {author} {\bibfnamefont {J.-F.}\ \bibnamefont {Juget}}, \bibinfo {author} {\bibfnamefont {P.}~\bibnamefont {Ninin}}, \bibinfo {author} {\bibfnamefont {R.}~\bibnamefont {Nunes}}, \bibinfo {author} {\bibfnamefont {T.}~\bibnamefont {Pettersson}}, \bibinfo {author} {\bibfnamefont {T.}~\bibnamefont {Riesco}}, \bibinfo {author} {\bibfnamefont {E.}~\bibnamefont {Sanchez-Corral~Mena}},\ and\ \bibinfo {author} {\bibfnamefont {G.}~\bibnamefont {Smith}},\ }\bibfield  {title} {\bibinfo
  {title} {{The LHC Access System}},\ }in\ \href {https://proceedings.jacow.org/icalepcs2009/papers/wep102.pdf} {\emph {\bibinfo {booktitle} {Proceedings of the 12th International Conference on Accelerator and Large Experimental Physics Control Systems (ICALEPCS'09)}}}\ (\bibinfo {address} {Kobe, Japan},\ \bibinfo {year} {2009})\ pp.\ \bibinfo {pages} {600--602},\ \bibinfo {note} {paper WEP102}\BibitemShut {NoStop}%
\bibitem [{\citenamefont {L{\"u}ders}\ \emph {et~al.}(2003)\citenamefont {L{\"u}ders}, \citenamefont {Flockhart}, \citenamefont {Morpurgo},\ and\ \citenamefont {Schmeling}}]{Luders:2003DSS}%
  \BibitemOpen
  \bibfield  {author} {\bibinfo {author} {\bibfnamefont {S.}~\bibnamefont {L{\"u}ders}}, \bibinfo {author} {\bibfnamefont {R.~B.}\ \bibnamefont {Flockhart}}, \bibinfo {author} {\bibfnamefont {G.}~\bibnamefont {Morpurgo}},\ and\ \bibinfo {author} {\bibfnamefont {S.~M.}\ \bibnamefont {Schmeling}},\ }\bibfield  {title} {\bibinfo {title} {{The CERN Detector Safety System for the LHC Experiments}},\ }in\ \href {https://proceedings.jacow.org/ica03/PAPERS/TH214.PDF} {\emph {\bibinfo {booktitle} {Proceedings of the 9th International Conference on Accelerator and Large Experimental Physics Control Systems (ICALEPCS'03)}}}\ (\bibinfo  {publisher} {JACoW Publishing},\ \bibinfo {address} {Gyeongju, Korea},\ \bibinfo {year} {2003})\ pp.\ \bibinfo {pages} {569--571},\ \bibinfo {note} {paper TH214},\ \Eprint {https://arxiv.org/abs/hep-ex/0305067} {hep-ex/0305067} \BibitemShut {NoStop}%
\bibitem [{\citenamefont {Bajko}\ \emph {et~al.}(2009)\citenamefont {Bajko}, \citenamefont {Bertinelli}, \citenamefont {Catalan-Lasheras}, \citenamefont {Claudet}, \citenamefont {Cruikshank}, \citenamefont {Dahlerup-Petersen}, \citenamefont {Denz}, \citenamefont {Fessia}, \citenamefont {Garion}, \citenamefont {Jimenez}, \citenamefont {Kirby}, \citenamefont {Lebrun}, \citenamefont {Le~Naour}, \citenamefont {Mess}, \citenamefont {Modena}, \citenamefont {Montabonnet}, \citenamefont {Nunes}, \citenamefont {Parma}, \citenamefont {Perin}, \citenamefont {de~Rijk}, \citenamefont {Rijllart}, \citenamefont {Rossi}, \citenamefont {Schmidt}, \citenamefont {Siemko}, \citenamefont {Strubin}, \citenamefont {Tavian}, \citenamefont {Thiesen}, \citenamefont {Tock}, \citenamefont {Todesco}, \citenamefont {Veness}, \citenamefont {Verweij}, \citenamefont {Walckiers}, \citenamefont {Van~Weelderen}, \citenamefont {Wolf}, \citenamefont {Feh{\'e}r}, \citenamefont {Flora}, \citenamefont {Koratzinos}, \citenamefont {Limon},\ and\
  \citenamefont {Strait}}]{Bajko:2009Incident}%
  \BibitemOpen
  \bibfield  {author} {\bibinfo {author} {\bibfnamefont {M.}~\bibnamefont {Bajko}}, \bibinfo {author} {\bibfnamefont {F.}~\bibnamefont {Bertinelli}}, \bibinfo {author} {\bibfnamefont {N.}~\bibnamefont {Catalan-Lasheras}}, \bibinfo {author} {\bibfnamefont {S.}~\bibnamefont {Claudet}}, \bibinfo {author} {\bibfnamefont {P.}~\bibnamefont {Cruikshank}}, \bibinfo {author} {\bibfnamefont {K.}~\bibnamefont {Dahlerup-Petersen}}, \bibinfo {author} {\bibfnamefont {R.}~\bibnamefont {Denz}}, \bibinfo {author} {\bibfnamefont {P.}~\bibnamefont {Fessia}}, \bibinfo {author} {\bibfnamefont {C.}~\bibnamefont {Garion}}, \bibinfo {author} {\bibfnamefont {J.~M.}\ \bibnamefont {Jimenez}}, \bibinfo {author} {\bibfnamefont {G.}~\bibnamefont {Kirby}}, \bibinfo {author} {\bibfnamefont {P.}~\bibnamefont {Lebrun}}, \bibinfo {author} {\bibfnamefont {S.}~\bibnamefont {Le~Naour}}, \bibinfo {author} {\bibfnamefont {K.-H.}\ \bibnamefont {Mess}}, \bibinfo {author} {\bibfnamefont {M.}~\bibnamefont {Modena}}, \bibinfo {author} {\bibfnamefont
  {V.}~\bibnamefont {Montabonnet}}, \bibinfo {author} {\bibfnamefont {R.}~\bibnamefont {Nunes}}, \bibinfo {author} {\bibfnamefont {V.}~\bibnamefont {Parma}}, \bibinfo {author} {\bibfnamefont {A.}~\bibnamefont {Perin}}, \bibinfo {author} {\bibfnamefont {G.}~\bibnamefont {de~Rijk}}, \bibinfo {author} {\bibfnamefont {A.}~\bibnamefont {Rijllart}}, \bibinfo {author} {\bibfnamefont {L.}~\bibnamefont {Rossi}}, \bibinfo {author} {\bibfnamefont {R.}~\bibnamefont {Schmidt}}, \bibinfo {author} {\bibfnamefont {A.}~\bibnamefont {Siemko}}, \bibinfo {author} {\bibfnamefont {P.}~\bibnamefont {Strubin}}, \bibinfo {author} {\bibfnamefont {L.}~\bibnamefont {Tavian}}, \bibinfo {author} {\bibfnamefont {H.}~\bibnamefont {Thiesen}}, \bibinfo {author} {\bibfnamefont {J.~P.}\ \bibnamefont {Tock}}, \bibinfo {author} {\bibfnamefont {E.}~\bibnamefont {Todesco}}, \bibinfo {author} {\bibfnamefont {R.}~\bibnamefont {Veness}}, \bibinfo {author} {\bibfnamefont {A.}~\bibnamefont {Verweij}}, \bibinfo {author} {\bibfnamefont {L.}~\bibnamefont
  {Walckiers}}, \bibinfo {author} {\bibfnamefont {R.}~\bibnamefont {Van~Weelderen}}, \bibinfo {author} {\bibfnamefont {R.}~\bibnamefont {Wolf}}, \bibinfo {author} {\bibfnamefont {S.}~\bibnamefont {Feh{\'e}r}}, \bibinfo {author} {\bibfnamefont {R.}~\bibnamefont {Flora}}, \bibinfo {author} {\bibfnamefont {M.}~\bibnamefont {Koratzinos}}, \bibinfo {author} {\bibfnamefont {P.}~\bibnamefont {Limon}},\ and\ \bibinfo {author} {\bibfnamefont {J.}~\bibnamefont {Strait}},\ }\href {https://cds.cern.ch/record/1168025/} {\emph {\bibinfo {title} {{Report of the Task Force on the Incident of 19th September 2008 at the LHC}}}},\ \bibinfo {type} {Tech. Rep.}\ \bibinfo {number} {CERN-LHC-PROJECT-REPORT-1168}\ (\bibinfo  {institution} {CERN},\ \bibinfo {address} {Geneva},\ \bibinfo {year} {2009})\BibitemShut {NoStop}%
\bibitem [{\citenamefont {Chorowski}\ \emph {et~al.}(2010)\citenamefont {Chorowski}, \citenamefont {Fydrych}, \citenamefont {Modlinski}, \citenamefont {Polinski}, \citenamefont {Tavian},\ and\ \citenamefont {Wach}}]{CryogenicRisk:2010}%
  \BibitemOpen
  \bibfield  {author} {\bibinfo {author} {\bibfnamefont {M.}~\bibnamefont {Chorowski}}, \bibinfo {author} {\bibfnamefont {J.}~\bibnamefont {Fydrych}}, \bibinfo {author} {\bibfnamefont {Z.}~\bibnamefont {Modlinski}}, \bibinfo {author} {\bibfnamefont {J.}~\bibnamefont {Polinski}}, \bibinfo {author} {\bibfnamefont {L.}~\bibnamefont {Tavian}},\ and\ \bibinfo {author} {\bibfnamefont {J.}~\bibnamefont {Wach}},\ }\href {https://cds.cern.ch/record/1277511/} {\emph {\bibinfo {title} {{Upgrade on Risk Analysis Following the 080919 Incident in the LHC Sector 3--4}}}},\ \bibinfo {type} {Tech. Rep.}\ \bibinfo {number} {CERN-ATS-Note-2010-033 TECH}\ (\bibinfo  {institution} {CERN},\ \bibinfo {address} {Geneva},\ \bibinfo {year} {2010})\BibitemShut {NoStop}%
\bibitem [{\citenamefont {Delille}\ \emph {et~al.}(2009)\citenamefont {Delille}, \citenamefont {Evrard}, \citenamefont {Inigo-Golfin}, \citenamefont {Lindell}, \citenamefont {Roy}, \citenamefont {Tavian}, \citenamefont {Thomas}, \citenamefont {Trant},\ and\ \citenamefont {V{\"o}llinger}}]{LHCSafetyTaskForce:2009}%
  \BibitemOpen
  \bibfield  {author} {\bibinfo {author} {\bibfnamefont {B.}~\bibnamefont {Delille}}, \bibinfo {author} {\bibfnamefont {S.}~\bibnamefont {Evrard}}, \bibinfo {author} {\bibfnamefont {J.}~\bibnamefont {Inigo-Golfin}}, \bibinfo {author} {\bibfnamefont {G.}~\bibnamefont {Lindell}}, \bibinfo {author} {\bibfnamefont {G.}~\bibnamefont {Roy}}, \bibinfo {author} {\bibfnamefont {L.}~\bibnamefont {Tavian}}, \bibinfo {author} {\bibfnamefont {E.}~\bibnamefont {Thomas}}, \bibinfo {author} {\bibfnamefont {R.}~\bibnamefont {Trant}},\ and\ \bibinfo {author} {\bibfnamefont {C.}~\bibnamefont {V{\"o}llinger}},\ }\href {https://cds.cern.ch/record/1184225/} {\emph {\bibinfo {title} {{Task Force Report: Safety of Personnel in LHC Underground Areas Following the Accident of 19th September 2008}}}},\ \bibinfo {type} {Tech. Rep.}\ \bibinfo {number} {CERN-ATS-2009-002}\ (\bibinfo  {institution} {CERN},\ \bibinfo {address} {Geneva},\ \bibinfo {year} {2009})\BibitemShut {NoStop}%
\bibitem [{\citenamefont {Gomez-Reino}\ \emph {et~al.}(2009)\citenamefont {Gomez-Reino}, \citenamefont {Bauer}, \citenamefont {Beccati}, \citenamefont {Behrens}, \citenamefont {Biery}, \citenamefont {Branson}, \citenamefont {Cano}, \citenamefont {Cheung}, \citenamefont {Ciganek}, \citenamefont {Cittolin}, \citenamefont {Coarasa}, \citenamefont {Deldicque}, \citenamefont {Erhan}, \citenamefont {Gigi}, \citenamefont {Glege}, \citenamefont {Gutleber}, \citenamefont {Hatton}, \citenamefont {Hwong}, \citenamefont {Laurens}, \citenamefont {Loizides}, \citenamefont {Lopez~Perez}, \citenamefont {Meijers}, \citenamefont {Meschi}, \citenamefont {Meyer}, \citenamefont {Mommsen}, \citenamefont {Moser}, \citenamefont {O'Dell}, \citenamefont {Orsini}, \citenamefont {Paus}, \citenamefont {Petrucci}, \citenamefont {Pieri}, \citenamefont {Racz}, \citenamefont {Sakulin}, \citenamefont {Sani}, \citenamefont {Schwick}, \citenamefont {Serrano~Margaleff}, \citenamefont {Shpakov}, \citenamefont {Simon}, \citenamefont {Sumorok},\
  and\ \citenamefont {Zanetti}}]{CMS:2009DCS}%
  \BibitemOpen
  \bibfield  {author} {\bibinfo {author} {\bibfnamefont {R.}~\bibnamefont {Gomez-Reino}}, \bibinfo {author} {\bibfnamefont {G.}~\bibnamefont {Bauer}}, \bibinfo {author} {\bibfnamefont {B.}~\bibnamefont {Beccati}}, \bibinfo {author} {\bibfnamefont {U.}~\bibnamefont {Behrens}}, \bibinfo {author} {\bibfnamefont {K.}~\bibnamefont {Biery}}, \bibinfo {author} {\bibfnamefont {J.}~\bibnamefont {Branson}}, \bibinfo {author} {\bibfnamefont {E.}~\bibnamefont {Cano}}, \bibinfo {author} {\bibfnamefont {H.}~\bibnamefont {Cheung}}, \bibinfo {author} {\bibfnamefont {M.}~\bibnamefont {Ciganek}}, \bibinfo {author} {\bibfnamefont {S.}~\bibnamefont {Cittolin}}, \bibinfo {author} {\bibfnamefont {J.~A.}\ \bibnamefont {Coarasa}}, \bibinfo {author} {\bibfnamefont {C.}~\bibnamefont {Deldicque}}, \bibinfo {author} {\bibfnamefont {S.}~\bibnamefont {Erhan}}, \bibinfo {author} {\bibfnamefont {D.}~\bibnamefont {Gigi}}, \bibinfo {author} {\bibfnamefont {F.}~\bibnamefont {Glege}}, \bibinfo {author} {\bibfnamefont {J.}~\bibnamefont
  {Gutleber}}, \bibinfo {author} {\bibfnamefont {D.}~\bibnamefont {Hatton}}, \bibinfo {author} {\bibfnamefont {Y.~L.}\ \bibnamefont {Hwong}}, \bibinfo {author} {\bibfnamefont {J.-F.}\ \bibnamefont {Laurens}}, \bibinfo {author} {\bibfnamefont {C.}~\bibnamefont {Loizides}}, \bibinfo {author} {\bibfnamefont {J.~A.}\ \bibnamefont {Lopez~Perez}}, \bibinfo {author} {\bibfnamefont {F.}~\bibnamefont {Meijers}}, \bibinfo {author} {\bibfnamefont {E.}~\bibnamefont {Meschi}}, \bibinfo {author} {\bibfnamefont {A.}~\bibnamefont {Meyer}}, \bibinfo {author} {\bibfnamefont {R.}~\bibnamefont {Mommsen}}, \bibinfo {author} {\bibfnamefont {R.}~\bibnamefont {Moser}}, \bibinfo {author} {\bibfnamefont {V.}~\bibnamefont {O'Dell}}, \bibinfo {author} {\bibfnamefont {L.}~\bibnamefont {Orsini}}, \bibinfo {author} {\bibfnamefont {C.}~\bibnamefont {Paus}}, \bibinfo {author} {\bibfnamefont {A.}~\bibnamefont {Petrucci}}, \bibinfo {author} {\bibfnamefont {M.}~\bibnamefont {Pieri}}, \bibinfo {author} {\bibfnamefont {A.}~\bibnamefont {Racz}},
  \bibinfo {author} {\bibfnamefont {H.}~\bibnamefont {Sakulin}}, \bibinfo {author} {\bibfnamefont {M.}~\bibnamefont {Sani}}, \bibinfo {author} {\bibfnamefont {C.}~\bibnamefont {Schwick}}, \bibinfo {author} {\bibfnamefont {J.~F.}\ \bibnamefont {Serrano~Margaleff}}, \bibinfo {author} {\bibfnamefont {D.}~\bibnamefont {Shpakov}}, \bibinfo {author} {\bibfnamefont {M.}~\bibnamefont {Simon}}, \bibinfo {author} {\bibfnamefont {K.}~\bibnamefont {Sumorok}},\ and\ \bibinfo {author} {\bibfnamefont {M.}~\bibnamefont {Zanetti}},\ }\bibfield  {title} {\bibinfo {title} {{The Compact Muon Solenoid Detector Control System}},\ }in\ \href {https://epaper.kek.jp/icalepcs2009/papers/mob005.pdf} {\emph {\bibinfo {booktitle} {Proceedings of the 12th International Conference on Accelerator and Large Experimental Physics Control Systems (ICALEPCS'09)}}}\ (\bibinfo {address} {Kobe, Japan},\ \bibinfo {year} {2009})\ pp.\ \bibinfo {pages} {10--12},\ \bibinfo {note} {paper MOB005}\BibitemShut {NoStop}%
\bibitem [{\citenamefont {Di~Calafiori}\ \emph {et~al.}(2019)\citenamefont {Di~Calafiori}, \citenamefont {Dissertori}, \citenamefont {Jim{\'e}nez~Estupi{\~n}{\'a}n}, \citenamefont {Zelepoukine}, \citenamefont {Lustermann}, \citenamefont {Tsirou},\ and\ \citenamefont {Verdini}}]{DiCalafiori:2020CMS_ECAL}%
  \BibitemOpen
  \bibfield  {author} {\bibinfo {author} {\bibfnamefont {D.}~\bibnamefont {Di~Calafiori}}, \bibinfo {author} {\bibfnamefont {G.}~\bibnamefont {Dissertori}}, \bibinfo {author} {\bibfnamefont {R.}~\bibnamefont {Jim{\'e}nez~Estupi{\~n}{\'a}n}}, \bibinfo {author} {\bibfnamefont {S.}~\bibnamefont {Zelepoukine}}, \bibinfo {author} {\bibfnamefont {W.}~\bibnamefont {Lustermann}}, \bibinfo {author} {\bibfnamefont {A.}~\bibnamefont {Tsirou}},\ and\ \bibinfo {author} {\bibfnamefont {P.~G.}\ \bibnamefont {Verdini}},\ }\bibfield  {title} {\bibinfo {title} {{The CMS ECAL Control and Safety Systems Upgrades During the CERN LHC Long Shutdown 2}},\ }in\ \href {https://doi.org/10.18429/JACoW-ICALEPCS2019-WEPHA041} {\emph {\bibinfo {booktitle} {Proceedings of the 17th International Conference on Accelerator and Large Experimental Physics Control Systems (ICALEPCS'19)}}}\ (\bibinfo  {publisher} {JACoW Publishing},\ \bibinfo {address} {New York, NY, USA},\ \bibinfo {year} {2019})\ pp.\ \bibinfo {pages} {1175--1179},\ \bibinfo
  {note} {paper WEPHA041}\BibitemShut {NoStop}%
\bibitem [{\citenamefont {Lebrun}(2000)}]{Lebrun:1999Cryogenics}%
  \BibitemOpen
  \bibfield  {author} {\bibinfo {author} {\bibfnamefont {P.}~\bibnamefont {Lebrun}},\ }\bibfield  {title} {\bibinfo {title} {{Cryogenics for the Large Hadron Collider}},\ }\href {https://doi.org/10.1109/77.828526} {\bibfield  {journal} {\bibinfo  {journal} {IEEE Trans. Appl. Supercond.}\ }\textbf {\bibinfo {volume} {10}},\ \bibinfo {pages} {1500} (\bibinfo {year} {2000})},\ \bibinfo {note} {presented at the 16th International Conference on Magnet Technology (MT-16), Tallahassee, Florida, 26 September--2 October 1999; CERN-LHC-Project-Report-338}\BibitemShut {NoStop}%
\bibitem [{\citenamefont {Tahir}\ \emph {et~al.}(2005)\citenamefont {Tahir}, \citenamefont {Goddard}, \citenamefont {Kain}, \citenamefont {Schmidt}, \citenamefont {Shutov}, \citenamefont {Lomonosov}, \citenamefont {Piriz}, \citenamefont {Temporal}, \citenamefont {Hoffmann},\ and\ \citenamefont {Fortov}}]{Tahir:2005LHCBeam}%
  \BibitemOpen
  \bibfield  {author} {\bibinfo {author} {\bibfnamefont {N.~A.}\ \bibnamefont {Tahir}}, \bibinfo {author} {\bibfnamefont {B.}~\bibnamefont {Goddard}}, \bibinfo {author} {\bibfnamefont {V.}~\bibnamefont {Kain}}, \bibinfo {author} {\bibfnamefont {R.}~\bibnamefont {Schmidt}}, \bibinfo {author} {\bibfnamefont {A.}~\bibnamefont {Shutov}}, \bibinfo {author} {\bibfnamefont {I.~V.}\ \bibnamefont {Lomonosov}}, \bibinfo {author} {\bibfnamefont {A.~R.}\ \bibnamefont {Piriz}}, \bibinfo {author} {\bibfnamefont {M.}~\bibnamefont {Temporal}}, \bibinfo {author} {\bibfnamefont {D.~H.~H.}\ \bibnamefont {Hoffmann}},\ and\ \bibinfo {author} {\bibfnamefont {V.~E.}\ \bibnamefont {Fortov}},\ }\bibfield  {title} {\bibinfo {title} {{Impact of 7-TeV/c Large Hadron Collider Proton Beam on a Copper Target}},\ }\href {https://doi.org/10.1063/1.1888031} {\bibfield  {journal} {\bibinfo  {journal} {J. Appl. Phys.}\ }\textbf {\bibinfo {volume} {97}},\ \bibinfo {pages} {083532} (\bibinfo {year} {2005})}\BibitemShut {NoStop}%
\bibitem [{\citenamefont {Tahir}\ \emph {et~al.}(2009)\citenamefont {Tahir}, \citenamefont {Schmidt}, \citenamefont {Brugger}, \citenamefont {Shutov}, \citenamefont {Lomonosov}, \citenamefont {Piriz},\ and\ \citenamefont {Hoffmann}}]{Tahir:2009Graphite}%
  \BibitemOpen
  \bibfield  {author} {\bibinfo {author} {\bibfnamefont {N.~A.}\ \bibnamefont {Tahir}}, \bibinfo {author} {\bibfnamefont {R.}~\bibnamefont {Schmidt}}, \bibinfo {author} {\bibfnamefont {M.}~\bibnamefont {Brugger}}, \bibinfo {author} {\bibfnamefont {A.}~\bibnamefont {Shutov}}, \bibinfo {author} {\bibfnamefont {I.~V.}\ \bibnamefont {Lomonosov}}, \bibinfo {author} {\bibfnamefont {A.~R.}\ \bibnamefont {Piriz}},\ and\ \bibinfo {author} {\bibfnamefont {D.~H.~H.}\ \bibnamefont {Hoffmann}},\ }\bibfield  {title} {\bibinfo {title} {{Simulations of Full Impact of the Large Hadron Collider Beam with a Solid Graphite Target}},\ }\href {https://doi.org/10.1017/S0263034609990206} {\bibfield  {journal} {\bibinfo  {journal} {Laser Part. Beams}\ }\textbf {\bibinfo {volume} {27}},\ \bibinfo {pages} {475} (\bibinfo {year} {2009})}\BibitemShut {NoStop}%
\bibitem [{\citenamefont {Claudet}\ \emph {et~al.}(2010)\citenamefont {Claudet}, \citenamefont {Lebrun}, \citenamefont {Tavian},\ and\ \citenamefont {Wagner}}]{Claudet:2010LHCcryo}%
  \BibitemOpen
  \bibfield  {author} {\bibinfo {author} {\bibfnamefont {S.}~\bibnamefont {Claudet}}, \bibinfo {author} {\bibfnamefont {P.}~\bibnamefont {Lebrun}}, \bibinfo {author} {\bibfnamefont {L.}~\bibnamefont {Tavian}},\ and\ \bibinfo {author} {\bibfnamefont {U.}~\bibnamefont {Wagner}},\ }\bibfield  {title} {\bibinfo {title} {{Exergy Analysis of the Cryogenic Helium Distribution System for the Large Hadron Collider (LHC)}},\ }\href {https://doi.org/10.1063/1.3422294} {\bibfield  {journal} {\bibinfo  {journal} {AIP Conf. Proc.}\ }\textbf {\bibinfo {volume} {1218}},\ \bibinfo {pages} {1267} (\bibinfo {year} {2010})},\ \bibinfo {note} {cERN-ATS-2010-020}\BibitemShut {NoStop}%
\bibitem [{\citenamefont {ten Kate}(2005)}]{tenKate:2005ATLASMagnets}%
  \BibitemOpen
  \bibfield  {author} {\bibinfo {author} {\bibfnamefont {H.~H.~J.}\ \bibnamefont {ten Kate}},\ }\bibfield  {title} {\bibinfo {title} {{ATLAS Superconducting Toroids and Solenoid}},\ }\href {https://doi.org/10.1109/TASC.2005.849560} {\bibfield  {journal} {\bibinfo  {journal} {IEEE Trans. Appl. Supercond.}\ }\textbf {\bibinfo {volume} {15}},\ \bibinfo {pages} {1267} (\bibinfo {year} {2005})},\ \bibinfo {note} {on behalf of the ATLAS Collaboration}\BibitemShut {NoStop}%
\bibitem [{\citenamefont {Sklet}(2006)}]{Sklet:2006SafetyBarriers}%
  \BibitemOpen
  \bibfield  {author} {\bibinfo {author} {\bibfnamefont {S.}~\bibnamefont {Sklet}},\ }\bibfield  {title} {\bibinfo {title} {Safety barriers: Definition, classification, and performance},\ }\href {https://doi.org/10.1016/j.jlp.2005.12.004} {\bibfield  {journal} {\bibinfo  {journal} {Journal of Loss Prevention in the Process Industries}\ }\textbf {\bibinfo {volume} {19}},\ \bibinfo {pages} {494} (\bibinfo {year} {2006})}\BibitemShut {NoStop}%
\bibitem [{\citenamefont {de~Dianous}\ and\ \citenamefont {Fi{\'e}vez}(2006)}]{DeDianousFievez:2006ARAMIS}%
  \BibitemOpen
  \bibfield  {author} {\bibinfo {author} {\bibfnamefont {V.}~\bibnamefont {de~Dianous}}\ and\ \bibinfo {author} {\bibfnamefont {C.}~\bibnamefont {Fi{\'e}vez}},\ }\bibfield  {title} {\bibinfo {title} {{ARAMIS} project: A more explicit demonstration of risk control through the use of bow-tie diagrams and the evaluation of safety barrier performance},\ }\href {https://doi.org/10.1016/j.jhazmat.2005.07.010} {\bibfield  {journal} {\bibinfo  {journal} {Journal of Hazardous Materials}\ }\textbf {\bibinfo {volume} {130}},\ \bibinfo {pages} {220} (\bibinfo {year} {2006})}\BibitemShut {NoStop}%
\bibitem [{\citenamefont {Schmidt}(2016)}]{Schmidt:2016MPS}%
  \BibitemOpen
  \bibfield  {author} {\bibinfo {author} {\bibfnamefont {R.}~\bibnamefont {Schmidt}},\ }\bibfield  {title} {\bibinfo {title} {{Machine Protection and Interlock Systems for Circular Machines---Example for LHC}},\ }in\ \href {https://doi.org/10.5170/CERN-2016-002.319} {\emph {\bibinfo {booktitle} {Proceedings of the 2014 Joint International Accelerator School: Beam Loss and Accelerator Protection}}}\ (\bibinfo  {publisher} {CERN},\ \bibinfo {address} {Newport Beach, CA, USA},\ \bibinfo {year} {2016})\ pp.\ \bibinfo {pages} {319--341},\ \Eprint {https://arxiv.org/abs/1608.03087} {arXiv:1608.03087 [physics.acc-ph]} \BibitemShut {NoStop}%
\bibitem [{\citenamefont {Cartier-Michaud}\ \emph {et~al.}(2021)\citenamefont {Cartier-Michaud}, \citenamefont {Apollonio}, \citenamefont {Blarasin}, \citenamefont {Todd},\ and\ \citenamefont {Uythoven}}]{CartierMichaud:2021RiskMatrices}%
  \BibitemOpen
  \bibfield  {author} {\bibinfo {author} {\bibfnamefont {T.}~\bibnamefont {Cartier-Michaud}}, \bibinfo {author} {\bibfnamefont {A.}~\bibnamefont {Apollonio}}, \bibinfo {author} {\bibfnamefont {G.}~\bibnamefont {Blarasin}}, \bibinfo {author} {\bibfnamefont {B.}~\bibnamefont {Todd}},\ and\ \bibinfo {author} {\bibfnamefont {J.}~\bibnamefont {Uythoven}},\ }\bibfield  {title} {\bibinfo {title} {{Data-Driven Risk Matrices for CERN's Accelerators}},\ }in\ \href {https://doi.org/10.18429/JACoW-IPAC2021-TUPAB325} {\emph {\bibinfo {booktitle} {Proceedings of the 12th International Particle Accelerator Conference (IPAC'21)}}}\ (\bibinfo  {publisher} {JACoW Publishing},\ \bibinfo {address} {Campinas, Brazil},\ \bibinfo {year} {2021})\ pp.\ \bibinfo {pages} {2260--2263},\ \bibinfo {note} {paper TUPAB325}\BibitemShut {NoStop}%
\bibitem [{\citenamefont {Kaplan}\ and\ \citenamefont {Garrick}(1981)}]{KaplanGarrick:1981}%
  \BibitemOpen
  \bibfield  {author} {\bibinfo {author} {\bibfnamefont {S.}~\bibnamefont {Kaplan}}\ and\ \bibinfo {author} {\bibfnamefont {B.~J.}\ \bibnamefont {Garrick}},\ }\bibfield  {title} {\bibinfo {title} {{On the Quantitative Definition of Risk}},\ }\href {https://doi.org/10.1111/j.1539-6924.1981.tb01350.x} {\bibfield  {journal} {\bibinfo  {journal} {Risk Analysis}\ }\textbf {\bibinfo {volume} {1}},\ \bibinfo {pages} {11} (\bibinfo {year} {1981})}\BibitemShut {NoStop}%
\bibitem [{\citenamefont {Aven}(2016)}]{Aven:2016Risk}%
  \BibitemOpen
  \bibfield  {author} {\bibinfo {author} {\bibfnamefont {T.}~\bibnamefont {Aven}},\ }\bibfield  {title} {\bibinfo {title} {{Risk Assessment and Risk Management: Review of Recent Advances on Their Foundation}},\ }\href {https://doi.org/10.1016/j.ejor.2015.12.023} {\bibfield  {journal} {\bibinfo  {journal} {European Journal of Operational Research}\ }\textbf {\bibinfo {volume} {253}},\ \bibinfo {pages} {1} (\bibinfo {year} {2016})}\BibitemShut {NoStop}%
\bibitem [{\citenamefont {Filippini}\ \emph {et~al.}(2005)\citenamefont {Filippini}, \citenamefont {Dehning}, \citenamefont {Guaglio}, \citenamefont {Rodriguez-Mateos}, \citenamefont {Schmidt}, \citenamefont {Todd}, \citenamefont {Uythoven}, \citenamefont {Vergara-Fern{\'a}ndez},\ and\ \citenamefont {Zerlauth}}]{Filippini:2005Reliability}%
  \BibitemOpen
  \bibfield  {author} {\bibinfo {author} {\bibfnamefont {R.}~\bibnamefont {Filippini}}, \bibinfo {author} {\bibfnamefont {B.}~\bibnamefont {Dehning}}, \bibinfo {author} {\bibfnamefont {G.}~\bibnamefont {Guaglio}}, \bibinfo {author} {\bibfnamefont {F.}~\bibnamefont {Rodriguez-Mateos}}, \bibinfo {author} {\bibfnamefont {R.}~\bibnamefont {Schmidt}}, \bibinfo {author} {\bibfnamefont {B.}~\bibnamefont {Todd}}, \bibinfo {author} {\bibfnamefont {J.}~\bibnamefont {Uythoven}}, \bibinfo {author} {\bibfnamefont {A.}~\bibnamefont {Vergara-Fern{\'a}ndez}},\ and\ \bibinfo {author} {\bibfnamefont {M.}~\bibnamefont {Zerlauth}},\ }\bibfield  {title} {\bibinfo {title} {{Reliability Assessment of the LHC Machine Protection System}},\ }in\ \href {https://doi.org/10.1109/PAC.2005.1590726} {\emph {\bibinfo {booktitle} {Proceedings of the 2005 Particle Accelerator Conference (PAC 2005)}}}\ (\bibinfo  {publisher} {IEEE},\ \bibinfo {address} {Knoxville, TN, USA},\ \bibinfo {year} {2005})\ pp.\ \bibinfo {pages} {1257--1259},\ \bibinfo
  {note} {cERN-LHC-Project-Report-812}\BibitemShut {NoStop}%
\bibitem [{\citenamefont {Yuan}\ \emph {et~al.}(2022)\citenamefont {Yuan}, \citenamefont {Yang}, \citenamefont {Reniers}, \citenamefont {Chen},\ and\ \citenamefont {Wu}}]{Yuan:2022SafetyBarriers}%
  \BibitemOpen
  \bibfield  {author} {\bibinfo {author} {\bibfnamefont {S.}~\bibnamefont {Yuan}}, \bibinfo {author} {\bibfnamefont {M.}~\bibnamefont {Yang}}, \bibinfo {author} {\bibfnamefont {G.}~\bibnamefont {Reniers}}, \bibinfo {author} {\bibfnamefont {C.}~\bibnamefont {Chen}},\ and\ \bibinfo {author} {\bibfnamefont {J.}~\bibnamefont {Wu}},\ }\bibfield  {title} {\bibinfo {title} {Safety barriers in the chemical process industries: A state-of-the-art review on their classification, assessment, and management},\ }\href {https://doi.org/10.1016/j.ssci.2021.105647} {\bibfield  {journal} {\bibinfo  {journal} {Safety Science}\ }\textbf {\bibinfo {volume} {148}},\ \bibinfo {pages} {105647} (\bibinfo {year} {2022})}\BibitemShut {NoStop}%
\bibitem [{\citenamefont {Rausand}(2014)}]{Rausand:2014SafetyCritical}%
  \BibitemOpen
  \bibfield  {author} {\bibinfo {author} {\bibfnamefont {M.}~\bibnamefont {Rausand}},\ }\href {https://doi.org/10.1002/9781118776353} {\emph {\bibinfo {title} {Reliability of Safety-Critical Systems: Theory and Applications}}}\ (\bibinfo  {publisher} {John Wiley \& Sons},\ \bibinfo {year} {2014})\BibitemShut {NoStop}%
\bibitem [{\citenamefont {Puccio}\ \emph {et~al.}(2010)\citenamefont {Puccio}, \citenamefont {Casta{\~n}eda~Serra}, \citenamefont {Kwiatkowski}, \citenamefont {Romera~Ramirez},\ and\ \citenamefont {Todd}}]{Puccio:2010BIS}%
  \BibitemOpen
  \bibfield  {author} {\bibinfo {author} {\bibfnamefont {B.}~\bibnamefont {Puccio}}, \bibinfo {author} {\bibfnamefont {A.}~\bibnamefont {Casta{\~n}eda~Serra}}, \bibinfo {author} {\bibfnamefont {M.}~\bibnamefont {Kwiatkowski}}, \bibinfo {author} {\bibfnamefont {I.}~\bibnamefont {Romera~Ramirez}},\ and\ \bibinfo {author} {\bibfnamefont {B.}~\bibnamefont {Todd}},\ }\bibfield  {title} {\bibinfo {title} {{The CERN Beam Interlock System: Principle and Operational Experience}},\ }in\ \href {https://proceedings.jacow.org/IPAC10/papers/wepeb073.pdf} {\emph {\bibinfo {booktitle} {Proceedings of the 1st International Particle Accelerator Conference (IPAC'10)}}}\ (\bibinfo {address} {Kyoto, Japan},\ \bibinfo {year} {2010})\ pp.\ \bibinfo {pages} {2866--2868},\ \bibinfo {note} {cERN-ATS-2010-128, paper WEPEB073}\BibitemShut {NoStop}%
\bibitem [{\citenamefont {Bartmann}\ \emph {et~al.}(2010)\citenamefont {Bartmann}, \citenamefont {Assmann}, \citenamefont {Bracco}, \citenamefont {Goddard}, \citenamefont {Kain}, \citenamefont {Rossi},\ and\ \citenamefont {Wollmann}}]{Bartmann:2010InjectionDump}%
  \BibitemOpen
  \bibfield  {author} {\bibinfo {author} {\bibfnamefont {W.}~\bibnamefont {Bartmann}}, \bibinfo {author} {\bibfnamefont {R.}~\bibnamefont {Assmann}}, \bibinfo {author} {\bibfnamefont {C.}~\bibnamefont {Bracco}}, \bibinfo {author} {\bibfnamefont {B.}~\bibnamefont {Goddard}}, \bibinfo {author} {\bibfnamefont {V.}~\bibnamefont {Kain}}, \bibinfo {author} {\bibfnamefont {A.}~\bibnamefont {Rossi}},\ and\ \bibinfo {author} {\bibfnamefont {D.}~\bibnamefont {Wollmann}},\ }\href {https://cds.cern.ch/record/1281630} {\emph {\bibinfo {title} {{LHC Injection and Dump Protection}}}},\ \bibinfo {type} {Tech. Rep.}\ \bibinfo {number} {CERN-ATS-2010-028}\ (\bibinfo  {institution} {CERN},\ \bibinfo {address} {Geneva},\ \bibinfo {year} {2010})\ \bibinfo {note} {contribution to the first LHC Commissioning Workshop, Evian-les-Bains, France, 2010}\BibitemShut {NoStop}%
\bibitem [{\citenamefont {Willeke}(2016)}]{Willeke:2016Reliability}%
  \BibitemOpen
  \bibfield  {author} {\bibinfo {author} {\bibfnamefont {F.~J.}\ \bibnamefont {Willeke}},\ }\bibfield  {title} {\bibinfo {title} {Reliability considerations for the operation of large accelerator user facilities},\ }in\ \href {https://doi.org/10.5170/CERN-2016-002.111} {\emph {\bibinfo {booktitle} {Proceedings of the 2014 Joint International Accelerator School: Beam Loss and Accelerator Protection}}},\ \bibinfo {series} {CERN Yellow Reports}, Vol.~\bibinfo {volume} {2},\ \bibinfo {editor} {edited by\ \bibinfo {editor} {\bibfnamefont {R.}~\bibnamefont {Schmidt}}}\ (\bibinfo  {publisher} {CERN},\ \bibinfo {address} {Geneva},\ \bibinfo {year} {2016})\ pp.\ \bibinfo {pages} {111--136}\BibitemShut {NoStop}%
\bibitem [{\citenamefont {Fern{\'a}ndez~Adiego}\ \emph {et~al.}(2023)\citenamefont {Fern{\'a}ndez~Adiego}, \citenamefont {Blanco~Vi{\~n}uela}, \citenamefont {Germinario}, \citenamefont {Mainaud~Durand},\ and\ \citenamefont {Sosin}}]{FernandezAdiego:2023ProtectionLayers}%
  \BibitemOpen
  \bibfield  {author} {\bibinfo {author} {\bibfnamefont {B.}~\bibnamefont {Fern{\'a}ndez~Adiego}}, \bibinfo {author} {\bibfnamefont {E.}~\bibnamefont {Blanco~Vi{\~n}uela}}, \bibinfo {author} {\bibfnamefont {A.}~\bibnamefont {Germinario}}, \bibinfo {author} {\bibfnamefont {H.}~\bibnamefont {Mainaud~Durand}},\ and\ \bibinfo {author} {\bibfnamefont {M.}~\bibnamefont {Sosin}},\ }\bibfield  {title} {\bibinfo {title} {Protection layers design for the high luminosity {LHC} full remote alignment system},\ }in\ \href {https://doi.org/10.18429/JACoW-ICALEPCS2023-TU2BCO02} {\emph {\bibinfo {booktitle} {Proceedings of the 19th International Conference on Accelerator and Large Experimental Physics Control Systems (ICALEPCS'23)}}}\ (\bibinfo {year} {2023})\ pp.\ \bibinfo {pages} {285--291}\BibitemShut {NoStop}%
\bibitem [{\citenamefont {Redaelli}(2016)}]{Redaelli:2016Collimation}%
  \BibitemOpen
  \bibfield  {author} {\bibinfo {author} {\bibfnamefont {S.}~\bibnamefont {Redaelli}},\ }\bibfield  {title} {\bibinfo {title} {{Beam Cleaning and Collimation Systems}},\ }in\ \href {https://doi.org/10.5170/CERN-2016-002.403} {\emph {\bibinfo {booktitle} {Proceedings of the 2014 Joint International Accelerator School: Beam Loss and Accelerator Protection}}}\ (\bibinfo  {publisher} {CERN},\ \bibinfo {address} {Newport Beach, CA, USA},\ \bibinfo {year} {2016})\ pp.\ \bibinfo {pages} {403--437},\ \Eprint {https://arxiv.org/abs/1608.03159} {arXiv:1608.03159 [physics.acc-ph]} \BibitemShut {NoStop}%
\bibitem [{\citenamefont {Dahlerup-Petersen}\ \emph {et~al.}(2004)\citenamefont {Dahlerup-Petersen}, \citenamefont {Denz}, \citenamefont {Milani}, \citenamefont {Rodr{\'i}guez-Mateos},\ and\ \citenamefont {Tegenfeldt}}]{RodriguezMateos:2004QPS}%
  \BibitemOpen
  \bibfield  {author} {\bibinfo {author} {\bibfnamefont {K.}~\bibnamefont {Dahlerup-Petersen}}, \bibinfo {author} {\bibfnamefont {R.}~\bibnamefont {Denz}}, \bibinfo {author} {\bibfnamefont {D.}~\bibnamefont {Milani}}, \bibinfo {author} {\bibfnamefont {F.}~\bibnamefont {Rodr{\'i}guez-Mateos}},\ and\ \bibinfo {author} {\bibfnamefont {F.}~\bibnamefont {Tegenfeldt}},\ }\bibfield  {title} {\bibinfo {title} {{The Quench Protection System for the LHC Test String 2}},\ }\href {https://doi.org/10.1109/TASC.2004.829063} {\bibfield  {journal} {\bibinfo  {journal} {IEEE Trans. Appl. Supercond.}\ }\textbf {\bibinfo {volume} {14}},\ \bibinfo {pages} {251} (\bibinfo {year} {2004})},\ \bibinfo {note} {cERN-LHC-PROJECT-REPORT-715}\BibitemShut {NoStop}%
\bibitem [{\citenamefont {Kramer}\ \emph {et~al.}(2008)\citenamefont {Kramer}, \citenamefont {Goddard}, \citenamefont {Gyr}, \citenamefont {Koschik}, \citenamefont {Uythoven},\ and\ \citenamefont {Weiler}}]{Kramer:2008LBDS}%
  \BibitemOpen
  \bibfield  {author} {\bibinfo {author} {\bibfnamefont {T.}~\bibnamefont {Kramer}}, \bibinfo {author} {\bibfnamefont {B.}~\bibnamefont {Goddard}}, \bibinfo {author} {\bibfnamefont {M.}~\bibnamefont {Gyr}}, \bibinfo {author} {\bibfnamefont {A.}~\bibnamefont {Koschik}}, \bibinfo {author} {\bibfnamefont {J.~A.}\ \bibnamefont {Uythoven}},\ and\ \bibinfo {author} {\bibfnamefont {T.}~\bibnamefont {Weiler}},\ }\bibfield  {title} {\bibinfo {title} {{Apertures in the LHC Beam Dump System and Beam Losses During Beam Abort}},\ }in\ \href@noop {} {\emph {\bibinfo {booktitle} {Proceedings of the 11th European Particle Accelerator Conference (EPAC'08)}}}\ (\bibinfo {address} {Genoa, Italy},\ \bibinfo {year} {2008})\ pp.\ \bibinfo {pages} {2665--2667},\ \bibinfo {note} {paper WEPP064; CERN-LHC-PROJECT-REPORT-1161}\BibitemShut {NoStop}%
\bibitem [{\citenamefont {Holzer}\ \emph {et~al.}(2012)\citenamefont {Holzer}, \citenamefont {Dehning}, \citenamefont {Effinger}, \citenamefont {Emery}, \citenamefont {Grishin}, \citenamefont {Hajdu}, \citenamefont {Jackson}, \citenamefont {Kurfuerst}, \citenamefont {Marsili}, \citenamefont {Misiowiec}, \citenamefont {Nagel}, \citenamefont {Nebot~del Busto}, \citenamefont {Nordt}, \citenamefont {Roderick}, \citenamefont {Sapinski},\ and\ \citenamefont {Zamantzas}}]{Holzer:2012BLM}%
  \BibitemOpen
  \bibfield  {author} {\bibinfo {author} {\bibfnamefont {E.~B.}\ \bibnamefont {Holzer}}, \bibinfo {author} {\bibfnamefont {B.}~\bibnamefont {Dehning}}, \bibinfo {author} {\bibfnamefont {E.}~\bibnamefont {Effinger}}, \bibinfo {author} {\bibfnamefont {J.}~\bibnamefont {Emery}}, \bibinfo {author} {\bibfnamefont {V.}~\bibnamefont {Grishin}}, \bibinfo {author} {\bibfnamefont {C.}~\bibnamefont {Hajdu}}, \bibinfo {author} {\bibfnamefont {S.}~\bibnamefont {Jackson}}, \bibinfo {author} {\bibfnamefont {C.}~\bibnamefont {Kurfuerst}}, \bibinfo {author} {\bibfnamefont {A.}~\bibnamefont {Marsili}}, \bibinfo {author} {\bibfnamefont {M.}~\bibnamefont {Misiowiec}}, \bibinfo {author} {\bibfnamefont {M.}~\bibnamefont {Nagel}}, \bibinfo {author} {\bibfnamefont {E.}~\bibnamefont {Nebot~del Busto}}, \bibinfo {author} {\bibfnamefont {A.}~\bibnamefont {Nordt}}, \bibinfo {author} {\bibfnamefont {C.}~\bibnamefont {Roderick}}, \bibinfo {author} {\bibfnamefont {M.}~\bibnamefont {Sapinski}},\ and\ \bibinfo {author} {\bibfnamefont
  {C.}~\bibnamefont {Zamantzas}},\ }\bibfield  {title} {\bibinfo {title} {Beam loss monitoring for lhc machine protection},\ }\href {https://doi.org/10.1016/j.phpro.2012.04.110} {\bibfield  {journal} {\bibinfo  {journal} {Physics Procedia}\ }\textbf {\bibinfo {volume} {37}},\ \bibinfo {pages} {2055} (\bibinfo {year} {2012})}\BibitemShut {NoStop}%
\bibitem [{\citenamefont {Lechner}\ \emph {et~al.}(2022)\citenamefont {Lechner}, \citenamefont {B{\'e}langer}, \citenamefont {Efthymiopoulos}, \citenamefont {Grob}, \citenamefont {Lindstrom}, \citenamefont {Schmidt},\ and\ \citenamefont {Wollmann}}]{Lechner:2022Dust}%
  \BibitemOpen
  \bibfield  {author} {\bibinfo {author} {\bibfnamefont {A.}~\bibnamefont {Lechner}}, \bibinfo {author} {\bibfnamefont {P.}~\bibnamefont {B{\'e}langer}}, \bibinfo {author} {\bibfnamefont {I.}~\bibnamefont {Efthymiopoulos}}, \bibinfo {author} {\bibfnamefont {L.}~\bibnamefont {Grob}}, \bibinfo {author} {\bibfnamefont {B.}~\bibnamefont {Lindstrom}}, \bibinfo {author} {\bibfnamefont {R.}~\bibnamefont {Schmidt}},\ and\ \bibinfo {author} {\bibfnamefont {D.}~\bibnamefont {Wollmann}},\ }\bibfield  {title} {\bibinfo {title} {Dust-induced beam losses in the cryogenic arcs of the cern large hadron collider},\ }\href {https://doi.org/10.1103/PhysRevAccelBeams.25.041001} {\bibfield  {journal} {\bibinfo  {journal} {Phys. Rev. Accel. Beams}\ }\textbf {\bibinfo {volume} {25}},\ \bibinfo {pages} {041001} (\bibinfo {year} {2022})},\ \Eprint {https://arxiv.org/abs/2112.11068} {arXiv:2112.11068 [physics.acc-ph]} \BibitemShut {NoStop}%
\bibitem [{\citenamefont {Burkhardt}\ \emph {et~al.}(2003)\citenamefont {Burkhardt}, \citenamefont {Goddard},\ and\ \citenamefont {Mertens}}]{Burkhardt:2003TransferProtection}%
  \BibitemOpen
  \bibfield  {author} {\bibinfo {author} {\bibfnamefont {H.}~\bibnamefont {Burkhardt}}, \bibinfo {author} {\bibfnamefont {B.}~\bibnamefont {Goddard}},\ and\ \bibinfo {author} {\bibfnamefont {V.}~\bibnamefont {Mertens}},\ }\bibfield  {title} {\bibinfo {title} {{Protection Devices in the Transfer Lines to the LHC}},\ }in\ \href {https://doi.org/10.1109/PAC.2003.1288648} {\emph {\bibinfo {booktitle} {Proceedings of the 2003 Particle Accelerator Conference (PAC 2003)}}},\ Vol.~\bibinfo {volume} {3}\ (\bibinfo  {publisher} {IEEE},\ \bibinfo {address} {Portland, OR, USA},\ \bibinfo {year} {2003})\ pp.\ \bibinfo {pages} {1721--1723},\ \bibinfo {note} {cERN-LHC-PROJECT-REPORT-641}\BibitemShut {NoStop}%
\bibitem [{\citenamefont {Kain}\ \emph {et~al.}(2005)\citenamefont {Kain}, \citenamefont {Goddard}, \citenamefont {Schmidt},\ and\ \citenamefont {Wenninger}}]{Kain:2005InjectionProtection}%
  \BibitemOpen
  \bibfield  {author} {\bibinfo {author} {\bibfnamefont {V.}~\bibnamefont {Kain}}, \bibinfo {author} {\bibfnamefont {B.}~\bibnamefont {Goddard}}, \bibinfo {author} {\bibfnamefont {R.}~\bibnamefont {Schmidt}},\ and\ \bibinfo {author} {\bibfnamefont {J.}~\bibnamefont {Wenninger}},\ }\bibfield  {title} {\bibinfo {title} {{Protection Level During Extraction, Transfer and Injection into the LHC}},\ }in\ \href {https://doi.org/10.1109/PAC.2005.1590815} {\emph {\bibinfo {booktitle} {Proceedings of the 2005 Particle Accelerator Conference (PAC 2005)}}}\ (\bibinfo  {publisher} {IEEE},\ \bibinfo {address} {Knoxville, TN, USA},\ \bibinfo {year} {2005})\ pp.\ \bibinfo {pages} {1505--1507},\ \bibinfo {note} {paper RPPE016; CERN-LHC-PROJECT-REPORT-851}\BibitemShut {NoStop}%
\bibitem [{\citenamefont {Holzer}\ \emph {et~al.}(2005)\citenamefont {Holzer}, \citenamefont {Dehning}, \citenamefont {Effinger}, \citenamefont {Emery}, \citenamefont {Ferioli}, \citenamefont {Gonzalez}, \citenamefont {Gschwendtner}, \citenamefont {Guaglio}, \citenamefont {Hodgson}, \citenamefont {Kramer}, \citenamefont {Leitner}, \citenamefont {Ponce}, \citenamefont {Prieto}, \citenamefont {Stockner},\ and\ \citenamefont {Zamantzas}}]{Holzer:2005BLM}%
  \BibitemOpen
  \bibfield  {author} {\bibinfo {author} {\bibfnamefont {E.~B.}\ \bibnamefont {Holzer}}, \bibinfo {author} {\bibfnamefont {B.}~\bibnamefont {Dehning}}, \bibinfo {author} {\bibfnamefont {E.}~\bibnamefont {Effinger}}, \bibinfo {author} {\bibfnamefont {J.}~\bibnamefont {Emery}}, \bibinfo {author} {\bibfnamefont {G.}~\bibnamefont {Ferioli}}, \bibinfo {author} {\bibfnamefont {J.~L.}\ \bibnamefont {Gonzalez}}, \bibinfo {author} {\bibfnamefont {E.}~\bibnamefont {Gschwendtner}}, \bibinfo {author} {\bibfnamefont {G.}~\bibnamefont {Guaglio}}, \bibinfo {author} {\bibfnamefont {M.}~\bibnamefont {Hodgson}}, \bibinfo {author} {\bibfnamefont {D.}~\bibnamefont {Kramer}}, \bibinfo {author} {\bibfnamefont {R.}~\bibnamefont {Leitner}}, \bibinfo {author} {\bibfnamefont {L.}~\bibnamefont {Ponce}}, \bibinfo {author} {\bibfnamefont {V.}~\bibnamefont {Prieto}}, \bibinfo {author} {\bibfnamefont {M.}~\bibnamefont {Stockner}},\ and\ \bibinfo {author} {\bibfnamefont {C.}~\bibnamefont {Zamantzas}},\ }\bibfield  {title} {\bibinfo {title}
  {{Beam Loss Monitoring System for the LHC}},\ }in\ \href {https://doi.org/10.1109/NSSMIC.2005.1596433} {\emph {\bibinfo {booktitle} {2005 IEEE Nuclear Science Symposium Conference Record}}},\ Vol.~\bibinfo {volume} {2}\ (\bibinfo  {publisher} {IEEE},\ \bibinfo {address} {San Juan, Puerto Rico},\ \bibinfo {year} {2005})\ pp.\ \bibinfo {pages} {1052--1056},\ \bibinfo {note} {cERN-AB-2006-009-BI}\BibitemShut {NoStop}%
\bibitem [{\citenamefont {Dehning}(2016)}]{Dehning:2016BLM}%
  \BibitemOpen
  \bibfield  {author} {\bibinfo {author} {\bibfnamefont {B.}~\bibnamefont {Dehning}},\ }\bibfield  {title} {\bibinfo {title} {Beam loss monitors at lhc},\ }in\ \href {https://doi.org/10.5170/CERN-2016-002.303} {\emph {\bibinfo {booktitle} {Proceedings of the 2014 Joint International Accelerator School: Beam Loss and Accelerator Protection}}},\ \bibinfo {series} {CERN Yellow Reports}, Vol.~\bibinfo {volume} {2}\ (\bibinfo  {publisher} {CERN},\ \bibinfo {address} {Geneva},\ \bibinfo {year} {2016})\ pp.\ \bibinfo {pages} {303--318},\ \Eprint {https://arxiv.org/abs/1608.03079} {arXiv:1608.03079 [physics.acc-ph]} \BibitemShut {NoStop}%
\bibitem [{\citenamefont {Blaszkiewicz}\ \emph {et~al.}(2023)\citenamefont {Blaszkiewicz}, \citenamefont {Apollonio}, \citenamefont {Bolton}, \citenamefont {Colinet}, \citenamefont {Felsberger}, \citenamefont {Romera}, \citenamefont {Secondo}, \citenamefont {Uythoven},\ and\ \citenamefont {Wollmann}}]{Blaszkiewicz:2023SMP}%
  \BibitemOpen
  \bibfield  {author} {\bibinfo {author} {\bibfnamefont {M.}~\bibnamefont {Blaszkiewicz}}, \bibinfo {author} {\bibfnamefont {A.}~\bibnamefont {Apollonio}}, \bibinfo {author} {\bibfnamefont {S.}~\bibnamefont {Bolton}}, \bibinfo {author} {\bibfnamefont {A.}~\bibnamefont {Colinet}}, \bibinfo {author} {\bibfnamefont {L.}~\bibnamefont {Felsberger}}, \bibinfo {author} {\bibfnamefont {I.}~\bibnamefont {Romera}}, \bibinfo {author} {\bibfnamefont {R.}~\bibnamefont {Secondo}}, \bibinfo {author} {\bibfnamefont {J.}~\bibnamefont {Uythoven}},\ and\ \bibinfo {author} {\bibfnamefont {D.}~\bibnamefont {Wollmann}},\ }\bibfield  {title} {\bibinfo {title} {{Reliability Studies for CERN's New Safe Machine Parameter System}},\ }in\ \href {https://doi.org/10.18429/JACoW-IPAC2023-THPA057} {\emph {\bibinfo {booktitle} {Proceedings of the 14th International Particle Accelerator Conference (IPAC'23)}}}\ (\bibinfo  {publisher} {JACoW Publishing},\ \bibinfo {address} {Venice, Italy},\ \bibinfo {year} {2023})\ pp.\ \bibinfo {pages}
  {4085--4088},\ \bibinfo {note} {paper THPA057}\BibitemShut {NoStop}%
\bibitem [{\citenamefont {Bolton}\ \emph {et~al.}(2024)\citenamefont {Bolton}, \citenamefont {Blaszkiewicz}, \citenamefont {Colinet}, \citenamefont {Felsberger}, \citenamefont {Guasch-Martinez}, \citenamefont {Martin}, \citenamefont {Romera}, \citenamefont {Secondo},\ and\ \citenamefont {Uythoven}}]{Bolton:2024SMP}%
  \BibitemOpen
  \bibfield  {author} {\bibinfo {author} {\bibfnamefont {S.}~\bibnamefont {Bolton}}, \bibinfo {author} {\bibfnamefont {M.}~\bibnamefont {Blaszkiewicz}}, \bibinfo {author} {\bibfnamefont {A.}~\bibnamefont {Colinet}}, \bibinfo {author} {\bibfnamefont {L.}~\bibnamefont {Felsberger}}, \bibinfo {author} {\bibfnamefont {J.}~\bibnamefont {Guasch-Martinez}}, \bibinfo {author} {\bibfnamefont {C.}~\bibnamefont {Martin}}, \bibinfo {author} {\bibfnamefont {I.}~\bibnamefont {Romera}}, \bibinfo {author} {\bibfnamefont {R.}~\bibnamefont {Secondo}},\ and\ \bibinfo {author} {\bibfnamefont {J.}~\bibnamefont {Uythoven}},\ }\bibfield  {title} {\bibinfo {title} {Development of a second-generation system for the reliable distribution of machine protection parameters},\ }in\ \href {https://doi.org/10.18429/JACoW-IPAC2024-THPG60} {\emph {\bibinfo {booktitle} {Proceedings of the 15th International Particle Accelerator Conference (IPAC'24)}}}\ (\bibinfo  {publisher} {JACoW Publishing},\ \bibinfo {address} {Geneva, Switzerland},\
  \bibinfo {year} {2024})\ pp.\ \bibinfo {pages} {3401--3404}\BibitemShut {NoStop}%
\bibitem [{\citenamefont {Barlow}\ \emph {et~al.}(2006)\citenamefont {Barlow}, \citenamefont {Bobbio}, \citenamefont {Carlier}, \citenamefont {Gr{\"a}wer}, \citenamefont {Voumard},\ and\ \citenamefont {Gjelsvik}}]{Barlow:2005BETS}%
  \BibitemOpen
  \bibfield  {author} {\bibinfo {author} {\bibfnamefont {R.~A.}\ \bibnamefont {Barlow}}, \bibinfo {author} {\bibfnamefont {P.}~\bibnamefont {Bobbio}}, \bibinfo {author} {\bibfnamefont {E.}~\bibnamefont {Carlier}}, \bibinfo {author} {\bibfnamefont {G.}~\bibnamefont {Gr{\"a}wer}}, \bibinfo {author} {\bibfnamefont {N.}~\bibnamefont {Voumard}},\ and\ \bibinfo {author} {\bibfnamefont {R.}~\bibnamefont {Gjelsvik}},\ }\href {https://cds.cern.ch/record/927409} {\emph {\bibinfo {title} {{The Beam Energy Tracking System of the LHC Beam Dumping System}}}},\ \bibinfo {type} {Tech. Rep.}\ \bibinfo {number} {CERN-AB-2006-006-BT}\ (\bibinfo  {institution} {CERN},\ \bibinfo {address} {Geneva},\ \bibinfo {year} {2006})\ \bibinfo {note} {presented at the 10th International Conference on Accelerator and Large Experimental Physics Control Systems (ICALEPCS 2005), Geneva, Switzerland, 10--14 October 2005, paper PO2.056-4}\BibitemShut {NoStop}%
\bibitem [{\citenamefont {Bracco}\ \emph {et~al.}(2020)\citenamefont {Bracco}, \citenamefont {Barnes},\ and\ \citenamefont {Lechner}}]{Bracco:2020InjectionDumping}%
  \BibitemOpen
  \bibfield  {author} {\bibinfo {author} {\bibfnamefont {C.}~\bibnamefont {Bracco}}, \bibinfo {author} {\bibfnamefont {M.~J.}\ \bibnamefont {Barnes}},\ and\ \bibinfo {author} {\bibfnamefont {A.}~\bibnamefont {Lechner}},\ }\bibfield  {title} {\bibinfo {title} {Beam injection and dumping systems},\ }in\ \href {https://doi.org/10.23731/CYRM-2020-0010.259} {\emph {\bibinfo {booktitle} {High-Luminosity Large Hadron Collider (HL-LHC): Technical Design Report}}},\ \bibinfo {series and number} {CERN Yellow Reports: Monographs}\ (\bibinfo  {publisher} {CERN},\ \bibinfo {address} {Geneva},\ \bibinfo {year} {2020})\ pp.\ \bibinfo {pages} {259--268}\BibitemShut {NoStop}%
\bibitem [{\citenamefont {Carbajo~Perez}\ \emph {et~al.}(2021)\citenamefont {Carbajo~Perez}, \citenamefont {Perillo-Marcone}, \citenamefont {Berthom{\'e}}, \citenamefont {Bertone}, \citenamefont {Biancacci}, \citenamefont {Bracco}, \citenamefont {Bregliozzi}, \citenamefont {Bulat}, \citenamefont {Cadiou}, \citenamefont {Calviani}, \citenamefont {Cattenoz}, \citenamefont {Cherif}, \citenamefont {Costa~Pinto}, \citenamefont {Dallocchio}, \citenamefont {Di~Castro}, \citenamefont {Fessia}, \citenamefont {Frankl}, \citenamefont {Fuchs}, \citenamefont {Garcia~Gavela}, \citenamefont {Gentini}, \citenamefont {Geisser}, \citenamefont {Gilardoni}, \citenamefont {Gonzalez De La Aleja~Cabana}, \citenamefont {Grenard}, \citenamefont {Joly}, \citenamefont {Lechner}, \citenamefont {Lendaro}, \citenamefont {Maestre}, \citenamefont {Page}, \citenamefont {Perez~Ornedo}, \citenamefont {Pugnat}, \citenamefont {Rigutto}, \citenamefont {Seidenbinder}, \citenamefont {Salvant}, \citenamefont {Sapountzis}, \citenamefont {Scibor},
  \citenamefont {Sola~Merino}, \citenamefont {Taborelli}, \citenamefont {Urrutia}, \citenamefont {Franqueira~Ximenes}, \citenamefont {Vieille}, \citenamefont {Vollinger},\ and\ \citenamefont {Yin~Vallgren}}]{CarbajoPerez:2021TDIS}%
  \BibitemOpen
  \bibfield  {author} {\bibinfo {author} {\bibfnamefont {D.}~\bibnamefont {Carbajo~Perez}}, \bibinfo {author} {\bibfnamefont {A.}~\bibnamefont {Perillo-Marcone}}, \bibinfo {author} {\bibfnamefont {E.}~\bibnamefont {Berthom{\'e}}}, \bibinfo {author} {\bibfnamefont {C.}~\bibnamefont {Bertone}}, \bibinfo {author} {\bibfnamefont {N.}~\bibnamefont {Biancacci}}, \bibinfo {author} {\bibfnamefont {C.}~\bibnamefont {Bracco}}, \bibinfo {author} {\bibfnamefont {G.}~\bibnamefont {Bregliozzi}}, \bibinfo {author} {\bibfnamefont {B.}~\bibnamefont {Bulat}}, \bibinfo {author} {\bibfnamefont {C.}~\bibnamefont {Cadiou}}, \bibinfo {author} {\bibfnamefont {M.}~\bibnamefont {Calviani}}, \bibinfo {author} {\bibfnamefont {G.}~\bibnamefont {Cattenoz}}, \bibinfo {author} {\bibfnamefont {A.}~\bibnamefont {Cherif}}, \bibinfo {author} {\bibfnamefont {P.}~\bibnamefont {Costa~Pinto}}, \bibinfo {author} {\bibfnamefont {A.}~\bibnamefont {Dallocchio}}, \bibinfo {author} {\bibfnamefont {M.}~\bibnamefont {Di~Castro}}, \bibinfo {author}
  {\bibfnamefont {P.}~\bibnamefont {Fessia}}, \bibinfo {author} {\bibfnamefont {M.~I.}\ \bibnamefont {Frankl}}, \bibinfo {author} {\bibfnamefont {J.-F.}\ \bibnamefont {Fuchs}}, \bibinfo {author} {\bibfnamefont {H.}~\bibnamefont {Garcia~Gavela}}, \bibinfo {author} {\bibfnamefont {L.}~\bibnamefont {Gentini}}, \bibinfo {author} {\bibfnamefont {J.-M.}\ \bibnamefont {Geisser}}, \bibinfo {author} {\bibfnamefont {S.}~\bibnamefont {Gilardoni}}, \bibinfo {author} {\bibfnamefont {M.~A.}\ \bibnamefont {Gonzalez De La Aleja~Cabana}}, \bibinfo {author} {\bibfnamefont {J.-L.}\ \bibnamefont {Grenard}}, \bibinfo {author} {\bibfnamefont {S.}~\bibnamefont {Joly}}, \bibinfo {author} {\bibfnamefont {A.}~\bibnamefont {Lechner}}, \bibinfo {author} {\bibfnamefont {J.}~\bibnamefont {Lendaro}}, \bibinfo {author} {\bibfnamefont {J.}~\bibnamefont {Maestre}}, \bibinfo {author} {\bibfnamefont {E.}~\bibnamefont {Page}}, \bibinfo {author} {\bibfnamefont {M.}~\bibnamefont {Perez~Ornedo}}, \bibinfo {author} {\bibfnamefont {D.}~\bibnamefont
  {Pugnat}}, \bibinfo {author} {\bibfnamefont {E.}~\bibnamefont {Rigutto}}, \bibinfo {author} {\bibfnamefont {R.}~\bibnamefont {Seidenbinder}}, \bibinfo {author} {\bibfnamefont {B.}~\bibnamefont {Salvant}}, \bibinfo {author} {\bibfnamefont {A.}~\bibnamefont {Sapountzis}}, \bibinfo {author} {\bibfnamefont {K.}~\bibnamefont {Scibor}}, \bibinfo {author} {\bibfnamefont {J.}~\bibnamefont {Sola~Merino}}, \bibinfo {author} {\bibfnamefont {M.}~\bibnamefont {Taborelli}}, \bibinfo {author} {\bibfnamefont {E.}~\bibnamefont {Urrutia}}, \bibinfo {author} {\bibfnamefont {R.}~\bibnamefont {Franqueira~Ximenes}}, \bibinfo {author} {\bibfnamefont {A.}~\bibnamefont {Vieille}}, \bibinfo {author} {\bibfnamefont {C.}~\bibnamefont {Vollinger}},\ and\ \bibinfo {author} {\bibfnamefont {C.}~\bibnamefont {Yin~Vallgren}},\ }\bibfield  {title} {\bibinfo {title} {New generation cern lhc injection dump---assembly and installation (tdis)},\ }in\ \href {https://doi.org/10.18429/JACoW-IPAC2021-WEPAB361} {\emph {\bibinfo {booktitle}
  {Proceedings of the 12th International Particle Accelerator Conference (IPAC'21)}}}\ (\bibinfo  {publisher} {JACoW Publishing},\ \bibinfo {year} {2021})\ pp.\ \bibinfo {pages} {3548--3551}\BibitemShut {NoStop}%
\bibitem [{\citenamefont {Wiesner}\ \emph {et~al.}(2018)\citenamefont {Wiesner}, \citenamefont {Bartmann}, \citenamefont {Bracco}, \citenamefont {Carlier}, \citenamefont {Ducimeti{\`e}re}, \citenamefont {Frankl}, \citenamefont {Fraser}, \citenamefont {Goddard}, \citenamefont {He{\ss}ler}, \citenamefont {Kramer}, \citenamefont {Lechner}, \citenamefont {Magnin}, \citenamefont {Senaj},\ and\ \citenamefont {Wollmann}}]{Wiesner:2018AsynchronousDump}%
  \BibitemOpen
  \bibfield  {author} {\bibinfo {author} {\bibfnamefont {C.}~\bibnamefont {Wiesner}}, \bibinfo {author} {\bibfnamefont {W.}~\bibnamefont {Bartmann}}, \bibinfo {author} {\bibfnamefont {C.}~\bibnamefont {Bracco}}, \bibinfo {author} {\bibfnamefont {E.}~\bibnamefont {Carlier}}, \bibinfo {author} {\bibfnamefont {L.}~\bibnamefont {Ducimeti{\`e}re}}, \bibinfo {author} {\bibfnamefont {M.~I.}\ \bibnamefont {Frankl}}, \bibinfo {author} {\bibfnamefont {M.~A.}\ \bibnamefont {Fraser}}, \bibinfo {author} {\bibfnamefont {B.}~\bibnamefont {Goddard}}, \bibinfo {author} {\bibfnamefont {C.}~\bibnamefont {He{\ss}ler}}, \bibinfo {author} {\bibfnamefont {T.}~\bibnamefont {Kramer}}, \bibinfo {author} {\bibfnamefont {A.}~\bibnamefont {Lechner}}, \bibinfo {author} {\bibfnamefont {N.}~\bibnamefont {Magnin}}, \bibinfo {author} {\bibfnamefont {V.}~\bibnamefont {Senaj}},\ and\ \bibinfo {author} {\bibfnamefont {D.}~\bibnamefont {Wollmann}},\ }\bibfield  {title} {\bibinfo {title} {Asynchronous beam dump tests at lhc},\ }in\ \href
  {https://doi.org/10.18429/JACoW-IPAC2018-MOPMF063} {\emph {\bibinfo {booktitle} {Proceedings of the 9th International Particle Accelerator Conference (IPAC'18)}}}\ (\bibinfo  {publisher} {JACoW Publishing},\ \bibinfo {year} {2018})\BibitemShut {NoStop}%
\bibitem [{\citenamefont {Chorowski}\ \emph {et~al.}(2000)\citenamefont {Chorowski}, \citenamefont {Konopka},\ and\ \citenamefont {Riddone}}]{Chorowski:2000Helium}%
  \BibitemOpen
  \bibfield  {author} {\bibinfo {author} {\bibfnamefont {M.}~\bibnamefont {Chorowski}}, \bibinfo {author} {\bibfnamefont {G.}~\bibnamefont {Konopka}},\ and\ \bibinfo {author} {\bibfnamefont {G.}~\bibnamefont {Riddone}},\ }\bibfield  {title} {\bibinfo {title} {{Helium Discharge and Dispersion in the LHC Accelerator Tunnel in Case of Cryogenic Failure}},\ }in\ \href {https://cds.cern.ch/record/449265} {\emph {\bibinfo {booktitle} {Proceedings of the 18th International Cryogenic Engineering Conference (ICEC 18)}}}\ (\bibinfo {address} {Mumbai, India},\ \bibinfo {year} {2000})\ pp.\ \bibinfo {pages} {231--234},\ \bibinfo {note} {cERN-LHC-Project-Report-381}\BibitemShut {NoStop}%
\bibitem [{\citenamefont {Chorowski}\ \emph {et~al.}(2001)\citenamefont {Chorowski}, \citenamefont {Erdt}, \citenamefont {Konopka},\ and\ \citenamefont {Riddone}}]{Chorowski:2001ODH}%
  \BibitemOpen
  \bibfield  {author} {\bibinfo {author} {\bibfnamefont {M.}~\bibnamefont {Chorowski}}, \bibinfo {author} {\bibfnamefont {W.~K.}\ \bibnamefont {Erdt}}, \bibinfo {author} {\bibfnamefont {G.}~\bibnamefont {Konopka}},\ and\ \bibinfo {author} {\bibfnamefont {G.}~\bibnamefont {Riddone}},\ }\href {https://cds.cern.ch/record/534110?ln=de} {\emph {\bibinfo {title} {{Oxygen Deficiency Hazard (ODH) Monitoring System in the LHC Tunnel}}}},\ \bibinfo {type} {Tech. Rep.}\ \bibinfo {number} {CERN-LHC-Project-Report-523}\ (\bibinfo  {institution} {CERN},\ \bibinfo {address} {Geneva},\ \bibinfo {year} {2001})\BibitemShut {NoStop}%
\bibitem [{\citenamefont {Sciver}(2012)}]{VanSciver:2012Helium}%
  \BibitemOpen
  \bibfield  {author} {\bibinfo {author} {\bibfnamefont {S.~W.~V.}\ \bibnamefont {Sciver}},\ }\href {https://doi.org/10.1007/978-1-4419-9979-5} {\emph {\bibinfo {title} {Helium Cryogenics}}},\ \bibinfo {edition} {2nd}\ ed.,\ International Cryogenics Monograph Series\ (\bibinfo  {publisher} {Springer},\ \bibinfo {address} {New York},\ \bibinfo {year} {2012})\BibitemShut {NoStop}%
\bibitem [{\citenamefont {Parma}\ and\ \citenamefont {Leclercq}(2017)}]{ParmaLeclercq:2017CryogenicSafety}%
  \BibitemOpen
  \bibfield  {author} {\bibinfo {author} {\bibfnamefont {V.}~\bibnamefont {Parma}}\ and\ \bibinfo {author} {\bibfnamefont {Y.}~\bibnamefont {Leclercq}},\ }\bibfield  {title} {\bibinfo {title} {Cryogenic safety in helium cryostats at cern},\ }\href {https://doi.org/10.1088/1757-899X/278/1/012171} {\bibfield  {journal} {\bibinfo  {journal} {IOP Conf. Ser. Mater. Sci. Eng.}\ }\textbf {\bibinfo {volume} {278}},\ \bibinfo {pages} {012171} (\bibinfo {year} {2017})}\BibitemShut {NoStop}%
\bibitem [{\citenamefont {Cruikshank}\ \emph {et~al.}(2010)\citenamefont {Cruikshank}, \citenamefont {Parma}, \citenamefont {Perin},\ and\ \citenamefont {Tavian}}]{Cruikshank:2010Relief}%
  \BibitemOpen
  \bibfield  {author} {\bibinfo {author} {\bibfnamefont {P.}~\bibnamefont {Cruikshank}}, \bibinfo {author} {\bibfnamefont {V.}~\bibnamefont {Parma}}, \bibinfo {author} {\bibfnamefont {A.}~\bibnamefont {Perin}},\ and\ \bibinfo {author} {\bibfnamefont {L.}~\bibnamefont {Tavian}},\ }\href {https://cds.cern.ch/record/1316475} {\emph {\bibinfo {title} {{New Protection Scheme and Pressure Relief-Valve Staging of the LHC Insulation Vacuum Enclosure Following the 19th September 2008 Incident}}}},\ \bibinfo {type} {Tech. Rep.}\ \bibinfo {number} {CERN-ATS-Note-2010-057 TECH}\ (\bibinfo  {institution} {CERN},\ \bibinfo {address} {Geneva},\ \bibinfo {year} {2010})\BibitemShut {NoStop}%
\bibitem [{\citenamefont {Chorowski}\ \emph {et~al.}(2011)\citenamefont {Chorowski}, \citenamefont {Fydrych}, \citenamefont {Modlinski}, \citenamefont {Polinski}, \citenamefont {Tavian},\ and\ \citenamefont {Wach}}]{Chorowski:2011RiskUpdate}%
  \BibitemOpen
  \bibfield  {author} {\bibinfo {author} {\bibfnamefont {M.}~\bibnamefont {Chorowski}}, \bibinfo {author} {\bibfnamefont {J.}~\bibnamefont {Fydrych}}, \bibinfo {author} {\bibfnamefont {Z.}~\bibnamefont {Modlinski}}, \bibinfo {author} {\bibfnamefont {J.}~\bibnamefont {Polinski}}, \bibinfo {author} {\bibfnamefont {L.}~\bibnamefont {Tavian}},\ and\ \bibinfo {author} {\bibfnamefont {J.}~\bibnamefont {Wach}},\ }\bibfield  {title} {\bibinfo {title} {{Risk Analysis Update of the LHC Cryogenic System Following the 19th September 2008 Incident}},\ }in\ \href {https://cds.cern.ch/record/1377044?ln=en} {\emph {\bibinfo {booktitle} {Proceedings of the 23rd International Cryogenic Engineering Conference and International Cryogenic Materials Conference 2010 (ICEC 23--ICMC 2010)}}}\ (\bibinfo  {publisher} {Oficyna Wydawnicza Politechniki Wroclawskiej},\ \bibinfo {address} {Wroclaw, Poland},\ \bibinfo {year} {2011})\ pp.\ \bibinfo {pages} {879--884},\ \bibinfo {note} {cERN-ATS-2011-019}\BibitemShut {NoStop}%
\bibitem [{\citenamefont {Boterenbrood}\ \emph {et~al.}(2004)\citenamefont {Boterenbrood}, \citenamefont {Burckhart}, \citenamefont {Cook}, \citenamefont {Filimonov}, \citenamefont {Hallgren}, \citenamefont {Heubers}, \citenamefont {Khomoutnikov}, \citenamefont {Ryabov},\ and\ \citenamefont {Varela}}]{Boterenbrood:2004ATLASDCS}%
  \BibitemOpen
  \bibfield  {author} {\bibinfo {author} {\bibfnamefont {H.}~\bibnamefont {Boterenbrood}}, \bibinfo {author} {\bibfnamefont {H.~J.}\ \bibnamefont {Burckhart}}, \bibinfo {author} {\bibfnamefont {J.}~\bibnamefont {Cook}}, \bibinfo {author} {\bibfnamefont {V.}~\bibnamefont {Filimonov}}, \bibinfo {author} {\bibfnamefont {B.}~\bibnamefont {Hallgren}}, \bibinfo {author} {\bibfnamefont {W.}~\bibnamefont {Heubers}}, \bibinfo {author} {\bibfnamefont {V.}~\bibnamefont {Khomoutnikov}}, \bibinfo {author} {\bibfnamefont {Y.}~\bibnamefont {Ryabov}},\ and\ \bibinfo {author} {\bibfnamefont {F.}~\bibnamefont {Varela}},\ }\bibfield  {title} {\bibinfo {title} {Design and implementation of the atlas detector control system},\ }\href {https://doi.org/10.1109/TNS.2004.828523} {\bibfield  {journal} {\bibinfo  {journal} {IEEE Trans. Nucl. Sci.}\ }\textbf {\bibinfo {volume} {51}},\ \bibinfo {pages} {495} (\bibinfo {year} {2004})}\BibitemShut {NoStop}%
\bibitem [{\citenamefont {Schmeling}\ \emph {et~al.}(2004)\citenamefont {Schmeling}, \citenamefont {Flockhart}, \citenamefont {L{\"u}ders},\ and\ \citenamefont {Morpurgo}}]{Schmeling:2004DSS}%
  \BibitemOpen
  \bibfield  {author} {\bibinfo {author} {\bibfnamefont {S.}~\bibnamefont {Schmeling}}, \bibinfo {author} {\bibfnamefont {R.~B.}\ \bibnamefont {Flockhart}}, \bibinfo {author} {\bibfnamefont {S.}~\bibnamefont {L{\"u}ders}},\ and\ \bibinfo {author} {\bibfnamefont {G.}~\bibnamefont {Morpurgo}},\ }\bibfield  {title} {\bibinfo {title} {The detector safety system for lhc experiments},\ }\href {https://doi.org/10.1109/TNS.2004.828631} {\bibfield  {journal} {\bibinfo  {journal} {IEEE Trans. Nucl. Sci.}\ }\textbf {\bibinfo {volume} {51}},\ \bibinfo {pages} {521} (\bibinfo {year} {2004})}\BibitemShut {NoStop}%
\bibitem [{\citenamefont {Bremer}(2000)}]{Bremer:2000ATLASLAr}%
  \BibitemOpen
  \bibfield  {author} {\bibinfo {author} {\bibfnamefont {J.}~\bibnamefont {Bremer}},\ }\bibfield  {title} {\bibinfo {title} {{The Cryogenic System for the ATLAS Liquid Argon Detector}},\ }in\ \href@noop {} {\emph {\bibinfo {booktitle} {Proceedings of the 18th International Cryogenic Engineering Conference (ICEC 18)}}}\ (\bibinfo {address} {Mumbai, India},\ \bibinfo {year} {2000})\ pp.\ \bibinfo {pages} {219--222},\ \bibinfo {note} {on behalf of the ATLAS Liquid Argon Collaboration; CERN-LHC-Project-Report-393}\BibitemShut {NoStop}%
\bibitem [{\citenamefont {Fabre}\ \emph {et~al.}(2009)\citenamefont {Fabre}, \citenamefont {Bremer}, \citenamefont {Chalifour}, \citenamefont {Gonidec}, \citenamefont {Passardi}, \citenamefont {Petit}, \citenamefont {Pezzetti},\ and\ \citenamefont {Wicek}}]{Fabre:2009ATLASLArCryo}%
  \BibitemOpen
  \bibfield  {author} {\bibinfo {author} {\bibfnamefont {C.}~\bibnamefont {Fabre}}, \bibinfo {author} {\bibfnamefont {J.}~\bibnamefont {Bremer}}, \bibinfo {author} {\bibfnamefont {M.}~\bibnamefont {Chalifour}}, \bibinfo {author} {\bibfnamefont {G.}~\bibnamefont {Gonidec}}, \bibinfo {author} {\bibfnamefont {G.}~\bibnamefont {Passardi}}, \bibinfo {author} {\bibfnamefont {P.}~\bibnamefont {Petit}}, \bibinfo {author} {\bibfnamefont {M.}~\bibnamefont {Pezzetti}},\ and\ \bibinfo {author} {\bibfnamefont {F.}~\bibnamefont {Wicek}},\ }\bibfield  {title} {\bibinfo {title} {{Design Principles and Operational Results of the Cryogenic System for the ATLAS Liquid Argon Calorimeter}},\ }in\ \href@noop {} {\emph {\bibinfo {booktitle} {Proceedings of the 22nd International Cryogenic Engineering Conference and International Cryogenic Materials Conference 2008 (ICEC 22--ICMC 2008)}}}\ (\bibinfo  {publisher} {Korea Institute of Applied Superconductivity and Cryogenics},\ \bibinfo {address} {Seoul, Republic of Korea},\ \bibinfo
  {year} {2009})\ pp.\ \bibinfo {pages} {787--792}\BibitemShut {NoStop}%
\bibitem [{\citenamefont {Cur{\'e}}\ \emph {et~al.}(2024)\citenamefont {Cur{\'e}}, \citenamefont {Le~Godec}, \citenamefont {Ostrega},\ and\ \citenamefont {Wagner}}]{Cure:2024CMSMagnet}%
  \BibitemOpen
  \bibfield  {author} {\bibinfo {author} {\bibfnamefont {B.}~\bibnamefont {Cur{\'e}}}, \bibinfo {author} {\bibfnamefont {G.}~\bibnamefont {Le~Godec}}, \bibinfo {author} {\bibfnamefont {M.}~\bibnamefont {Ostrega}},\ and\ \bibinfo {author} {\bibfnamefont {U.}~\bibnamefont {Wagner}},\ }\bibfield  {title} {\bibinfo {title} {Fifteen years of operation of the compact muon solenoid detector superconducting magnet},\ }\href {https://doi.org/10.1109/TASC.2023.3336838} {\bibfield  {journal} {\bibinfo  {journal} {IEEE Trans. Appl. Supercond.}\ }\textbf {\bibinfo {volume} {34}},\ \bibinfo {pages} {1} (\bibinfo {year} {2024})}\BibitemShut {NoStop}%
\bibitem [{\citenamefont {{ATLAS Collaboration}}(2020)}]{ATLAS:2020DataQuality}%
  \BibitemOpen
  \bibfield  {author} {\bibinfo {author} {\bibnamefont {{ATLAS Collaboration}}},\ }\bibfield  {title} {\bibinfo {title} {{ATLAS} data quality operations and performance for 2015--2018 data-taking},\ }\href {https://doi.org/10.1088/1748-0221/15/04/P04003} {\bibfield  {journal} {\bibinfo  {journal} {JINST}\ }\textbf {\bibinfo {volume} {15}}\bibfield  {number} {\bibinfo  {number} { (04)},\ \bibinfo {pages} {P04003}},\ }\Eprint {https://arxiv.org/abs/1911.04632} {arXiv:1911.04632 [physics.ins-det]} \BibitemShut {NoStop}%
\bibitem [{\citenamefont {Buchanan}\ \emph {et~al.}(2008)\citenamefont {Buchanan}, \citenamefont {Chen}, \citenamefont {Gingrich}, \citenamefont {Liu}, \citenamefont {Chen}, \citenamefont {Damazio}, \citenamefont {Densing}, \citenamefont {Duffin}, \citenamefont {Farrell}, \citenamefont {Kandasamy}, \citenamefont {Kierstead}, \citenamefont {Lanni}, \citenamefont {Lissauer}, \citenamefont {Ma}, \citenamefont {Makowiecki}, \citenamefont {Muller}, \citenamefont {Radeka}, \citenamefont {Rescia}, \citenamefont {Ruggiero}, \citenamefont {Takai}, \citenamefont {Wolniewicz}, \citenamefont {Ghazlane}, \citenamefont {Hoummada}, \citenamefont {Hervas}, \citenamefont {Hott}, \citenamefont {Wilkens}, \citenamefont {Ban}, \citenamefont {Boettcher}, \citenamefont {Brooijmans}, \citenamefont {Chi}, \citenamefont {Caughron}, \citenamefont {Cooke}, \citenamefont {Copic}, \citenamefont {Dannheim}, \citenamefont {Gara}, \citenamefont {Haas}, \citenamefont {Katsanos}, \citenamefont {Parsons}, \citenamefont {Simion}, \citenamefont
  {Sippach}, \citenamefont {Zhang}, \citenamefont {Zhou}, \citenamefont {Eckstein}, \citenamefont {Kobel}, \citenamefont {Ladygin}, \citenamefont {Auge}, \citenamefont {Bernier}, \citenamefont {Bouchel}, \citenamefont {Bozzone}, \citenamefont {Breton}, \citenamefont {de~La~Taille}, \citenamefont {Falleau}, \citenamefont {Fournier}, \citenamefont {Imbert}, \citenamefont {Martin-Chassard}, \citenamefont {Perus}, \citenamefont {Richer}, \citenamefont {Seguin~Moreau}, \citenamefont {Serin}, \citenamefont {Tocut}, \citenamefont {Veillet}, \citenamefont {Zerwas}, \citenamefont {Colas}, \citenamefont {Dumont-Dayot}, \citenamefont {Massol}, \citenamefont {Perrodo}, \citenamefont {Perrot}, \citenamefont {Wingerter-Seez}, \citenamefont {Escalier}, \citenamefont {Hubaut}, \citenamefont {Laforge}, \citenamefont {Le~Dortz}, \citenamefont {Schwemling}, \citenamefont {Collot}, \citenamefont {Dzahini}, \citenamefont {Gallin-Martel}, \citenamefont {Martin}, \citenamefont {Cwienk}, \citenamefont {Fent}, \citenamefont
  {Kurchaninov}, \citenamefont {Citterio}, \citenamefont {Mazzanti}, \citenamefont {Tartarelli}, \citenamefont {Bansal}, \citenamefont {Boulahouache}, \citenamefont {Cleland}, \citenamefont {Liu}, \citenamefont {McDonald}, \citenamefont {Paolone}, \citenamefont {Rabel}, \citenamefont {Savinov}, \citenamefont {Zuk}, \citenamefont {Benslama}, \citenamefont {Borgeaud}, \citenamefont {de~La~Bro{\"i}se}, \citenamefont {Delagnes}, \citenamefont {Le~Coguie}, \citenamefont {Mansouli{\'e}},\ and\ \citenamefont {Pascual}}]{ATLAS:LArElectronics}%
  \BibitemOpen
  \bibfield  {author} {\bibinfo {author} {\bibfnamefont {N.~J.}\ \bibnamefont {Buchanan}}, \bibinfo {author} {\bibfnamefont {L.}~\bibnamefont {Chen}}, \bibinfo {author} {\bibfnamefont {D.~M.}\ \bibnamefont {Gingrich}}, \bibinfo {author} {\bibfnamefont {S.}~\bibnamefont {Liu}}, \bibinfo {author} {\bibfnamefont {H.}~\bibnamefont {Chen}}, \bibinfo {author} {\bibfnamefont {D.}~\bibnamefont {Damazio}}, \bibinfo {author} {\bibfnamefont {F.}~\bibnamefont {Densing}}, \bibinfo {author} {\bibfnamefont {S.}~\bibnamefont {Duffin}}, \bibinfo {author} {\bibfnamefont {J.}~\bibnamefont {Farrell}}, \bibinfo {author} {\bibfnamefont {S.}~\bibnamefont {Kandasamy}}, \bibinfo {author} {\bibfnamefont {J.}~\bibnamefont {Kierstead}}, \bibinfo {author} {\bibfnamefont {F.}~\bibnamefont {Lanni}}, \bibinfo {author} {\bibfnamefont {D.}~\bibnamefont {Lissauer}}, \bibinfo {author} {\bibfnamefont {H.}~\bibnamefont {Ma}}, \bibinfo {author} {\bibfnamefont {D.}~\bibnamefont {Makowiecki}}, \bibinfo {author} {\bibfnamefont {T.}~\bibnamefont
  {Muller}}, \bibinfo {author} {\bibfnamefont {V.}~\bibnamefont {Radeka}}, \bibinfo {author} {\bibfnamefont {S.}~\bibnamefont {Rescia}}, \bibinfo {author} {\bibfnamefont {R.}~\bibnamefont {Ruggiero}}, \bibinfo {author} {\bibfnamefont {H.}~\bibnamefont {Takai}}, \bibinfo {author} {\bibfnamefont {K.}~\bibnamefont {Wolniewicz}}, \bibinfo {author} {\bibfnamefont {H.}~\bibnamefont {Ghazlane}}, \bibinfo {author} {\bibfnamefont {A.}~\bibnamefont {Hoummada}}, \bibinfo {author} {\bibfnamefont {L.}~\bibnamefont {Hervas}}, \bibinfo {author} {\bibfnamefont {T.}~\bibnamefont {Hott}}, \bibinfo {author} {\bibfnamefont {H.~G.}\ \bibnamefont {Wilkens}}, \bibinfo {author} {\bibfnamefont {J.}~\bibnamefont {Ban}}, \bibinfo {author} {\bibfnamefont {S.}~\bibnamefont {Boettcher}}, \bibinfo {author} {\bibfnamefont {G.}~\bibnamefont {Brooijmans}}, \bibinfo {author} {\bibfnamefont {C.~Y.}\ \bibnamefont {Chi}}, \bibinfo {author} {\bibfnamefont {S.}~\bibnamefont {Caughron}}, \bibinfo {author} {\bibfnamefont {M.}~\bibnamefont {Cooke}},
  \bibinfo {author} {\bibfnamefont {K.}~\bibnamefont {Copic}}, \bibinfo {author} {\bibfnamefont {D.}~\bibnamefont {Dannheim}}, \bibinfo {author} {\bibfnamefont {A.}~\bibnamefont {Gara}}, \bibinfo {author} {\bibfnamefont {A.}~\bibnamefont {Haas}}, \bibinfo {author} {\bibfnamefont {I.}~\bibnamefont {Katsanos}}, \bibinfo {author} {\bibfnamefont {J.~A.}\ \bibnamefont {Parsons}}, \bibinfo {author} {\bibfnamefont {S.}~\bibnamefont {Simion}}, \bibinfo {author} {\bibfnamefont {W.}~\bibnamefont {Sippach}}, \bibinfo {author} {\bibfnamefont {L.}~\bibnamefont {Zhang}}, \bibinfo {author} {\bibfnamefont {N.}~\bibnamefont {Zhou}}, \bibinfo {author} {\bibfnamefont {P.}~\bibnamefont {Eckstein}}, \bibinfo {author} {\bibfnamefont {M.}~\bibnamefont {Kobel}}, \bibinfo {author} {\bibfnamefont {E.}~\bibnamefont {Ladygin}}, \bibinfo {author} {\bibfnamefont {E.}~\bibnamefont {Auge}}, \bibinfo {author} {\bibfnamefont {R.}~\bibnamefont {Bernier}}, \bibinfo {author} {\bibfnamefont {M.}~\bibnamefont {Bouchel}}, \bibinfo {author}
  {\bibfnamefont {A.}~\bibnamefont {Bozzone}}, \bibinfo {author} {\bibfnamefont {D.}~\bibnamefont {Breton}}, \bibinfo {author} {\bibfnamefont {C.}~\bibnamefont {de~La~Taille}}, \bibinfo {author} {\bibfnamefont {I.}~\bibnamefont {Falleau}}, \bibinfo {author} {\bibfnamefont {D.}~\bibnamefont {Fournier}}, \bibinfo {author} {\bibfnamefont {P.}~\bibnamefont {Imbert}}, \bibinfo {author} {\bibfnamefont {G.}~\bibnamefont {Martin-Chassard}}, \bibinfo {author} {\bibfnamefont {A.}~\bibnamefont {Perus}}, \bibinfo {author} {\bibfnamefont {J.~P.}\ \bibnamefont {Richer}}, \bibinfo {author} {\bibfnamefont {N.}~\bibnamefont {Seguin~Moreau}}, \bibinfo {author} {\bibfnamefont {L.}~\bibnamefont {Serin}}, \bibinfo {author} {\bibfnamefont {V.}~\bibnamefont {Tocut}}, \bibinfo {author} {\bibfnamefont {J.~J.}\ \bibnamefont {Veillet}}, \bibinfo {author} {\bibfnamefont {D.}~\bibnamefont {Zerwas}}, \bibinfo {author} {\bibfnamefont {J.}~\bibnamefont {Colas}}, \bibinfo {author} {\bibfnamefont {N.}~\bibnamefont {Dumont-Dayot}}, \bibinfo
  {author} {\bibfnamefont {N.}~\bibnamefont {Massol}}, \bibinfo {author} {\bibfnamefont {P.}~\bibnamefont {Perrodo}}, \bibinfo {author} {\bibfnamefont {G.}~\bibnamefont {Perrot}}, \bibinfo {author} {\bibfnamefont {I.}~\bibnamefont {Wingerter-Seez}}, \bibinfo {author} {\bibfnamefont {M.}~\bibnamefont {Escalier}}, \bibinfo {author} {\bibfnamefont {F.}~\bibnamefont {Hubaut}}, \bibinfo {author} {\bibfnamefont {B.}~\bibnamefont {Laforge}}, \bibinfo {author} {\bibfnamefont {O.}~\bibnamefont {Le~Dortz}}, \bibinfo {author} {\bibfnamefont {P.}~\bibnamefont {Schwemling}}, \bibinfo {author} {\bibfnamefont {J.}~\bibnamefont {Collot}}, \bibinfo {author} {\bibfnamefont {D.}~\bibnamefont {Dzahini}}, \bibinfo {author} {\bibfnamefont {M.~L.}\ \bibnamefont {Gallin-Martel}}, \bibinfo {author} {\bibfnamefont {P.}~\bibnamefont {Martin}}, \bibinfo {author} {\bibfnamefont {W.~D.}\ \bibnamefont {Cwienk}}, \bibinfo {author} {\bibfnamefont {J.}~\bibnamefont {Fent}}, \bibinfo {author} {\bibfnamefont {L.}~\bibnamefont {Kurchaninov}},
  \bibinfo {author} {\bibfnamefont {M.}~\bibnamefont {Citterio}}, \bibinfo {author} {\bibfnamefont {M.}~\bibnamefont {Mazzanti}}, \bibinfo {author} {\bibfnamefont {F.}~\bibnamefont {Tartarelli}}, \bibinfo {author} {\bibfnamefont {V.}~\bibnamefont {Bansal}}, \bibinfo {author} {\bibfnamefont {C.}~\bibnamefont {Boulahouache}}, \bibinfo {author} {\bibfnamefont {W.}~\bibnamefont {Cleland}}, \bibinfo {author} {\bibfnamefont {B.}~\bibnamefont {Liu}}, \bibinfo {author} {\bibfnamefont {J.}~\bibnamefont {McDonald}}, \bibinfo {author} {\bibfnamefont {V.}~\bibnamefont {Paolone}}, \bibinfo {author} {\bibfnamefont {J.}~\bibnamefont {Rabel}}, \bibinfo {author} {\bibfnamefont {V.}~\bibnamefont {Savinov}}, \bibinfo {author} {\bibfnamefont {G.}~\bibnamefont {Zuk}}, \bibinfo {author} {\bibfnamefont {K.}~\bibnamefont {Benslama}}, \bibinfo {author} {\bibfnamefont {P.}~\bibnamefont {Borgeaud}}, \bibinfo {author} {\bibfnamefont {X.}~\bibnamefont {de~La~Bro{\"i}se}}, \bibinfo {author} {\bibfnamefont {E.}~\bibnamefont {Delagnes}},
  \bibinfo {author} {\bibfnamefont {A.}~\bibnamefont {Le~Coguie}}, \bibinfo {author} {\bibfnamefont {B.}~\bibnamefont {Mansouli{\'e}}},\ and\ \bibinfo {author} {\bibfnamefont {J.}~\bibnamefont {Pascual}},\ }\bibfield  {title} {\bibinfo {title} {{ATLAS} liquid argon calorimeter front end electronics},\ }\href {https://doi.org/10.1088/1748-0221/3/09/P09003} {\bibfield  {journal} {\bibinfo  {journal} {JINST}\ }\textbf {\bibinfo {volume} {3}}\bibinfo  {number} { (09)},\ \bibinfo {pages} {P09003}}\BibitemShut {NoStop}%
\bibitem [{\citenamefont {{Tracker Group of the CMS Collaboration}}(2021{\natexlab{a}})}]{CMS:Phase1Pixel}%
  \BibitemOpen
\bibfield  {number} {  }\bibfield  {author} {\bibinfo {author} {\bibnamefont {{Tracker Group of the CMS Collaboration}}},\ }\bibfield  {title} {\bibinfo {title} {The {CMS} phase-1 pixel detector upgrade},\ }\href {https://doi.org/10.1088/1748-0221/16/02/P02027} {\bibfield  {journal} {\bibinfo  {journal} {JINST}\ }\textbf {\bibinfo {volume} {16}}\bibfield  {number} {\bibinfo  {number} { (02)},\ \bibinfo {pages} {P02027}},\ }\Eprint {https://arxiv.org/abs/2012.14304} {arXiv:2012.14304 [physics.ins-det]} \BibitemShut {NoStop}%
\bibitem [{\citenamefont {Kello}(2025)}]{CMS:TrackerRun3}%
  \BibitemOpen
  \bibfield  {author} {\bibinfo {author} {\bibfnamefont {T.}~\bibnamefont {Kello}} (\bibinfo {collaboration} {CMS Collaboration}),\ }\bibfield  {title} {\bibinfo {title} {The {CMS} tracker performance in {LHC} run 3},\ }\href {https://doi.org/10.22323/1.468.0013} {\bibfield  {journal} {\bibinfo  {journal} {PoS}\ }\textbf {\bibinfo {volume} {TIPP2023}},\ \bibinfo {pages} {013} (\bibinfo {year} {2025})}\BibitemShut {NoStop}%
\bibitem [{\citenamefont {{CMS Collaboration}}(2017{\natexlab{a}})}]{CMS:DAQPhase2TDR}%
  \BibitemOpen
  \bibfield  {author} {\bibinfo {author} {\bibnamefont {{CMS Collaboration}}},\ }\href {https://cds.cern.ch/record/2283193} {\emph {\bibinfo {title} {The Phase-2 Upgrade of the {CMS} {DAQ}: Interim Technical Design Report}}},\ \bibinfo {type} {Tech. Rep.}\ \bibinfo {number} {CERN-LHCC-2017-014, CMS-TDR-018}\ (\bibinfo  {institution} {CERN},\ \bibinfo {address} {Geneva},\ \bibinfo {year} {2017})\BibitemShut {NoStop}%
\bibitem [{\citenamefont {{ATLAS Collaboration}}(2024)}]{ATLAS:2024Run3}%
  \BibitemOpen
  \bibfield  {author} {\bibinfo {author} {\bibnamefont {{ATLAS Collaboration}}},\ }\bibfield  {title} {\bibinfo {title} {The {ATLAS} experiment at the {CERN} large hadron collider: a description of the detector configuration for run 3},\ }\href {https://doi.org/10.1088/1748-0221/19/05/P05063} {\bibfield  {journal} {\bibinfo  {journal} {JINST}\ }\textbf {\bibinfo {volume} {19}}\bibfield  {number} {\bibinfo  {number} { (05)},\ \bibinfo {pages} {P05063}},\ }\Eprint {https://arxiv.org/abs/2305.16623} {arXiv:2305.16623 [physics.ins-det]} \BibitemShut {NoStop}%
\bibitem [{\citenamefont {{Tracker Group of the CMS Collaboration}}(2021{\natexlab{b}})}]{CMS:2020PixelUpgrade}%
  \BibitemOpen
  \bibfield  {author} {\bibinfo {author} {\bibnamefont {{Tracker Group of the CMS Collaboration}}},\ }\bibfield  {title} {\bibinfo {title} {The {CMS} phase-1 pixel detector upgrade},\ }\href {https://doi.org/10.1088/1748-0221/16/02/P02027} {\bibfield  {journal} {\bibinfo  {journal} {JINST}\ }\textbf {\bibinfo {volume} {16}}\bibfield  {number} {\bibinfo  {number} { (02)},\ \bibinfo {pages} {P02027}},\ }\Eprint {https://arxiv.org/abs/2012.14304} {arXiv:2012.14304 [physics.ins-det]} \BibitemShut {NoStop}%
\bibitem [{\citenamefont {Butz}(2015)}]{CMS:DryGasPlant}%
  \BibitemOpen
  \bibfield  {author} {\bibinfo {author} {\bibfnamefont {E.}~\bibnamefont {Butz}},\ }\bibfield  {title} {\bibinfo {title} {{CMS} strip detector: Operational experience and run1 to run2 transition},\ }\href {https://doi.org/10.22323/1.227.0004} {\bibfield  {journal} {\bibinfo  {journal} {PoS}\ }\textbf {\bibinfo {volume} {Vertex2014}},\ \bibinfo {pages} {004} (\bibinfo {year} {2015})},\ \bibinfo {note} {on behalf of the CMS Collaboration}\BibitemShut {NoStop}%
\bibitem [{\citenamefont {{CMS Collaboration}}(2017{\natexlab{b}})}]{CMS:TrackerThermal}%
  \BibitemOpen
  \bibfield  {author} {\bibinfo {author} {\bibnamefont {{CMS Collaboration}}},\ }\bibfield  {title} {\bibinfo {title} {Mechanical stability of the {CMS} strip tracker measured with a laser alignment system},\ }\href {https://doi.org/10.1088/1748-0221/12/04/P04023} {\bibfield  {journal} {\bibinfo  {journal} {JINST}\ }\textbf {\bibinfo {volume} {12}}\bibfield  {number} {\bibinfo  {number} { (04)},\ \bibinfo {pages} {P04023}},\ }\Eprint {https://arxiv.org/abs/1701.02022} {arXiv:1701.02022 [physics.ins-det]} \BibitemShut {NoStop}%
\bibitem [{\citenamefont {Ramos}\ \emph {et~al.}(2026)\citenamefont {Ramos}, \citenamefont {Tytgat}, \citenamefont {Mota~Amarilo}, \citenamefont {Samalan}, \citenamefont {Skovpen}, \citenamefont {Alves}, \citenamefont {Alves~Coelho}, \citenamefont {Marujo~da Silva}, \citenamefont {Barroso Ferreira~Filho}, \citenamefont {Da~Costa}, \citenamefont {De~Jesus~Damiao}, \citenamefont {Ferreira}, \citenamefont {Fonseca De~Souza}, \citenamefont {Mundim}, \citenamefont {Nogima}, \citenamefont {Pinheiro}, \citenamefont {Santoro}, \citenamefont {Thiel}, \citenamefont {Gomes De~Souza}, \citenamefont {Aleksandrov}, \citenamefont {Hadjiiska}, \citenamefont {Iaydjiev}, \citenamefont {Shopova}, \citenamefont {Sultanov}, \citenamefont {Dimitrov}, \citenamefont {Litov}, \citenamefont {Pavlov}, \citenamefont {Petkov}, \citenamefont {Petrov}, \citenamefont {Shumka}, \citenamefont {Cao}, \citenamefont {Diao}, \citenamefont {Gong}, \citenamefont {Hou}, \citenamefont {Kou}, \citenamefont {Liu}, \citenamefont {Song}, \citenamefont
  {Wang}, \citenamefont {Zhao}, \citenamefont {Qian}, \citenamefont {Avila}, \citenamefont {Barbosa~Trujillo}, \citenamefont {Cabrera}, \citenamefont {Florez}, \citenamefont {Reyes~Vega}, \citenamefont {Aly}, \citenamefont {Radi}, \citenamefont {Assran}, \citenamefont {Crotty}, \citenamefont {Mahmoud}, \citenamefont {Balleyguier}, \citenamefont {Chen}, \citenamefont {Combaret}, \citenamefont {Galbit}, \citenamefont {Gouzevitch}, \citenamefont {Grenier}, \citenamefont {Laktineh}, \citenamefont {Luciol}, \citenamefont {Mirabito}, \citenamefont {Tromeur}, \citenamefont {Bagaturia}, \citenamefont {Kemularia}, \citenamefont {Lomidze}, \citenamefont {Tsamalaidze}, \citenamefont {Amoozegar}, \citenamefont {Boghrati}, \citenamefont {Ebrahimi}, \citenamefont {Esfandi}, \citenamefont {Hosseini}, \citenamefont {Mohammadi~Najafabadi}, \citenamefont {Zareian}, \citenamefont {Abbrescia}, \citenamefont {De~Filippis}, \citenamefont {Iaselli}, \citenamefont {Loddo}, \citenamefont {Pugliese}, \citenamefont {Benussi},
  \citenamefont {Bianco}, \citenamefont {Meola}, \citenamefont {Piccolo}, \citenamefont {Buontempo}, \citenamefont {Carnevali}, \citenamefont {Lista}, \citenamefont {Paolucci}, \citenamefont {Fienga}, \citenamefont {Braghieri}, \citenamefont {Montagna}, \citenamefont {Riccardi}, \citenamefont {Salvini}, \citenamefont {Vitulo}, \citenamefont {Kim}, \citenamefont {Asilar}, \citenamefont {Ryou}, \citenamefont {Choi}, \citenamefont {Hong}, \citenamefont {Lee}, \citenamefont {Goh}, \citenamefont {Shin}, \citenamefont {Lee}, \citenamefont {Pedraza}, \citenamefont {Uribe~Estrada}, \citenamefont {Castilla-Valdez}, \citenamefont {Lopez-Fernandez}, \citenamefont {Sanchez~Hernandez}, \citenamefont {Ramirez~Garcia}, \citenamefont {Ramirez~Guadarrama}, \citenamefont {Shah}, \citenamefont {Vazquez}, \citenamefont {Zaganidis}, \citenamefont {Ahmad}, \citenamefont {Asghar}, \citenamefont {Hoorani}, \citenamefont {Muhammad},\ and\ \citenamefont {Eysermans}}]{Ramos:2026CMSRPC}%
  \BibitemOpen
  \bibfield  {author} {\bibinfo {author} {\bibfnamefont {D.}~\bibnamefont {Ramos}}, \bibinfo {author} {\bibfnamefont {M.}~\bibnamefont {Tytgat}}, \bibinfo {author} {\bibfnamefont {K.}~\bibnamefont {Mota~Amarilo}}, \bibinfo {author} {\bibfnamefont {A.}~\bibnamefont {Samalan}}, \bibinfo {author} {\bibfnamefont {K.}~\bibnamefont {Skovpen}}, \bibinfo {author} {\bibfnamefont {G.~A.}\ \bibnamefont {Alves}}, \bibinfo {author} {\bibfnamefont {E.}~\bibnamefont {Alves~Coelho}}, \bibinfo {author} {\bibfnamefont {F.}~\bibnamefont {Marujo~da Silva}}, \bibinfo {author} {\bibfnamefont {M.}~\bibnamefont {Barroso Ferreira~Filho}}, \bibinfo {author} {\bibfnamefont {E.~M.}\ \bibnamefont {Da~Costa}}, \bibinfo {author} {\bibfnamefont {D.}~\bibnamefont {De~Jesus~Damiao}}, \bibinfo {author} {\bibfnamefont {B.~C.}\ \bibnamefont {Ferreira}}, \bibinfo {author} {\bibfnamefont {S.}~\bibnamefont {Fonseca De~Souza}}, \bibinfo {author} {\bibfnamefont {L.}~\bibnamefont {Mundim}}, \bibinfo {author} {\bibfnamefont {H.}~\bibnamefont {Nogima}},
  \bibinfo {author} {\bibfnamefont {J.~P.}\ \bibnamefont {Pinheiro}}, \bibinfo {author} {\bibfnamefont {A.}~\bibnamefont {Santoro}}, \bibinfo {author} {\bibfnamefont {M.}~\bibnamefont {Thiel}}, \bibinfo {author} {\bibfnamefont {R.}~\bibnamefont {Gomes De~Souza}}, \bibinfo {author} {\bibfnamefont {A.}~\bibnamefont {Aleksandrov}}, \bibinfo {author} {\bibfnamefont {R.}~\bibnamefont {Hadjiiska}}, \bibinfo {author} {\bibfnamefont {P.}~\bibnamefont {Iaydjiev}}, \bibinfo {author} {\bibfnamefont {M.}~\bibnamefont {Shopova}}, \bibinfo {author} {\bibfnamefont {G.}~\bibnamefont {Sultanov}}, \bibinfo {author} {\bibfnamefont {A.}~\bibnamefont {Dimitrov}}, \bibinfo {author} {\bibfnamefont {L.}~\bibnamefont {Litov}}, \bibinfo {author} {\bibfnamefont {B.}~\bibnamefont {Pavlov}}, \bibinfo {author} {\bibfnamefont {P.}~\bibnamefont {Petkov}}, \bibinfo {author} {\bibfnamefont {A.}~\bibnamefont {Petrov}}, \bibinfo {author} {\bibfnamefont {E.}~\bibnamefont {Shumka}}, \bibinfo {author} {\bibfnamefont {P.}~\bibnamefont {Cao}},
  \bibinfo {author} {\bibfnamefont {W.}~\bibnamefont {Diao}}, \bibinfo {author} {\bibfnamefont {W.}~\bibnamefont {Gong}}, \bibinfo {author} {\bibfnamefont {Q.}~\bibnamefont {Hou}}, \bibinfo {author} {\bibfnamefont {H.}~\bibnamefont {Kou}}, \bibinfo {author} {\bibfnamefont {Z.-A.}\ \bibnamefont {Liu}}, \bibinfo {author} {\bibfnamefont {J.}~\bibnamefont {Song}}, \bibinfo {author} {\bibfnamefont {N.}~\bibnamefont {Wang}}, \bibinfo {author} {\bibfnamefont {J.}~\bibnamefont {Zhao}}, \bibinfo {author} {\bibfnamefont {S.~J.}\ \bibnamefont {Qian}}, \bibinfo {author} {\bibfnamefont {C.}~\bibnamefont {Avila}}, \bibinfo {author} {\bibfnamefont {D.~A.}\ \bibnamefont {Barbosa~Trujillo}}, \bibinfo {author} {\bibfnamefont {A.}~\bibnamefont {Cabrera}}, \bibinfo {author} {\bibfnamefont {C.~A.}\ \bibnamefont {Florez}}, \bibinfo {author} {\bibfnamefont {J.~A.}\ \bibnamefont {Reyes~Vega}}, \bibinfo {author} {\bibfnamefont {R.}~\bibnamefont {Aly}}, \bibinfo {author} {\bibfnamefont {A.}~\bibnamefont {Radi}}, \bibinfo {author}
  {\bibfnamefont {Y.}~\bibnamefont {Assran}}, \bibinfo {author} {\bibfnamefont {I.}~\bibnamefont {Crotty}}, \bibinfo {author} {\bibfnamefont {M.~A.}\ \bibnamefont {Mahmoud}}, \bibinfo {author} {\bibfnamefont {L.}~\bibnamefont {Balleyguier}}, \bibinfo {author} {\bibfnamefont {X.}~\bibnamefont {Chen}}, \bibinfo {author} {\bibfnamefont {C.}~\bibnamefont {Combaret}}, \bibinfo {author} {\bibfnamefont {G.}~\bibnamefont {Galbit}}, \bibinfo {author} {\bibfnamefont {M.}~\bibnamefont {Gouzevitch}}, \bibinfo {author} {\bibfnamefont {G.}~\bibnamefont {Grenier}}, \bibinfo {author} {\bibfnamefont {I.~B.}\ \bibnamefont {Laktineh}}, \bibinfo {author} {\bibfnamefont {A.}~\bibnamefont {Luciol}}, \bibinfo {author} {\bibfnamefont {L.}~\bibnamefont {Mirabito}}, \bibinfo {author} {\bibfnamefont {W.}~\bibnamefont {Tromeur}}, \bibinfo {author} {\bibfnamefont {I.}~\bibnamefont {Bagaturia}}, \bibinfo {author} {\bibfnamefont {O.}~\bibnamefont {Kemularia}}, \bibinfo {author} {\bibfnamefont {I.}~\bibnamefont {Lomidze}}, \bibinfo {author}
  {\bibfnamefont {Z.}~\bibnamefont {Tsamalaidze}}, \bibinfo {author} {\bibfnamefont {V.}~\bibnamefont {Amoozegar}}, \bibinfo {author} {\bibfnamefont {B.}~\bibnamefont {Boghrati}}, \bibinfo {author} {\bibfnamefont {M.}~\bibnamefont {Ebrahimi}}, \bibinfo {author} {\bibfnamefont {F.}~\bibnamefont {Esfandi}}, \bibinfo {author} {\bibfnamefont {Y.}~\bibnamefont {Hosseini}}, \bibinfo {author} {\bibfnamefont {M.}~\bibnamefont {Mohammadi~Najafabadi}}, \bibinfo {author} {\bibfnamefont {E.}~\bibnamefont {Zareian}}, \bibinfo {author} {\bibfnamefont {M.}~\bibnamefont {Abbrescia}}, \bibinfo {author} {\bibfnamefont {N.}~\bibnamefont {De~Filippis}}, \bibinfo {author} {\bibfnamefont {G.}~\bibnamefont {Iaselli}}, \bibinfo {author} {\bibfnamefont {F.}~\bibnamefont {Loddo}}, \bibinfo {author} {\bibfnamefont {G.}~\bibnamefont {Pugliese}}, \bibinfo {author} {\bibfnamefont {L.}~\bibnamefont {Benussi}}, \bibinfo {author} {\bibfnamefont {S.}~\bibnamefont {Bianco}}, \bibinfo {author} {\bibfnamefont {S.}~\bibnamefont {Meola}}, \bibinfo
  {author} {\bibfnamefont {D.}~\bibnamefont {Piccolo}}, \bibinfo {author} {\bibfnamefont {S.}~\bibnamefont {Buontempo}}, \bibinfo {author} {\bibfnamefont {F.}~\bibnamefont {Carnevali}}, \bibinfo {author} {\bibfnamefont {L.}~\bibnamefont {Lista}}, \bibinfo {author} {\bibfnamefont {P.}~\bibnamefont {Paolucci}}, \bibinfo {author} {\bibfnamefont {F.}~\bibnamefont {Fienga}}, \bibinfo {author} {\bibfnamefont {A.}~\bibnamefont {Braghieri}}, \bibinfo {author} {\bibfnamefont {P.}~\bibnamefont {Montagna}}, \bibinfo {author} {\bibfnamefont {C.}~\bibnamefont {Riccardi}}, \bibinfo {author} {\bibfnamefont {P.}~\bibnamefont {Salvini}}, \bibinfo {author} {\bibfnamefont {P.}~\bibnamefont {Vitulo}}, \bibinfo {author} {\bibfnamefont {T.~J.}\ \bibnamefont {Kim}}, \bibinfo {author} {\bibfnamefont {E.}~\bibnamefont {Asilar}}, \bibinfo {author} {\bibfnamefont {Y.}~\bibnamefont {Ryou}}, \bibinfo {author} {\bibfnamefont {S.}~\bibnamefont {Choi}}, \bibinfo {author} {\bibfnamefont {B.}~\bibnamefont {Hong}}, \bibinfo {author}
  {\bibfnamefont {K.~S.}\ \bibnamefont {Lee}}, \bibinfo {author} {\bibfnamefont {J.}~\bibnamefont {Goh}}, \bibinfo {author} {\bibfnamefont {J.}~\bibnamefont {Shin}}, \bibinfo {author} {\bibfnamefont {Y.}~\bibnamefont {Lee}}, \bibinfo {author} {\bibfnamefont {I.}~\bibnamefont {Pedraza}}, \bibinfo {author} {\bibfnamefont {C.}~\bibnamefont {Uribe~Estrada}}, \bibinfo {author} {\bibfnamefont {H.}~\bibnamefont {Castilla-Valdez}}, \bibinfo {author} {\bibfnamefont {R.}~\bibnamefont {Lopez-Fernandez}}, \bibinfo {author} {\bibfnamefont {A.}~\bibnamefont {Sanchez~Hernandez}}, \bibinfo {author} {\bibfnamefont {M.}~\bibnamefont {Ramirez~Garcia}}, \bibinfo {author} {\bibfnamefont {D.~L.}\ \bibnamefont {Ramirez~Guadarrama}}, \bibinfo {author} {\bibfnamefont {M.~A.}\ \bibnamefont {Shah}}, \bibinfo {author} {\bibfnamefont {E.}~\bibnamefont {Vazquez}}, \bibinfo {author} {\bibfnamefont {N.}~\bibnamefont {Zaganidis}}, \bibinfo {author} {\bibfnamefont {A.}~\bibnamefont {Ahmad}}, \bibinfo {author} {\bibfnamefont {M.~I.}\
  \bibnamefont {Asghar}}, \bibinfo {author} {\bibfnamefont {H.~R.}\ \bibnamefont {Hoorani}}, \bibinfo {author} {\bibfnamefont {S.}~\bibnamefont {Muhammad}},\ and\ \bibinfo {author} {\bibfnamefont {J.}~\bibnamefont {Eysermans}},\ }\bibfield  {title} {\bibinfo {title} {New {RPC} gas mixtures for sustainable operation in the {CMS} experiment},\ }\href {https://doi.org/10.1016/j.nima.2025.170969} {\bibfield  {journal} {\bibinfo  {journal} {Nucl. Instrum. Meth. A}\ }\textbf {\bibinfo {volume} {1082}},\ \bibinfo {pages} {170969} (\bibinfo {year} {2026})},\ \bibinfo {note} {on behalf of the CMS Muon Group}\BibitemShut {NoStop}%
\bibitem [{\citenamefont {{CMS Collaboration}}(2024)}]{CMS:2024Run3}%
  \BibitemOpen
  \bibfield  {author} {\bibinfo {author} {\bibnamefont {{CMS Collaboration}}},\ }\bibfield  {title} {\bibinfo {title} {Development of the {CMS} detector for the {CERN} {LHC} run 3},\ }\href {https://doi.org/10.1088/1748-0221/19/05/P05064} {\bibfield  {journal} {\bibinfo  {journal} {JINST}\ }\textbf {\bibinfo {volume} {19}}\bibfield  {number} {\bibinfo  {number} { (05)},\ \bibinfo {pages} {P05064}},\ }\Eprint {https://arxiv.org/abs/2309.05466} {arXiv:2309.05466 [physics.ins-det]} \BibitemShut {NoStop}%
\bibitem [{\citenamefont {Rao}(2014)}]{CMS:FireFoam}%
  \BibitemOpen
  \bibfield  {author} {\bibinfo {author} {\bibfnamefont {A.}~\bibnamefont {Rao}},\ }\href {https://cds.cern.ch/record/1998534} {\bibinfo {title} {{CMS} to improve underground fire-fighting system}},\ \bibinfo {howpublished} {CERN / CMS Experiment} (\bibinfo {year} {2014}),\ \bibinfo {note} {published 24 February 2014}\BibitemShut {NoStop}%
\bibitem [{\citenamefont {{CERN}}(2006)}]{CMS:FireFoamTest}%
  \BibitemOpen
  \bibfield  {author} {\bibinfo {author} {\bibnamefont {{CERN}}},\ }\bibfield  {title} {\bibinfo {title} {{CMS} installations are put to the test},\ }\href {https://cds.cern.ch/record/955085} {\bibfield  {journal} {\bibinfo  {journal} {CERN Bulletin}\ }\textbf {\bibinfo {volume} {23/2006--24/2006}},\ \bibinfo {pages} {1} (\bibinfo {year} {2006})}\BibitemShut {NoStop}%
\bibitem [{\citenamefont {Forkel-Wirth}\ \emph {et~al.}(2013)\citenamefont {Forkel-Wirth}, \citenamefont {Roesler}, \citenamefont {Silari}, \citenamefont {Streit-Bianchi}, \citenamefont {Theis}, \citenamefont {Vincke},\ and\ \citenamefont {Vincke}}]{CERN:RadiationProtectionCode}%
  \BibitemOpen
  \bibfield  {author} {\bibinfo {author} {\bibfnamefont {D.}~\bibnamefont {Forkel-Wirth}}, \bibinfo {author} {\bibfnamefont {S.}~\bibnamefont {Roesler}}, \bibinfo {author} {\bibfnamefont {M.}~\bibnamefont {Silari}}, \bibinfo {author} {\bibfnamefont {M.}~\bibnamefont {Streit-Bianchi}}, \bibinfo {author} {\bibfnamefont {C.}~\bibnamefont {Theis}}, \bibinfo {author} {\bibfnamefont {H.}~\bibnamefont {Vincke}},\ and\ \bibinfo {author} {\bibfnamefont {H.}~\bibnamefont {Vincke}},\ }\bibfield  {title} {\bibinfo {title} {{Radiation Protection at CERN}},\ }in\ \href {https://doi.org/10.5170/CERN-2013-001.415} {\emph {\bibinfo {booktitle} {Proceedings of the CERN Accelerator School: Course on High Power Hadron Machines}}}\ (\bibinfo  {publisher} {CERN},\ \bibinfo {address} {Bilbao, Spain},\ \bibinfo {year} {2013})\ pp.\ \bibinfo {pages} {415--436},\ \bibinfo {note} {cERN Yellow Report CERN-2013-001; CERN Accelerator School held in Bilbao, Spain, 24 May--2 June 2011},\ \Eprint {https://arxiv.org/abs/1303.6519}
  {arXiv:1303.6519 [physics.acc-ph]} \BibitemShut {NoStop}%
\bibitem [{\citenamefont {Harrison}\ \emph {et~al.}(2021)\citenamefont {Harrison}, \citenamefont {Balonov}, \citenamefont {Bochud}, \citenamefont {Martin}, \citenamefont {Menzel}, \citenamefont {Ortiz-L{\'o}pez}, \citenamefont {Smith-Bindman}, \citenamefont {Simmonds},\ and\ \citenamefont {Wakeford}}]{ICRP:2021Publication147}%
  \BibitemOpen
  \bibfield  {author} {\bibinfo {author} {\bibfnamefont {J.~D.}\ \bibnamefont {Harrison}}, \bibinfo {author} {\bibfnamefont {M.}~\bibnamefont {Balonov}}, \bibinfo {author} {\bibfnamefont {F.}~\bibnamefont {Bochud}}, \bibinfo {author} {\bibfnamefont {C.}~\bibnamefont {Martin}}, \bibinfo {author} {\bibfnamefont {H.-G.}\ \bibnamefont {Menzel}}, \bibinfo {author} {\bibfnamefont {P.}~\bibnamefont {Ortiz-L{\'o}pez}}, \bibinfo {author} {\bibfnamefont {R.}~\bibnamefont {Smith-Bindman}}, \bibinfo {author} {\bibfnamefont {J.~R.}\ \bibnamefont {Simmonds}},\ and\ \bibinfo {author} {\bibfnamefont {R.}~\bibnamefont {Wakeford}},\ }\bibfield  {title} {\bibinfo {title} {Use of dose quantities in radiological protection},\ }\bibfield  {journal} {\bibinfo  {journal} {Annals of the ICRP}\ }\textbf {\bibinfo {volume} {50}},\ \href {https://doi.org/10.1177/0146645320911864} {10.1177/0146645320911864} (\bibinfo {year} {2021}),\ \bibinfo {note} {iCRP Publication 147}\BibitemShut {NoStop}%
\bibitem [{\citenamefont {Otto}(2024)}]{Otto:2024ICRU95}%
  \BibitemOpen
  \bibfield  {author} {\bibinfo {author} {\bibfnamefont {T.}~\bibnamefont {Otto}},\ }\bibfield  {title} {\bibinfo {title} {{ICRU} report 95: New operational quantities for external radiation exposure},\ }\href {https://doi.org/10.1177/01466453241283931k} {\bibfield  {journal} {\bibinfo  {journal} {Annals of the ICRP}\ }\textbf {\bibinfo {volume} {53}},\ \bibinfo {pages} {74} (\bibinfo {year} {2024})}\BibitemShut {NoStop}%
\bibitem [{\citenamefont {Fass{\`o}}\ and\ \citenamefont {Rokni}(2009)}]{FassoRokni:2009ALARA}%
  \BibitemOpen
  \bibfield  {author} {\bibinfo {author} {\bibfnamefont {A.}~\bibnamefont {Fass{\`o}}}\ and\ \bibinfo {author} {\bibfnamefont {S.}~\bibnamefont {Rokni}},\ }\bibfield  {title} {\bibinfo {title} {Operational radiation protection in high-energy physics accelerators: implementation of {ALARA} in design and operation of accelerators},\ }\href {https://doi.org/10.1093/rpd/ncp189} {\bibfield  {journal} {\bibinfo  {journal} {Radiation Protection Dosimetry}\ }\textbf {\bibinfo {volume} {137}},\ \bibinfo {pages} {94} (\bibinfo {year} {2009})}\BibitemShut {NoStop}%
\bibitem [{\citenamefont {Leveson}(2004)}]{Leveson:2004AccidentModel}%
  \BibitemOpen
  \bibfield  {author} {\bibinfo {author} {\bibfnamefont {N.}~\bibnamefont {Leveson}},\ }\bibfield  {title} {\bibinfo {title} {A new accident model for engineering safer systems},\ }\href {https://doi.org/10.1016/S0925-7535(03)00047-X} {\bibfield  {journal} {\bibinfo  {journal} {Safety Science}\ }\textbf {\bibinfo {volume} {42}},\ \bibinfo {pages} {237} (\bibinfo {year} {2004})}\BibitemShut {NoStop}%
\bibitem [{\citenamefont {Perrow}(1999)}]{Perrow:1999NormalAccidents}%
  \BibitemOpen
  \bibfield  {author} {\bibinfo {author} {\bibfnamefont {C.}~\bibnamefont {Perrow}},\ }\href@noop {} {\emph {\bibinfo {title} {Normal Accidents: Living with High-Risk Technologies}}},\ \bibinfo {edition} {updated}\ ed.\ (\bibinfo  {publisher} {Princeton University Press},\ \bibinfo {address} {Princeton},\ \bibinfo {year} {1999})\BibitemShut {NoStop}%
\bibitem [{\citenamefont {Leveson}(2012)}]{Leveson:2012SaferWorld}%
  \BibitemOpen
  \bibfield  {author} {\bibinfo {author} {\bibfnamefont {N.~G.}\ \bibnamefont {Leveson}},\ }\href {https://doi.org/10.7551/mitpress/8179.001.0001} {\emph {\bibinfo {title} {{Engineering a Safer World: Systems Thinking Applied to Safety}}}}\ (\bibinfo  {publisher} {MIT Press},\ \bibinfo {address} {Cambridge, Massachusetts},\ \bibinfo {year} {2012})\BibitemShut {NoStop}%
\bibitem [{\citenamefont {Rasmussen}(1997)}]{Rasmussen:1997RiskManagement}%
  \BibitemOpen
  \bibfield  {author} {\bibinfo {author} {\bibfnamefont {J.}~\bibnamefont {Rasmussen}},\ }\bibfield  {title} {\bibinfo {title} {Risk management in a dynamic society: A modelling problem},\ }\href {https://doi.org/10.1016/S0925-7535(97)00052-0} {\bibfield  {journal} {\bibinfo  {journal} {Safety Science}\ }\textbf {\bibinfo {volume} {27}},\ \bibinfo {pages} {183} (\bibinfo {year} {1997})}\BibitemShut {NoStop}%
\bibitem [{\citenamefont {Hollnagel}\ \emph {et~al.}(2006)\citenamefont {Hollnagel}, \citenamefont {Woods},\ and\ \citenamefont {Leveson}}]{Hollnagel:2006Resilience}%
  \BibitemOpen
  \bibinfo {editor} {\bibfnamefont {E.}~\bibnamefont {Hollnagel}}, \bibinfo {editor} {\bibfnamefont {D.~D.}\ \bibnamefont {Woods}},\ and\ \bibinfo {editor} {\bibfnamefont {N.}~\bibnamefont {Leveson}},\ eds.,\ \href@noop {} {\emph {\bibinfo {title} {Resilience Engineering: Concepts and Precepts}}}\ (\bibinfo  {publisher} {Ashgate},\ \bibinfo {address} {Aldershot},\ \bibinfo {year} {2006})\BibitemShut {NoStop}%
\bibitem [{\citenamefont {Beltramello}\ \emph {et~al.}(2009)\citenamefont {Beltramello}, \citenamefont {Burckhart}, \citenamefont {Franz}, \citenamefont {Jaekel}, \citenamefont {Jeckel}, \citenamefont {L{\"u}ders}, \citenamefont {Morpurgo}, \citenamefont {dos Santos~Pedrosa}, \citenamefont {Pommes},\ and\ \citenamefont {Sandaker}}]{Beltramello:2009DSS}%
  \BibitemOpen
  \bibfield  {author} {\bibinfo {author} {\bibfnamefont {O.}~\bibnamefont {Beltramello}}, \bibinfo {author} {\bibfnamefont {H.~J.}\ \bibnamefont {Burckhart}}, \bibinfo {author} {\bibfnamefont {S.}~\bibnamefont {Franz}}, \bibinfo {author} {\bibfnamefont {M.}~\bibnamefont {Jaekel}}, \bibinfo {author} {\bibfnamefont {M.}~\bibnamefont {Jeckel}}, \bibinfo {author} {\bibfnamefont {S.}~\bibnamefont {L{\"u}ders}}, \bibinfo {author} {\bibfnamefont {G.}~\bibnamefont {Morpurgo}}, \bibinfo {author} {\bibfnamefont {F.}~\bibnamefont {dos Santos~Pedrosa}}, \bibinfo {author} {\bibfnamefont {K.}~\bibnamefont {Pommes}},\ and\ \bibinfo {author} {\bibfnamefont {H.}~\bibnamefont {Sandaker}},\ }\bibfield  {title} {\bibinfo {title} {The detector safety system of the atlas experiment},\ }\href {https://doi.org/10.1088/1748-0221/4/09/P09012} {\bibfield  {journal} {\bibinfo  {journal} {Journal of Instrumentation}\ }\textbf {\bibinfo {volume} {4}},\ \bibinfo {pages} {P09012}}\BibitemShut {NoStop}%
\bibitem [{\citenamefont {Garino}\ \emph {et~al.}(2013)\citenamefont {Garino}, \citenamefont {Daudin}, \citenamefont {De~Jonghe}, \citenamefont {Dorsival}, \citenamefont {Dos Santos~Pedrosa}, \citenamefont {Dumont}, \citenamefont {Foraz},\ and\ \citenamefont {Reguero~Fuentes}}]{Garino:2013IMPACT}%
  \BibitemOpen
  \bibfield  {author} {\bibinfo {author} {\bibfnamefont {C.}~\bibnamefont {Garino}}, \bibinfo {author} {\bibfnamefont {B.}~\bibnamefont {Daudin}}, \bibinfo {author} {\bibfnamefont {J.}~\bibnamefont {De~Jonghe}}, \bibinfo {author} {\bibfnamefont {A.}~\bibnamefont {Dorsival}}, \bibinfo {author} {\bibfnamefont {F.~B.}\ \bibnamefont {Dos Santos~Pedrosa}}, \bibinfo {author} {\bibfnamefont {G.}~\bibnamefont {Dumont}}, \bibinfo {author} {\bibfnamefont {K.}~\bibnamefont {Foraz}},\ and\ \bibinfo {author} {\bibfnamefont {E.}~\bibnamefont {Reguero~Fuentes}},\ }\bibfield  {title} {\bibinfo {title} {{Intervention Management from Operation to Shutdown}},\ }in\ \href {https://proceedings.jacow.org/IPAC2013/papers/thpwa035.pdf} {\emph {\bibinfo {booktitle} {Proceedings of the 4th International Particle Accelerator Conference (IPAC'13)}}}\ (\bibinfo  {publisher} {JACoW},\ \bibinfo {address} {Shanghai, China},\ \bibinfo {year} {2013})\ pp.\ \bibinfo {pages} {3705--3707},\ \bibinfo {note} {paper THPWA035; CDS Record
  2010151}\BibitemShut {NoStop}%
\bibitem [{\citenamefont {{CERN}}(2024)}]{CERN:AnnualReport2023}%
  \BibitemOpen
  \bibfield  {author} {\bibinfo {author} {\bibnamefont {{CERN}}},\ }\href {https://doi.org/10.17181/AnnualReport2023} {\emph {\bibinfo {title} {{CERN Annual Report 2023}}}},\ \bibinfo {series} {Annual Report of the European Organization for Nuclear Research}, Vol.~\bibinfo {volume} {69}\ (\bibinfo  {publisher} {CERN},\ \bibinfo {address} {Geneva},\ \bibinfo {year} {2024})\BibitemShut {NoStop}%
\bibitem [{\citenamefont {{CERN}}(2025)}]{CERN:ProgressReport2024}%
  \BibitemOpen
  \bibfield  {author} {\bibinfo {author} {\bibnamefont {{CERN}}},\ }\href {https://cds.cern.ch/record/2941416} {\emph {\bibinfo {title} {{Annual Progress Report 2024}}}},\ \bibinfo {type} {Tech. Rep.}\ \bibinfo {number} {CERN/SPC/1252/Rev., CERN/FC/6883/Rev., CERN/3886/Rev.}\ (\bibinfo  {institution} {CERN},\ \bibinfo {address} {Geneva},\ \bibinfo {year} {2025})\ \bibinfo {note} {approved by the CERN Council, June 2025}\BibitemShut {NoStop}%
\bibitem [{\citenamefont {Tock}\ \emph {et~al.}(2025)\citenamefont {Tock}, \citenamefont {Herty}, \citenamefont {Borglund}, \citenamefont {Duchemin}, \citenamefont {Bedel}, \citenamefont {Dos Santos~Pedrosa}, \citenamefont {Riddone}, \citenamefont {Etheridge}, \citenamefont {Fernandez}, \citenamefont {Bernardini}, \citenamefont {Modena}, \citenamefont {Fessia}, \citenamefont {Steerenberg}, \citenamefont {Blanchard}, \citenamefont {Roesler}, \citenamefont {Vargas~Jorba},\ and\ \citenamefont {Marsh}}]{Tock:2025LS3}%
  \BibitemOpen
  \bibfield  {author} {\bibinfo {author} {\bibfnamefont {J.-P.}\ \bibnamefont {Tock}}, \bibinfo {author} {\bibfnamefont {A.}~\bibnamefont {Herty}}, \bibinfo {author} {\bibfnamefont {A.}~\bibnamefont {Borglund}}, \bibinfo {author} {\bibfnamefont {C.}~\bibnamefont {Duchemin}}, \bibinfo {author} {\bibfnamefont {C.}~\bibnamefont {Bedel}}, \bibinfo {author} {\bibfnamefont {F.}~\bibnamefont {Dos Santos~Pedrosa}}, \bibinfo {author} {\bibfnamefont {G.}~\bibnamefont {Riddone}}, \bibinfo {author} {\bibfnamefont {J.}~\bibnamefont {Etheridge}}, \bibinfo {author} {\bibfnamefont {J.}~\bibnamefont {Fernandez}}, \bibinfo {author} {\bibfnamefont {M.}~\bibnamefont {Bernardini}}, \bibinfo {author} {\bibfnamefont {M.}~\bibnamefont {Modena}}, \bibinfo {author} {\bibfnamefont {P.}~\bibnamefont {Fessia}}, \bibinfo {author} {\bibfnamefont {R.}~\bibnamefont {Steerenberg}}, \bibinfo {author} {\bibfnamefont {S.}~\bibnamefont {Blanchard}}, \bibinfo {author} {\bibfnamefont {S.}~\bibnamefont {Roesler}}, \bibinfo {author} {\bibfnamefont
  {M.}~\bibnamefont {Vargas~Jorba}},\ and\ \bibinfo {author} {\bibfnamefont {S.}~\bibnamefont {Marsh}},\ }\bibfield  {title} {\bibinfo {title} {{The Third Long Shutdown (LS3) of the CERN Accelerator Complex}},\ }in\ \href {https://doi.org/10.18429/JACoW-IPAC2025-TUBN1} {\emph {\bibinfo {booktitle} {Proceedings of the 16th International Particle Accelerator Conference (IPAC'25)}}}\ (\bibinfo  {publisher} {JACoW Publishing},\ \bibinfo {address} {Taipei, Taiwan},\ \bibinfo {year} {2025})\ pp.\ \bibinfo {pages} {891--894},\ \bibinfo {note} {paper TUBN1}\BibitemShut {NoStop}%
\bibitem [{\citenamefont {Evrard}\ \emph {et~al.}(2025)\citenamefont {Evrard}, \citenamefont {Contini}, \citenamefont {Dole}, \citenamefont {Fernandez}, \citenamefont {Godineau}, \citenamefont {Hakulinen}, \citenamefont {Havart}, \citenamefont {Krautsztrung}, \citenamefont {Ladzinski}, \citenamefont {Lang}, \citenamefont {Martel}, \citenamefont {Morand}, \citenamefont {Munoz~Codoceo}, \citenamefont {Ninin}, \citenamefont {Nunes}, \citenamefont {Olivet~Dessent}, \citenamefont {Raffourt}, \citenamefont {Rios}, \citenamefont {Sanchez-Corral~Mena},\ and\ \citenamefont {Suwalska}}]{Evrard:2025CERNPersonnelSafety}%
  \BibitemOpen
  \bibfield  {author} {\bibinfo {author} {\bibfnamefont {S.}~\bibnamefont {Evrard}}, \bibinfo {author} {\bibfnamefont {L.}~\bibnamefont {Contini}}, \bibinfo {author} {\bibfnamefont {M.}~\bibnamefont {Dole}}, \bibinfo {author} {\bibfnamefont {J.}~\bibnamefont {Fernandez}}, \bibinfo {author} {\bibfnamefont {G.}~\bibnamefont {Godineau}}, \bibinfo {author} {\bibfnamefont {T.}~\bibnamefont {Hakulinen}}, \bibinfo {author} {\bibfnamefont {F.}~\bibnamefont {Havart}}, \bibinfo {author} {\bibfnamefont {I.}~\bibnamefont {Krautsztrung}}, \bibinfo {author} {\bibfnamefont {T.}~\bibnamefont {Ladzinski}}, \bibinfo {author} {\bibfnamefont {T.}~\bibnamefont {Lang}}, \bibinfo {author} {\bibfnamefont {P.}~\bibnamefont {Martel}}, \bibinfo {author} {\bibfnamefont {B.}~\bibnamefont {Morand}}, \bibinfo {author} {\bibfnamefont {M.}~\bibnamefont {Munoz~Codoceo}}, \bibinfo {author} {\bibfnamefont {P.}~\bibnamefont {Ninin}}, \bibinfo {author} {\bibfnamefont {R.}~\bibnamefont {Nunes}}, \bibinfo {author} {\bibfnamefont {G.}~\bibnamefont
  {Olivet~Dessent}}, \bibinfo {author} {\bibfnamefont {D.}~\bibnamefont {Raffourt}}, \bibinfo {author} {\bibfnamefont {O.}~\bibnamefont {Rios}}, \bibinfo {author} {\bibfnamefont {E.}~\bibnamefont {Sanchez-Corral~Mena}},\ and\ \bibinfo {author} {\bibfnamefont {A.}~\bibnamefont {Suwalska}},\ }\bibfield  {title} {\bibinfo {title} {{Consolidation of Personnel Safety Systems at CERN}},\ }in\ \href {https://doi.org/10.18429/JACoW-IPAC2025-THPS022} {\emph {\bibinfo {booktitle} {Proceedings of the 16th International Particle Accelerator Conference (IPAC'25)}}}\ (\bibinfo  {publisher} {JACoW Publishing},\ \bibinfo {address} {Taipei, Taiwan},\ \bibinfo {year} {2025})\ pp.\ \bibinfo {pages} {3000--3002},\ \bibinfo {note} {paper THPS022}\BibitemShut {NoStop}%
\bibitem [{\citenamefont {Apollinari}\ \emph {et~al.}(2017)\citenamefont {Apollinari}, \citenamefont {B{\'e}jar~Alonso}, \citenamefont {Br{\"u}ning}, \citenamefont {Fessia}, \citenamefont {Lamont}, \citenamefont {Rossi},\ and\ \citenamefont {Tavian}}]{Apollinari:2017HLLHC}%
  \BibitemOpen
  \bibinfo {editor} {\bibfnamefont {G.}~\bibnamefont {Apollinari}}, \bibinfo {editor} {\bibfnamefont {I.}~\bibnamefont {B{\'e}jar~Alonso}}, \bibinfo {editor} {\bibfnamefont {O.}~\bibnamefont {Br{\"u}ning}}, \bibinfo {editor} {\bibfnamefont {P.}~\bibnamefont {Fessia}}, \bibinfo {editor} {\bibfnamefont {M.}~\bibnamefont {Lamont}}, \bibinfo {editor} {\bibfnamefont {L.}~\bibnamefont {Rossi}},\ and\ \bibinfo {editor} {\bibfnamefont {L.}~\bibnamefont {Tavian}},\ eds.,\ \href {https://doi.org/10.23731/CYRM-2017-004} {\emph {\bibinfo {title} {{High-Luminosity Large Hadron Collider (HL-LHC): Technical Design Report V. 0.1}}}},\ \bibinfo {series} {CERN Yellow Reports: Monographs}, Vol.\ \bibinfo {volume} {4/2017}\ (\bibinfo  {publisher} {CERN},\ \bibinfo {address} {Geneva},\ \bibinfo {year} {2017})\BibitemShut {NoStop}%
\bibitem [{\citenamefont {Adorisio}\ \emph {et~al.}(2015)\citenamefont {Adorisio}, \citenamefont {B{\'e}jar~Alonso}, \citenamefont {Gascon}, \citenamefont {Otto},\ and\ \citenamefont {Roesler}}]{Adorisio:2015HLLHCSafety}%
  \BibitemOpen
  \bibfield  {author} {\bibinfo {author} {\bibfnamefont {C.}~\bibnamefont {Adorisio}}, \bibinfo {author} {\bibfnamefont {I.}~\bibnamefont {B{\'e}jar~Alonso}}, \bibinfo {author} {\bibfnamefont {J.~C.}\ \bibnamefont {Gascon}}, \bibinfo {author} {\bibfnamefont {T.}~\bibnamefont {Otto}},\ and\ \bibinfo {author} {\bibfnamefont {S.}~\bibnamefont {Roesler}},\ }\bibfield  {title} {\bibinfo {title} {{Safety}},\ }in\ \href {https://doi.org/10.5170/CERN-2015-005.249} {\emph {\bibinfo {booktitle} {{High-Luminosity Large Hadron Collider (HL-LHC): Preliminary Design Report}}}},\ \bibinfo {series and number} {\bibinfo {series} {CERN Yellow Reports}\ No.\ \bibinfo {number} {CERN-2015-005}},\ \bibinfo {editor} {edited by\ \bibinfo {editor} {\bibfnamefont {G.}~\bibnamefont {Apollinari}}, \bibinfo {editor} {\bibfnamefont {I.}~\bibnamefont {B{\'e}jar~Alonso}}, \bibinfo {editor} {\bibfnamefont {O.}~\bibnamefont {Br{\"u}ning}}, \bibinfo {editor} {\bibfnamefont {M.}~\bibnamefont {Lamont}},\ and\ \bibinfo {editor} {\bibfnamefont
  {L.}~\bibnamefont {Rossi}}}\ (\bibinfo  {publisher} {CERN},\ \bibinfo {address} {Geneva},\ \bibinfo {year} {2015})\ pp.\ \bibinfo {pages} {249--256},\ \Eprint {https://arxiv.org/abs/1705.09512} {arXiv:1705.09512 [physics.acc-ph]} \BibitemShut {NoStop}%
\bibitem [{\citenamefont {Benedikt}\ \emph {et~al.}(2025)\citenamefont {Benedikt}, \citenamefont {Zimmermann}, \citenamefont {Auchmann}, \citenamefont {Bartmann}, \citenamefont {Burnet}, \citenamefont {Carli}, \citenamefont {Chanc{\'e}}, \citenamefont {Craievich}, \citenamefont {Giovannozzi}, \citenamefont {Grojean}, \citenamefont {Gutleber}, \citenamefont {Hanke}, \citenamefont {Henriques}, \citenamefont {Janot}, \citenamefont {Louren{\c c}o}, \citenamefont {Mangano}, \citenamefont {Otto}, \citenamefont {Poole}, \citenamefont {Rajagopalan}, \citenamefont {Raubenheimer}, \citenamefont {Todesco} \emph {et~al.}}]{Benedikt:2025FCCFeasibility}%
  \BibitemOpen
  \bibfield  {author} {\bibinfo {author} {\bibfnamefont {M.}~\bibnamefont {Benedikt}}, \bibinfo {author} {\bibfnamefont {F.}~\bibnamefont {Zimmermann}}, \bibinfo {author} {\bibfnamefont {B.}~\bibnamefont {Auchmann}}, \bibinfo {author} {\bibfnamefont {W.}~\bibnamefont {Bartmann}}, \bibinfo {author} {\bibfnamefont {J.~P.}\ \bibnamefont {Burnet}}, \bibinfo {author} {\bibfnamefont {C.}~\bibnamefont {Carli}}, \bibinfo {author} {\bibfnamefont {A.}~\bibnamefont {Chanc{\'e}}}, \bibinfo {author} {\bibfnamefont {P.}~\bibnamefont {Craievich}}, \bibinfo {author} {\bibfnamefont {M.}~\bibnamefont {Giovannozzi}}, \bibinfo {author} {\bibfnamefont {C.}~\bibnamefont {Grojean}}, \bibinfo {author} {\bibfnamefont {J.}~\bibnamefont {Gutleber}}, \bibinfo {author} {\bibfnamefont {K.}~\bibnamefont {Hanke}}, \bibinfo {author} {\bibfnamefont {A.}~\bibnamefont {Henriques}}, \bibinfo {author} {\bibfnamefont {P.}~\bibnamefont {Janot}}, \bibinfo {author} {\bibfnamefont {C.}~\bibnamefont {Louren{\c c}o}}, \bibinfo {author} {\bibfnamefont
  {M.}~\bibnamefont {Mangano}}, \bibinfo {author} {\bibfnamefont {T.}~\bibnamefont {Otto}}, \bibinfo {author} {\bibfnamefont {J.}~\bibnamefont {Poole}}, \bibinfo {author} {\bibfnamefont {S.}~\bibnamefont {Rajagopalan}}, \bibinfo {author} {\bibfnamefont {T.}~\bibnamefont {Raubenheimer}}, \bibinfo {author} {\bibfnamefont {E.}~\bibnamefont {Todesco}}, \emph {et~al.},\ }\bibfield  {title} {\bibinfo {title} {{Future Circular Collider Feasibility Study Report: Volume 2 Accelerators, Technical Infrastructure and Safety}},\ }\href {https://doi.org/10.1140/epjs/s11734-025-01967-4} {\bibfield  {journal} {\bibinfo  {journal} {Eur. Phys. J. Spec. Top.}\ }\textbf {\bibinfo {volume} {234}},\ \bibinfo {pages} {5713} (\bibinfo {year} {2025})},\ \bibinfo {note} {cERN-FCC-ACC-2025-0004; CERN DOI 10.17181/CERN.EBAY.7W4X},\ \Eprint {https://arxiv.org/abs/2505.00274} {arXiv:2505.00274 [physics.acc-ph]} \BibitemShut {NoStop}%
\bibitem [{\citenamefont {{International Muon Collider Collaboration}}(2024)}]{Accettura:2024MuonCollider}%
  \BibitemOpen
  \bibfield  {author} {\bibinfo {author} {\bibnamefont {{International Muon Collider Collaboration}}},\ }\href {https://doi.org/10.23731/CYRM-2024-002} {\emph {\bibinfo {title} {{Interim Report for the International Muon Collider Collaboration}}}},\ \bibinfo {series} {CERN Yellow Reports: Monographs}, Vol.\ \bibinfo {volume} {2/2024}\ (\bibinfo  {publisher} {CERN},\ \bibinfo {address} {Geneva},\ \bibinfo {year} {2024})\ \Eprint {https://arxiv.org/abs/2407.12450} {arXiv:2407.12450 [physics.acc-ph]} \BibitemShut {NoStop}%
\bibitem [{\citenamefont {Wollmann}\ \emph {et~al.}(2020)\citenamefont {Wollmann}, \citenamefont {Denz}, \citenamefont {Lindstr{\"o}m}, \citenamefont {Ravaioli}, \citenamefont {Rodr{\'i}guez~Mateos}, \citenamefont {Siemko}, \citenamefont {Uythoven}, \citenamefont {Verweij}, \citenamefont {Will},\ and\ \citenamefont {Zerlauth}}]{Wollmann:2020HLLHCMPS}%
  \BibitemOpen
  \bibfield  {author} {\bibinfo {author} {\bibfnamefont {D.}~\bibnamefont {Wollmann}}, \bibinfo {author} {\bibfnamefont {R.}~\bibnamefont {Denz}}, \bibinfo {author} {\bibfnamefont {B.}~\bibnamefont {Lindstr{\"o}m}}, \bibinfo {author} {\bibfnamefont {E.}~\bibnamefont {Ravaioli}}, \bibinfo {author} {\bibfnamefont {F.}~\bibnamefont {Rodr{\'i}guez~Mateos}}, \bibinfo {author} {\bibfnamefont {A.}~\bibnamefont {Siemko}}, \bibinfo {author} {\bibfnamefont {J.}~\bibnamefont {Uythoven}}, \bibinfo {author} {\bibfnamefont {A.}~\bibnamefont {Verweij}}, \bibinfo {author} {\bibfnamefont {A.}~\bibnamefont {Will}},\ and\ \bibinfo {author} {\bibfnamefont {M.}~\bibnamefont {Zerlauth}},\ }\bibfield  {title} {\bibinfo {title} {{Machine Protection}},\ }in\ \href {https://doi.org/10.23731/CYRM-2020-0010.153} {\emph {\bibinfo {booktitle} {{High-Luminosity Large Hadron Collider (HL-LHC): Technical Design Report}}}},\ \bibinfo {series} {CERN Yellow Reports: Monographs}, Vol.~\bibinfo {volume} {10}\ (\bibinfo  {publisher} {CERN},\
  \bibinfo {address} {Geneva},\ \bibinfo {year} {2020})\ pp.\ \bibinfo {pages} {153--168}\BibitemShut {NoStop}%
\bibitem [{\citenamefont {Redaelli}\ \emph {et~al.}(2020)\citenamefont {Redaelli}, \citenamefont {Bruce}, \citenamefont {Lechner},\ and\ \citenamefont {Mereghetti}}]{Redaelli:2020HLLHCCollimation}%
  \BibitemOpen
  \bibfield  {author} {\bibinfo {author} {\bibfnamefont {S.}~\bibnamefont {Redaelli}}, \bibinfo {author} {\bibfnamefont {R.}~\bibnamefont {Bruce}}, \bibinfo {author} {\bibfnamefont {A.}~\bibnamefont {Lechner}},\ and\ \bibinfo {author} {\bibfnamefont {A.}~\bibnamefont {Mereghetti}},\ }\bibfield  {title} {\bibinfo {title} {{Collimation System}},\ }in\ \href {https://doi.org/10.23731/CYRM-2020-0010.87} {\emph {\bibinfo {booktitle} {{High-Luminosity Large Hadron Collider (HL-LHC): Technical Design Report}}}},\ \bibinfo {series} {CERN Yellow Reports: Monographs}, Vol.~\bibinfo {volume} {10}\ (\bibinfo  {publisher} {CERN},\ \bibinfo {address} {Geneva},\ \bibinfo {year} {2020})\ pp.\ \bibinfo {pages} {87--114}\BibitemShut {NoStop}%
\bibitem [{\citenamefont {Tygier}\ \emph {et~al.}(2019)\citenamefont {Tygier}, \citenamefont {Appleby}, \citenamefont {Bruce}, \citenamefont {Mirarchi}, \citenamefont {Redaelli},\ and\ \citenamefont {Valloni}}]{Tygier:2019Collimation}%
  \BibitemOpen
  \bibfield  {author} {\bibinfo {author} {\bibfnamefont {S.}~\bibnamefont {Tygier}}, \bibinfo {author} {\bibfnamefont {R.~B.}\ \bibnamefont {Appleby}}, \bibinfo {author} {\bibfnamefont {R.}~\bibnamefont {Bruce}}, \bibinfo {author} {\bibfnamefont {D.}~\bibnamefont {Mirarchi}}, \bibinfo {author} {\bibfnamefont {S.}~\bibnamefont {Redaelli}},\ and\ \bibinfo {author} {\bibfnamefont {A.}~\bibnamefont {Valloni}},\ }\bibfield  {title} {\bibinfo {title} {{Performance of the Large Hadron Collider Cleaning System During the Squeeze: Simulations and Measurements}},\ }\href {https://doi.org/10.1103/PhysRevAccelBeams.22.023001} {\bibfield  {journal} {\bibinfo  {journal} {Phys. Rev. Accel. Beams}\ }\textbf {\bibinfo {volume} {22}},\ \bibinfo {pages} {023001} (\bibinfo {year} {2019})}\BibitemShut {NoStop}%
\bibitem [{\citenamefont {Apsimon}\ \emph {et~al.}(2019)\citenamefont {Apsimon}, \citenamefont {Burt}, \citenamefont {Dexter}, \citenamefont {Shipman}, \citenamefont {Castilla}, \citenamefont {Macpherson}, \citenamefont {Ness~Sjobak}, \citenamefont {Santamaria~Garcia}, \citenamefont {Stapley}, \citenamefont {Alekou},\ and\ \citenamefont {Appleby}}]{Apsimon:2019CrabQuench}%
  \BibitemOpen
  \bibfield  {author} {\bibinfo {author} {\bibfnamefont {R.}~\bibnamefont {Apsimon}}, \bibinfo {author} {\bibfnamefont {G.}~\bibnamefont {Burt}}, \bibinfo {author} {\bibfnamefont {A.}~\bibnamefont {Dexter}}, \bibinfo {author} {\bibfnamefont {N.}~\bibnamefont {Shipman}}, \bibinfo {author} {\bibfnamefont {A.}~\bibnamefont {Castilla}}, \bibinfo {author} {\bibfnamefont {A.}~\bibnamefont {Macpherson}}, \bibinfo {author} {\bibfnamefont {K.}~\bibnamefont {Ness~Sjobak}}, \bibinfo {author} {\bibfnamefont {A.}~\bibnamefont {Santamaria~Garcia}}, \bibinfo {author} {\bibfnamefont {N.}~\bibnamefont {Stapley}}, \bibinfo {author} {\bibfnamefont {A.}~\bibnamefont {Alekou}},\ and\ \bibinfo {author} {\bibfnamefont {R.~B.}\ \bibnamefont {Appleby}},\ }\bibfield  {title} {\bibinfo {title} {{Prediction of Beam Losses During Crab Cavity Quenches at the High Luminosity LHC}},\ }\href {https://doi.org/10.1103/PhysRevAccelBeams.22.061001} {\bibfield  {journal} {\bibinfo  {journal} {Phys. Rev. Accel. Beams}\ }\textbf {\bibinfo {volume}
  {22}},\ \bibinfo {pages} {061001} (\bibinfo {year} {2019})}\BibitemShut {NoStop}%
\bibitem [{\citenamefont {Claudet}\ \emph {et~al.}(2020)\citenamefont {Claudet}, \citenamefont {Ferlin}, \citenamefont {Monneret}, \citenamefont {Perin}, \citenamefont {Pirotte}, \citenamefont {Sisti},\ and\ \citenamefont {Van~Weelderen}}]{Claudet:2020HLLHCCryogenics}%
  \BibitemOpen
  \bibfield  {author} {\bibinfo {author} {\bibfnamefont {S.}~\bibnamefont {Claudet}}, \bibinfo {author} {\bibfnamefont {G.}~\bibnamefont {Ferlin}}, \bibinfo {author} {\bibfnamefont {E.}~\bibnamefont {Monneret}}, \bibinfo {author} {\bibfnamefont {A.}~\bibnamefont {Perin}}, \bibinfo {author} {\bibfnamefont {O.}~\bibnamefont {Pirotte}}, \bibinfo {author} {\bibfnamefont {M.}~\bibnamefont {Sisti}},\ and\ \bibinfo {author} {\bibfnamefont {R.}~\bibnamefont {Van~Weelderen}},\ }\bibfield  {title} {\bibinfo {title} {{Cryogenics for the HL-LHC}},\ }in\ \href {https://doi.org/10.23731/CYRM-2020-0010.189} {\emph {\bibinfo {booktitle} {{High-Luminosity Large Hadron Collider (HL-LHC): Technical Design Report}}}},\ \bibinfo {series} {CERN Yellow Reports: Monographs}, Vol.~\bibinfo {volume} {10}\ (\bibinfo  {publisher} {CERN},\ \bibinfo {address} {Geneva},\ \bibinfo {year} {2020})\ pp.\ \bibinfo {pages} {189--198}\BibitemShut {NoStop}%
\bibitem [{\citenamefont {Buttar}(2025)}]{Buttar:2025ATLASITk}%
  \BibitemOpen
  \bibfield  {author} {\bibinfo {author} {\bibfnamefont {C.}~\bibnamefont {Buttar}},\ }\bibfield  {title} {\bibinfo {title} {{ATLAS ITk Pixel Detector Overview}},\ }\href {https://doi.org/10.1016/j.nima.2024.169978} {\bibfield  {journal} {\bibinfo  {journal} {Nucl. Instrum. Meth. A}\ }\textbf {\bibinfo {volume} {1070}},\ \bibinfo {pages} {169978} (\bibinfo {year} {2025})}\BibitemShut {NoStop}%
\bibitem [{\citenamefont {Kuo}(2025)}]{Kuo:2025CMSInnerTracker}%
  \BibitemOpen
  \bibfield  {author} {\bibinfo {author} {\bibfnamefont {C.-C.}\ \bibnamefont {Kuo}},\ }\bibfield  {title} {\bibinfo {title} {{CMS Inner Tracker Upgrade for the HL-LHC: Design, Development, and Production Status}},\ }\href {https://doi.org/10.1016/j.nima.2025.170696} {\bibfield  {journal} {\bibinfo  {journal} {Nucl. Instrum. Meth. A}\ }\textbf {\bibinfo {volume} {1080}},\ \bibinfo {pages} {170696} (\bibinfo {year} {2025})}\BibitemShut {NoStop}%
\bibitem [{\citenamefont {{ATLAS Collaboration}}(2015)}]{ATLAS:2015PhaseIIScoping}%
  \BibitemOpen
  \bibfield  {author} {\bibinfo {author} {\bibnamefont {{ATLAS Collaboration}}},\ }\href {https://doi.org/10.17181/CERN.7CRX.AJHP} {\emph {\bibinfo {title} {{ATLAS Phase-II Upgrade Scoping Document}}}},\ \bibinfo {type} {Tech. Rep.}\ (\bibinfo  {institution} {CERN},\ \bibinfo {year} {2015})\BibitemShut {NoStop}%
\bibitem [{\citenamefont {Lorenzon}\ \emph {et~al.}(2024)\citenamefont {Lorenzon}, \citenamefont {Bozzato}, \citenamefont {Froeschl},\ and\ \citenamefont {Kouskoura}}]{Lorenzon:2024CMSActivation}%
  \BibitemOpen
  \bibfield  {author} {\bibinfo {author} {\bibfnamefont {T.}~\bibnamefont {Lorenzon}}, \bibinfo {author} {\bibfnamefont {D.}~\bibnamefont {Bozzato}}, \bibinfo {author} {\bibfnamefont {R.}~\bibnamefont {Froeschl}},\ and\ \bibinfo {author} {\bibfnamefont {V.}~\bibnamefont {Kouskoura}},\ }\bibfield  {title} {\bibinfo {title} {{Radiation Protection Aspects of Long Shutdown 3 at the CERN LHC Experiments: A Methodology for Residual Activation Assessments Applied to the CMS Case Study}},\ }in\ \href {https://www.oecd-nea.org/upload/docs/application/pdf/2024-05/satif15-session-5.pdf} {\emph {\bibinfo {booktitle} {{SATIF-15: 15th Workshop on Shielding Aspects of Accelerators, Targets and Irradiation Facilities}}}}\ (\bibinfo  {publisher} {OECD Nuclear Energy Agency},\ \bibinfo {year} {2024})\BibitemShut {NoStop}%
\bibitem [{\citenamefont {Bozzato}\ \emph {et~al.}(2023)\citenamefont {Bozzato}, \citenamefont {Froeschl},\ and\ \citenamefont {Kouskoura}}]{Bozzato:2023LHCExperiments}%
  \BibitemOpen
  \bibfield  {author} {\bibinfo {author} {\bibfnamefont {D.}~\bibnamefont {Bozzato}}, \bibinfo {author} {\bibfnamefont {R.}~\bibnamefont {Froeschl}},\ and\ \bibinfo {author} {\bibfnamefont {V.}~\bibnamefont {Kouskoura}},\ }\bibfield  {title} {\bibinfo {title} {Operational radiation protection challenges for the lhc experiments},\ }\href {https://doi.org/10.1093/rpd/ncad126} {\bibfield  {journal} {\bibinfo  {journal} {Radiation Protection Dosimetry}\ }\textbf {\bibinfo {volume} {199}},\ \bibinfo {pages} {785} (\bibinfo {year} {2023})}\BibitemShut {NoStop}%
\bibitem [{\citenamefont {Brugger}\ and\ \citenamefont {Roesler}(2005)}]{BruggerRoesler:2005CollimatorDose}%
  \BibitemOpen
  \bibfield  {author} {\bibinfo {author} {\bibfnamefont {M.}~\bibnamefont {Brugger}}\ and\ \bibinfo {author} {\bibfnamefont {S.}~\bibnamefont {Roesler}},\ }\bibfield  {title} {\bibinfo {title} {Remanent dose rates around the collimators of the lhc beam cleaning insertions},\ }\href {https://doi.org/10.1093/rpd/nci053} {\bibfield  {journal} {\bibinfo  {journal} {Radiation Protection Dosimetry}\ }\textbf {\bibinfo {volume} {115}},\ \bibinfo {pages} {470} (\bibinfo {year} {2005})}\BibitemShut {NoStop}%
\bibitem [{\citenamefont {Kershaw}\ \emph {et~al.}(2013)\citenamefont {Kershaw}, \citenamefont {Feral}, \citenamefont {Grenard}, \citenamefont {Feniet}, \citenamefont {De~Man}, \citenamefont {Hazelaar-Bal}, \citenamefont {Bertone},\ and\ \citenamefont {Ingo}}]{Kershaw:2013RemoteHandling}%
  \BibitemOpen
  \bibfield  {author} {\bibinfo {author} {\bibfnamefont {K.}~\bibnamefont {Kershaw}}, \bibinfo {author} {\bibfnamefont {B.}~\bibnamefont {Feral}}, \bibinfo {author} {\bibfnamefont {J.-L.}\ \bibnamefont {Grenard}}, \bibinfo {author} {\bibfnamefont {T.}~\bibnamefont {Feniet}}, \bibinfo {author} {\bibfnamefont {S.}~\bibnamefont {De~Man}}, \bibinfo {author} {\bibfnamefont {C.}~\bibnamefont {Hazelaar-Bal}}, \bibinfo {author} {\bibfnamefont {C.}~\bibnamefont {Bertone}},\ and\ \bibinfo {author} {\bibfnamefont {R.}~\bibnamefont {Ingo}},\ }\bibfield  {title} {\bibinfo {title} {{Remote Inspection, Measurement and Handling for Maintenance and Operation at CERN}},\ }\href {https://doi.org/10.5772/56849} {\bibfield  {journal} {\bibinfo  {journal} {International Journal of Advanced Robotic Systems}\ }\textbf {\bibinfo {volume} {10}},\ \bibinfo {pages} {382} (\bibinfo {year} {2013})}\BibitemShut {NoStop}%
\bibitem [{\citenamefont {Infantino}\ \emph {et~al.}(2023)\citenamefont {Infantino}, \citenamefont {Bernardes}, \citenamefont {Calviani}, \citenamefont {Dumont}, \citenamefont {El-Idrissi}, \citenamefont {Mouret}, \citenamefont {Tromel},\ and\ \citenamefont {Vincke}}]{Infantino:2023BeamDumpRP}%
  \BibitemOpen
  \bibfield  {author} {\bibinfo {author} {\bibfnamefont {A.}~\bibnamefont {Infantino}}, \bibinfo {author} {\bibfnamefont {A.-P.}\ \bibnamefont {Bernardes}}, \bibinfo {author} {\bibfnamefont {M.}~\bibnamefont {Calviani}}, \bibinfo {author} {\bibfnamefont {G.}~\bibnamefont {Dumont}}, \bibinfo {author} {\bibfnamefont {S.}~\bibnamefont {El-Idrissi}}, \bibinfo {author} {\bibfnamefont {R.}~\bibnamefont {Mouret}}, \bibinfo {author} {\bibfnamefont {C.}~\bibnamefont {Tromel}},\ and\ \bibinfo {author} {\bibfnamefont {H.}~\bibnamefont {Vincke}},\ }\bibfield  {title} {\bibinfo {title} {Radiation protection challenges in the upgrade, autopsy and disposal of the lhc beam dump},\ }\href {https://doi.org/10.1093/rpd/ncad118} {\bibfield  {journal} {\bibinfo  {journal} {Radiation Protection Dosimetry}\ }\textbf {\bibinfo {volume} {199}},\ \bibinfo {pages} {891} (\bibinfo {year} {2023})}\BibitemShut {NoStop}%
\bibitem [{\citenamefont {Otto}\ \emph {et~al.}(2020)\citenamefont {Otto}, \citenamefont {Adorisio}, \citenamefont {Gaignant}, \citenamefont {Infantino},\ and\ \citenamefont {Maietta}}]{Otto:2020HLLHCSafety}%
  \BibitemOpen
  \bibfield  {author} {\bibinfo {author} {\bibfnamefont {T.}~\bibnamefont {Otto}}, \bibinfo {author} {\bibfnamefont {C.}~\bibnamefont {Adorisio}}, \bibinfo {author} {\bibfnamefont {C.}~\bibnamefont {Gaignant}}, \bibinfo {author} {\bibfnamefont {A.}~\bibnamefont {Infantino}},\ and\ \bibinfo {author} {\bibfnamefont {M.}~\bibnamefont {Maietta}},\ }\bibfield  {title} {\bibinfo {title} {{Safety}},\ }in\ \href {https://doi.org/10.23731/CYRM-2020-0010.329} {\emph {\bibinfo {booktitle} {{High-Luminosity Large Hadron Collider (HL-LHC): Technical Design Report}}}},\ \bibinfo {series} {CERN Yellow Reports: Monographs}, Vol.~\bibinfo {volume} {10}\ (\bibinfo  {publisher} {CERN},\ \bibinfo {address} {Geneva},\ \bibinfo {year} {2020})\ pp.\ \bibinfo {pages} {329--342}\BibitemShut {NoStop}%
\bibitem [{\citenamefont {Elie}\ \emph {et~al.}(2023)\citenamefont {Elie}, \citenamefont {Dyrcz}, \citenamefont {Infantino}, \citenamefont {Tromel},\ and\ \citenamefont {Vincke}}]{Elie:2023HLRadiationProtection}%
  \BibitemOpen
  \bibfield  {author} {\bibinfo {author} {\bibfnamefont {L.}~\bibnamefont {Elie}}, \bibinfo {author} {\bibfnamefont {P.}~\bibnamefont {Dyrcz}}, \bibinfo {author} {\bibfnamefont {A.}~\bibnamefont {Infantino}}, \bibinfo {author} {\bibfnamefont {C.}~\bibnamefont {Tromel}},\ and\ \bibinfo {author} {\bibfnamefont {H.}~\bibnamefont {Vincke}},\ }\bibfield  {title} {\bibinfo {title} {{Radiation Protection Challenges for the Large Hadron Collider Upgrade}},\ }\href {https://doi.org/10.1093/rpd/ncad119} {\bibfield  {journal} {\bibinfo  {journal} {Radiation Protection Dosimetry}\ }\textbf {\bibinfo {volume} {199}},\ \bibinfo {pages} {900} (\bibinfo {year} {2023})}\BibitemShut {NoStop}%
\bibitem [{\citenamefont {{FCC Collaboration}}(2025)}]{Benedikt:2025FCCCivilEngineering}%
  \BibitemOpen
  \bibfield  {author} {\bibinfo {author} {\bibnamefont {{FCC Collaboration}}},\ }\bibfield  {title} {\bibinfo {title} {{Future Circular Collider Feasibility Study Report: Volume 3, Civil Engineering, Implementation and Sustainability}},\ }\href {https://doi.org/10.1140/epjs/s11734-025-01958-5} {\bibfield  {journal} {\bibinfo  {journal} {Eur. Phys. J. Spec. Top.}\ }\textbf {\bibinfo {volume} {234}},\ \bibinfo {pages} {5113} (\bibinfo {year} {2025})},\ \Eprint {https://arxiv.org/abs/2505.00273} {arXiv:2505.00273 [physics.acc-ph]} \BibitemShut {NoStop}%
\bibitem [{\citenamefont {Delprat}\ \emph {et~al.}(2025)\citenamefont {Delprat}, \citenamefont {Naydenov}, \citenamefont {Bradu},\ and\ \citenamefont {Brodzinski}}]{Delprat:2025FCCCryogenics}%
  \BibitemOpen
  \bibfield  {author} {\bibinfo {author} {\bibfnamefont {L.}~\bibnamefont {Delprat}}, \bibinfo {author} {\bibfnamefont {B.}~\bibnamefont {Naydenov}}, \bibinfo {author} {\bibfnamefont {B.}~\bibnamefont {Bradu}},\ and\ \bibinfo {author} {\bibfnamefont {K.}~\bibnamefont {Brodzinski}},\ }\bibfield  {title} {\bibinfo {title} {{Status of the FCC Cryogenics Feasibility Study}},\ }\href {https://doi.org/10.1088/1757-899X/1327/1/012001} {\bibfield  {journal} {\bibinfo  {journal} {IOP Conf. Ser. Mater. Sci. Eng.}\ }\textbf {\bibinfo {volume} {1327}},\ \bibinfo {pages} {012001} (\bibinfo {year} {2025})},\ \bibinfo {note} {proceedings of the 29th International Cryogenic Engineering Conference and International Cryogenic Materials Conference (ICEC29--ICMC2024), Geneva, Switzerland, 22--26 July 2024}\BibitemShut {NoStop}%
\bibitem [{\citenamefont {Manczak}\ \emph {et~al.}(2024)\citenamefont {Manczak}, \citenamefont {Frasca}, \citenamefont {Lechner}, \citenamefont {Ahdida}, \citenamefont {Schulte}, \citenamefont {Calzolari}, \citenamefont {Bottura}, \citenamefont {Calviani}, \citenamefont {Franqueira~Ximenes}, \citenamefont {Portone}, \citenamefont {Rogers},\ and\ \citenamefont {Saura~Esteban}}]{Manczak:2024MuonTargetRadiation}%
  \BibitemOpen
  \bibfield  {author} {\bibinfo {author} {\bibfnamefont {J.}~\bibnamefont {Manczak}}, \bibinfo {author} {\bibfnamefont {A.}~\bibnamefont {Frasca}}, \bibinfo {author} {\bibfnamefont {A.}~\bibnamefont {Lechner}}, \bibinfo {author} {\bibfnamefont {C.}~\bibnamefont {Ahdida}}, \bibinfo {author} {\bibfnamefont {D.}~\bibnamefont {Schulte}}, \bibinfo {author} {\bibfnamefont {D.}~\bibnamefont {Calzolari}}, \bibinfo {author} {\bibfnamefont {L.}~\bibnamefont {Bottura}}, \bibinfo {author} {\bibfnamefont {M.}~\bibnamefont {Calviani}}, \bibinfo {author} {\bibfnamefont {R.}~\bibnamefont {Franqueira~Ximenes}}, \bibinfo {author} {\bibfnamefont {A.}~\bibnamefont {Portone}}, \bibinfo {author} {\bibfnamefont {C.}~\bibnamefont {Rogers}},\ and\ \bibinfo {author} {\bibfnamefont {F.}~\bibnamefont {Saura~Esteban}},\ }\bibfield  {title} {\bibinfo {title} {{Radiation Load Studies for the Proton Target Area of a Multi-TeV Muon Collider}},\ }in\ \href {https://doi.org/10.18429/JACoW-IPAC2024-WEPR28} {\emph {\bibinfo {booktitle}
  {Proceedings of the 15th International Particle Accelerator Conference (IPAC2024)}}}\ (\bibinfo  {publisher} {JACoW Publishing},\ \bibinfo {address} {Geneva},\ \bibinfo {year} {2024})\ pp.\ \bibinfo {pages} {2544--2547}\BibitemShut {NoStop}%
\bibitem [{\citenamefont {{MICE Collaboration}}(2024)}]{MICE:2024IonizationCooling}%
  \BibitemOpen
  \bibfield  {author} {\bibinfo {author} {\bibnamefont {{MICE Collaboration}}},\ }\bibfield  {title} {\bibinfo {title} {{Transverse Emittance Reduction in Muon Beams by Ionization Cooling}},\ }\href {https://doi.org/10.1038/s41567-024-02547-4} {\bibfield  {journal} {\bibinfo  {journal} {Nature Physics}\ }\textbf {\bibinfo {volume} {20}},\ \bibinfo {pages} {1558} (\bibinfo {year} {2024})}\BibitemShut {NoStop}%
\bibitem [{\citenamefont {Lechner}\ \emph {et~al.}(2024)\citenamefont {Lechner}, \citenamefont {Frasca}, \citenamefont {Carli}, \citenamefont {Schulte}, \citenamefont {Calzolari}, \citenamefont {Amorim}, \citenamefont {Salvat~Pujol}, \citenamefont {Lerner}, \citenamefont {Manczak}, \citenamefont {Ferreira~Somoza}, \citenamefont {Skoufaris}, \citenamefont {Bottura}, \citenamefont {Borges~de Sousa}, \citenamefont {van Weelderen}, \citenamefont {Caiffi},\ and\ \citenamefont {Mariotto}}]{Lechner:2024MuonMagnetShielding}%
  \BibitemOpen
  \bibfield  {author} {\bibinfo {author} {\bibfnamefont {A.}~\bibnamefont {Lechner}}, \bibinfo {author} {\bibfnamefont {A.}~\bibnamefont {Frasca}}, \bibinfo {author} {\bibfnamefont {C.}~\bibnamefont {Carli}}, \bibinfo {author} {\bibfnamefont {D.}~\bibnamefont {Schulte}}, \bibinfo {author} {\bibfnamefont {D.}~\bibnamefont {Calzolari}}, \bibinfo {author} {\bibfnamefont {D.}~\bibnamefont {Amorim}}, \bibinfo {author} {\bibfnamefont {F.}~\bibnamefont {Salvat~Pujol}}, \bibinfo {author} {\bibfnamefont {G.}~\bibnamefont {Lerner}}, \bibinfo {author} {\bibfnamefont {J.}~\bibnamefont {Manczak}}, \bibinfo {author} {\bibfnamefont {J.}~\bibnamefont {Ferreira~Somoza}}, \bibinfo {author} {\bibfnamefont {K.}~\bibnamefont {Skoufaris}}, \bibinfo {author} {\bibfnamefont {L.}~\bibnamefont {Bottura}}, \bibinfo {author} {\bibfnamefont {P.}~\bibnamefont {Borges~de Sousa}}, \bibinfo {author} {\bibfnamefont {R.}~\bibnamefont {van Weelderen}}, \bibinfo {author} {\bibfnamefont {B.}~\bibnamefont {Caiffi}},\ and\ \bibinfo {author}
  {\bibfnamefont {S.}~\bibnamefont {Mariotto}},\ }\bibfield  {title} {\bibinfo {title} {{Radiation Shielding Studies for Superconducting Magnets in Multi-TeV Muon Colliders}},\ }in\ \href {https://doi.org/10.18429/JACoW-IPAC2024-WEPR26} {\emph {\bibinfo {booktitle} {Proceedings of the 15th International Particle Accelerator Conference (IPAC2024)}}}\ (\bibinfo  {publisher} {JACoW Publishing},\ \bibinfo {address} {Geneva},\ \bibinfo {year} {2024})\ pp.\ \bibinfo {pages} {2536--2539}\BibitemShut {NoStop}%
\bibitem [{\citenamefont {Lucchesi}\ \emph {et~al.}(2024)\citenamefont {Lucchesi}, \citenamefont {Bartosik}, \citenamefont {Calzolari}, \citenamefont {Castelli},\ and\ \citenamefont {Lechner}}]{Lucchesi:2024MuonMDI}%
  \BibitemOpen
  \bibfield  {author} {\bibinfo {author} {\bibfnamefont {D.}~\bibnamefont {Lucchesi}}, \bibinfo {author} {\bibfnamefont {N.}~\bibnamefont {Bartosik}}, \bibinfo {author} {\bibfnamefont {D.}~\bibnamefont {Calzolari}}, \bibinfo {author} {\bibfnamefont {L.}~\bibnamefont {Castelli}},\ and\ \bibinfo {author} {\bibfnamefont {A.}~\bibnamefont {Lechner}},\ }\bibfield  {title} {\bibinfo {title} {{Machine-Detector Interface for Multi-TeV Muon Collider}},\ }\href {https://doi.org/10.22323/1.449.0630} {\bibfield  {journal} {\bibinfo  {journal} {PoS}\ }\textbf {\bibinfo {volume} {EPS-HEP2023}},\ \bibinfo {pages} {630} (\bibinfo {year} {2024})}\BibitemShut {NoStop}%
\bibitem [{\citenamefont {Lerner}\ \emph {et~al.}(2024)\citenamefont {Lerner}, \citenamefont {Frasca}, \citenamefont {Lechner}, \citenamefont {Carli}, \citenamefont {Ahdida},\ and\ \citenamefont {Manczak}}]{Lerner:2024MuonNeutrinoDose}%
  \BibitemOpen
  \bibfield  {author} {\bibinfo {author} {\bibfnamefont {G.}~\bibnamefont {Lerner}}, \bibinfo {author} {\bibfnamefont {A.}~\bibnamefont {Frasca}}, \bibinfo {author} {\bibfnamefont {A.}~\bibnamefont {Lechner}}, \bibinfo {author} {\bibfnamefont {C.}~\bibnamefont {Carli}}, \bibinfo {author} {\bibfnamefont {C.}~\bibnamefont {Ahdida}},\ and\ \bibinfo {author} {\bibfnamefont {J.}~\bibnamefont {Manczak}},\ }\bibfield  {title} {\bibinfo {title} {{FLUKA Simulations of Neutrino-Induced Effective Dose at a Muon Collider}},\ }in\ \href {https://doi.org/10.18429/JACoW-IPAC2024-WEPR27} {\emph {\bibinfo {booktitle} {Proceedings of the 15th International Particle Accelerator Conference (IPAC2024)}}}\ (\bibinfo  {publisher} {JACoW Publishing},\ \bibinfo {address} {Geneva},\ \bibinfo {year} {2024})\BibitemShut {NoStop}%
\bibitem [{\citenamefont {Wied}\ \emph {et~al.}(2020)\citenamefont {Wied}, \citenamefont {Oehmen},\ and\ \citenamefont {Welo}}]{Wied:2020Resilience}%
  \BibitemOpen
  \bibfield  {author} {\bibinfo {author} {\bibfnamefont {M.}~\bibnamefont {Wied}}, \bibinfo {author} {\bibfnamefont {J.}~\bibnamefont {Oehmen}},\ and\ \bibinfo {author} {\bibfnamefont {T.}~\bibnamefont {Welo}},\ }\bibfield  {title} {\bibinfo {title} {Conceptualizing resilience in engineering systems: An analysis of the literature},\ }\href {https://doi.org/10.1002/sys.21491} {\bibfield  {journal} {\bibinfo  {journal} {Systems Engineering}\ }\textbf {\bibinfo {volume} {23}},\ \bibinfo {pages} {3} (\bibinfo {year} {2020})}\BibitemShut {NoStop}%
\bibitem [{\citenamefont {Mentges}\ \emph {et~al.}(2023)\citenamefont {Mentges}, \citenamefont {Halekotte}, \citenamefont {Schneider}, \citenamefont {Demmer},\ and\ \citenamefont {Lichte}}]{Mentges:2023ResilienceGlossary}%
  \BibitemOpen
  \bibfield  {author} {\bibinfo {author} {\bibfnamefont {A.}~\bibnamefont {Mentges}}, \bibinfo {author} {\bibfnamefont {L.}~\bibnamefont {Halekotte}}, \bibinfo {author} {\bibfnamefont {M.}~\bibnamefont {Schneider}}, \bibinfo {author} {\bibfnamefont {T.}~\bibnamefont {Demmer}},\ and\ \bibinfo {author} {\bibfnamefont {D.}~\bibnamefont {Lichte}},\ }\bibfield  {title} {\bibinfo {title} {A resilience glossary shaped by context: Reviewing resilience-related terms for critical infrastructures},\ }\href {https://doi.org/10.1016/j.ijdrr.2023.103893} {\bibfield  {journal} {\bibinfo  {journal} {International Journal of Disaster Risk Reduction}\ }\textbf {\bibinfo {volume} {96}},\ \bibinfo {pages} {103893} (\bibinfo {year} {2023})}\BibitemShut {NoStop}%
\bibitem [{\citenamefont {{International Atomic Energy Agency}}(2014)}]{IAEA:2014Decommissioning}%
  \BibitemOpen
  \bibfield  {author} {\bibinfo {author} {\bibnamefont {{International Atomic Energy Agency}}},\ }\href {https://www.iaea.org/publications/10676/decommissioning-of-facilities} {\emph {\bibinfo {title} {{Decommissioning of Facilities}}}},\ \bibinfo {type} {IAEA Safety Standards Series}\ \bibinfo {number} {GSR Part 6}\ (\bibinfo  {institution} {International Atomic Energy Agency},\ \bibinfo {address} {Vienna},\ \bibinfo {year} {2014})\BibitemShut {NoStop}%
\bibitem [{\citenamefont {{International Atomic Energy Agency}}(2019)}]{IAEA:2019ResearchDecommissioning}%
  \BibitemOpen
  \bibfield  {author} {\bibinfo {author} {\bibnamefont {{International Atomic Energy Agency}}},\ }\href {https://www.iaea.org/publications/12291/decommissioning-of-medical-industrial-and-research-facilities} {\emph {\bibinfo {title} {{Decommissioning of Medical, Industrial and Research Facilities}}}},\ \bibinfo {type} {IAEA Safety Standards Series}\ \bibinfo {number} {SSG-49}\ (\bibinfo  {institution} {International Atomic Energy Agency},\ \bibinfo {address} {Vienna},\ \bibinfo {year} {2019})\BibitemShut {NoStop}%
\end{thebibliography}
